\PassOptionsToPackage{labelfont=bf,labelsep=period}{caption}
\PassOptionsToPackage{colorlinks=true,linkcolor=SaddleBrown,urlcolor=SeaGreen,citecolor=Crimson,hyperfigures,hyperfootnotes}{hyperref}
\documentclass[10pt,a4paper,english,twoside]{llncs}

\usepackage{etoolbox}
\usepackage{imakeidx}
\usepackage{enumerate}
\usepackage[vlined]{algorithm2e}
\usepackage[utf8]{inputenc}
\usepackage[T1]{fontenc}
\usepackage{multirow}
\usepackage{url}
\usepackage{amsmath,amssymb}
\usepackage{epsfig}
\usepackage[usenames,svgnames]{xcolor}
\usepackage{listings}
\usepackage{acronym}
\usepackage{wrapfig}
\usepackage{graphicx}
\usepackage{rotating}
\usepackage{colortbl}
\usepackage{pifont}
\usepackage{flafter}
\usepackage{tabularx}
\usepackage{nth}
\usepackage{fourier}
\usepackage{marvosym}
\usepackage{lettrine}
\usepackage{xkeyval}
\usepackage{makecell}
\usepackage{bookmark}
\usepackage{subfig}
\usepackage{fancyhdr}
\usepackage[strict]{changepage}
\usepackage{eso-pic}
\usepackage{lastpage}
\usepackage{trfsigns}
\usepackage{hyperref}
\usepackage[nomain,acronym,shortcuts]{glossaries} 
\usepackage[all]{hypcap}
\usepackage{cleveref}
\usepackage{caption}

\noist

\definecolor{notparticipating}{RGB}{225,225,225}

\definecolor{nofailure}{RGB}{0,0,0}
\definecolor{txtnofailure}{RGB}{255,255,255}

\definecolor{bestscore}{RGB}{90,90,90}
\definecolor{txtbestscore}{RGB}{255,255,255}

\definecolor{ednf}{RGB}{255,255,255}
\definecolor{txtednf}{RGB}{0,0,0}

\definecolor{movf}{RGB}{255,255,255}
\definecolor{txtmovf}{RGB}{0,0,0}

\definecolor{sovf}{RGB}{255,255,255}
\definecolor{txtsovf}{RGB}{100,100,100}

\input{includes/macros-al}

\date{}
\title{
  Raw Report on the Model Checking Contest
\\
  at Petri Nets 2012
}
\author{
     F. Kordon\inst{1}
\and A. Linard\inst{2}
\and
    \\
     D. Buchs\inst{2}
\and M. Colange\inst{1}
\and S. Evangelista\inst{3}
\and
    \\
     L. Fronc\inst{4}
\and L.M. Hillah\inst{1}
\and N. Lohmann\inst{5}
\and
    \\
     E. Paviot-Adet\inst{1}
\and F Pommereau\inst{4}
\and C. Rohr\inst{6}
\and \\
     Y. Thierry-Mieg\inst{1}
\and H. Wimmel\inst{5}
\and K. Wolf\inst{5}
}

\institute{
     LIP6, CNRS UMR 7606, Université P. \& M. Curie~--~Paris~6 \\
     4, place Jussieu, F-75252 Paris Cedex 05, France\\
     \email{Fabrice.Kordon@lip6.fr},
     \email{Maximilien.Colange@lip6.fr},
     \email{Lom-Messan.Hillah@lip6.fr},
     \email{Emmanuel.Paviot-Adet@lip6.fr},
     \email{Yann.Thierry-Mieg@lip6.fr} \\
\and
     Centre Universitaire d'Informatique, Université de Genève \\
     7, route de Drize, CH-1227 Carouge, Switzerland \\
     \email{Alban.Linard@unige.ch},
     \email{Didier.Buchs@unige.ch} \\
\and
     LIPN, CNRS UMR 7030, Université Paris 13 \\
     99, av. J-B Clément, 93430 Villetaneuse, France \\
     \email{sami.evangelista@lipn.univ-paris13.fr} \\
\and
     IBISC, Université d'Évry Val d'Essonne \\
     22 Boulevard de France, 91037 Évry Cedex France\\
     \email{lfronc@ibisc.univ-evry.fr}, \email{franck.pommereau@ibisc.univ-evry.fr} \\
\and
     Universität Rostock, 
     18051 Rostock, Germany \\
     \email{niels.lohmann@uni-rostock.de},
     \email{Wimmel@Informatik.Uni-Oldenburg.de}, \email{karsten.wolf@uni-rostock.de}
\and
     Brandenburg University of Technology at Cottbus \\
     Postbox 10 13 44, 03013 Cottbus, Germany \\
     \email{rohrch@tu-cottbus.de}
}

\crefname{algocf}{Algorithm}{Alg.}
\DontPrintSemicolon

\lstdefinelanguage{command}%
{ basicstyle=\normalsize\ttfamily
, flexiblecolumns=true
}%

\lstdefinelanguage{title}%
{ basicstyle=\normalsize\ttfamily
, flexiblecolumns=true
}%
\lstdefinelanguage{body}%
{ basicstyle=\footnotesize\ttfamily
, flexiblecolumns=true
}%

\makeindex[options=-s mcc, intoc, columns=2]
\makeglossaries

\newacronym[user1=MCC]{MCC}{%
  Model Checking Contest%
}{%
  Model Checking Contest @ Petri nets%
}
\newacronym[user1=MCC!2011]{MCC2011}{%
  MCC'2011%
}{%
  Model Checking Contest 2011%
}
\newacronym[user1=MCC!2012]{MCC2012}{%
  MCC'2012%
}{%
  Model Checking Contest 2012%
}
\newacronym[user1=MCC!2013]{MCC2013}{%
  MCC'2013%
}{%
  Model Checking Contest 2013%
}

\newacronym{SUMo}{SUMo}{International Workshop on Scalable and Usable Model Checking for Petri nets and other models of concurrency}
\newacronym{DD}{DD}{decision diagram}
\newacronym{IDD}{IDD}{Interval Decision Diagram}
\newacronym{SigmaDD}{\ensuremath{\Sigma}DD}{\ensuremath{\Sigma}~Decision Diagram}
\newacronym{ZDD}{ZDD}{Zero-Suppressed Decision Diagram}
\newacronym{ZMDD}{ZMDD}{Zero-Suppressed Multi-Terminal Decision Diagram}
\newacronym{BDD}{BDD}{Binary Decision Diagram}
\newacronym{DDD}{DDD}{Data Decision Diagram}
\newacronym{SDD}{SDD}{Set Decision Diagram}
\newacronym{PN}{Petri net}{Petri net}
\newacronym{GSPN}{GSPN}{Generalized Stochastic Petri Nets}
\newacronym{PT}{P/T}{Place/Transition net}
\newacronym{SN}{SN}{Symmetric net}
\newacronym{SNB}{SNB}{Symmetric nets with Bags}
\newacronym{CN}{Colored}{Colored net}
\newacronym{APN}{APN}{Algebraic Petri net}
\newacronym{HLPN}{HLPN}{High Level Petri net}
\newacronym{ITS}{ITS}{Instantiable Transition Systems}
\newacronym{AADT}{AADT}{Abstract Algebraic Data Type}
\newacronym{CTL}{CTL}{Computation Tree Logic}
\newacronym{LTL}{LTL}{Linear Temporal Logic}
\newacronym[user1=PNML]{PNML}{PNML}{Petri Net Markup Language}
\newacronym[user1=Model!cs\_repetitions]{CS-Repetitions}{%
  cs\_repetitions%
}{%
  Client/Server with repetitions%
}

\newacronym[user1=Model!echo]{Echo}{%
  echo%
}{%
  Echo algorithm%
}

\newacronym[user1=Model!eratosthenes]{Eratosthenes}{%
  eratosthenes%
}{%
  Eratosthenes' sieve%
}

\newacronym[user1=Model!FMS]{FMS}{%
  FMS%
}{%
  Flexible Manufacturing System%
}

\newacronym[user1=Model!galloc\_res]{Galloc}{%
  galloc\_res%
}{%
  Global Allocation Resource Management%
}

\newacronym[user1=Model!Kanban]{Kanban}{%
  Kanban%
}{%
  Kanban System%
}

\newacronym[user1=Model!lamport\_fmea]{Lamport}{%
  lamport\_fmea%
}{%
  Lamport's fast mutual exclusion algorithm%
}

\newacronym[user1=Model!MAPK]{MAPK}{%
  MAPK%
}{%
  Mitogen-Activated Protein Kinase Kascade%
}

\newacronym[user1=Model!neo-election]{Neo-Election}{%
  neo-election%
}{%
  Neo Election Protocol%
}

\newacronym[user1=Model!Peterson]{Peterson}{%
  Peterson%
}{%
  Peterson's mutual exclusion algorithm%
}

\newacronym[user1=Model!philo\_dyn]{Dynamic-Philosophers}{%
  philo\_dyn%
}{%
  Dynamic Dining Philosophers%
}

\newacronym[user1=Model!Philosophers]{Philosophers}{%
  Philosophers%
}{%
  Dining Philosophers%
}

\newacronym[user1=Model!planning]{Planning}{%
  planning%
}{%
  AI Planning%
}

\newacronym[user1=Model!railroad]{Railroad}{%
  railroad%
}{%
  Railroad%
}
\newacronym[user1=Model!ring]{Ring}{%
  ring%
}{%
  Three-Module Ring%
}

\newacronym[user1=Model!rwmutex]{RW-Mutex}{%
  rwmutex%
}{%
  Reader/Writer Mutual Exclusion%
}

\newacronym[user1=Model!SharedMemory]{Shared-Memory}{%
  SharedMemory%
}{%
  Shared Memory%
}

\newacronym[user1=Model!simple\_lbs]{Simple-LBS}{%
  simple\_lbs%
}{%
  Simple Load Balancing System%
}

\newacronym[user1=Model!TokenRing]{Token-Ring}{%
  TokenRing%
}{%
  Token Ring%
}

\newacronym[user1=Tool!ACTIVITY-LOCAL]{ACTIVITY-LOCAL}{%
  \lstinline!ACTIVITY-LOCAL!%
}{%
  \lstinline!ACTIVITY-LOCAL!%
}

\newacronym[user1=Tool!AlPiNA]{AlPiNA}{%
  \lstinline!AlPiNA!%
}{%
  \lstinline!AlPiNA!%
}

\newacronym[user1=Tool!Crocodile]{Crocodile}{%
  \lstinline!Crocodile!%
}{%
  \lstinline!Crocodile!%
}

\newacronym[user1=Tool!GreatSPN]{GreatSPN}{%
  \lstinline!GreatSPN!%
}{%
  \lstinline!GreatSPN!%
}

\newacronym[user1=Tool!Helena]{Helena}{%
  \lstinline!Helena!%
}{%
  \lstinline!Helena!%
}

\newacronym[user1=Tool!ITS-Tools]{ITS-Tools}{%
  \lstinline!ITS-Tools!%
}{%
  \lstinline!ITS-Tools!%
}

\newacronym[user1=Tool!LoLA-binstore]{LoLA-binstore}{%
  \lstinline!LoLa-binstore!%
}{%
  \lstinline!LoLa-binstore!%
}

\newacronym[user1=Tool!LoLA-bloom]{LoLA-bloom}{%
  \lstinline!LoLa-bloom!%
}{%
  \lstinline!LoLa-bloom!%
}

\newacronym[user1=Tool!LoLA]{LoLA}{%
  \lstinline!LoLA!%
}{%
  \lstinline!LoLA!%
}

\newacronym[user1=Tool!Marcie]{Marcie}{%
  \lstinline!Marcie!%
}{%
  \lstinline!Marcie!%
}

\newacronym[user1=Tool!Neco]{Neco}{%
  \lstinline!Neco!%
}{%
  \lstinline!Neco!%
}

\newacronym[user1=Tool!PNXDD]{PNXDD}{%
  \lstinline!PNXDD!%
}{%
  \lstinline!PNXDD!%
}

\newacronym[user1=Tool!PeTe]{PeTe}{%
  \lstinline!PeTe!%
}{%
  \lstinline!PeTe!%
}

\newacronym[user1=Tool!Sara]{Sara}{%
  \lstinline!Sara!%
}{%
  \lstinline!Sara!%
}

\newacronym[user1=Tool!SMART]{SMART}{%
  \lstinline!SMART!%
}{%
  \lstinline!SMART!%
}

\newacronym[user1=Tool!YASPA]{YASPA}{%
  \lstinline!YASPA!%
}{%
  \lstinline!YASPA!%
}

\newacronym[user1=Tool!UPPAAL]{UPPAAL}{%
  \lstinline!UPPAAL!%
}{%
  \lstinline!UPPAAL!%
}

\renewcommand\subsubsection{\@startsection{subsubsection}{3}{\z@}%
 {1em}%
 {.5em}%
 {\reset@font\normalsize\bfseries}}

\makeatletter
\renewcommand\section{\@startsection{section}{1}{\z@}%
 {1em}%
 {.5em}%
 {\reset@font\large\bfseries}}
\renewcommand\subsection{\@startsection{subsection}{2}{\z@}%
 {1em}%
 {.5em}%
 {\reset@font\normalsize\bfseries}}
\renewcommand\subsubsection{\@startsection{subsubsection}{3}{\z@}%
  {1em}%
  {-1em}%
  {\normalfont\normalsize\bfseries}}
\makeatother

\makeatletter
\apptocmd{\@gls@link}{\index{\glsentryuseri{#2}}}{}{}
\makeatother

\begin{document}
\pagenumbering{roman}
\thispagestyle{empty}
\input Starburst.fd
\newcommand\CoverPicture{
\put(-5,-100){
\parbox[b][\paperheight]{\paperwidth}{%
\vfill
\centering
\includegraphics[width=23.5cm,keepaspectratio]{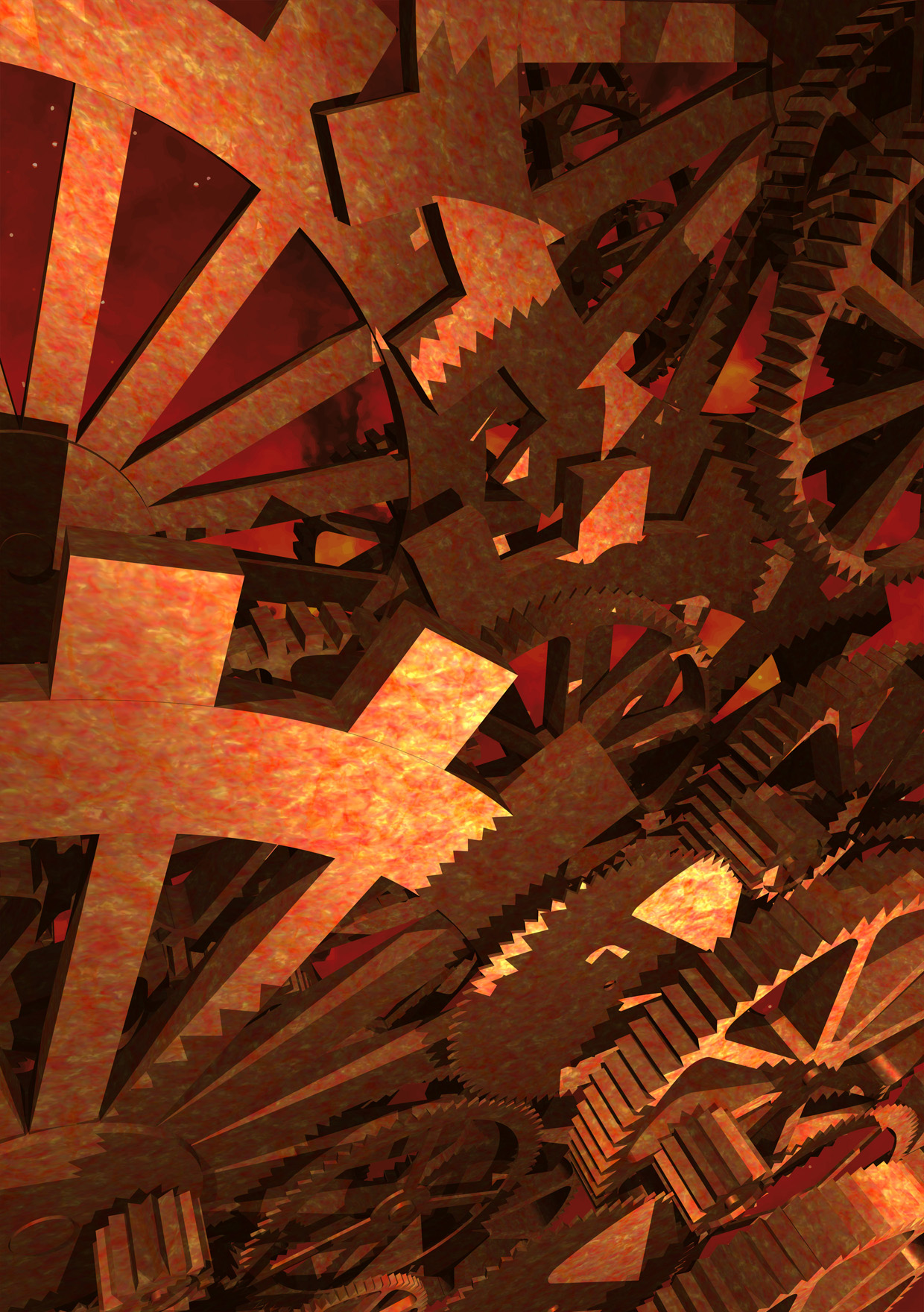}%
\vfill
}}}
\AddToShipoutPicture{\CoverPicture}
\begin{center}
\color{white}
{\bfseries\usefont{U}{Starburst}{xl}{n}
\mbox{}
\vfill
\Large
\textsf{* * *}
\vfill 
\huge
Model Checking Contest
\\
\textsf{@} Petri Nets
\vfill
Report on the 2012 edition
\vskip 1.0cm}
\huge
September 2012
\vskip 2.0cm
\begin{minipage}{.8\textwidth}
\Large\centering
     F. Kordon,
     A. Linard,
     \par
     D. Buchs,
     M. Colange,
     S. Evangelista,
     L. Fronc,
     \par
     L.M. Hillah,
     N. Lohmann,
     E. Paviot-Adet,
     F Pommereau,
     \par
     C. Rohr,
     Y. Thierry-Mieg,
     H. Wimmel,
     K. Wolf
\end{minipage}
\vfill
\textsf{\Large* * *}
\vfill
\vskip 9.0cm
\mbox{}
\end{center}
\clearpage
\renewcommand\CoverPicture{}
\thispagestyle{empty}
\cleardoublepage

\setcounter{tocdepth}{2}
\tableofcontents
\cleardoublepage

\cfoot{\thepage}
\pagenumbering{arabic}
\maketitle

\begin{abstract}

This article presents the results of the Model Checking Contest held at Petri
Nets 2012 in Hambourg. This contest aimed at a fair and experimental
evaluation of the performances of model checking techniques applied to Petri
nets. This is the second edition after a successful one in 2011~\cite{mcc2011}.

The participating tools were compared on several examinations (state space
generation and evaluation of several types of formul{\ae} -- structural,
reachability, LTL, CTL) run on a set of common models (Place/Transition and
Symmetric Petri nets).

After a short overview of the contest, this paper provides the raw results
from the context, model per model and examination per examination.

\end{abstract}

\section{Introduction}

\label{sec:introduction}

When verifying by model checking a system with formal methods, such as
\aclp{PN}, one may have several questions such as:

\begin{quotation}
``When creating the model of a system,
should we use structural analysis or an explicit model checker to debug the model?''

``When verifying the final model of a highly concurrent system,
should we use a symmetry-based or a partial order reduction-based model checker?''

``When updating a model with large variable domains,
should we use a decision diagram-based or an abstraction-based model checker?''
\end{quotation}

Results that help to answer these questions are spread among numerous papers
in numerous conferences. The choice of the models and tools used in benchmarks
is rarely sufficient to answer these questions. Benchmark results are
available a long time after their publication, even if the computer
architecture has changed a lot. Moreover, as they are executed over several
platforms and composed of different models, conclusions are not easy.

The objective of the \acl{MCC} is to compare the efficiency of verification
techniques according to the characteristics of the models. To do so, the
\acs{MCC} compares tools on several classes of models with scaling
capabilities, \emph{e.g.}, values that set up the ``size'' of the associated
state space.

Through a benchmark, our goal is to identify the techniques that can tackle a
given type of problem identified in a ``typical model'', for a given class of
problem (\emph{e.g.}, state space generation, evaluation of reachability or
temporal formula{\ae}, etc.).

The second edition of the \acl{MCC} took place within the context of the Petri
Nets and ACSD 2012 conferences, in Hamburg, Germany\footnote{A First edition
took place in Newcastle, UK, withing the context of the SUMo workshop,
associated to the Petri Nets et ACSD 2011 conferences~\cite{mcc2011}.}. The
original submission procedure was published early March 2012 and submissions
gathered by mid-May 2012. After some tuning of the execution environment, the
evaluation procedure was operated on a cluster early June. Results were
presented during the SUMo workshop, on June \nth{26}, 2012.

\bigskip
The goal of this paper is to report the raw data provided by this second
edition of the Model Checking Contest. It reflects the vision of
\acs{MCC2012}
organizers, as it was first presented in Hamburg. All tool
developers are listed in~\Cref{sec:conclusion}.

\bigskip
The article is structured as follows.
\Cref{sec:models}~presents the models proposed in this second edition.
Then, \Cref{sec:tools}~lists some information on the participating tools.
\Cref{sec:models}~provides an overview of the evaluation methodology.
Finally, \Cref{sec:statespace,sec:structformulae,sec:reachformulae}
present the raw data we collected from~\acs{MCC2012}.

\cleardoublepage
\section{Selected Models}
\label{sec:models}

A first innovation in \acs{MCC2012} was to add a ``call for
model''. The idea was to gather benchmarks from the community. Then, on top of
the 7 models proposed in the first edition, we got 12 models from various
institutes, usually coming from some tool benchmarks. When models were
colored, we requested submitters to provide both the colored version and the
Place/Transition (P/T) equivalent version.

The models were provided in \acs{PNML} format~\cite{iso159092,pnmlorg}. Tools developers could use
the \acs{PNML} description to build the one of their tool.
\acs{PNML} was then used as the
reference. 17 out of the 19 models had scaling parameters. Thus, we could
process them with various values and then see how tools could scale up with
these models.

We provide below a brief description of models. Moreover, each one has a
data-sheet available on \acs{MCC2012} web site: \url{http://mcc.lip6.fr/2012}.

\subsection{Models from \acs{MCC2011}} These first seven models are from the
\acs{MCC2011}. In this set, \acs{MAPK} is the only model coming from a case study (biology).

\subsubsection{\Acs{FMS}} belongs to the \acs{GreatSPN} and
\acs{SMART}~\cite{smart} benchmarks. It models a
\acl{FMS}~\cite{fms:ciardo:1993}. The scaling parameter corresponds to the
number of initial tokens held in three places. The following values were used:
$2, 5, 10, 20, 50, 100, 200, 500$.

\subsubsection{\acs{Kanban}} \cite{kanban:ciardo:1996} models a
Kanban system. The scaling parameter corresponds to the number of initial
tokens held in four places. The following values were used: $5, 10, 20, 50,
100, 200, 500, 1\,000$.

\subsubsection{\Acs{MAPK}} models a biological system: the
\acl{MAPK}~\cite{mapk:heiner:2008}. The scaling parameter changes the initial
number of tokens held in seven places. The following values were used: $8, 20,
40, 80, 160, 320$.

\subsubsection{\acs{Peterson}} models
\acl{Peterson}~\cite{Peterson81} in its generalized version for $N$ processes.
This algorithm is based on shared memory communication and uses a loop with
$N-1$ iterations, each iteration is in charge of stopping one of the competing
processes. The scaling parameter is the number of involved processes. The
following values were used: $2, 3, 4, 5, 6$.

\subsubsection{\acs{Philosophers}} models the famous
\acl{Philosophers} problem introduced by E.W. Dijkstra in
1965~\cite{philo:dijkstra65} to illustrate an inappropriate use of shared
resources, thus generating deadlocks or starvation. The scaling parameter is
the number of philosophers. The following values were used: $5$, $10$, $20$,
$50$, $100$, $500$, $1\,000$, $5\,000$, $10\,000$, $50\,000$, $100\,000$.

\subsubsection{\acs{Shared-Memory}} is a model taken from the
\acs{GreatSPN} benchmarks~\cite{sharedmemory}. It models a system composed of
$P$ processors competing for the access to a shared memory (built with their
local memory) using a unique shared bus. The scaling parameter is the number
of processors. The following values were used: $5, 10, 20, 50, 100, 200, 500,
1\,000, 2\,000, 5\,000, 10\,000, 20\,000, 50\,000$.

\subsubsection{\acs{Token-Ring}} is another problem proposed by
E.W. Dijkstra~\cite{TokenRing}. It models a system where a set of machines is
placed in a ring, numbered $0$ to $N-1$. Each machine $i$ only knows its own
state and the state of its left neighbor, i.e., machine $(i-1) \mod (N)$.
Machine number 0 plays a special role, and it is called the ``bottom
machine''. A protocol ensuring non-starvation determines which machine has a
``privilege'' (e.g. the right to access a resource). The scaling parameter is
the number of machines. The following values were used: $5, 10, 15, 20, 30,
40, 50, 100, 200, 300, 400, 500$.

\subsection{New Models in \acs{MCC2012}} These twelve models were
submitted by the community for \acs{MCC2012}. In this new set, several models
are coming from larger case studies: \acs{Neo-Election}, \acs{Planning}, and
\acs{Ring}.

\subsubsection{\acs{CS-Repetitions}} models a client/server application
with $C$ clients and $S$ servers. Communication from clients to servers is not
reliable, with requests stored in a buffer of size $B$. Communication from
servers to clients are reliable. A client send its message until it receives
an answer. The scaling parameter is a function of $C$ for a fixed number of
severs. The following values were used: $25, 49, 100, 225,400,625,900$.

\subsubsection{\acs{Echo}} This file specifies the Echo Algorithm
(see~\cite{reisig98}) for grid like networks. It is a protocol for propagation
of information with feedback in a network. A distinguished agent (initiator),
starts the distribution of a message by sending it to all its neighbors. On
receiving some first message, every other agent forwards the message to all
its neighbors, except the one it received its first message from. Then it
awaits messages from all recipients of its forwards (regardless whether these
messages had been intended as forwards or acknowledgments) and replies to the
agent where it received its first message from. As soon as the initiator
receives a message from all its neighbors, the protocol terminates. In this
example, agents are arranged in a hypercube that can be scaled in two values:
$D$, the number of \emph{dimensions} and $R$, the number of \emph{agents per
dimensions}. The scaling parameter is a combination of $D$ and $R$. The
following values were used: $d2r11$, $d2r13$, $d2r15$, $d2r17$, $d2r19$,
$d2r9$, $d3r3$, $d3r5$, $d3r7$, $d4r3$, $d5r3$.

\subsubsection{\acs{Eratosthenes}} This model implements the sieve of
Eratosthenes~\cite{eratosthenessieve}. The scaling parameter is the size of
the sieve. The following values were used: $5, 10, 20, 50 ,100, 200, 500$.

\subsubsection{\acs{Galloc}} It models the deadlock-free management
of mutually exclusive resources known as the ``global allocation
strategy''~\cite{krakoviak88}. When a process enters a critical section, it
locks all the resources needed to be used in the critical section (in the
model, 4 max). Then, it can release a subset of these resources (max 2 in the
model) at a time (and then stay in the critical section) or exit the critical
section, thus releasing all the remaining resources it locks. The scaling
parameter is a value $N$ for $N$ processes and $N\times 2$ resources. The
following values were used: $3, 5, 6, 7, 9, 10, 11$.

\subsubsection{\acs{Lamport}} It models Lamport's fast mutual
exclusion algorithm designed for multi-processor architectures with a shared
memory and was studied in~\cite{lamportFMEA}. The scaling parameter is the
number of processes competing for the critical section. The following values
were used: $2,3,4,5,6,7,8$.

\subsubsection{\acs{Neo-Election}} The Neo protocol aims at managing
large distributed databases on clusters of workstations. The machines on the
cluster may have several roles. This model focusses on master nodes which
handle the communications between all nodes, and in particular requests for
accessing database objects. Prior to that all master nodes agree on a primary
master which will be the operating one, the other master nodes being
secondary, waiting to replace the primary master if needed. This model
specifies this election algorithm~\cite{neoppod-ATPN10}. The scaling parameter
is the number of master nodes. The following values were used: $2, 3, 4, 5, 6,
7, 8$.

\subsubsection{\acs{Dynamic-Philosophers}} is a variation of the Dining Philosophers
where philosophers can join or quit the table~\cite{philodyn}. Each
philosopher has its own fork, as in the usual version. The interesting point
is that identifiers of left and right for each philosopher must be computed or
stored somewhere. A philosopher can enter the table only if the two forks
around his position are available. He can leave if his fork is free, and he is
thinking. The scaling parameter is the maximum number of philosophers. The
following values were used: $2,3,10,20,50, 80$.

\subsubsection{\acs{Planning}} It models the equipment (displays,
canvases, documents, and lamps) of a smart conference room of the University
of Rostock. It was derived from a proprietary description format that was used
by an AI planning tool to generated plans to bring the room in a desired
state, for instance displaying a document on a certain canvas while switching
off the lights. This problem can be expressed as a reachability problem.
This model has no scaling parameter.

\subsubsection{\acs{Railroad}} it corresponds to the Petri nets semantics
of an ABCD model of a railroad crossing system. t has three components: a gate
sub-net, a controller sub-net and $n$ tracks sub-nets that differ only by an
identifier $k$ in $\{0,\dots,n-1\}$. These components communicate through
shared places, some being low-level places to exchange signals, others being
integer-valued places to exchange tracks identifiers. The controller also has
a place to count the number of trains at a given time. The scaling parameter
is the number of tracks. The following values were used: $5, 10, 20, 50, 100$.

\subsubsection{\acs{Ring}} It models a three-module ring
architecture~\cite{ringref}. The communication architecture contains as many
channels as there are modules. It tests the occurrence of global deadlock
arising from a local one. It uses stoppable clocking scheme on arbitrated
input and output channels. This model has no scaling parameter.

\subsubsection{\acs{RW-Mutex}} It models a system with readers and
writers~\cite{reisig98}. Reading can be conducted concurrently whereas writing
has to be done exclusively. This is modeled by a number of semaphores (one for
each reader) that need to be collected by a writer prior to writing. The
scaling parameter a combination of $R$, the number of readers and $W$, the
number of writers. The following values were used: $r10w10$, $r10w20$,
$r10w50$, $r10w100$, $r10w500$, $r10w1000$, $r10w2000$, $r20w10$, $r100w10$,
$r500w10$, $r1000w10$, $r2000w10$.

\subsubsection{\acs{Simple-LBS}} models a simple load balancing system
composed of a set of clients, two servers, and between these, a load balancer
process. The scaling parameter is the number of clients to be balanced over
the servers. The following values were used: $2$, $3$, $4$, $5$, $6$, $7$,
$8$, $9$, $10$, $15$, $20$.

\clearpage
\section{Participating Tools}
\label{sec:tools}

We got 10 tool submissions summarized in~\Cref{tab:submissions}:
\acs{AlPiNA},
\acs{Crocodile},
\acs{Helena},
\acs{ITS-Tools},
\acs{LoLA-binstore},
\acs{LoLA-bloom},
\acs{Marcie},
\acs{Neco},
\acs{PNXDD},
and \acs{Sara}.
Their descriptions, written by the tool developers,
are given in this section.

\begin{table}[b]
   \centering
   {\small
   \begin{tabular}{|r||c|c|c|c|}
      \hline
      \textbf{Tool Name} & \textbf{Team} & \textbf{Institution} & \textbf{Country} & \textbf{Contact Name} \\
      \hline
      \hline
      \acs{AlPiNA} & {CUI/SMV} & {Univ. Geneva} & {Switzerland} & {D. Buchs} \\
      \hline
      \acs{Crocodile} & {LIP6/MoVe} & {UPMC} & {France} & {M. Colange} \\
      \hline
      \acs{Helena} & {LIPN/LCR} & {Univ. Paris 13} & {France} & {S. Evangelista} \\
      \hline
      \acs{ITS-Tools} & {LIP6/MoVe} & {UPMC} & {France} & {Y. Thierry-Mieg} \\
      \hline
      \acs{LoLA} & {Team Rostock} & {Univ. Rostock} & {Germany} & {N. Lohmann \& K. Wolf} \\
      \hline
      \acs{Marcie} & {DSSZ} & {Univ. Cottbus} & {Germany} & {M. Heiner \& C. Rohr} \\
      \hline
      \acs{Neco} & {IBISC} & {Univ. Evry} & {France} & {L. Fronc} \\
      \hline
      \acs{PNXDD} & {LIP6/MoVe} & {UPMC} & {France} & {E. Paviot-Adet} \\
      \hline
      \acs{Sara} & {Team Rostock} & {Univ. Rostock} & {Germany} & {H. Wimmel \& K. Wolf} \\
      \hline
    \end{tabular}}
    \caption{Summary of data on participating tools\label{tab:submissions}}
\end{table}

\subsubsection{\acs{AlPiNA}}%
\footnote{Tool is available at \url{http://alpina.unige.ch}.}
\cite{2011-hostettler-0} stands for
\aclp{APN} Analyzer and is a symbolic model checker for \aclp{APN}. It can
verify various state properties expressed in a first order logic property
language.

\aclp{APN} (\acsp{APN}) (\aclp{PN} + \aclp{AADT}) is a powerful formalism to
model concurrent systems in a compact way. Usually, concurrent systems have
very large sets of states, that grow very fast in relation to the system size.
Symbolic Model Checking (\acs{DD}-based one) is a proven technique to handle
the State Space Explosion for simpler formalisms such as \aclp{PT}.
\acs{AlPiNA} extend these concepts to handle algebraic values that can be
located in net places.

For this purpose \acs{AlPiNA} uses enhanced \acsp{DD} such as \aclp{DDD} and
\aclp{SDD} for representing the place vectors and
\aclp{SigmaDD}~\cite{buchs:termgraph:09} for representing the values of the
\acs{APN}. It also allows to specify both algebraic and topological clusters
to group states together and thus to reduce the memory footprint. Particular
care has been taken to let users freely model their systems in \acsp{APN} and
in a second independent step to tune the optimization parameters such as
unfolding rules, variable order, and algebraic clustering. Compared to
\acl{CN} approaches, \acs{AlPiNA}~\cite{buchs:tacas:2010} solves problems
related to the unbounded nature of data types and uses the inductive nature of
\aclp{AADT} to generalize the unfolding and clustering techniques to any kind
of data structure.

\acs{AlPiNA}'s additional goal is to provide a user friendly suite of tools
for checking models based on the \aclp{APN} formalism. In order to provide
great user experience, it leverages the Eclipse platform.

\subsubsection{\acs{Crocodile}}%
\footnote{Tool is available at \url{http://move.lip6.fr/software/Crocodile}.}
\cite{PN2011} was initially
designed as a demonstration tool for the so-called symbolic/symbolic
approach~\cite{symbolic2}. It combines two techniques for handling the
combinatorial explosion of the state space that are both called
``symbolic''.

The first ``symbolic'' technique concerns the reduction of the reachability
graph of a system by its symmetries. The method used in \acs{Crocodile} was
first introduced in~\cite{chiola91} for the \aclp{SN}, and was then extended
to the \ac{SNB} in~\cite{acc2009}. A symbolic reachability graph (also called
quotient graph) can be built for such types of \aclp{PN}, thus dramatically
reducing the size of the state space.

The second ``symbolic'' technique consists in storing the reachability graph
using \aclp{DD}, leading to a symbolic memory encoding. \acs{Crocodile} relies
on Hierarchical \aclp{SDD}~\cite{FORTE2005}. These present several interesting
features, such as hierarchy, and the ability to define inductive operations.

Still under development, \acs{Crocodile} essentially generates the state space
of a~\acs{SNB} and then processes reachability properties. The version
submitted for this second edition of the \acs{MCC} includes new strategies for the
management of bags~\cite{crocodile2}.

\subsubsection{\acs{Helena}}%
\footnote{Tool is available at \url{http://helena-mc.sourceforge.net}.}
\cite{Helena} is an explicit state model checker for \aclp{HLPN}.
It is a command-line tool available available under the GNU GPL.

\acs{Helena} tackles the state explosion problem mostly through a reduction of
parallelism operated at two stages of the verification process. First, static
reduction rules are applied on the model in order to produce a smaller net
that -- provided some structural conditions are verified -- is equivalent to
the original one but has a smaller reachability graph. Second, during the
search, partial order reduction is employed to limit, as much as possible, the
exploration of redundant paths in the reachability graph. This reduction is
based on the detection of independent transitions of the net at a given
marking. Other reduction techniques are also implemented by \acs{Helena},
e.g., state compression, but were disabled during the contest due to their
inadequacy with the proposed models.

\subsubsection{\acs{ITS-Tools}}%
\footnote{Tool is available at \url{http://ddd.lip6.fr}.}
\cite{TPHK} are a set of tools to analyze
\acl{ITS}, introduced in~\cite{TPHK}. This formalism allows compositional
specification using a notion of type and instance inspired by component
oriented models. The basic elementary types are labeled automata structures,
or labeled \aclp{PN} with some classical extensions (inhibitor arcs, reset
arcs$\dots$). The instances are composed using event-based label
synchronization.

The main strength of \acs{ITS-Tools} is that they rely on Hierarchical
\aclp{SDD}~\cite{FORTE2005} to perform analysis. These \aclp{DD} support
hierarchy, allowing to share representation of states for some subsystems.
When the system is very regular or symmetric, recursive encodings~\cite{TPHK}
may even allow to reach logarithmic overall complexity when performing
analysis. Within the contest, the \acs{Philosophers} and \acs{Token-Ring}
examples proved to be tractable using this recursive folding feature.

\aclp{SDD} also offer support for automatically enabling the ``saturation''
algorithm for symbolic least fixpoint computations~\cite{FI2009}, a feature
allowing to drastically reduce time and memory consumption. This feature was
used in all computations.

\subsubsection{\acs{LoLA}} \footnote{Tool is
available at \url{http://www.informatik.uni-rostock.de/tpp/lola}.}
\cite{Wolf_2007_atpn} is an explicit \acl{PN} state space verification tools.
It can verify a variety of properties ranging from questions regarding
single \acl{PN} nodes (\emph{e.g.}, boundedness of a place or quasiliveness of
a transition), reachability of a given state or a state predicate, typical
questions related to a whole \acl{PN} (\emph{e.g.}, deadlock freedom,
reversibility, or boundedness), and the validity of temporal logical formulae
such as \acs{CTL}. It has been successfully used in case studies from
various domains, including asynchronous circuits, biochemical reaction chains,
services, business processes, and parameterized Boolean programs.

For each property, \acs{LoLA} provides tailored versions of state space
reduction techniques such as stubborn sets, symmetry reduction, coverability
graph generation, or methods involving the \acl{PN} invariant calculus.
Depending on the property to be preserved, these techniques can also be used
in combination to only generate a small portion of the state space.


Two versions of \acs{LoLA} were submitted in 2012.
\acs{LoLA-binstore} is a complete reimplementation of the tool since 2011.
\acs{LoLA-bloom} uses a bit hashing technique.

\subsubsection{\acs{Marcie}}%
\footnote{Tool is available at \url{http://www-dssz.informatik.tu-cottbus.de/DSSZ/Software/Marcie}.}
\cite{srh2011}
is a tool for the analysis of \acl{GSPN}, supporting
qualitative and quantitative analyses including model checking facilities.
Particular features are symbolic state space analysis including efficient
saturation-based state space generation, evaluation of standard \acl{PN}
properties as well as \acs{CTL} model checking.

Most of \acs{Marcie}'s features are realized on top of an \acl{IDD}
implementation~\cite{tovchigrechko2008}. \Acsp{IDD} are used to efficiently
encode interval logic functions representing marking sets of bounded
\aclp{PN}. This allows to efficiently support qualitative state space based
analysis techniques~\cite{schwarick2010}. Further, \acs{Marcie} applies heuristics
for the computation of static variable orders to achieve small \acl{DD}
representations.

For quantitative analysis \acs{Marcie} implements a multi-threaded on-the-fly
computation of the underlying CTMC. It is thus less sensitive to the number of
distinct rate values than approaches based on, e.g., Multi-Terminal Decision
Diagrams. Further it offers symbolic CSRL model checking and permits to
compute reward expectations. Additionally \acs{Marcie} provides simulative and
explicit approximative numerical analysis techniques.

\subsubsection{\acs{Neco}}%
\footnote{Tool is available at \url{http://code.google.com/p/neco-net-compiler/}.}
is a coloured Petri nets
compiler which produces libraries for explicit model-checking. These libraries
can be used to build state spaces. It is a command-line tool available under
the GNU Lesser GPL.

\acs{Neco} is based on the SNAKES~\cite{SNAKES} toolkit and handles coloured Petri
nets annotated with arbitrary Python~\cite{Python} objects. This allows for a
high degree of expressivity. Extracting information from models, \acs{Neco} can
identify object types and produce optimized Python or C exploration code. The
later is done using the Cython~\cite{Cython} language. Moreover, if a part of
the model cannot be compiled efficiently a Python fallback is used to handle
this part of the model.

\acs{Neco} also performs model based optimizations using place bounds~\cite{FP11}
and control flow places for more efficient firing functions. However, these
optimizations are closely related to a modelling language we use which allows
them to be assumed by construction. Because the models from the contest were
not provided with such properties, these optimizations could not be used.

\subsubsection{\acs{PNXDD}}%
\footnote{Tool is available at \url{https://srcdev.lip6.fr/trac/research/NEOPPOD/wiki/pnxdd}.}
generates
the state-space of \aclp{PT}. When \aclp{CN} are used in the \acs{MCC},
equivalent \acsp{PT} are obtained after an ``optimized''
unfolding~\cite{p2006linar} (unused places and transitions are detected and
suppressed).

State space storage relies on Hierarchical
\aclp{SDD}~\cite{FORTE2005}~(\acsp{SDD}). These are \aclp{DD} with any data
type associated to arcs (see \emph{e.g.},~\cite{polyDD} for an overview of
\acs{DD}-like structures). If the associated data type is another \acs{SDD},
hierarchical structures can be constructed.

Since \acs{PNXDD} exploits hierarchy, a state is seen as a tree, where the
leaves correspond to places marking. This particular structure offers greater
sharing opportunities than a, for instance, vector-based representation. The
conception of such a tree is critical to reach good performances and
heuristics are being elaborated for this purpose~\cite{2011-hong-0}. The one
used for the \acs{MCC} is based on~\cite{force:2003}: for colored models that
do scale via the size of color types, \acs{PNXDD} uses a tree-like version of
this heuristic, while the original version is kept when colored models only
scale via the number of tokens in the initial marking.

\subsubsection{\acs{Sara}}%
\footnote{Tool is available at \url{http://www.service-technology.org/tools/download}.}
uses the state
equation, known to be a necessary criterion for reachability and in a modified
way also for other properties like coverability, to avoid enumerating all
possible states of a system~\cite{narrowing01}. A minimal solution of the
state equation in form of a transition vector is transformed into a tree of
possible firing sequences for this solution. A firing sequence using all the
transitions given in the solution (with the correct multiplicity) reaches the
goal.

For tree pruning, partial order reduction is used, especially in the form of
stubborn sets~\cite{Schmidt_1999_atpn,ksv06}. If the goal cannot be reached
using the obtained solution, places that do not get enough tokens are
computed. Constraints are built and added to the state equation (in a
CEGAR-like fashion~\cite{cegar00}). These constraints modify the former
solution by adding transition invariants, temporarily allowing for additional
tokens on the undermarked places.

\acs{Sara} detects unreachable goals either from an unsatisfiable state
equation or by cutting the solution space to finite size when the repeated
addition of transition invariants is known not to move towards the goal. A
more involved explanation of the algorithm behind \acs{Sara} can be found
in~\cite{sara11}.

\clearpage
\section{Evaluation Methodology}
\label{sec:methodology}

Roughly, the evaluation methodology was the same as for \acs{MCC2011}
(it is presented in~\cite{mcc2011}).
The main differences were the following:

\begin{enumerate}
\item we changed some of the examinations,
      to allow checking of more properties;
\item we monitored tools by means of virtual machines, which is more complex
to operate but much less intrusive, thus leading to more accurate results.
\end{enumerate}

\subsection{Examinations} There were two types of examinations in
\acs{MCC2011}: state space
generation and evaluation of reachability formul{\ae}.
We enriched this second examination by dividing it with subclasses of
formul{\ae}:

\begin{itemize}

\item \emph{Structural}: these only refer to structural aspects of the net
such as existence and value of a place bound,
level of transition liveness, and the existence of a deadlock;

\item \emph{Reachability}: these only refer to a combination of profile over
states by comparing token values in places and/or computing transition
fireability;

\item \emph{LTL}: these formul{\ae} contain LTL operators
mixed with reachability ones;

\item \emph{CTL}: these contain CTL operators
mixed with reachability ones.

\end{itemize}

All details on the formul{\ae} and the associated language is described
in~\cite{propmanual}.

We decided to operate a set of formul{\ae} per model and per scaling parameter
when it existed. Since models were proposed in both colored and P/T versions,
we also proposed the formul{\ae} in both types.

These two choices made the preparation of the contest quite difficult. First,
computing equivalent formul{\ae} for colored nets and their P/T equivalent is
sometimes not possible. Second, we had no time to check that formul{\ae} could
be satisfied or unsatisfied as for the \acs{MCC2012}. This impede the results
of the formul{\ae} examinations as we state in~\Cref{sec:conclusion}.

\subsection{Use of Virtual Machines} Tools were submitted within a disk
image that was provided with a Linux distribution (Windows could have been
proposed but no tool requested it).

Then, we processed the examination on a 23-nodes cluster equipped with
PowerEdge R410 (6 ports gigabits) and 1.5To local disks. Each node operated an
Intel Xeon E5645@2.40GHz (6 cores, 12 threads) and had 8GB of memory (DDR3,
1333).

We considered one examination on one model for a given scaling parameter (when
there was one) as a \emph{run}. The contest required 2419 run in total (639
for the state space examination and 1780 for the structural and reachability
examinations). This does not include the necessary trials required to evaluate
the procedure and check (with their developers) that tools were operating
appropriately.

\begin{algorithm}
	\KwIn{$M$, a set of scalable models to be processed}
	\ForEach{$m \in M$}{
		launch the virtual machine $m$\;
		\ForEach {$v$, scaling values for $m$}{
			\ForEach {$e$, in StateSpace, StrctFormulae, ReachFormulae, CTLFormulae, LTLFormulae}{
			connect to the virtual machine\;
			request $e$ on $m$ for value $v$\;
			\If{Virtual machine reports confinement error}
				{continue to next model}
			}
		}
		Operate an epilogue for $m$\;
	}
	\caption{Actions performed for each tool by the invocation script
  \label{algo:mainscript}}
\end{algorithm}

The ``examinations'' requested for the contest were performed thanks to an
invocation script that iterated invocation of each tool over models instances.
This invocation script, adapted from the one of the first edition, is
presented in~\Cref{algo:mainscript}. The major improvement is to stop
executing the tool as soon as one instance fails. Another concern (that makes
the script more complex without changing its principle) was to increase the
use of parallelism over the nodes of the cluster.

\clearpage
\section{Raw Data for the State Space Generation}
\label{sec:statespace}
\index{State Space!Tool results (summary)}
\begin{table}[h]
	\centering
  \lstset{
    basicstyle=\scriptsize\fontfamily{fvm}\selectfont
  }
  \footnotesize
  \renewcommand\cellalign{cc}
  \setlength\rotheadsize{5em}
  \hspace{-1em}
  \begin{tabular}{|c|c||m{5em}|m{3.5em}|m{3.5em}|m{6em}|m{2.5em}|m{2.5em}|m{5em}|m{5em}|m{3.5em}|m{1.5em}|}
    \cline{3-12}
    \multicolumn{2}{c|}{}
  & \rothead{\textbf{AlPiNA}}
  & \rothead{\textbf{Crocodile}}
  & \rothead{\textbf{Helena}}
  & \rothead{\textbf{ITS-Tools}}
  & \rothead{\textbf{LoLA (binstore)}}
  & \rothead{\textbf{LoLA (bloom)}}
  & \rothead{\textbf{Marcie}}
  & \rothead{\textbf{Neco}}
  & \rothead{\textbf{PNXDD}}
  & \rothead{\textbf{Sara}}
    \lstset{
      basicstyle=\footnotesize\fontfamily{fvm}\selectfont
    }
  \\
  \hline
  \hline
    \multirow{7}{*}{\begin{sideways}\textbf{MCC 2011 models}\end{sideways}}
  & \textbf{FMS}
  & \result[reached=some,typepn=pt]{20} 
  & \result[reached=nc]{} 
  & \result[reached=nc]{} 
  & \result[reached=best,typepn=pt]{100} 
  & \result[reached=nc]{} 
  & \result[reached=nc]{} 
  & \result[reached=best,typepn=pt]{100} 
  & \result[reached=nc]{} 
  & \result[reached=some,typepn=pt]{50} 
  & \result[reached=nc]{} 
  \\
  & \textbf{Kanban}
  & \result[reached=some,typepn=pt]{20} 
  & \result[reached=nc]{} 
  & \result[reached=nc]{} 
  & \result[reached=max,typepn=pt]{1\,000} 
  & \result[reached=nc]{} 
  & \result[reached=nc]{} 
  & \result[reached=some,typepn=pt]{100} 
  & \result[reached=some,typepn=pt]{5} 
  & \result[reached=some,typepn=pt]{50} 
  & \result[reached=nc]{} 
  \\
  & \textbf{MAPK}
  & \result[reached=none]{} 
  & \result[reached=nc]{} 
  & \result[reached=nc]{} 
  & \result[reached=best,typepn=pt]{80} 
  & \result[reached=nc]{} 
  & \result[reached=nc]{} 
  & \result[reached=best,typepn=pt]{80} 
  & \result[reached=best,typepn=pt]{8} 
  & \result[reached=best,typepn=pt]{80} 
  & \result[reached=nc]{} 
  \\
  & \textbf{Peterson}
  & \result[reached=some,typepn=cn]{3} 
  & \result[reached=nc]{} 
  & \result[reached=none]{} 
  & \result[reached=nc]{} 
  & \result[reached=nc]{} 
  & \result[reached=nc]{} 
  & \result[reached=some,typepn=pt]{2} 
  & \result[reached=some,typepn=pt]{2} 
  & \result[reached=best,typepn=pt]{4} 
  & \result[reached=nc]{} 
  \\
  & \textbf{Philosophers}
  & \result[reached=some,typepn=cn]{500} 
  & \result[reached=nc]{} 
  & \result[reached=some,typepn=cn]{20} 
  & \result[reached=max,typepn=pt]{100\,000} 
  & \result[reached=nc]{} 
  & \result[reached=nc]{} 
  & \result[reached=some,typepn=pt]{2000} 
  & \result[reached=nc]{} 
  & \result[reached=some,typepn=pt]{100} 
  & \result[reached=nc]{} 
  \\
  & \textbf{Shared-Memory}
  & \result[reached=some,typepn=cn]{20} 
  & \result[reached=some,typepn=cn]{100} 
  & \result[reached=some,typepn=cn]{20} 
  & \result[reached=some,typepn=pt]{20} 
  & \result[reached=nc]{} 
  & \result[reached=nc]{} 
  & \result[reached=best,typepn=pt]{200} 
  & \result[reached=nc]{} 
  & \result[reached=some,typepn=pt]{20} 
  & \result[reached=nc]{} 
  \\
  & \textbf{TokenRing}
  & \result[reached=some,typepn=cn]{5} 
  & \result[reached=nc]{} 
  & \result[reached=best,typepn=cn]{20} 
  & \result[reached=best,typepn=pt]{20} 
  & \result[reached=nc]{} 
  & \result[reached=nc]{} 
  & \result[reached=best,typepn=pt]{20} 
  & \result[reached=nc]{} 
  & \result[reached=some,typepn=pt]{10} 
  & \result[reached=nc]{} 
  \\
  \hline
  \hline
    \multirow{12}{*}{\begin{sideways}\textbf{new models from MCC 2012}\end{sideways}}
  & \textbf{Cs\_repetitions}
  & \result[reached=none,typepn=no]{} 
  & \result[reached=none,typepn=no]{} 
  & \result[reached=nc]{} 
  & \result[reached=nc]{} 
  & \result[reached=nc]{} 
  & \result[reached=nc]{} 
  & \result[reached=none,typepn=no]{} 
  & \result[reached=nc]{} 
  & \result[reached=none,typepn=no]{} 
  & \result[reached=nc]{} 
  \\
  & \textbf{Echo}
  & \result[reached=none,typepn=no]{} 
  & \result[reached=nc]{} 
  & \result[reached=nc]{} 
  & \result[reached=none,typepn=no]{} 
  & \result[reached=nc]{} 
  & \result[reached=nc]{} 
  & \result[reached=some,typepn=pt]{d2r11} 
  & \result[reached=nc]{} 
  & \result[reached=nc]{} 
  & \result[reached=nc]{} 
  \\
  & \textbf{Eratosthenes}
  & \result[reached=max,typepn=cn]{500} 
  & \result[reached=nc]{} 
  & \result[reached=none]{} 
  & \result[reached=max,typepn=pt]{500} 
  & \result[reached=nc]{} 
  & \result[reached=nc]{} 
  & \result[reached=max,typepn=pt]{500} 
  & \result[reached=some,typepn=cn]{20} 
  & \result[reached=nc]{} 
  & \result[reached=nc]{} 
  \\
  & \textbf{Galloc\_res}
  & \result[reached=none]{} 
  & \result[reached=best,typepn=cn]{3} 
  & \result[reached=nc]{} 
  & \result[reached=best,typepn=pt]{3} 
  & \result[reached=nc]{} 
  & \result[reached=nc]{} 
  & \result[reached=best,typepn=pt]{3} 
  & \result[reached=nc]{} 
  & \result[reached=best,typepn=pt]{3} 
  & \result[reached=nc]{} 
  \\
  & \textbf{Lamport\_fmea}
  & \result[reached=some,typepn=cn]{3} 
  & \result[reached=nc]{} 
  & \result[reached=none]{} 
  & \result[reached=best,typepn=pt]{5} 
  & \result[reached=nc]{} 
  & \result[reached=nc]{} 
  & \result[reached=some,typepn=pt]{3} 
  & \result[reached=nc]{} 
  & \result[reached=nc]{} 
  & \result[reached=nc]{} 
  \\
  & \textbf{NEO-election}
  & \result[reached=none]{} 
  & \result[reached=none]{} 
  & \result[reached=none]{} 
  & \result[reached=none]{} 
  & \result[reached=nc]{} 
  & \result[reached=nc]{} 
  & \result[reached=none]{} 
  & \result[reached=nc]{} 
  & \result[reached=nc]{} 
  & \result[reached=nc]{} 
  \\
  & \textbf{Philo\_dyn}
  & \result[reached=some,typepn=cn]{3} 
  & \result[reached=nc]{} 
  & \result[reached=best,typepn=cn]{50} 
  & \result[reached=some,typepn=pt]{3} 
  & \result[reached=nc]{} 
  & \result[reached=nc]{} 
  & \result[reached=some,typepn=pt]{3} 
  & \result[reached=some,typepn=pt]{3} 
  & \result[reached=some,typepn=pt]{3} 
  & \result[reached=nc]{} 
  \\
  & \textbf{Planning}
  & \result[reached=none]{} 
  & \result[reached=nc]{} 
  & \result[reached=nc]{} 
  & \result[reached=none]{} 
  & \result[reached=nc]{} 
  & \result[reached=nc]{} 
  & \result[reached=none]{} 
  & \result[reached=nc]{} 
  & \result[reached=nc]{} 
  & \result[reached=nc]{} 
  \\
  & \textbf{Railroad}
  & \result[reached=some,typepn=cn]{5} 
  & \result[reached=nc]{} 
  & \result[reached=nc]{} 
  & \result[reached=best,typepn=pt]{10} 
  & \result[reached=nc]{} 
  & \result[reached=nc]{} 
  & \result[reached=best,typepn=pt]{10} 
  & \result[reached=nc]{} 
  & \result[reached=nc]{} 
  & \result[reached=nc]{} 
   \\
  & \textbf{Ring}
  & \result[reached=none]{} 
  & \result[reached=nc]{} 
  & \result[reached=nc]{} 
  & \result[reached=max,typepn=pt]{--} 
  & \result[reached=nc]{} 
  & \result[reached=nc]{} 
  & \result[reached=max,typepn=pt]{--} 
  & \result[reached=nc]{} 
  & \result[reached=nc]{} 
  & \result[reached=nc]{} 
  \\
  & \textbf{Rw\_mutex}
  & \result[reached=some,typepn=pt]{r10w100} 
  & \result[reached=nc]{} 
  & \result[reached=nc]{} 
  & \result[reached=best,typepn=pt]{r10w2000} 
  & \result[reached=nc]{} 
  & \result[reached=nc]{} 
  & \result[reached=some,typepn=pt]{r10w100} 
  & \result[reached=some,typepn=pt]{r10w100} 
  & \result[reached=nc]{} 
  & \result[reached=nc]{} 
  \\
  & \textbf{Simple\_lbs}
  & \result[reached=some,typepn=pt]{2} 
  & \result[reached=nc]{} 
  & \result[reached=nc]{} 
  & \result[reached=max,typepn=pt]{20} 
  & \result[reached=nc]{} 
  & \result[reached=nc]{} 
  & \result[reached=some,typepn=pt]{5} 
  & \result[reached=some,typepn=pt]{5} 
  & \result[reached=max,typepn=pt]{20} 
  & \result[reached=nc]{} 
  \\
  \hline
  \end{tabular}
  \caption{Results for the state space generation examination}
  \label{tab:rg}
\end{table}

This section shows the raw results of the state space generation examination.
\Cref{tab:rg}~summarizes the highest scaling parameter reached by the tools
for each model. Then Charts generated from the data collected for this
examination are provided.

This table, as well as~\Cref{tab:fs,tab:fr},
should be interpreted using the
legend below (a value shows the maximum scaling parameter and, on the
rightmost part of the cell, the type of Petri net used by the tool is
indicated -- P for P/T nets or C for Colored ones):

\label{position:legend}\legend

\medskip
\Cref{sec:sttatespacesizes}~presents the data computed by tools.
\Cref{sec:ssprocessedmodels}~shows how models have been handled
by tools and
\Cref{sec:ss:bymodels,sec:ss:bytools}~summarize how tools did cope
with models. 

Then, \Cref{sec:ssmodechart}~presents the charts showing the evolution of
memory and CPU according to the scaling parameter (if any). Only models
handled by at least one tool are displayed.

Finally, \Crefrange{sec:sstoolfirst}{sec:sstoollast} show the
evolution of memory and CPU consumption while tools are performing
the state space generation for a model.

\medskip\noindent\textbf{No comment is provided intentionally, the objective
of this document mainly being to report all the data we collected this year.}

\subsection{Computed sizes for the State Spaces}
\label{sec:sttatespacesizes}
\index{State Space!Computed data}

\begin{table}
\centering
\begin{tabular}{|c|c||c|c||c|c||c|c|}
	\hline
	\textbf{scale} & \textbf{|S|} & \textbf{scale} & \textbf{|S|} & \textbf{scale} & \textbf{|S|} & \textbf{scale} & \textbf{|S|}\\
	\hline
	\hline
	\multicolumn{8}{|c|}{\textbf{cs\_repetitions}}\\
	25 & ? & \cellcolor{gray!50} & \cellcolor{gray!50} & \cellcolor{gray!50} & \cellcolor{gray!50} & \cellcolor{gray!50} & \cellcolor{gray!50}\\
	\hline
	\multicolumn{8}{|c|}{\textbf{echo}}\\
	d2r11 & 9\,615 & d2r13 & ? & \cellcolor{gray!50} & \cellcolor{gray!50} & \cellcolor{gray!50} & \cellcolor{gray!50}\\
	\hline
	\multicolumn{8}{|c|}{\textbf{eratosthenes}}\\
	5 & 2 & 20 & 2\,048 & 100 & 1.889$\times10^{22}$ & 500 & 4.132$\times10^{121}$\\
	10 & 32 & 50 & 1.718$\times10^{10}$ & 200 & 1.142$\times10^{46}$ & \cellcolor{gray!50} & \cellcolor{gray!50} \\
	\hline
	\multicolumn{8}{|c|}{\textbf{FMS}}\\
	2 & 3\,444 & 10 & 2.501$\times10^{9}$ & 50 & 4.240$\times10^{17}$ & 200 & ?\\
	5 & 2.895$\times10^{6}$ & 20 & 6.029$\times10^{12}$ & 100 & 2.703$\times10^{21}$ & \cellcolor{gray!50} & \cellcolor{gray!50} \\
	\hline
	\multicolumn{8}{|c|}{\textbf{galloc\_res}}\\
	3 & 6\,320 & 5 & ? & \cellcolor{gray!50} & \cellcolor{gray!50} & \cellcolor{gray!50} & \cellcolor{gray!50} \\
	\hline
	\multicolumn{8}{|c|}{\textbf{Kanban}}\\
	5 & 2.546$\times10^{6}$ & 20 & 8.054$\times10^{11}$ & 100 & 1.726$\times10^{19}$ & 500 & 7.086$\times10^{26}$\\
	10 & 1.006$\times10^{9}$ & 50 & 1.043$\times10^{16}$ & 200 & 3.173$\times10^{22}$ & 1000 & 1.420$\times10^{30}$\\
	\hline
	\multicolumn{8}{|c|}{\textbf{lamport\_fmea}}\\
	2 & 380 & 4 & 1.915$\times10^{6}$ & 6 & ? & \cellcolor{gray!50} & \cellcolor{gray!50} \\
	3 & 19\,742 & 5 & 5.307$\times10^{8}$ & \cellcolor{gray!50} & \cellcolor{gray!50} & \cellcolor{gray!50} & \cellcolor{gray!50} \\
	\hline
	\multicolumn{8}{|c|}{\textbf{MAPK}}\\
	8 & 6.111$\times10^{6}$ & 40 & 4.783$\times10^{14}$ & 160 & ? & \cellcolor{gray!50} & \cellcolor{gray!50} \\
	20 & 8.813$\times10^{10}$ & 80 & 5.635$\times10^{18}$ & \cellcolor{gray!50} & \cellcolor{gray!50} & \cellcolor{gray!50} & \cellcolor{gray!50} \\
	\hline
	\multicolumn{8}{|c|}{\textbf{neo-election}}\\
	2 & ? & \cellcolor{gray!50} & \cellcolor{gray!50} & \cellcolor{gray!50} & \cellcolor{gray!50} & \cellcolor{gray!50} & \cellcolor{gray!50} \\
	\hline
	\multicolumn{8}{|c|}{\textbf{Peterson}}\\
	2 & 20\,754 & 3 & 3.408$\times10^{6}$ & 4 & 6.299$\times10^{8}$ & 5 & ?\\
	\hline
	\multicolumn{8}{|c|}{\textbf{philo\_dyn}}\\
	2 & 27 & 10 & unsafe$^{(1)}$ & 50 & unsafe$^{(1)}$ & \cellcolor{gray!50} & \cellcolor{gray!50} \\
	3 & 7\,251 & 20 & unsafe$^{(1)}$ & 80 & ? & \cellcolor{gray!50} & \cellcolor{gray!50} \\
	\hline
	\multicolumn{8}{|c|}{\textbf{Philosophers}}\\
	5 & 243 & 50 & 7.179$\times10^{23}$ & 500 & 3.640$\times10^{238}$ & 5000 & 4.039$\times10^{2385}$\\
	10 & 59\,049 & 100 & 5.150$\times10^{47}$ & 1000 & 1.322$\times10^{477}$ & 10000 & 1.631$\times10^{4771}$\\
	20 & 3.487$\times10^{9}$ & 200 & 2.660$\times10^{95}$ & 2000 & 1.748$\times10^{954}$ & $\geq$50000 & $\infty$\\
	\hline
	\multicolumn{8}{|c|}{\textbf{planning}}\\
	--- & ? & \cellcolor{gray!50} & \cellcolor{gray!50} & \cellcolor{gray!50} & \cellcolor{gray!50} & \cellcolor{gray!50} & \cellcolor{gray!50} \\
	\hline
	\multicolumn{8}{|c|}{\textbf{railroad}}\\
	5 & 1\,838 & 10 & 2.038$\times10^{6}$ & 20 & ? & \cellcolor{gray!50} & \cellcolor{gray!50} \\
	\hline
	\multicolumn{8}{|c|}{\textbf{ring}}\\
	--- & unsafe$^{(2)}$ & \cellcolor{gray!50} & \cellcolor{gray!50} & \cellcolor{gray!50} & \cellcolor{gray!50} & \cellcolor{gray!50} & \cellcolor{gray!50} \\
	\hline
	\multicolumn{8}{|c|}{\textbf{rwmutex}}\\
	r10w10 & 1\,034 & r10w50 & 1\,074 & r10w500 & 1\,524 & r10w2000 & 3\,024\\
	r10w20 & 1\,044 & r10w100 & 1\,124 & r10w1000 & 2\,024 & r20w10 & ?\\
	\hline
	\multicolumn{8}{|c|}{\textbf{SharedMemory}}\\
	5 & 1\,863 & 20 & 4.451$\times10^{11}$ & 100 & unsafe$^{(3)}$ & 500 & unsafe$^{(3)}$\\
	10 & 1.831$\times10^{6}$ & 50 & unsafe$^{(3)}$ & 200 & unsafe$^{(3)}$ & \cellcolor{gray!50} & \cellcolor{gray!50} \\
	\hline
	\multicolumn{8}{|c|}{\textbf{simple\_lbs}}\\
	2 & 832 & 10 & 4.060$\times10^{8}$ & 20 & 4.583$\times10^{15}$ & \cellcolor{gray!50} & \cellcolor{gray!50} \\
	5 & 116\,176 & 15 & 1.374$\times10^{12}$ & \cellcolor{gray!50} & \cellcolor{gray!50} & \cellcolor{gray!50} & \cellcolor{gray!50} \\
	\hline
	\multicolumn{8}{|c|}{\textbf{TokenRing}}\\
	5 & 166 & 10 & 58\,905 & 20 & 2.447$\times10^{10}$ & 50 & ? \\
\hline
\end{tabular}
\caption{Size of computed state spaces\label{tab:result:RG}}
\end{table}

This section presents the data computed by tools for the various state spaces
that were proposed. This information is summarized in~\Cref{tab:result:RG}: columns ``scale'' shows the value of the scaling
parameter and the column ``|S|'' aside shows the state space size
corresponding to this value. ``?'' are displayed when the value is unknown
(i.e. no tool could compute the value). In one situation, we display $\infty$
since it was reported that the big numbers C++ library fails to compute the
value.

In some cases, the value returned by tools differ. For \acs{Crocodile}, this is
normal because it counts symbolic states (i.e. the ones of the quotient graph)
in the sense of~\cite{chiola97symbolic} while other tools count explicit
states.

This also concerns other tools but it can be due to the use of specific
techniques preventing the computation of some ``useless states'' like the use
of partial order techniques. So we only provide the values when tools agree or
when only one tool reaches a value, when the values that are also computed by
some other tools are confirmed.

We thus have three ``unsafe'' situations remaining where we prefer not to
display any value:

\begin{enumerate}

\item \acs{Dynamic-Philosophers}: for the two first values of the scaling parameter,
\acs{PNXDD}, \acs{Marcie} and \acs{ITS-Tools} agree on the size of the state
space (\acs{AlPiNA} and \acs{Helena} do not) ;
but, for higher values, only \acs{Helena} is able to compute the state
space.

\item \acs{Ring}: here, only \acs{ITS-Tools} and \acs{Marcie} are able to cope with this model
but they disagree.

\item \acs{Shared-Memory}: for the values $\geq 50$, only \acs{Crocodile}
and \acs{Marcie} are
able to build the state space. However, \acs{Marcie} disagree with other tools on
the first results and \acs{Crocodile} does not count the same type of states. 

\end{enumerate}

\subsection{Processed Models}
\label{sec:ssprocessedmodels}
\index{State Space!Processed models}

This section summarizes how models were processed by tools.
Let us first note that no tool succeeded in this examination for the following
models:
\begin{itemize} 
\item \acs{CS-Repetitions},
\item \acs{Neo-Election}, 
\item \acs{Planning}.
\end{itemize} 

These models constitute challenges for the next edition.

\subsection{Radars by models}
\label{sec:ss:bymodels}

\begin{figure}[p]
\centering
\begin{adjustwidth}{-2em}{-2em}
\noindent
\includegraphics[scale=.35]{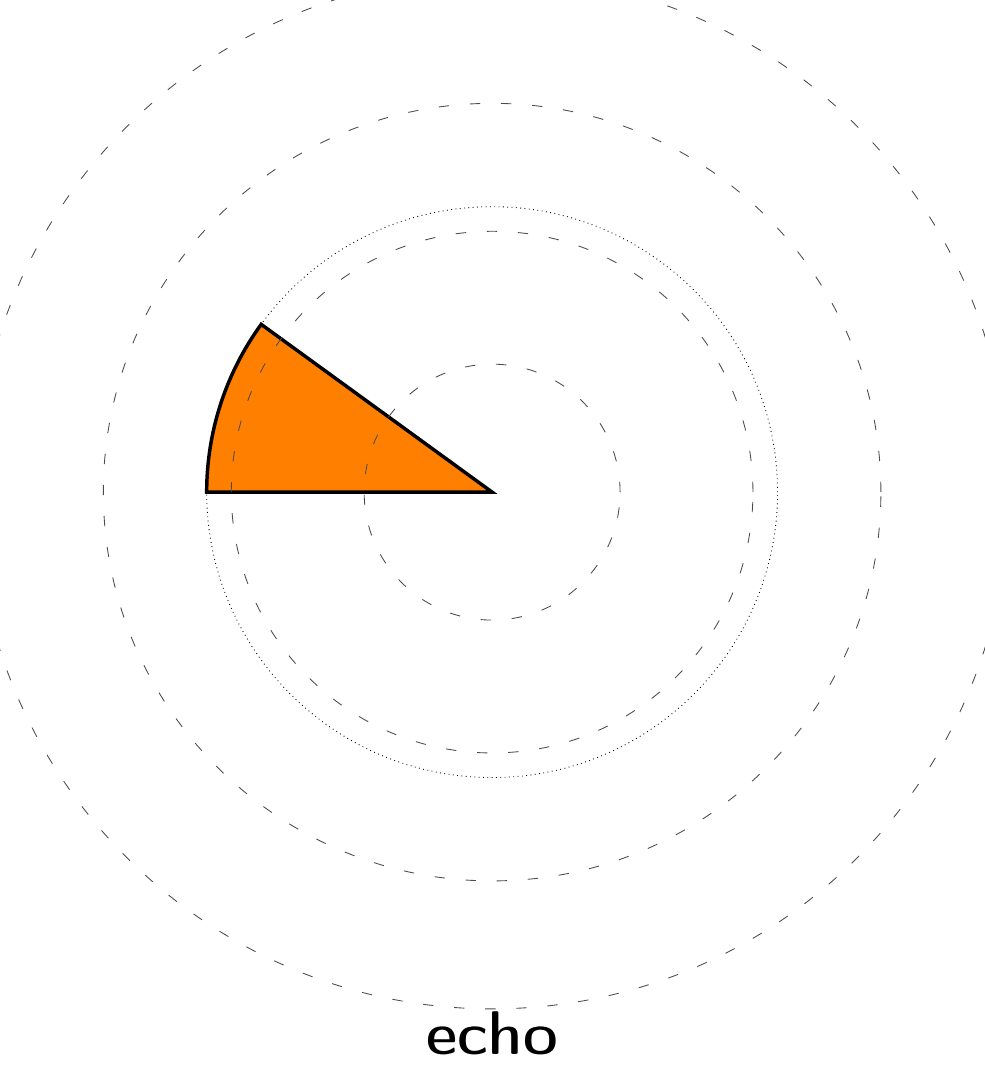}
\hfill
\includegraphics[scale=.35]{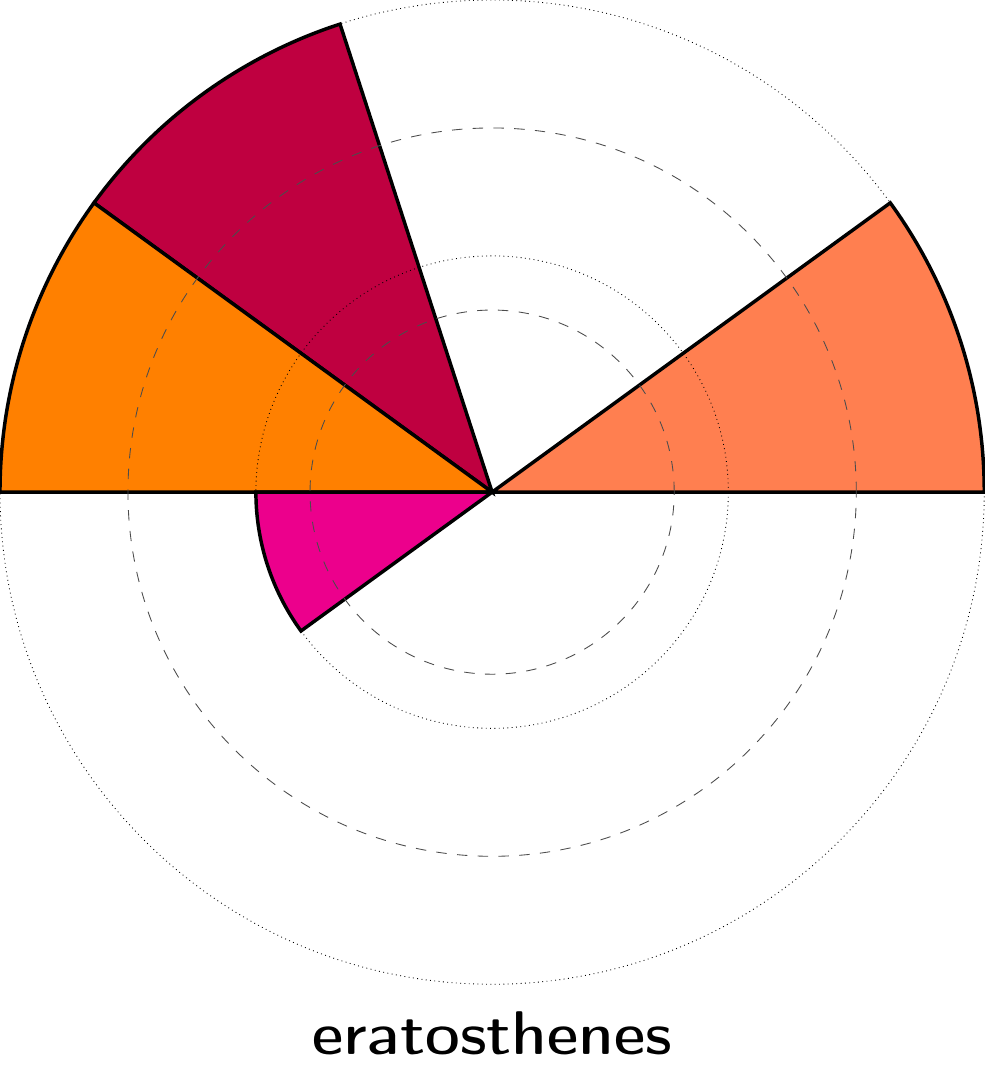}
\hfill
\includegraphics[scale=.35]{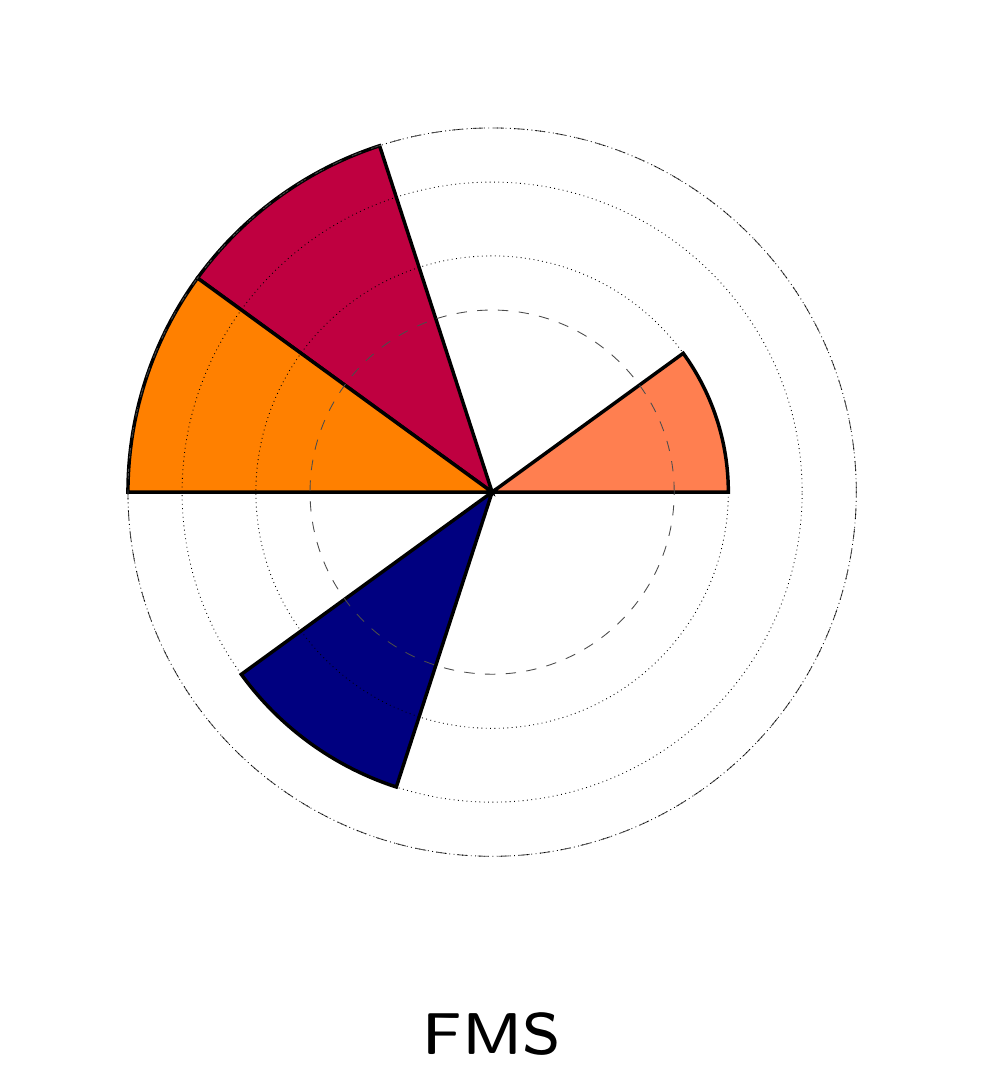}
\hfill
\includegraphics[scale=.35]{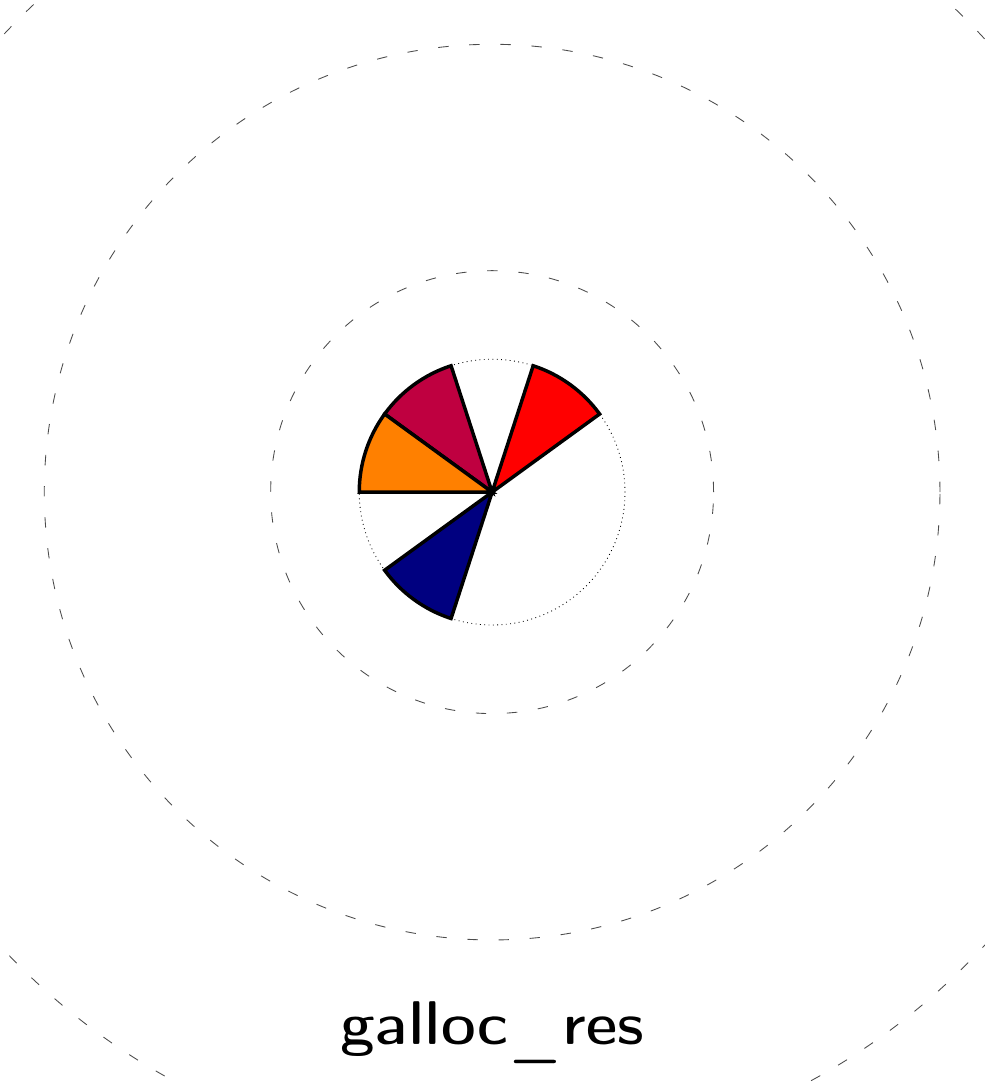}
\\
\medskip
\includegraphics[scale=.35]{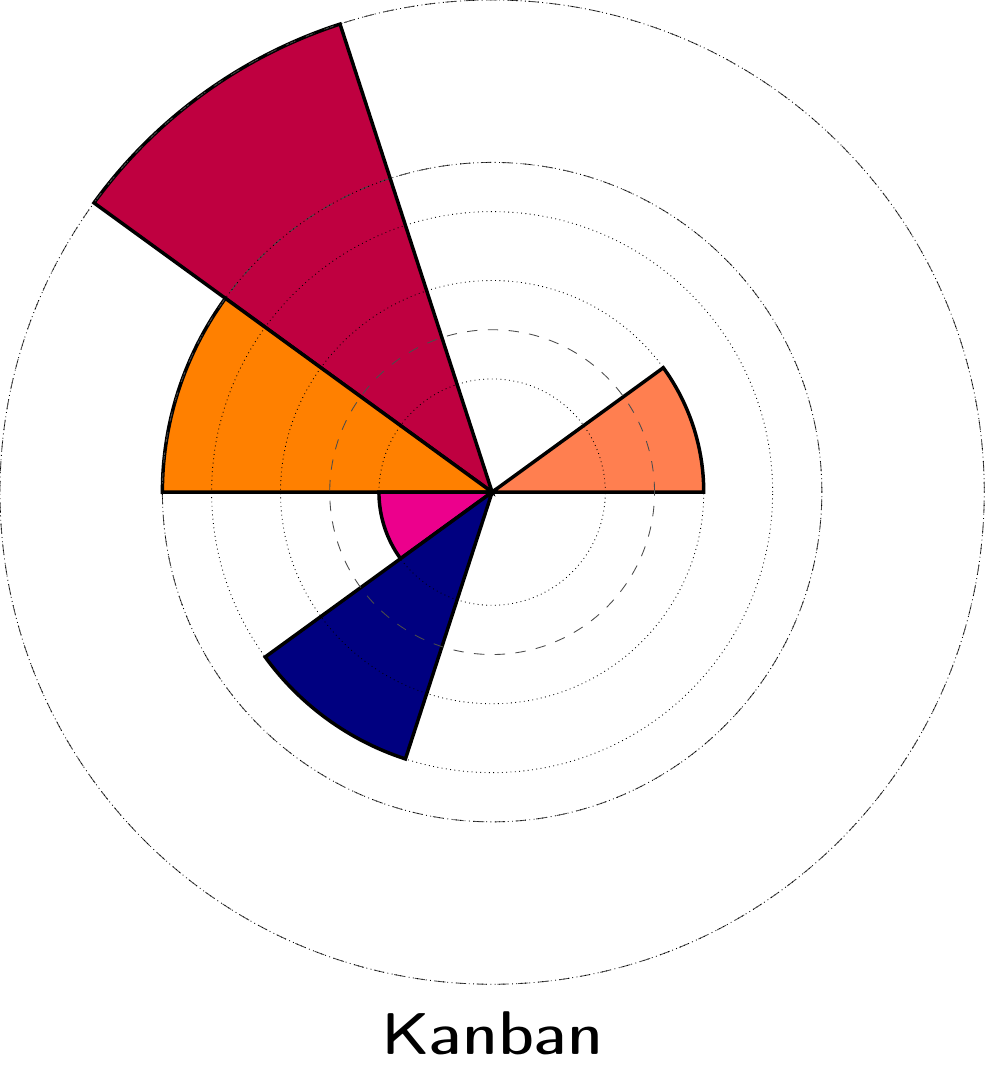}
\hfill
\includegraphics[scale=.35]{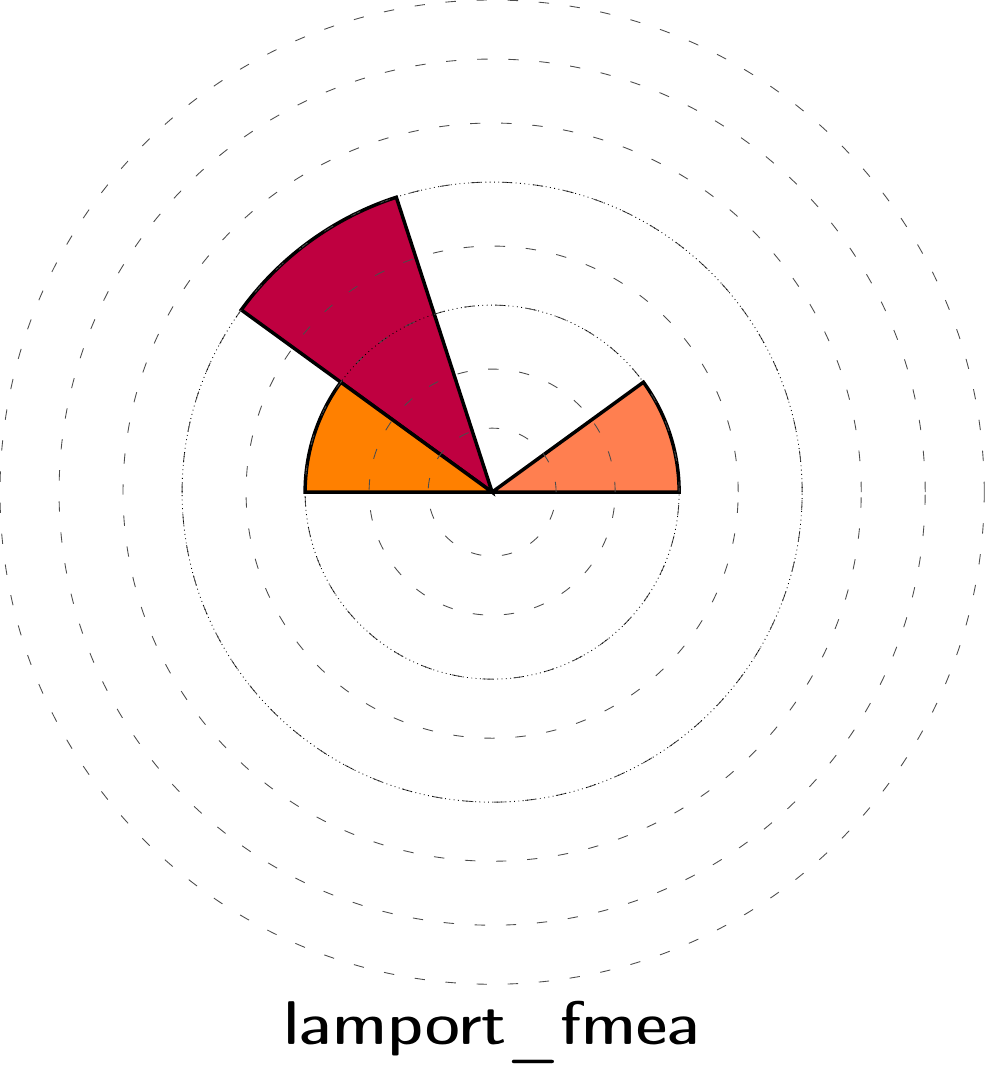}
\hfill
\includegraphics[scale=.35]{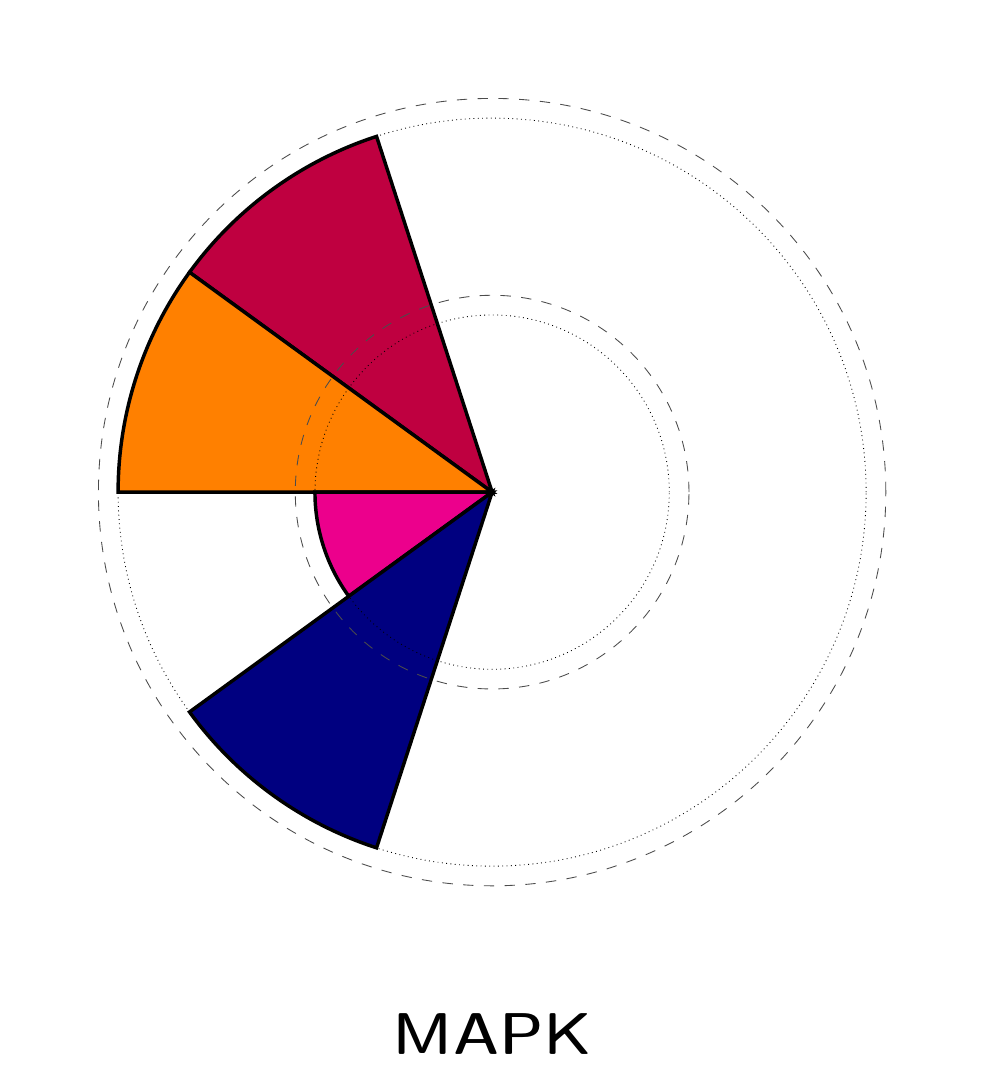}
\hfill
\includegraphics[scale=.35]{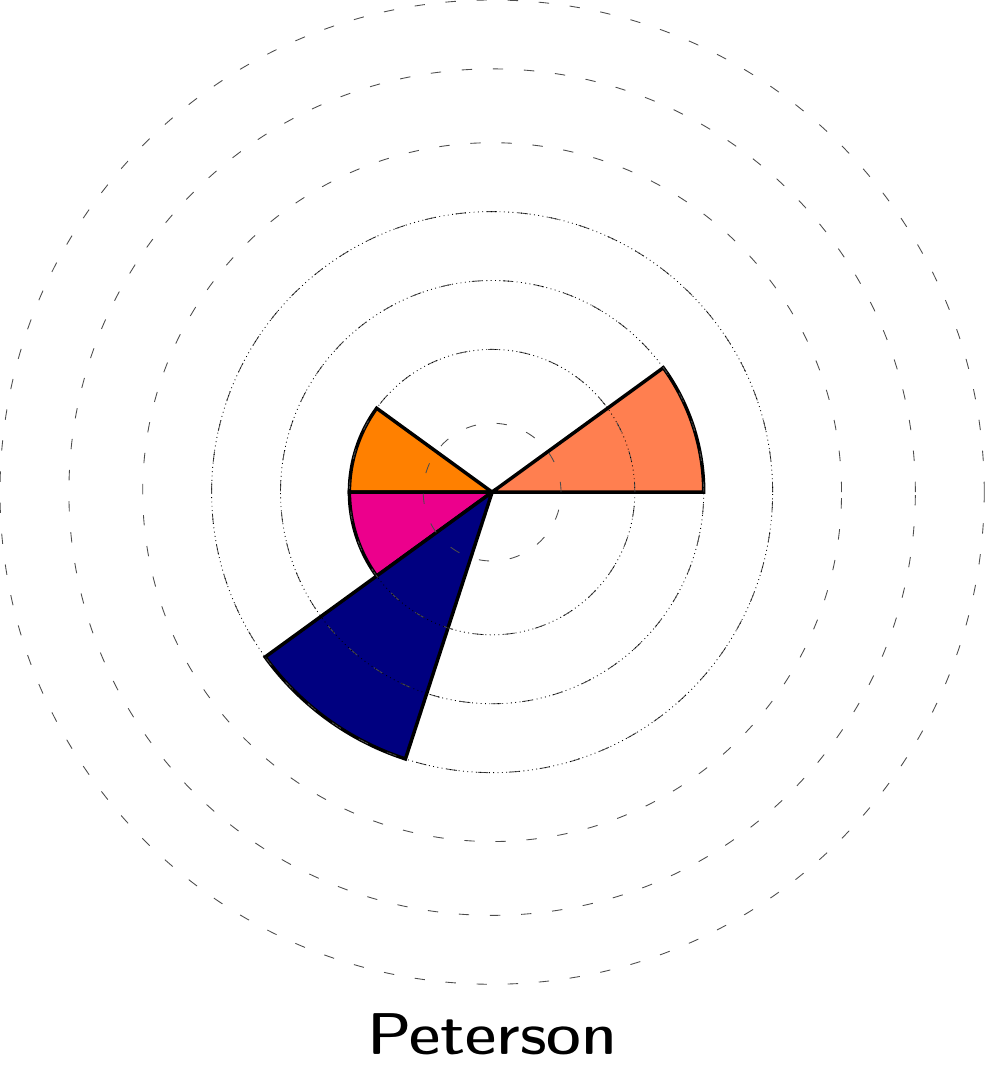}
\\
\medskip
\includegraphics[scale=.35]{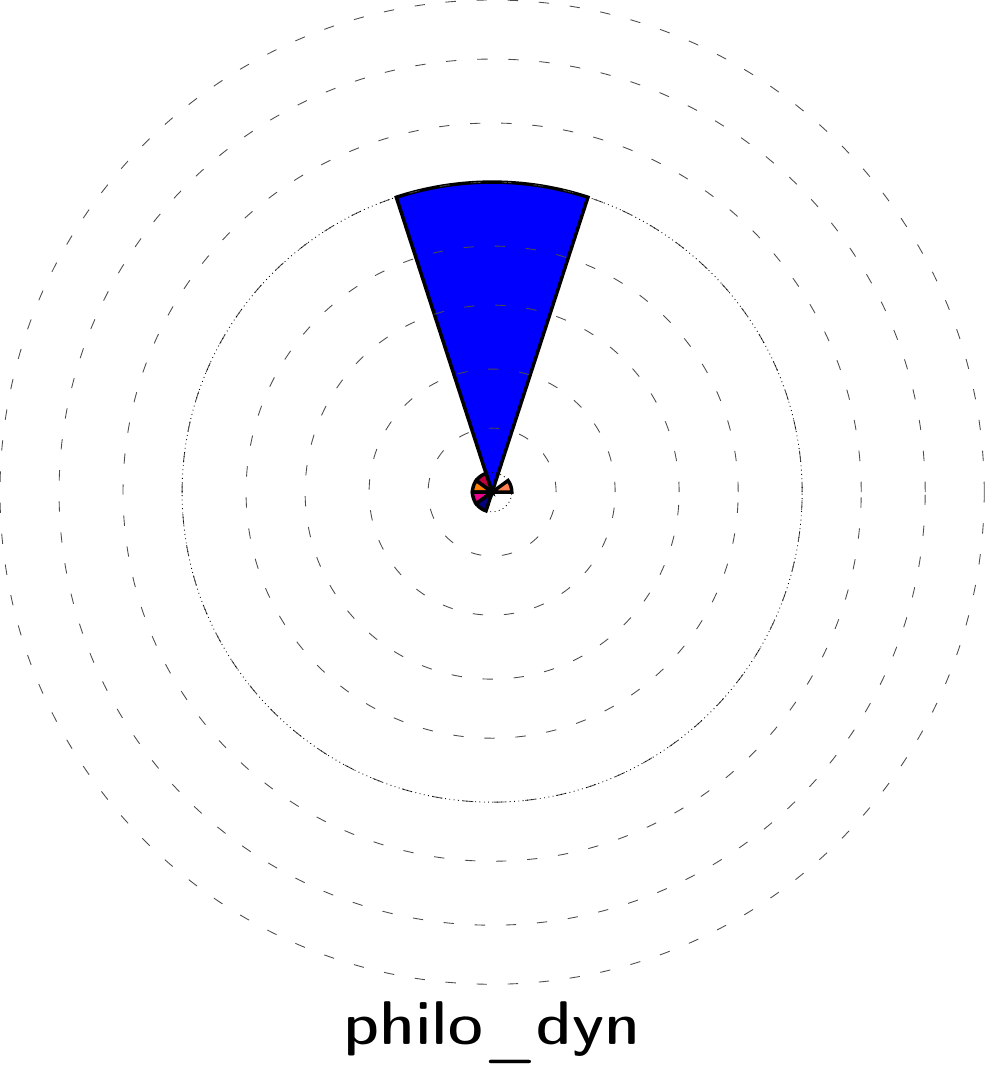}
\hfill
\includegraphics[scale=.35]{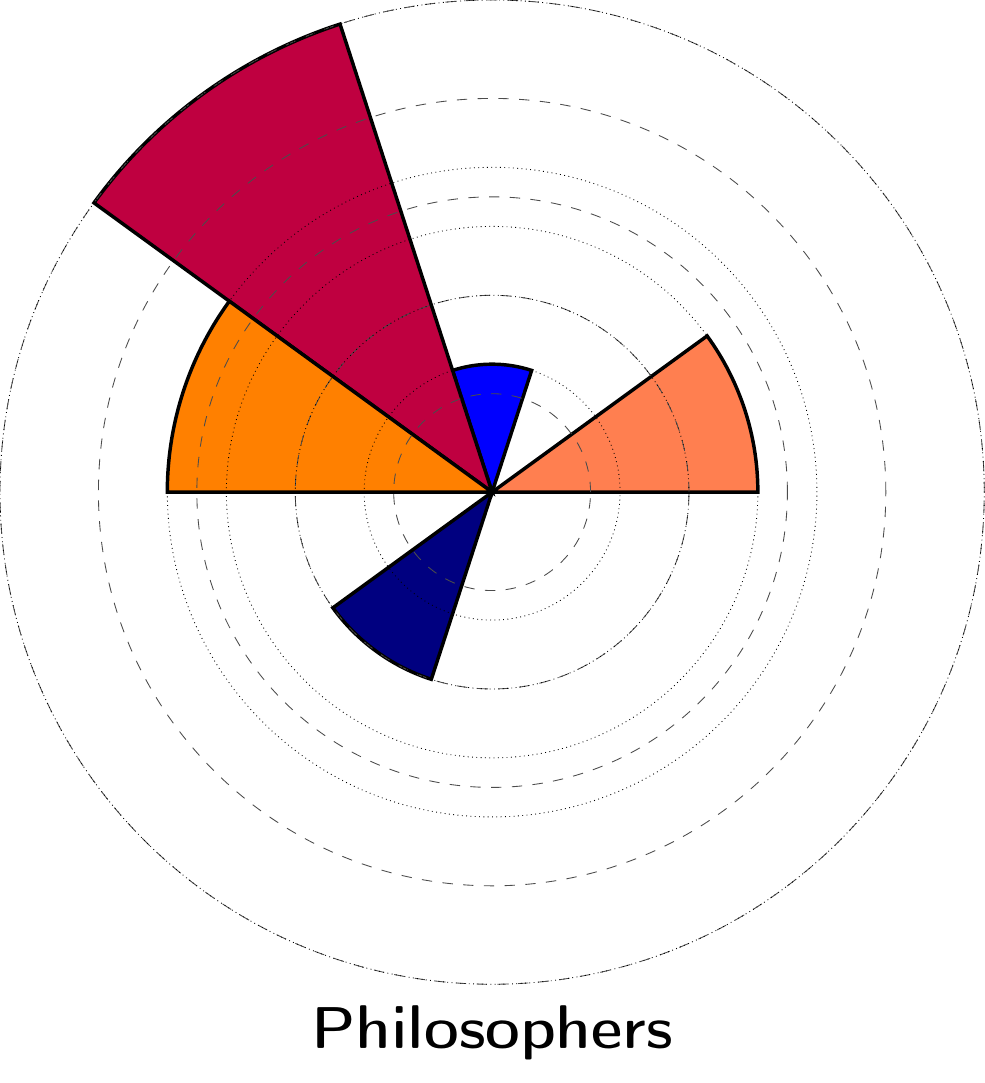}
\hfill
\includegraphics[scale=.35]{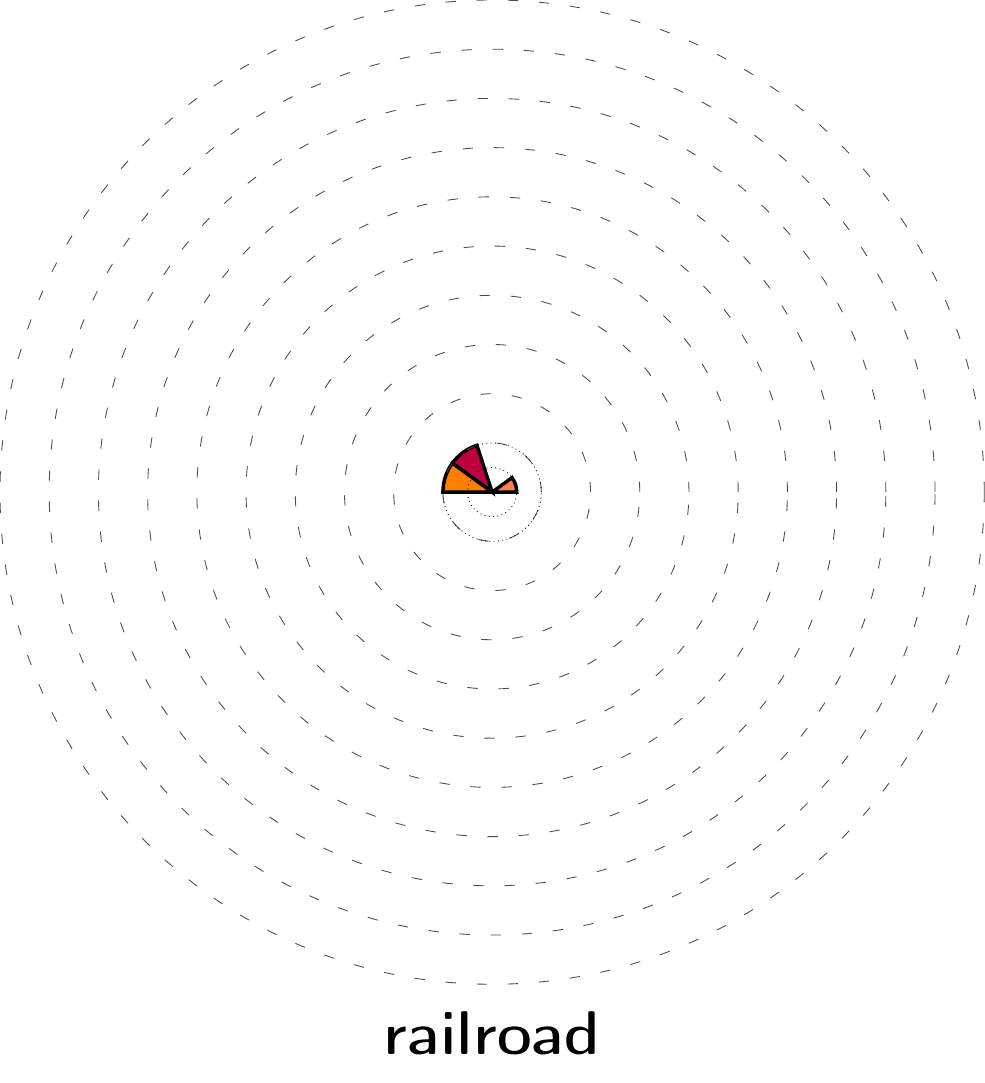}
\hfill
\includegraphics[scale=.35]{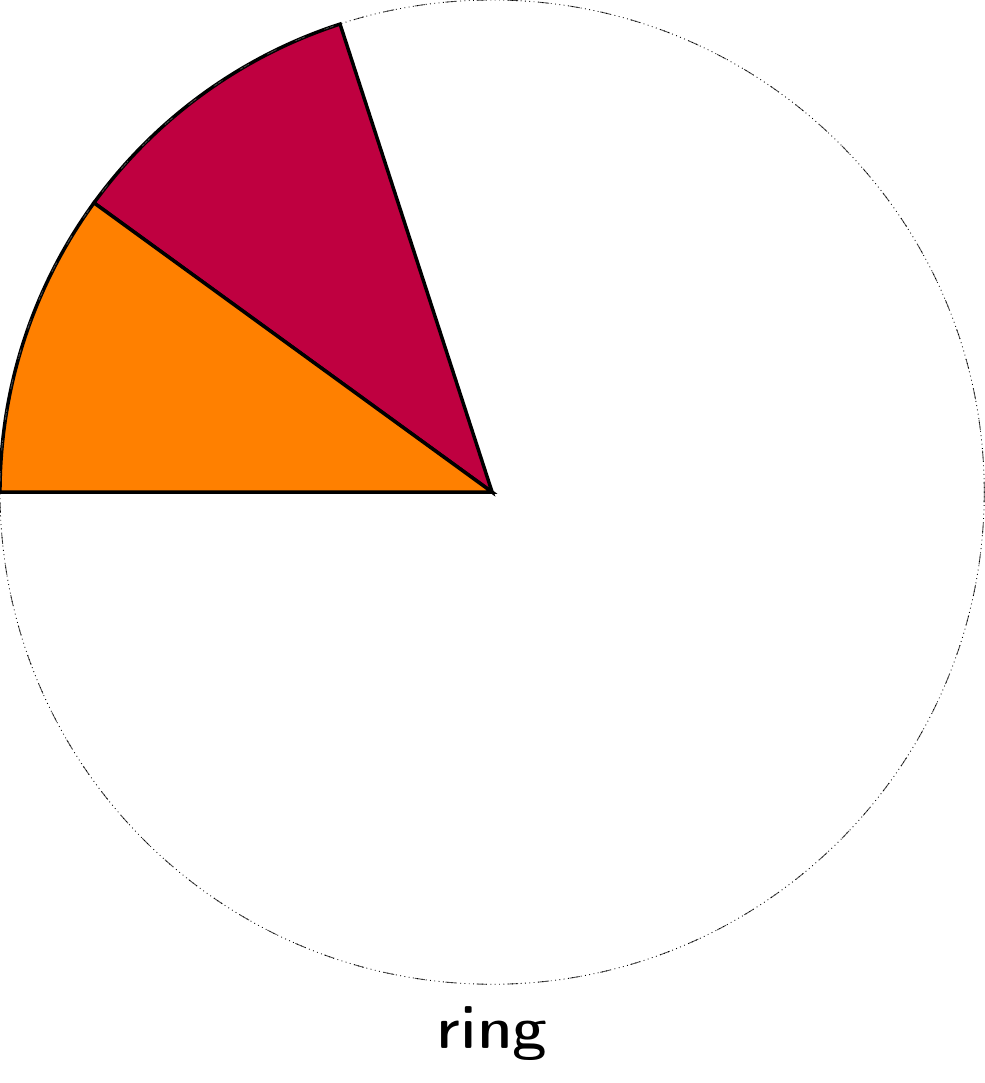}
\\
\medskip
\includegraphics[scale=.35]{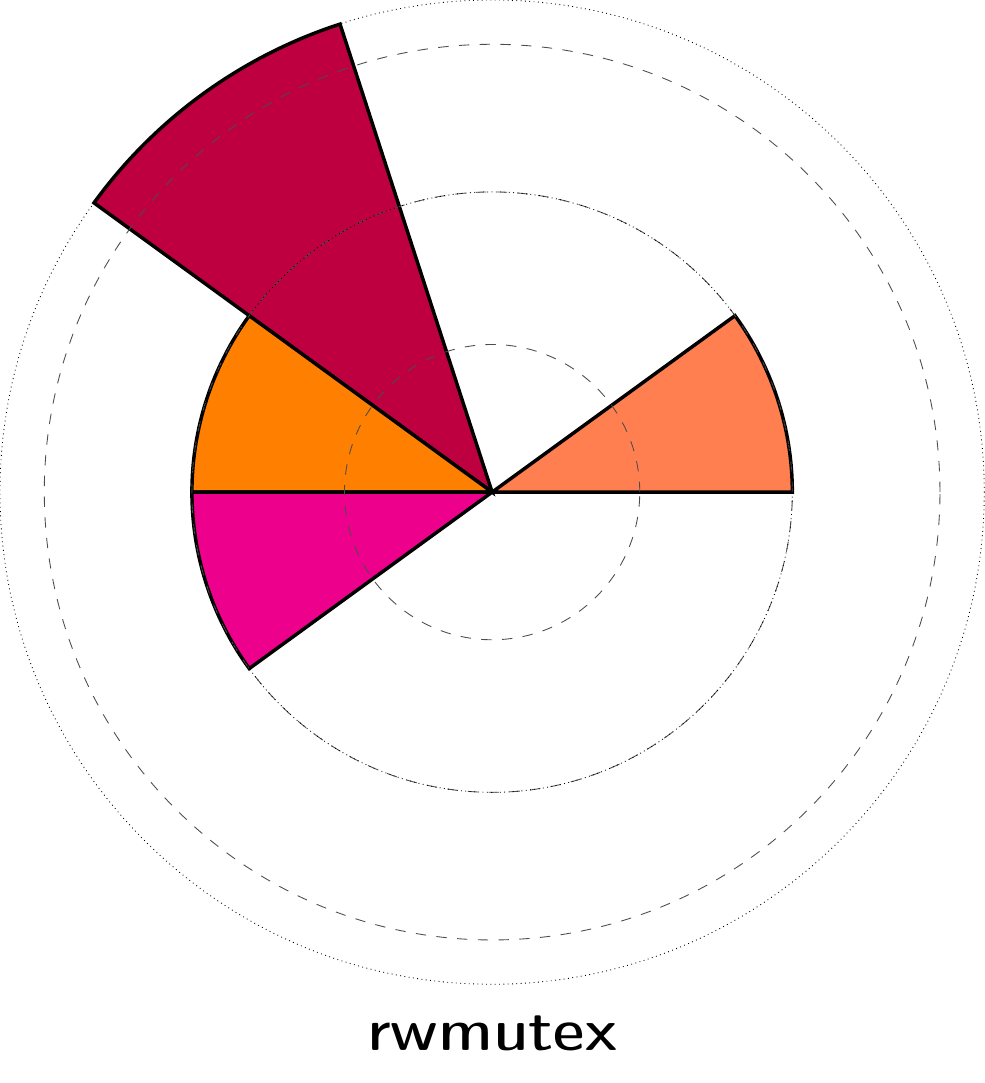}
\hfill
\includegraphics[scale=.35]{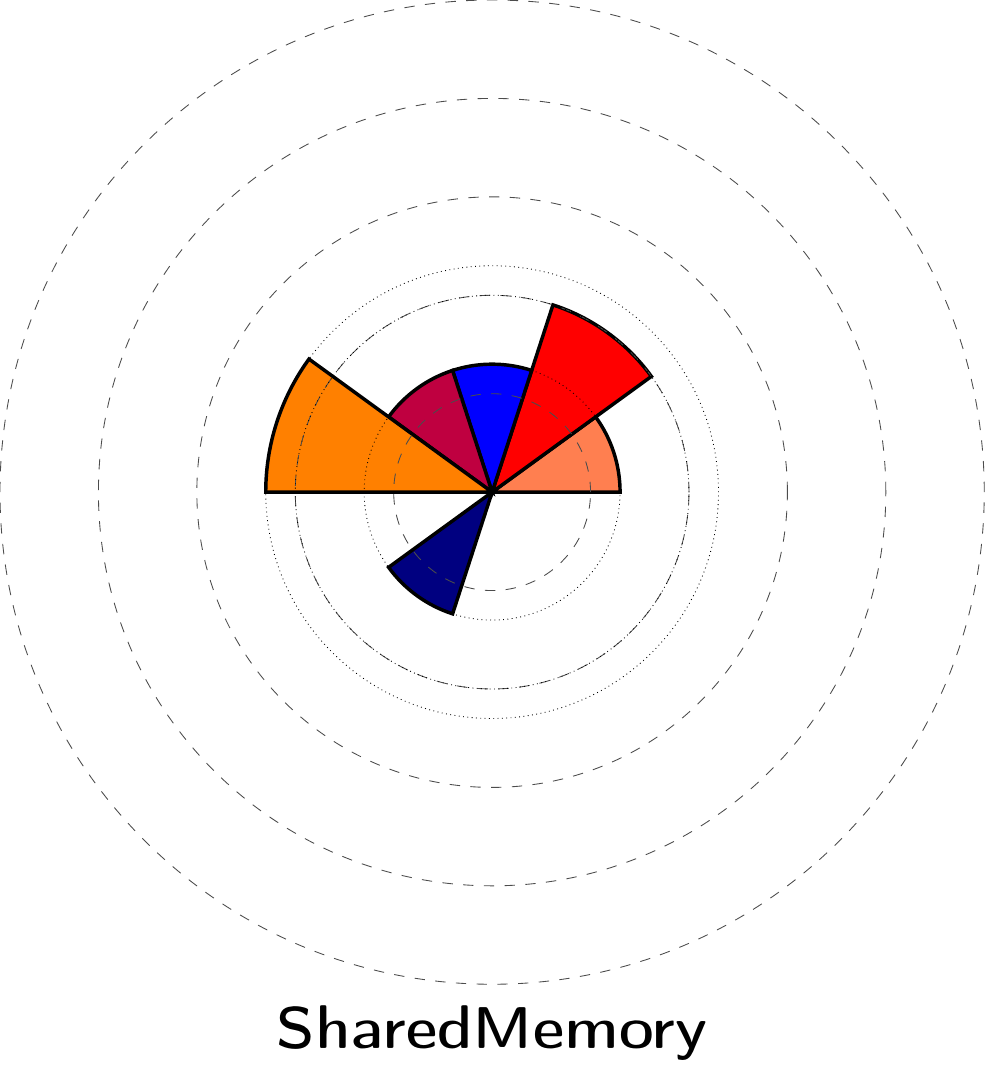}
\hfill
\includegraphics[scale=.35]{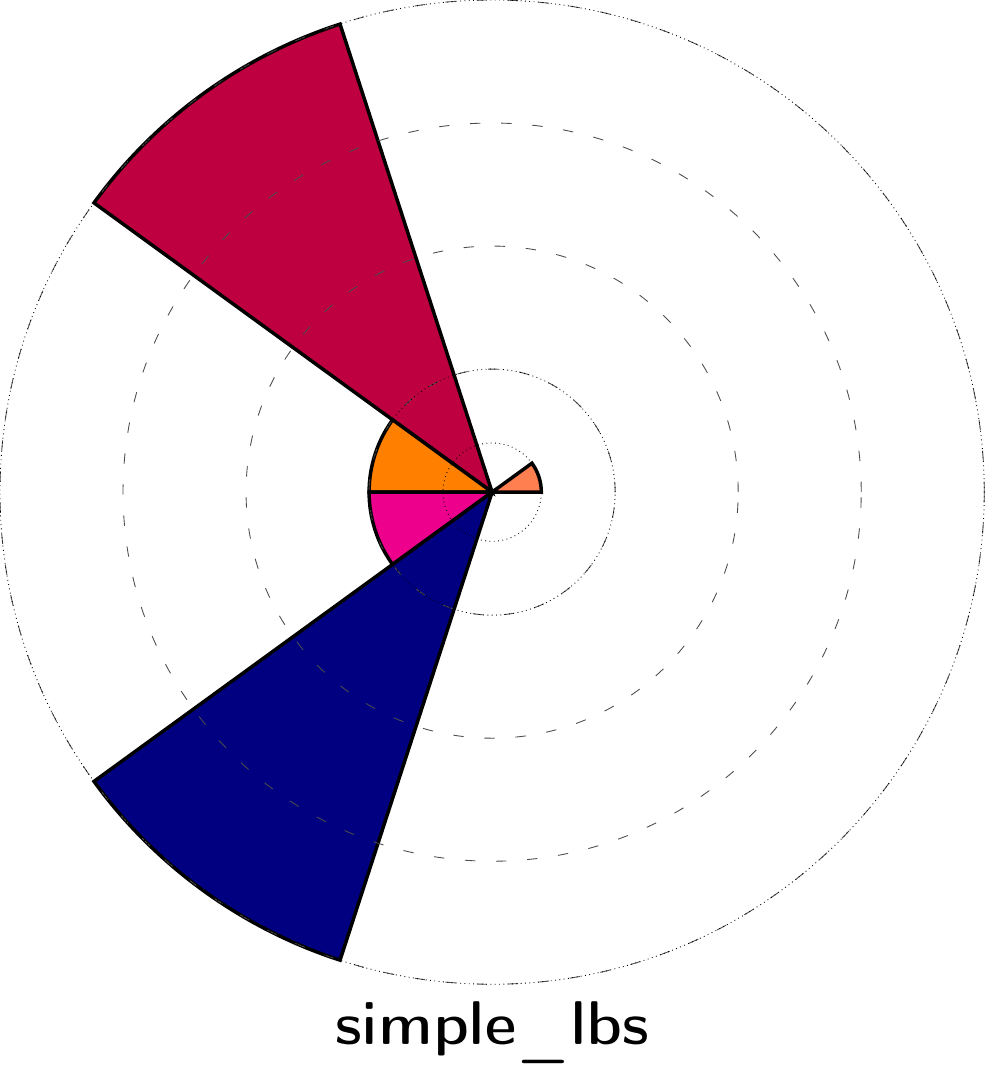}
\hfill
\includegraphics[scale=.35]{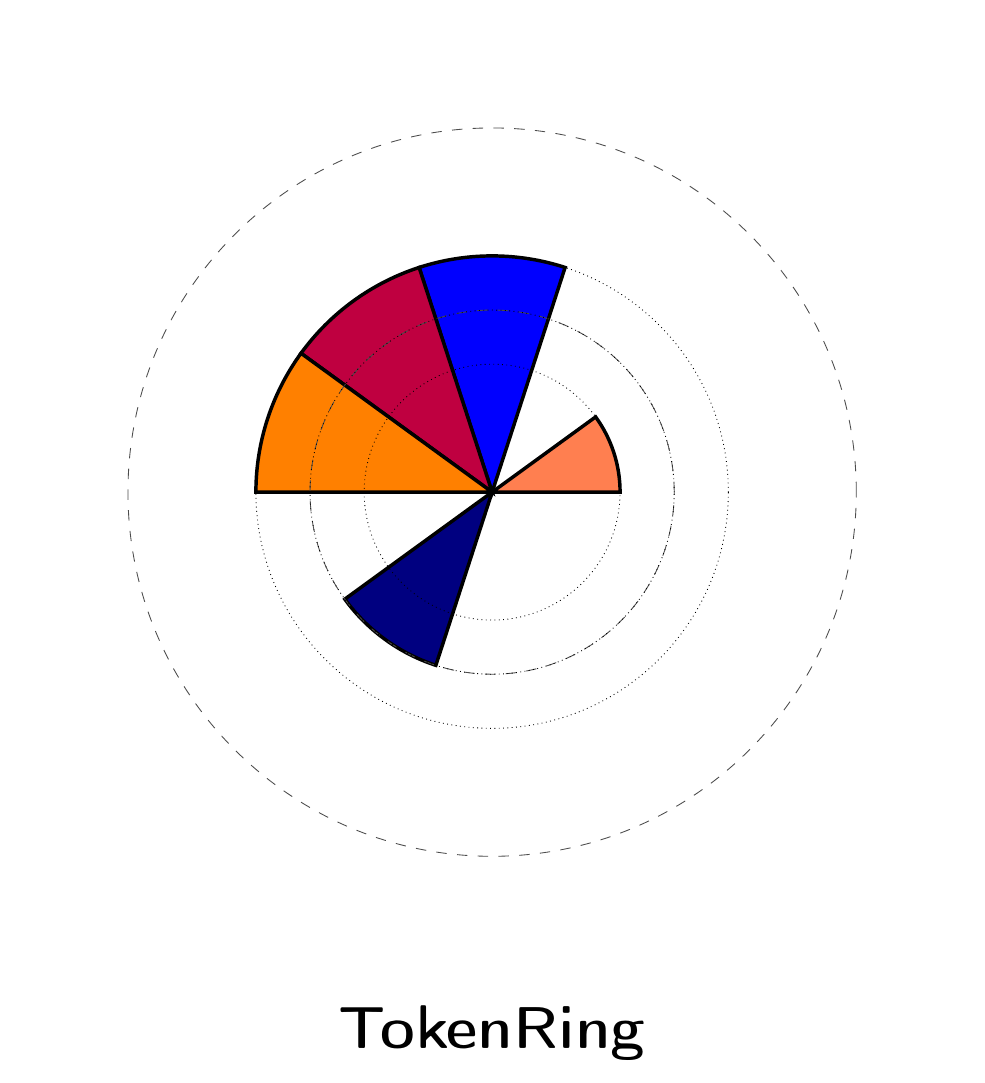}
\\
\bigskip
\mbox{}
\hfill
\includegraphics[scale=.7]{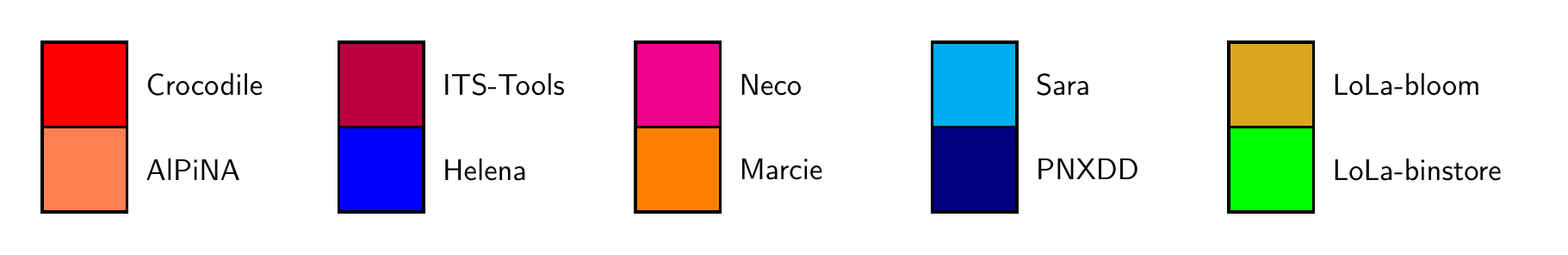}
\hfill
\mbox{}
\end{adjustwidth}
\caption{Highest parameter reached for each model,
         in state space generation}
\label{fig:ss:radar:models}
\end{figure}

\Cref{fig:ss:radar:models}~represents graphically through a set of radar
diagrams the highest parameter reached by the tools, for each model. Each
diagram corresponds to one model, \emph{e.g.}, \acs{Echo} or
\acs{Kanban}. Each diagram is
divided in ten slices, one for each competing tool, always at the same
position.

The length of the slice corresponds to the highest parameter reached by the
tool. When a slice does not appear, the tool could not process even the
smallest parameter. For instance, \acs{AlPiNA} handles some parameters for
\acs{Eratosthenes} and \acs{FMS}, but is not able to handle the smallest
parameter for \acs{MAPK}. The figure also shows that \acs{Marcie} handles all
parameters for \acs{Eratosthenes} and \acs{FMS}, but does not reach the
highest one for \acs{Lamport}.

Note that the scale depends on the model : when the parameters of a model vary
within a small range (less than~$100$), a linear scale is used (showed using
loosely dashed circles as in the \acs{Peterson} model), whereas a logarithmic
scale is used for larger parameter values (showed using densely dashed circles
as in the \acs{Philosophers} model). We also show dotted circles for the
results of the tools, in order to allow easier comparison.

Some models (\acs{RW-Mutex} and \acs{Echo}) have complex parameters, built
from two values. Tools were only able to handle variation of the second
parameter, so we only represent it in the figure, in order to show an integer
value.

\subsection{Radars by tools}
\label{sec:ss:bytools}
\index{State Space!Handling of models}

\begin{figure}[p]
\centering
\begin{adjustwidth}{0em}{0em}
\noindent
\includegraphics[scale=.4]{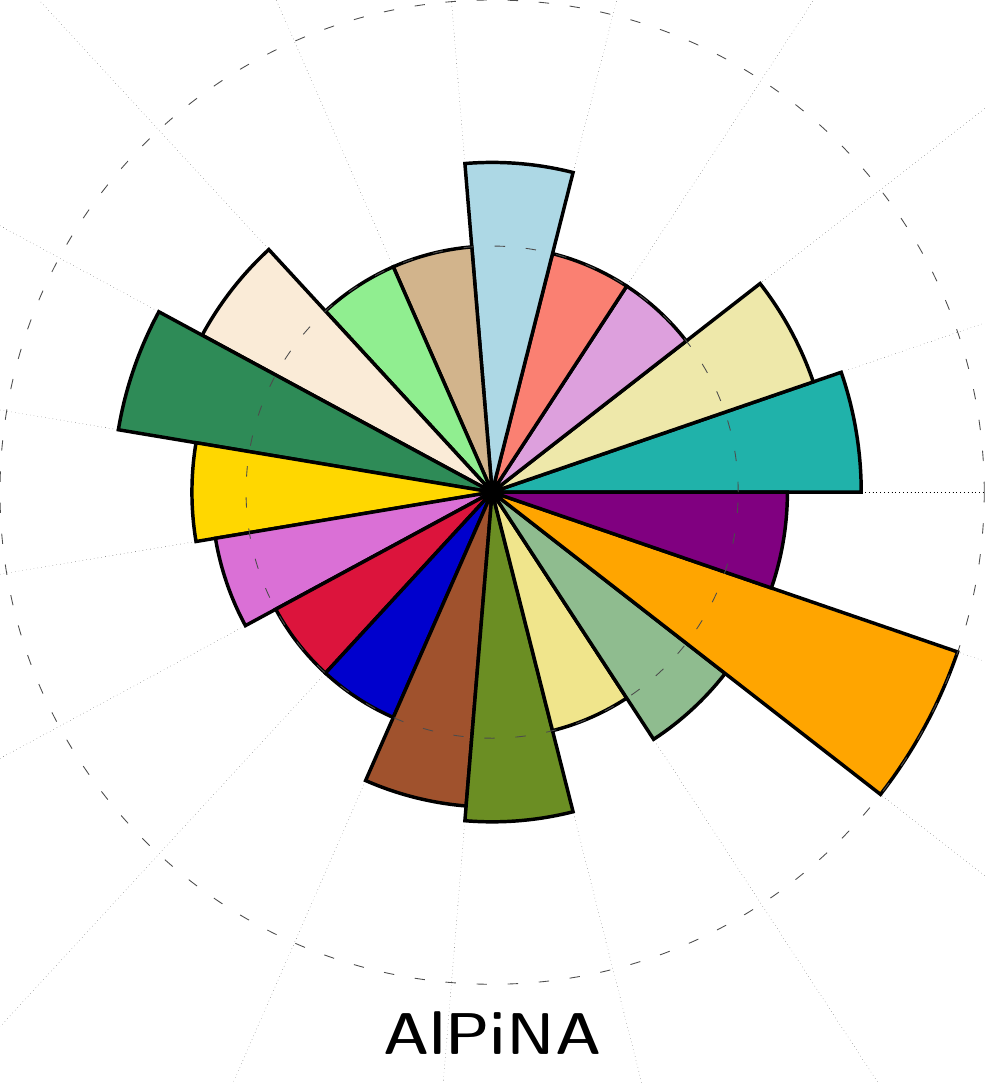}
\hfill
\includegraphics[scale=.4]{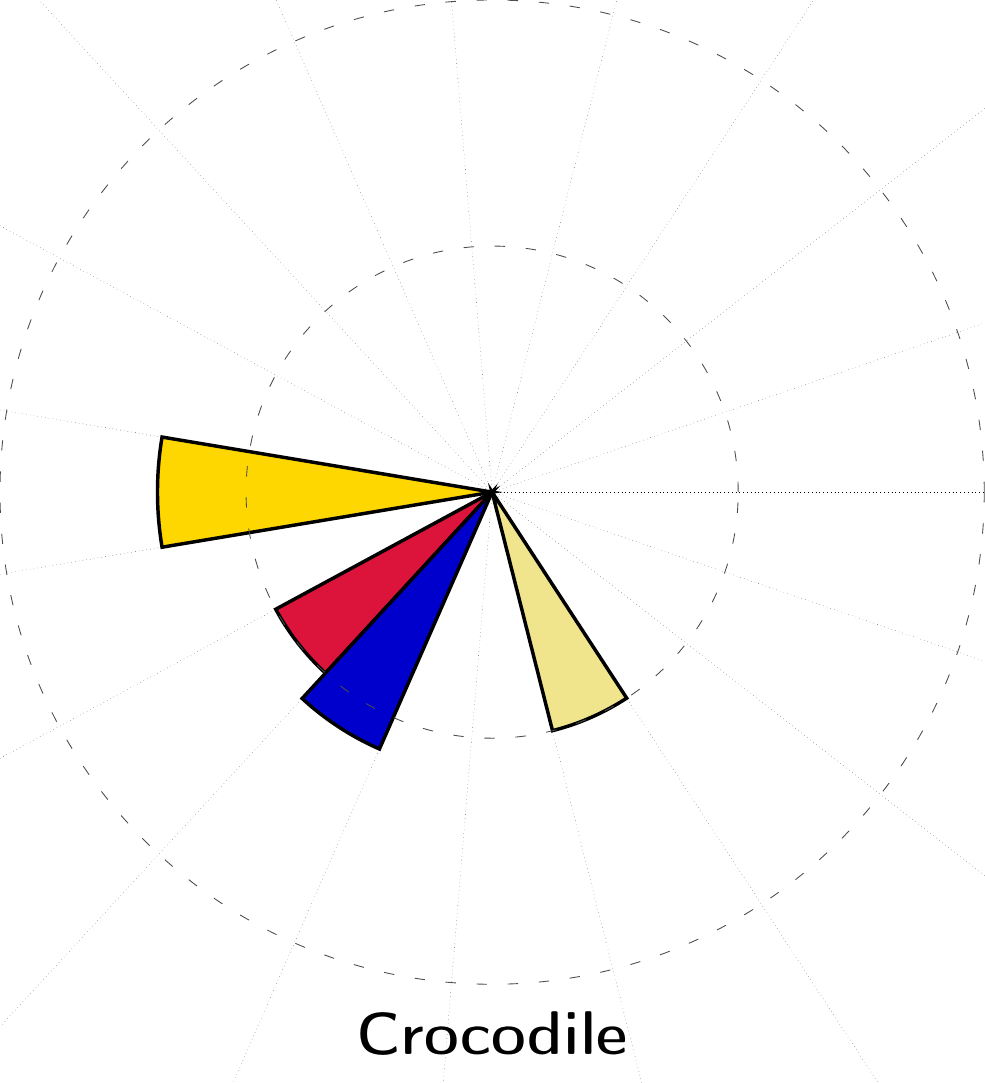}
\hfill
\includegraphics[scale=.4]{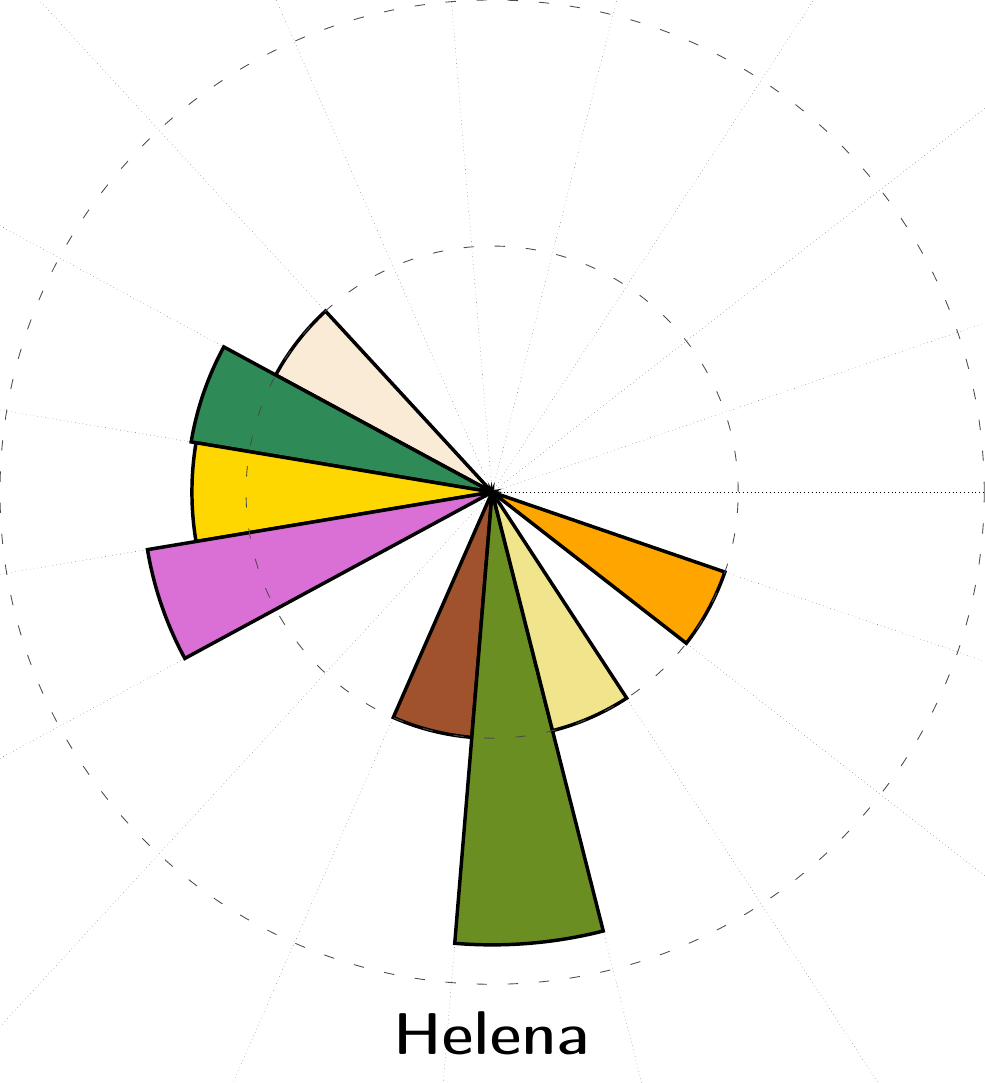}
\\
\medskip
\includegraphics[scale=.4]{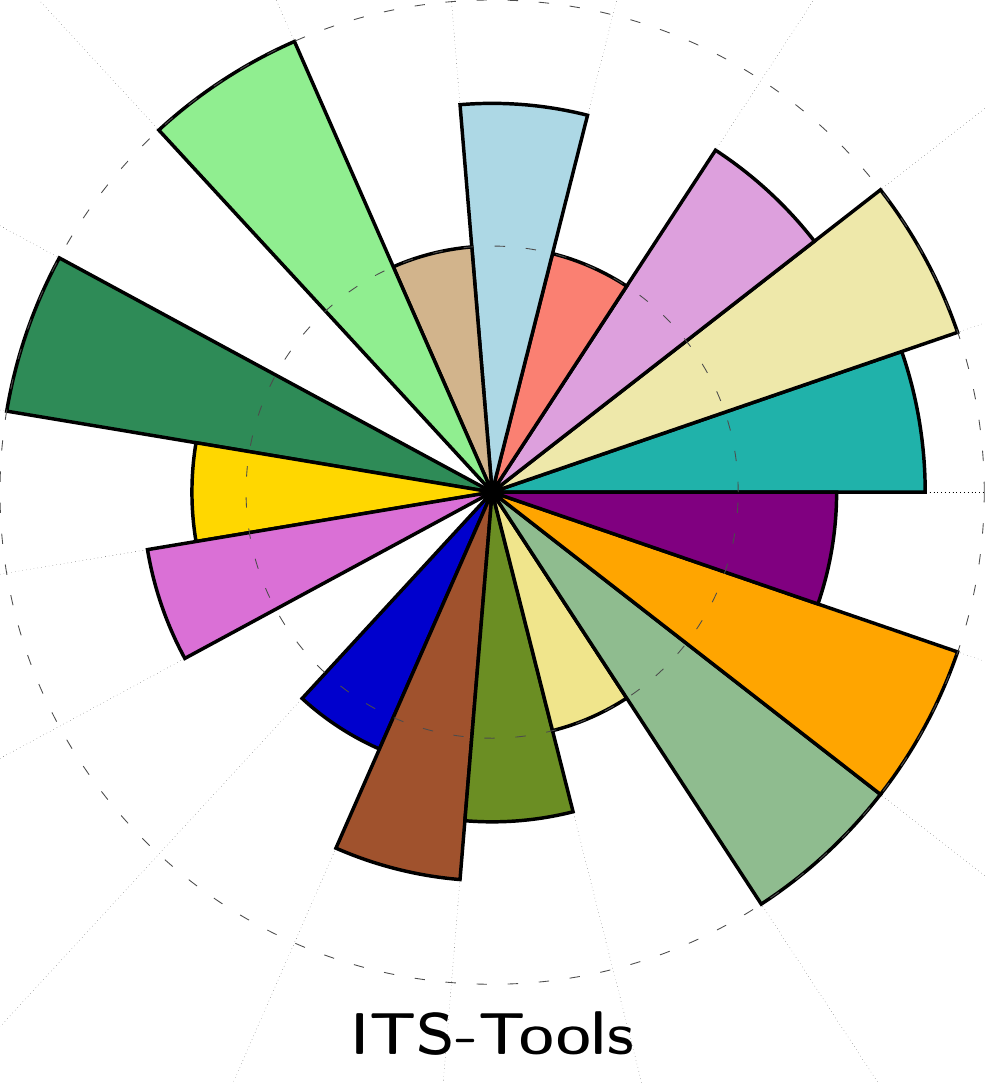}
\hfill
\includegraphics[scale=.4]{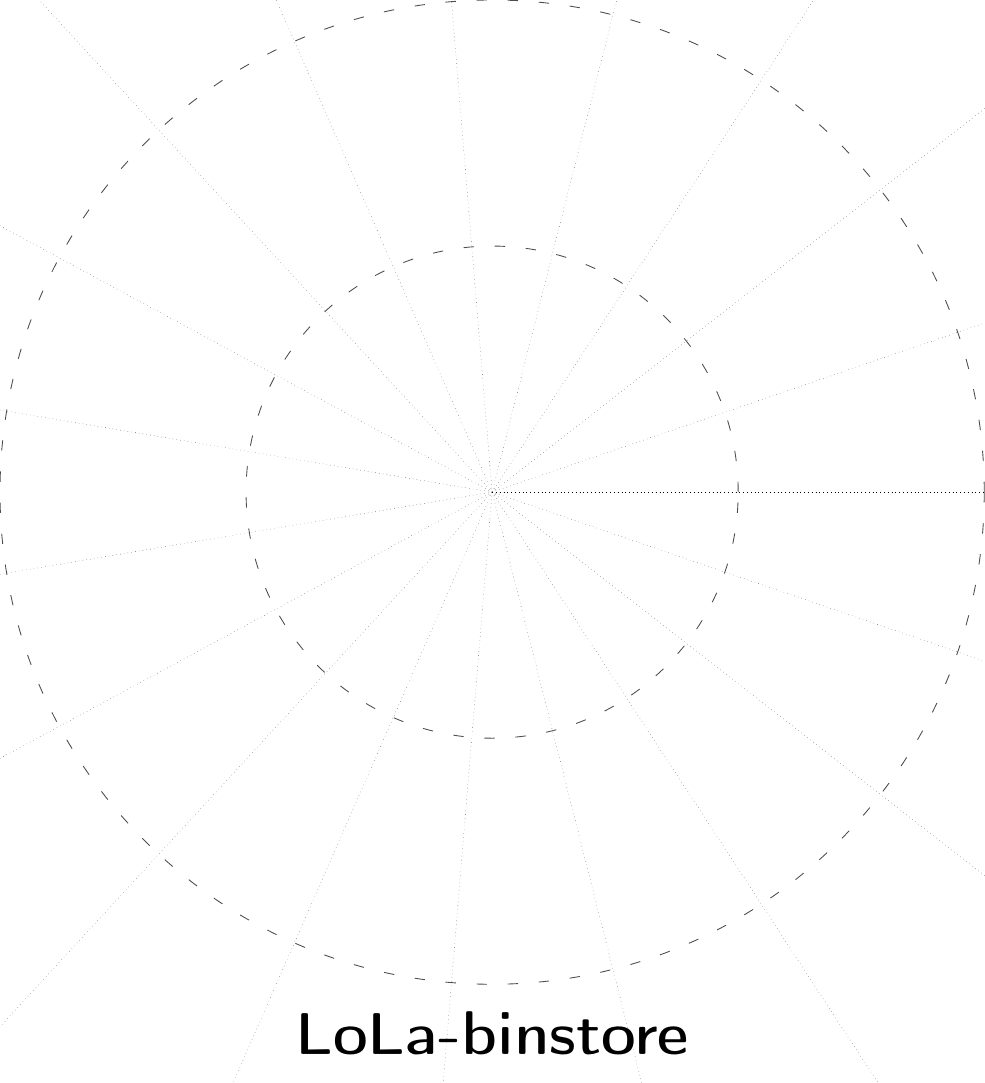}
\hfill
\includegraphics[scale=.4]{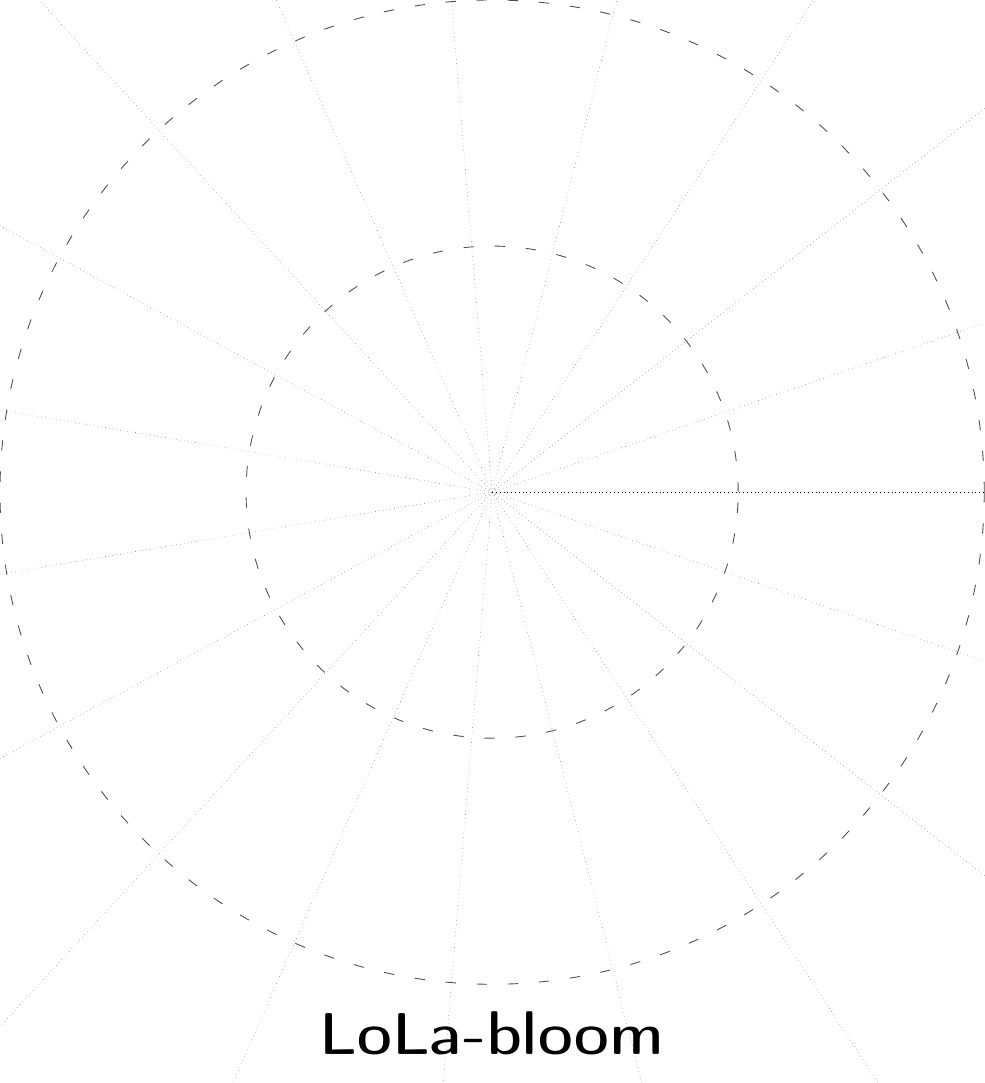}
\\
\medskip
\includegraphics[scale=.4]{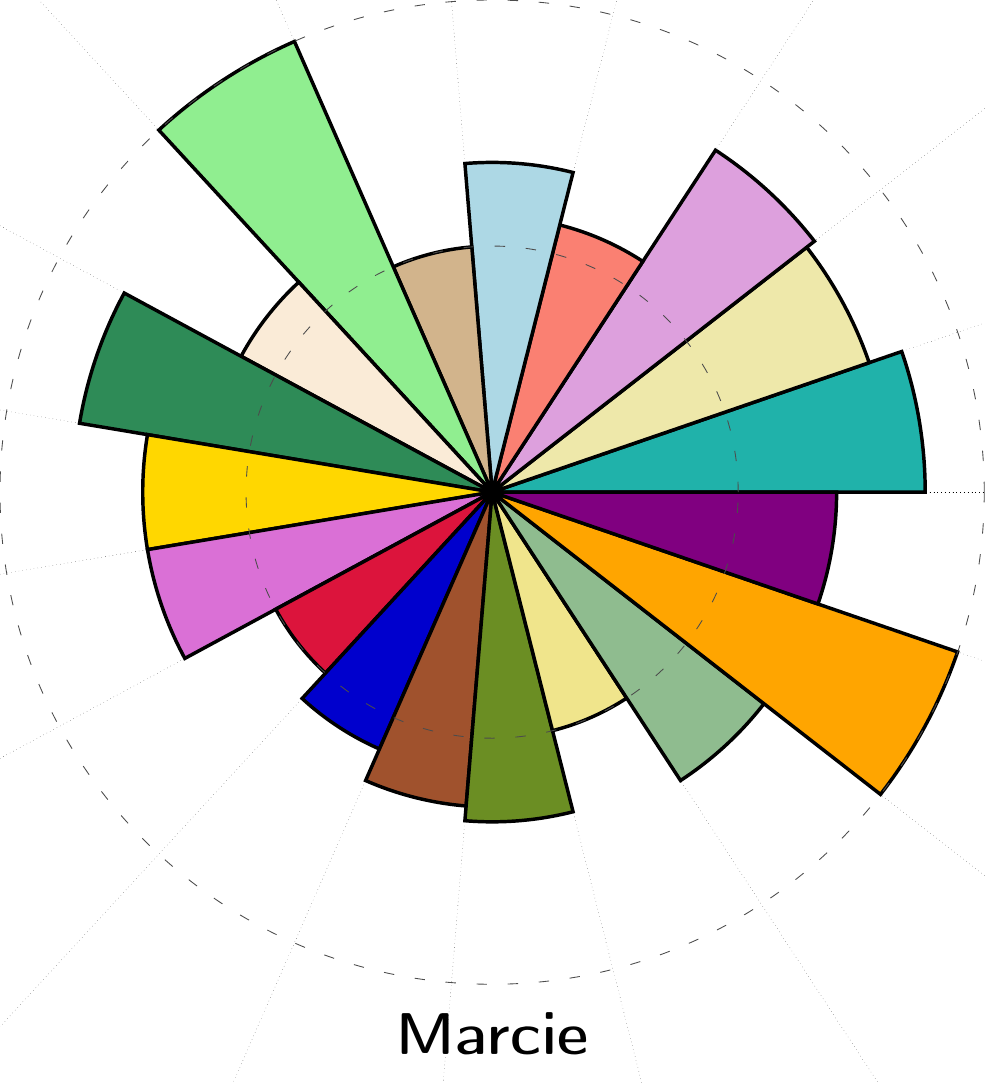}
\hfill
\includegraphics[scale=.4]{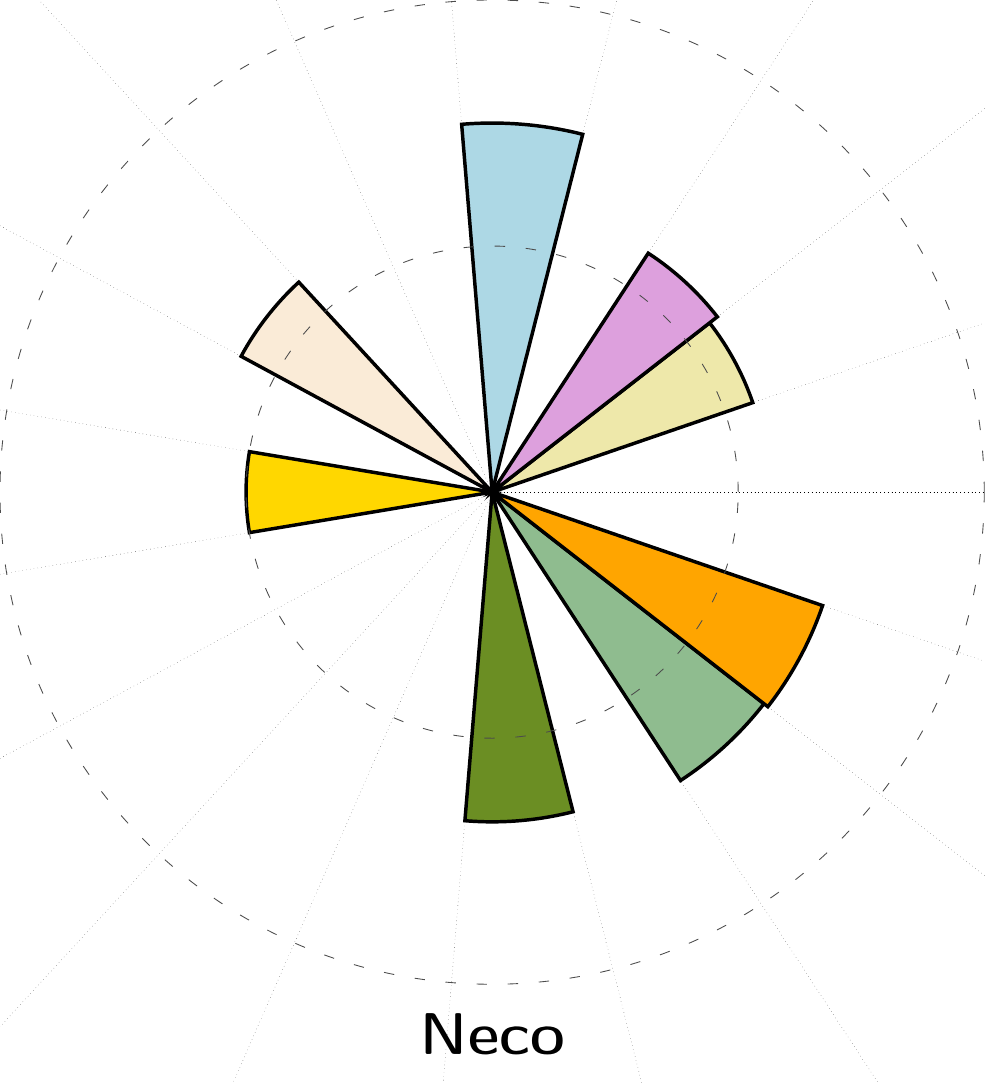}
\hfill
\includegraphics[scale=.4]{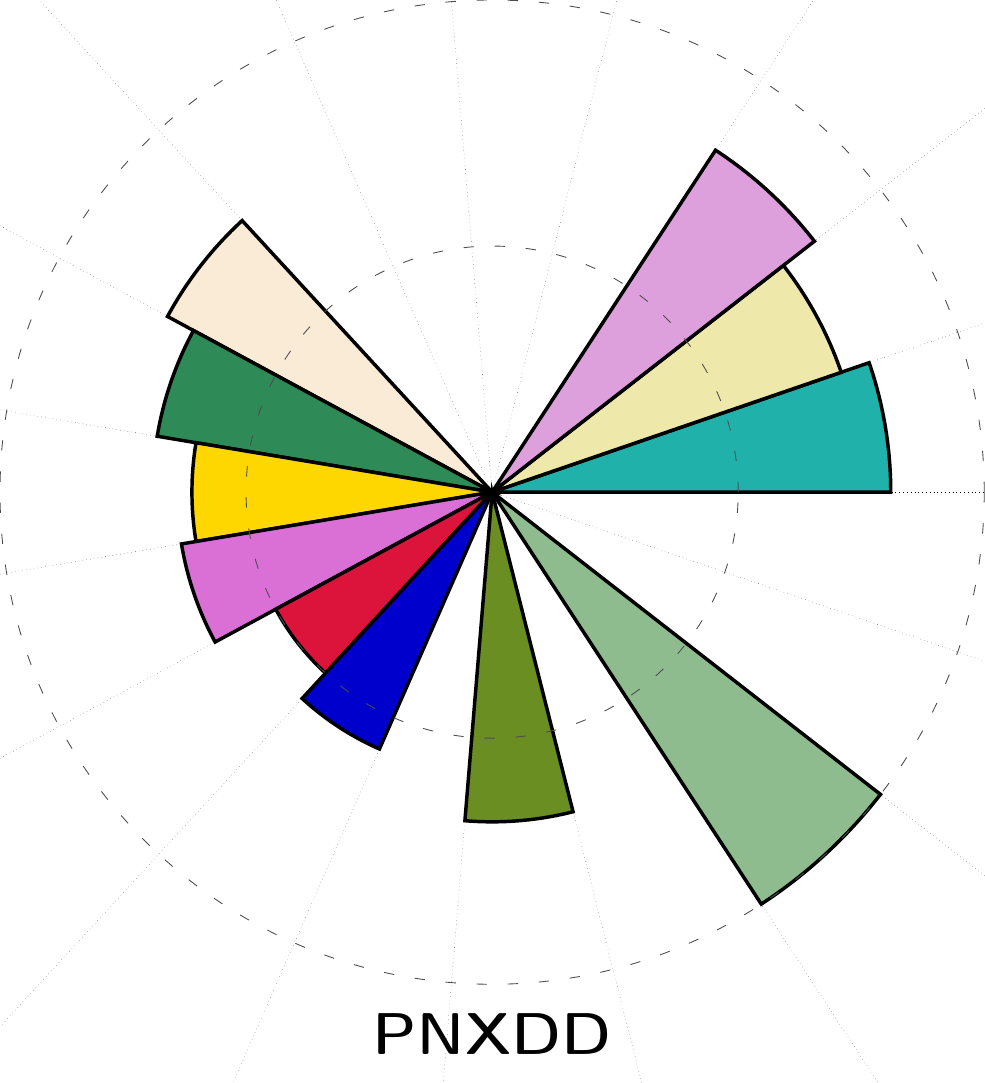}
\\
\medskip
\mbox{}
\hfill
\includegraphics[scale=.4]{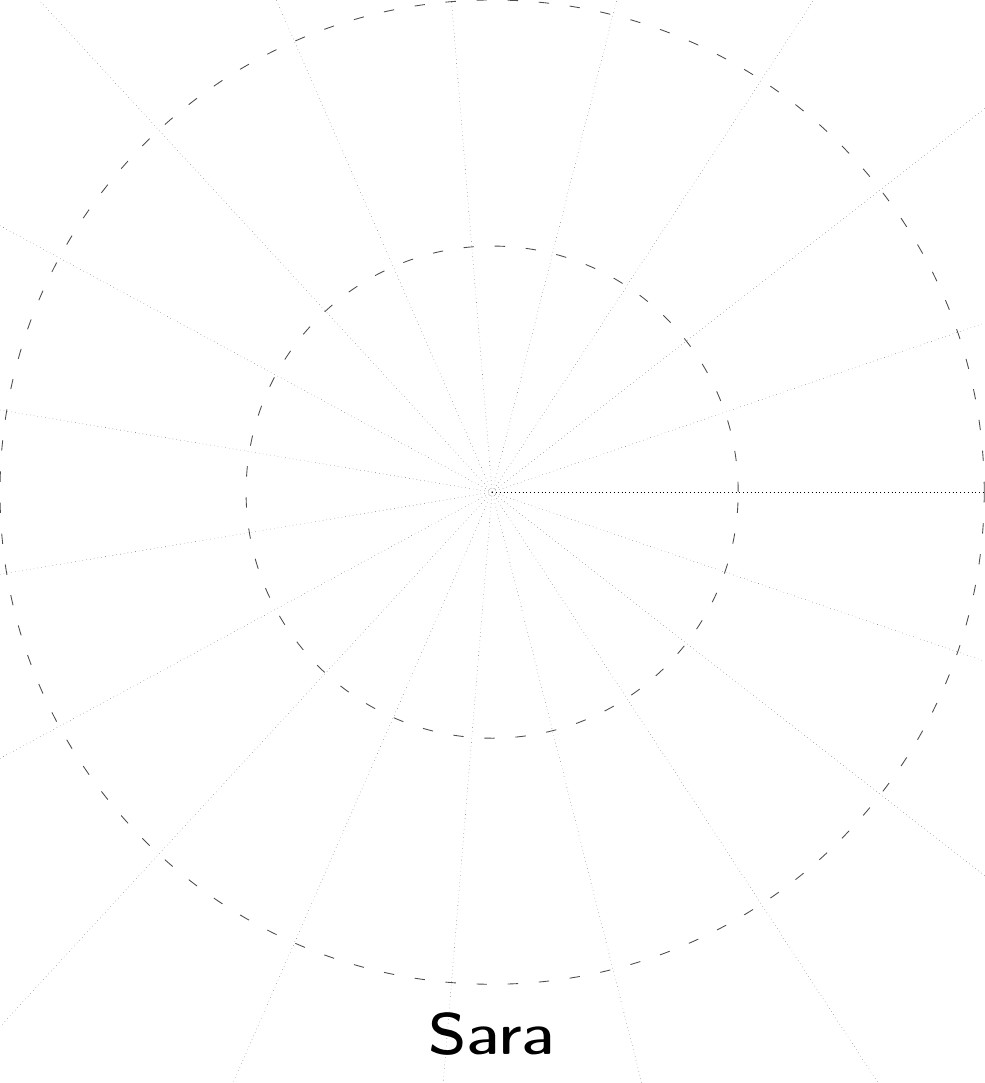}
\hfill
\mbox{}
\\
\bigskip
\mbox{}
\hfill
\includegraphics[scale=.7]{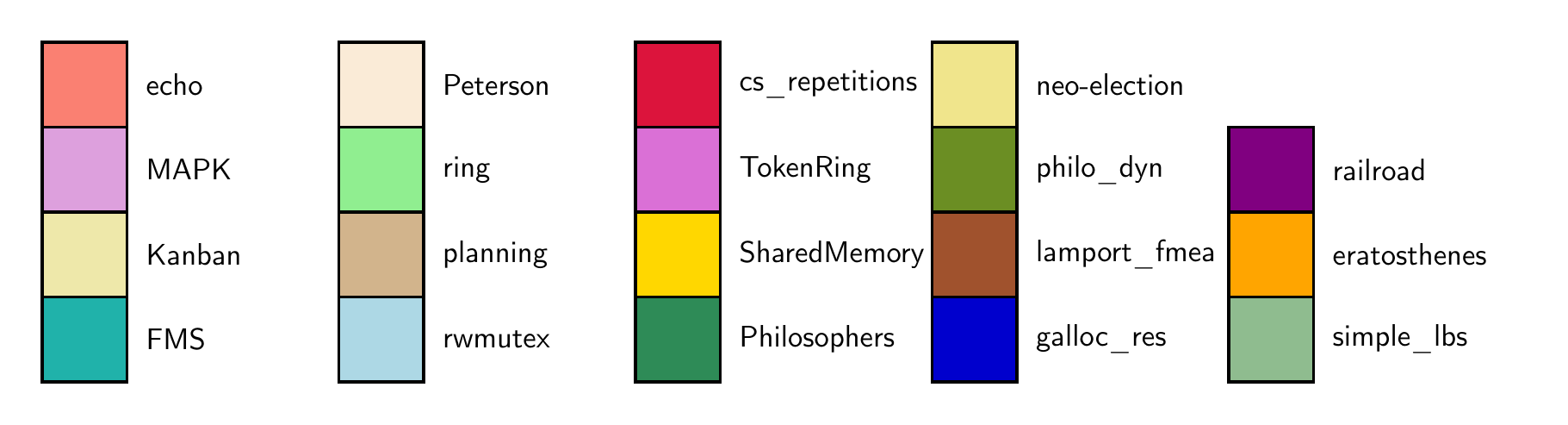}
\hfill
\mbox{}
\end{adjustwidth}
\caption{Handled parameters for each tool,
         in state space generation}
\label{fig:ss:radar:tools}
\end{figure}

\Cref{fig:ss:radar:tools}~presents also radar diagrams showing
graphically the participation and results by tool.
Each slice in the radars represents a model, always at the same position.

These diagrams differ from those in~\Cref{fig:ss:radar:models}.
Each diagram contains two circles:
a slice reaches the inner circle if the tool participates
to the state space competition for this model, but fails.
The size of the slice between the inner and the outer circles represents
how many parameters are handled.
Its length is the ratio between the number of handled parameters
(without failure) and the total number of parameters.
Thus, we do not take into account the parameter values.

Two interpretations can rise from the observation of these diagrams:

\begin{itemize}

\item First, the colored surface in the inner circle shows how many models
could be read by the tool. For instance, \acs{AlPiNA}, \acs{ITS-Tools} and
\acs{Marcie} seem to be non-specialized tools, contrary to \acs{Crocodile} or
\acs{Neco}. This interpretation is biased, as a small surface can also mean
that the tool developer did not have time to handle all models.

\item Second, the surface between the inner and outer circles shows how well
the tool can handle a model. Here, we see that \acs{ITS-Tools} is very
efficient for five models, where it handles all parameters.

\end{itemize}

Note that \acs{LoLA-binstore}, \acs{LoLA-bloom} and \acs{Sara}
did not participate to the state space examination.

\subsection{Charts for Models}
\label{sec:ssmodechart}

The charts below correspond to the way tools cope with state space
generation for the model that were supported.

\medskip
\noindent
\textbf{Charts for echo}
\nopagebreak[4]\\
\makebox[\textwidth]{
\index{State Space Charts!echo}
\includegraphics[width=.5\textwidth]{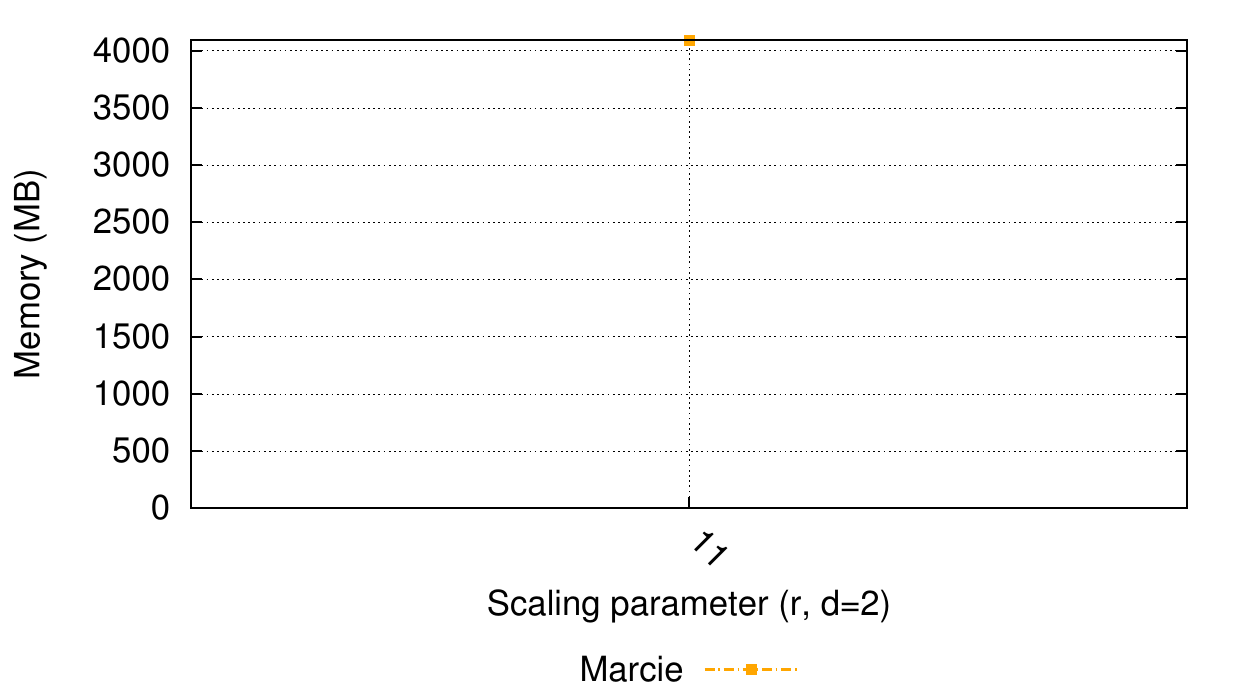}
\includegraphics[width=.5\textwidth]{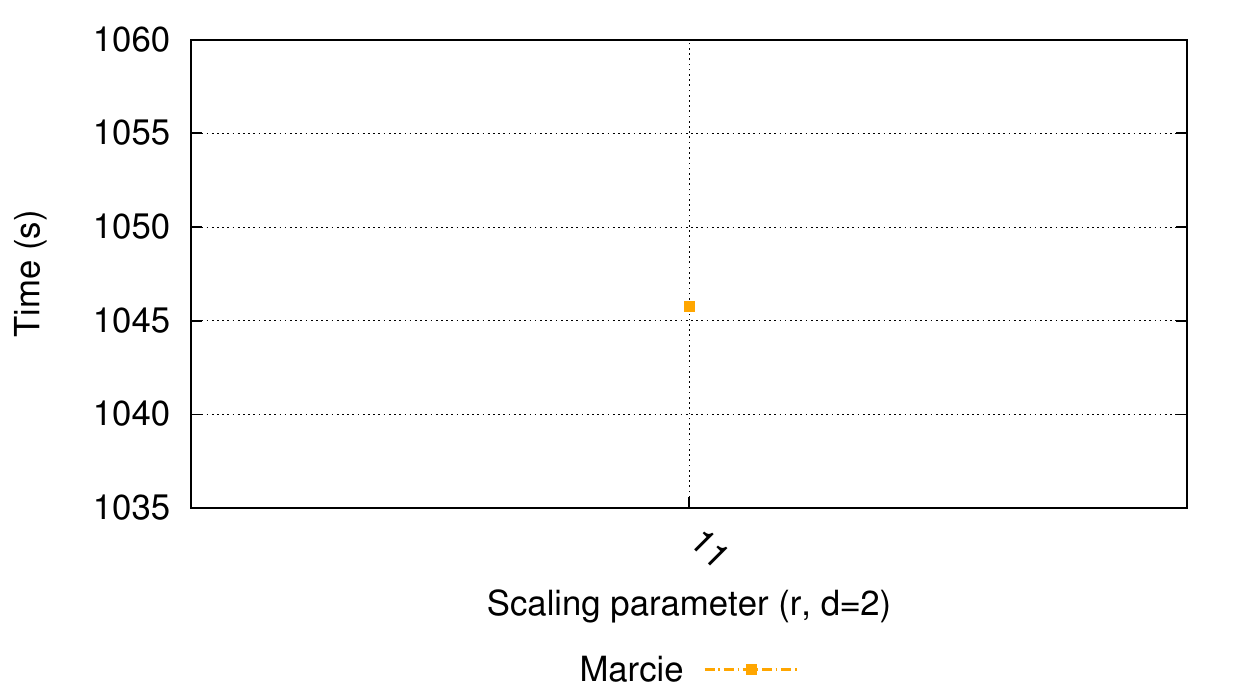}
}

\medskip
\noindent
\textbf{Charts for eratosthenes}
\nopagebreak[4]\\
\makebox[\textwidth]{
\index{State Space Charts!eratosthenes}
\includegraphics[width=.5\textwidth]{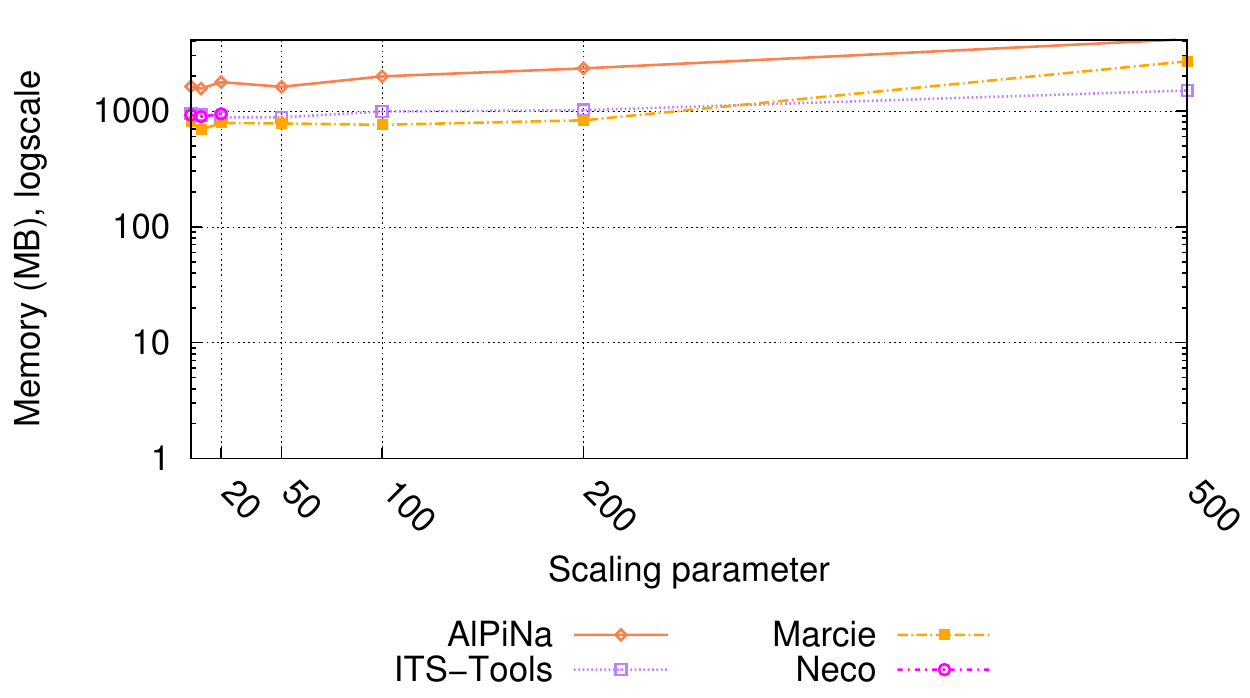}
\includegraphics[width=.5\textwidth]{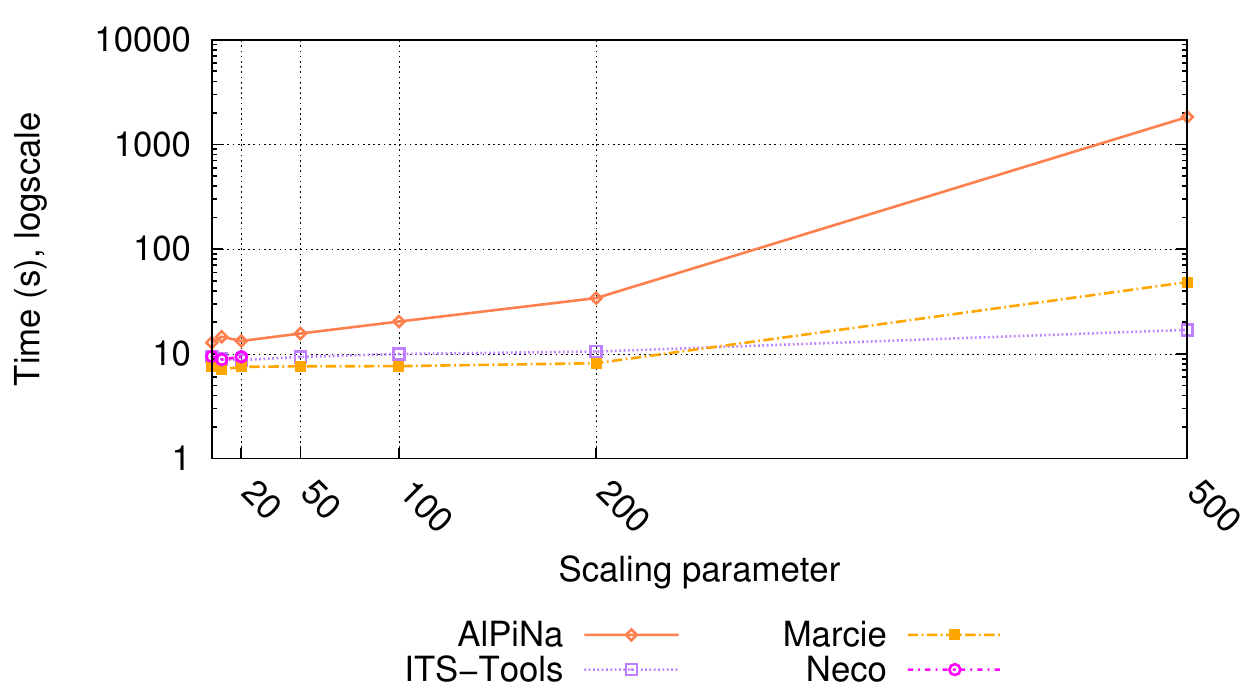}
}

\medskip
\noindent
\textbf{Charts for FMS}
\nopagebreak[4]\\
\makebox[\textwidth]{
\index{State Space Charts!FMS}
\includegraphics[width=.5\textwidth]{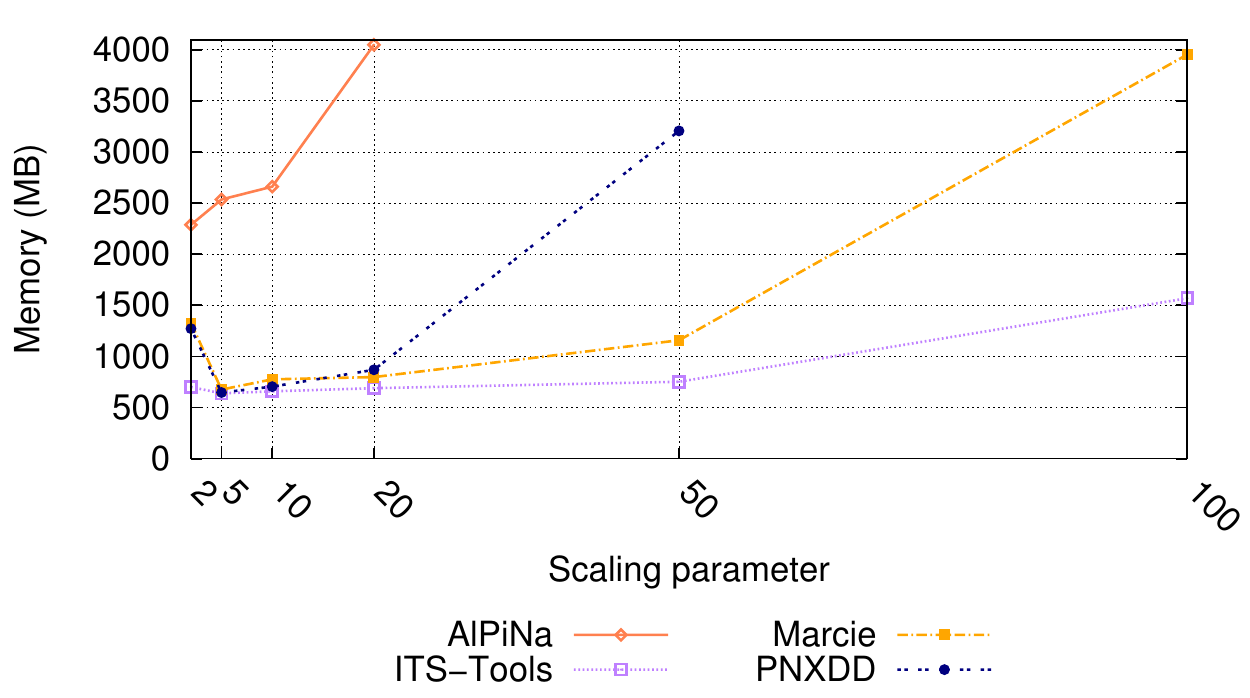}
\includegraphics[width=.5\textwidth]{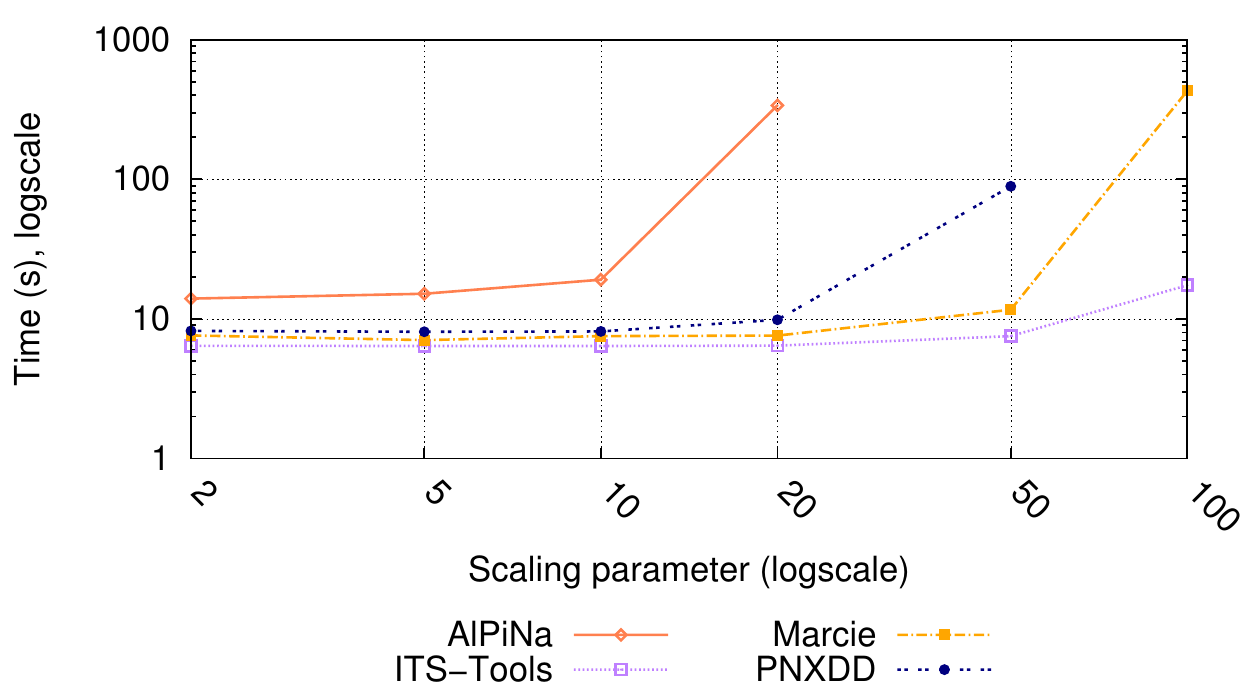}
}

\medskip
\noindent
\textbf{Charts for galloc\_res}
\nopagebreak[4]\\
\makebox[\textwidth]{
\index{State Space Charts!galloc\_res}
\includegraphics[width=.5\textwidth]{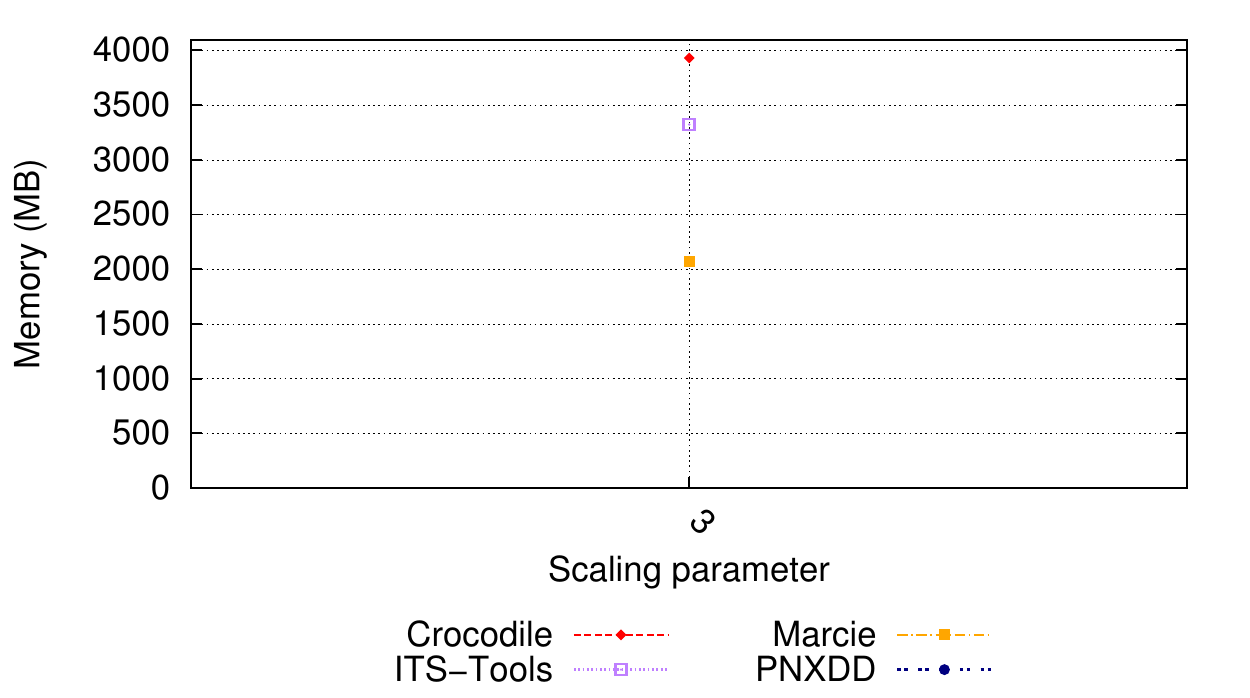}
\includegraphics[width=.5\textwidth]{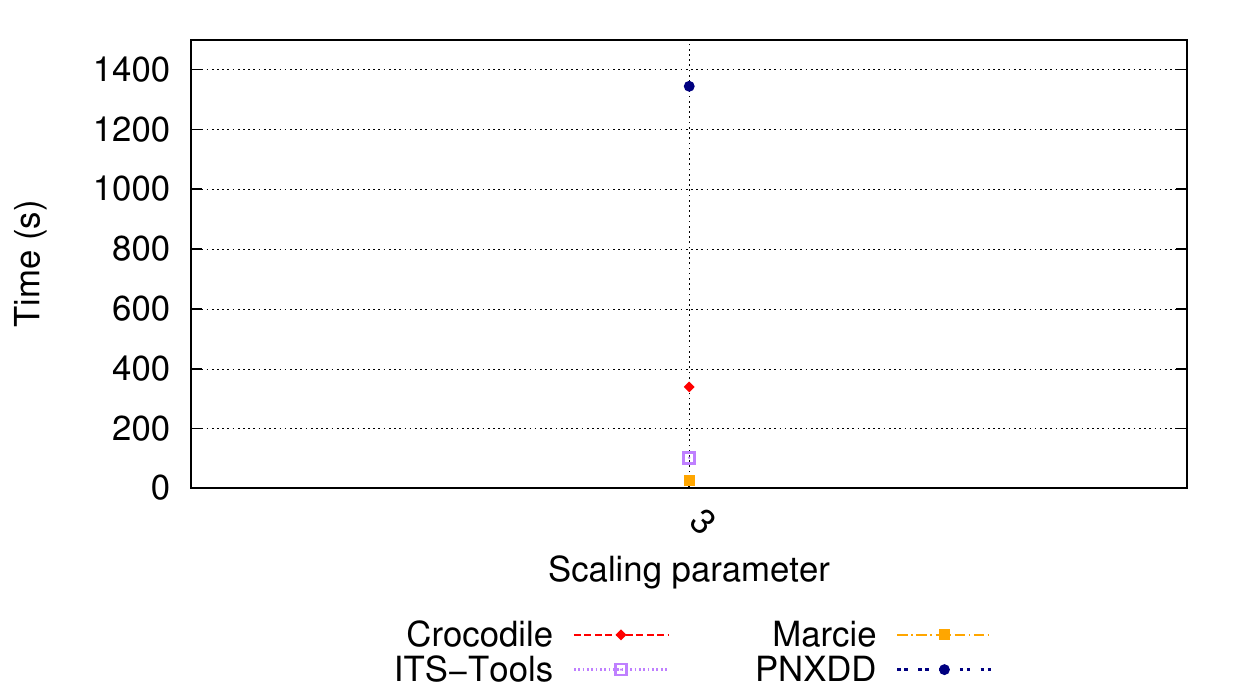}
}

\medskip
\noindent
\textbf{Charts for Kanban}
\nopagebreak[4]\\
\makebox[\textwidth]{
\index{State Space Charts!Kanban}
\includegraphics[width=.5\textwidth]{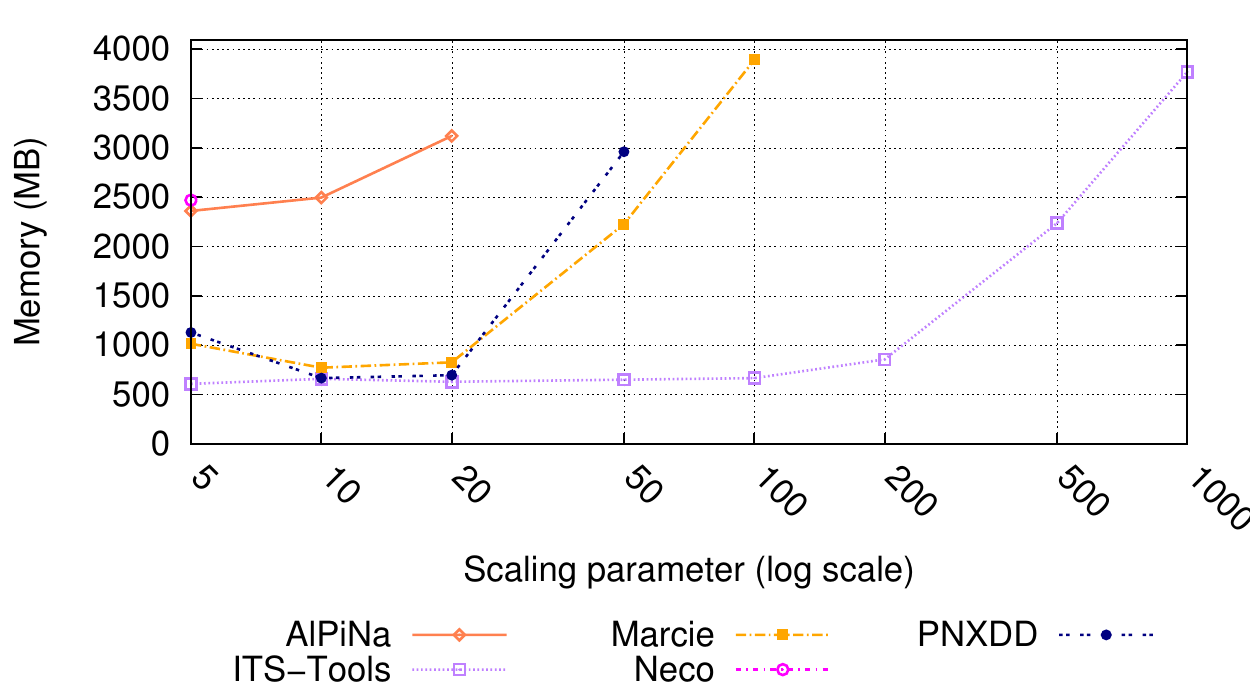}
\includegraphics[width=.5\textwidth]{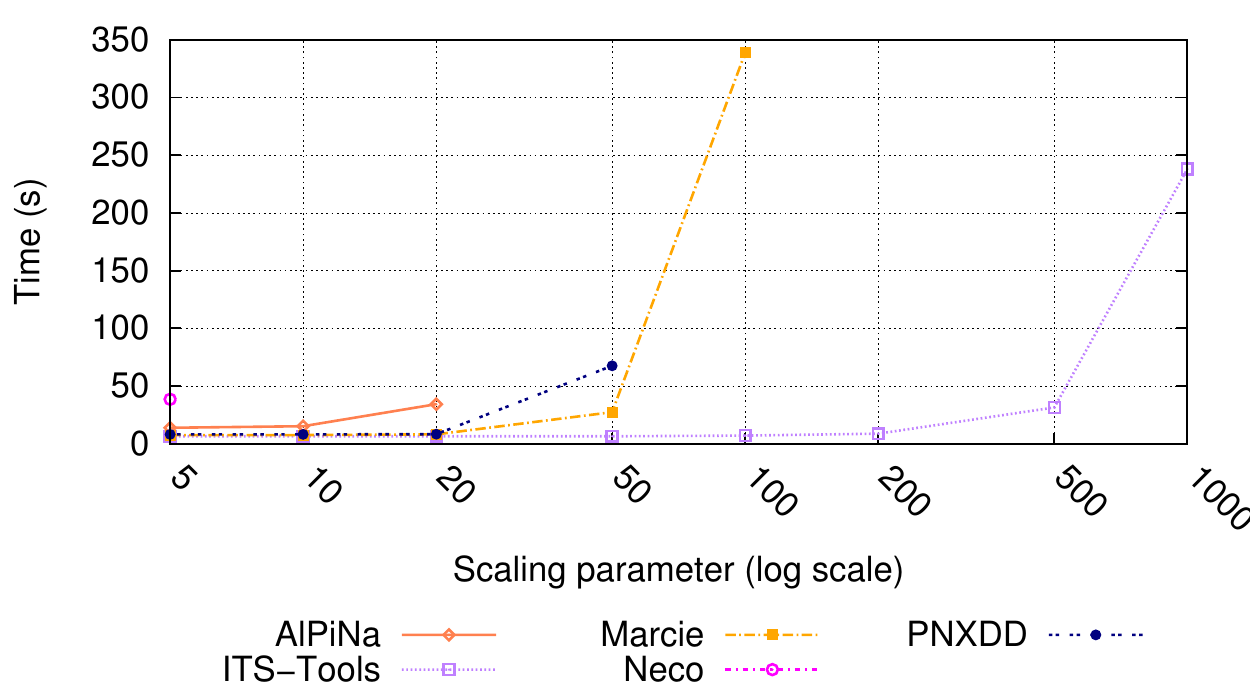}
}

\medskip
\noindent
\textbf{Charts for lamport\_fmea}
\index{State Space Charts!lamport\_fmea}
\nopagebreak[4]\\
\makebox[\textwidth]{
\includegraphics[width=.5\textwidth]{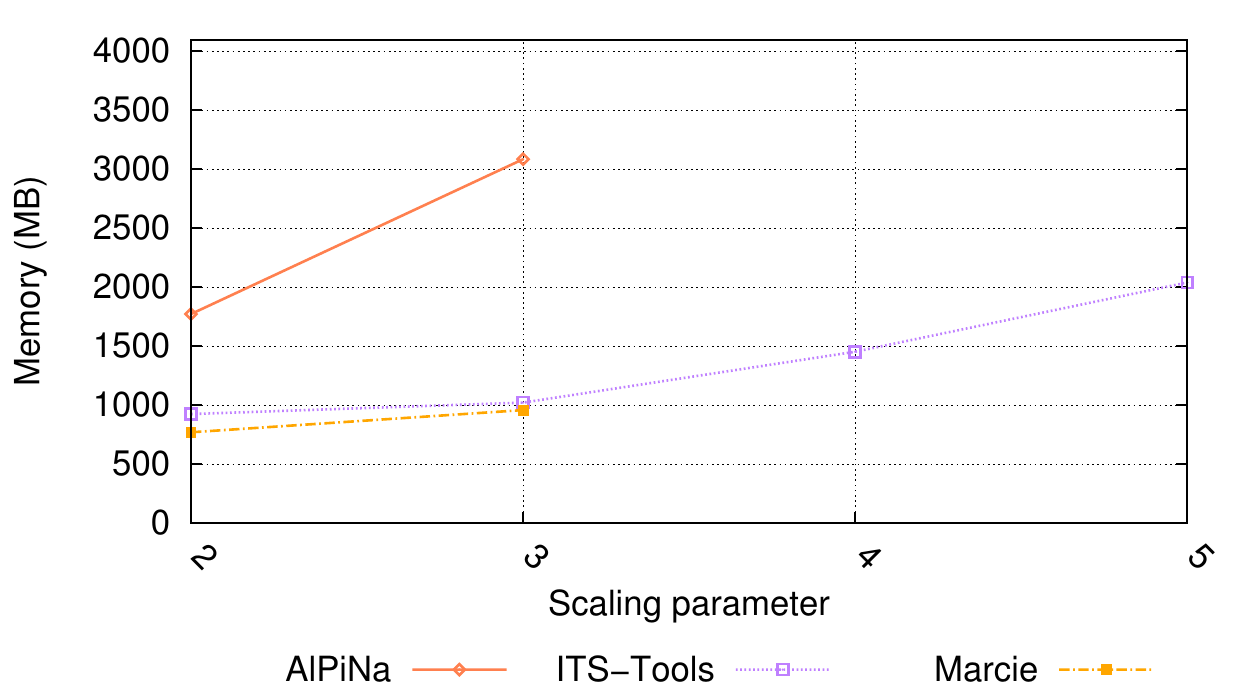}
\includegraphics[width=.5\textwidth]{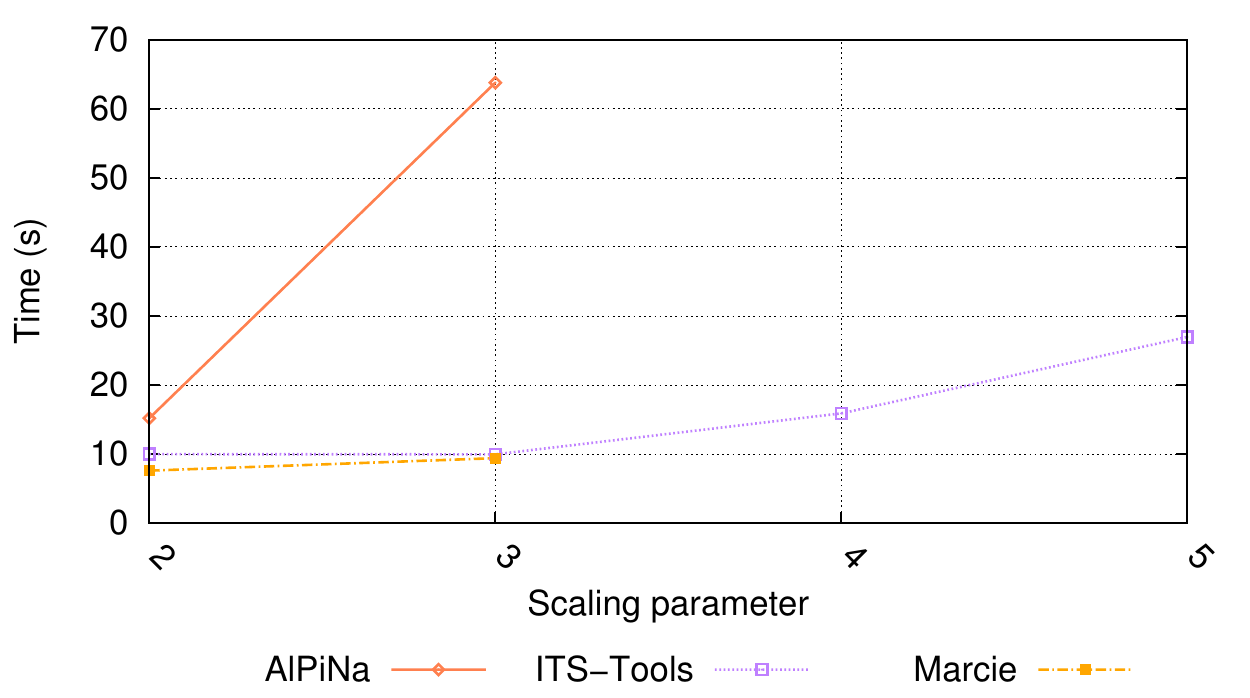}
}

\medskip
\noindent
\textbf{Charts for MAPK}
\nopagebreak[4]\\
\makebox[\textwidth]{
\index{State Space Charts!MAPK}
\includegraphics[width=.5\textwidth]{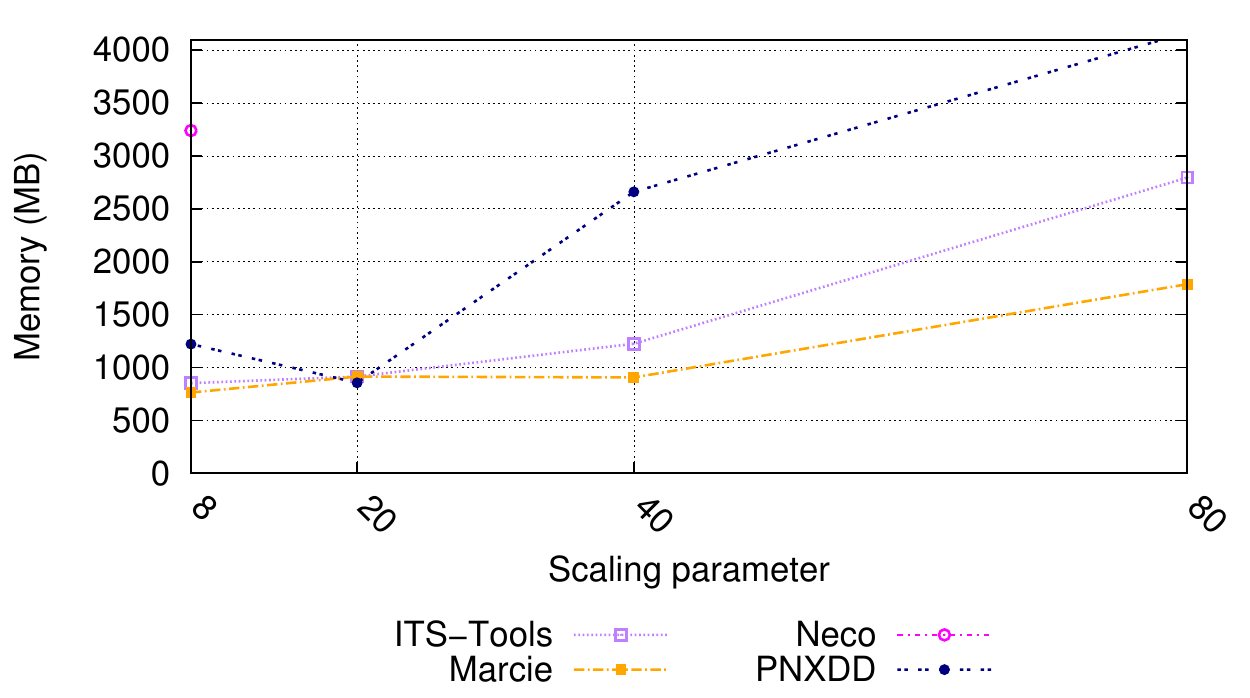}
\includegraphics[width=.5\textwidth]{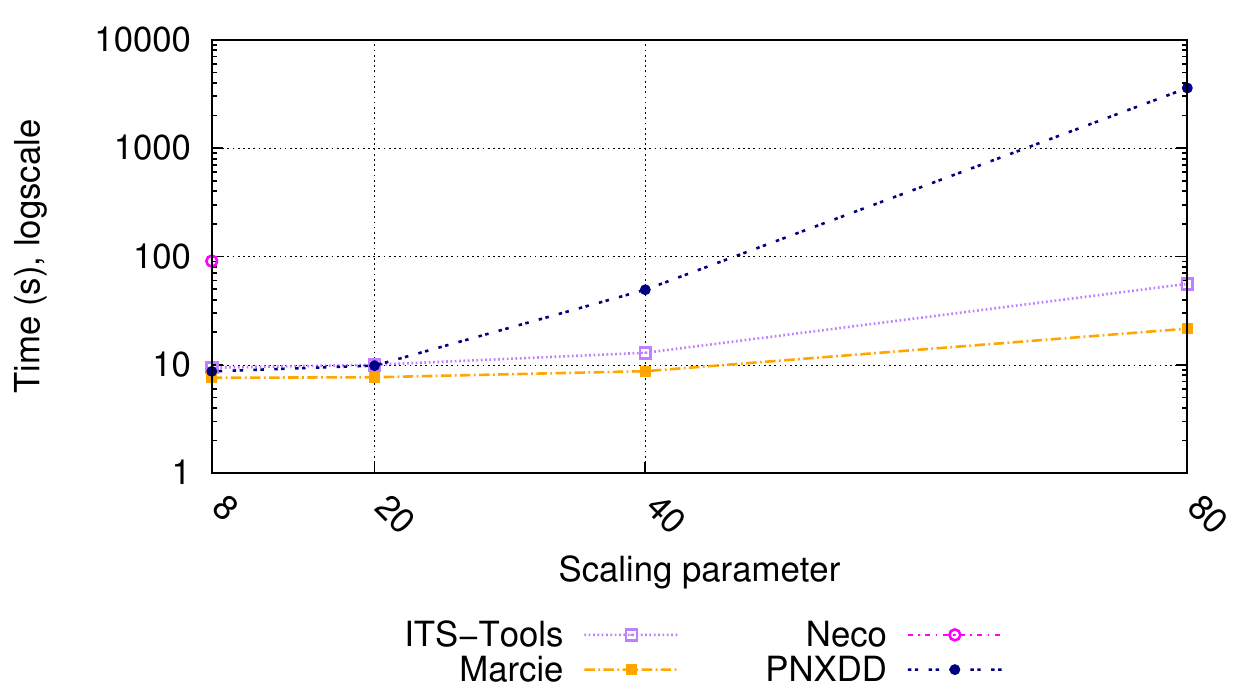}
}

\medskip
\noindent
\textbf{Charts for Peterson}
\nopagebreak[4]\\
\makebox[\textwidth]{
\index{State Space Charts!Peterson}
\includegraphics[width=.5\textwidth]{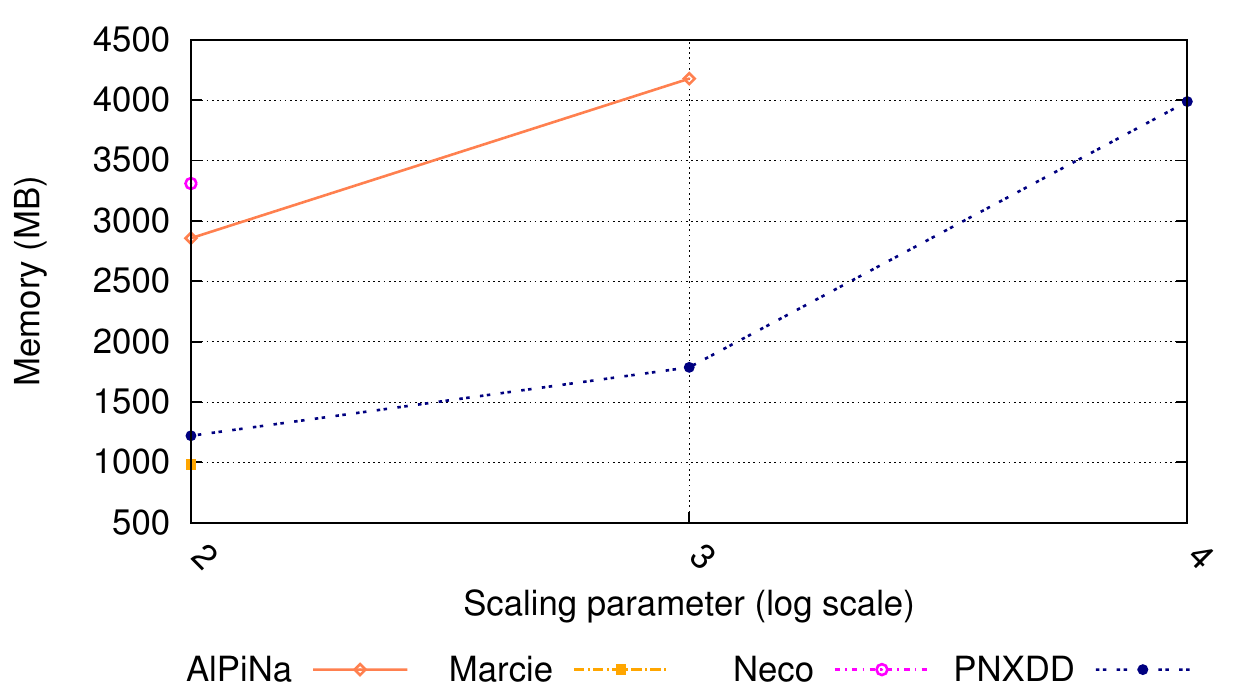}
\includegraphics[width=.5\textwidth]{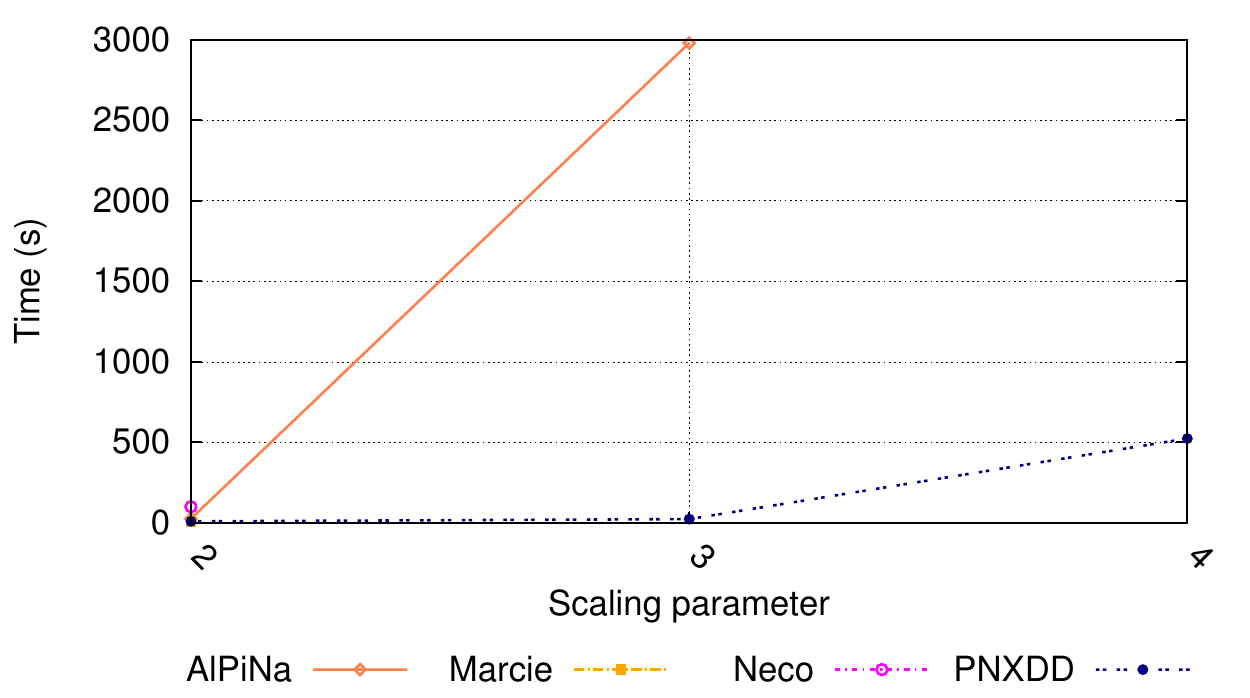}
}

\medskip
\noindent
\textbf{Charts for philo\_dyn}
\nopagebreak[4]\\
\makebox[\textwidth]{
\index{State Space Charts!philo\_dyn}
\includegraphics[width=.5\textwidth]{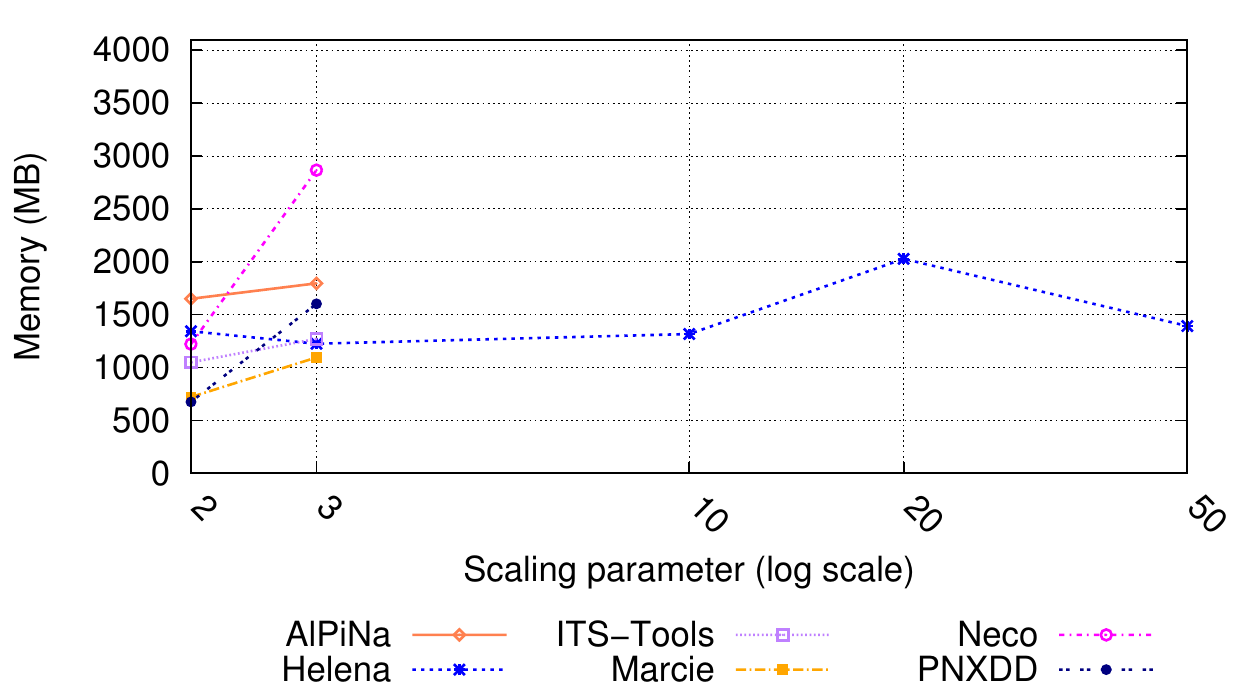}
\includegraphics[width=.5\textwidth]{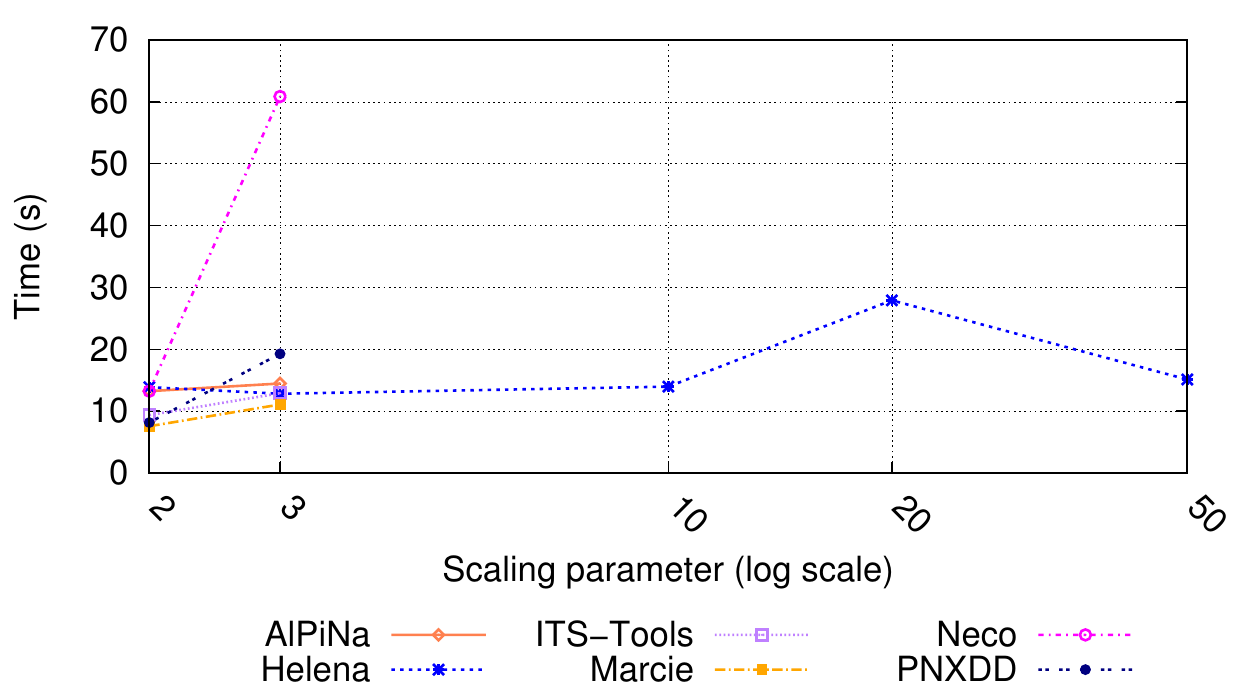}
}

\medskip
\noindent
\textbf{Charts for Philosophers}
\nopagebreak[4]\\
\makebox[\textwidth]{
\index{State Space Charts!Philosophers}
\includegraphics[width=.5\textwidth]{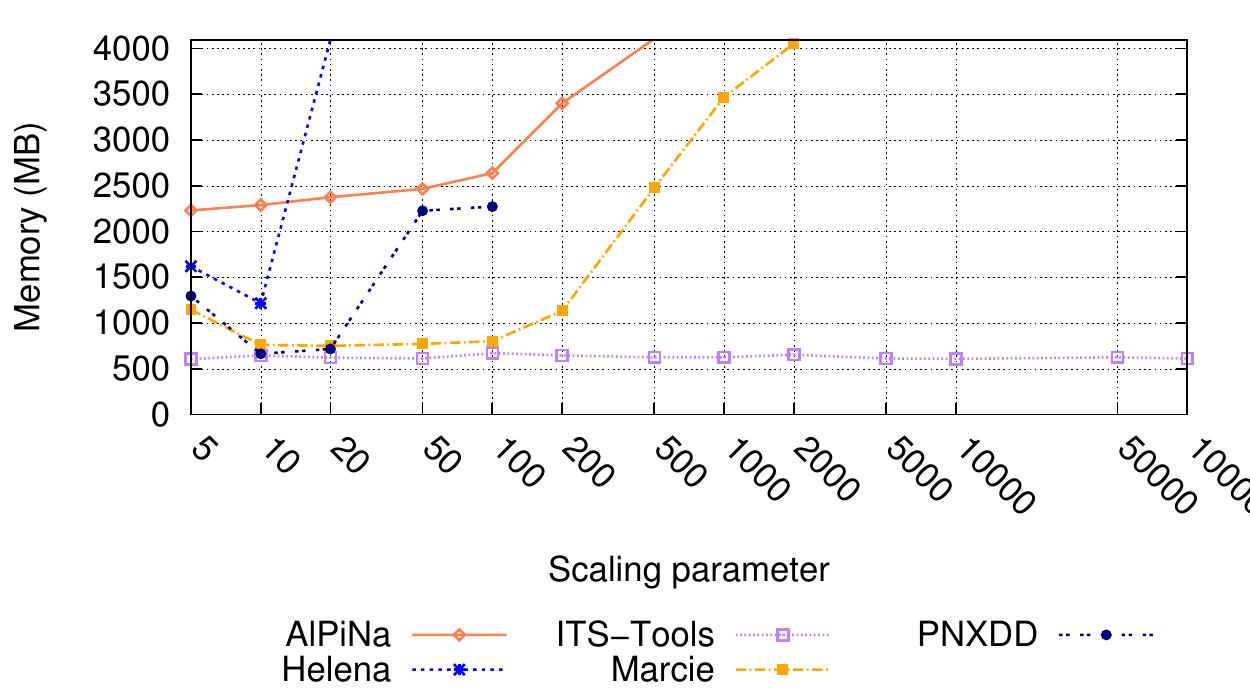}
\includegraphics[width=.5\textwidth]{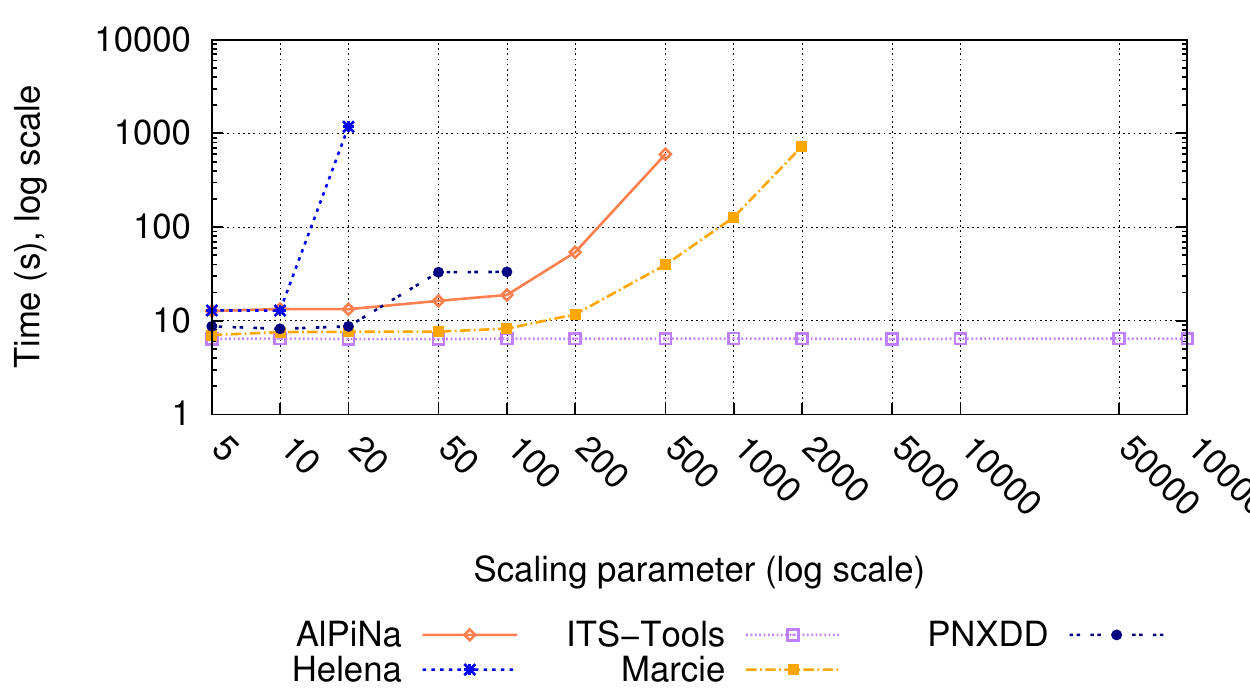}
}

\medskip
\noindent
\textbf{Charts for railroad}
\nopagebreak[4]\\
\makebox[\textwidth]{
\index{State Space Charts!railroad}
\includegraphics[width=.5\textwidth]{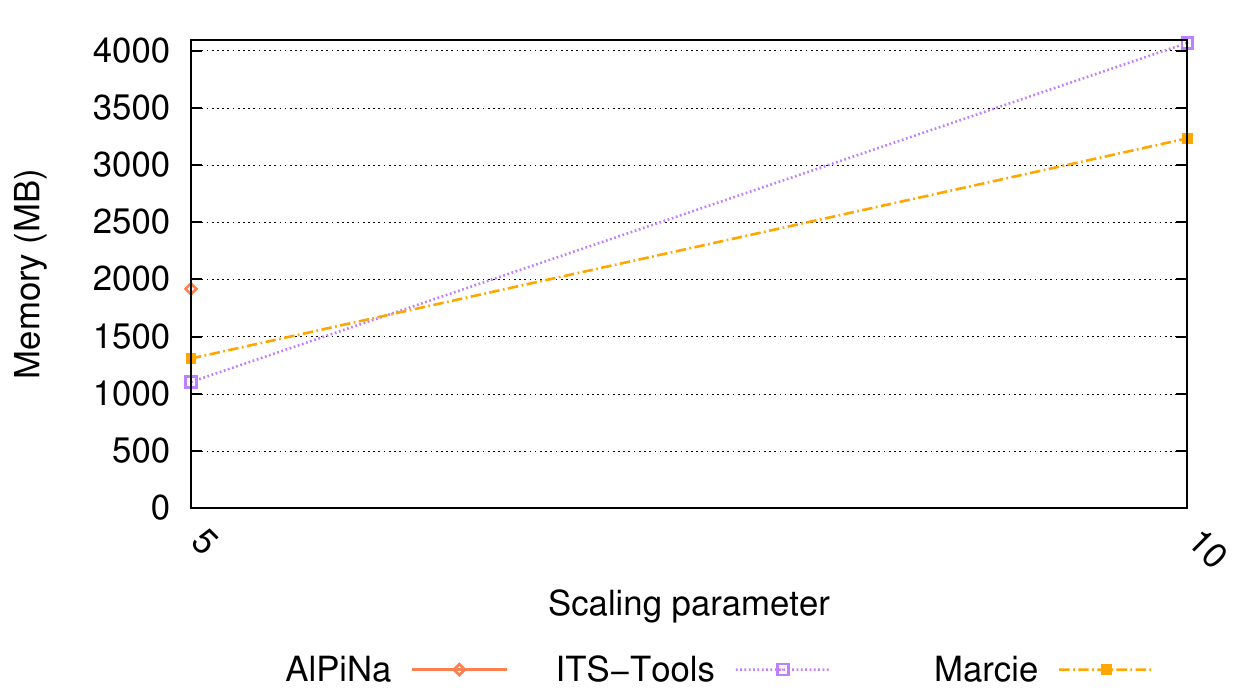}
\includegraphics[width=.5\textwidth]{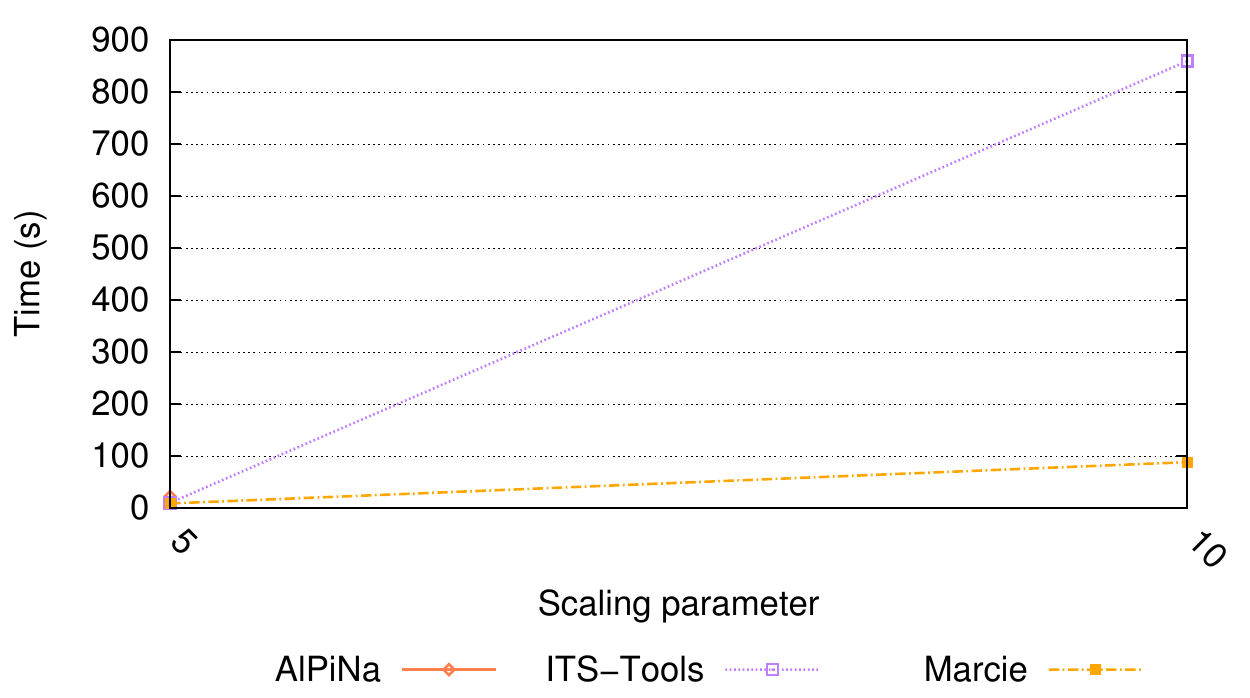}
}

\medskip
\noindent
\textbf{Charts for ring}
\nopagebreak[4]\\
\makebox[\textwidth]{
\index{State Space Charts!ring}
\includegraphics[width=.5\textwidth]{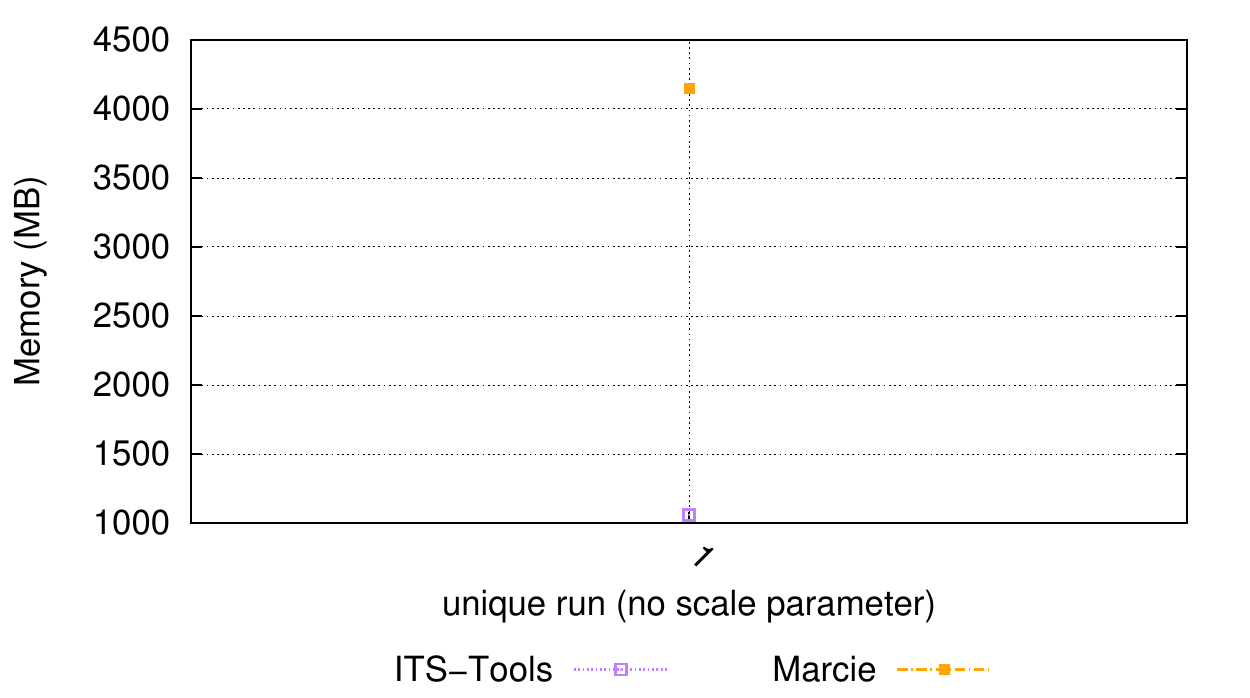}
\includegraphics[width=.5\textwidth]{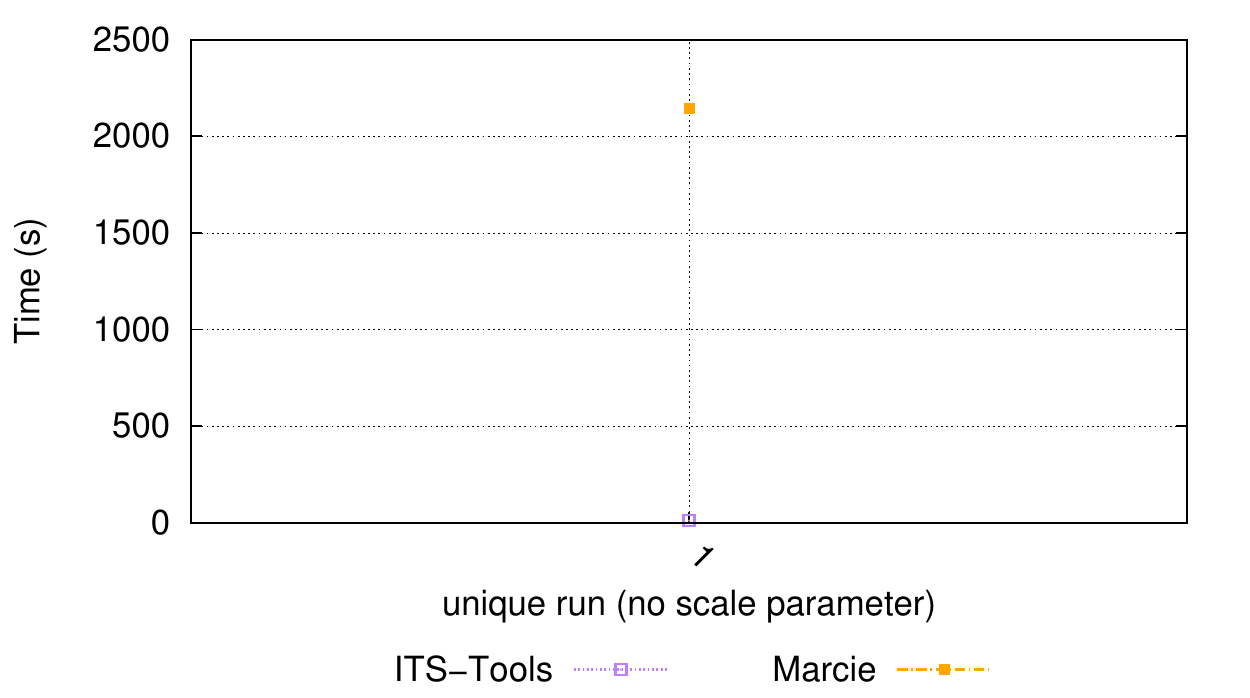}
}

\medskip
\noindent
\textbf{Charts for rwmutex}
\nopagebreak[4]\\
\makebox[\textwidth]{
\index{State Space Charts!rwmutex}
\includegraphics[width=.5\textwidth]{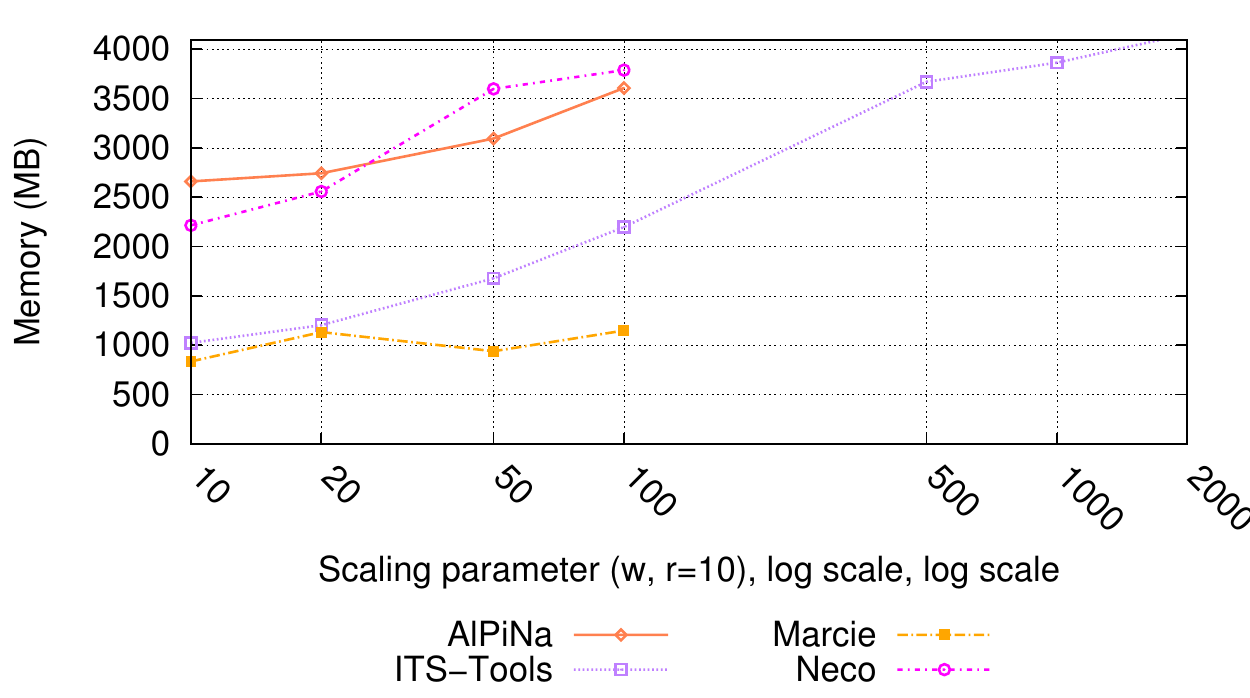}
\includegraphics[width=.5\textwidth]{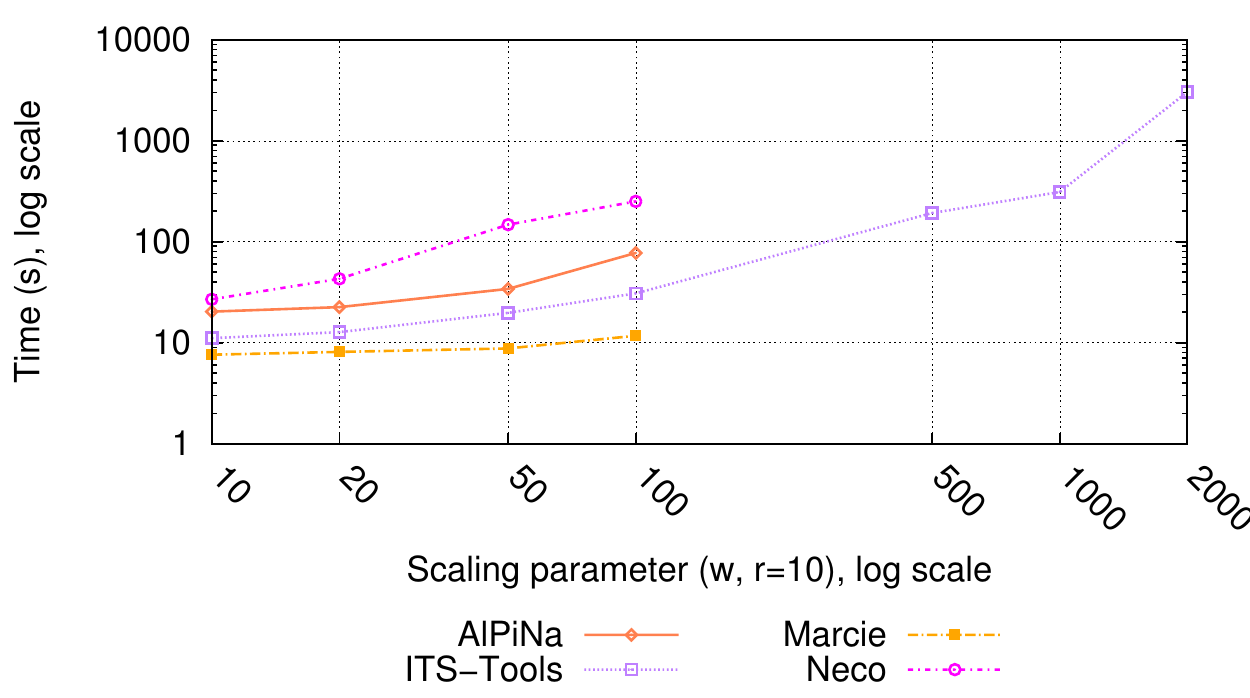}
}

\medskip
\noindent
\textbf{Charts for SharedMemory}
\nopagebreak[4]\\
\makebox[\textwidth]{
\index{State Space Charts!SharedMemory}
\includegraphics[width=.5\textwidth]{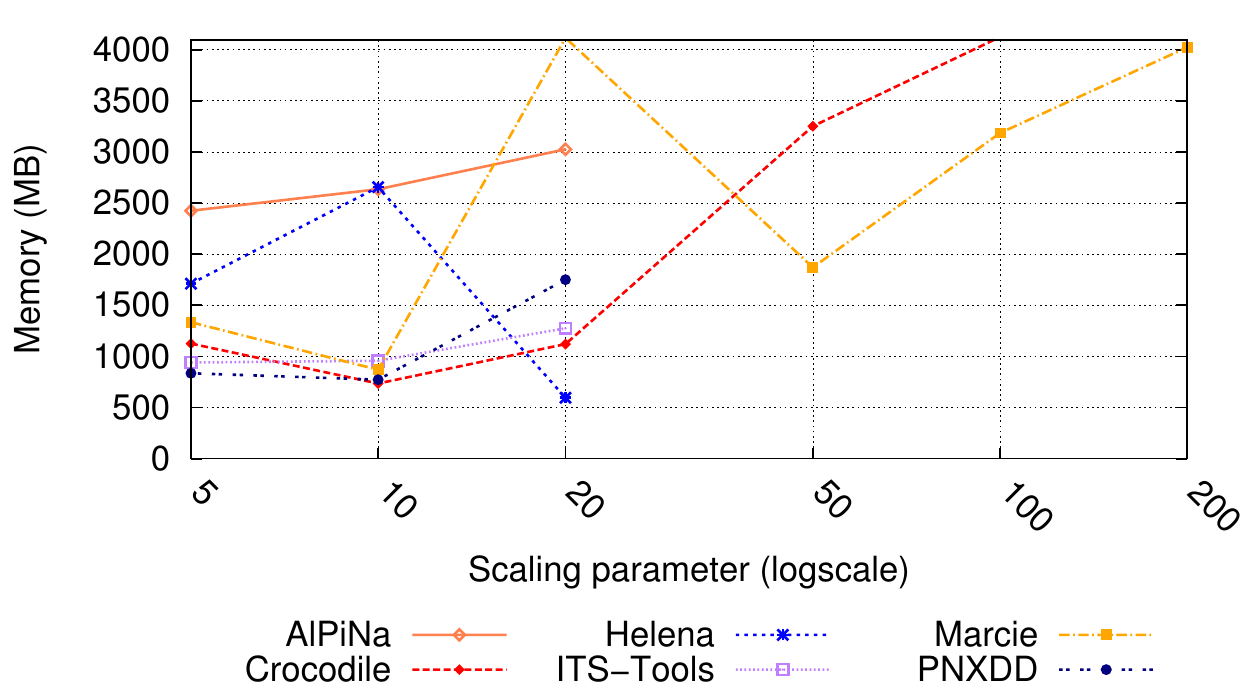}
\includegraphics[width=.5\textwidth]{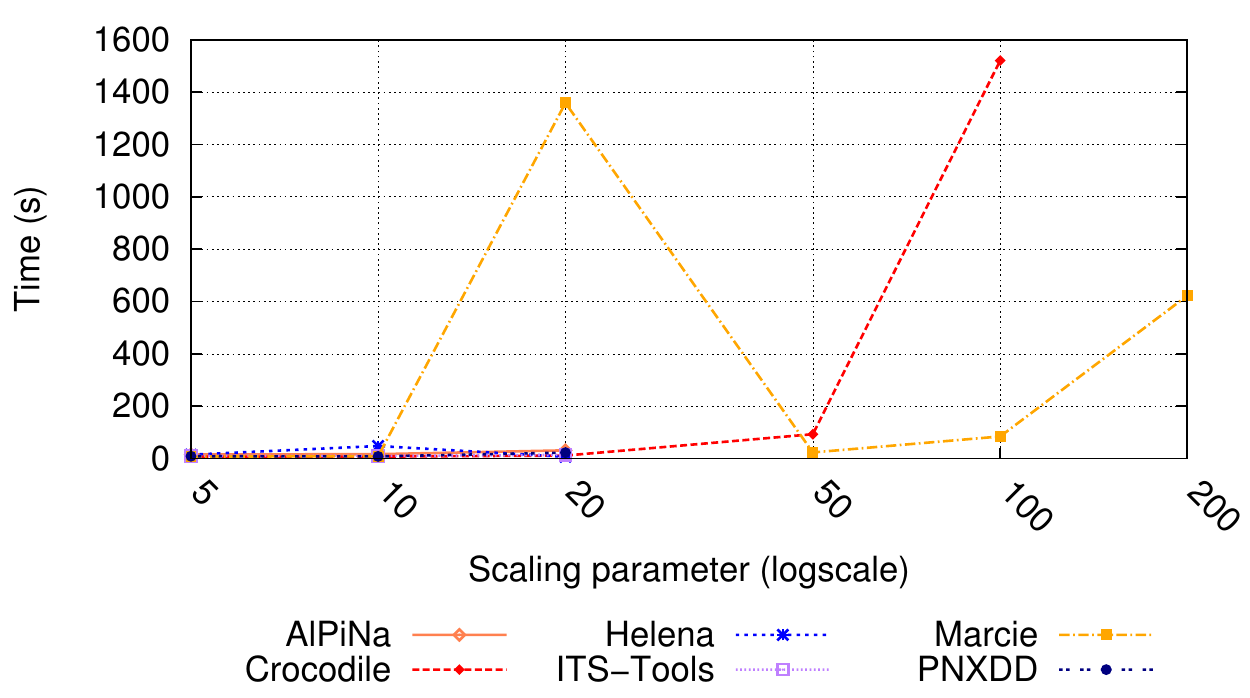}
}

\medskip
\noindent
\textbf{Charts for simple\_lbs}
\nopagebreak[4]\\
\makebox[\textwidth]{
\index{State Space Charts!simple\_lbs}
\includegraphics[width=.5\textwidth]{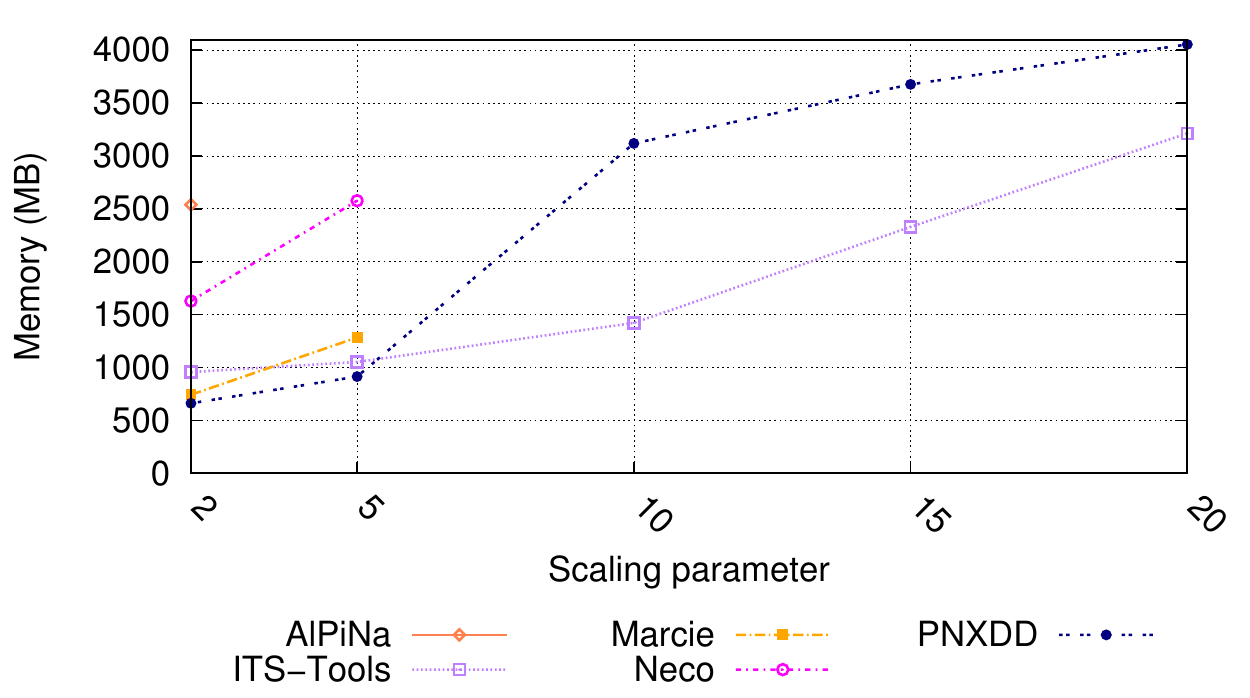}
\includegraphics[width=.5\textwidth]{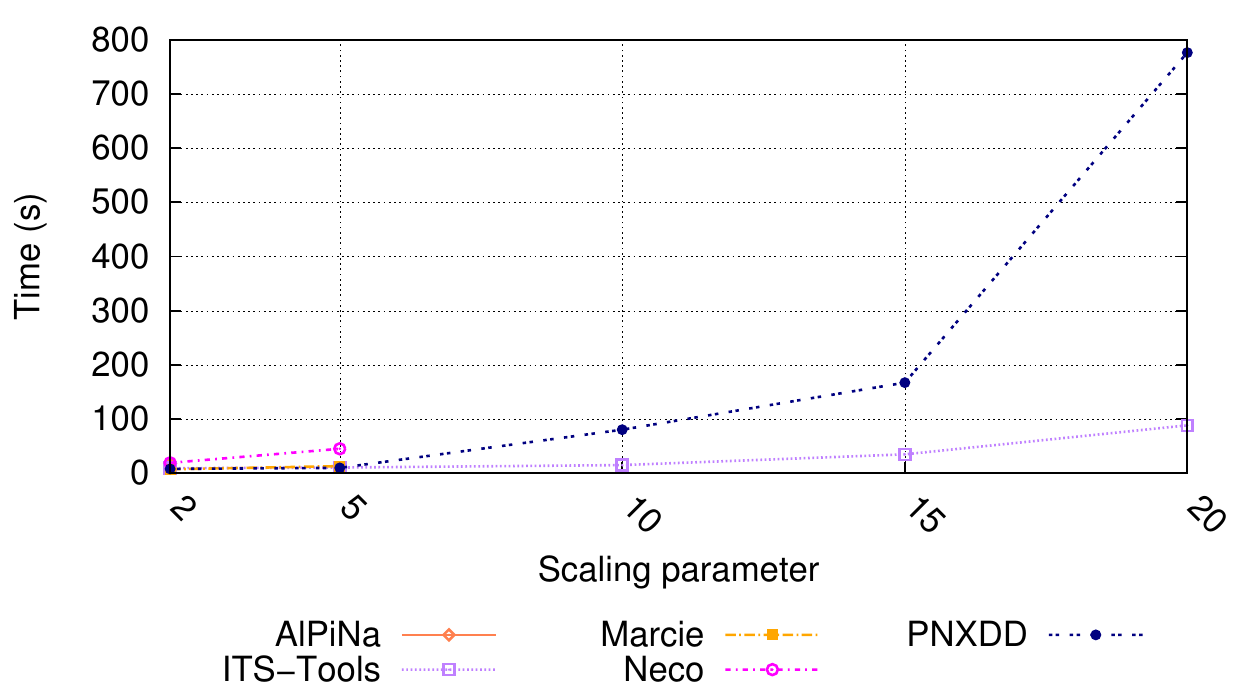}
}

\medskip
\noindent
\textbf{Charts for TokenRing}
\nopagebreak[4]\\
\makebox[\textwidth]{
\index{State Space Charts!TokenRing}
\includegraphics[width=.5\textwidth]{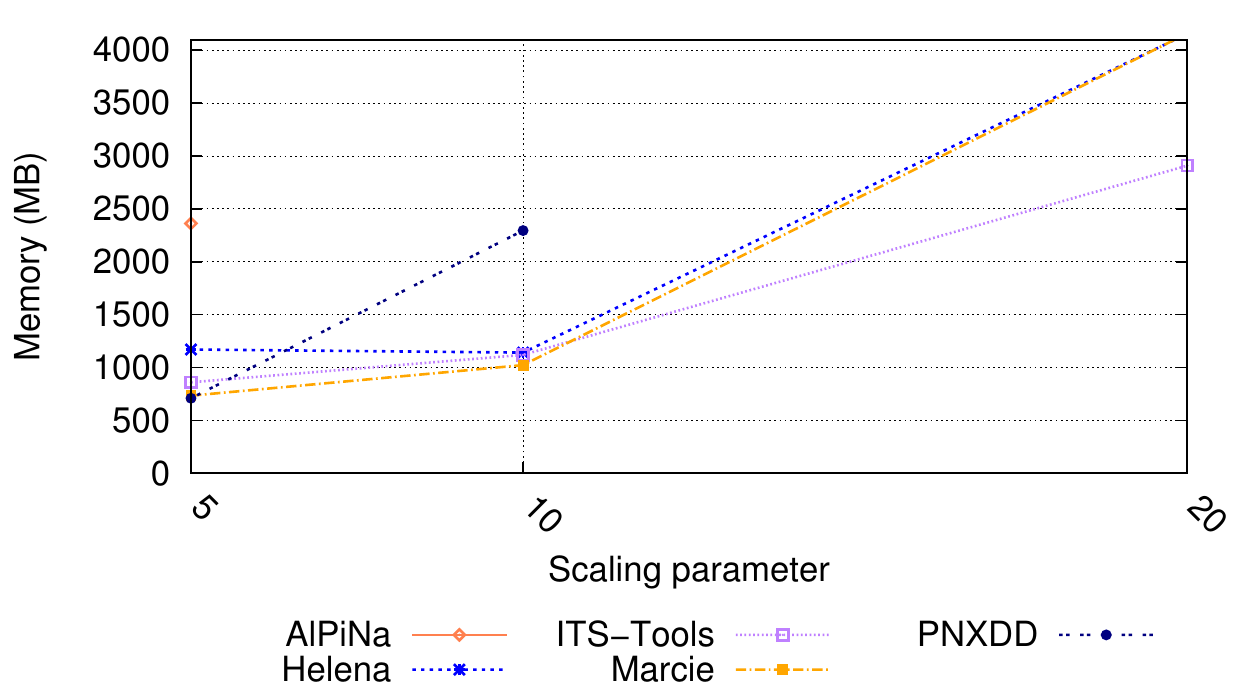}
\includegraphics[width=.5\textwidth]{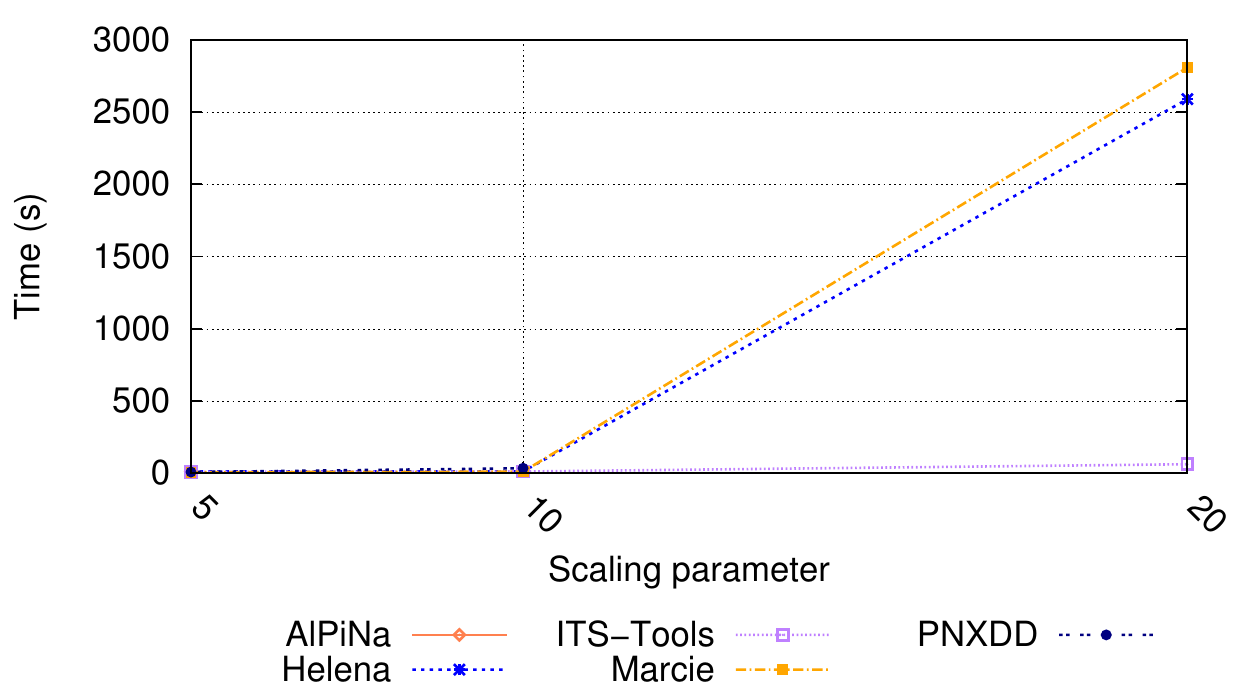}
}

\subsection{Execution Charts for AlPiNA}
\label{sec:sstoolfirst}

We provide here the execution charts observed for AlPiNA over
the models it could compete with.

\subsubsection{Executions for cs\_repetitions}
1 chart has been generated.
\index{Execution (by tool)!AlPiNA}
\index{Execution (by model)!cs\_repetitions!AlPiNA}

\noindent\includegraphics[width=.5\textwidth]{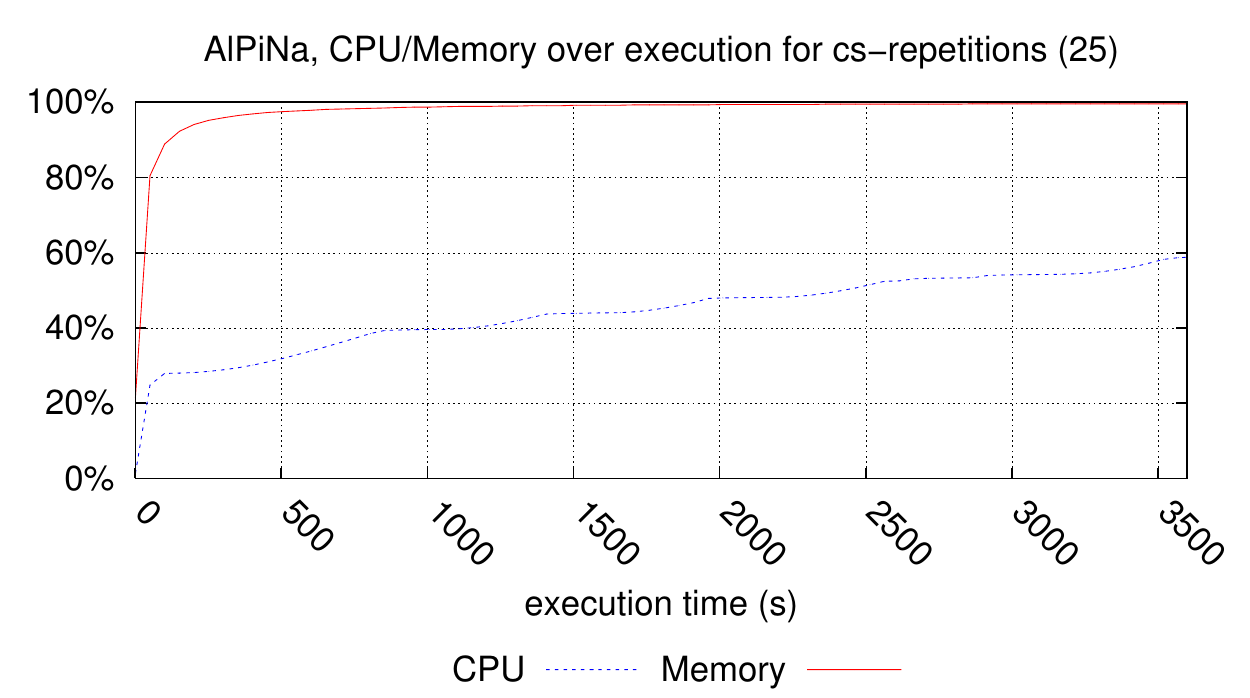}

\subsubsection{Executions for echo}
1 chart has been generated.
\index{Execution (by tool)!AlPiNA}
\index{Execution (by model)!echo!AlPiNA}

\noindent\includegraphics[width=.5\textwidth]{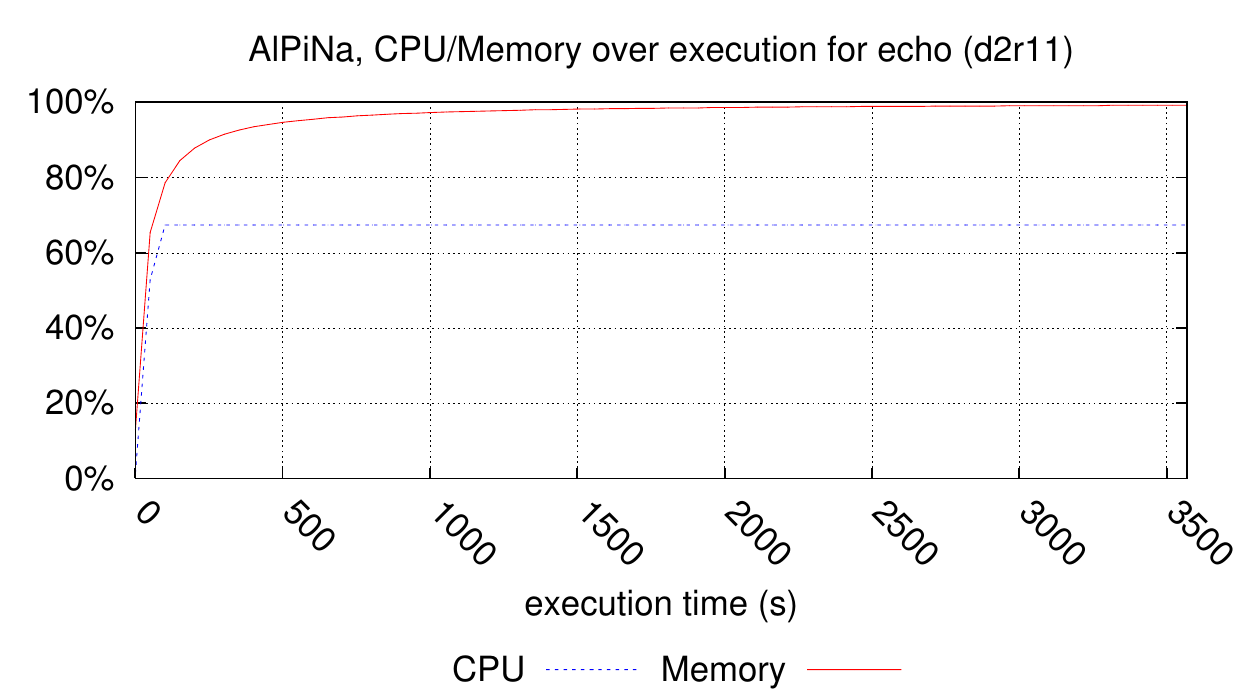}

\subsubsection{Executions for eratosthenes}
7 charts have been generated.
\index{Execution (by tool)!AlPiNA}
\index{Execution (by model)!eratosthenes!AlPiNA}

\noindent\includegraphics[width=.5\textwidth]{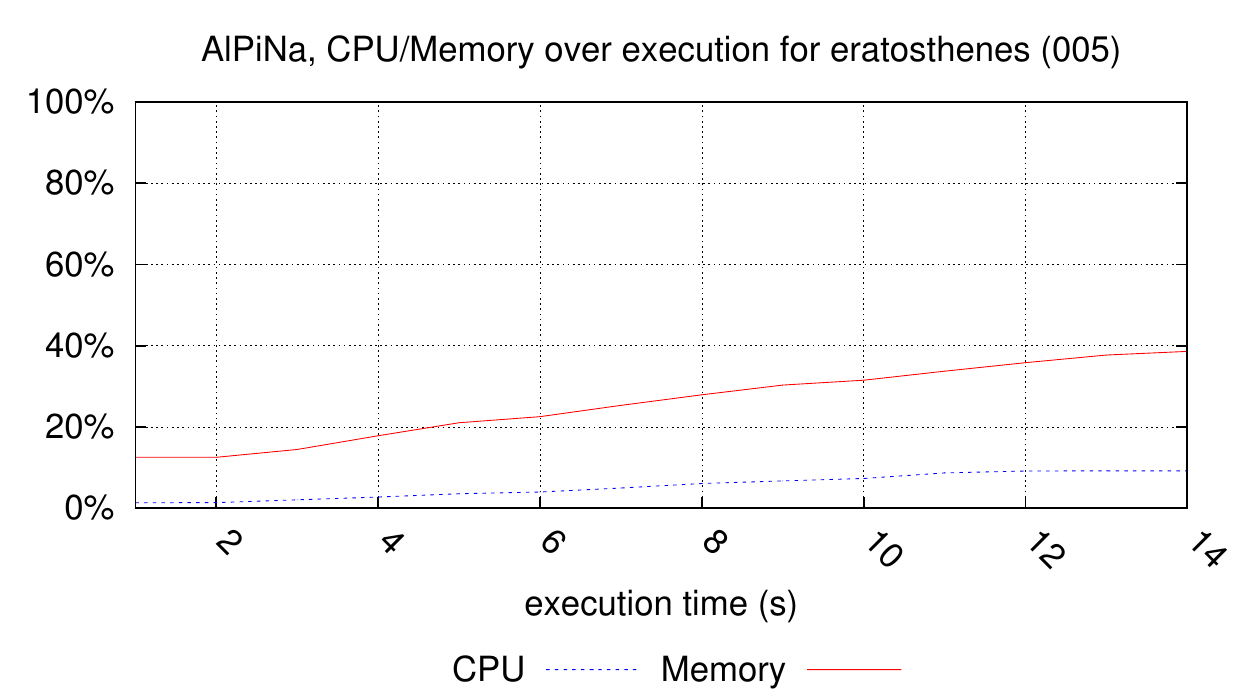}
\includegraphics[width=.5\textwidth]{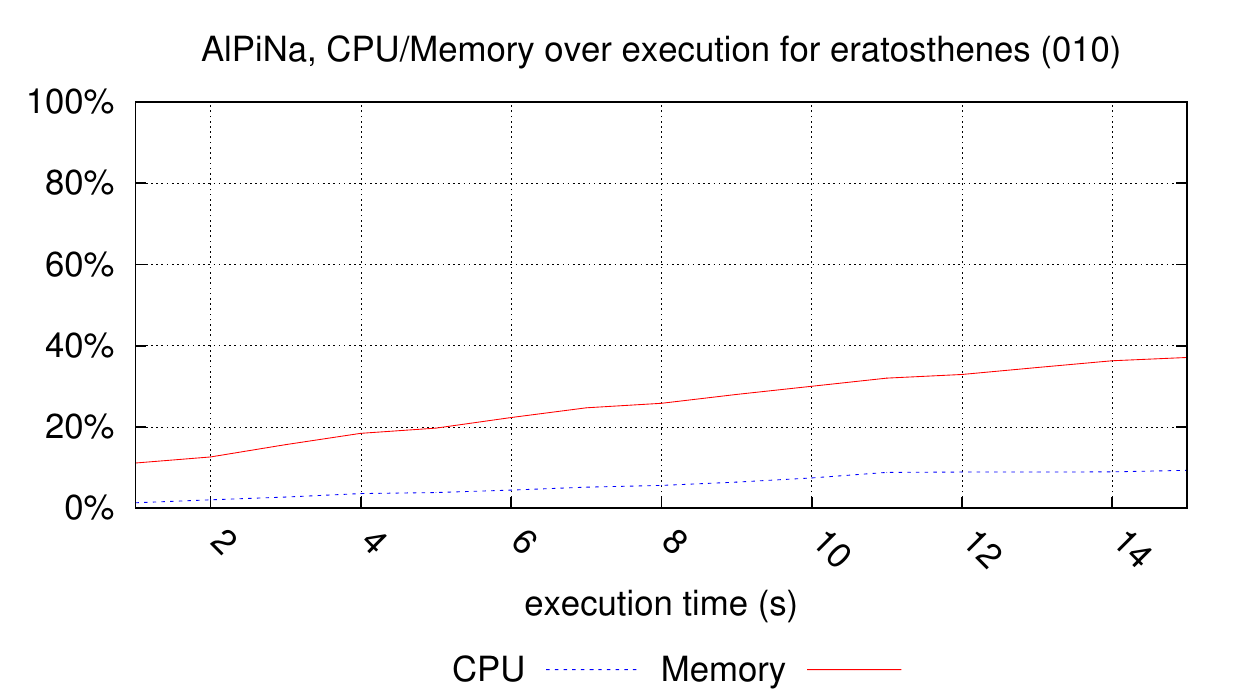}

\noindent\includegraphics[width=.5\textwidth]{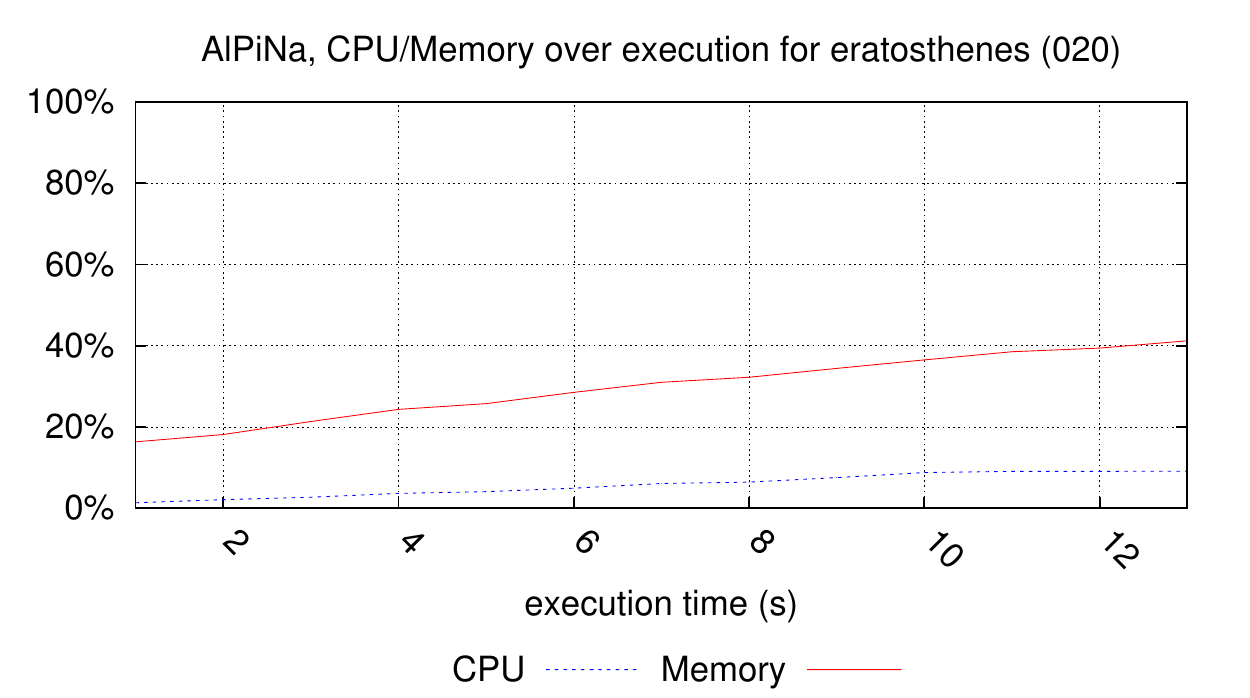}
\includegraphics[width=.5\textwidth]{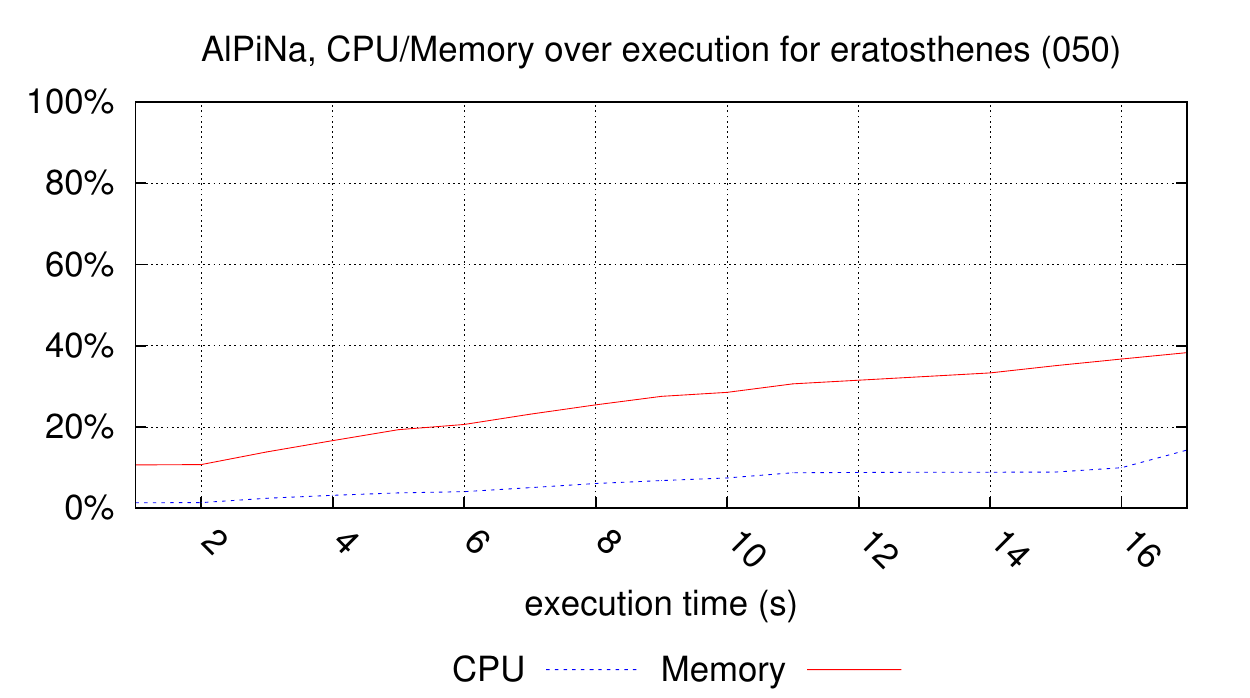}

\noindent\includegraphics[width=.5\textwidth]{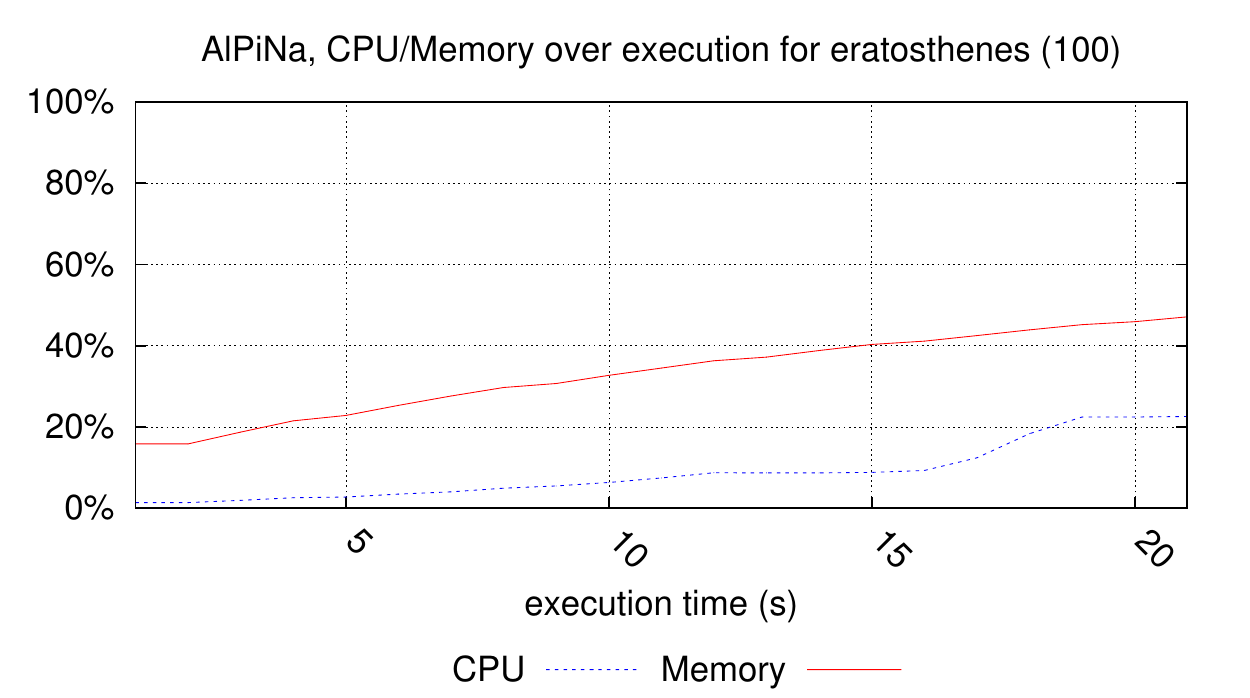}
\includegraphics[width=.5\textwidth]{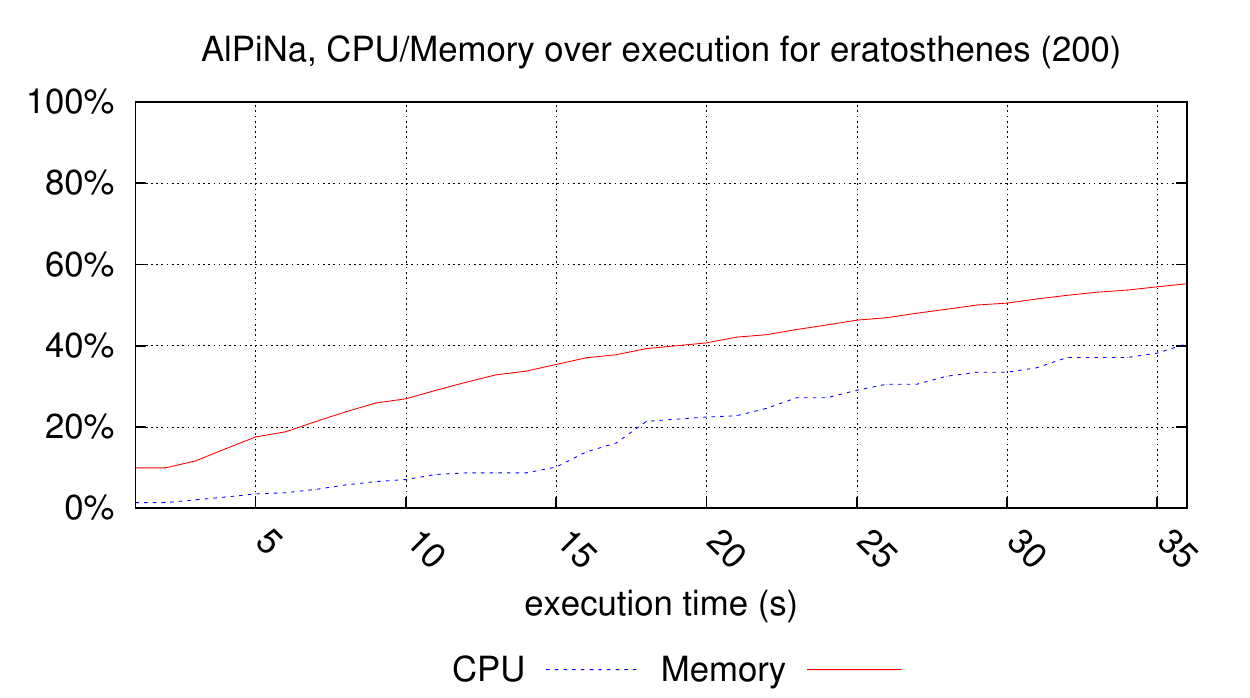}

\noindent\includegraphics[width=.5\textwidth]{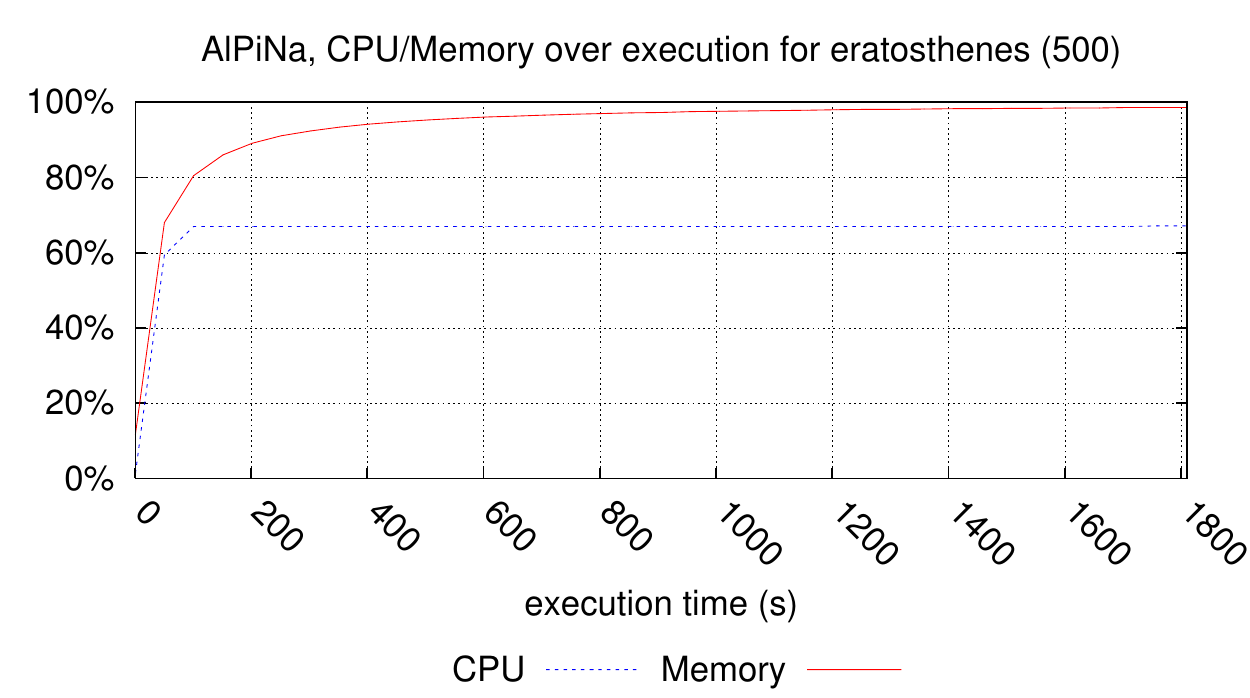}

\subsubsection{Executions for FMS}
5 charts have been generated.
\index{Execution (by tool)!AlPiNA}
\index{Execution (by model)!FMS!AlPiNA}

\noindent\includegraphics[width=.5\textwidth]{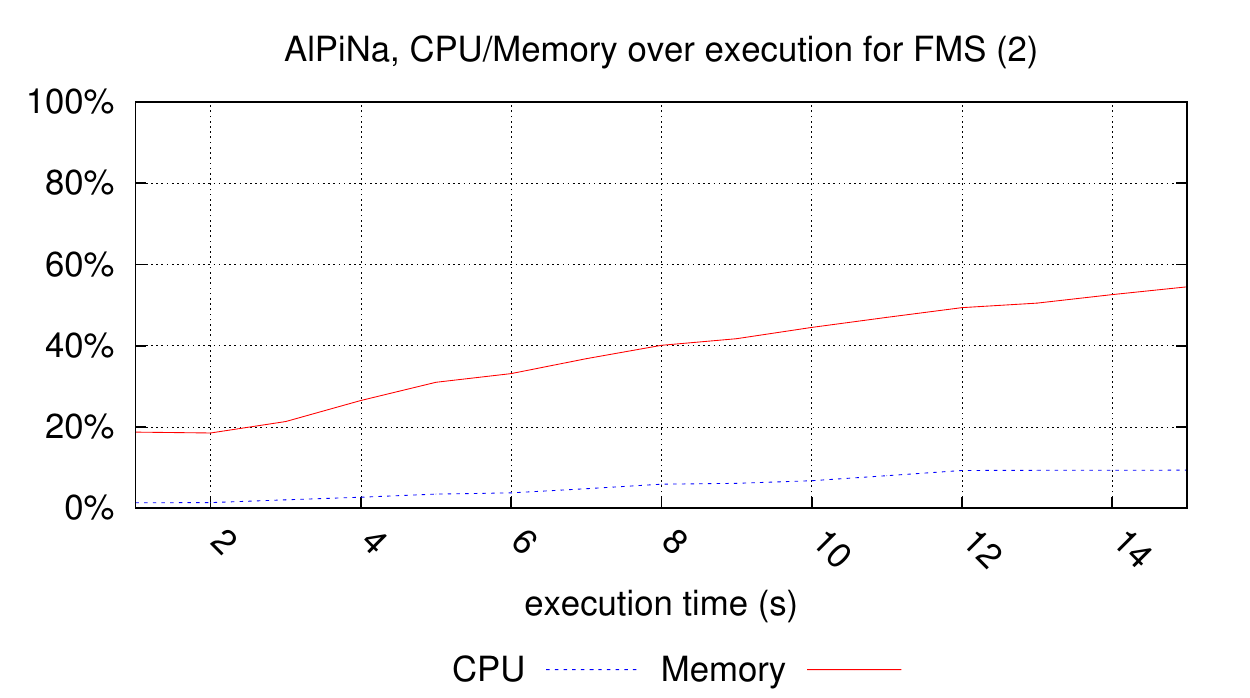}
\includegraphics[width=.5\textwidth]{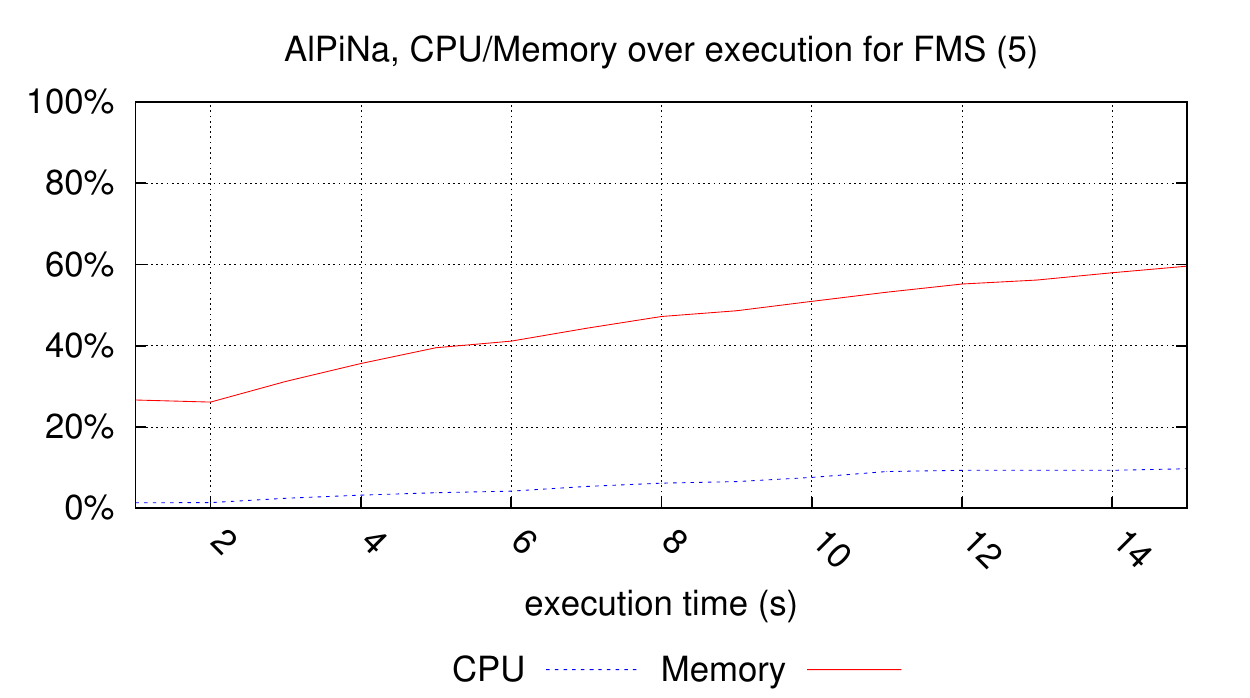}

\noindent\includegraphics[width=.5\textwidth]{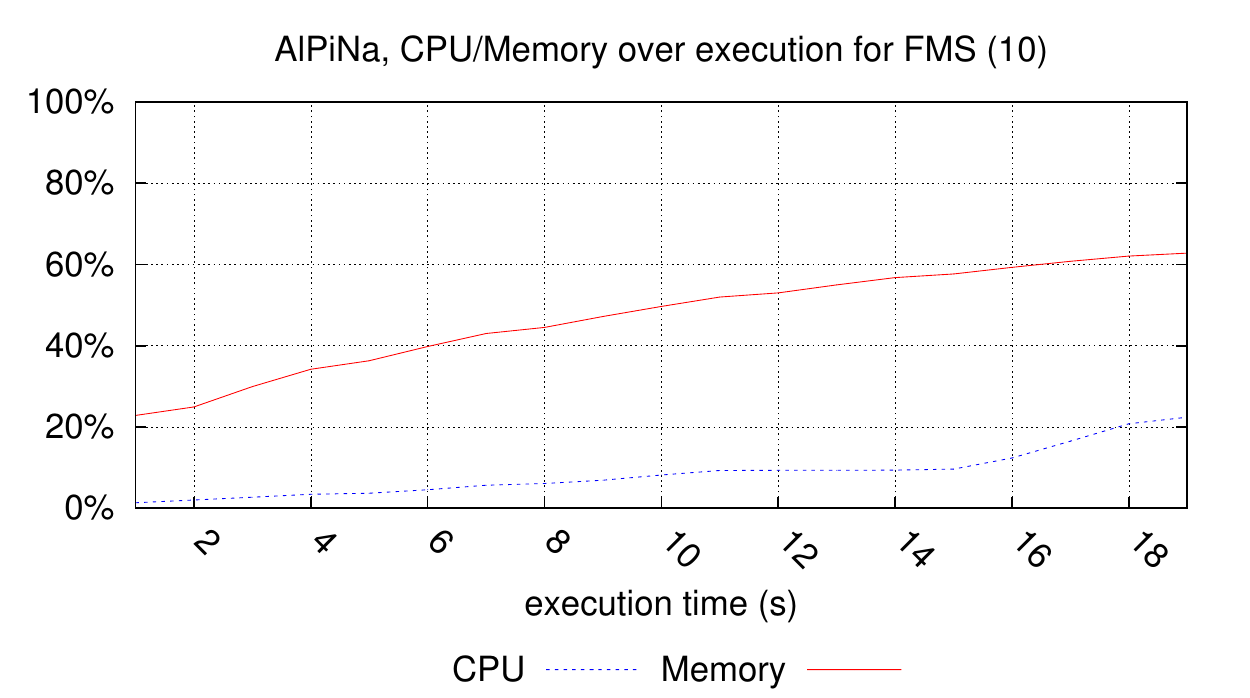}
\includegraphics[width=.5\textwidth]{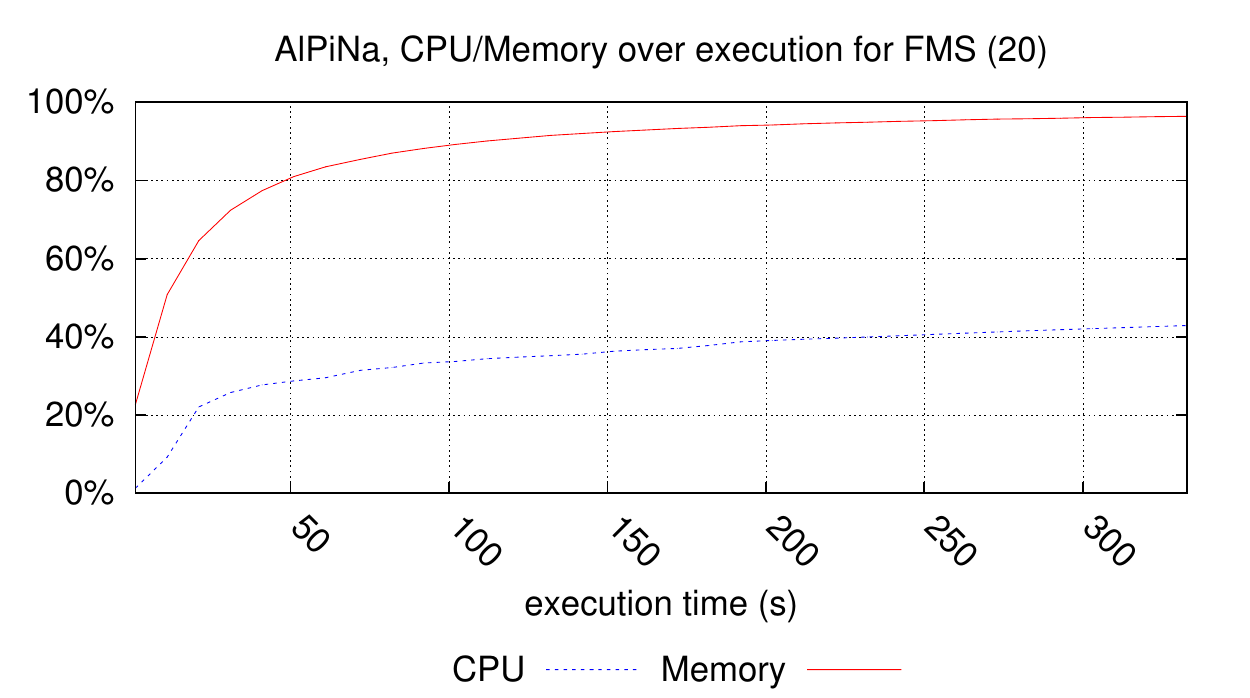}

\noindent\includegraphics[width=.5\textwidth]{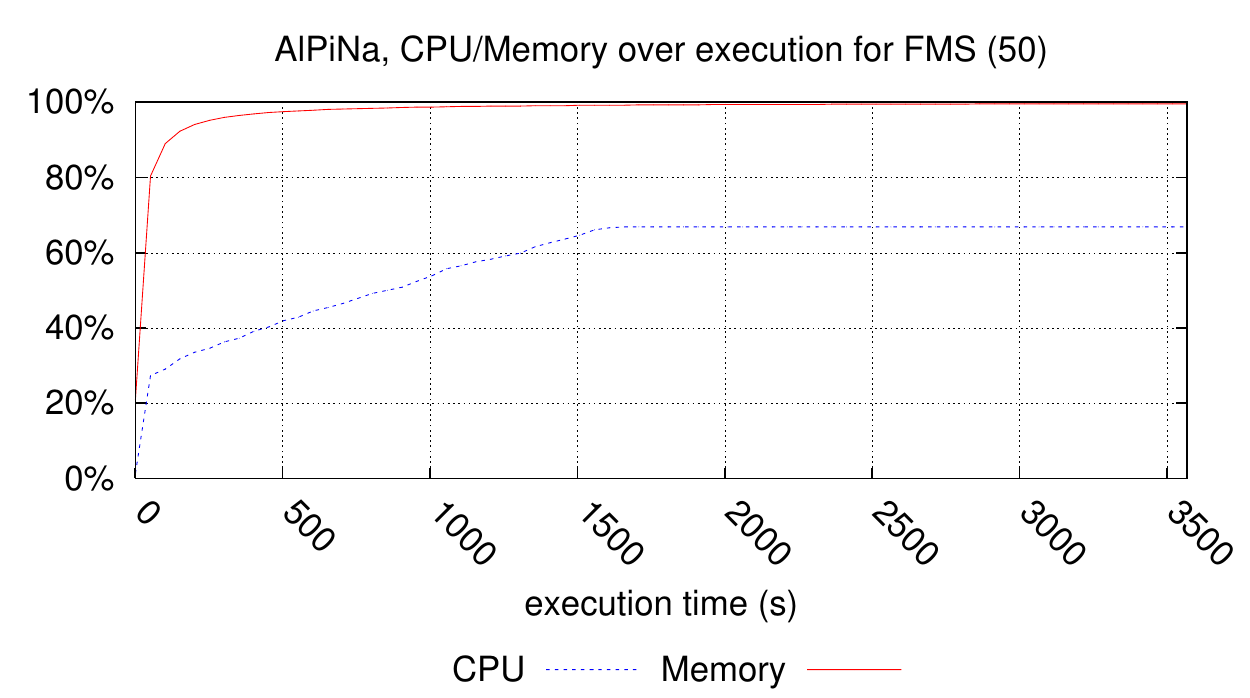}

\subsubsection{Executions for galloc\_res}
1 chart has been generated.
\index{Execution (by tool)!AlPiNA}
\index{Execution (by model)!galloc\_res!AlPiNA}

\noindent\includegraphics[width=.5\textwidth]{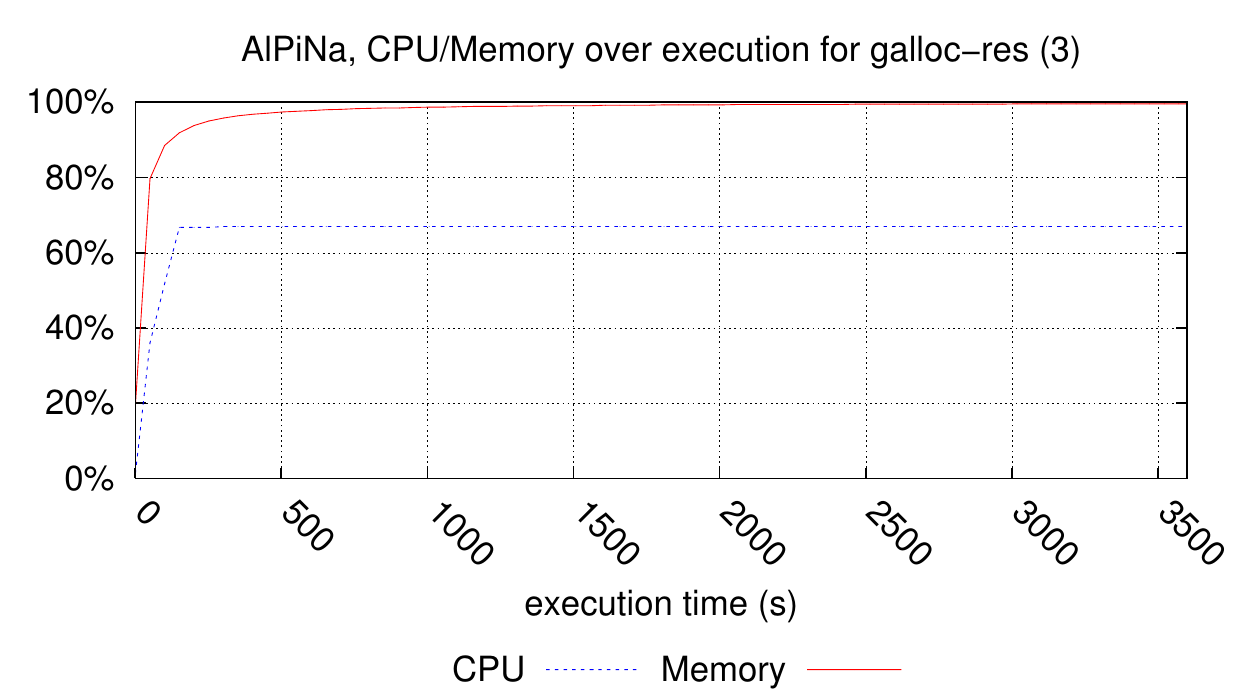}

\subsubsection{Executions for Kanban}
4 charts have been generated.
\index{Execution (by tool)!AlPiNA}
\index{Execution (by model)!Kanban!AlPiNA}

\noindent\includegraphics[width=.5\textwidth]{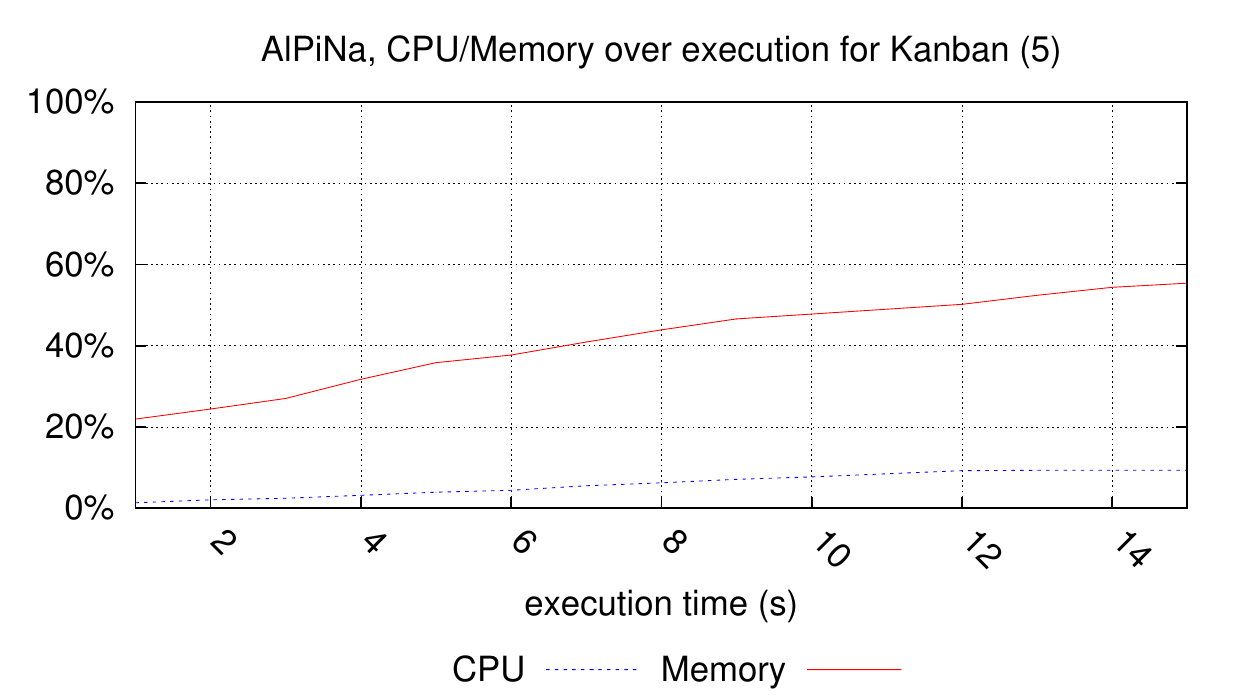}
\includegraphics[width=.5\textwidth]{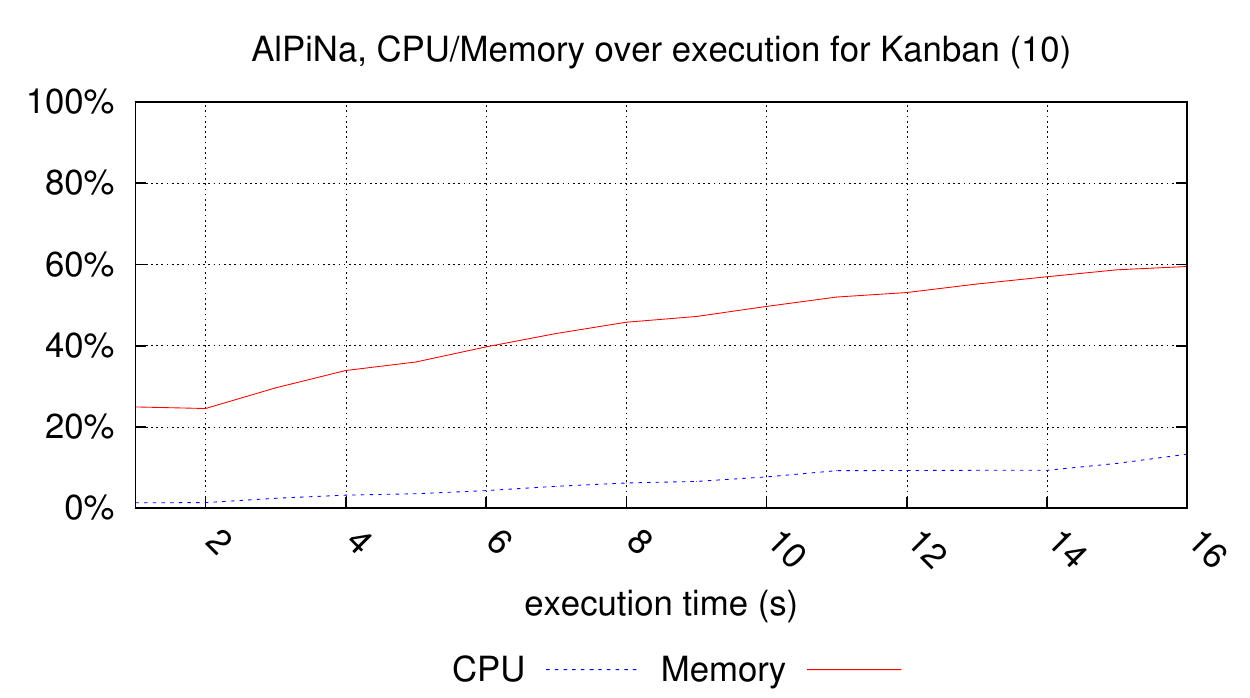}

\noindent\includegraphics[width=.5\textwidth]{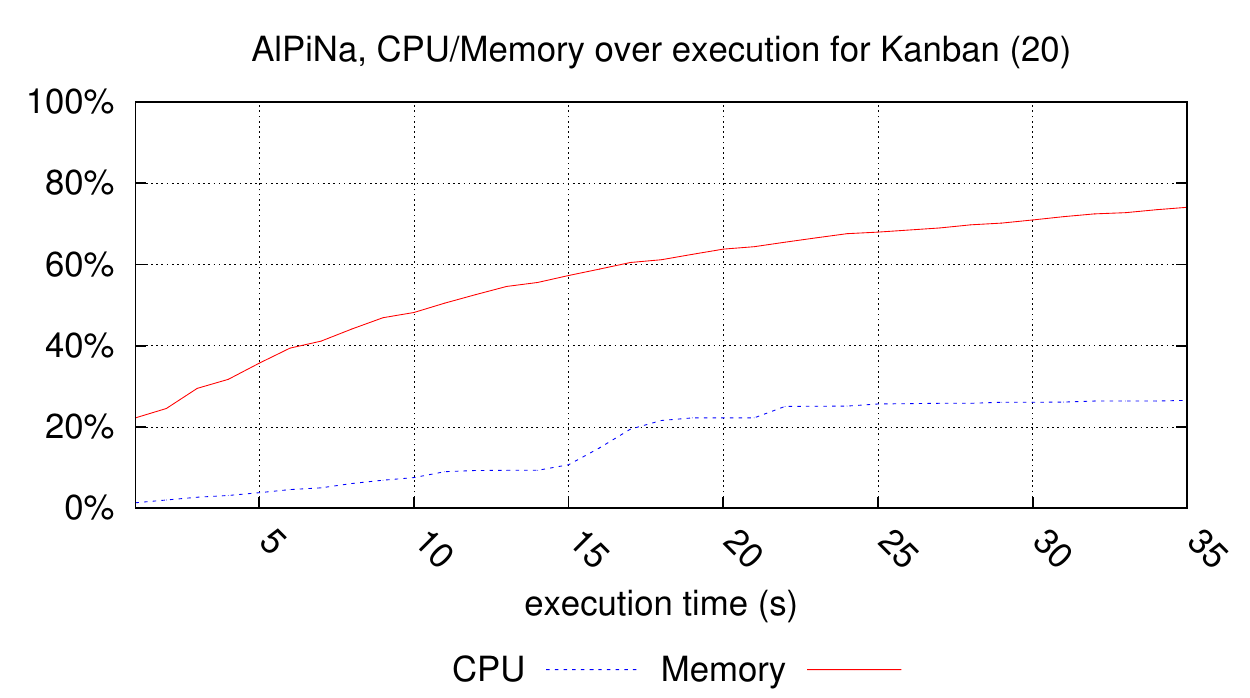}
\includegraphics[width=.5\textwidth]{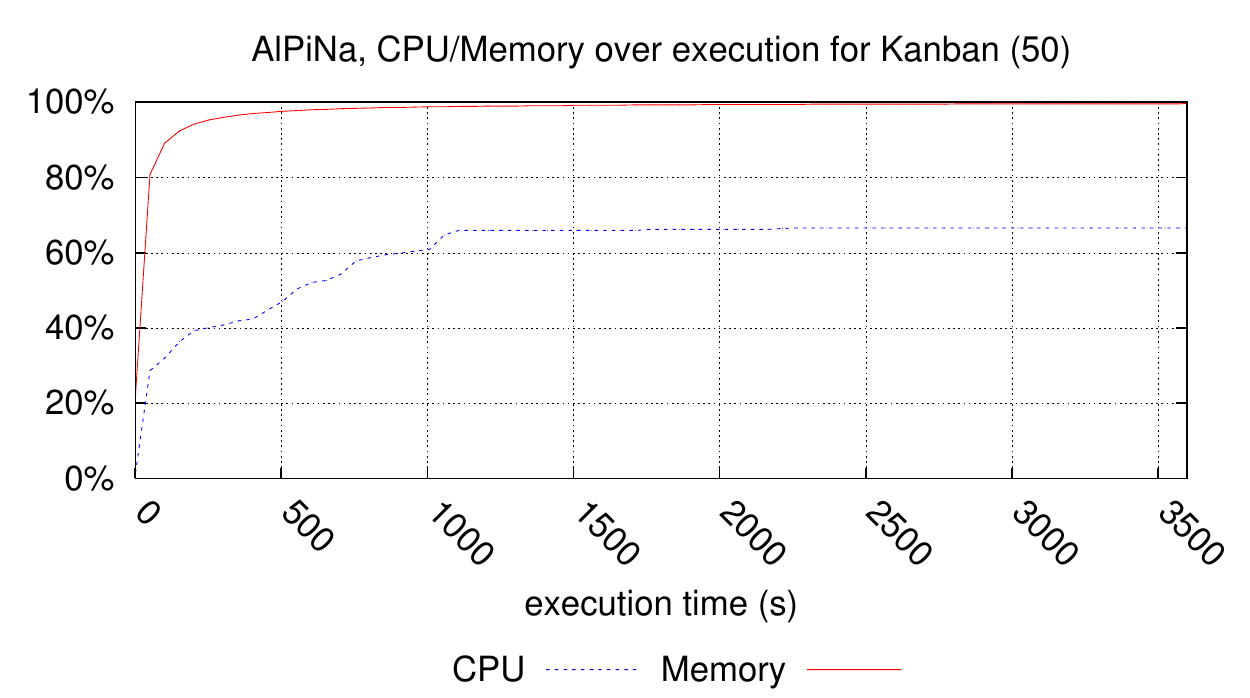}

\vfill\eject
\subsubsection{Executions for lamport\_fmea}
3 charts have been generated.
\index{Execution (by tool)!AlPiNA}
\index{Execution (by model)!lamport\_fmea!AlPiNA}

\noindent\includegraphics[width=.5\textwidth]{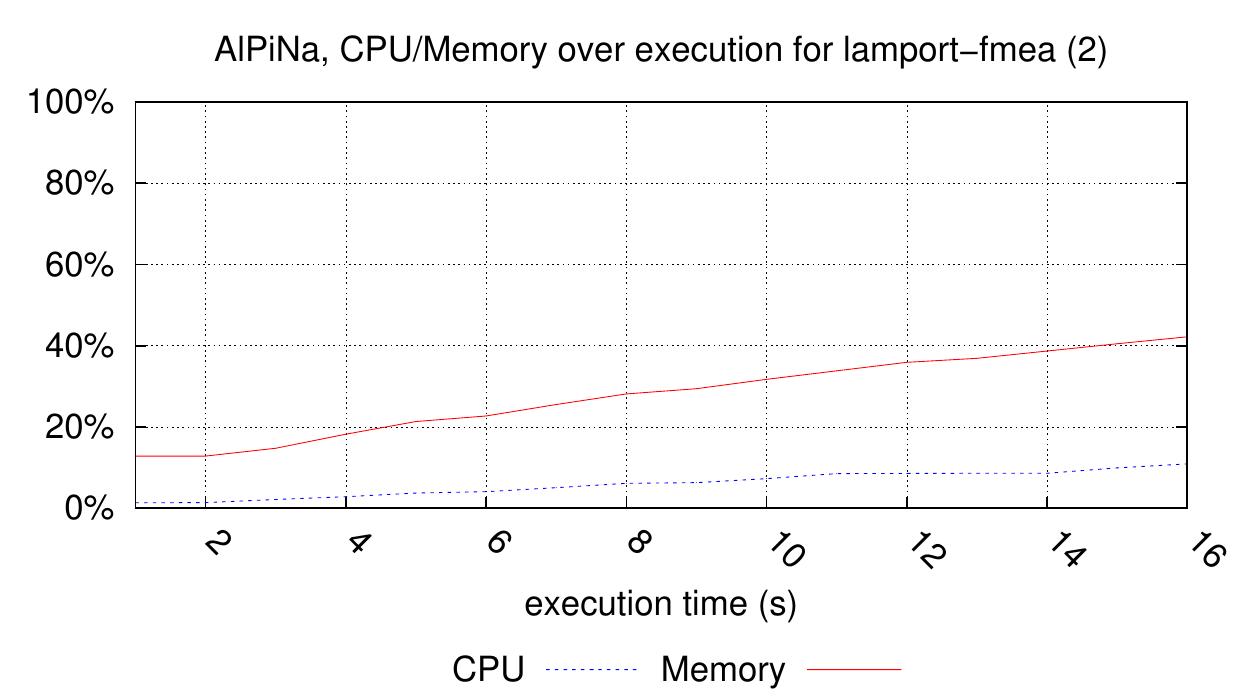}
\includegraphics[width=.5\textwidth]{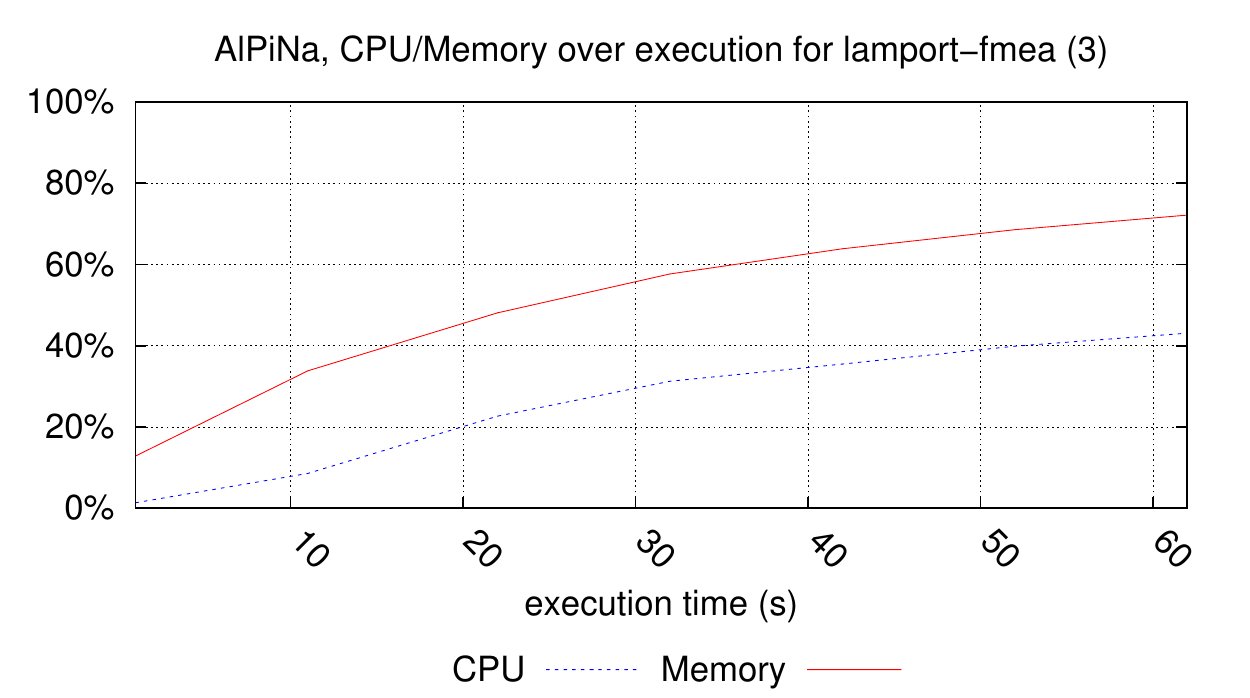}

\noindent\includegraphics[width=.5\textwidth]{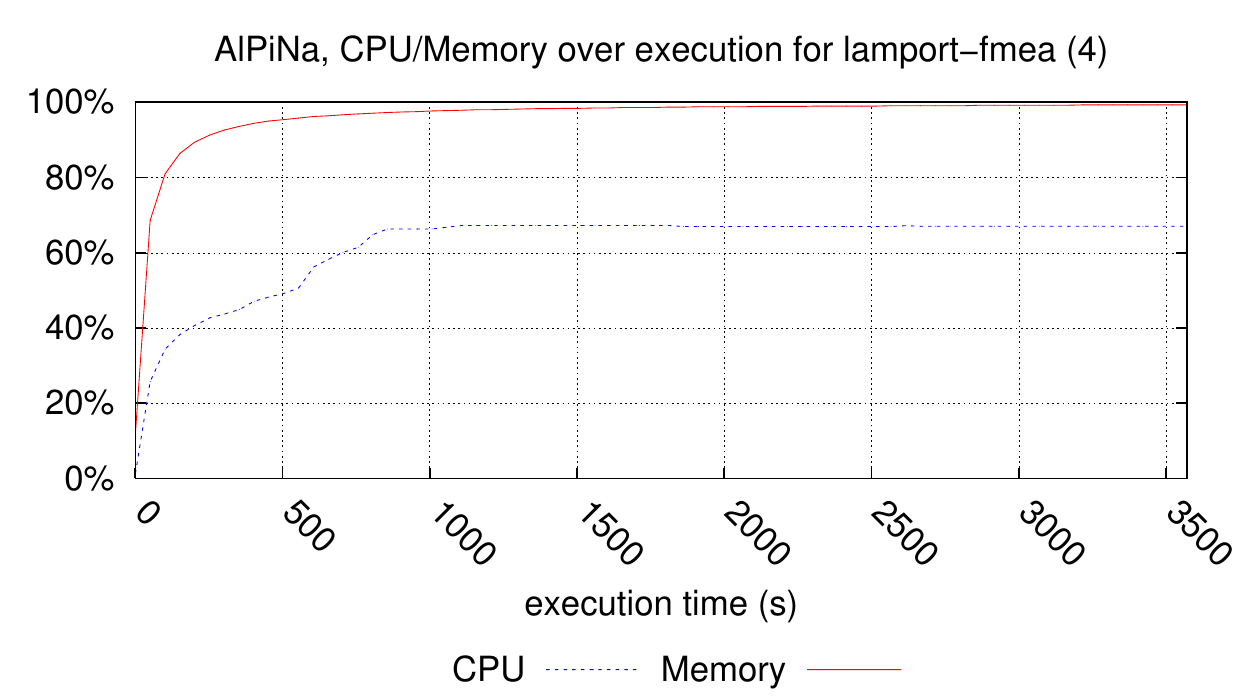}

\subsubsection{Executions for MAPK}
1 chart has been generated.
\index{Execution (by tool)!AlPiNA}
\index{Execution (by model)!MAPK!AlPiNA}

\noindent\includegraphics[width=.5\textwidth]{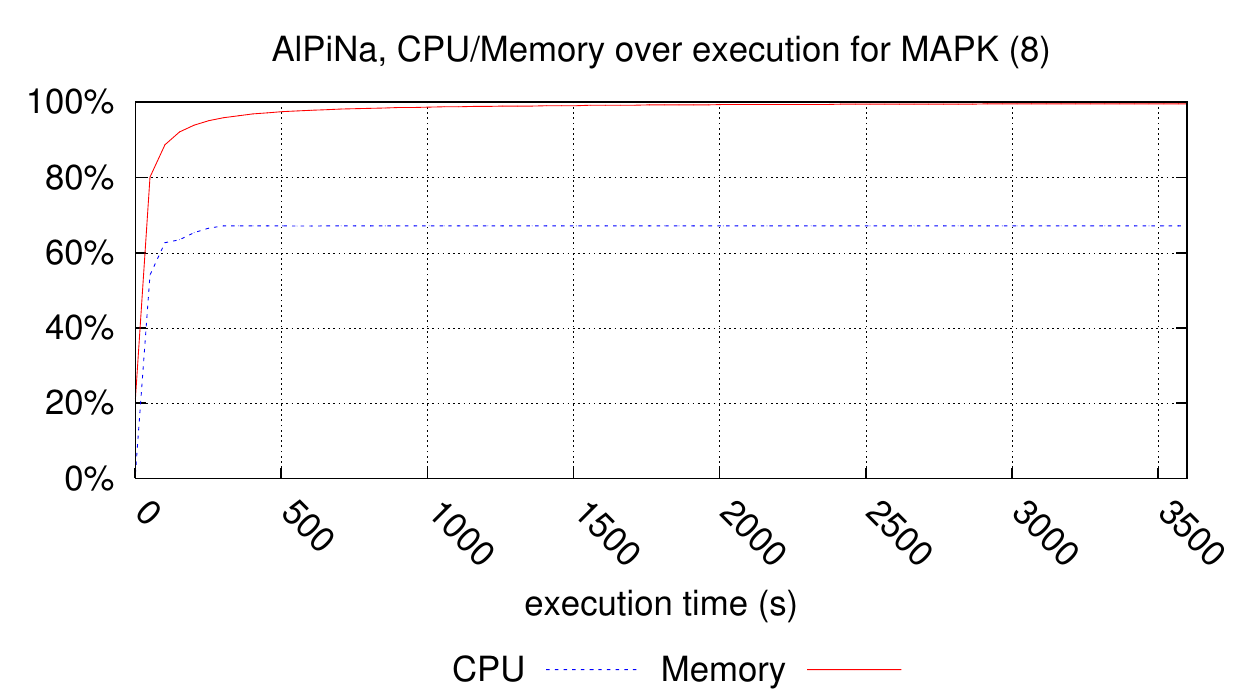}

\subsubsection{Executions for neo-election}
1 chart has been generated.
\index{Execution (by tool)!AlPiNA}
\index{Execution (by model)!neo-election!AlPiNA}

\noindent\includegraphics[width=.5\textwidth]{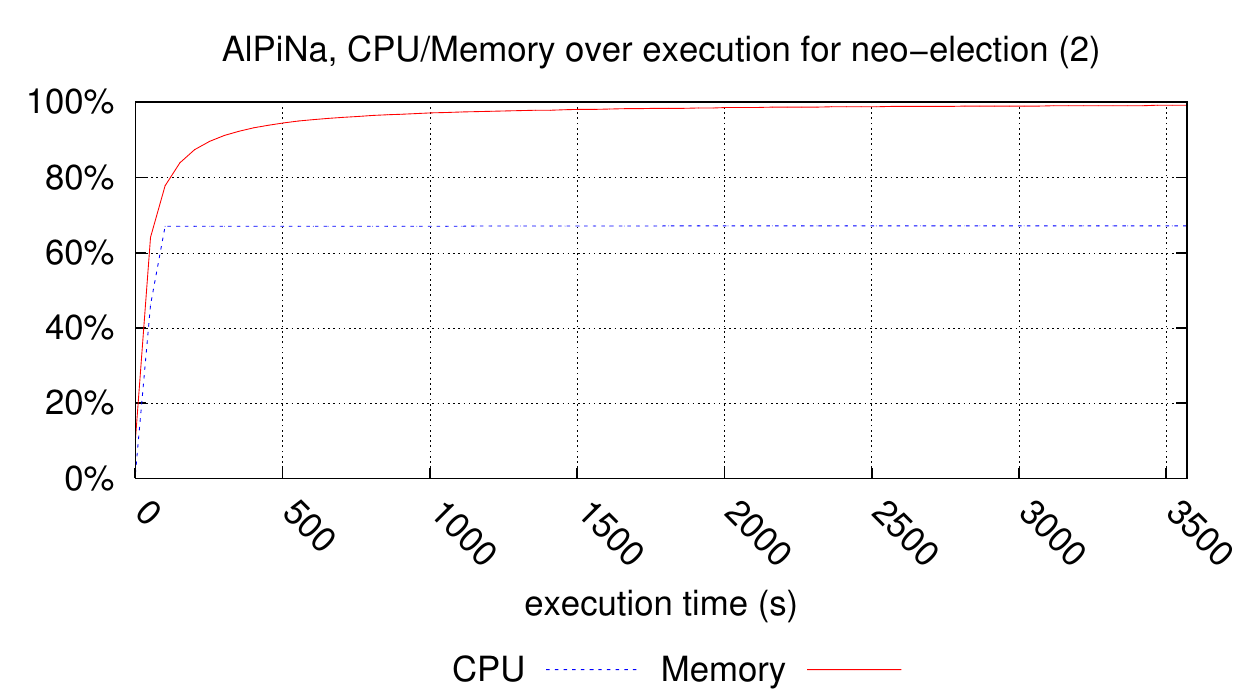}

\vfill\eject
\subsubsection{Executions for Peterson}
3 charts have been generated.
\index{Execution (by tool)!AlPiNA}
\index{Execution (by model)!Peterson!AlPiNA}

\noindent\includegraphics[width=.5\textwidth]{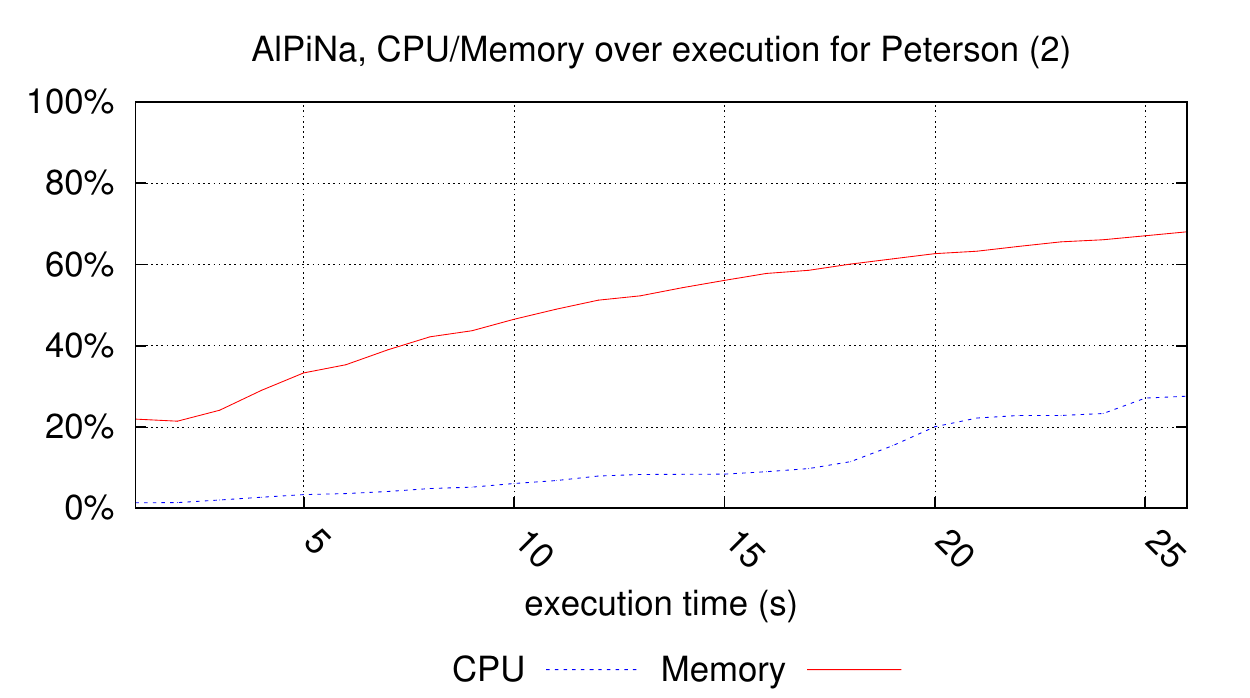}
\includegraphics[width=.5\textwidth]{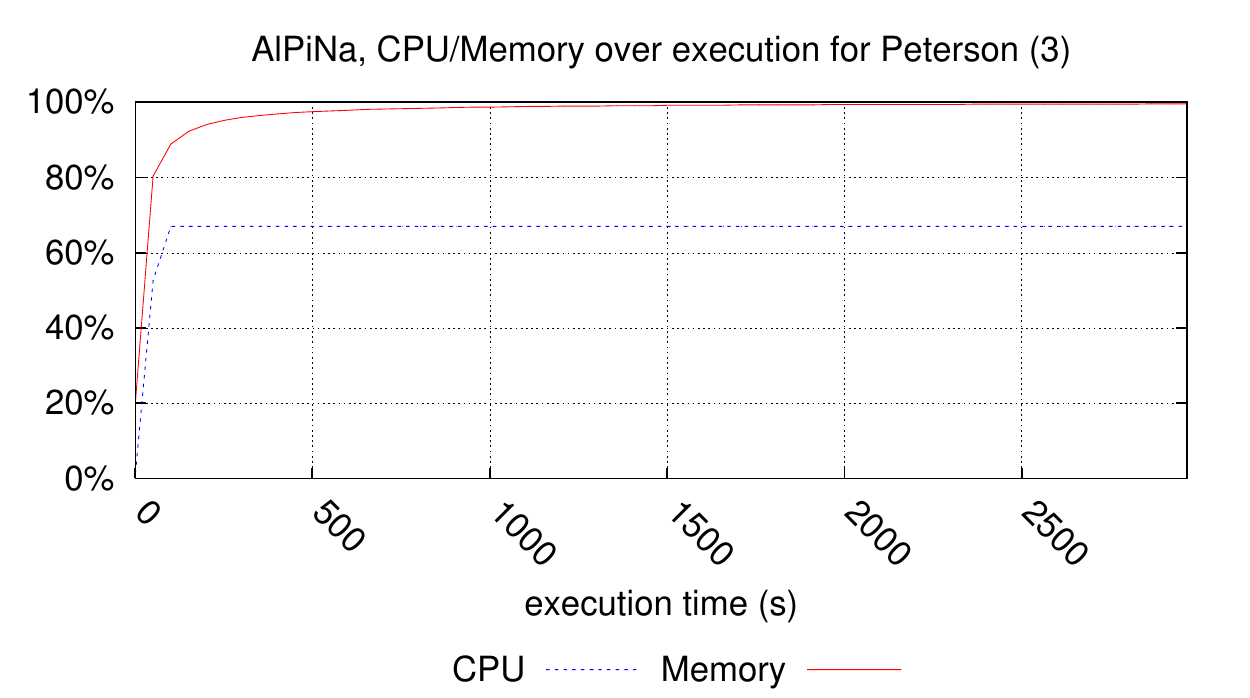}

\noindent\includegraphics[width=.5\textwidth]{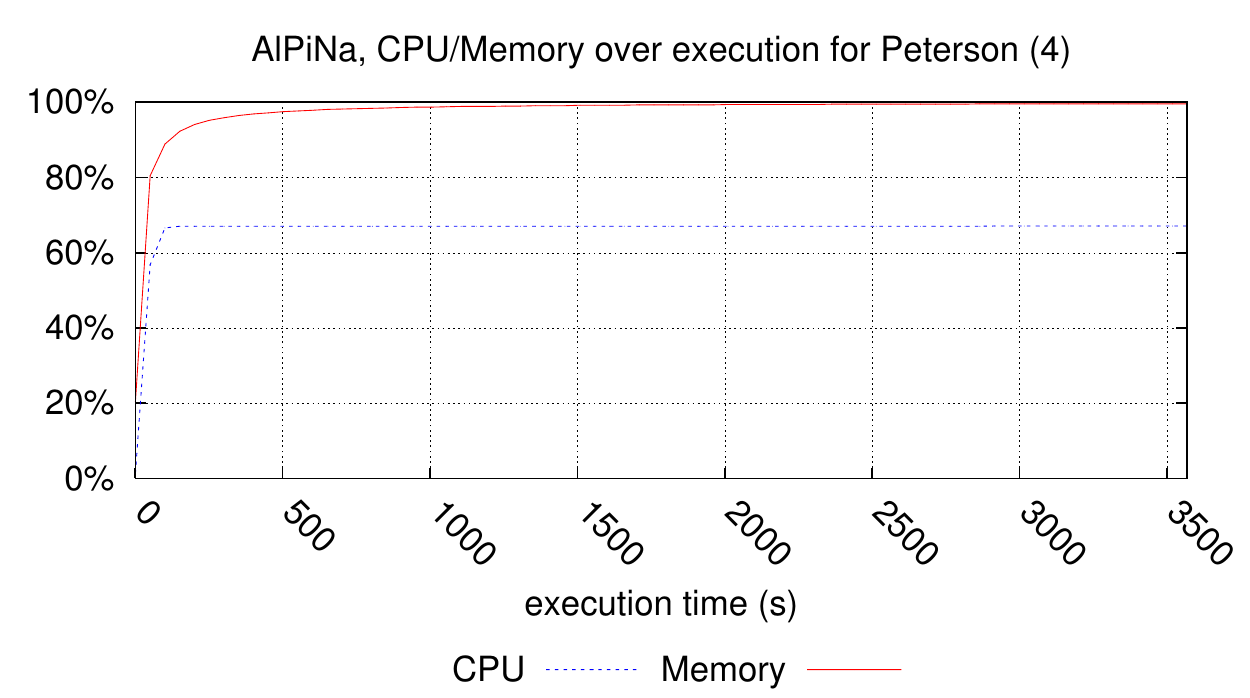}

\subsubsection{Executions for philo\_dyn}
2 charts have been generated.
\index{Execution (by tool)!AlPiNA}
\index{Execution (by model)!philo\_dyn!AlPiNA}

\noindent\includegraphics[width=.5\textwidth]{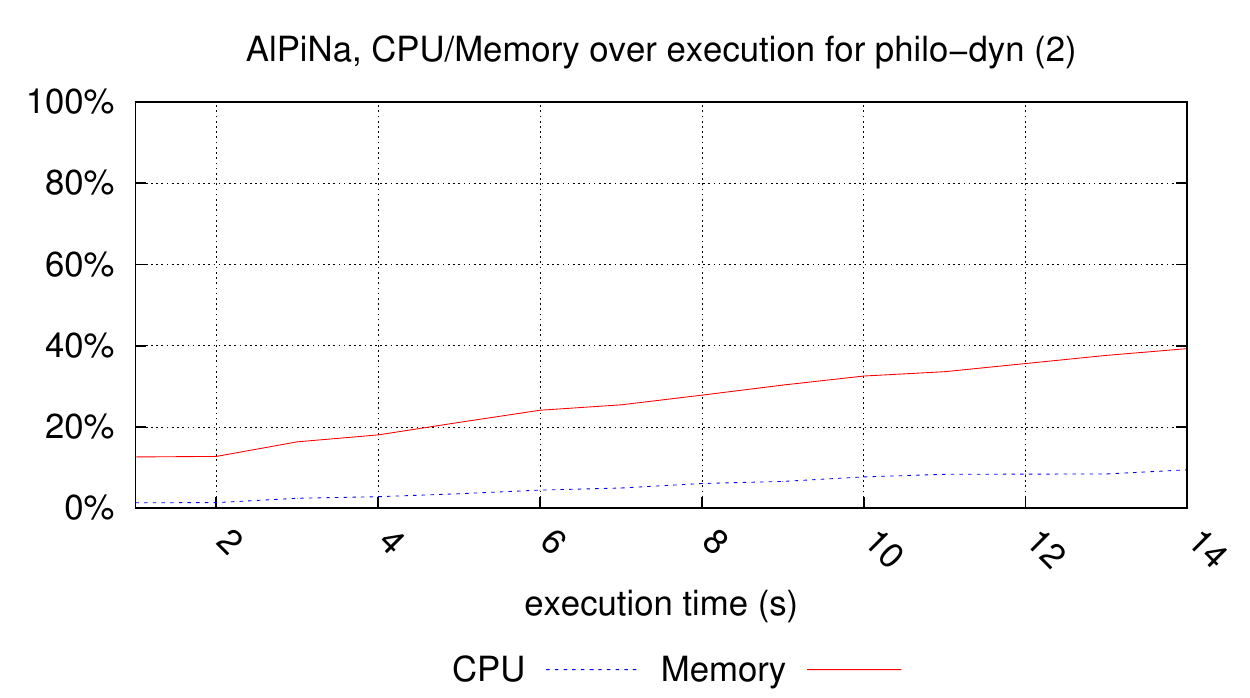}
\includegraphics[width=.5\textwidth]{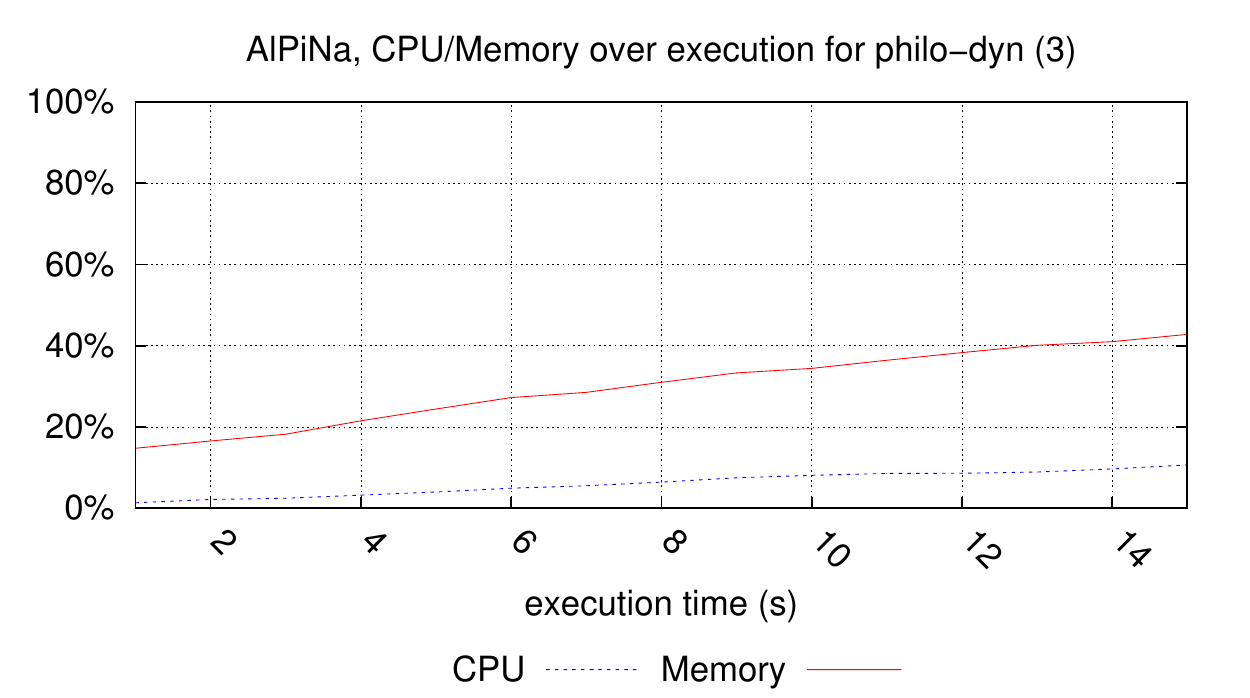}

\noindent\includegraphics[width=.5\textwidth]{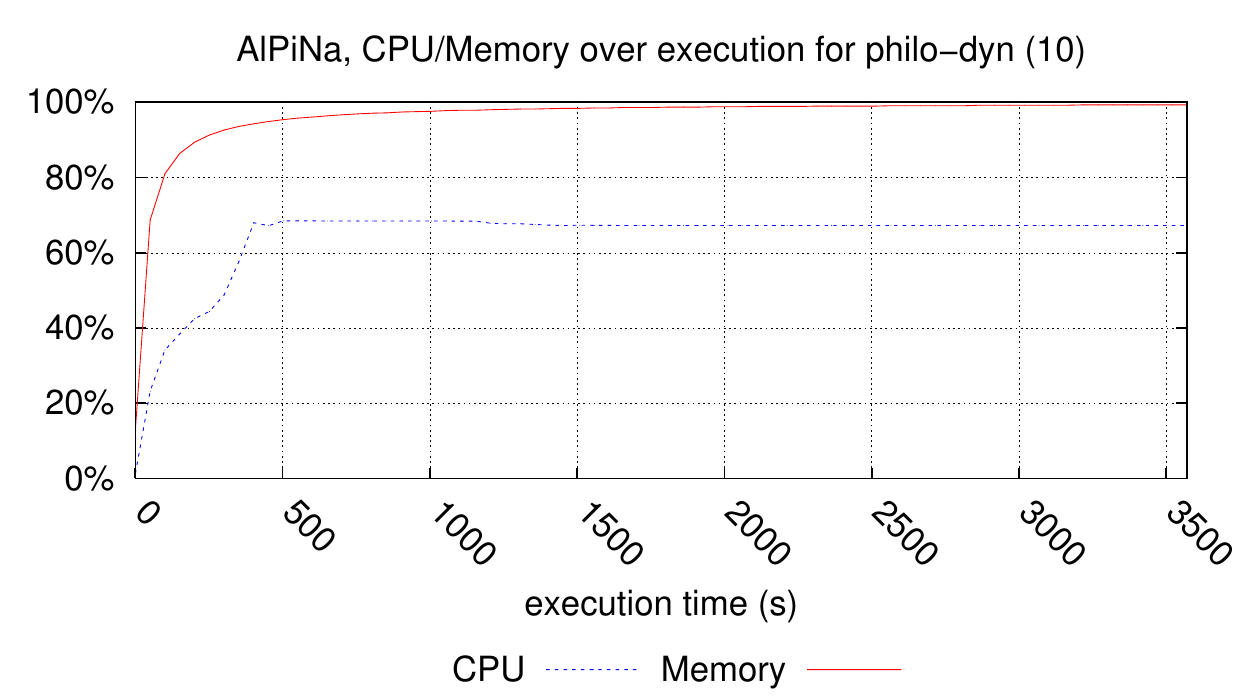}

\vfill\eject
\subsubsection{Executions for Philosophers}
8 charts have been generated.
\index{Execution (by tool)!AlPiNA}
\index{Execution (by model)!Philosophers!AlPiNA}

\noindent\includegraphics[width=.5\textwidth]{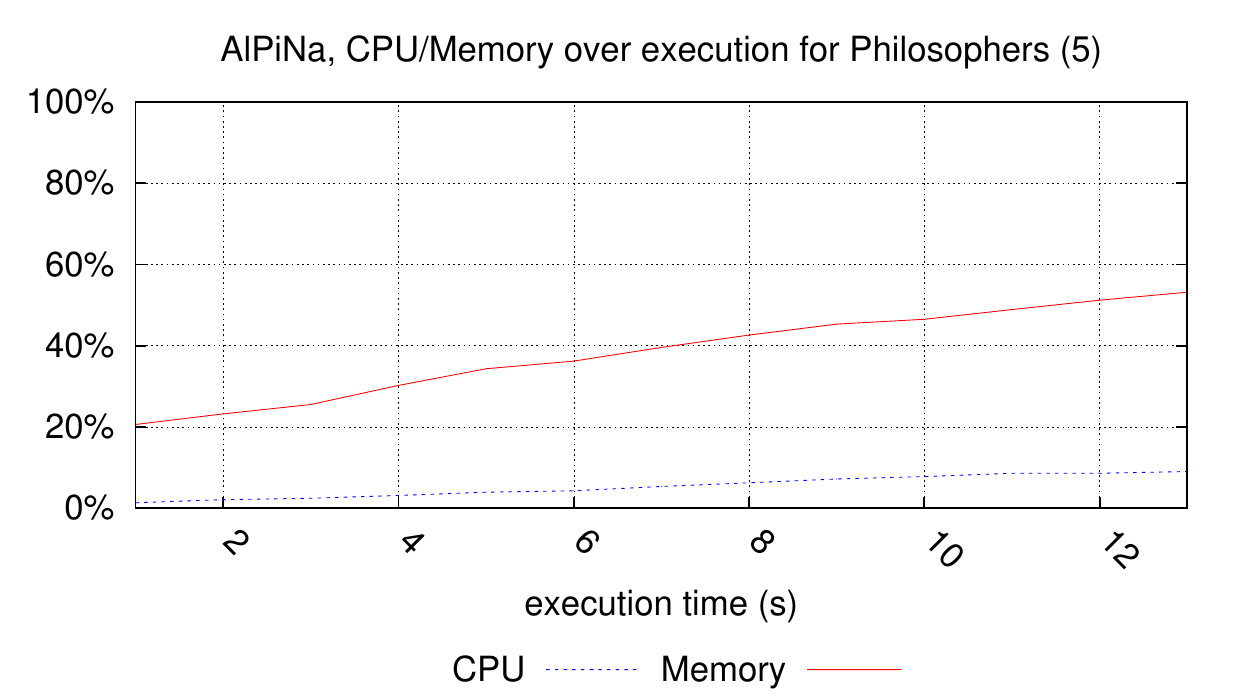}
\includegraphics[width=.5\textwidth]{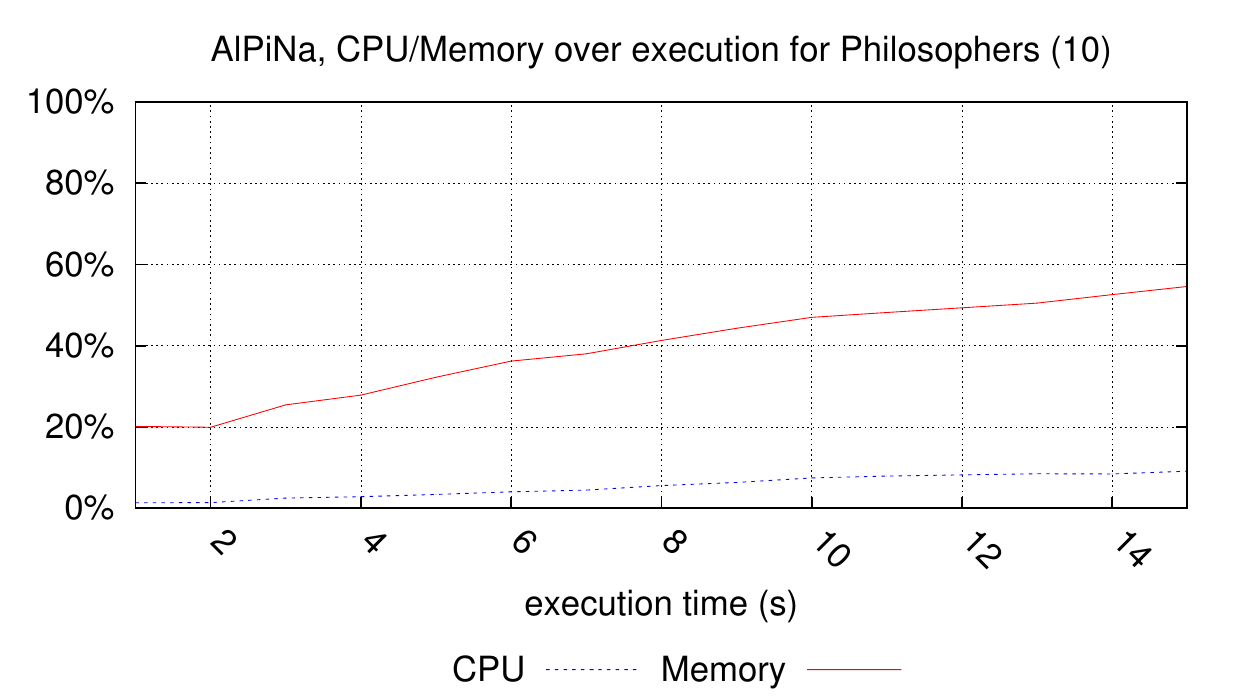}

\noindent\includegraphics[width=.5\textwidth]{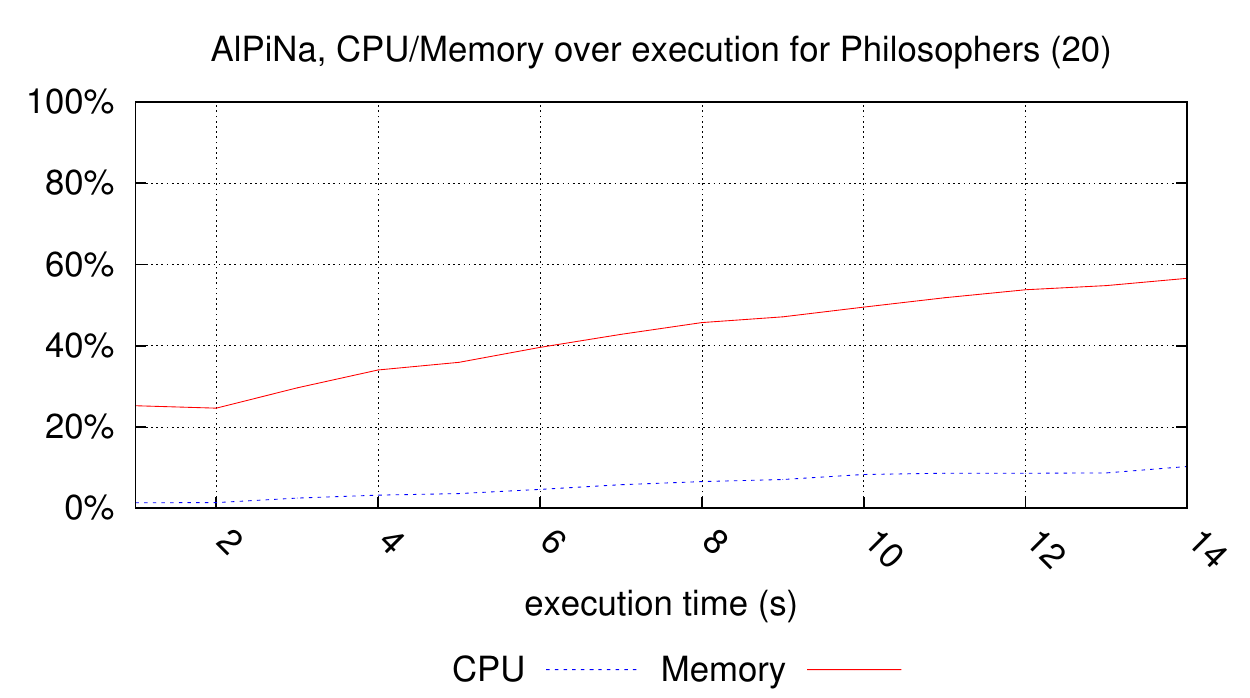}
\includegraphics[width=.5\textwidth]{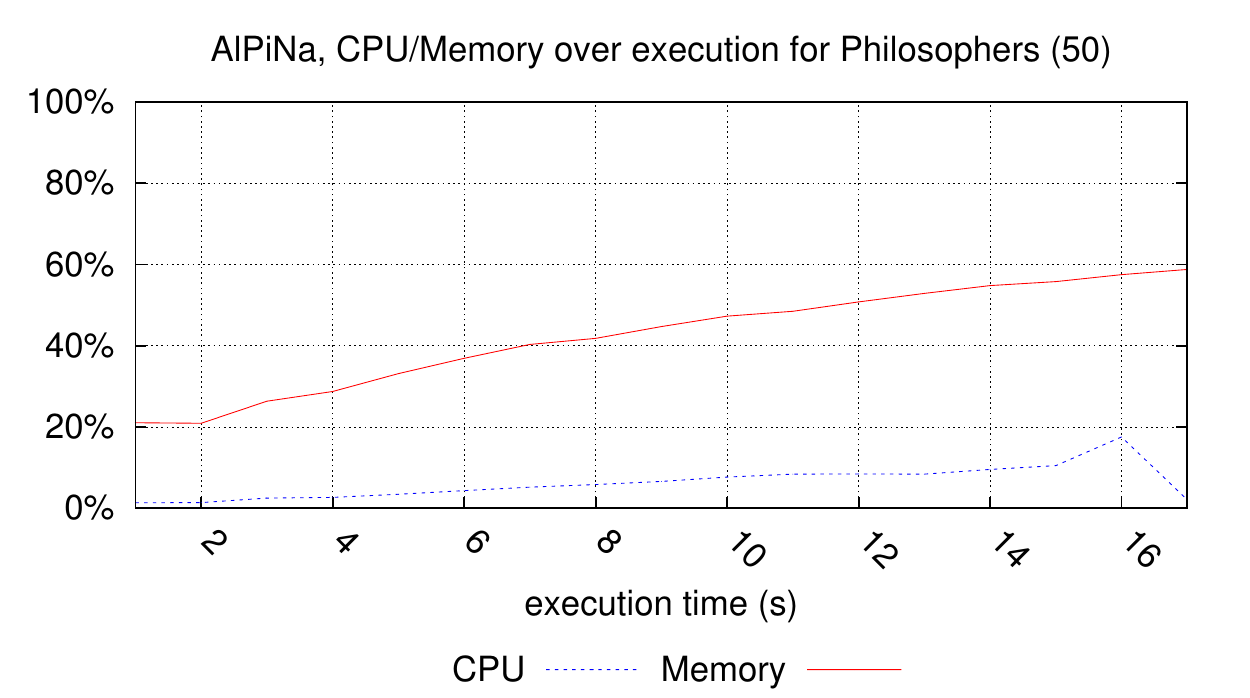}

\noindent\includegraphics[width=.5\textwidth]{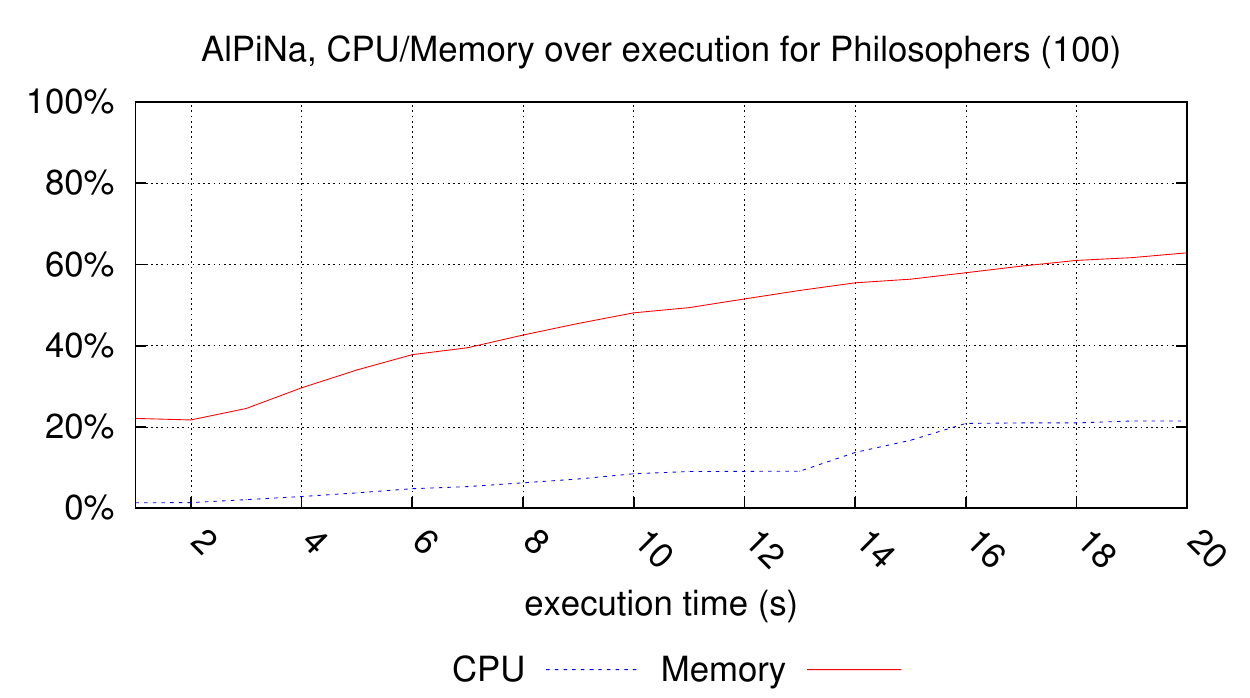}
\includegraphics[width=.5\textwidth]{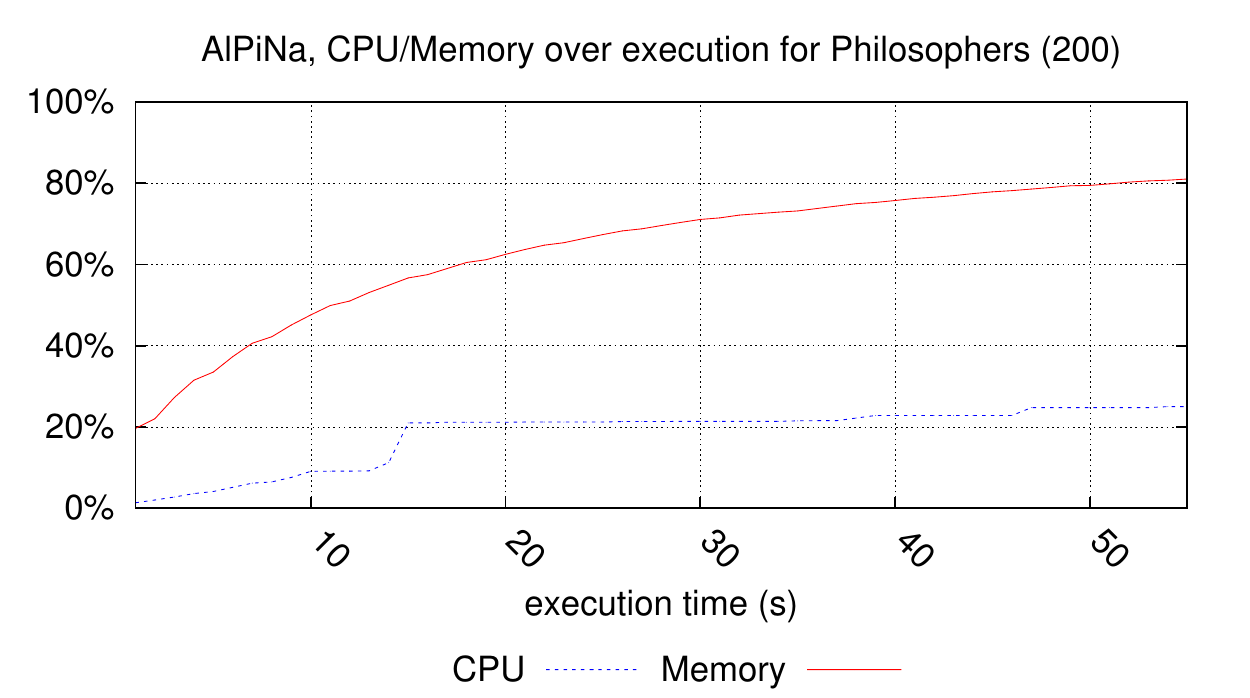}

\noindent\includegraphics[width=.5\textwidth]{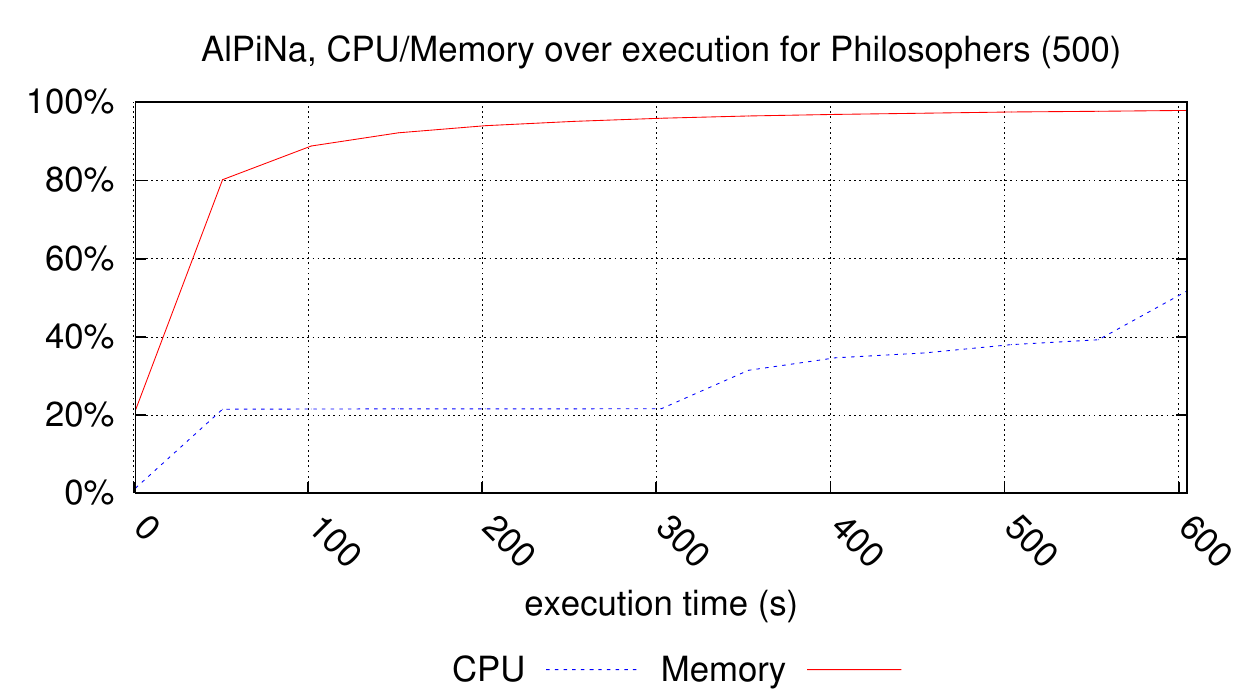}
\includegraphics[width=.5\textwidth]{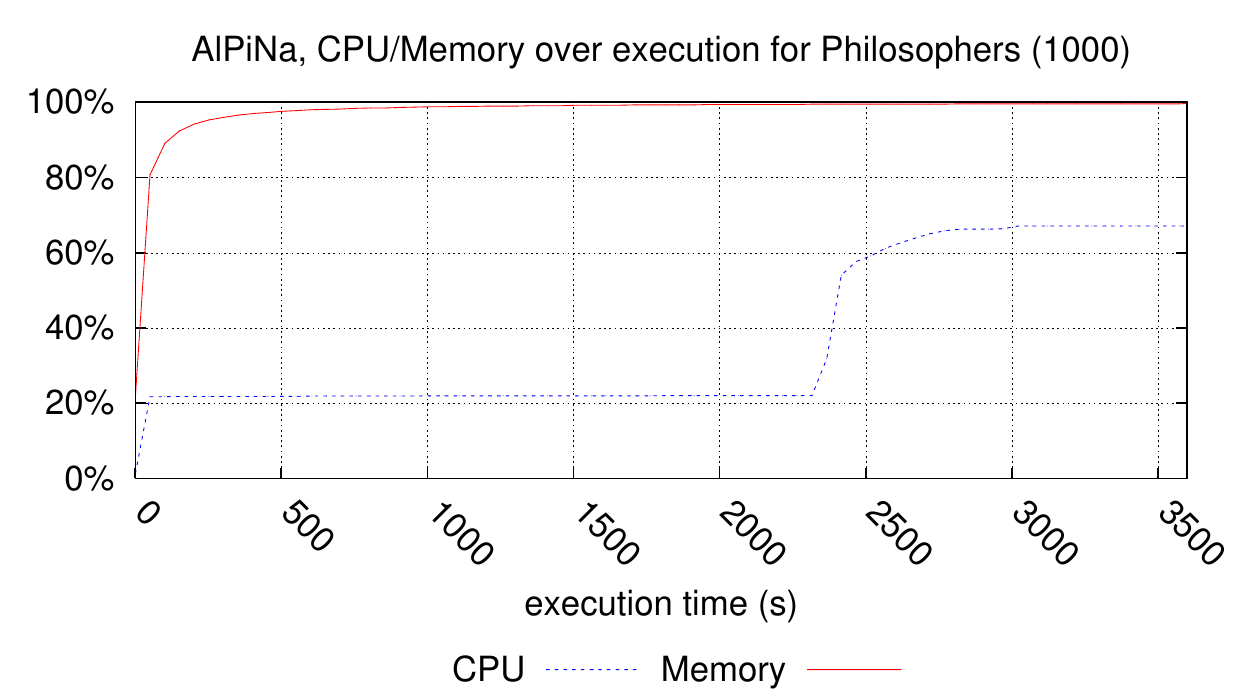}

\vfill\eject
\subsubsection{Executions for planning}
1 chart has been generated.
\index{Execution (by tool)!AlPiNA}
\index{Execution (by model)!planning!AlPiNA}

\noindent\includegraphics[width=.5\textwidth]{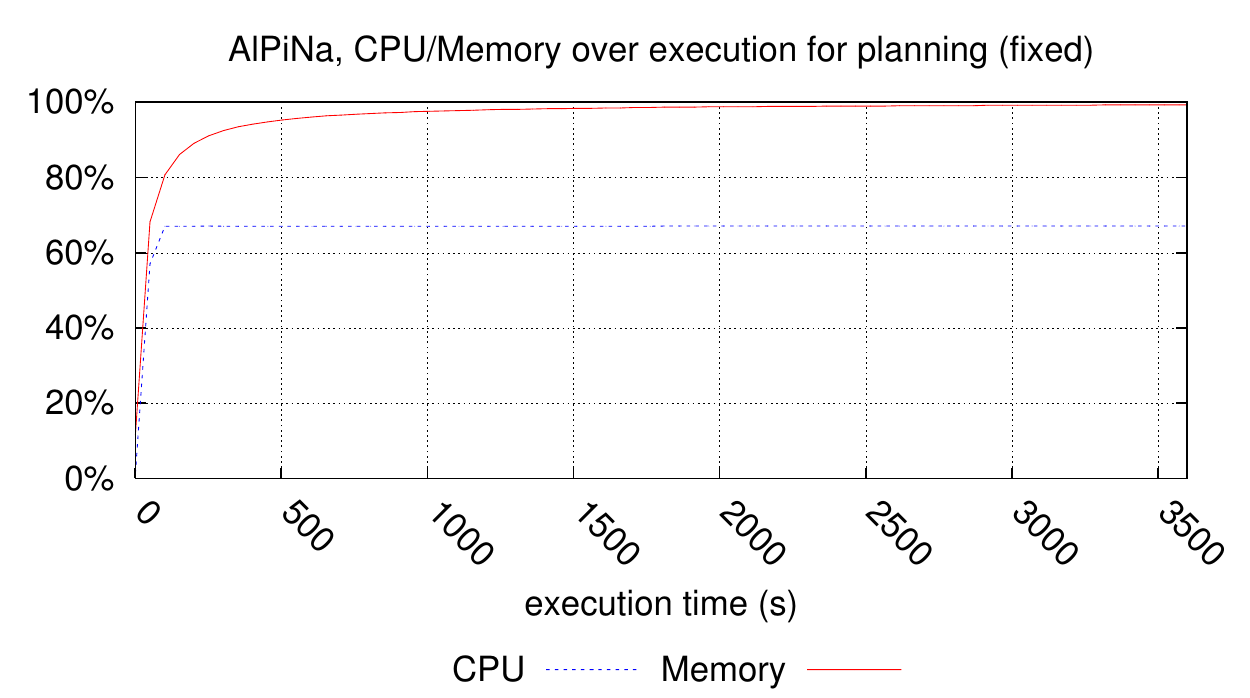}

\subsubsection{Executions for railroad}
2 charts have been generated.
\index{Execution (by tool)!AlPiNA}
\index{Execution (by model)!railroad!AlPiNA}

\noindent\includegraphics[width=.5\textwidth]{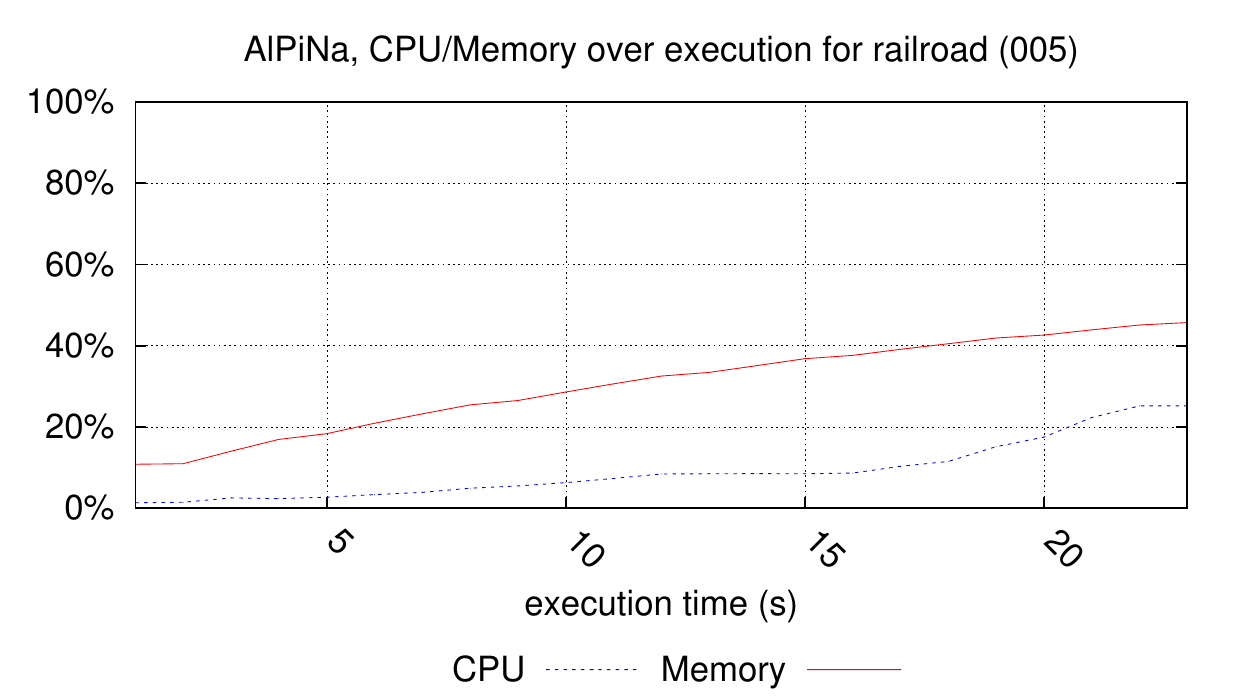}
\includegraphics[width=.5\textwidth]{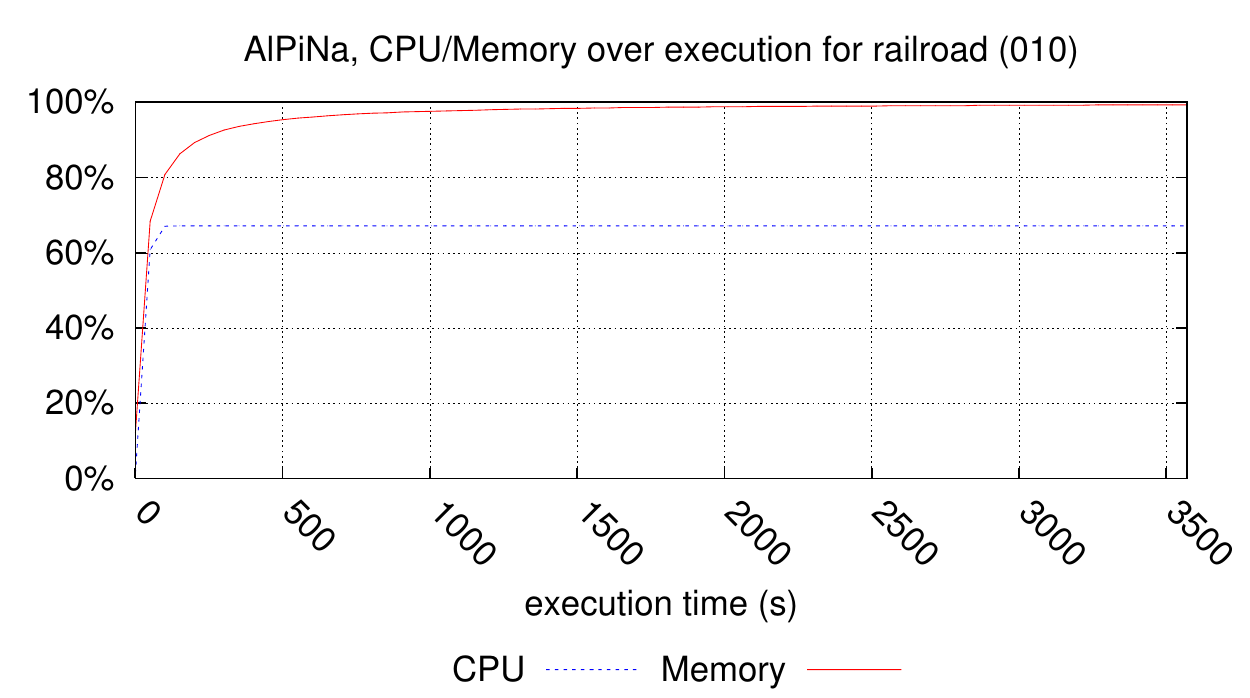}

\subsubsection{Executions for ring}
1 chart has been generated.
\index{Execution (by tool)!AlPiNA}
\index{Execution (by model)!ring!AlPiNA}

\noindent\includegraphics[width=.5\textwidth]{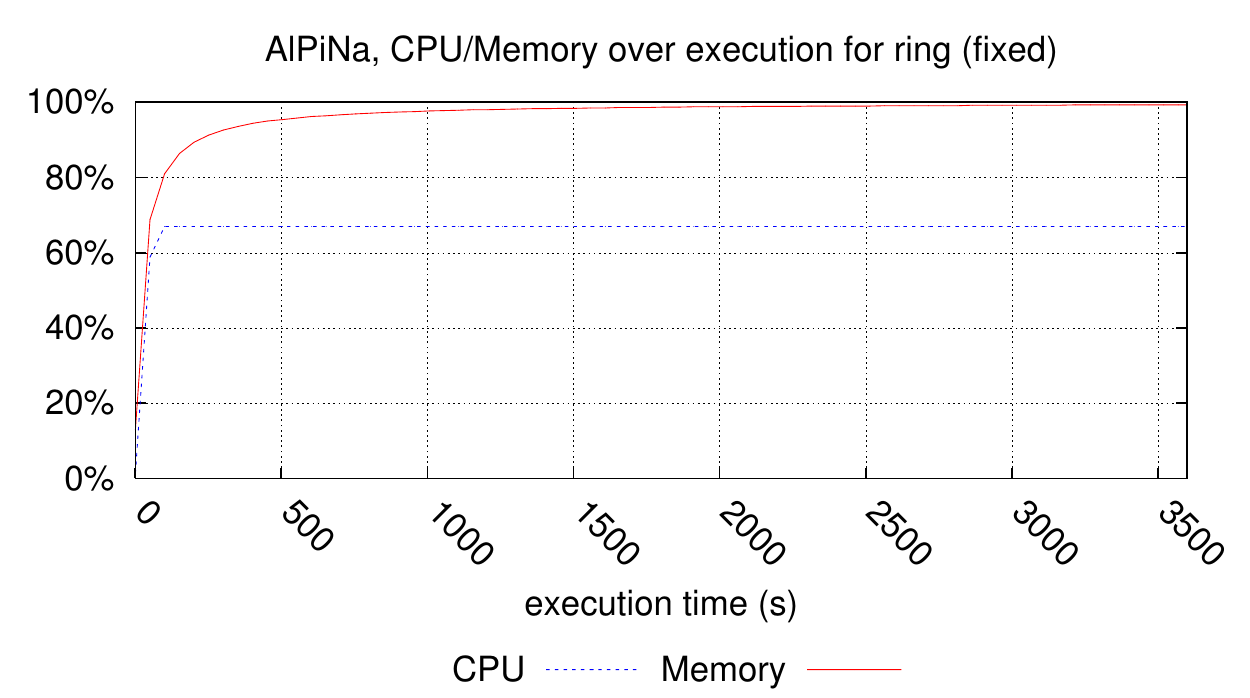}

\subsubsection{Executions for rw\_mutex}
5 charts have been generated.
\index{Execution (by tool)!AlPiNA}
\index{Execution (by model)!rw\_mutex!AlPiNA}

\noindent\includegraphics[width=.5\textwidth]{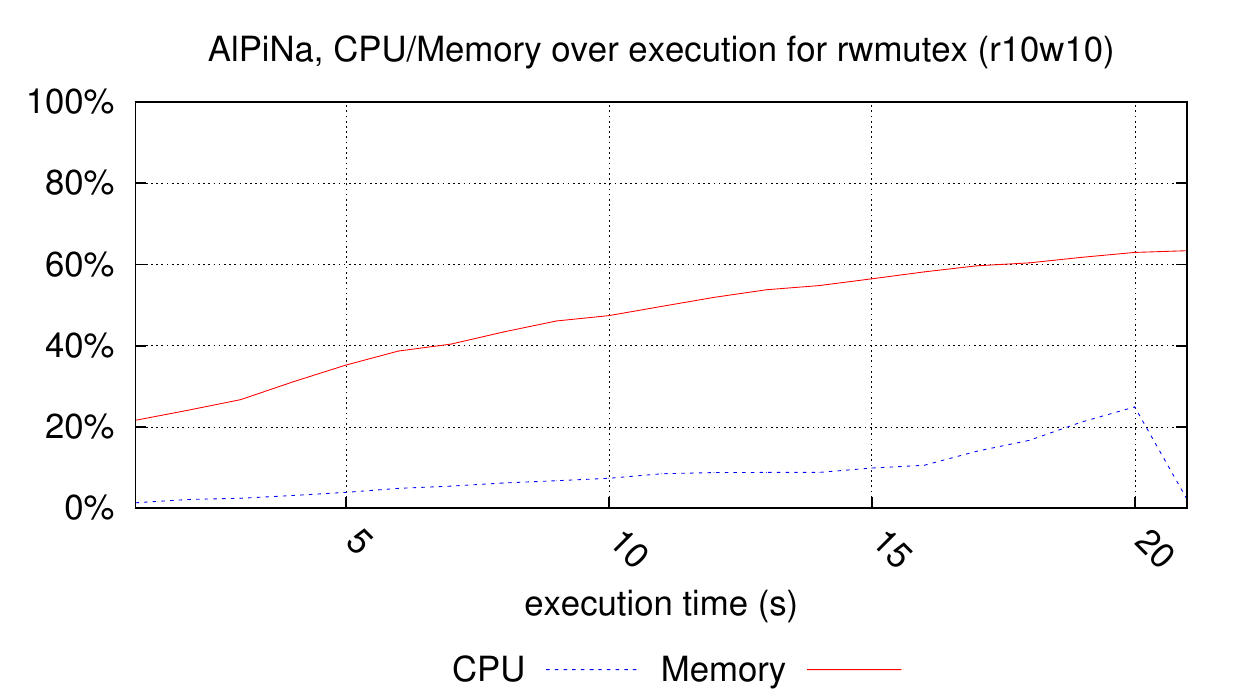}
\includegraphics[width=.5\textwidth]{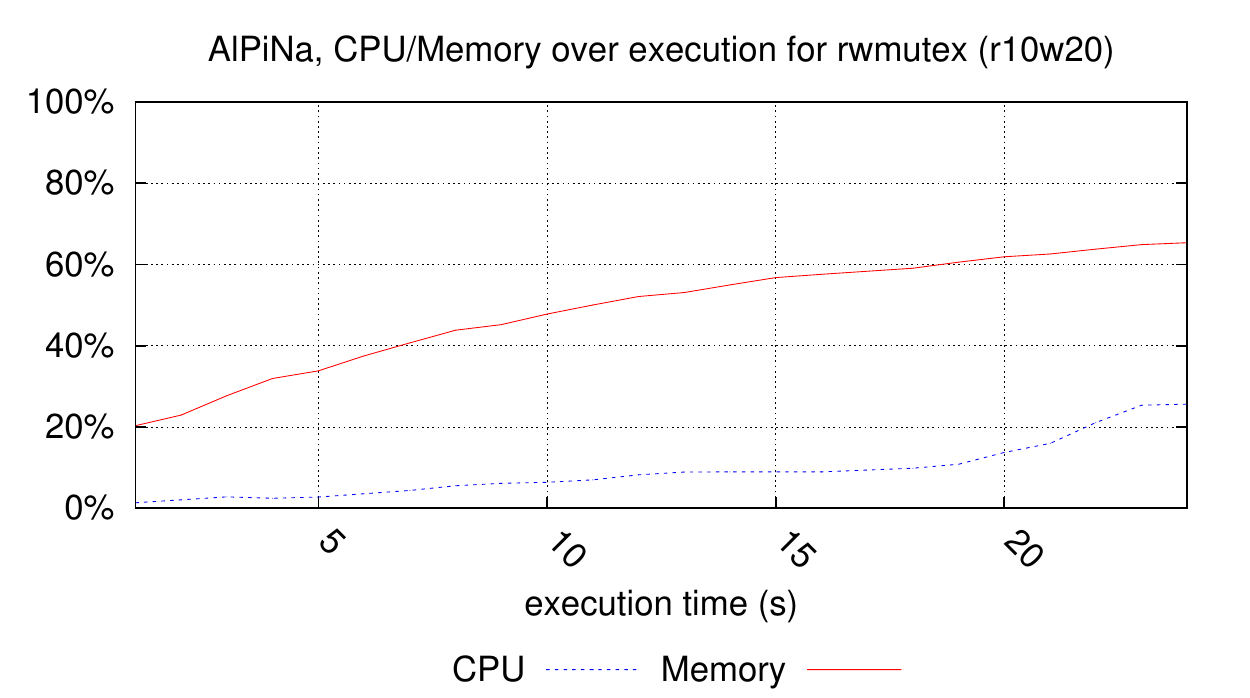}

\noindent\includegraphics[width=.5\textwidth]{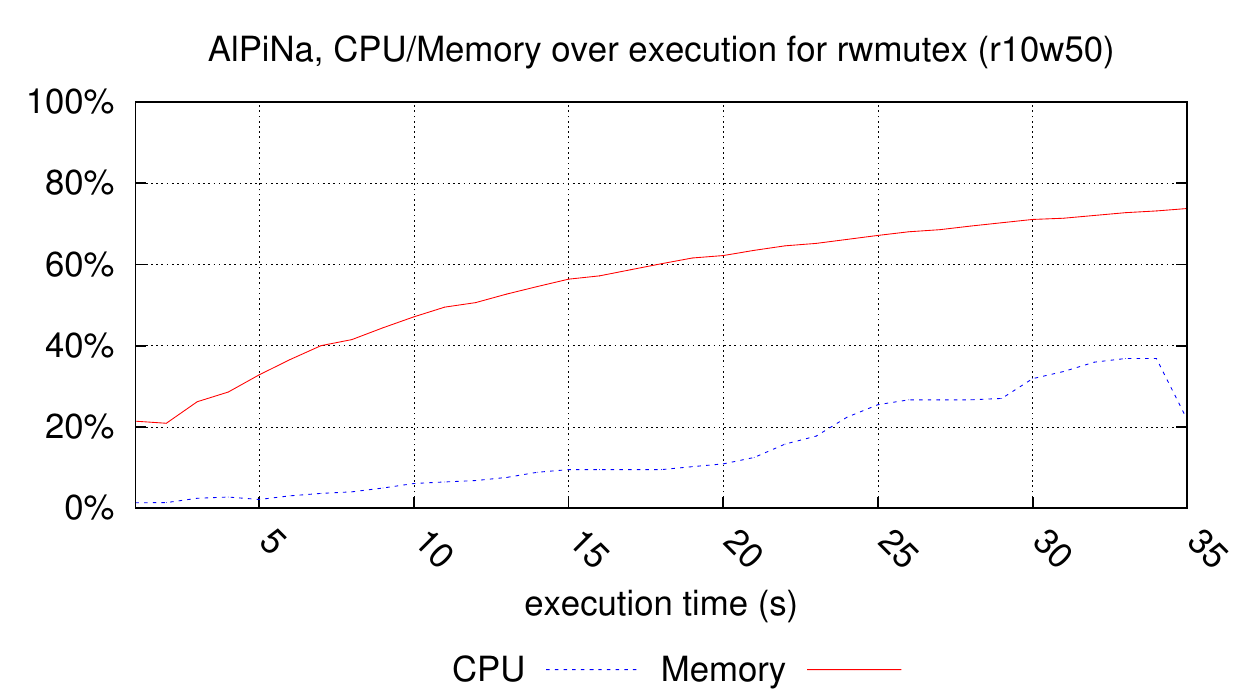}
\includegraphics[width=.5\textwidth]{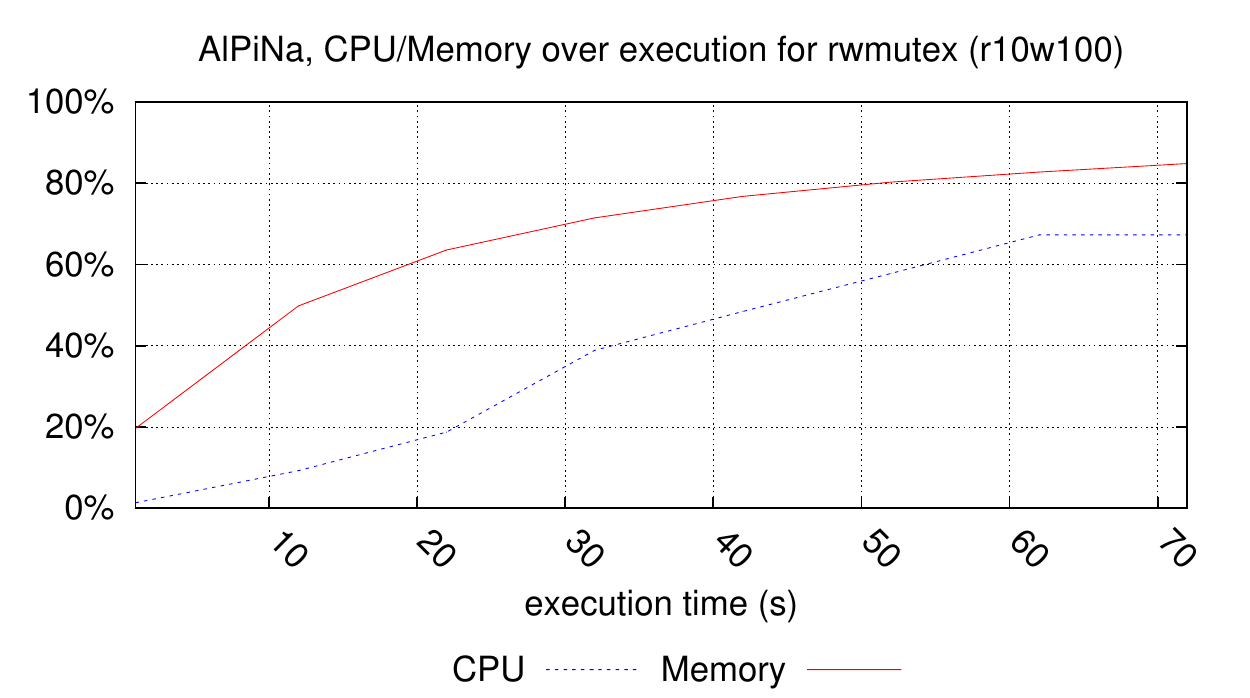}

\noindent\includegraphics[width=.5\textwidth]{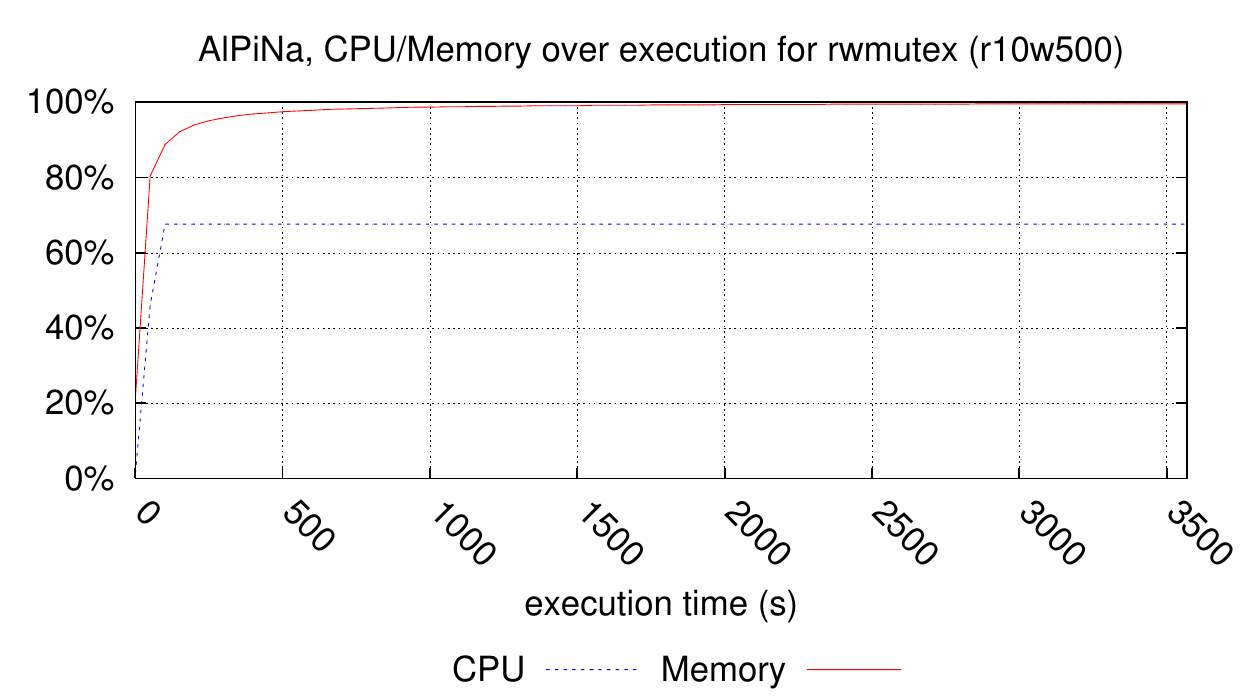}

\subsubsection{Executions for simple\_lbs}
2 charts have been generated.
\index{Execution (by tool)!AlPiNA}
\index{Execution (by model)!simple\_lbs!AlPiNA}

\noindent\includegraphics[width=.5\textwidth]{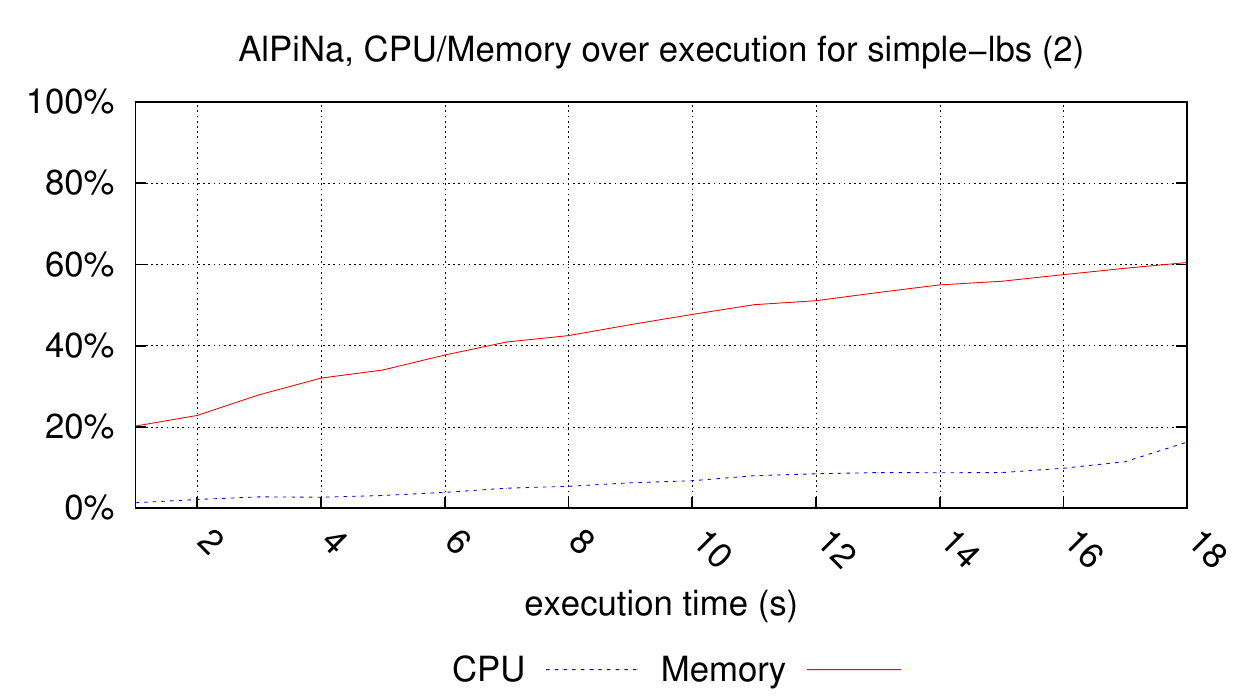}
\includegraphics[width=.5\textwidth]{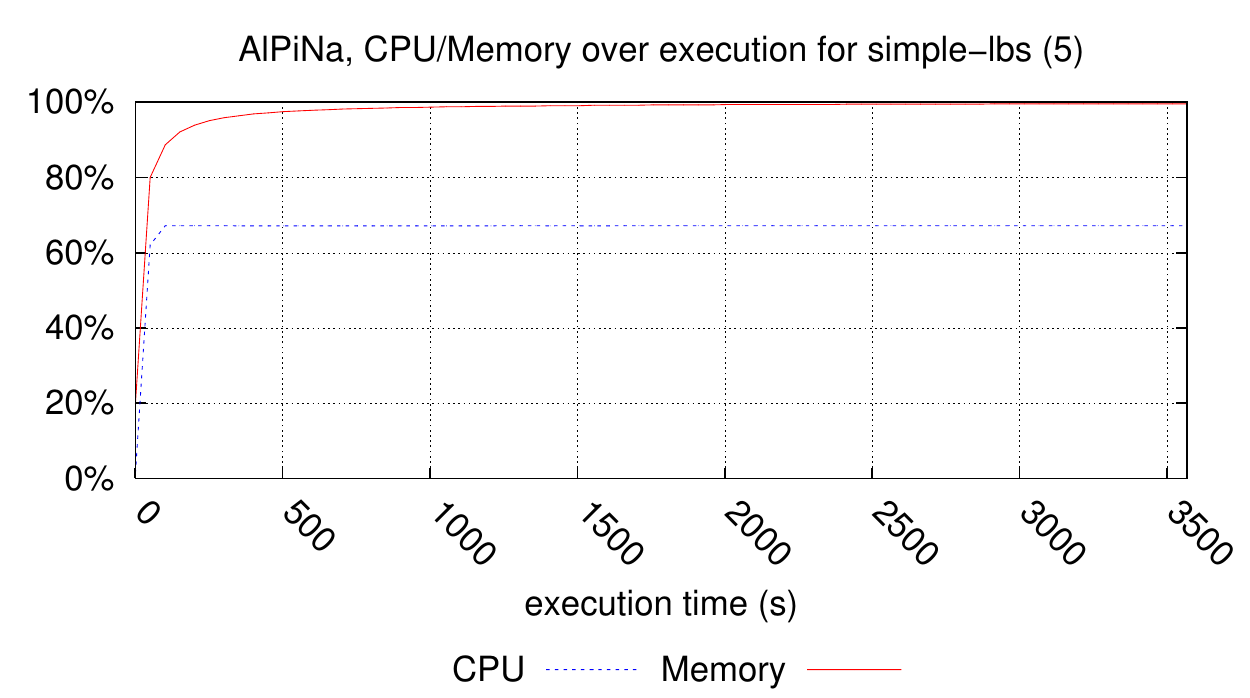}

\subsubsection{Executions for SharedMemory}
4 charts have been generated.
\index{Execution (by tool)!AlPiNA}
\index{Execution (by model)!SharedMemory!AlPiNA}

\noindent\includegraphics[width=.5\textwidth]{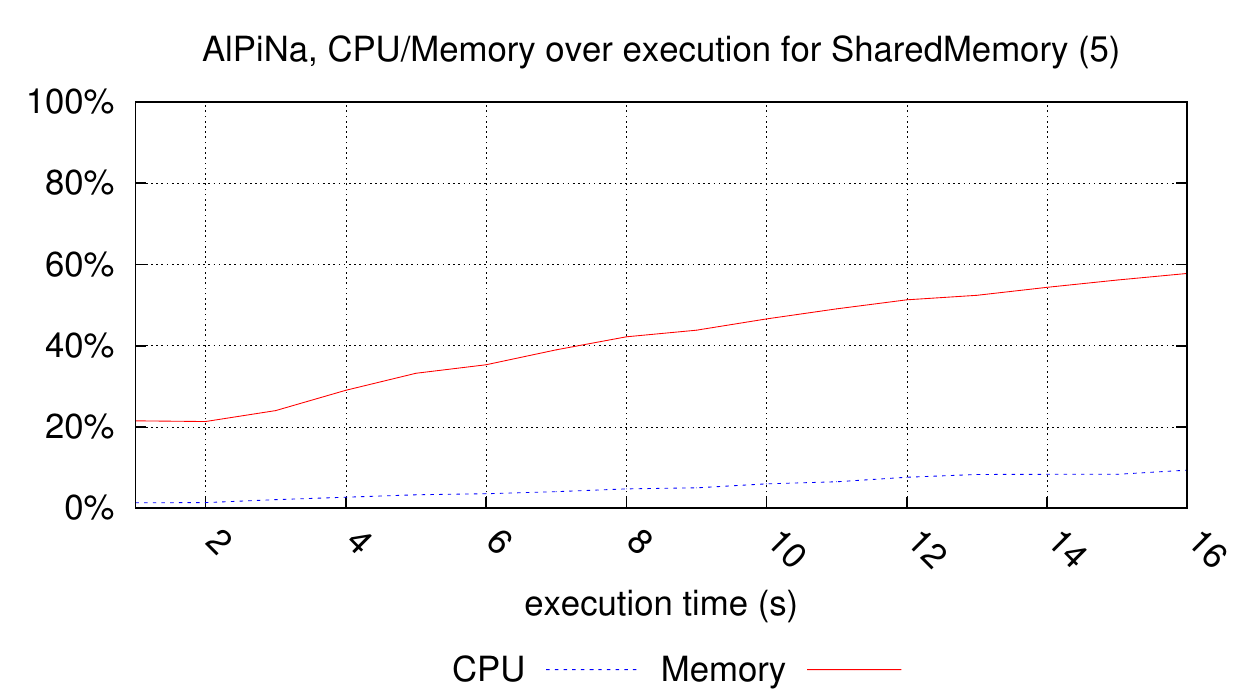}
\includegraphics[width=.5\textwidth]{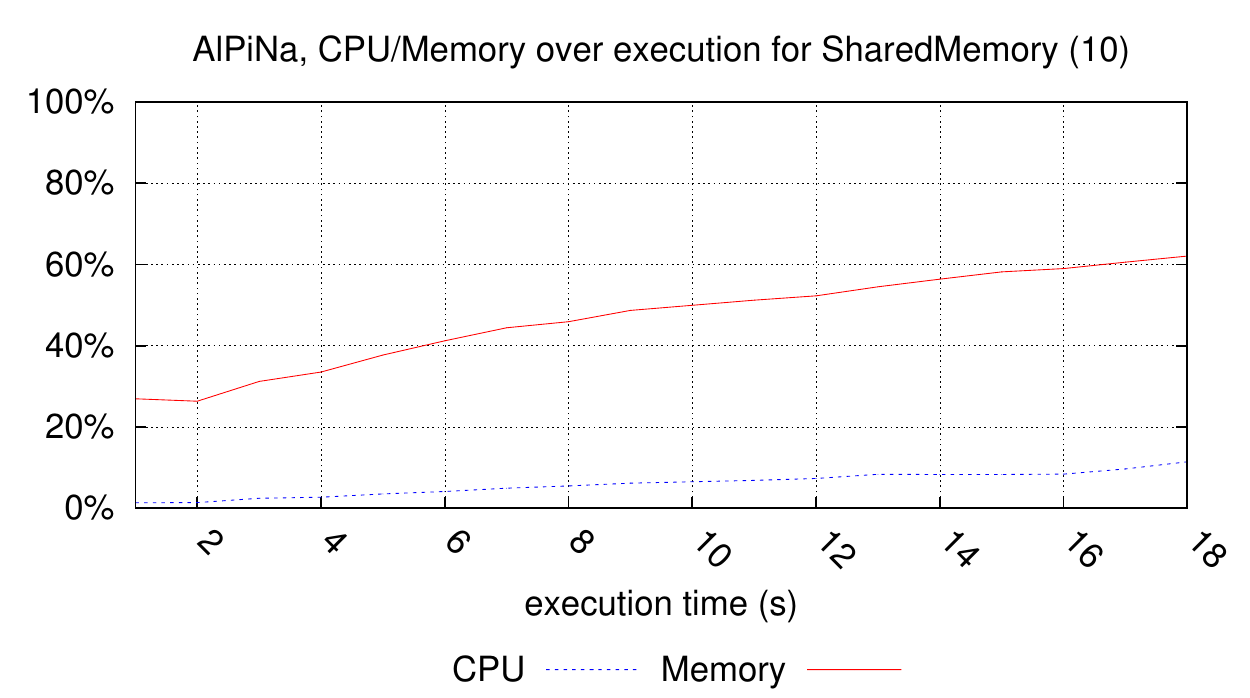}

\noindent\includegraphics[width=.5\textwidth]{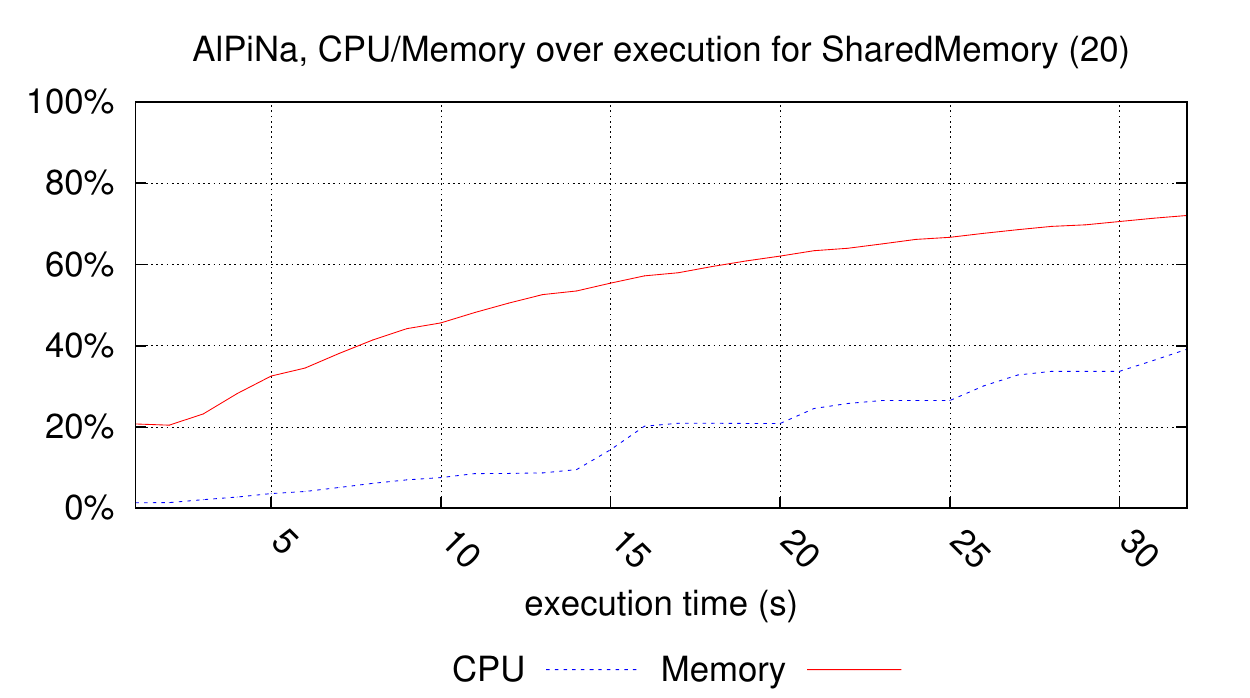}
\includegraphics[width=.5\textwidth]{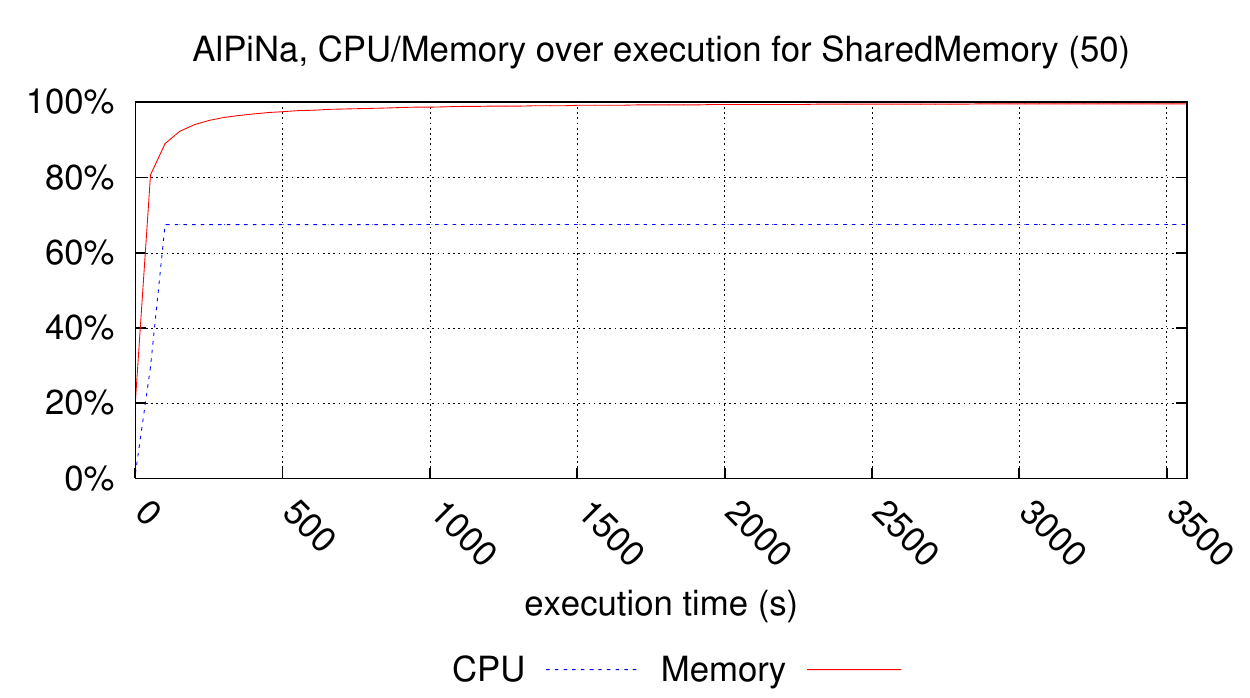}

\subsubsection{Executions for ToKenRing}
2 charts have been generated.
\index{Execution (by tool)!AlPiNA}
\index{Execution (by model)!SharedMemory!AlPiNA}

\noindent\includegraphics[width=.5\textwidth]{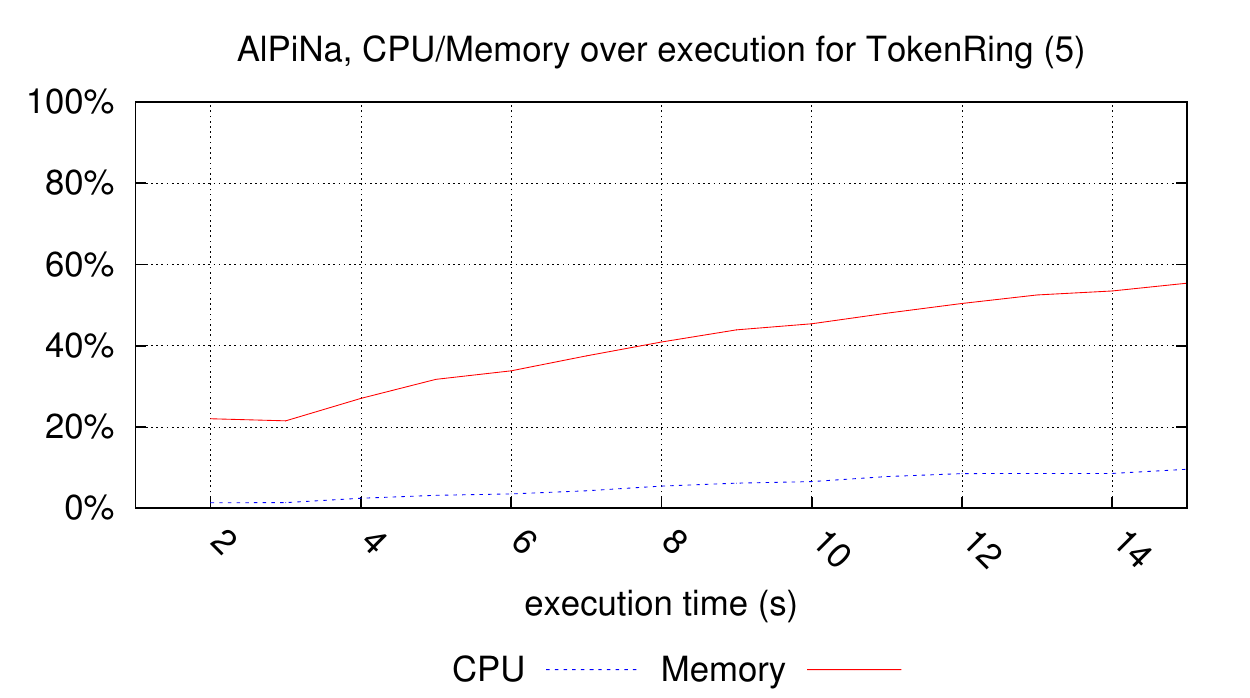}
\includegraphics[width=.5\textwidth]{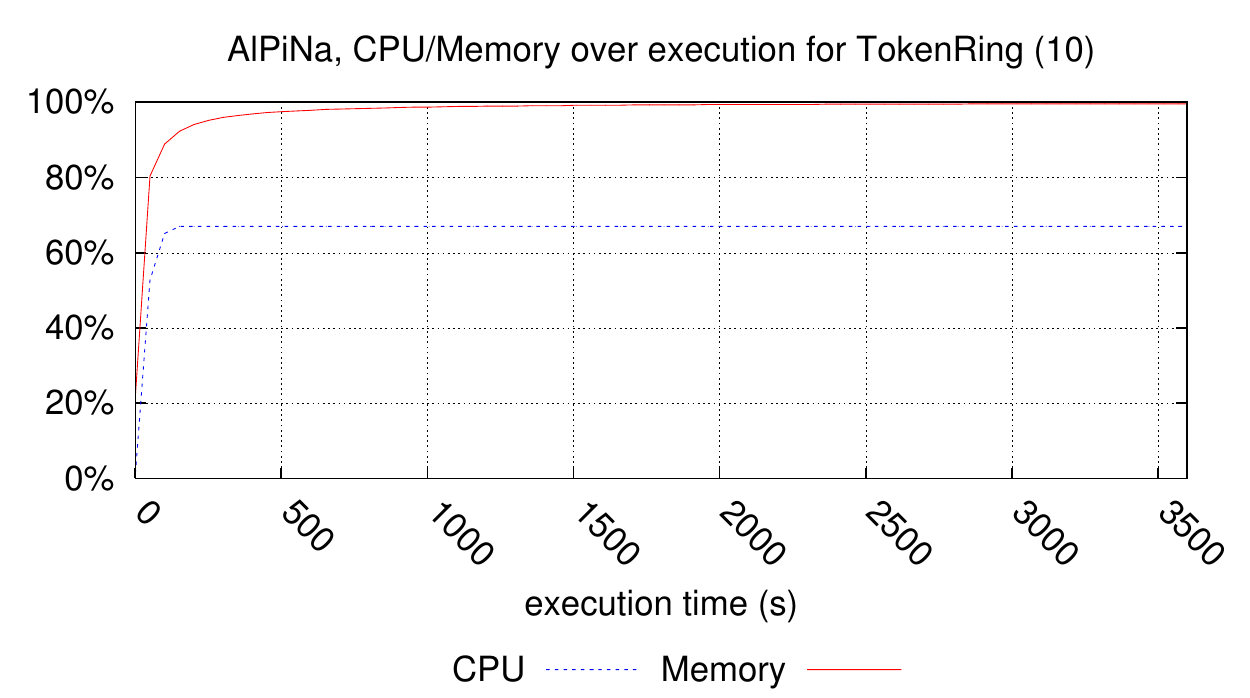}

\subsection{Execution Charts for Crocodile}

We provide here the execution charts observed for Crocodile over
the models it could compete with.

\subsubsection{Executions for cs\_repetitions}
1 chart has been generated.
\index{Execution (by tool)!Crocodile}
\index{Execution (by model)!cs\_repetitions!Crocodile}

\noindent\includegraphics[width=.5\textwidth]{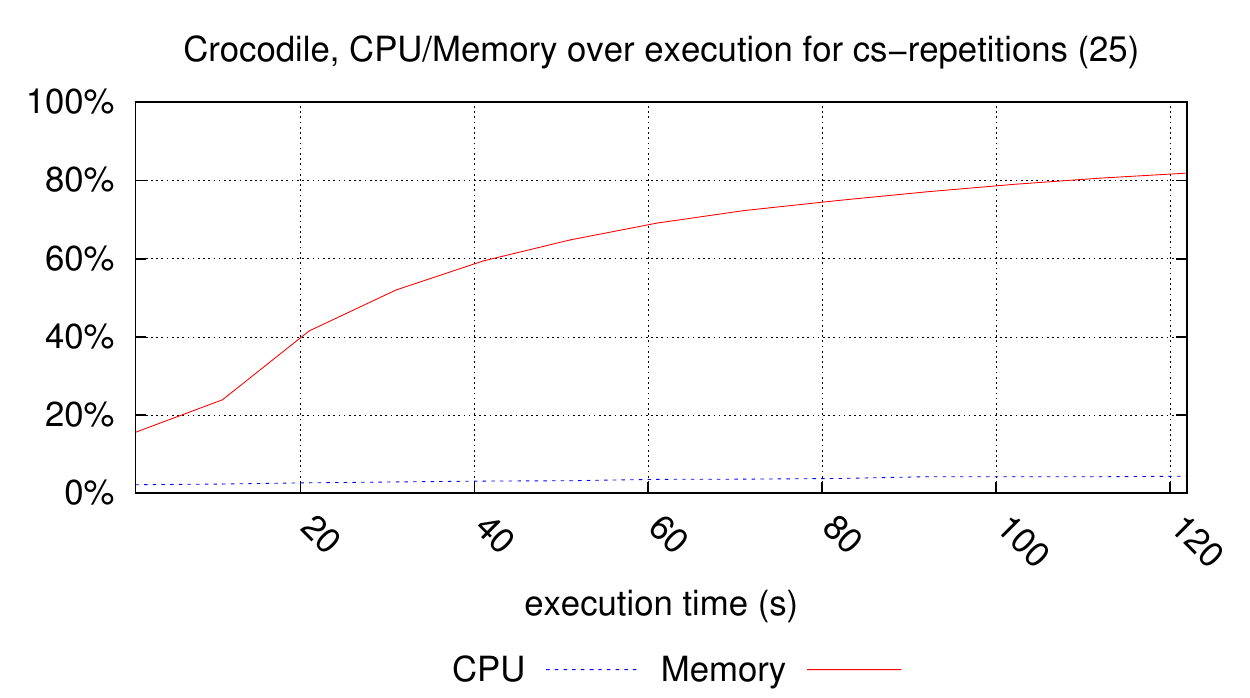}

\subsubsection{Executions for galloc\_res}
2 charts have been generated.
\index{Execution (by tool)!Crocodile}
\index{Execution (by model)!galloc\_res!Crocodile}

\noindent\includegraphics[width=.5\textwidth]{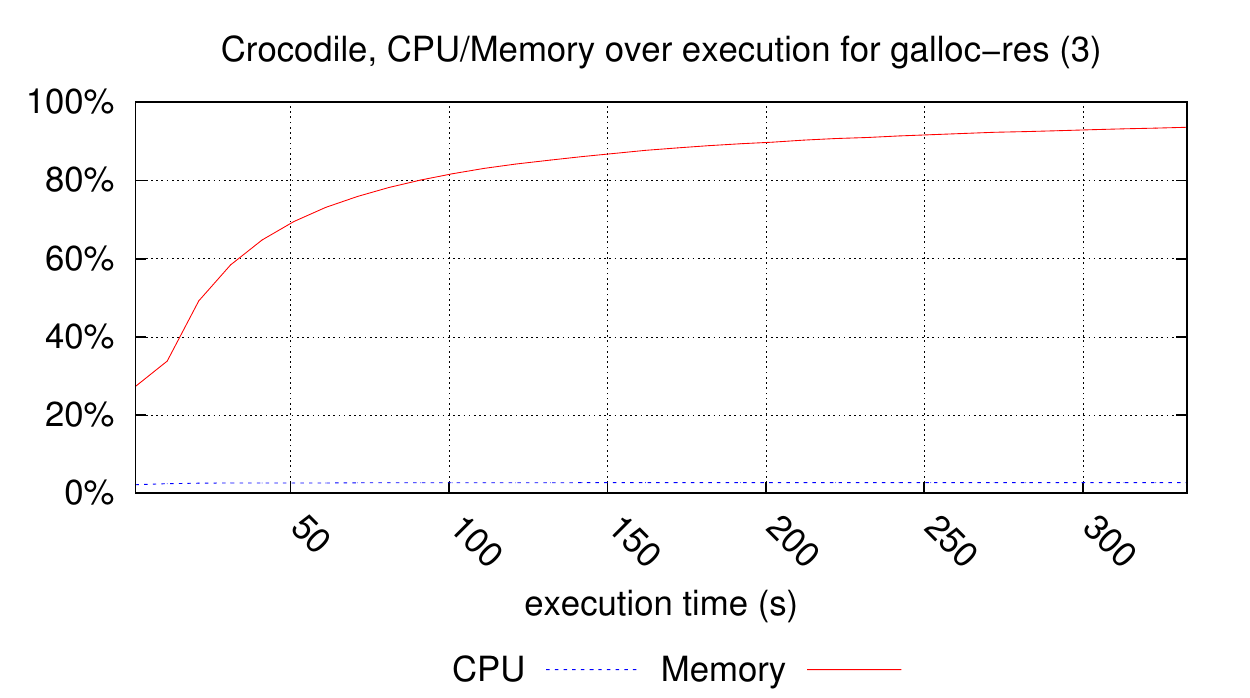}
\includegraphics[width=.5\textwidth]{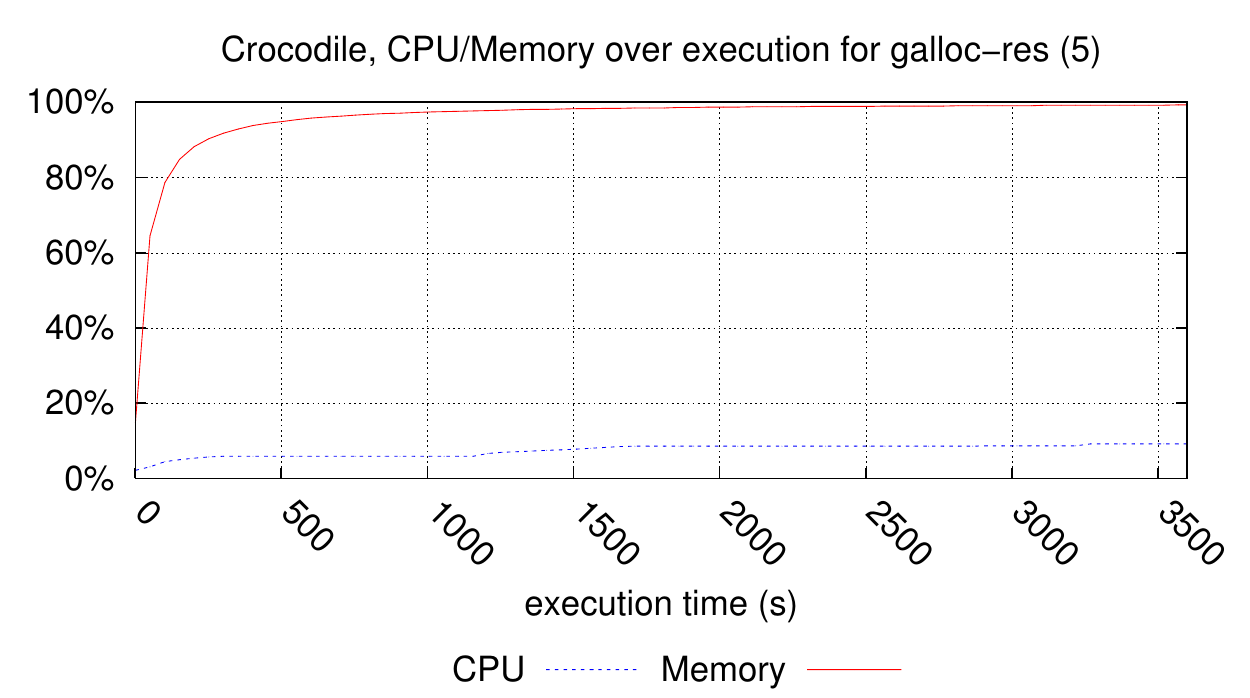}

\vfill\eject
\subsubsection{Executions for SharedMemory}
6 charts have been generated.
\index{Execution (by tool)!Crocodile}
\index{Execution (by model)!SharedMemory!Crocodile}

\noindent\includegraphics[width=.5\textwidth]{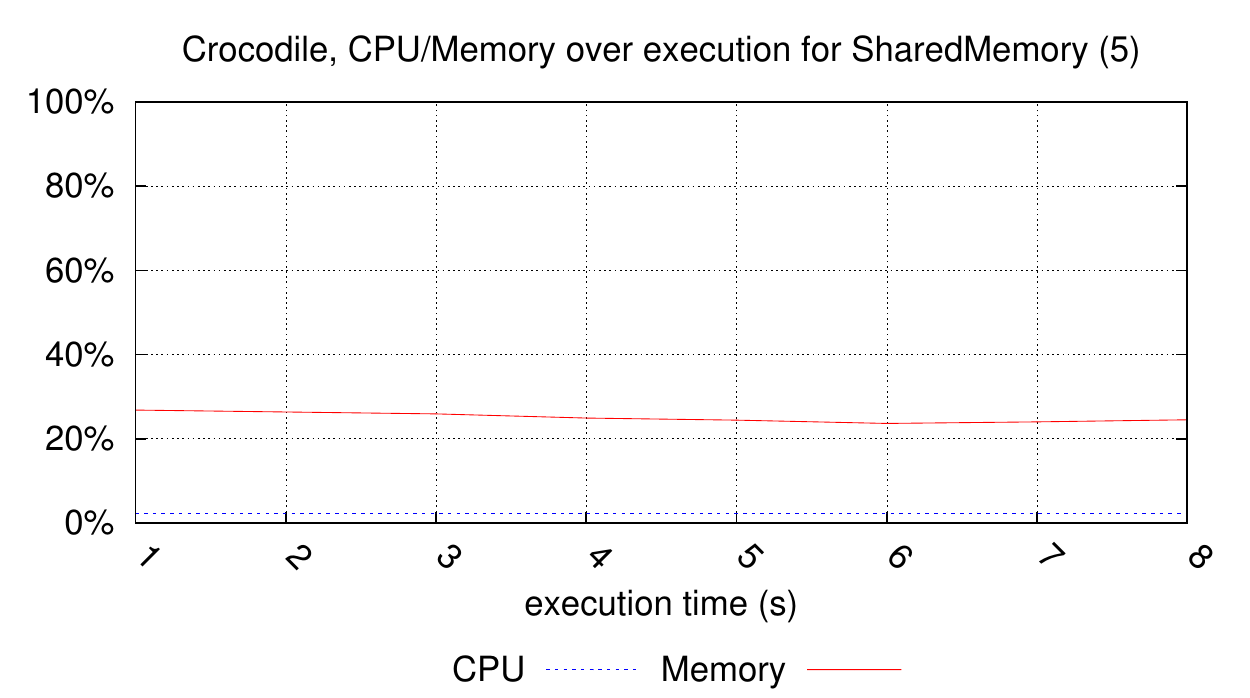}
\includegraphics[width=.5\textwidth]{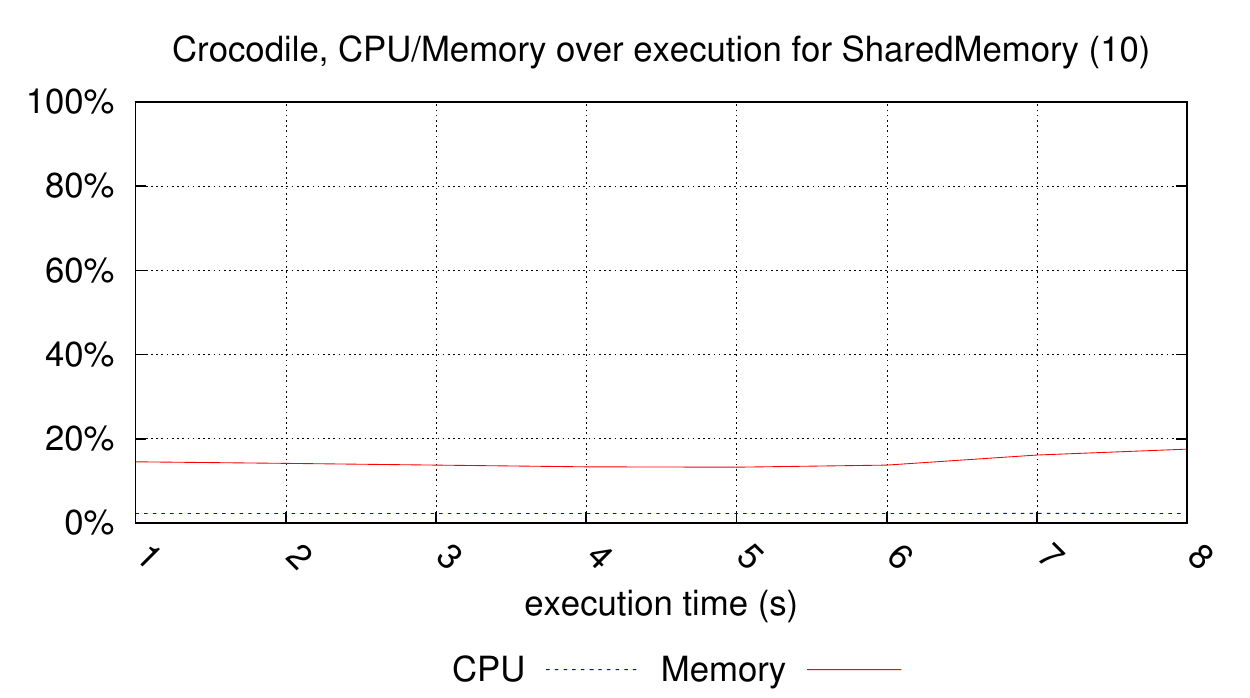}

\noindent\includegraphics[width=.5\textwidth]{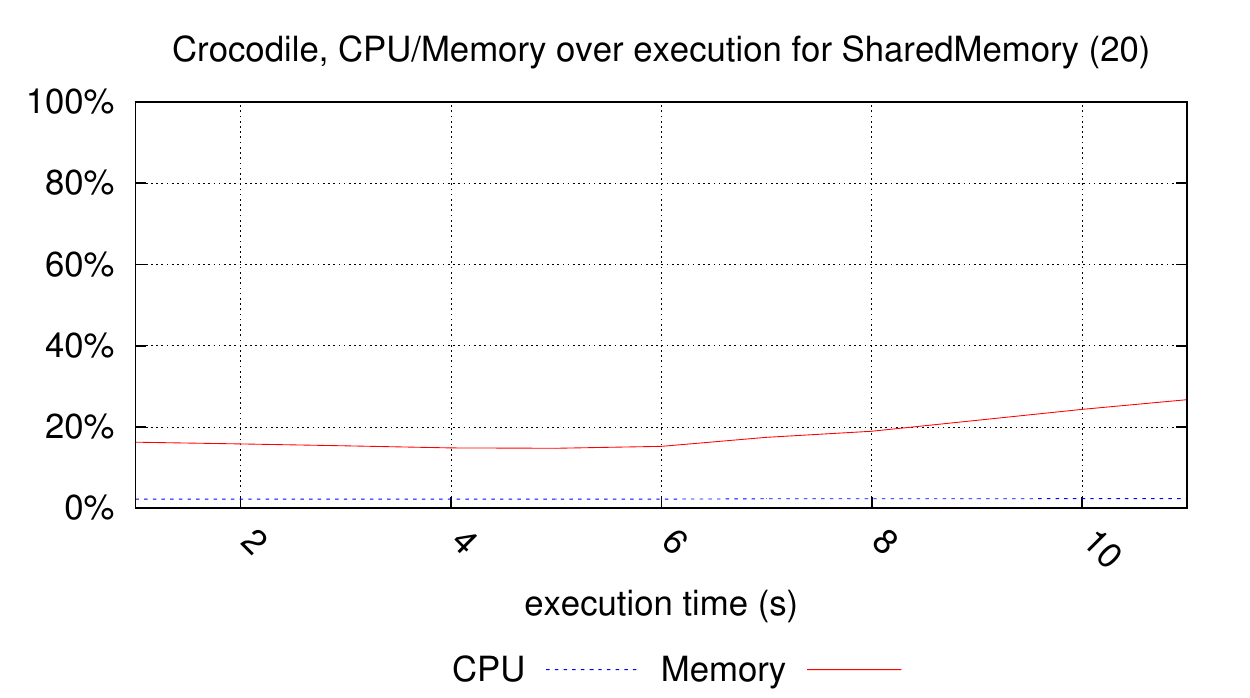}
\includegraphics[width=.5\textwidth]{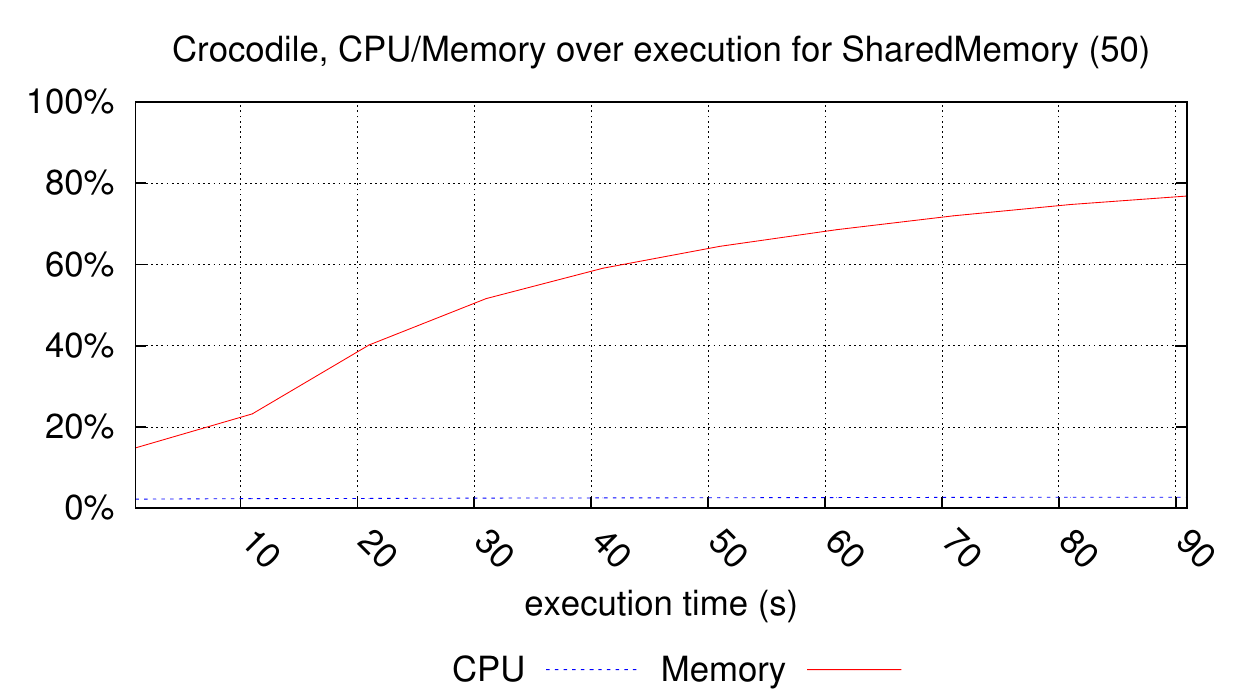}

\noindent\includegraphics[width=.5\textwidth]{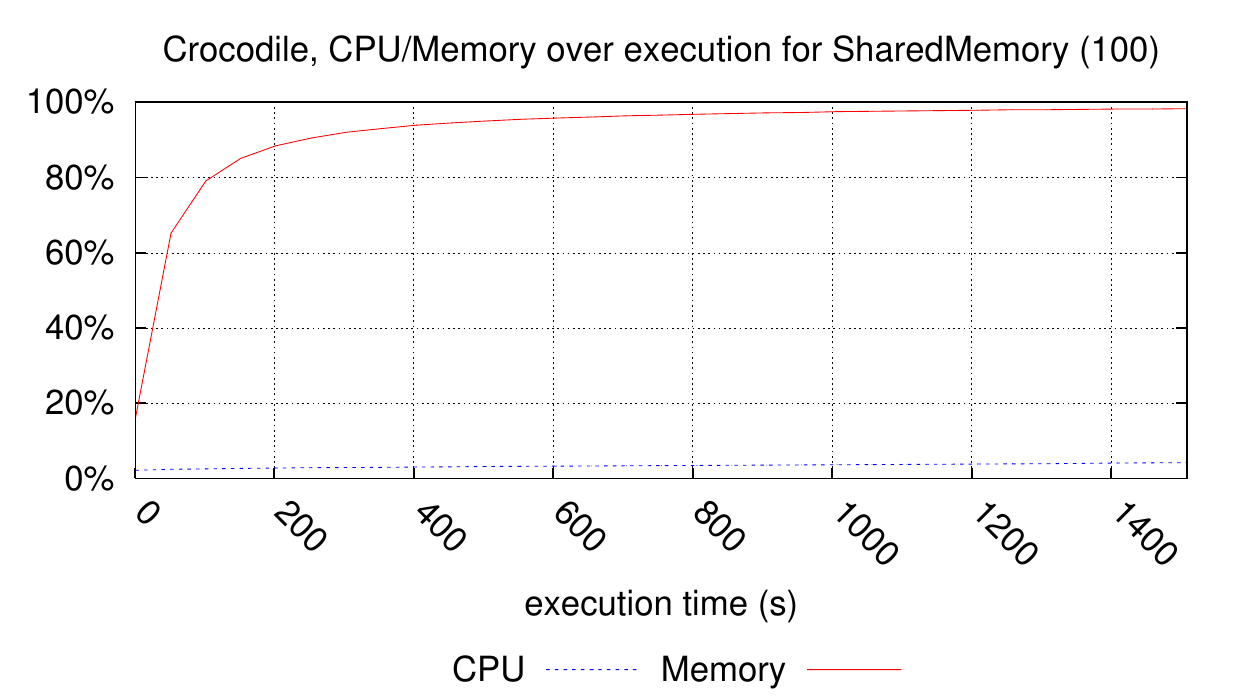}
\includegraphics[width=.5\textwidth]{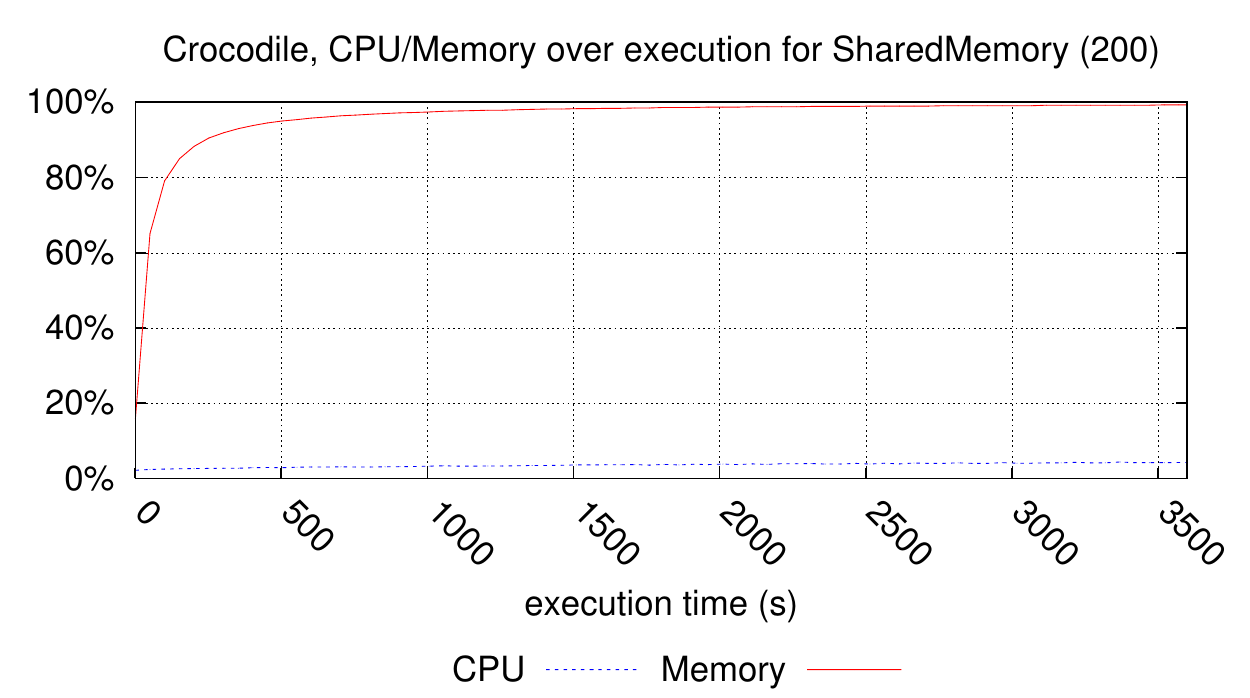}

\subsection{Execution Charts for Helena}

We provide here the execution charts observed for Helena over
the models it could compete with.

\subsubsection{Executions for lamport\_fmea}
1 chart has been generated.
\index{Execution (by tool)!Helena}
\index{Execution (by model)!lamport\_fmea!Helena}

\noindent\includegraphics[width=.5\textwidth]{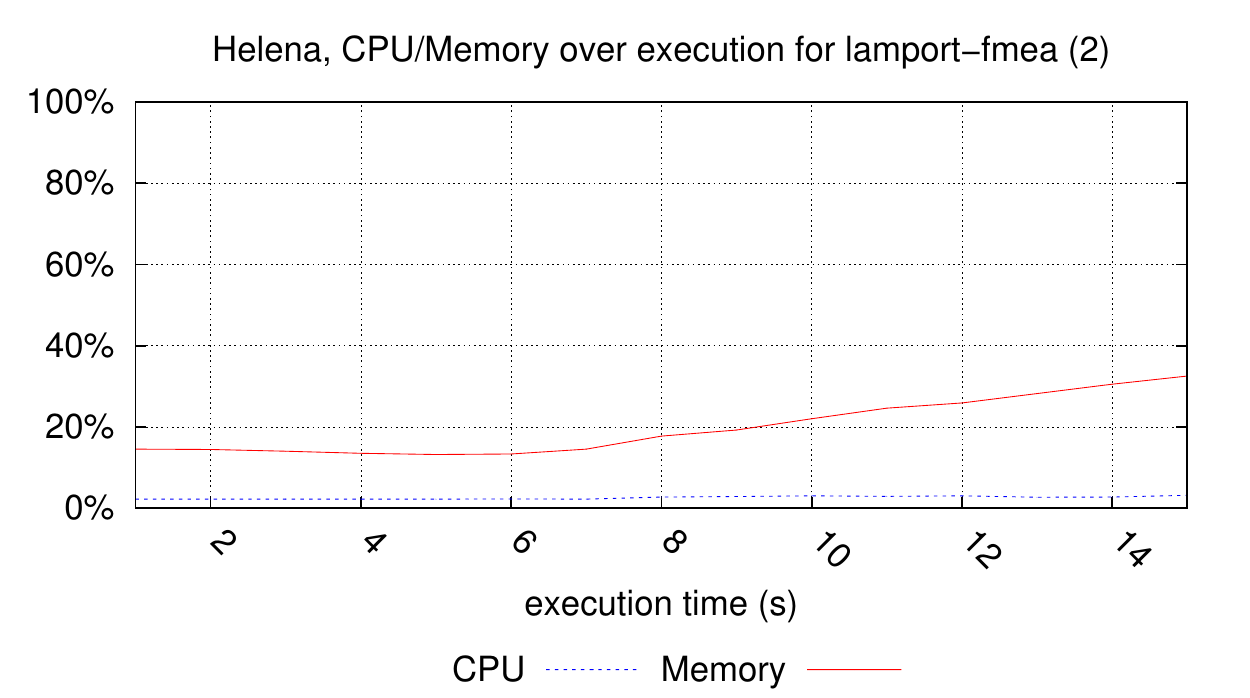}

\vfill\eject
\subsubsection{Executions for neo-election}
1 chart has been generated.
\index{Execution (by tool)!Helena}
\index{Execution (by model)!neo-election!Helena}

\noindent\includegraphics[width=.5\textwidth]{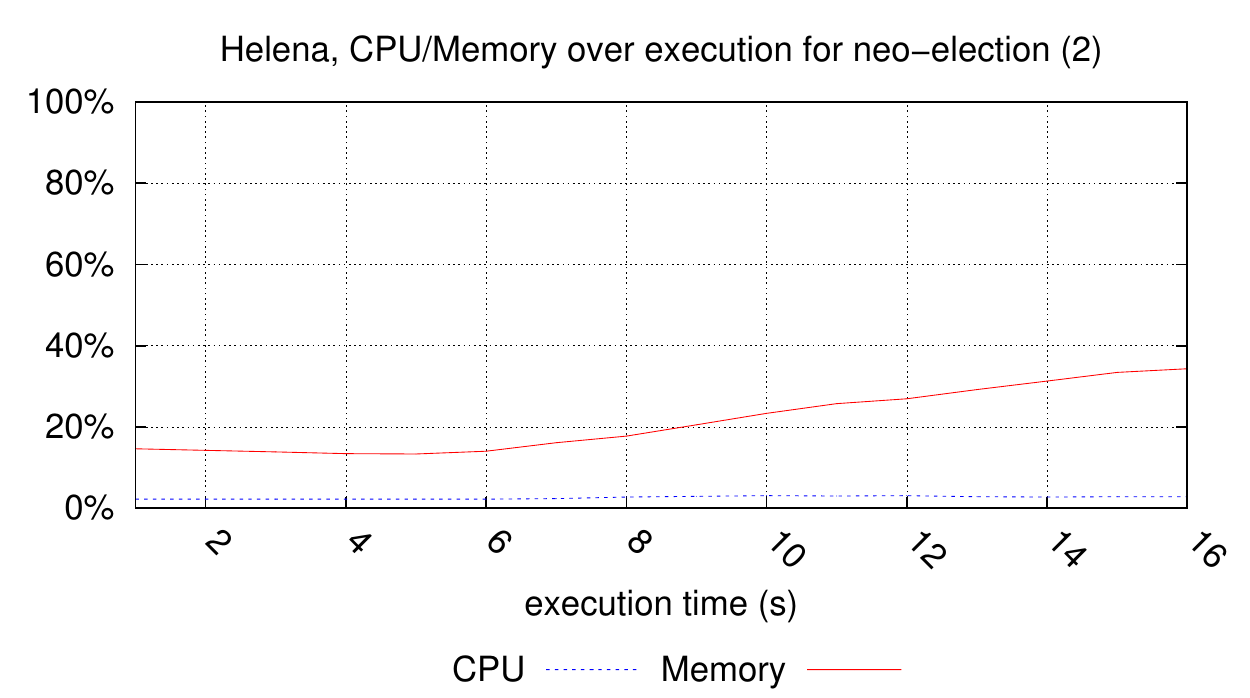}

\subsubsection{Executions for philo\_dyn}
6 charts have been generated.
\index{Execution (by tool)!Helena}
\index{Execution (by model)!philo\_dyn!Helena}

\noindent\includegraphics[width=.5\textwidth]{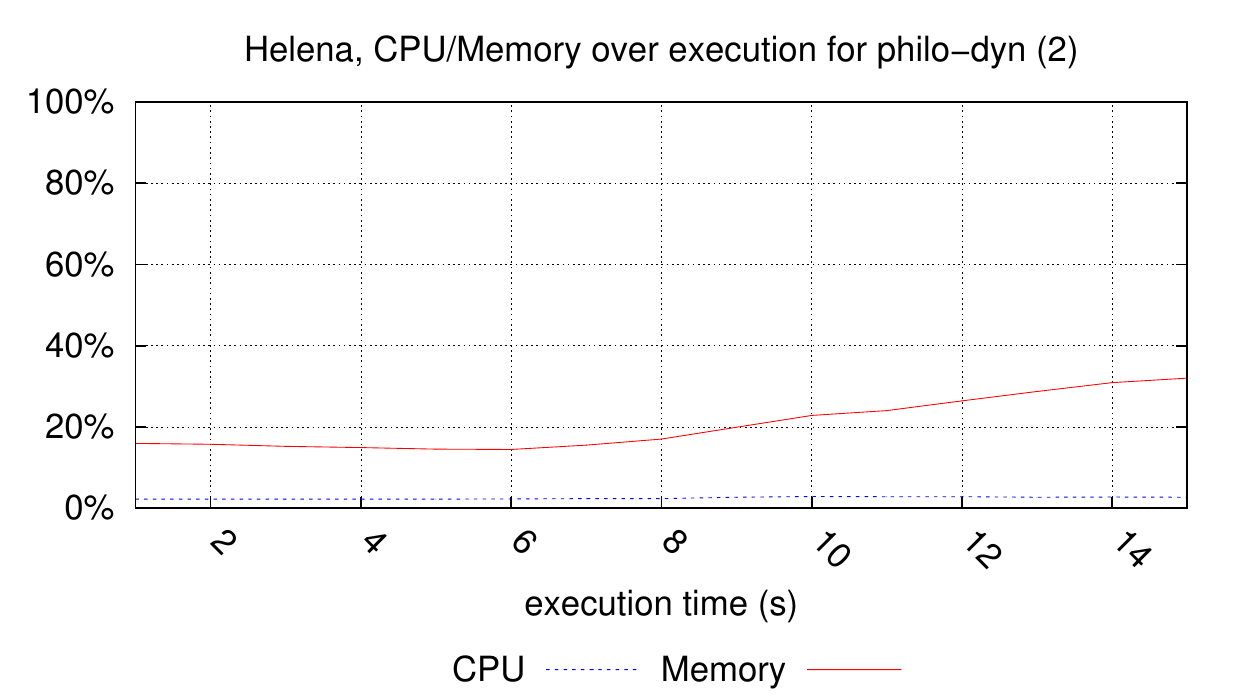}
\includegraphics[width=.5\textwidth]{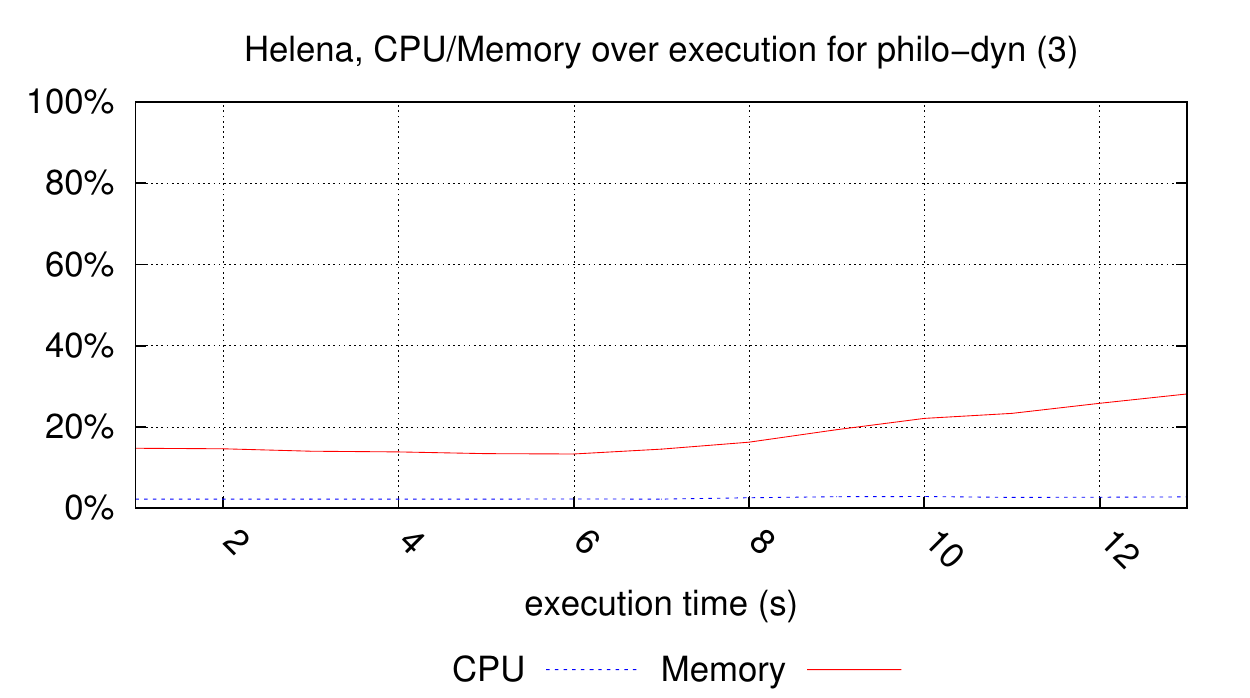}

\noindent\includegraphics[width=.5\textwidth]{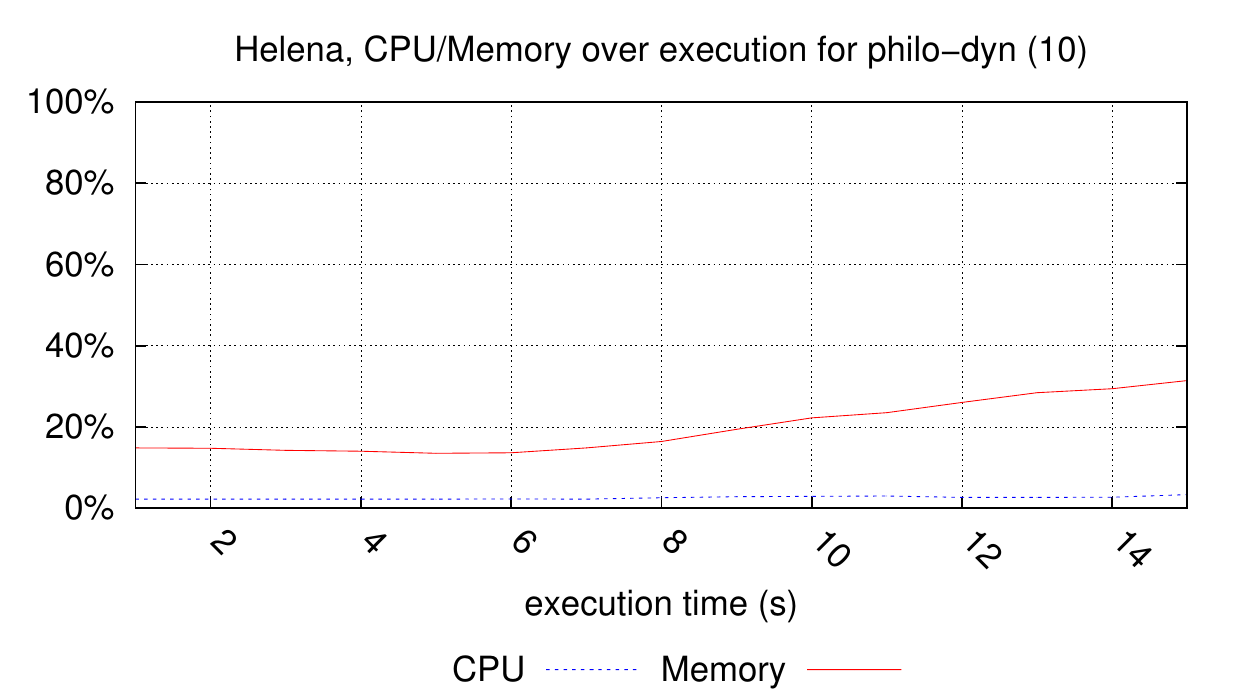}
\includegraphics[width=.5\textwidth]{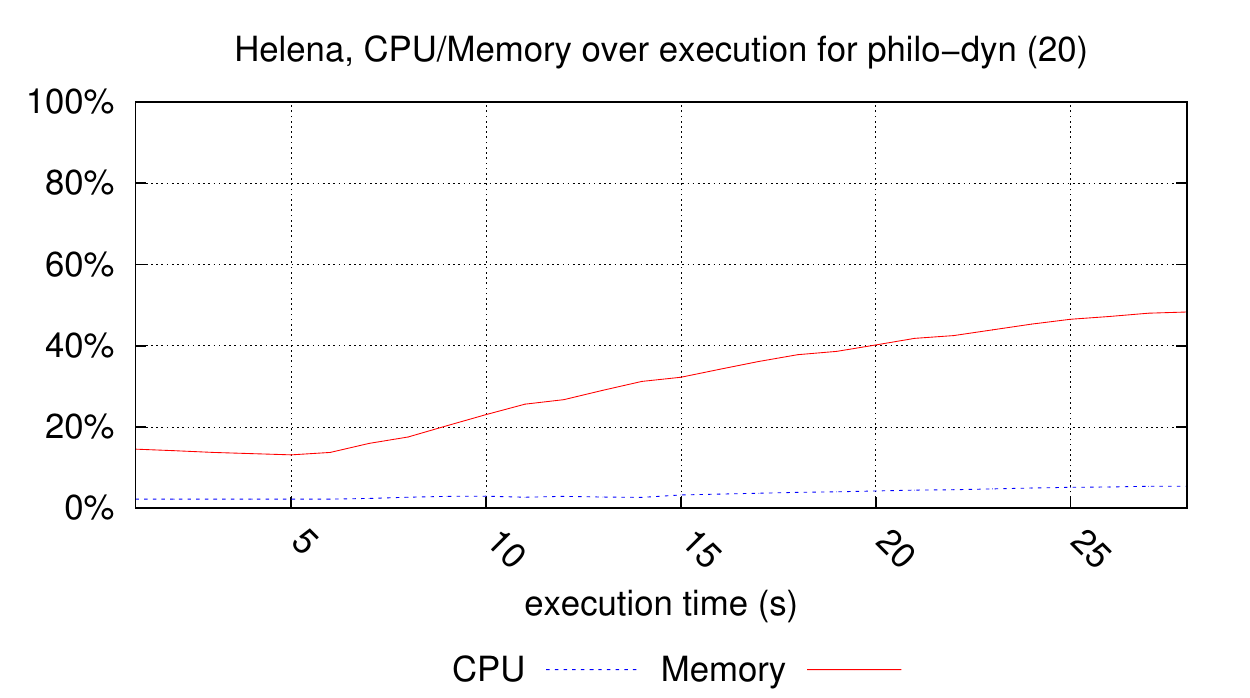}

\noindent\includegraphics[width=.5\textwidth]{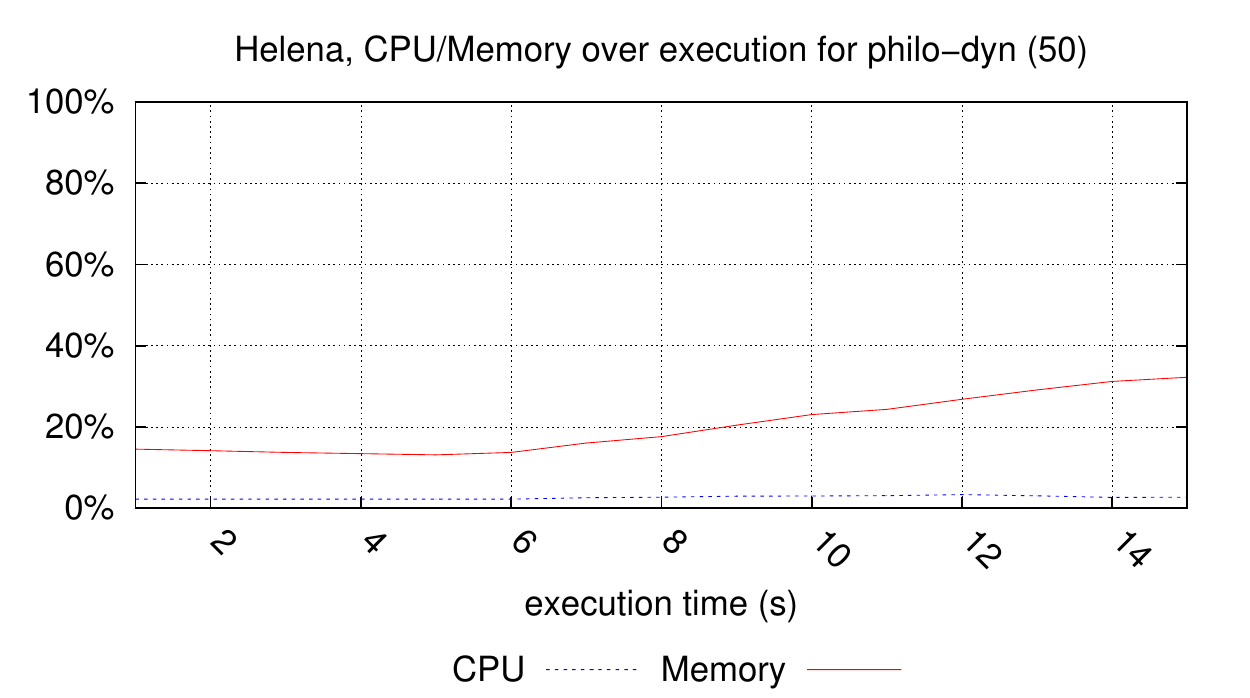}
\includegraphics[width=.5\textwidth]{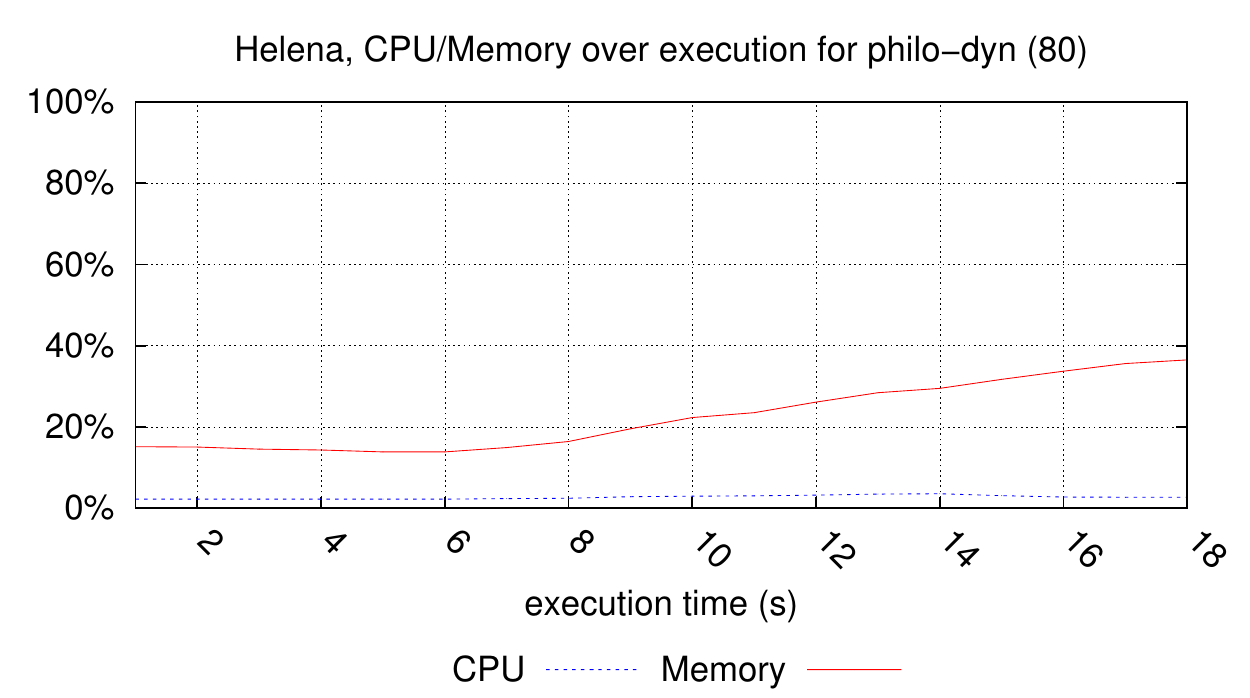}

\vfill\eject
\subsubsection{Executions for Philosophers}
4 charts have been generated.
\index{Execution (by tool)!Helena}
\index{Execution (by model)!Philosophers!Helena}

\noindent\includegraphics[width=.5\textwidth]{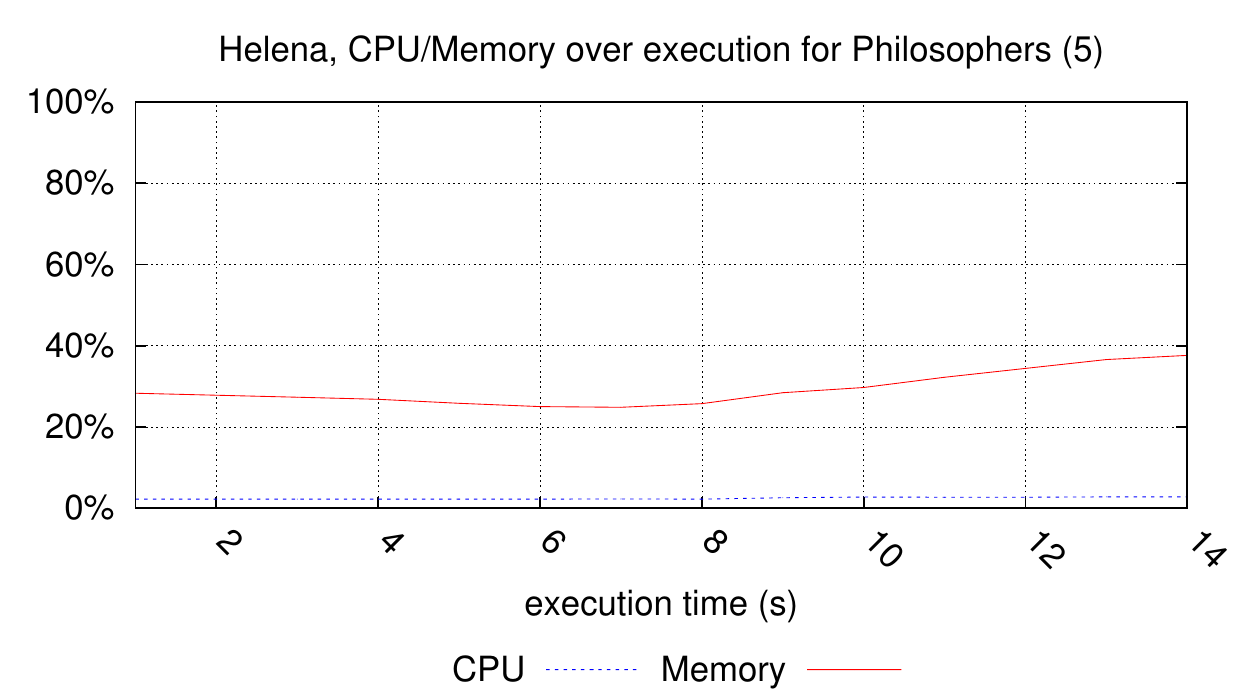}
\includegraphics[width=.5\textwidth]{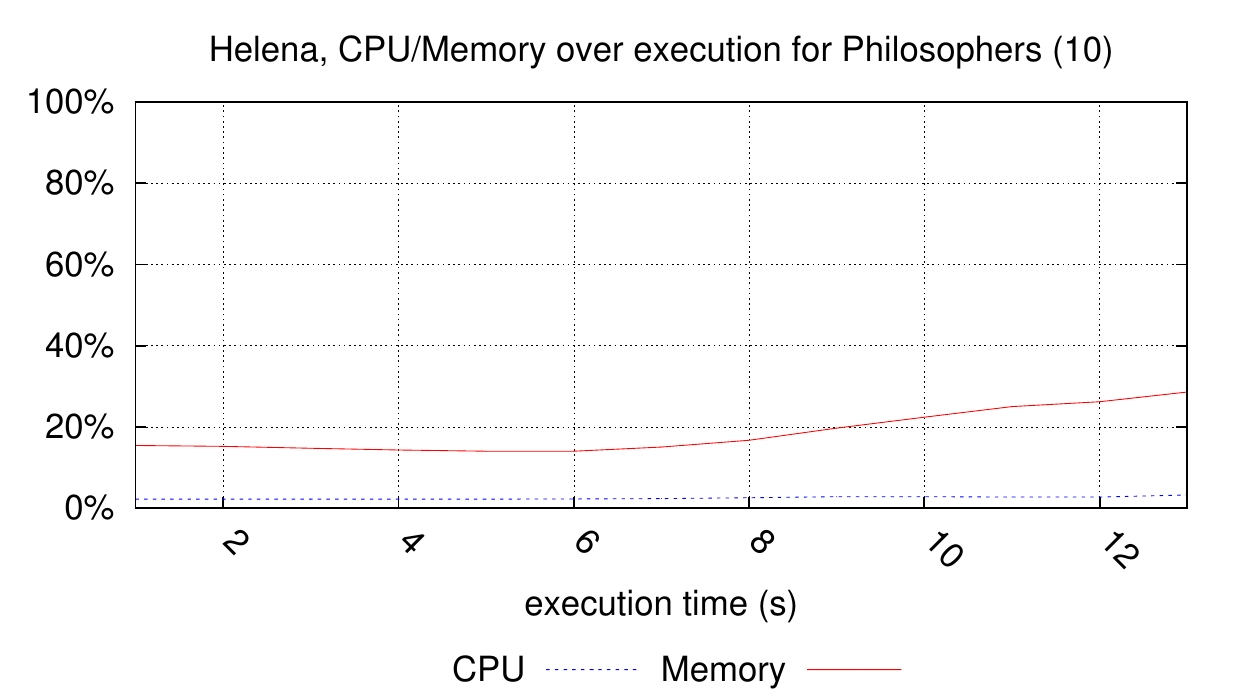}

\noindent\includegraphics[width=.5\textwidth]{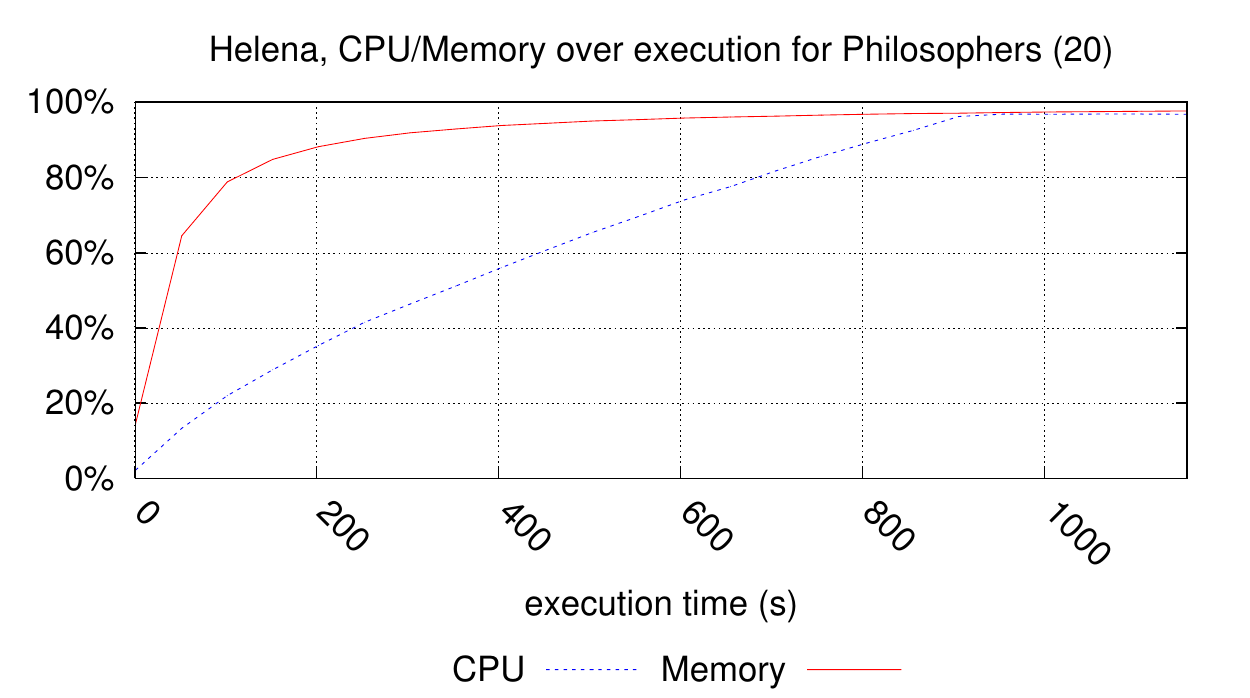}
\includegraphics[width=.5\textwidth]{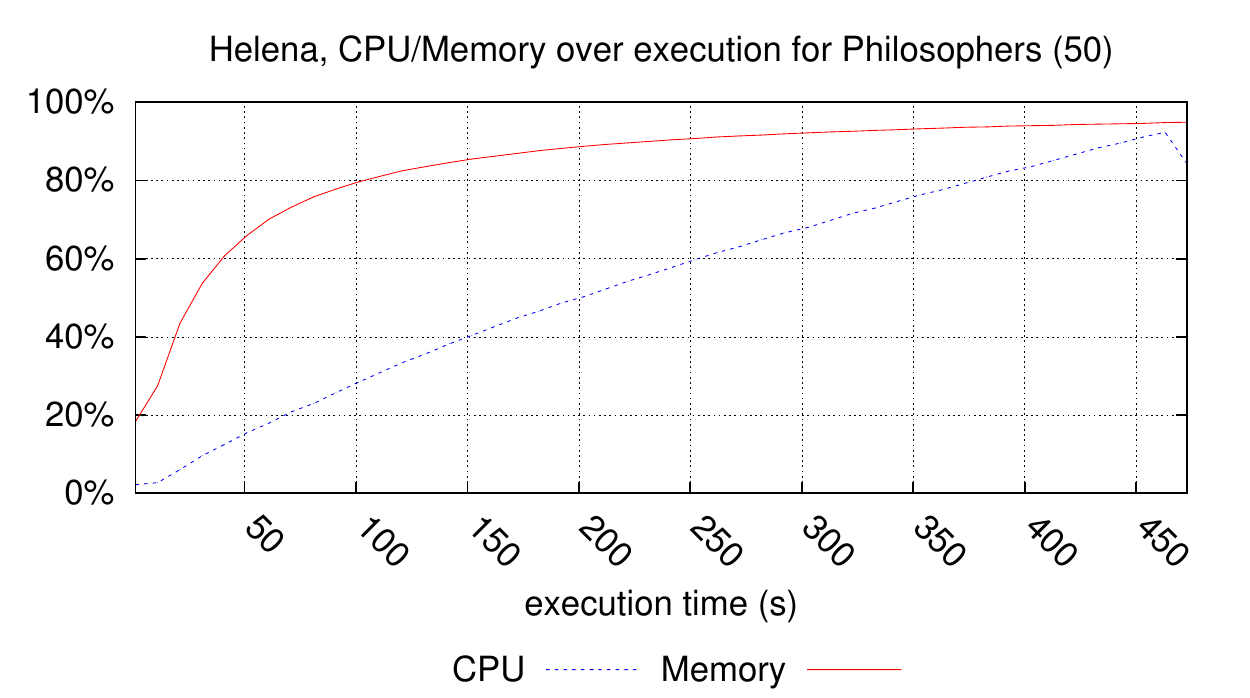}

\subsubsection{Executions for SharedMemory}
4 charts have been generated.
\index{Execution (by tool)!Helena}
\index{Execution (by model)!SharedMemory!Helena}

\noindent\includegraphics[width=.5\textwidth]{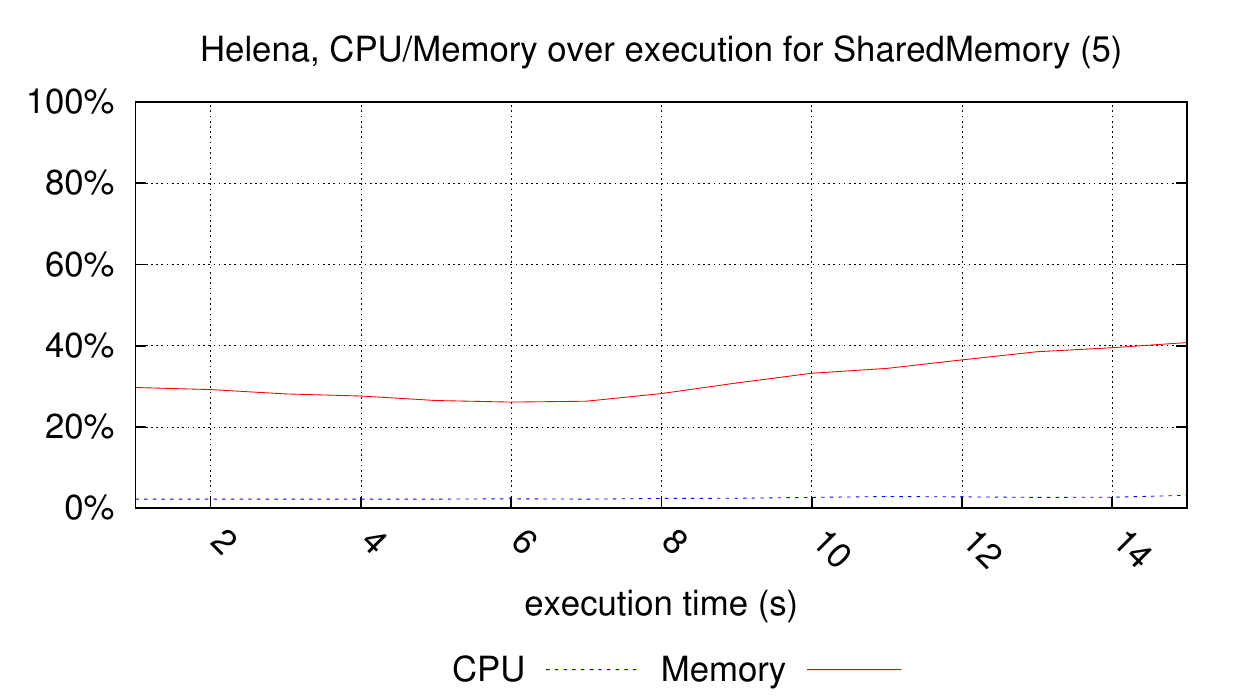}
\includegraphics[width=.5\textwidth]{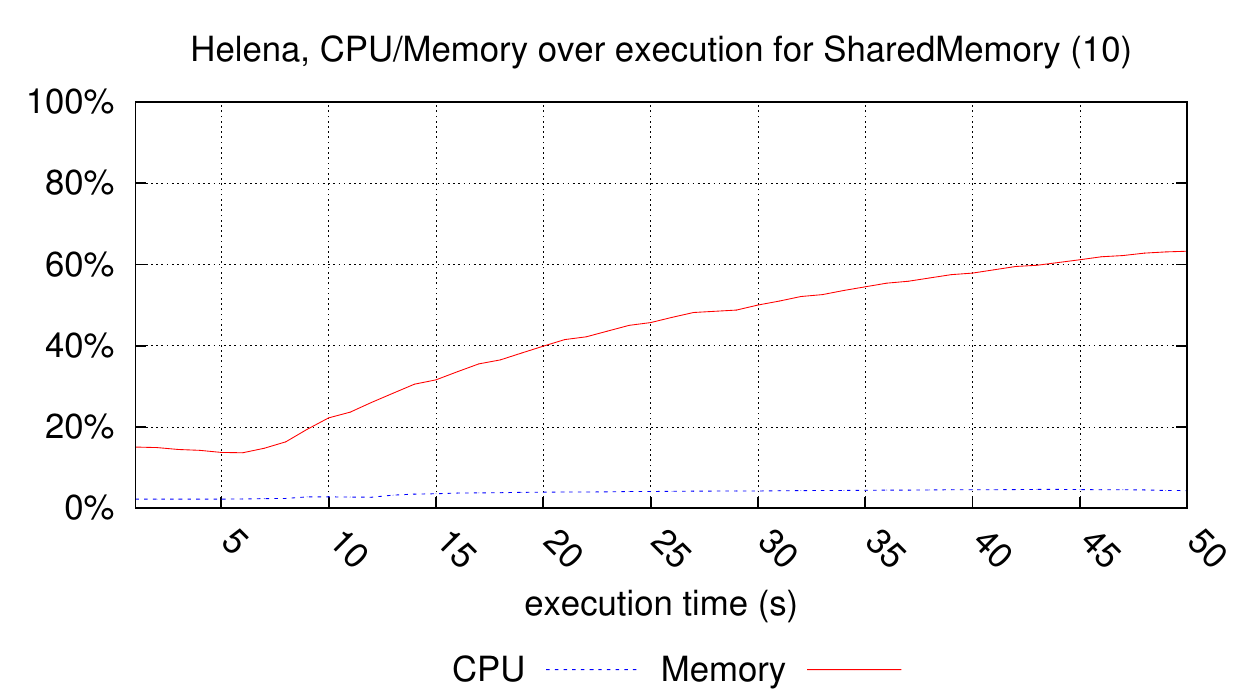}

\noindent\includegraphics[width=.5\textwidth]{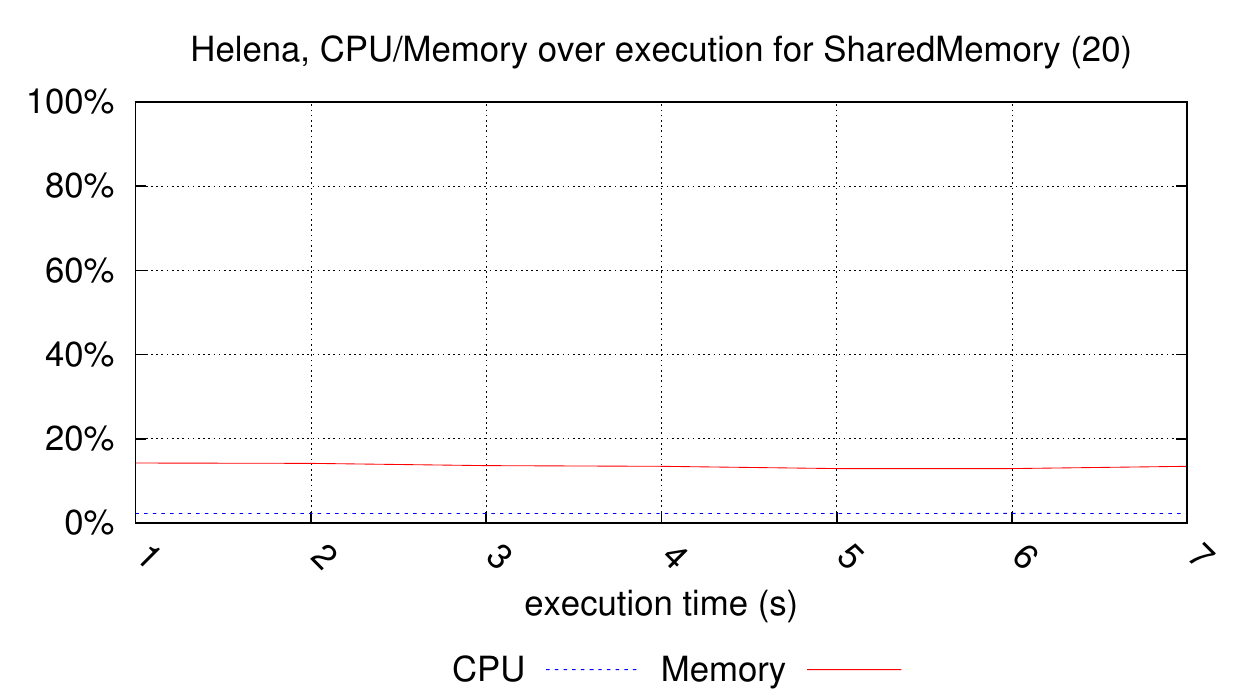}
\includegraphics[width=.5\textwidth]{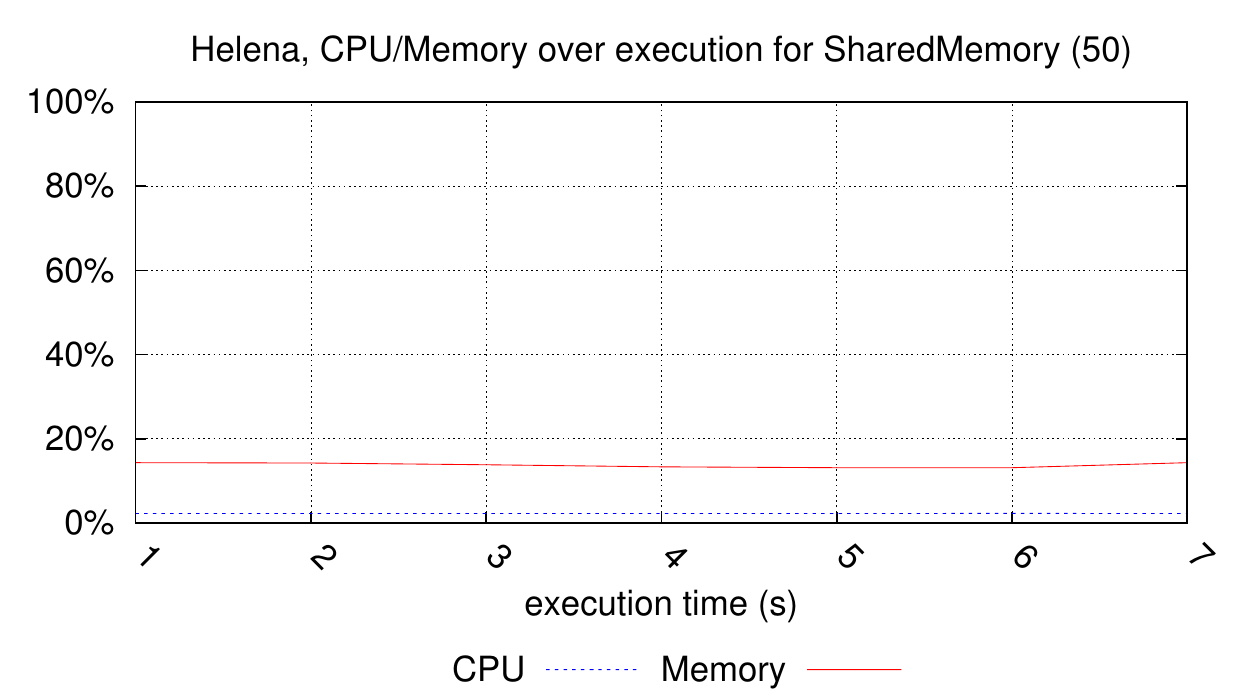}

\vfill\eject
\subsubsection{Executions for TokenRing}
4 charts have been generated.
\index{Execution (by tool)!Helena}
\index{Execution (by model)!TokenRing!Helena}

\noindent\includegraphics[width=.5\textwidth]{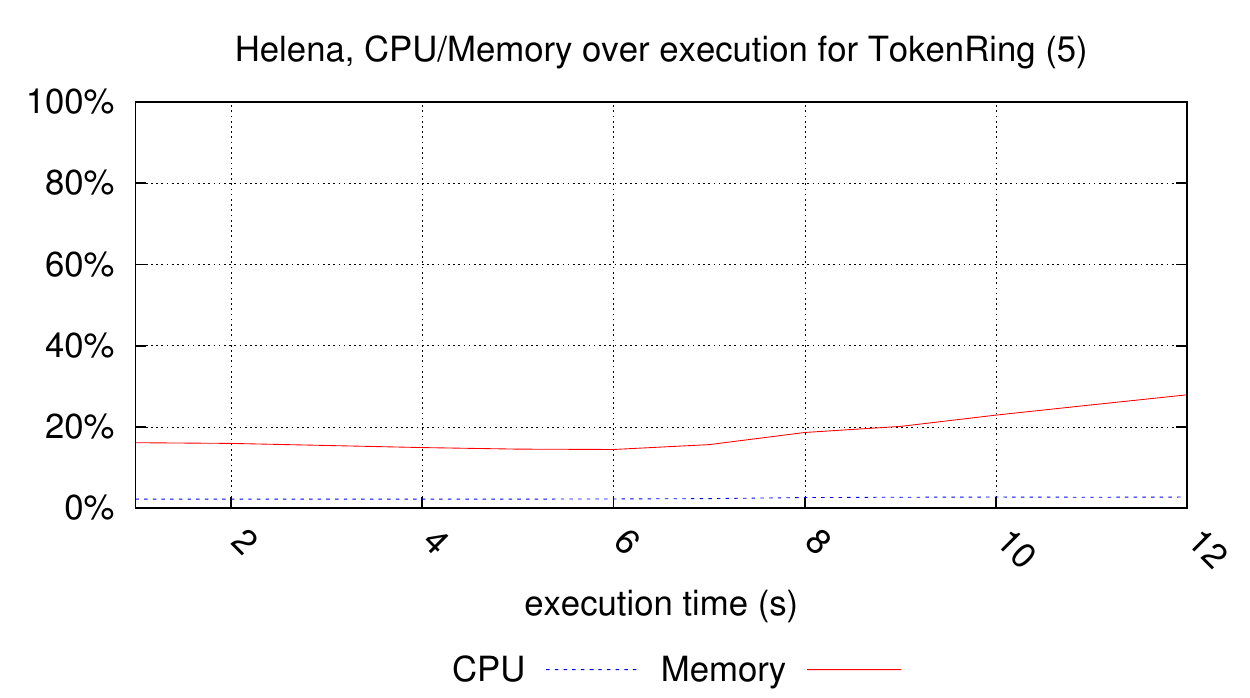}
\includegraphics[width=.5\textwidth]{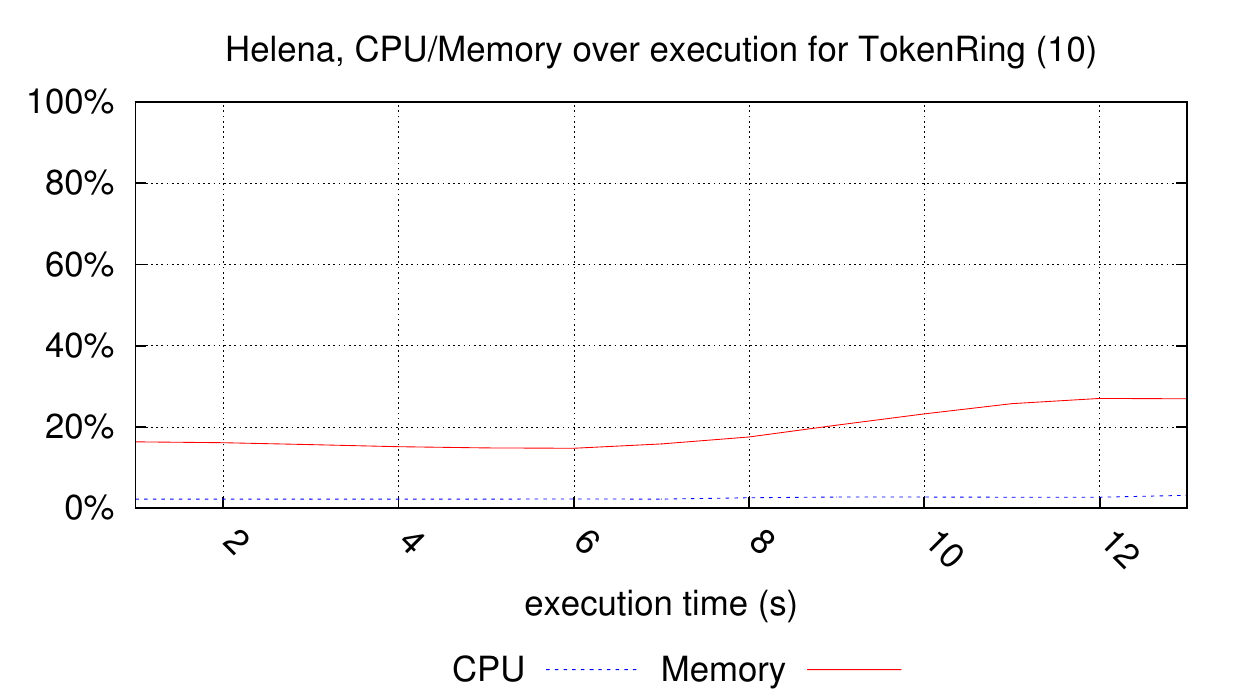}

\noindent\includegraphics[width=.5\textwidth]{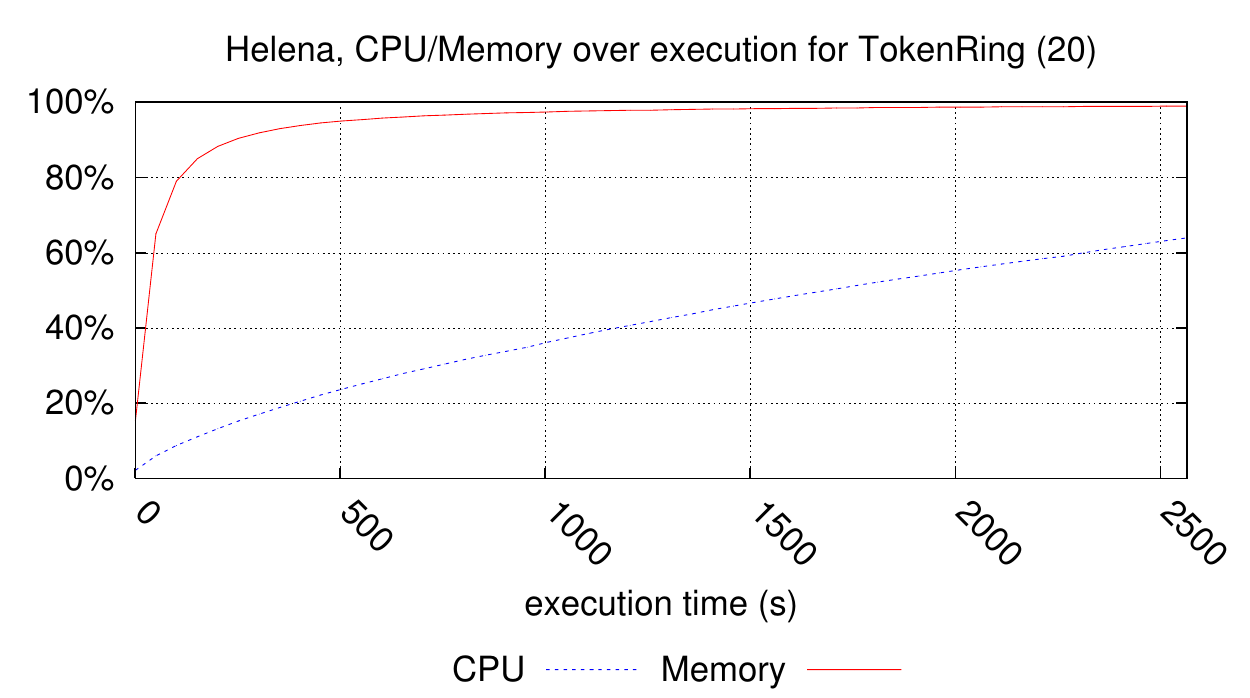}
\includegraphics[width=.5\textwidth]{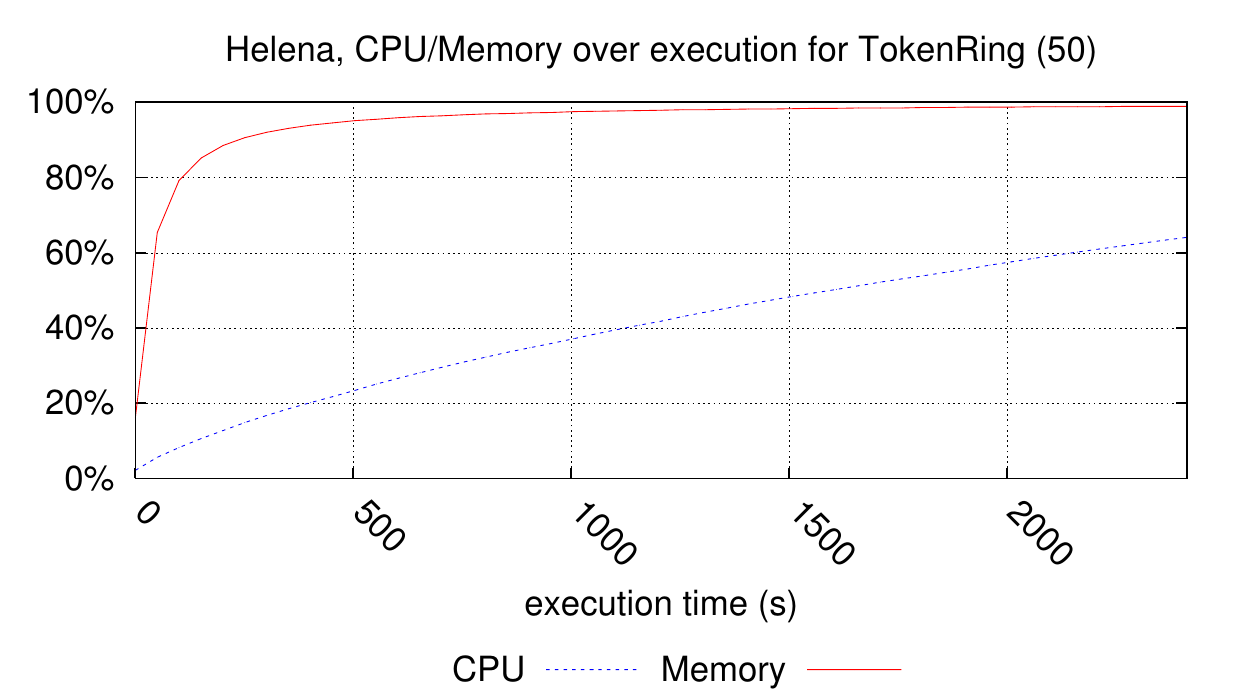}

\subsection{Execution Charts for ITS-Tools}

We provide here the execution charts observed for ITS-Tools over
the models it could compete with.

\subsubsection{Executions for echo}
1 chart has been generated.
\index{Execution (by tool)!ITS-Tools}
\index{Execution (by model)!echo!ITS-Tools}

\noindent\includegraphics[width=.5\textwidth]{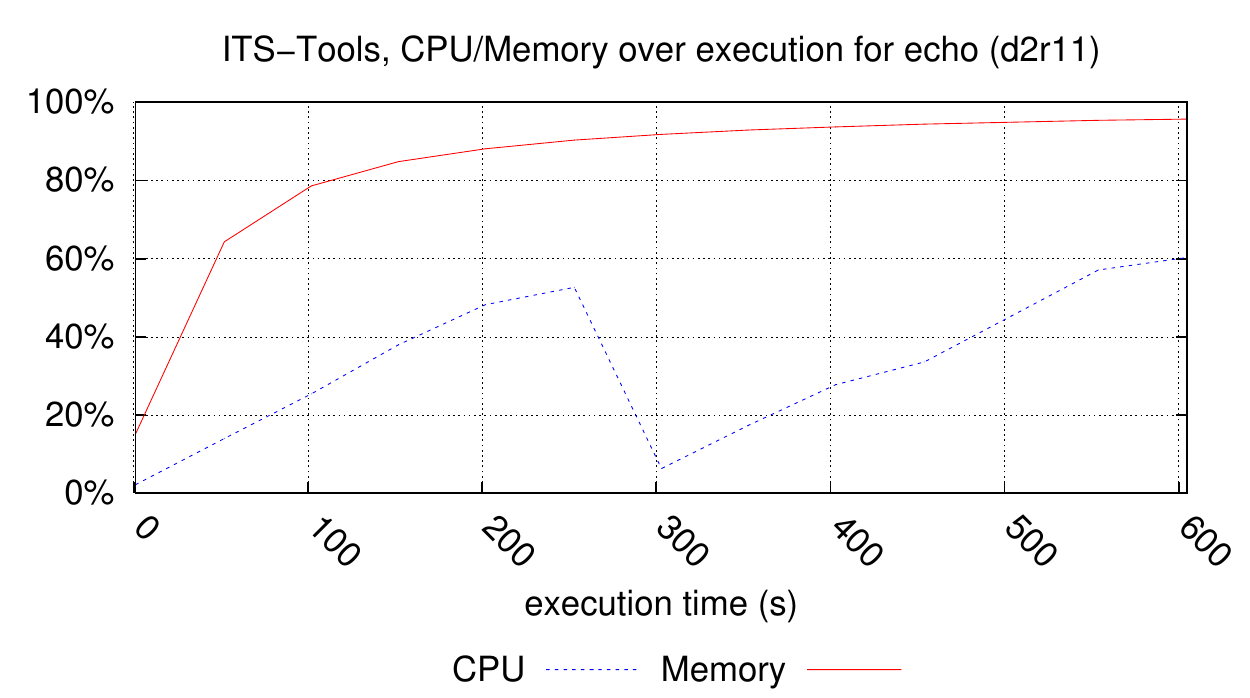}

\subsubsection{Executions for eratosthenes}
7 charts have been generated.
\index{Execution (by tool)!ITS-Tools}
\index{Execution (by model)!eratosthenes!ITS-Tools}

\noindent\includegraphics[width=.5\textwidth]{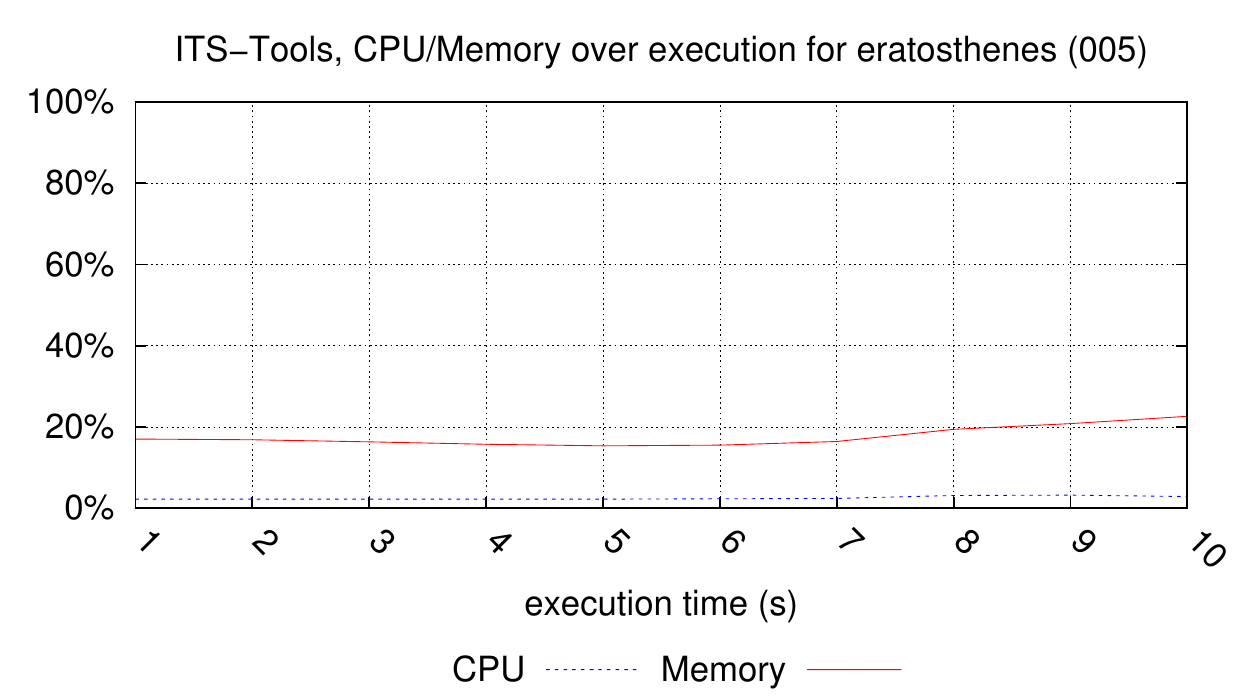}
\includegraphics[width=.5\textwidth]{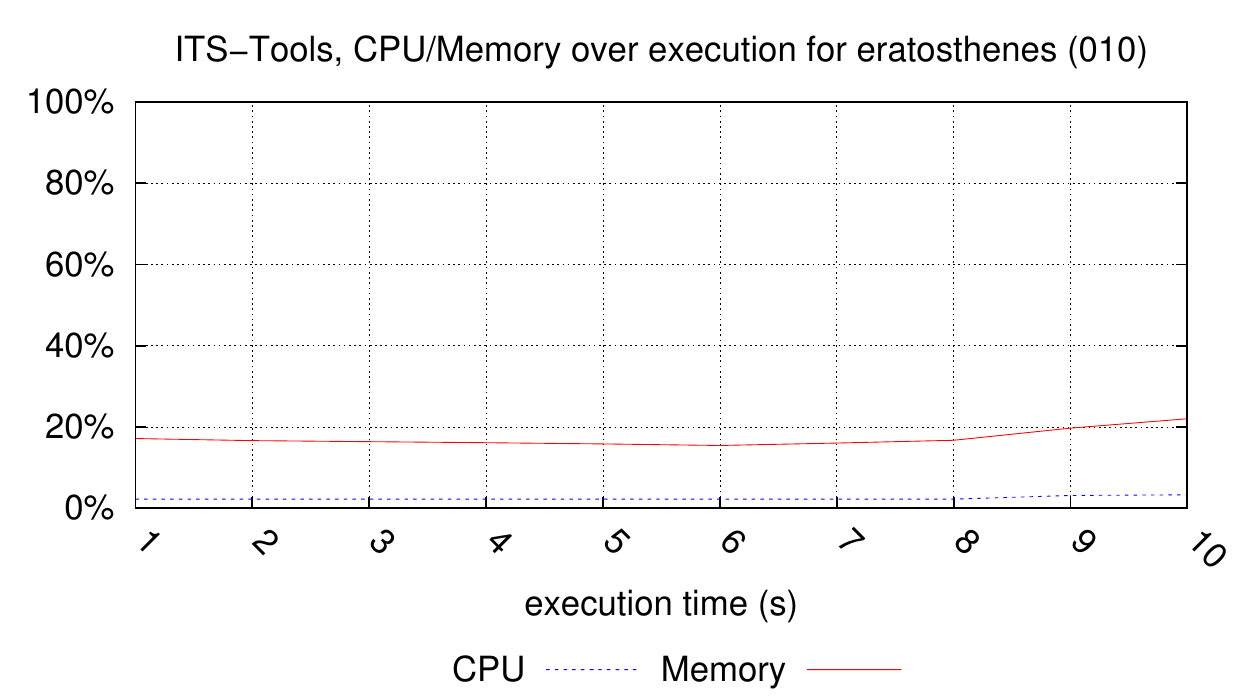}

\noindent\includegraphics[width=.5\textwidth]{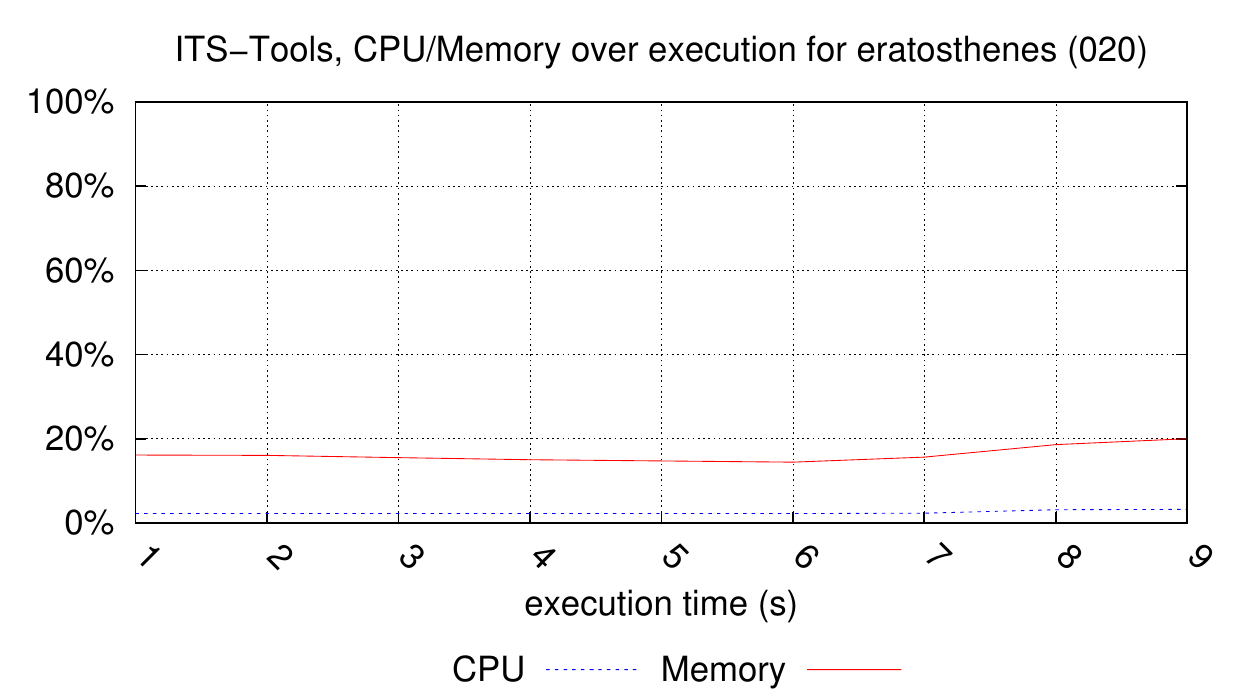}
\includegraphics[width=.5\textwidth]{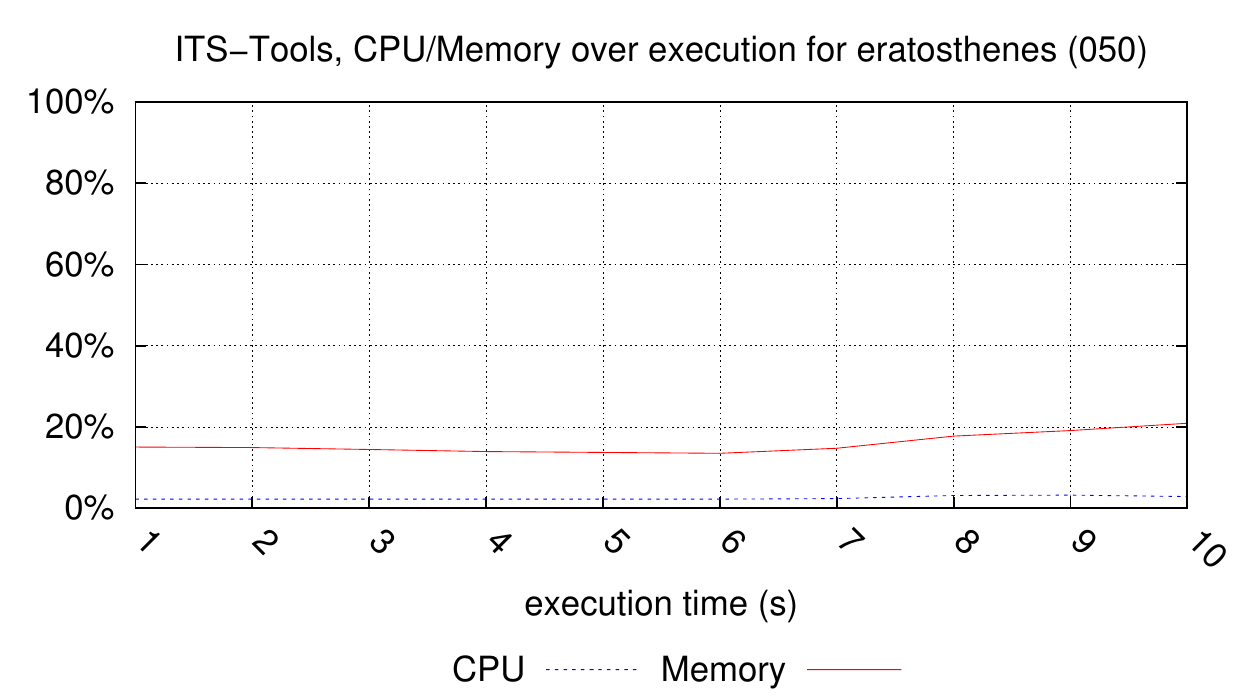}

\noindent\includegraphics[width=.5\textwidth]{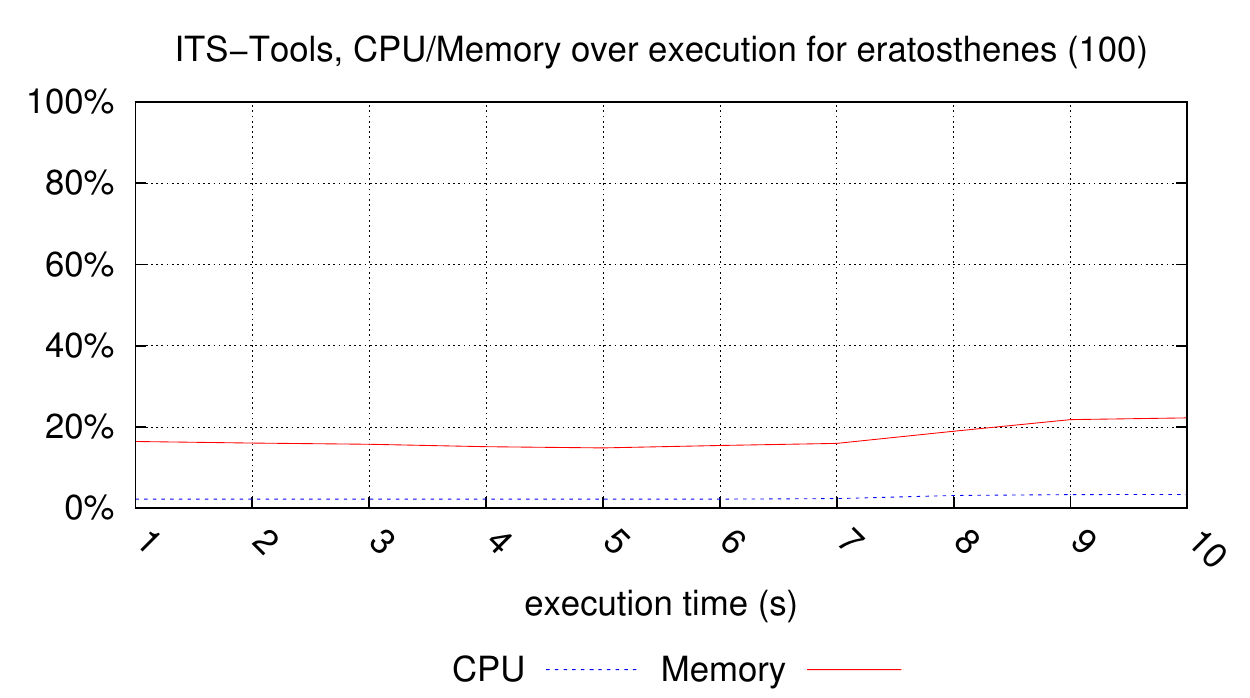}
\includegraphics[width=.5\textwidth]{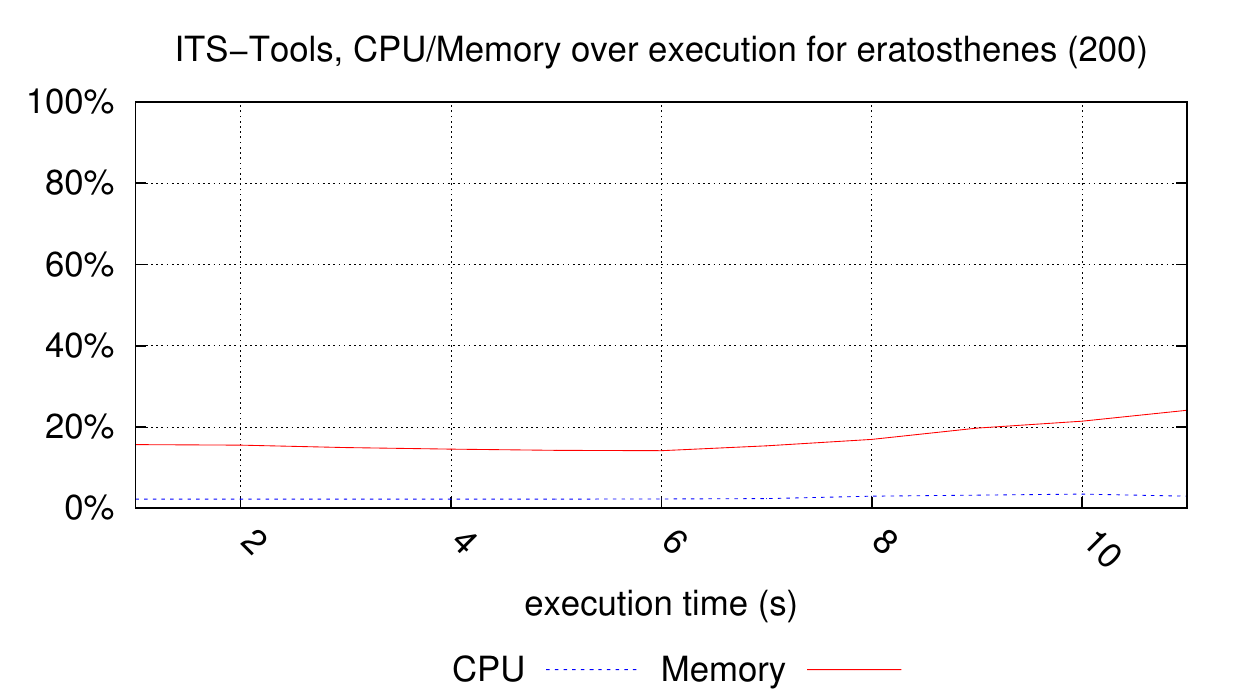}

\noindent\includegraphics[width=.5\textwidth]{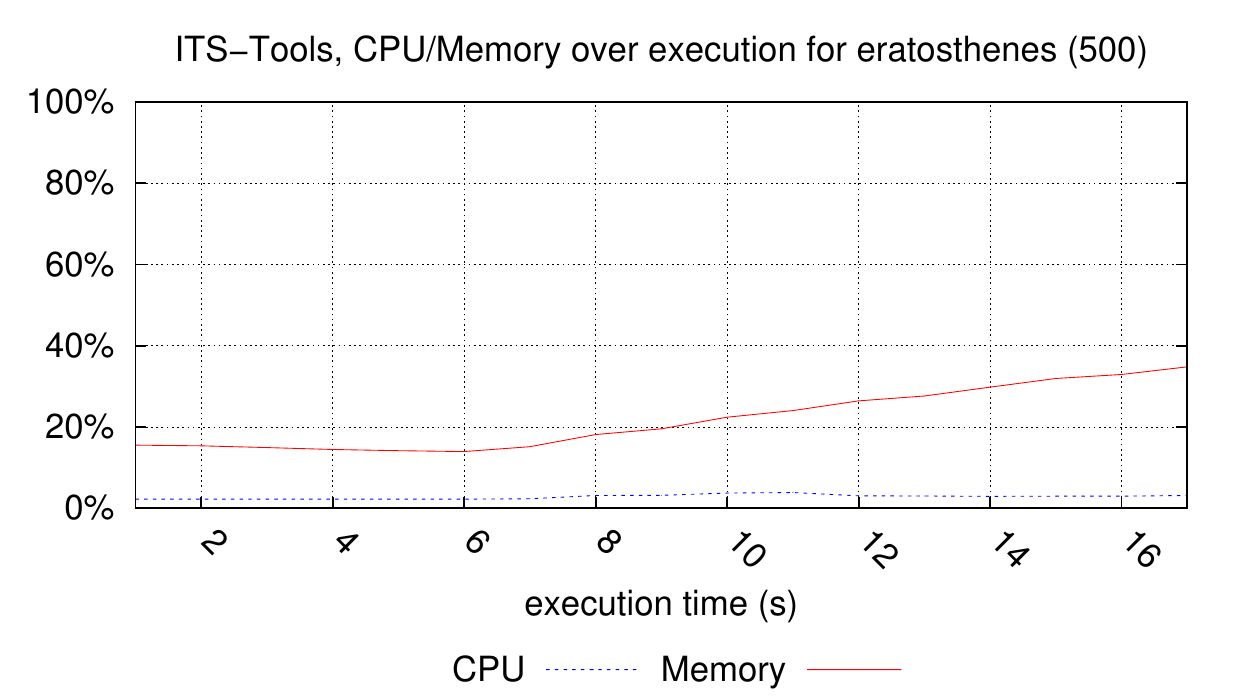}

\subsubsection{Executions for FMS}
7 charts have been generated.
\index{Execution (by tool)!ITS-Tools}
\index{Execution (by model)!FMS!ITS-Tools}

\noindent\includegraphics[width=.5\textwidth]{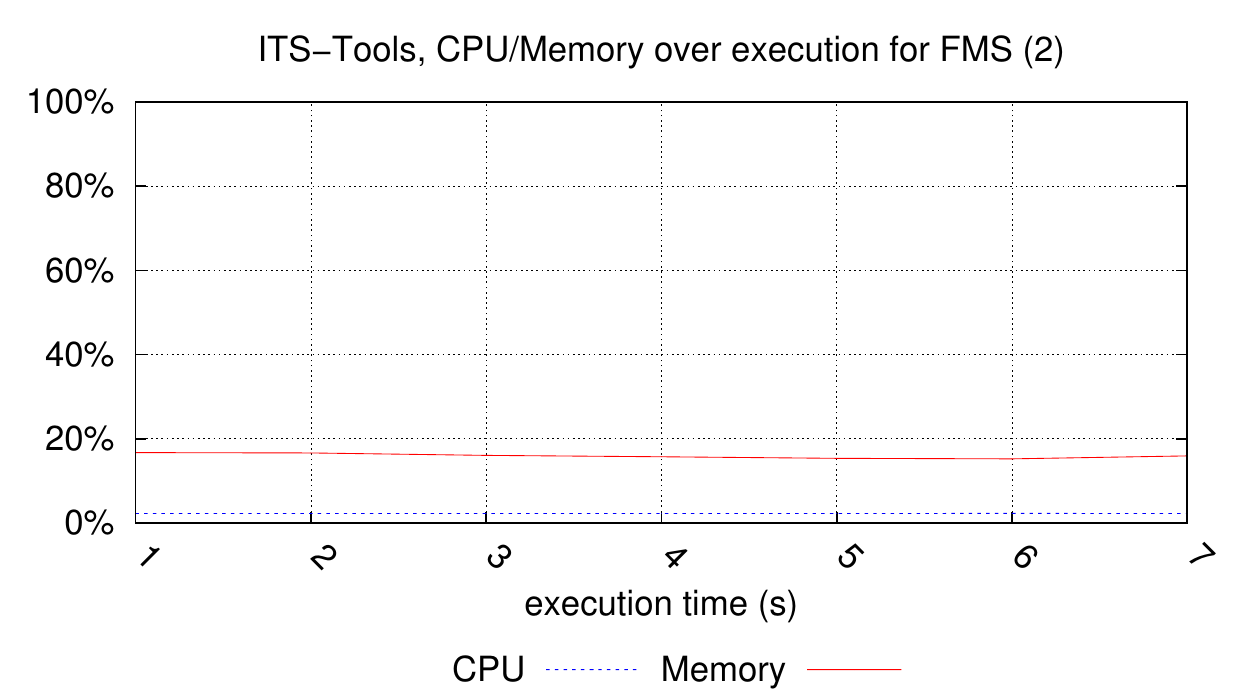}
\includegraphics[width=.5\textwidth]{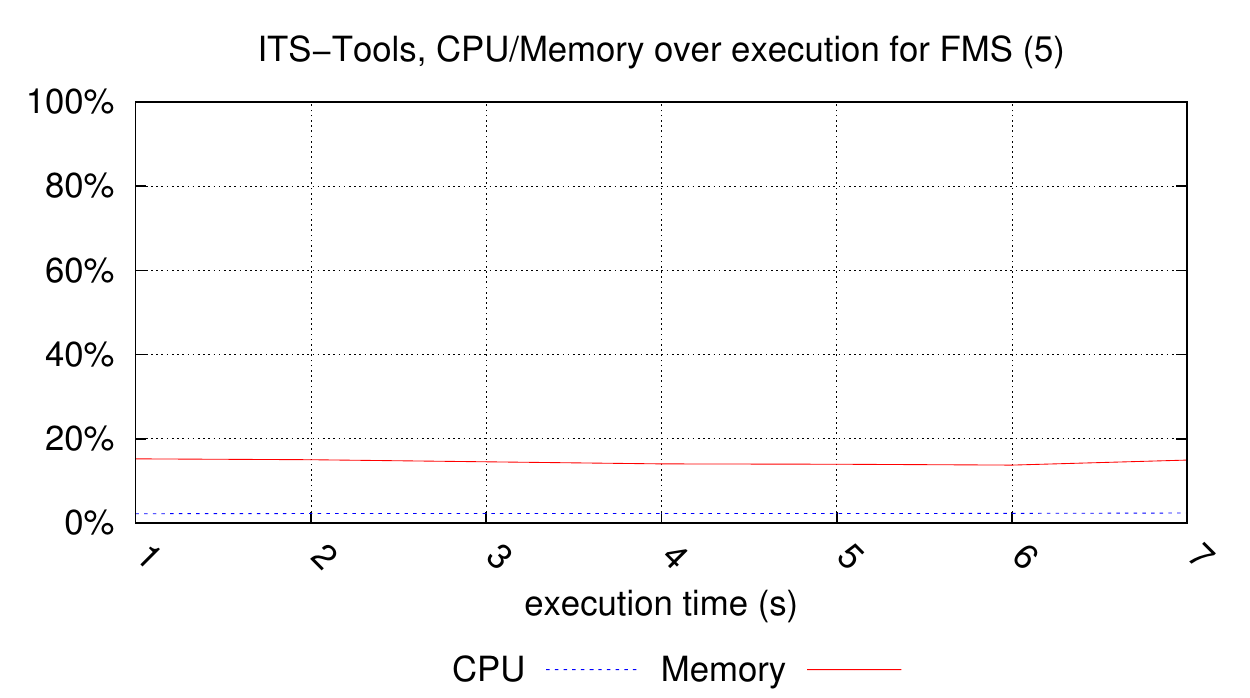}

\noindent\includegraphics[width=.5\textwidth]{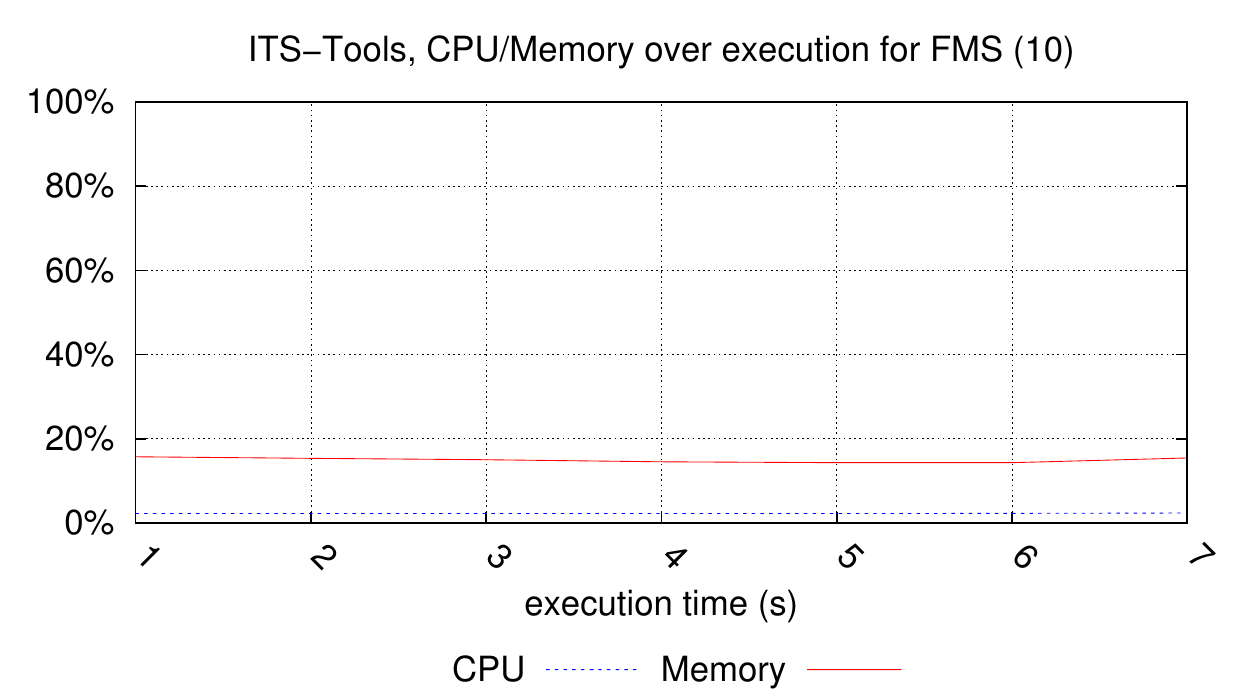}
\includegraphics[width=.5\textwidth]{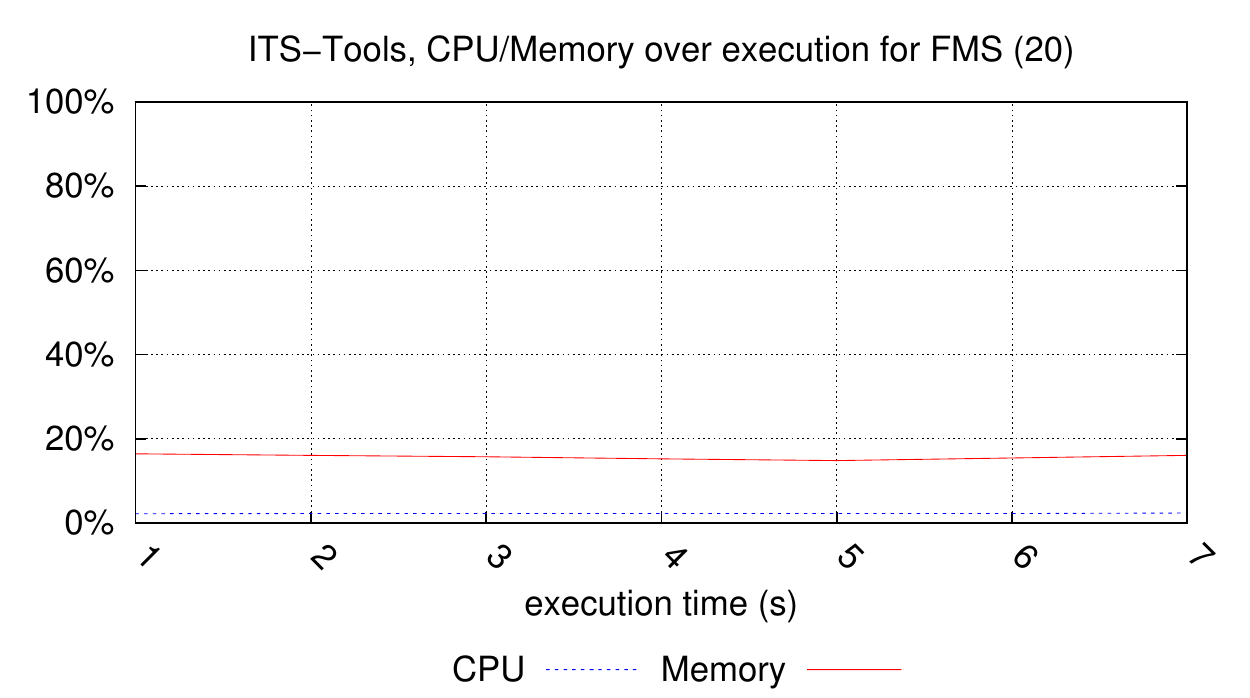}

\noindent\includegraphics[width=.5\textwidth]{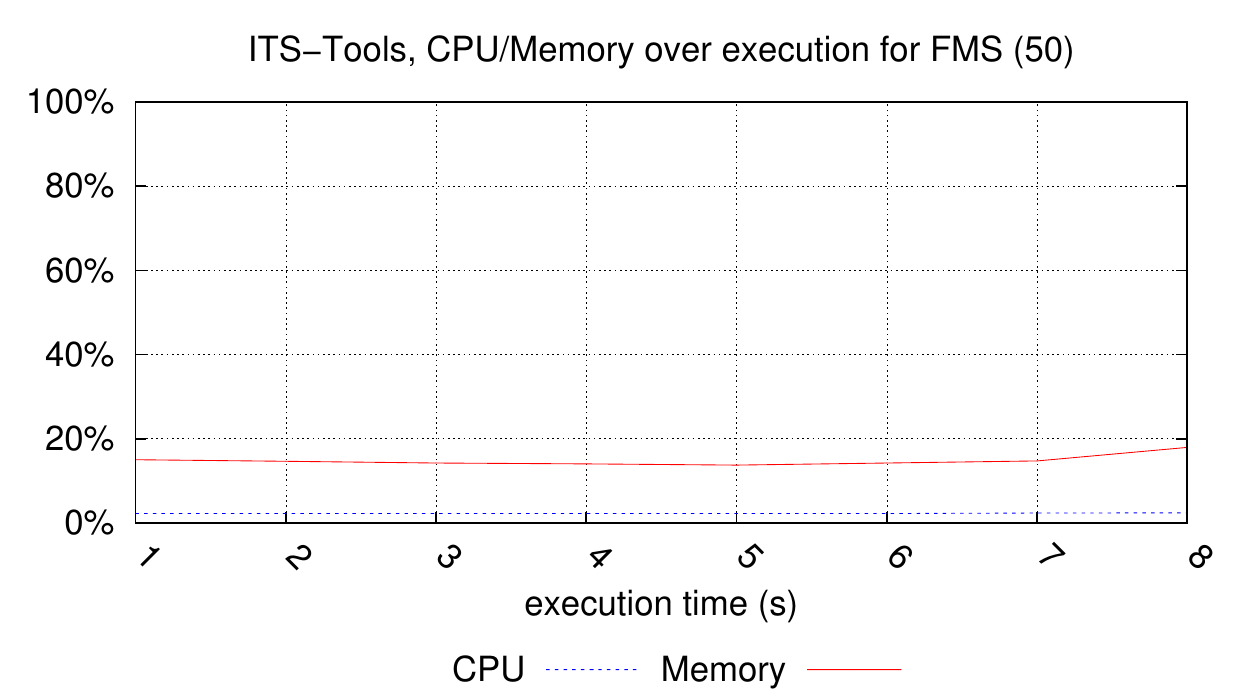}
\includegraphics[width=.5\textwidth]{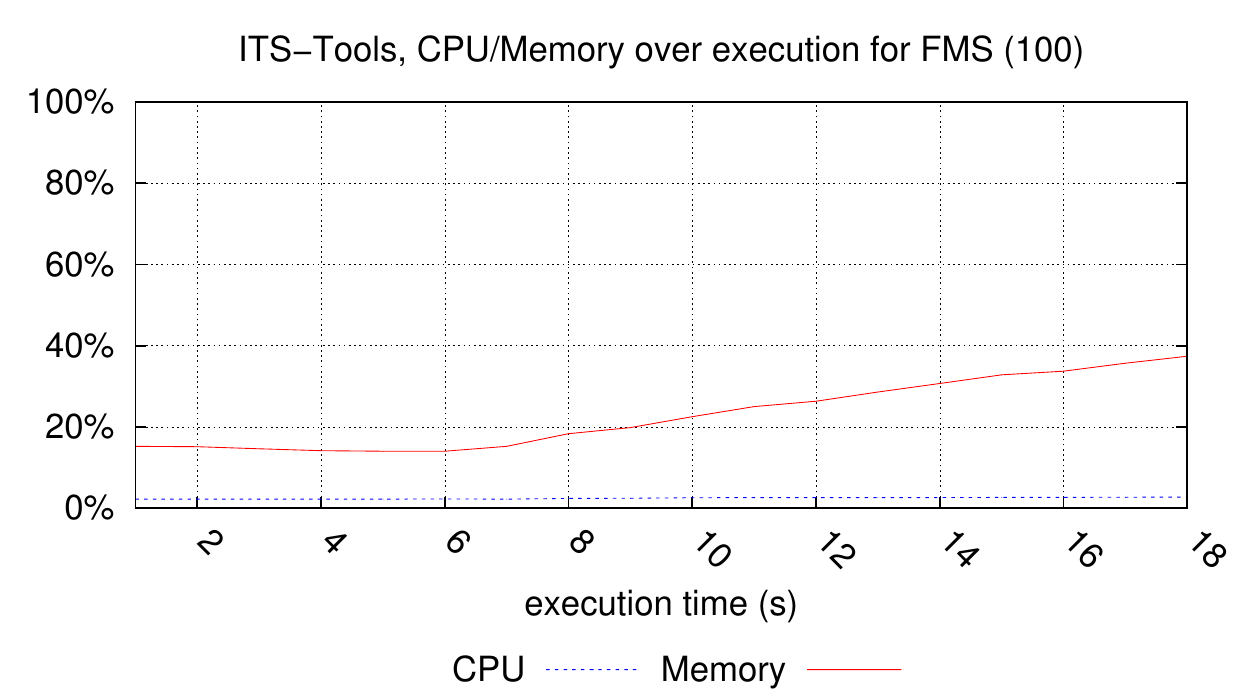}

\noindent\includegraphics[width=.5\textwidth]{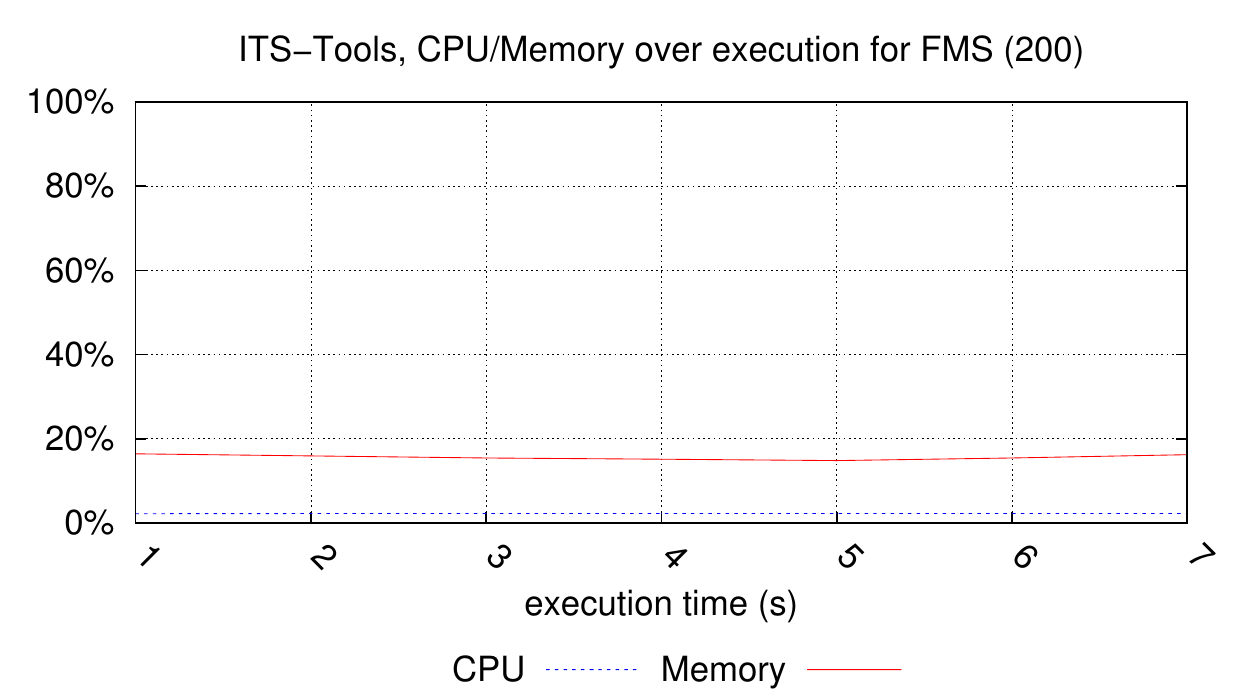}

\subsubsection{Executions for galloc\_res}
2 charts have been generated.
\index{Execution (by tool)!ITS-Tools}
\index{Execution (by model)!galloc\_res!ITS-Tools}

\noindent\includegraphics[width=.5\textwidth]{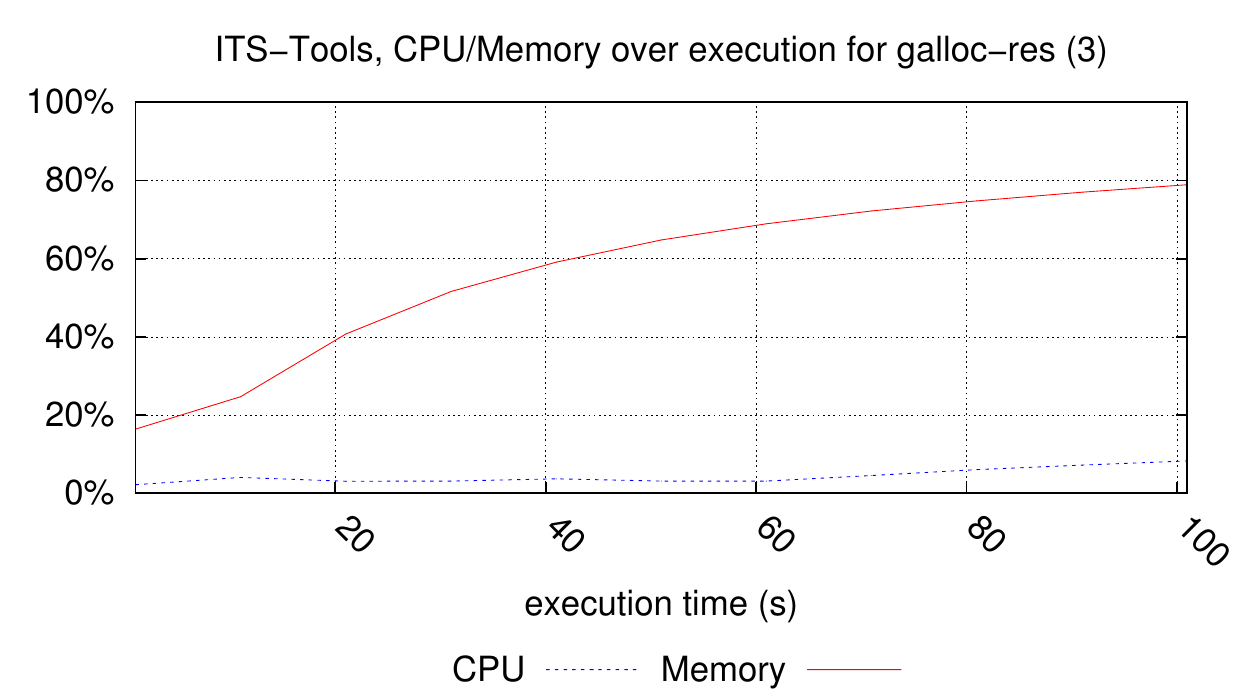}
\includegraphics[width=.5\textwidth]{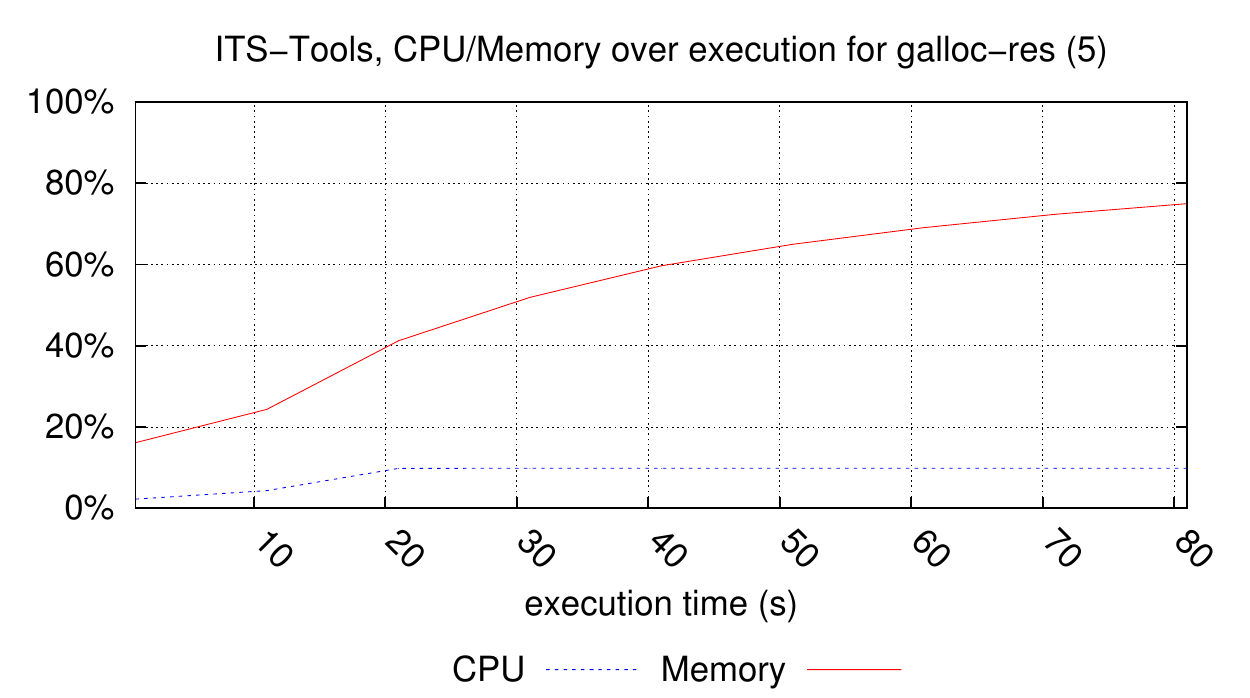}

\subsubsection{Executions for Kanban}
8 charts have been generated.
\index{Execution (by tool)!ITS-Tools}
\index{Execution (by model)!Kanban!ITS-Tools}

\noindent\includegraphics[width=.5\textwidth]{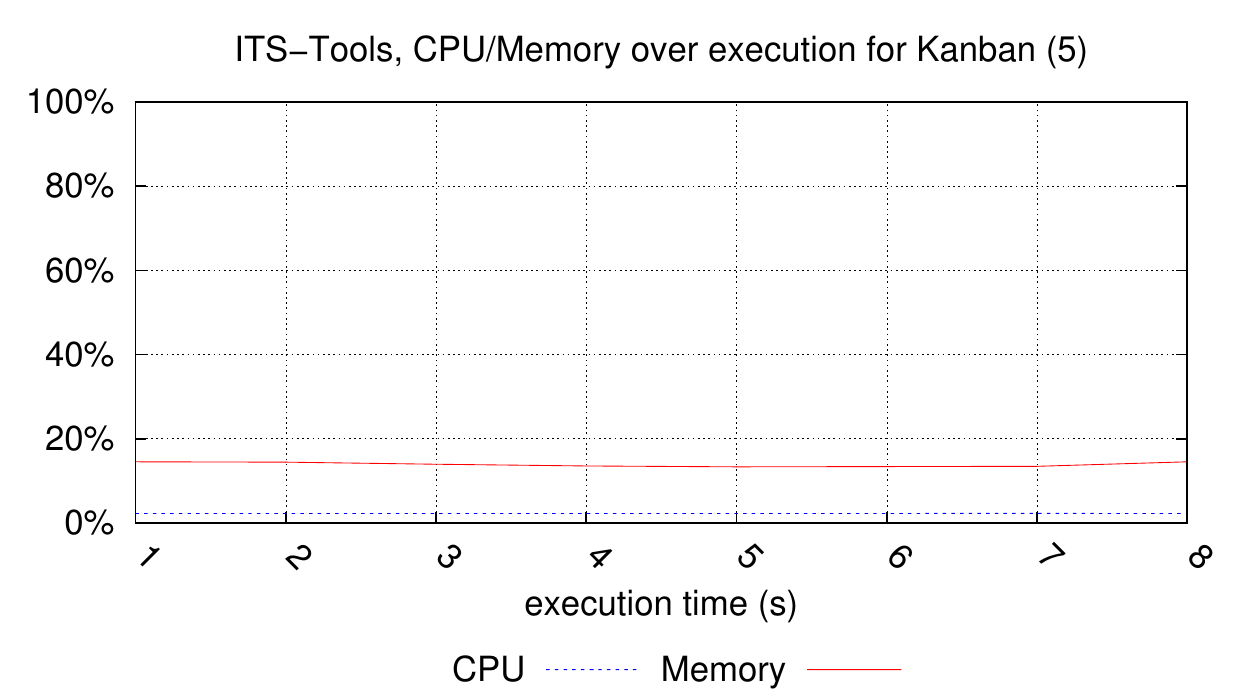}
\includegraphics[width=.5\textwidth]{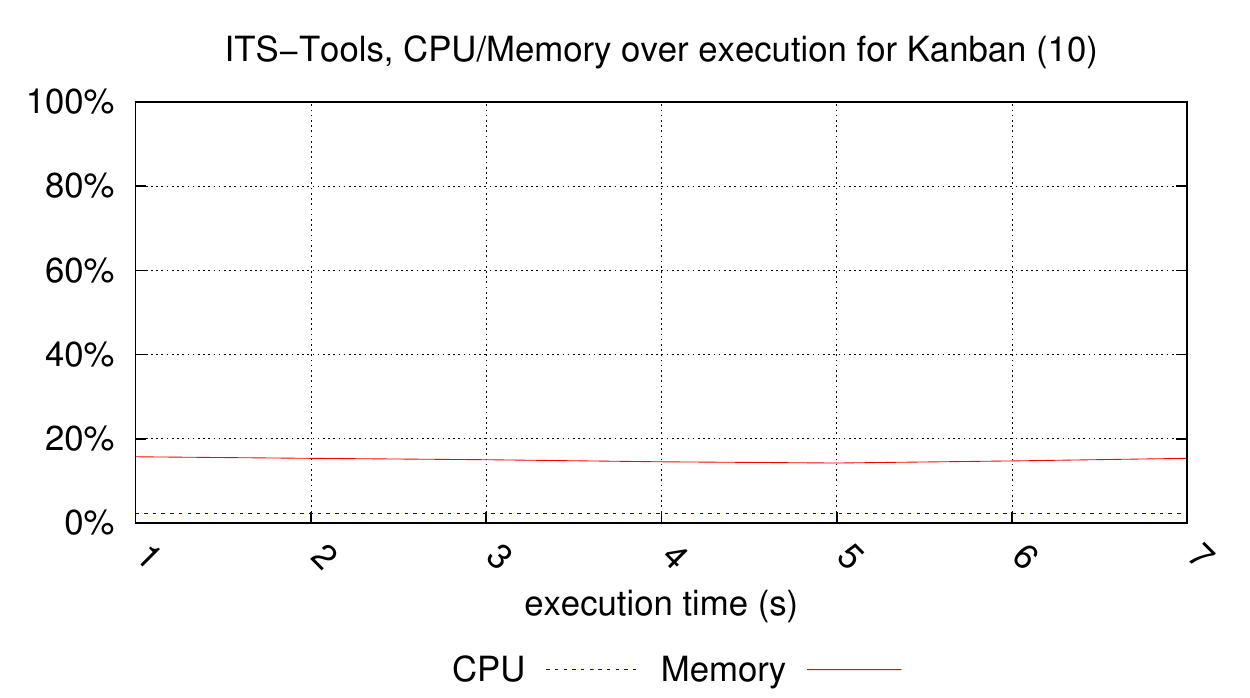}

\noindent\includegraphics[width=.5\textwidth]{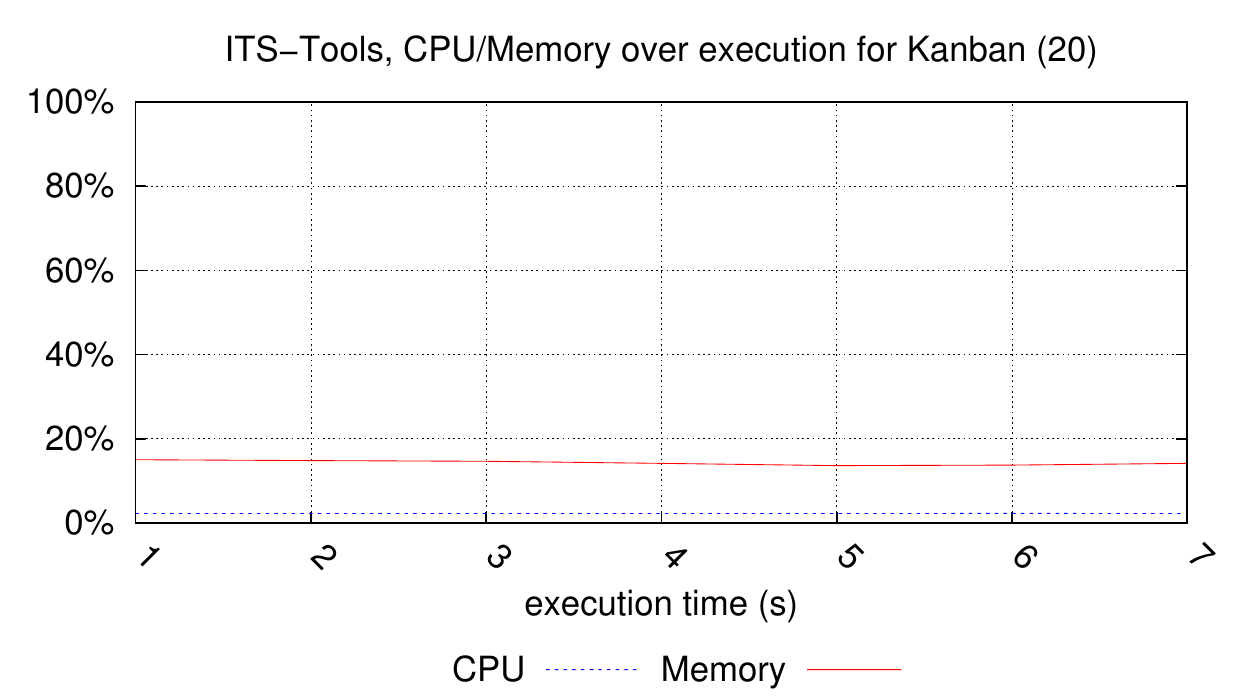}
\includegraphics[width=.5\textwidth]{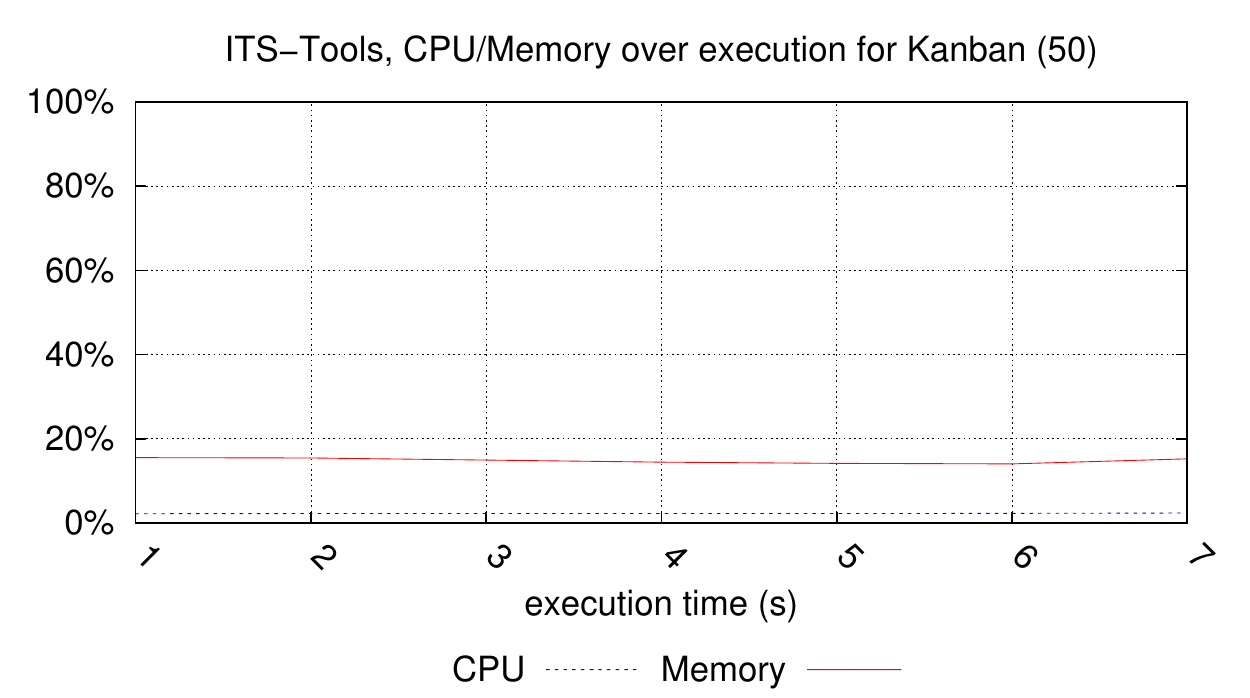}

\noindent\includegraphics[width=.5\textwidth]{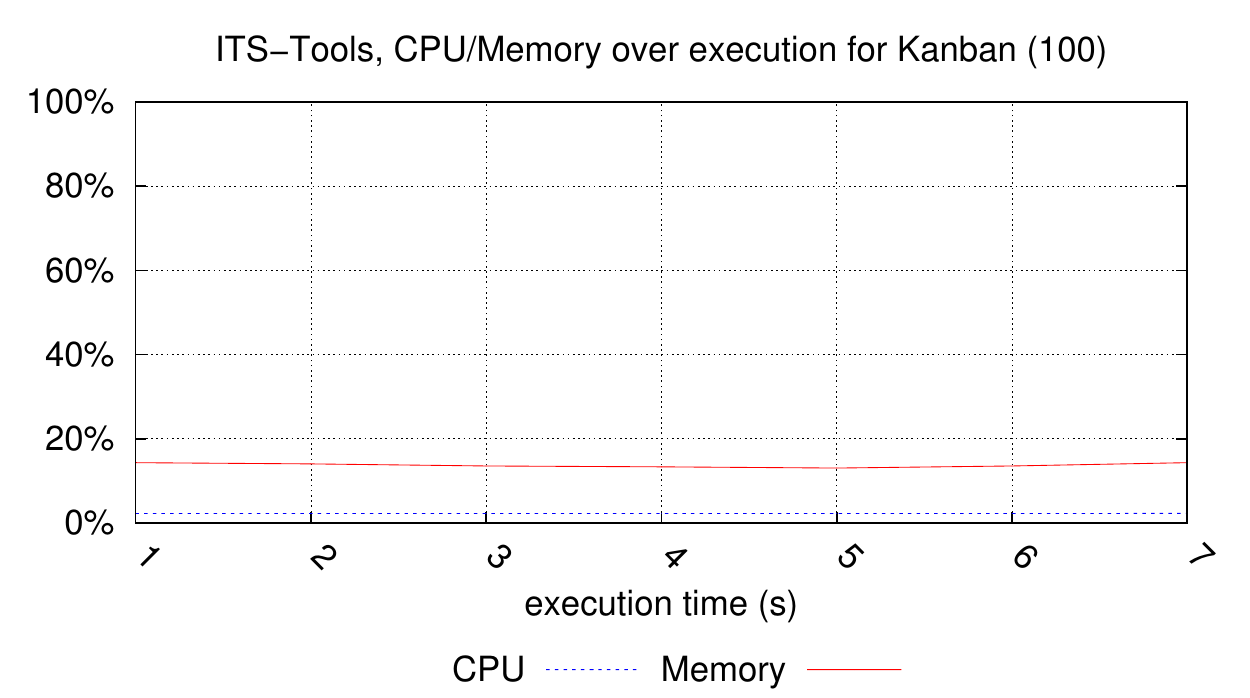}
\includegraphics[width=.5\textwidth]{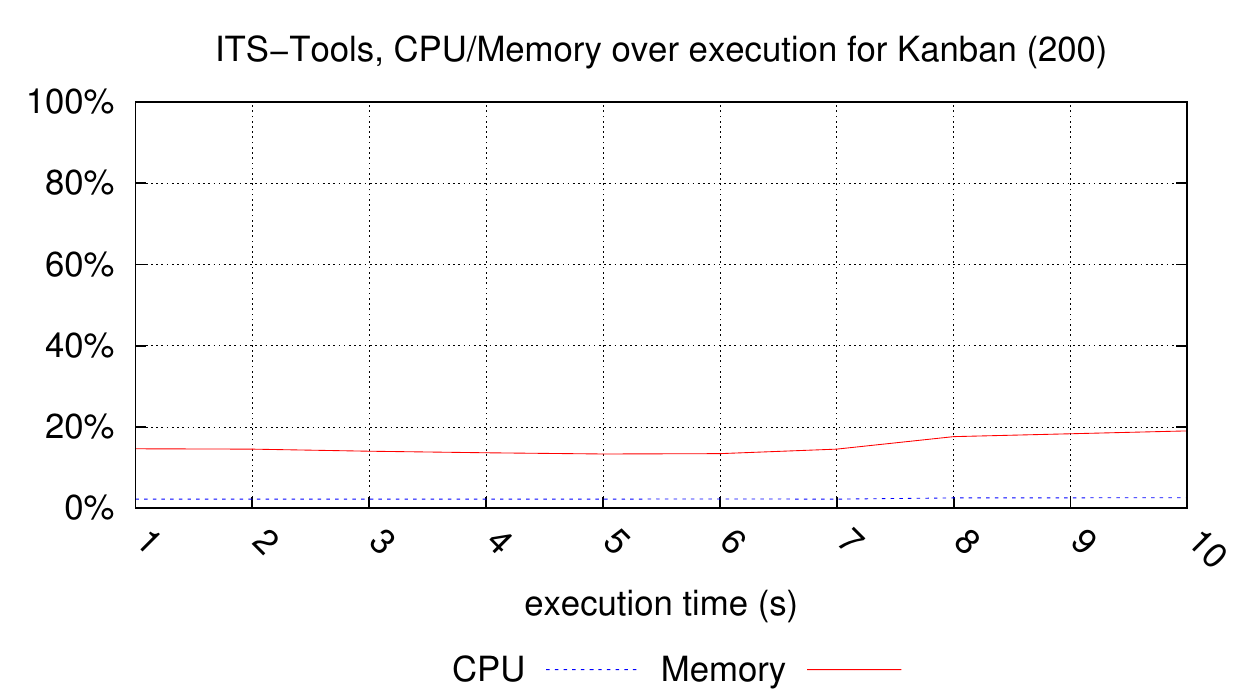}

\noindent\includegraphics[width=.5\textwidth]{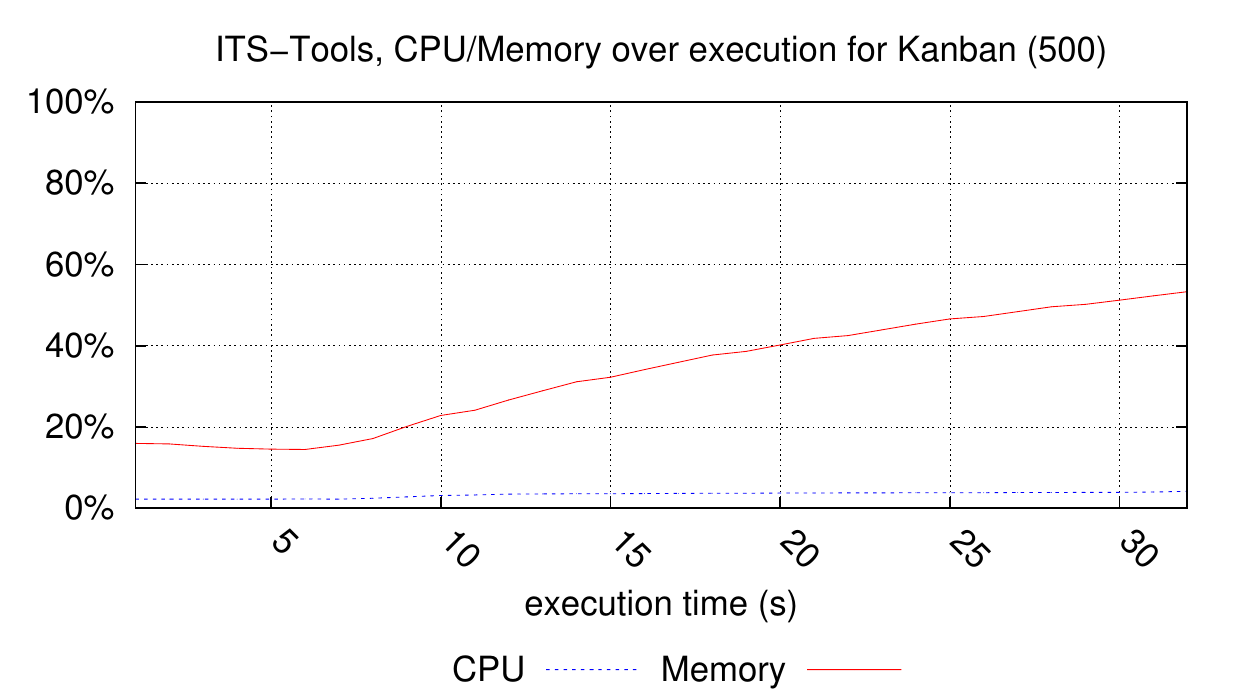}
\includegraphics[width=.5\textwidth]{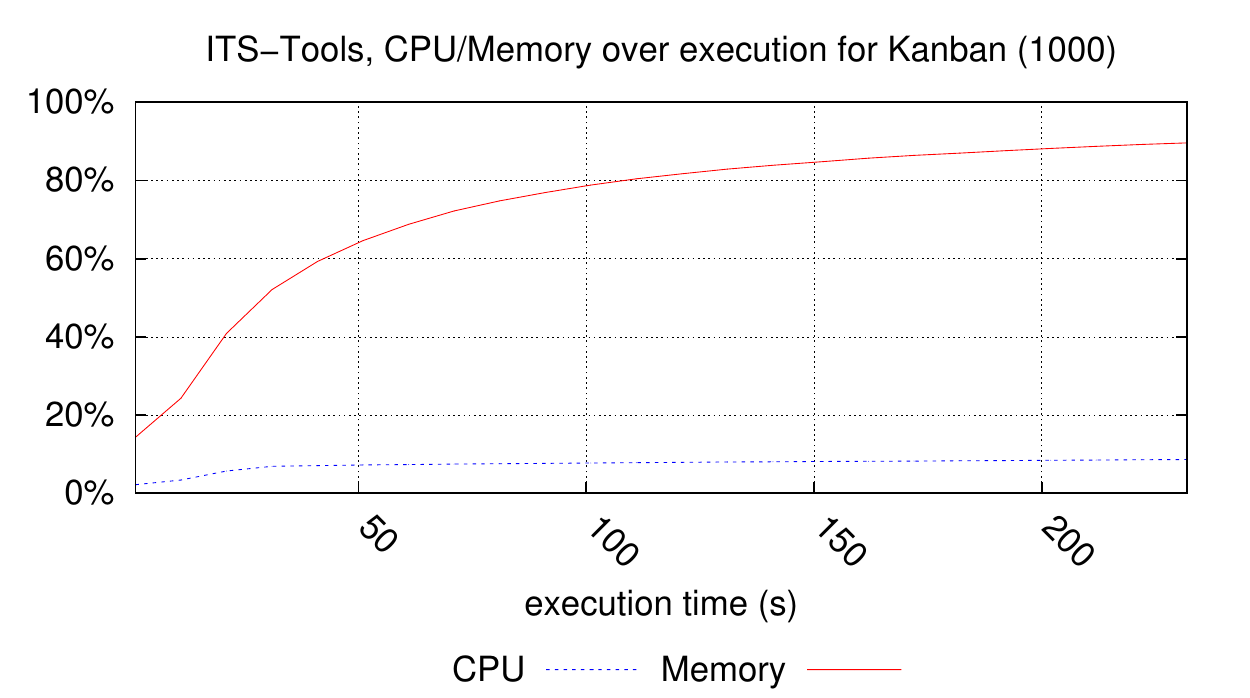}

\subsubsection{Executions for lamport\_fmea}
5 charts have been generated.
\index{Execution (by tool)!ITS-Tools}
\index{Execution (by model)!lamport\_fmea!ITS-Tools}

\noindent\includegraphics[width=.5\textwidth]{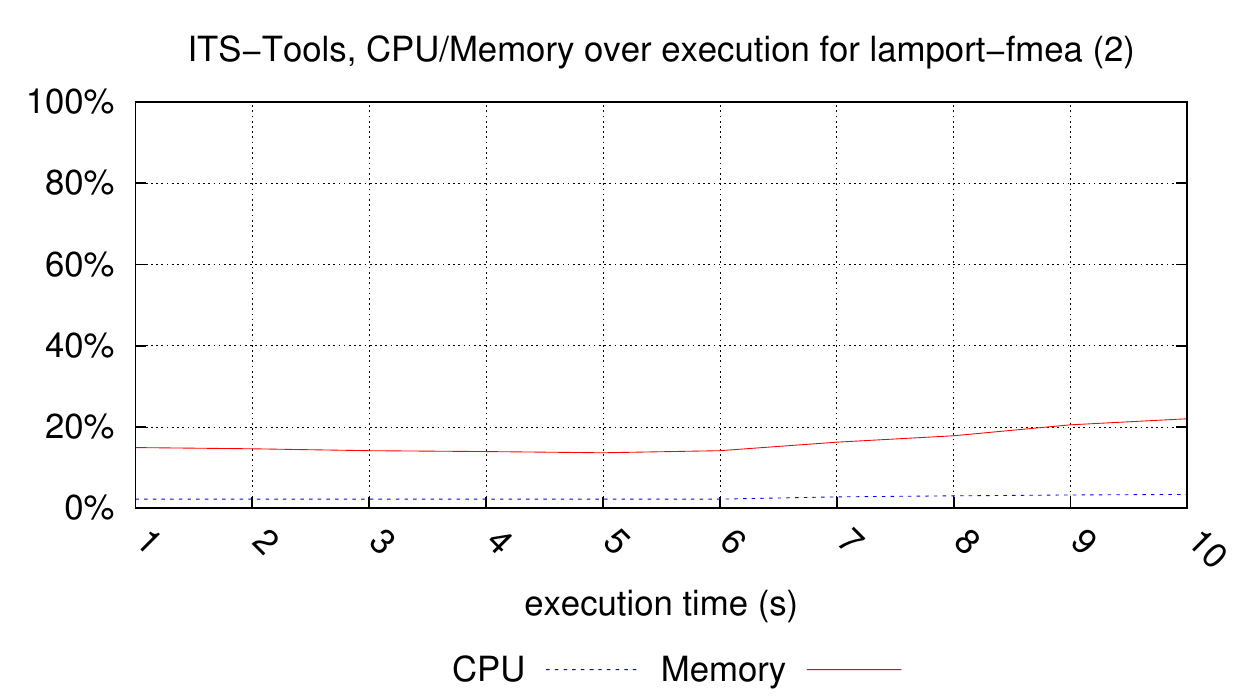}
\includegraphics[width=.5\textwidth]{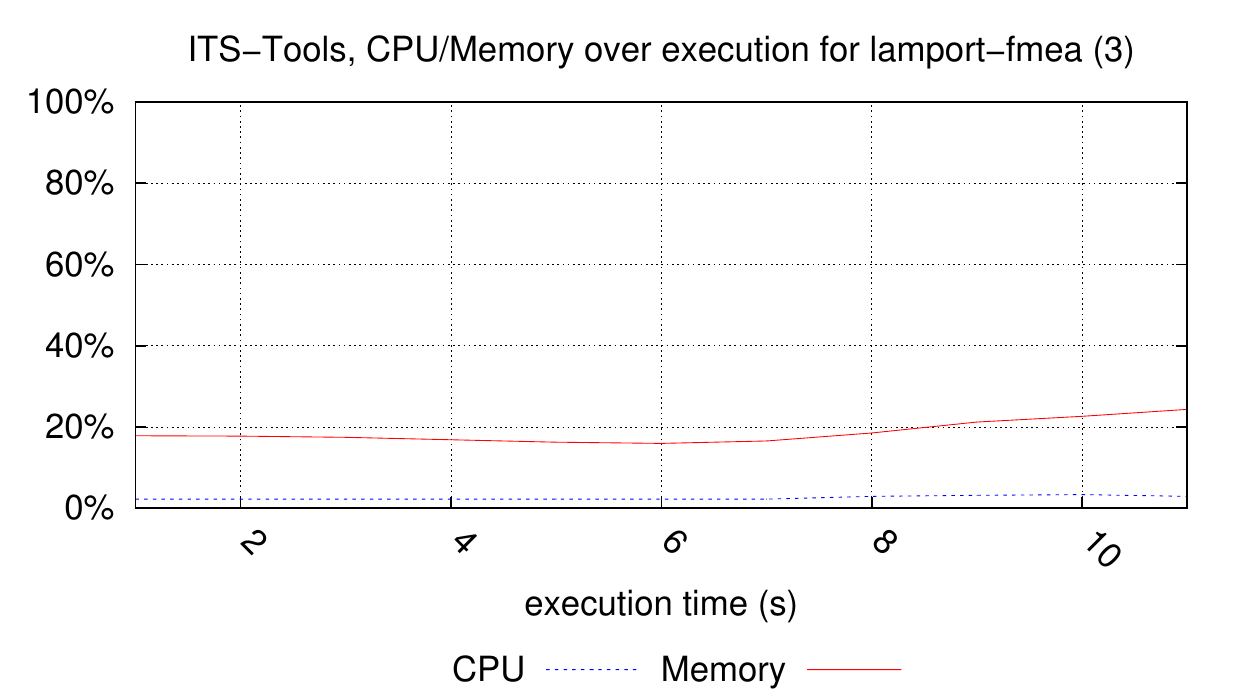}

\noindent\includegraphics[width=.5\textwidth]{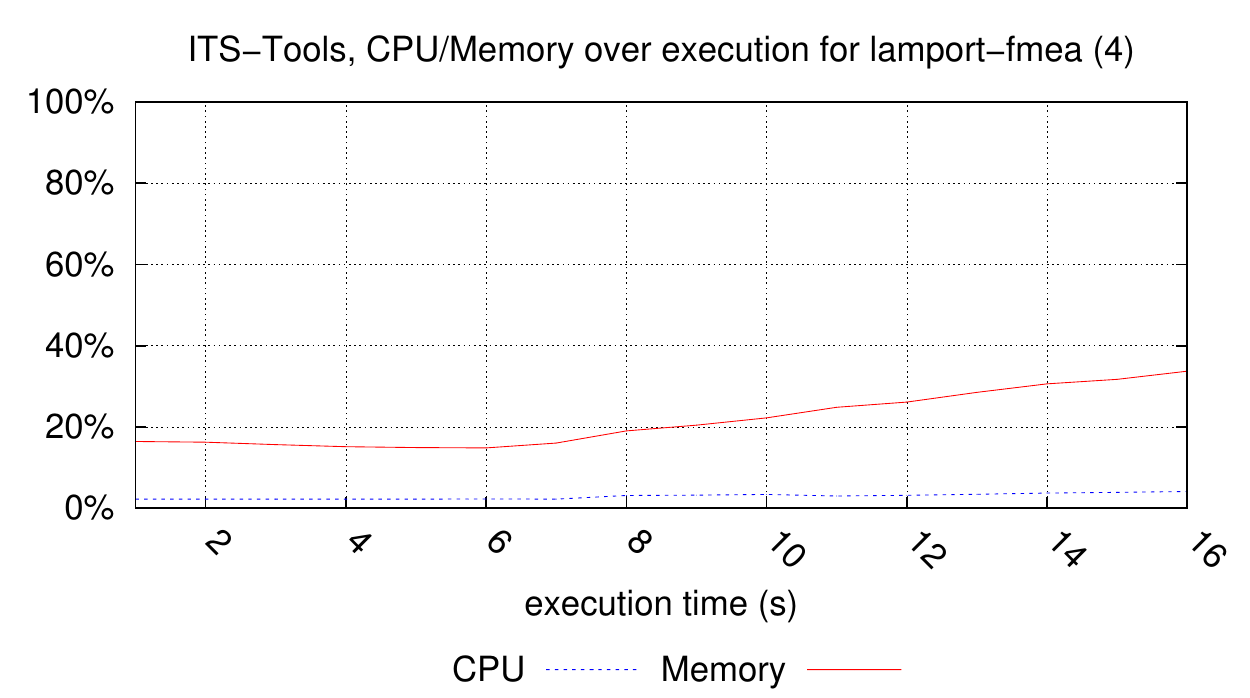}
\includegraphics[width=.5\textwidth]{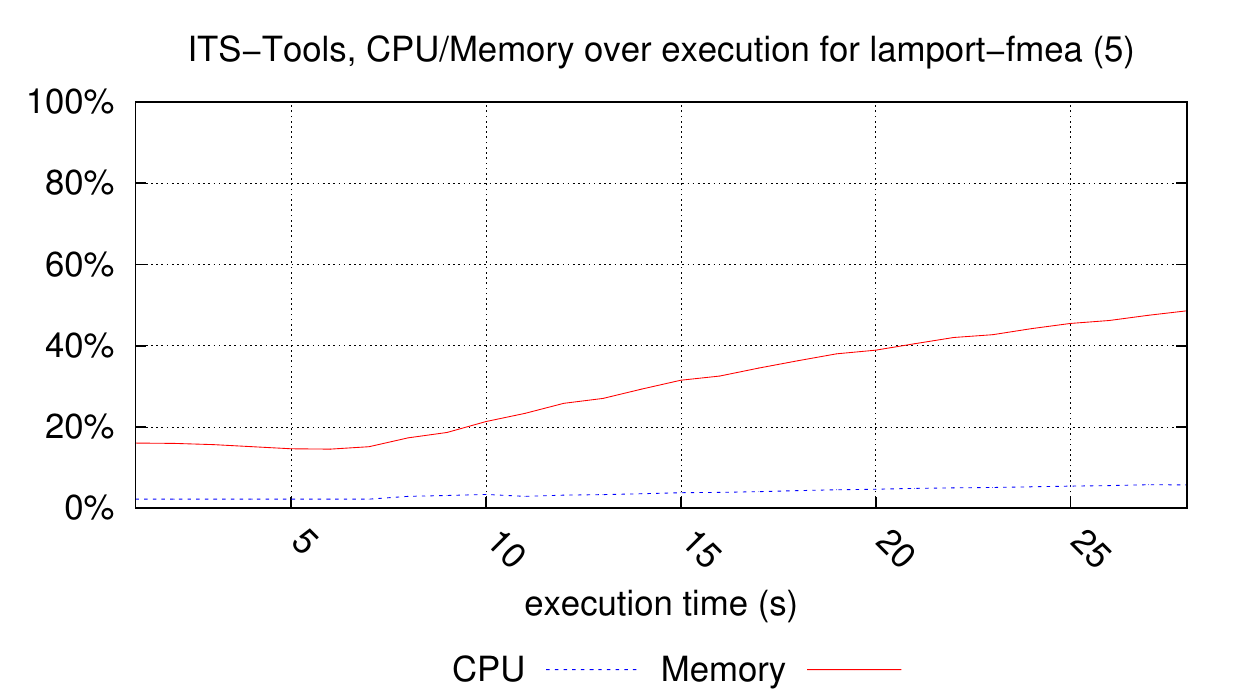}

\noindent\includegraphics[width=.5\textwidth]{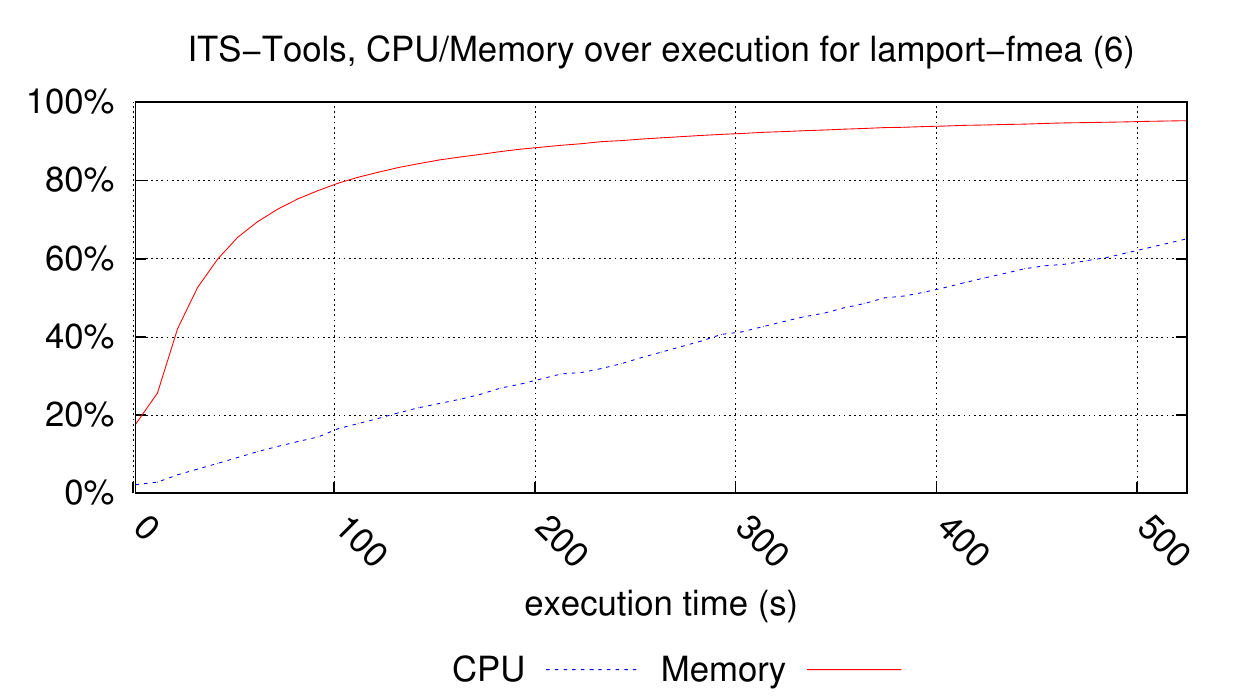}

\subsubsection{Executions for MAPK}
5 charts have been generated.
\index{Execution (by tool)!ITS-Tools}
\index{Execution (by model)!MAPK!ITS-Tools}

\noindent\includegraphics[width=.5\textwidth]{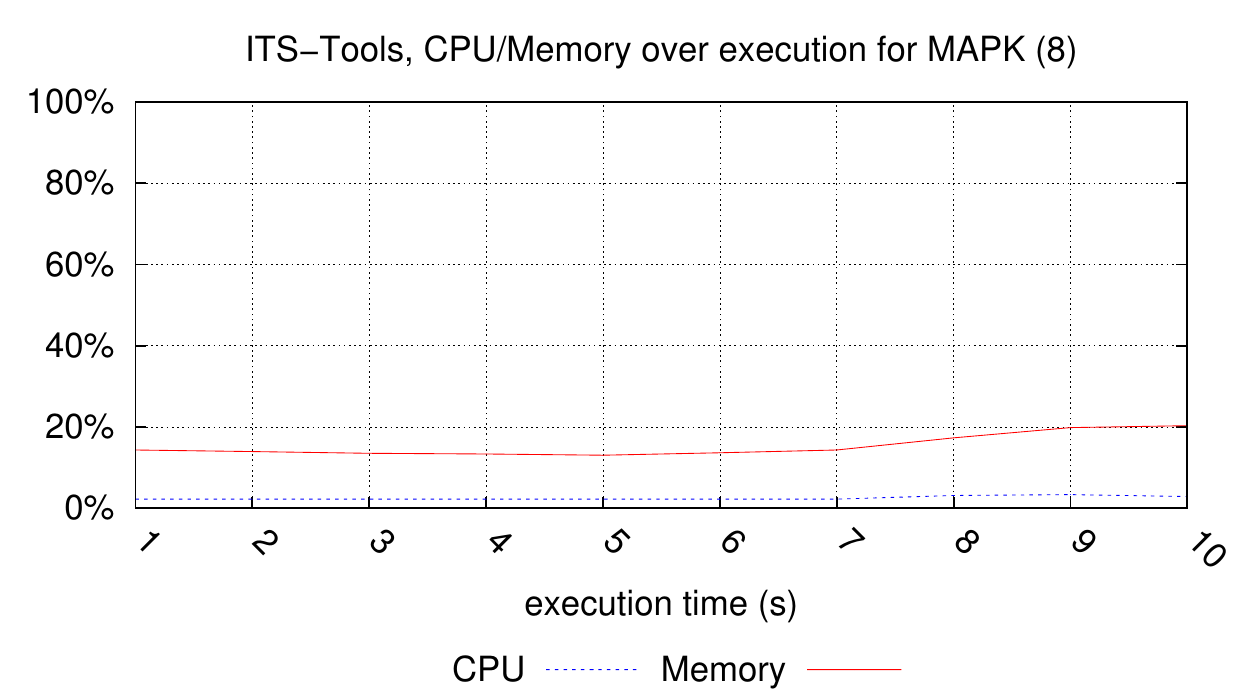}
\includegraphics[width=.5\textwidth]{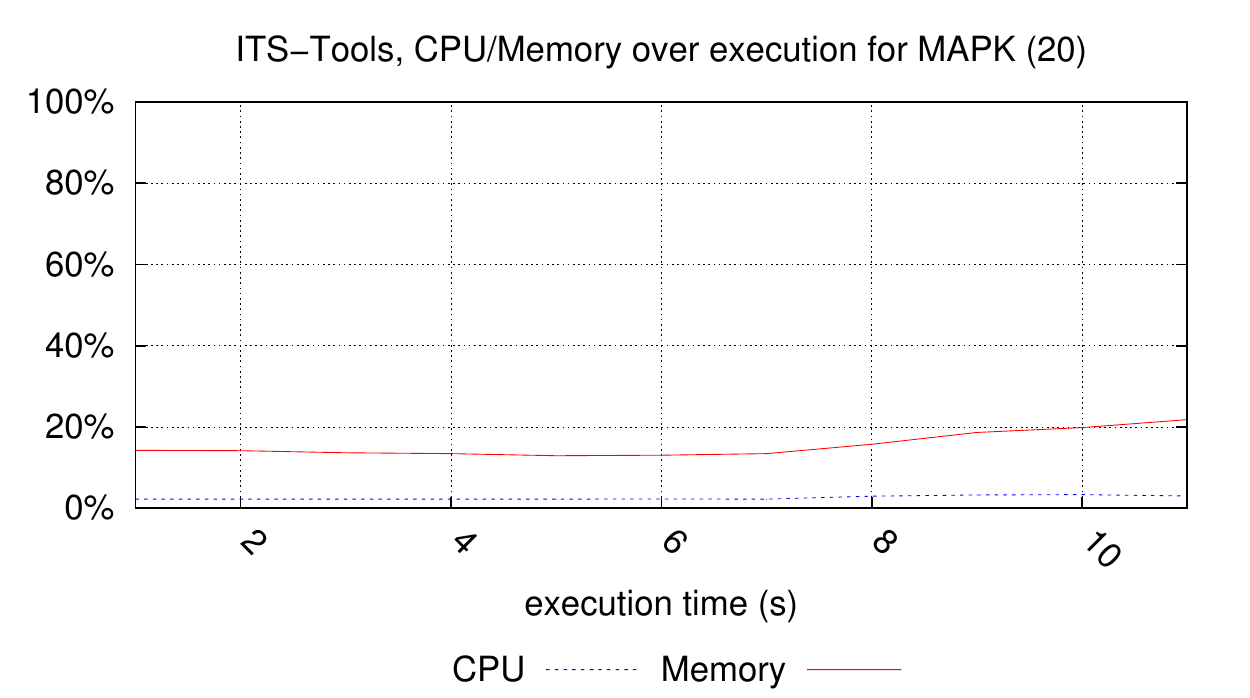}

\noindent\includegraphics[width=.5\textwidth]{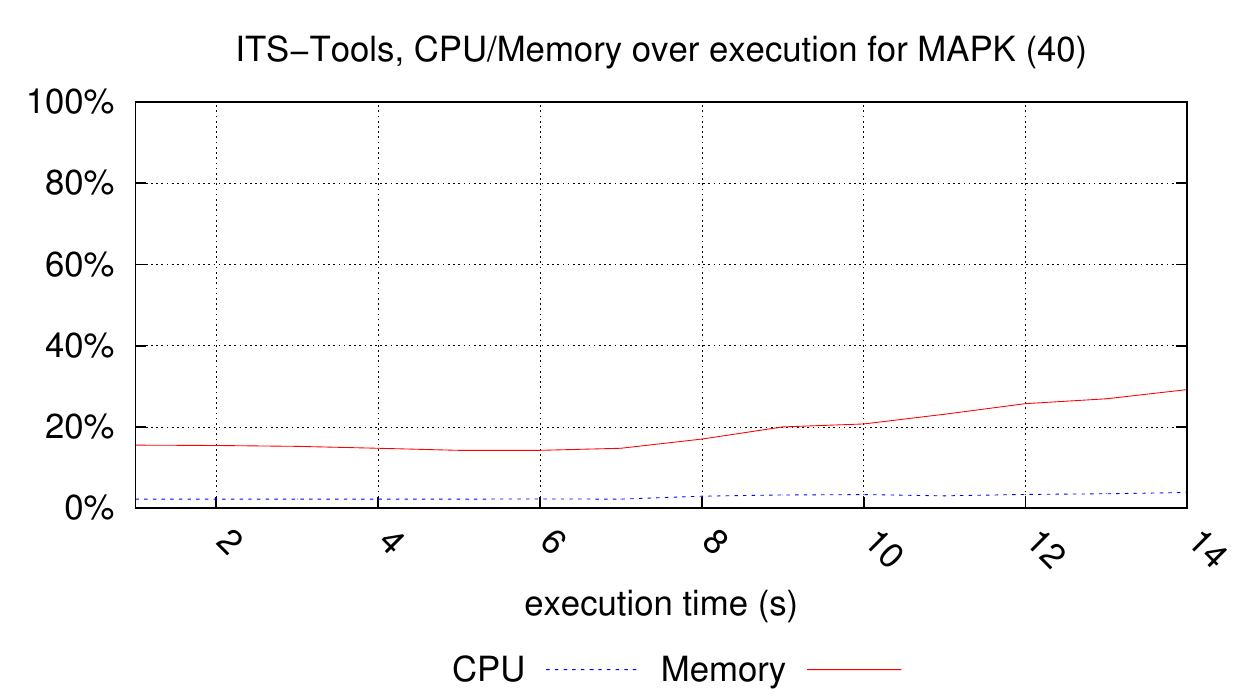}
\includegraphics[width=.5\textwidth]{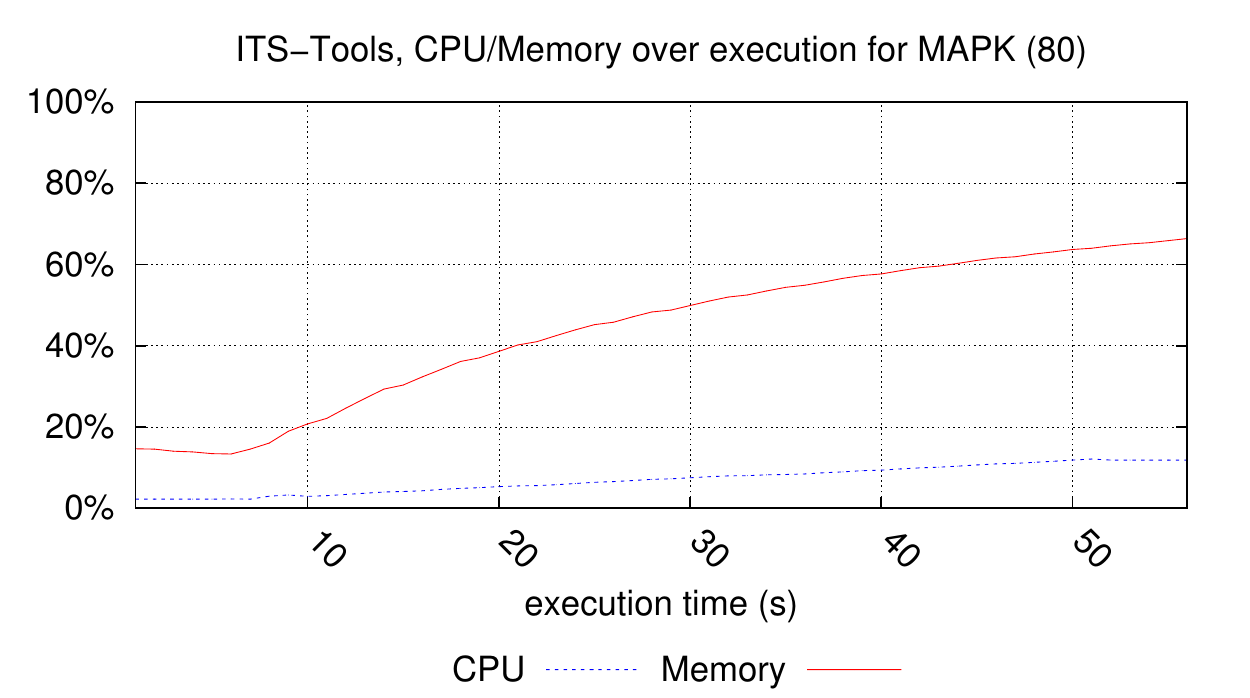}

\noindent\includegraphics[width=.5\textwidth]{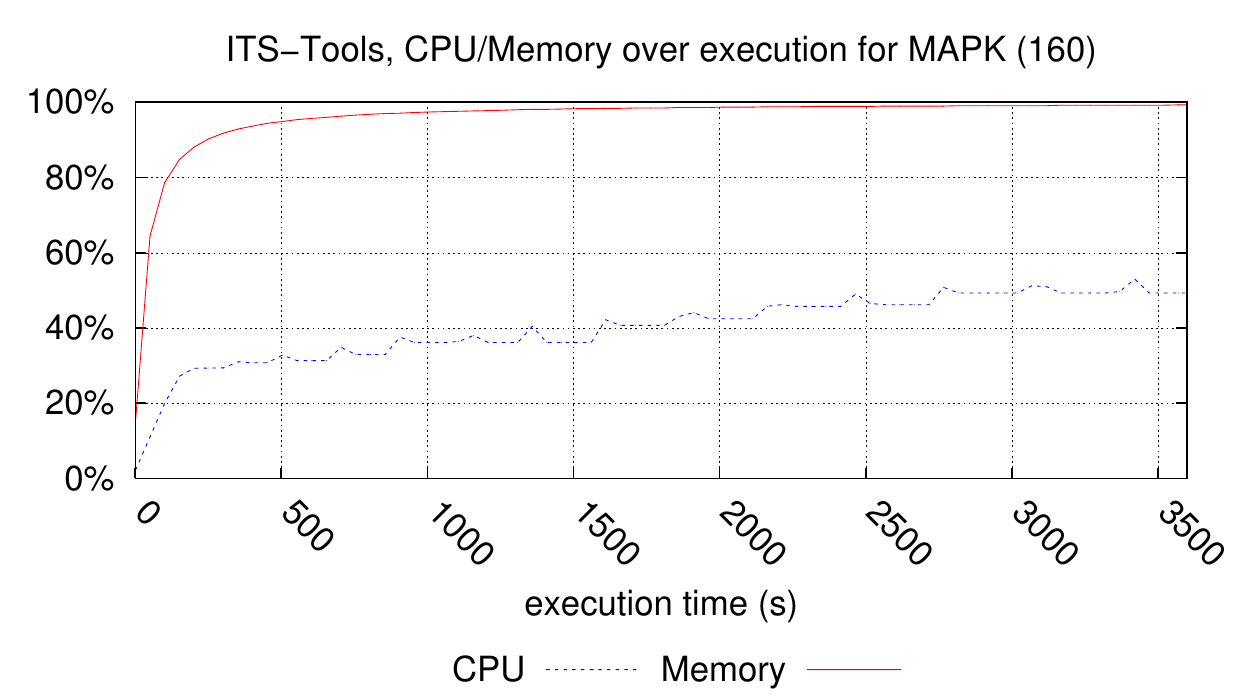}

\vfill\eject
\subsubsection{Executions for neo-election}
1 chart has been generated.
\index{Execution (by tool)!ITS-Tools}
\index{Execution (by model)!neo-election!ITS-Tools}

\noindent\includegraphics[width=.5\textwidth]{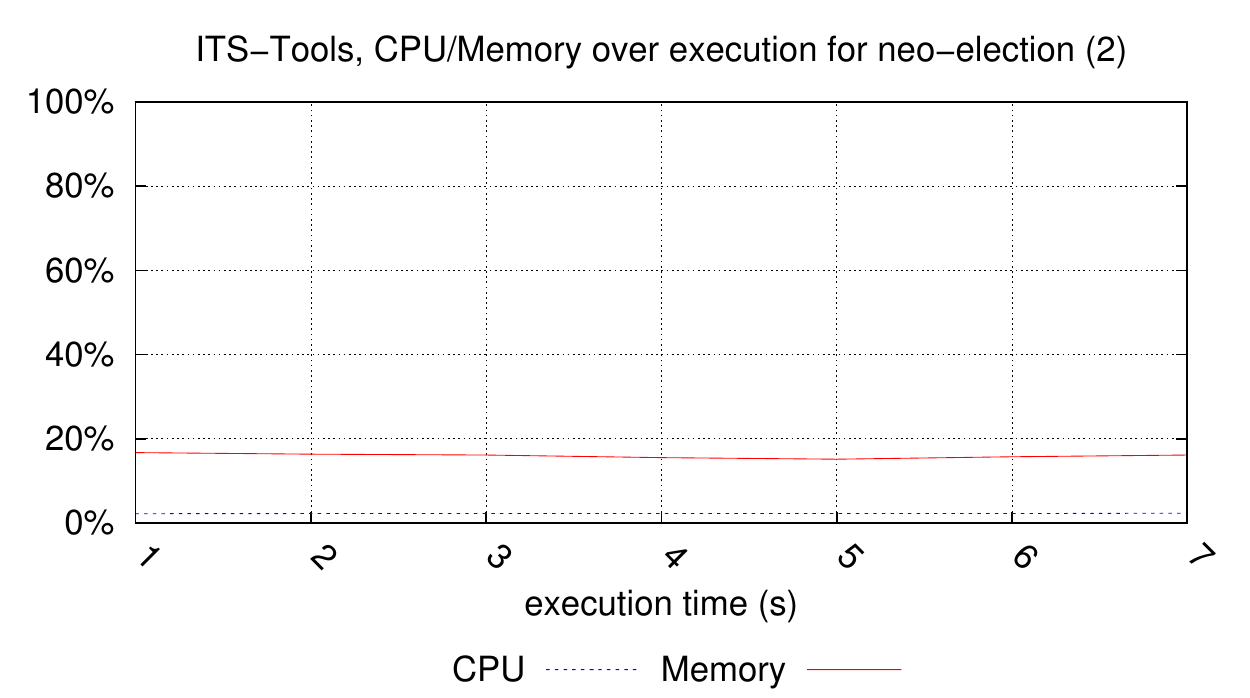}

\subsubsection{Executions for philo\_dyn}
6 charts have been generated.
\index{Execution (by tool)!ITS-Tools}
\index{Execution (by model)!philo\_dyn!ITS-Tools}

\noindent\includegraphics[width=.5\textwidth]{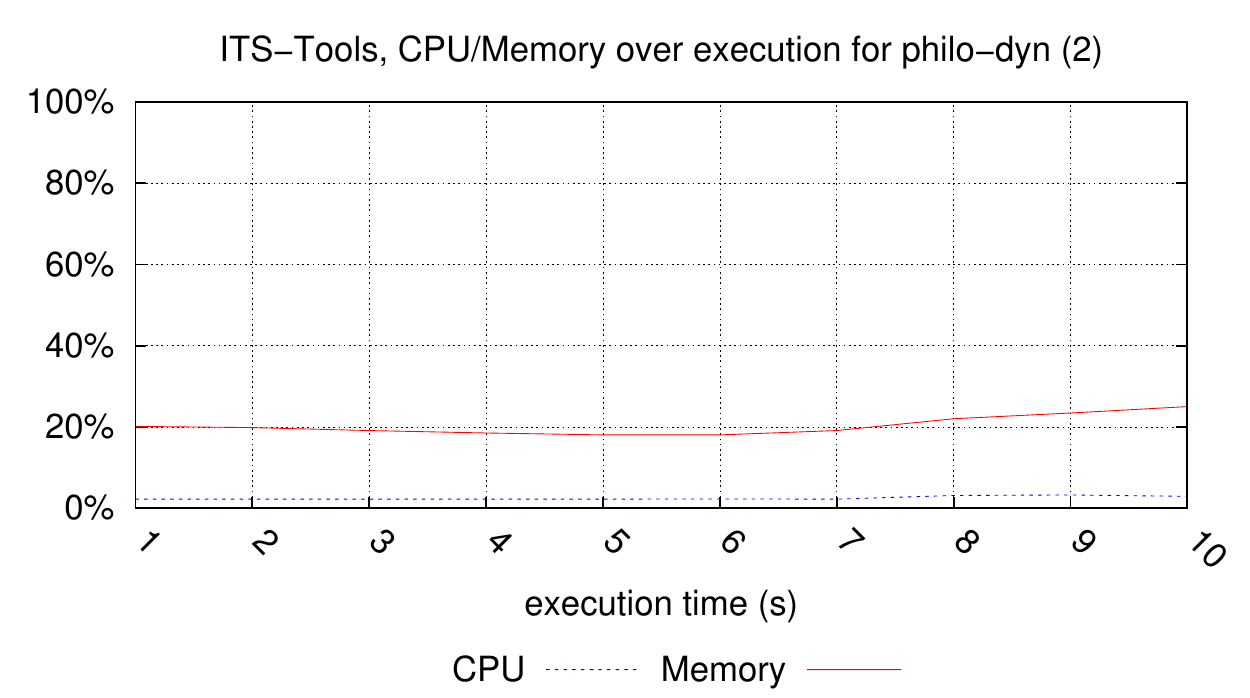}
\includegraphics[width=.5\textwidth]{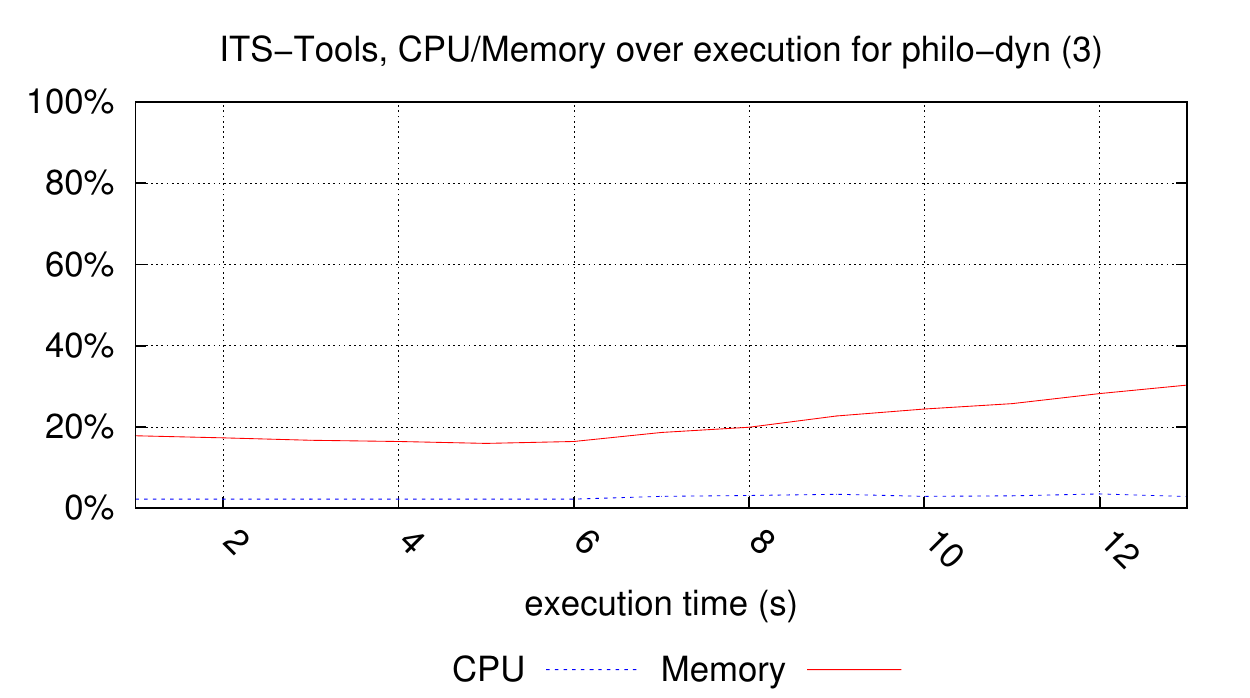}

\noindent\includegraphics[width=.5\textwidth]{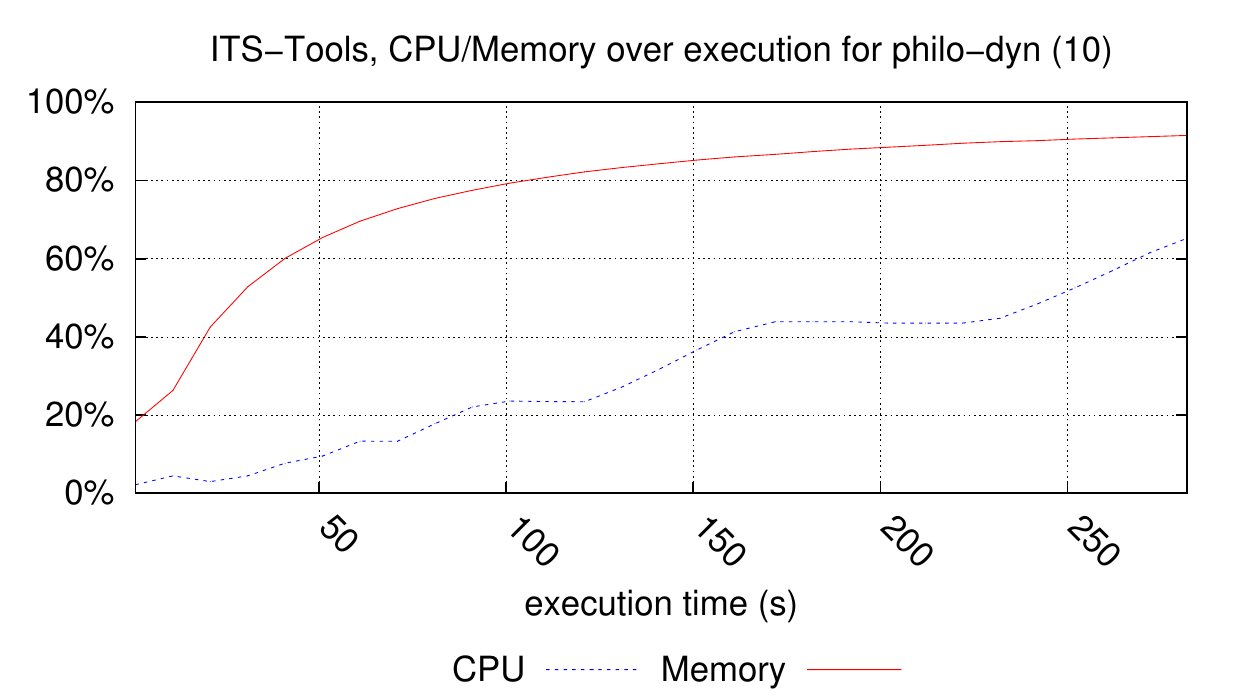}

\subsubsection{Executions for Philosophers}
13 charts have been generated.
\index{Execution (by tool)!ITS-Tools}
\index{Execution (by model)!Philosophers!ITS-Tools}

\noindent\includegraphics[width=.5\textwidth]{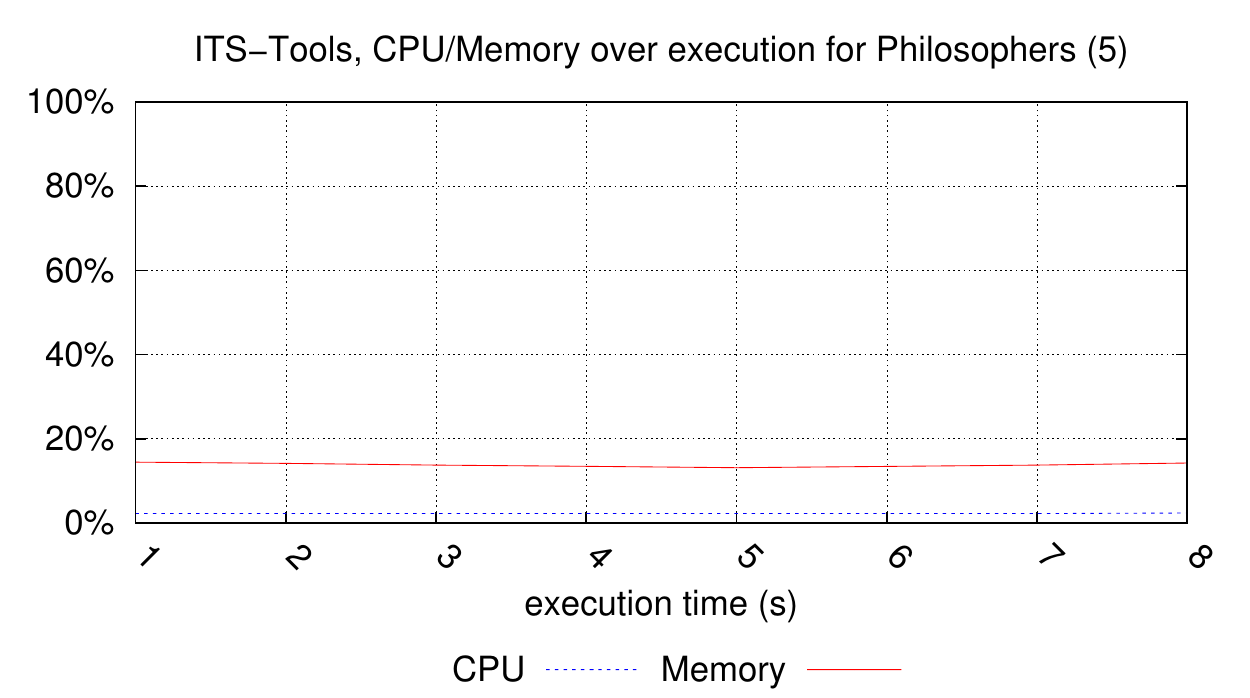}
\includegraphics[width=.5\textwidth]{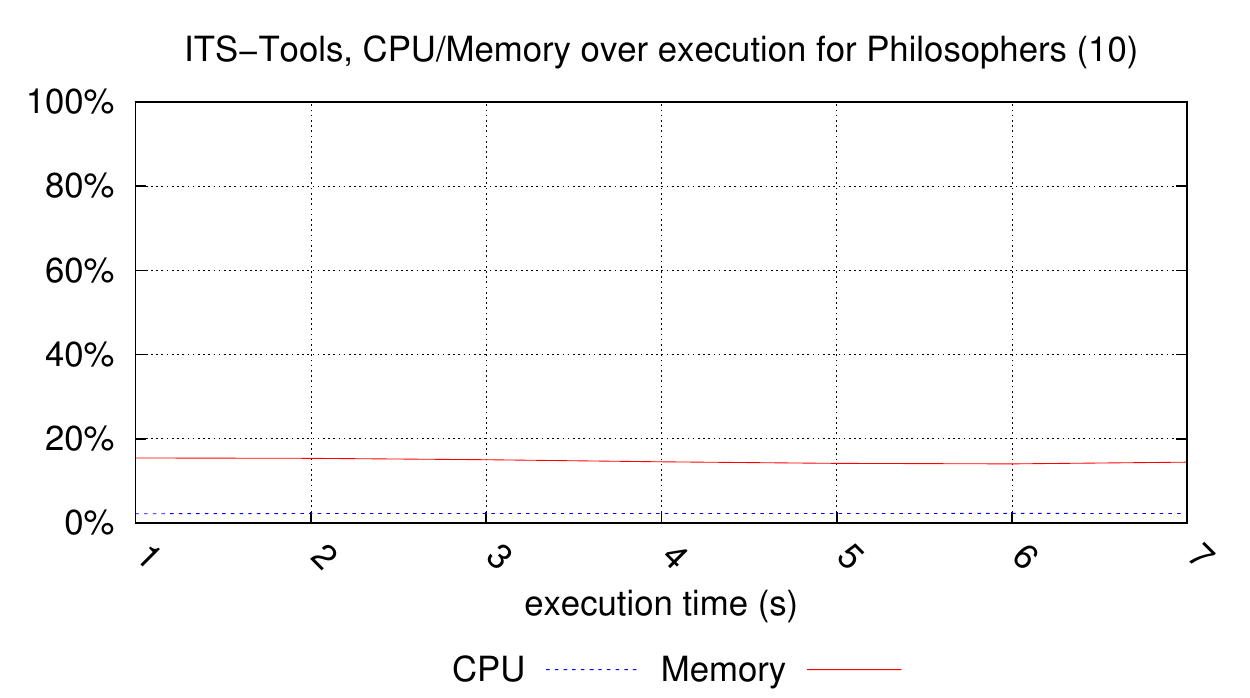}

\noindent\includegraphics[width=.5\textwidth]{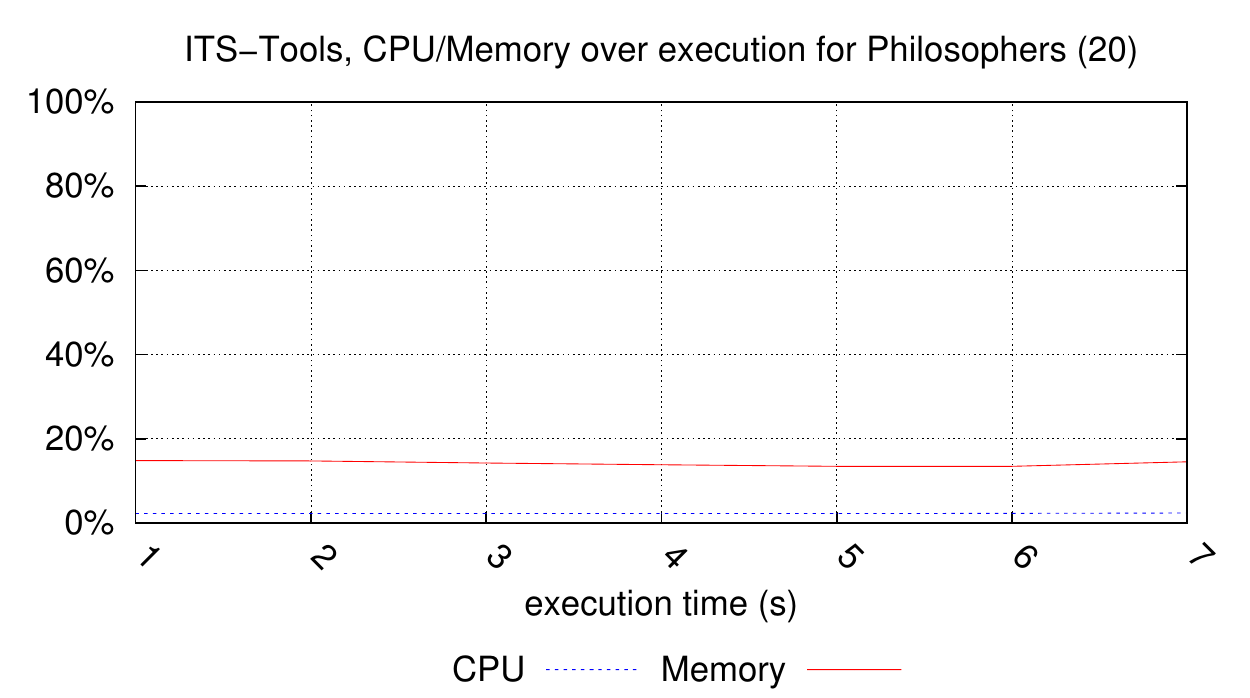}
\includegraphics[width=.5\textwidth]{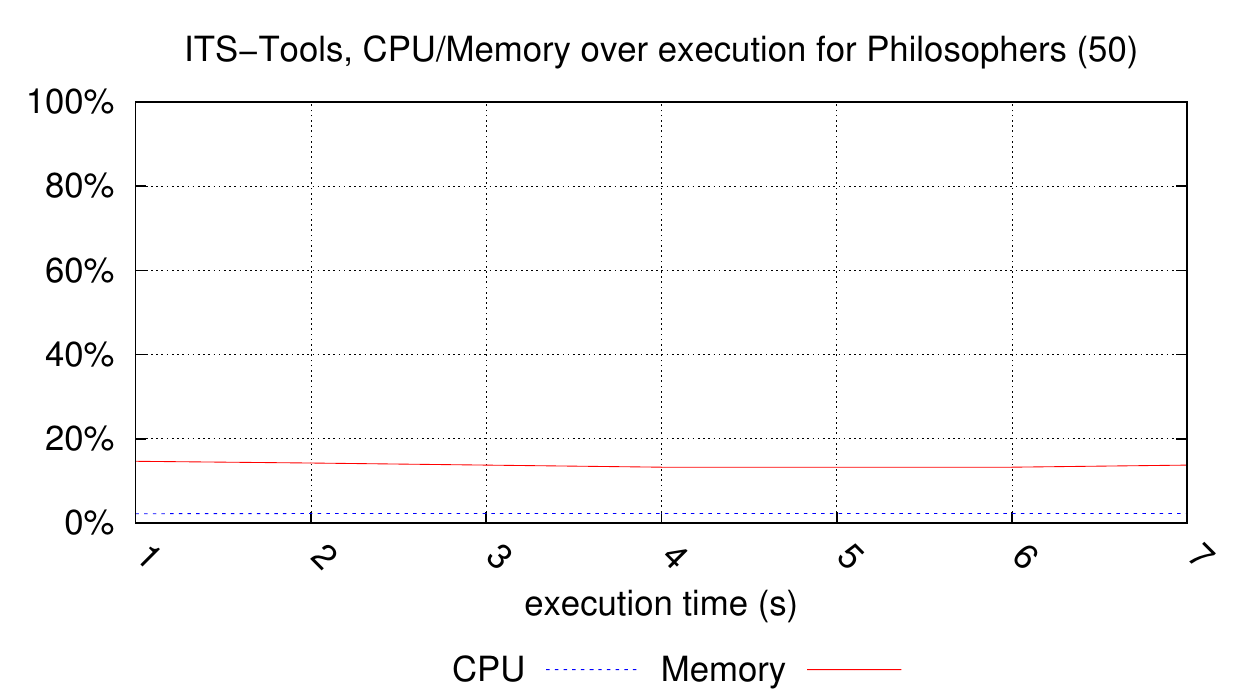}

\noindent\includegraphics[width=.5\textwidth]{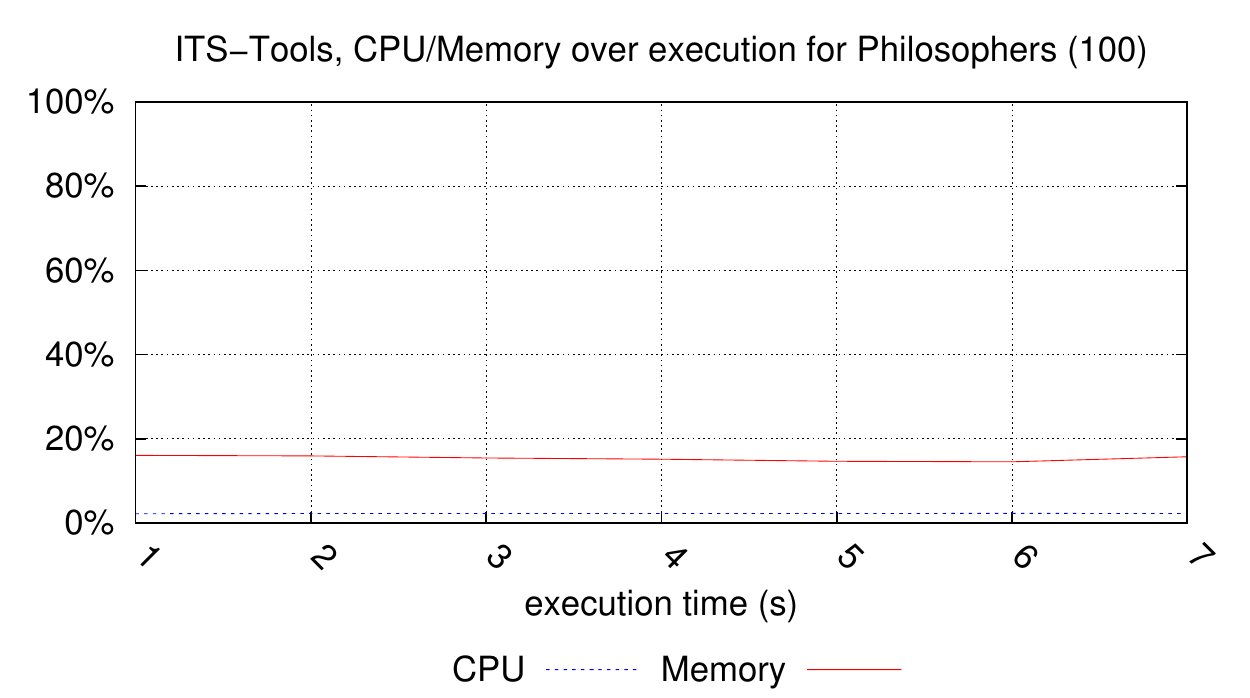}
\includegraphics[width=.5\textwidth]{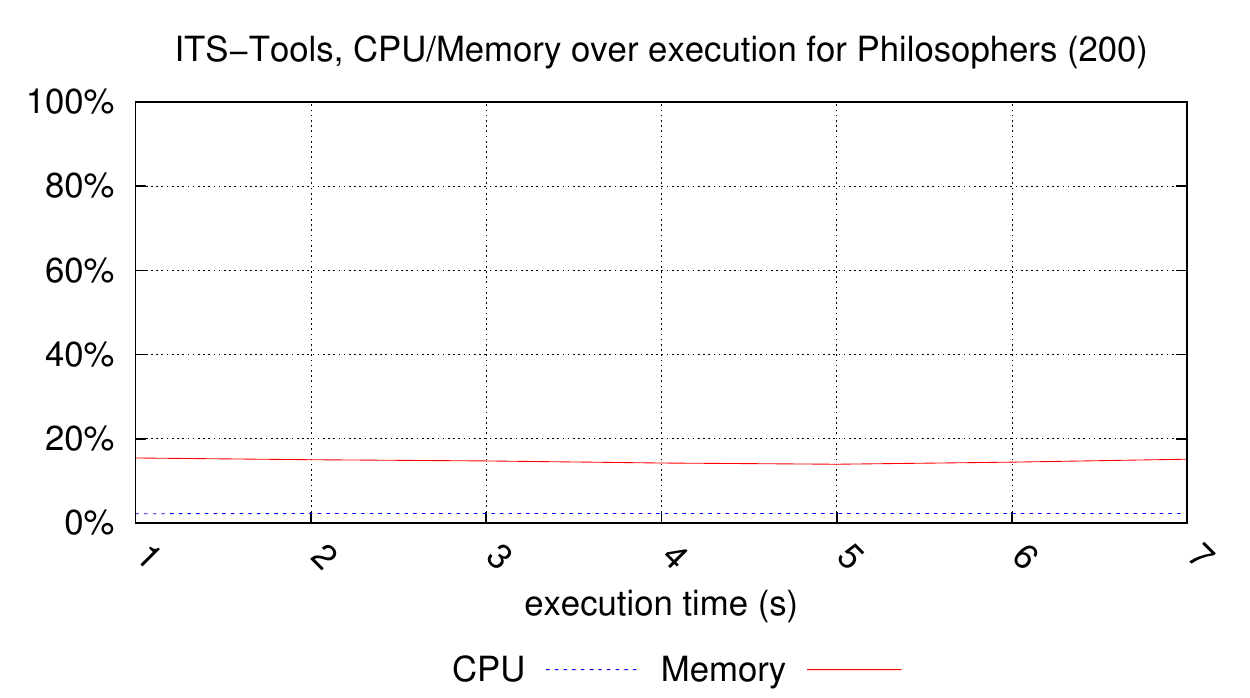}

\noindent\includegraphics[width=.5\textwidth]{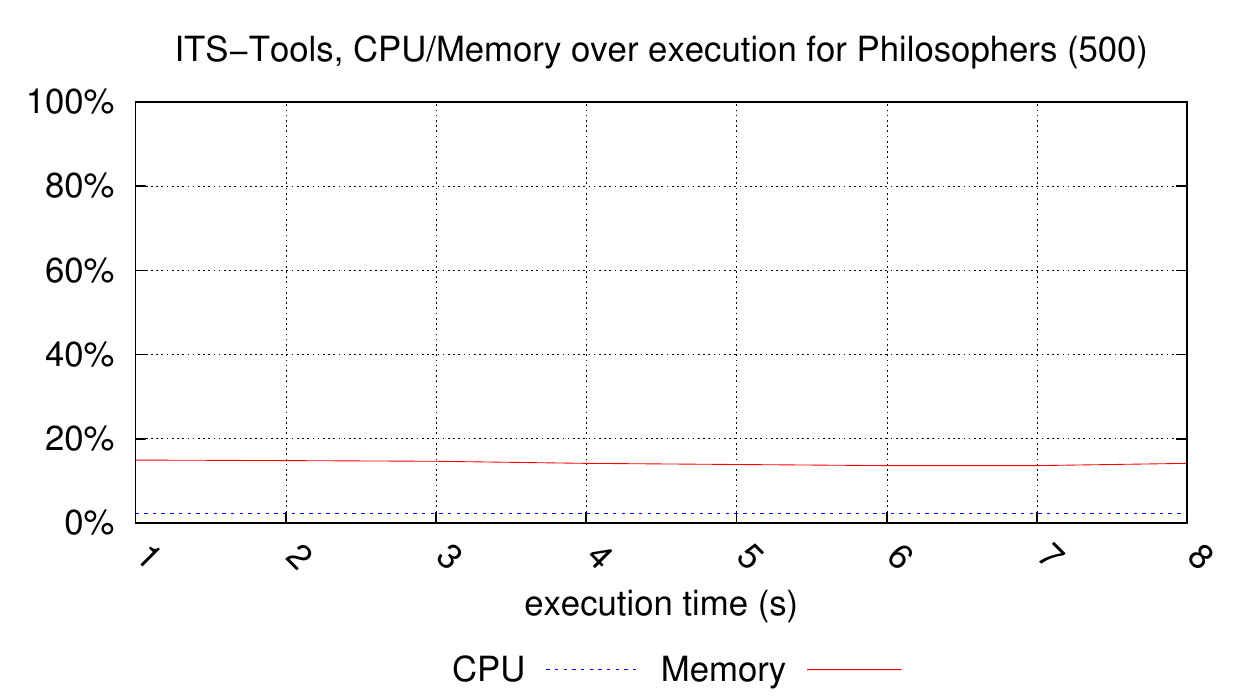}
\includegraphics[width=.5\textwidth]{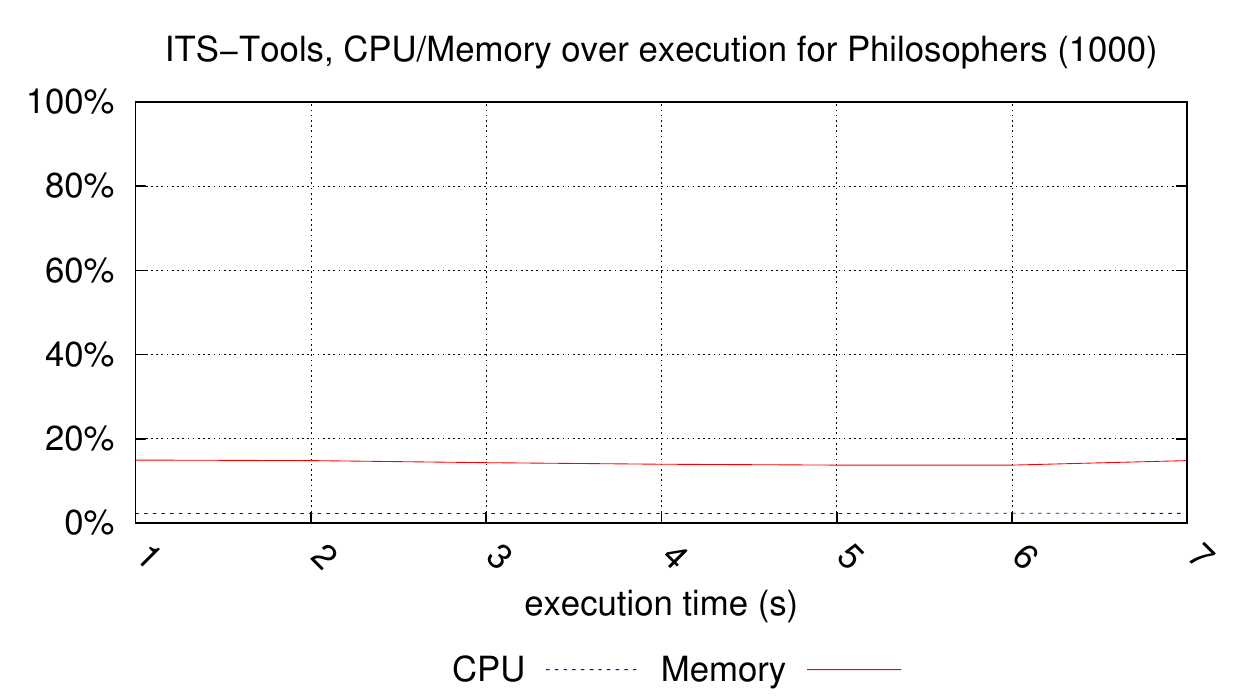}

\index{Execution (by tool)!ITS-Tools}
\noindent\includegraphics[width=.5\textwidth]{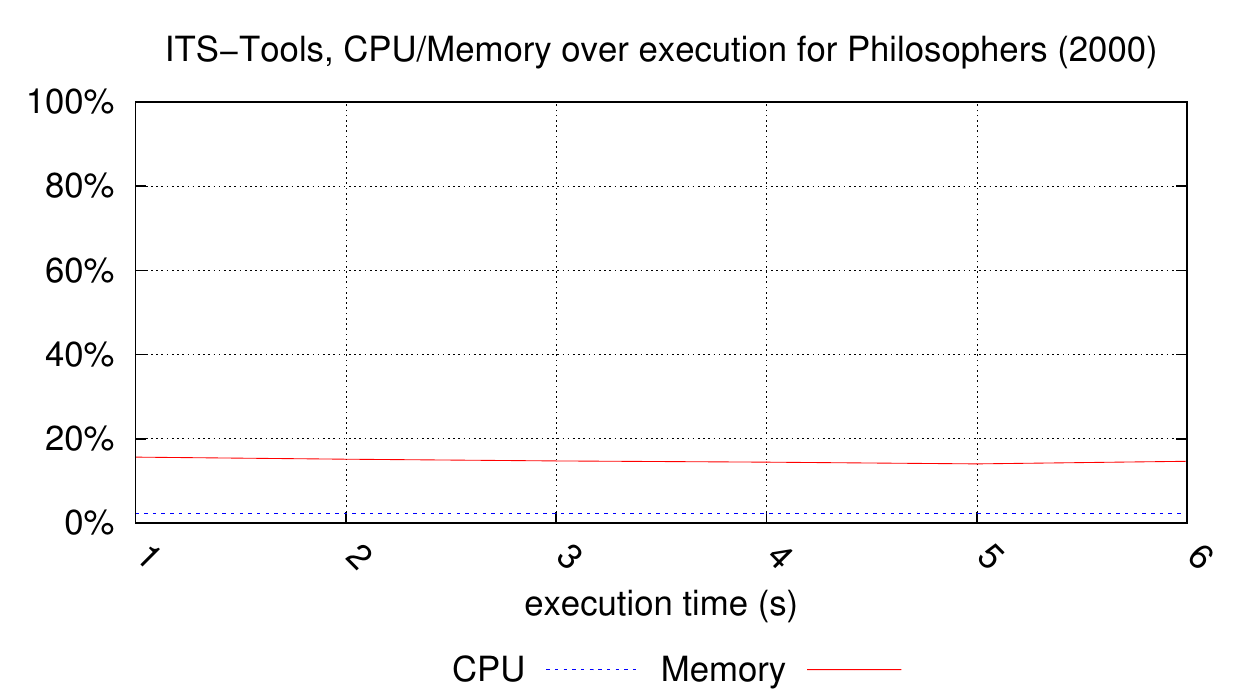}
\includegraphics[width=.5\textwidth]{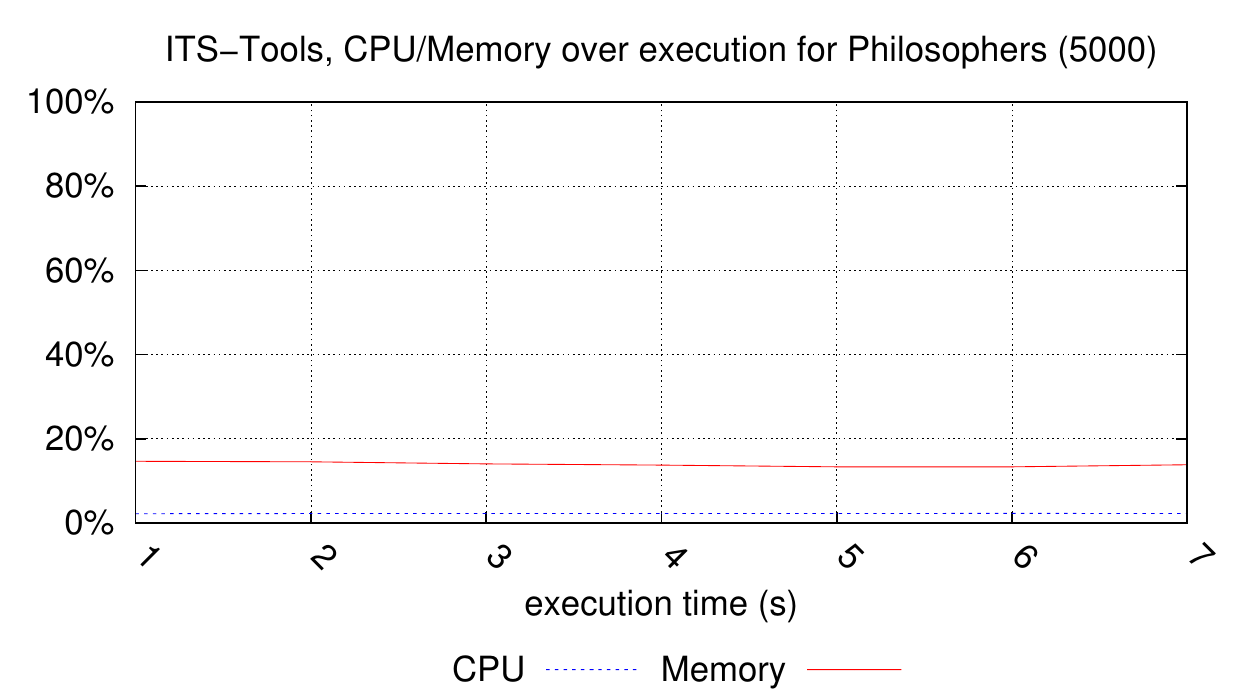}

\noindent\includegraphics[width=.5\textwidth]{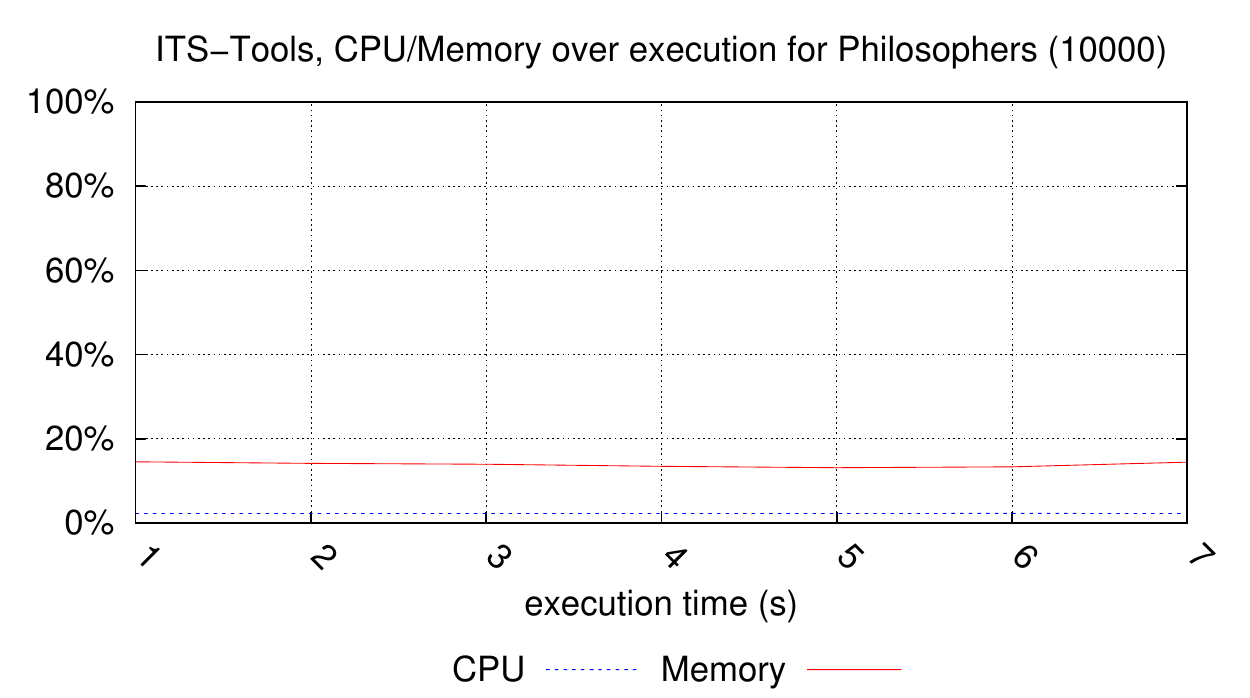}
\includegraphics[width=.5\textwidth]{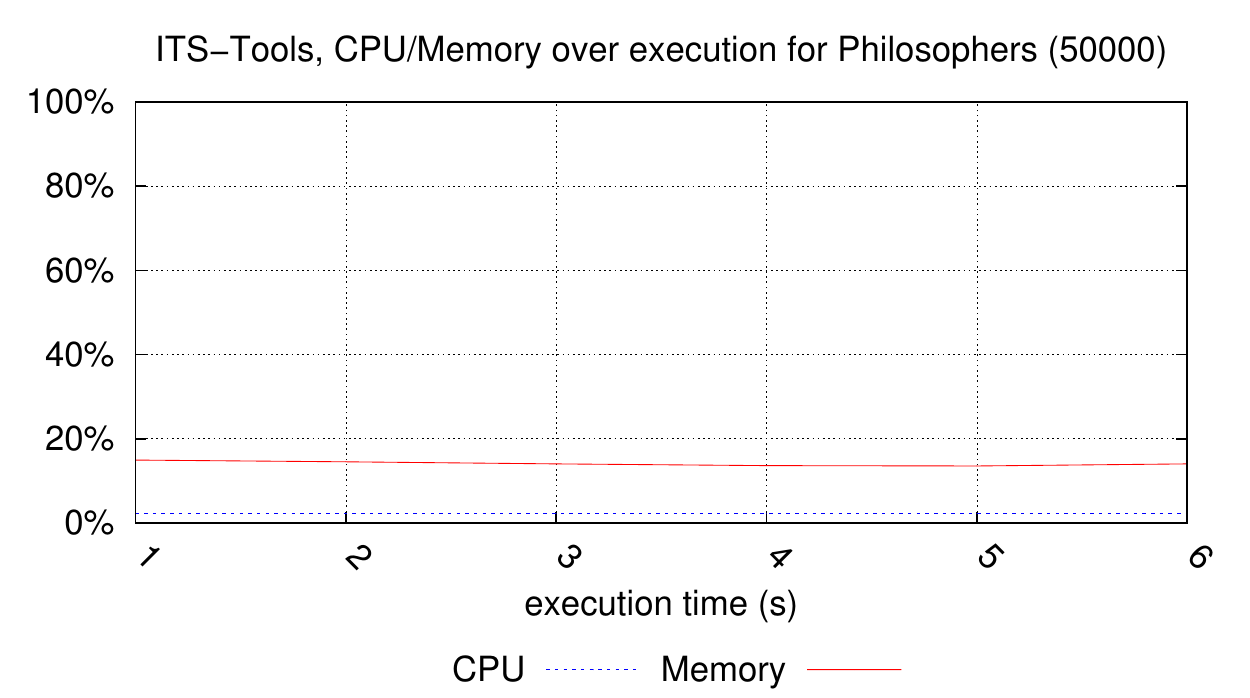}

\noindent\includegraphics[width=.5\textwidth]{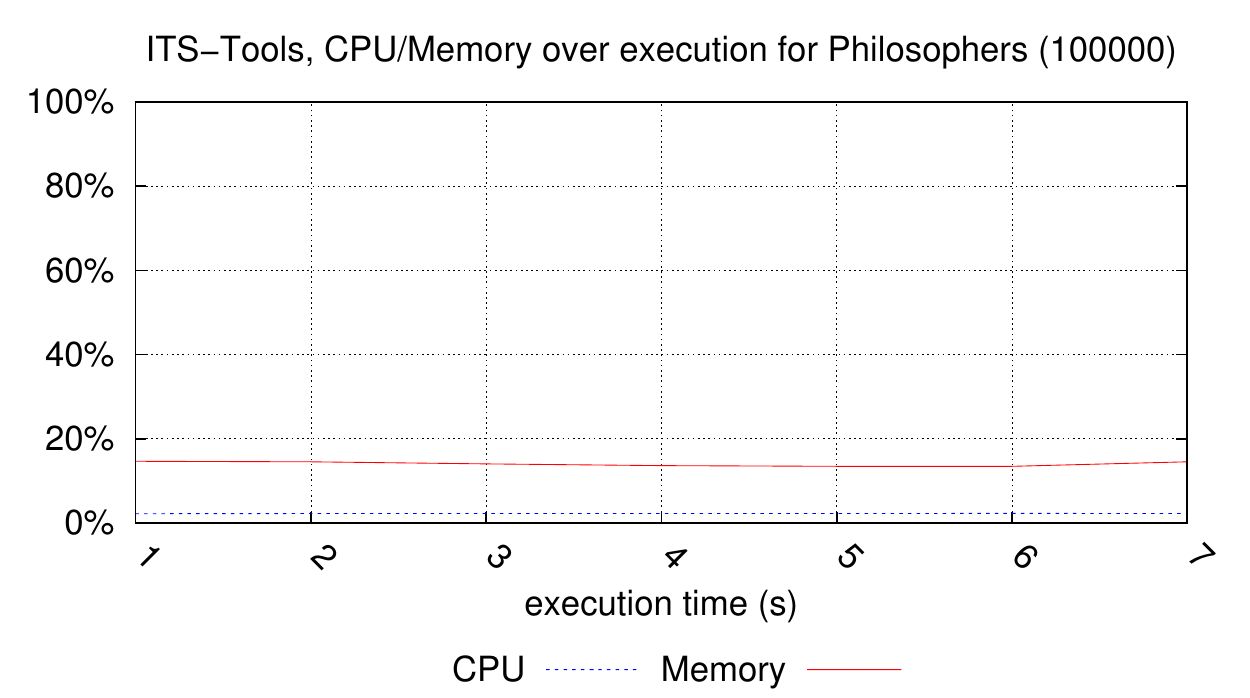}

\subsubsection{Executions for planning}
1 chart has been generated.
\index{Execution (by tool)!ITS-Tools}
\index{Execution (by model)!planning!ITS-Tools}

\noindent\includegraphics[width=.5\textwidth]{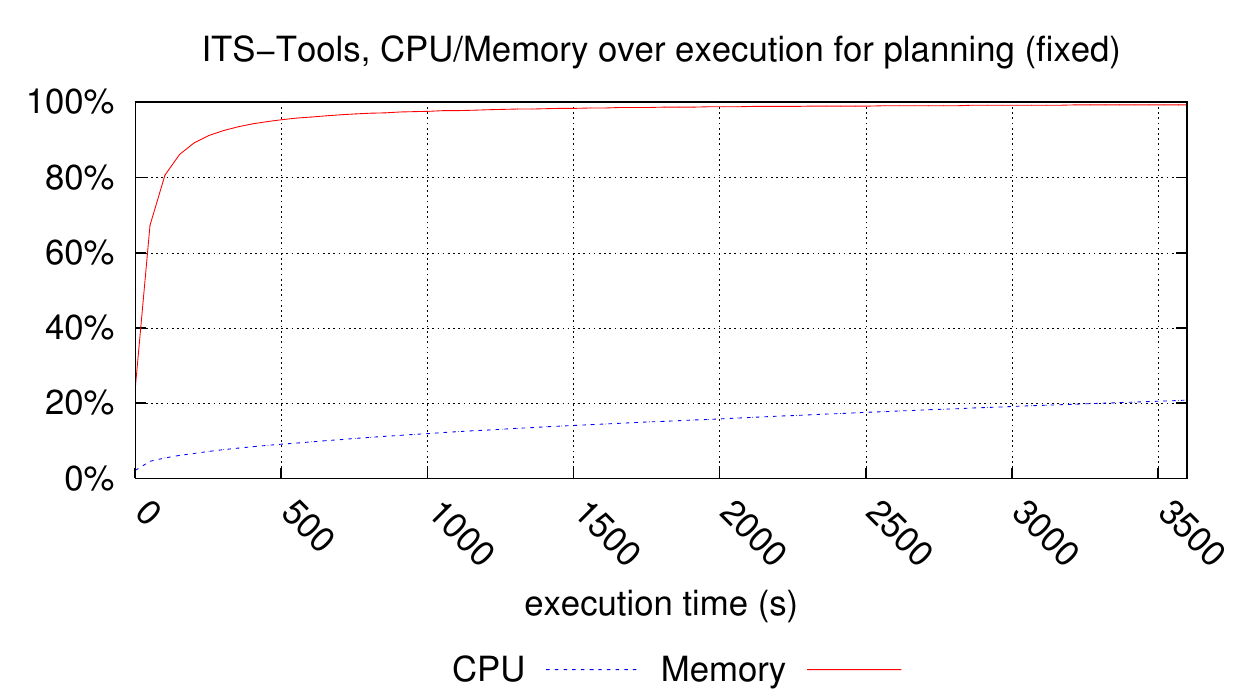}

\subsubsection{Executions for railroad}
3 charts have been generated.
\index{Execution (by tool)!ITS-Tools}
\index{Execution (by model)!railroad!ITS-Tools}

\noindent\includegraphics[width=.5\textwidth]{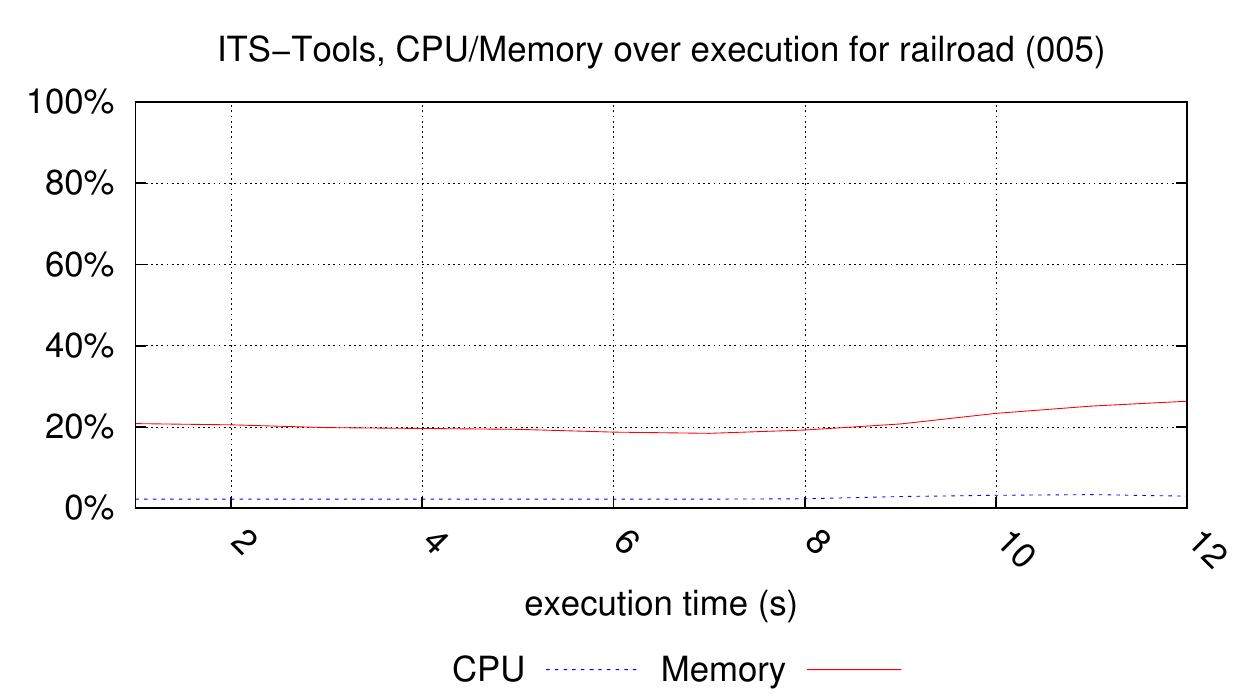}
\includegraphics[width=.5\textwidth]{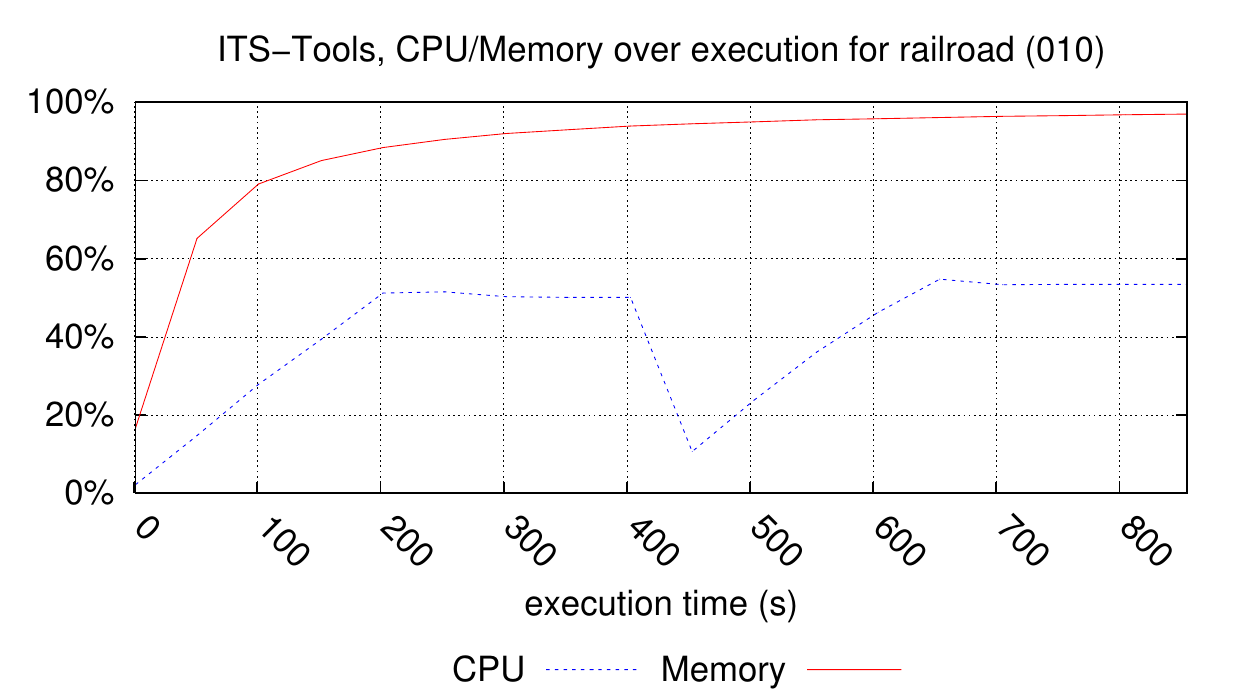}

\noindent\includegraphics[width=.5\textwidth]{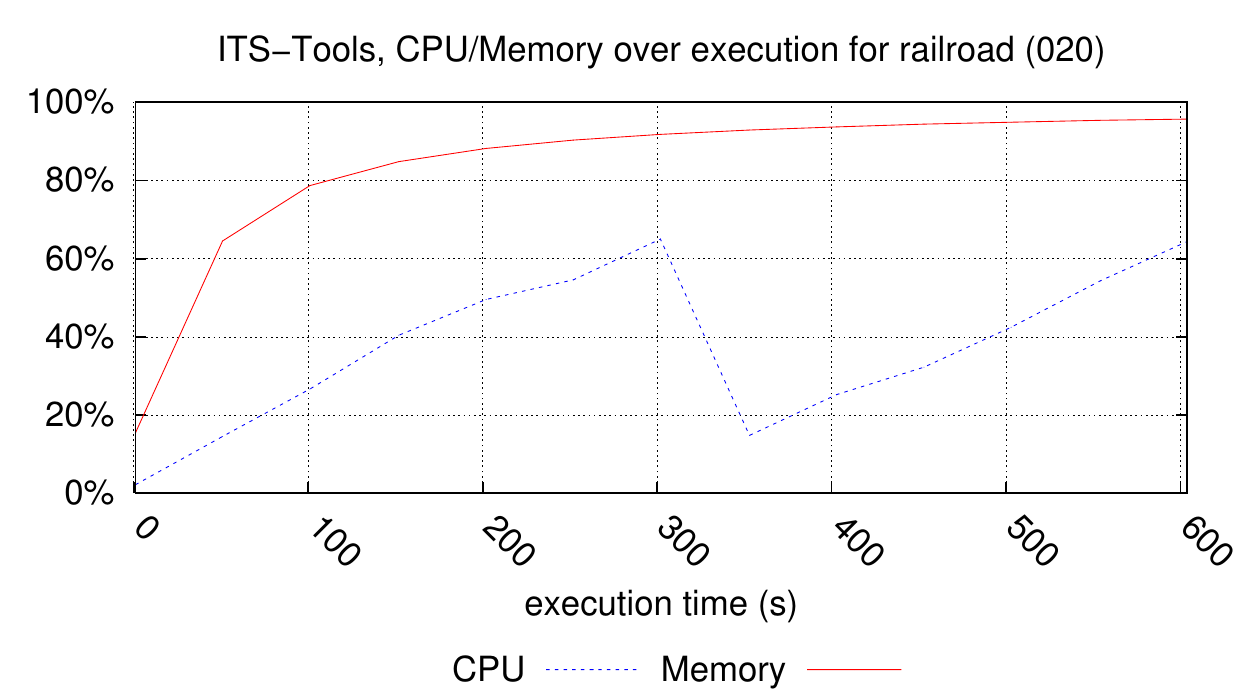}

\vfill\eject
\subsubsection{Executions for ring}
1 chart has been generated.
\index{Execution (by tool)!ITS-Tools}
\index{Execution (by model)!ring!ITS-Tools}

\noindent\includegraphics[width=.5\textwidth]{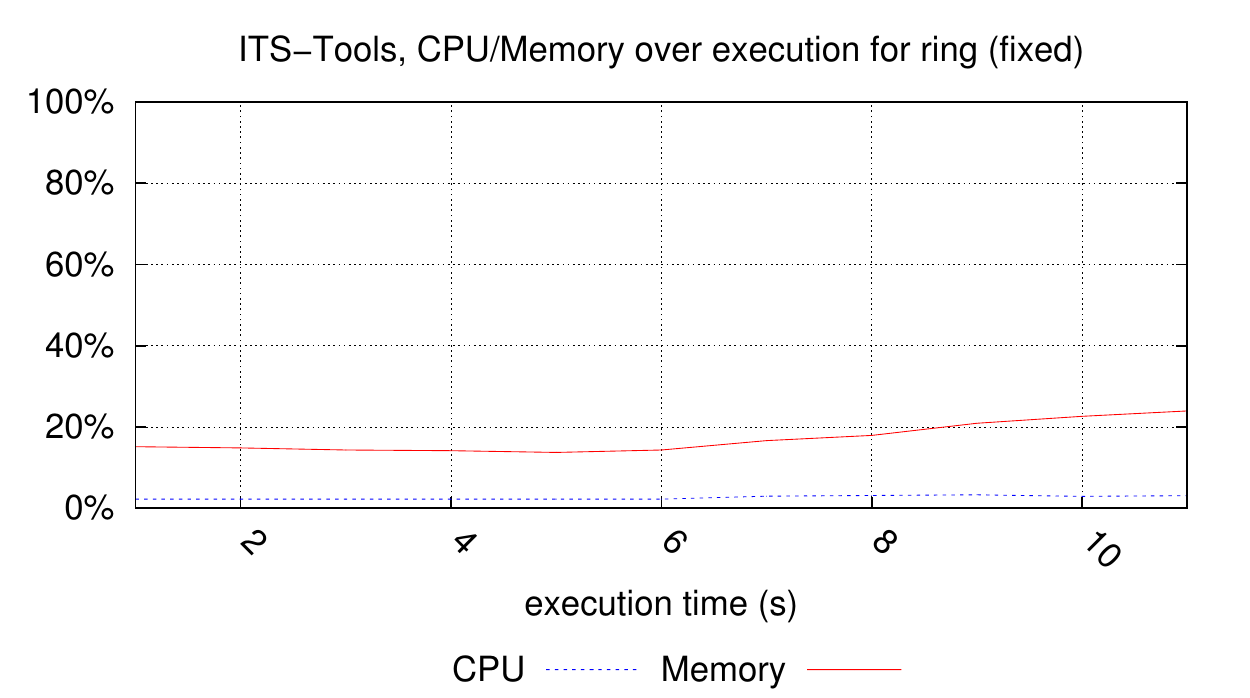}

\subsubsection{Executions for rw\_mutex}
4 charts have been generated.
\index{Execution (by tool)!ITS-Tools}
\index{Execution (by model)!rw\_mutex!ITS-Tools}

\noindent\includegraphics[width=.5\textwidth]{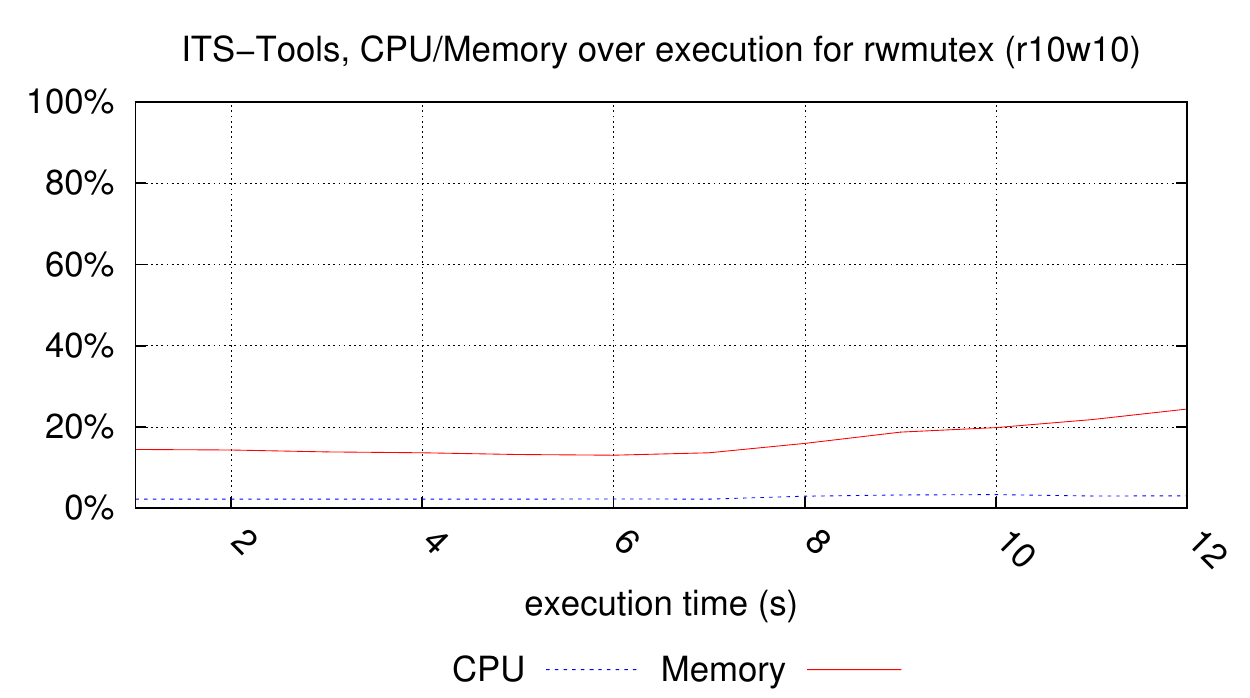}
\includegraphics[width=.5\textwidth]{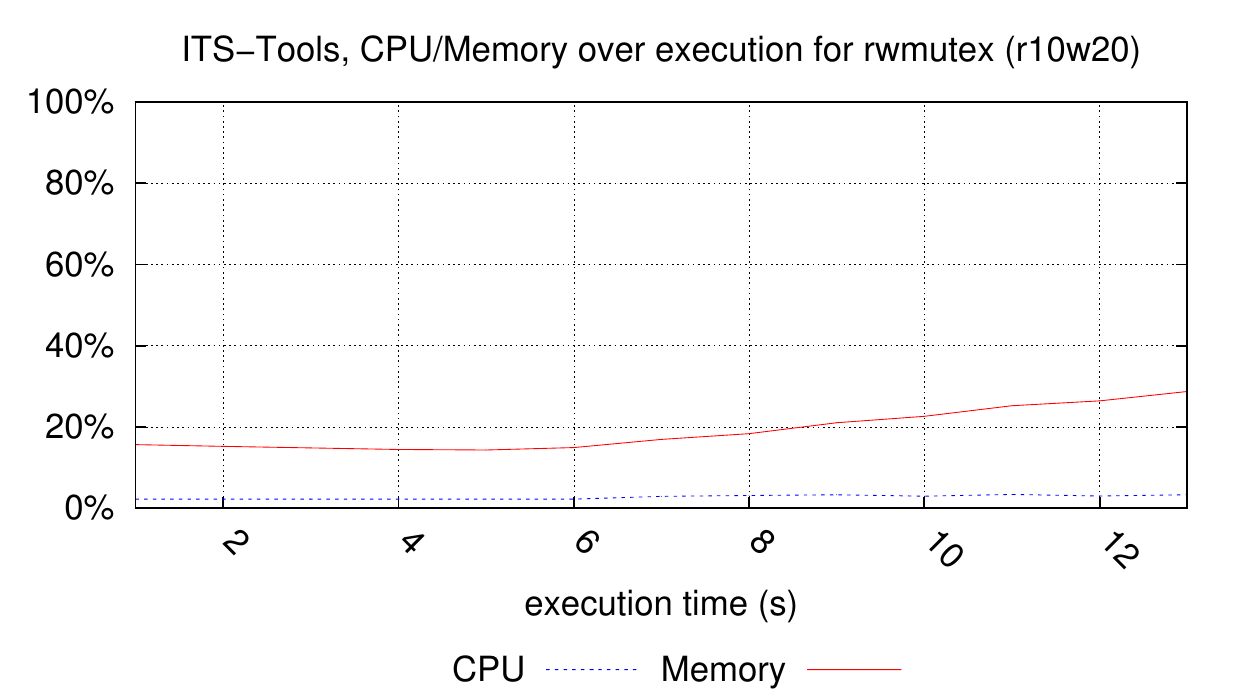}

\noindent\includegraphics[width=.5\textwidth]{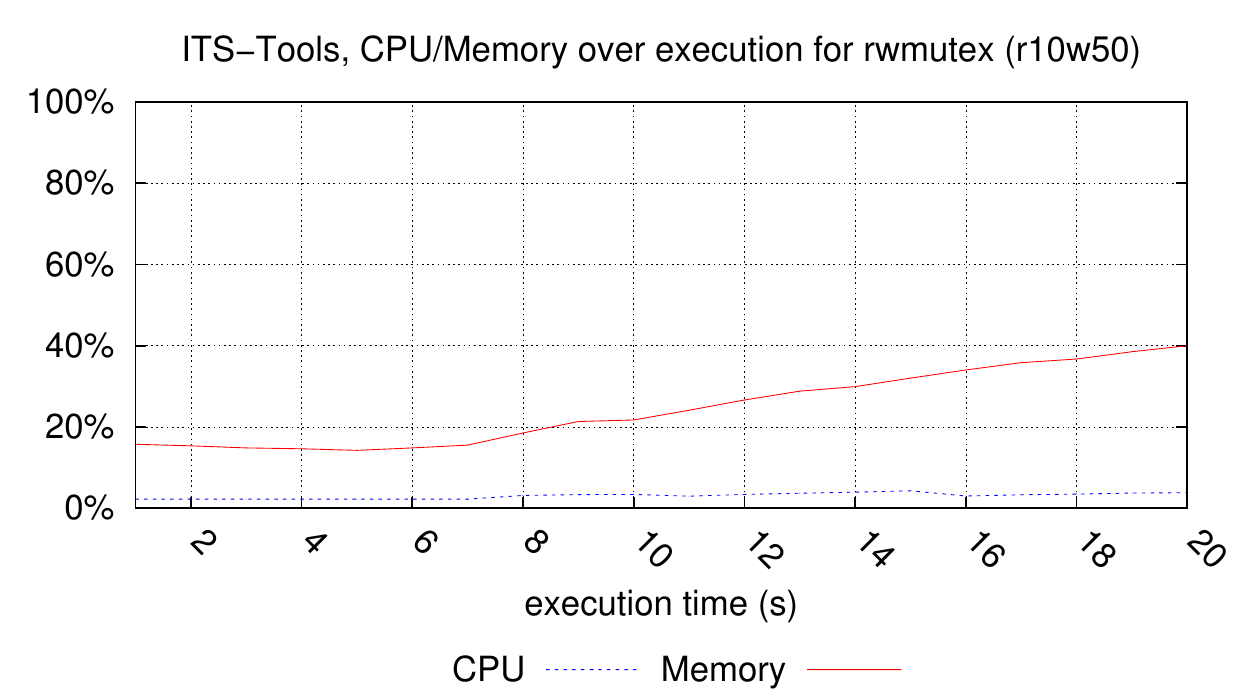}
\includegraphics[width=.5\textwidth]{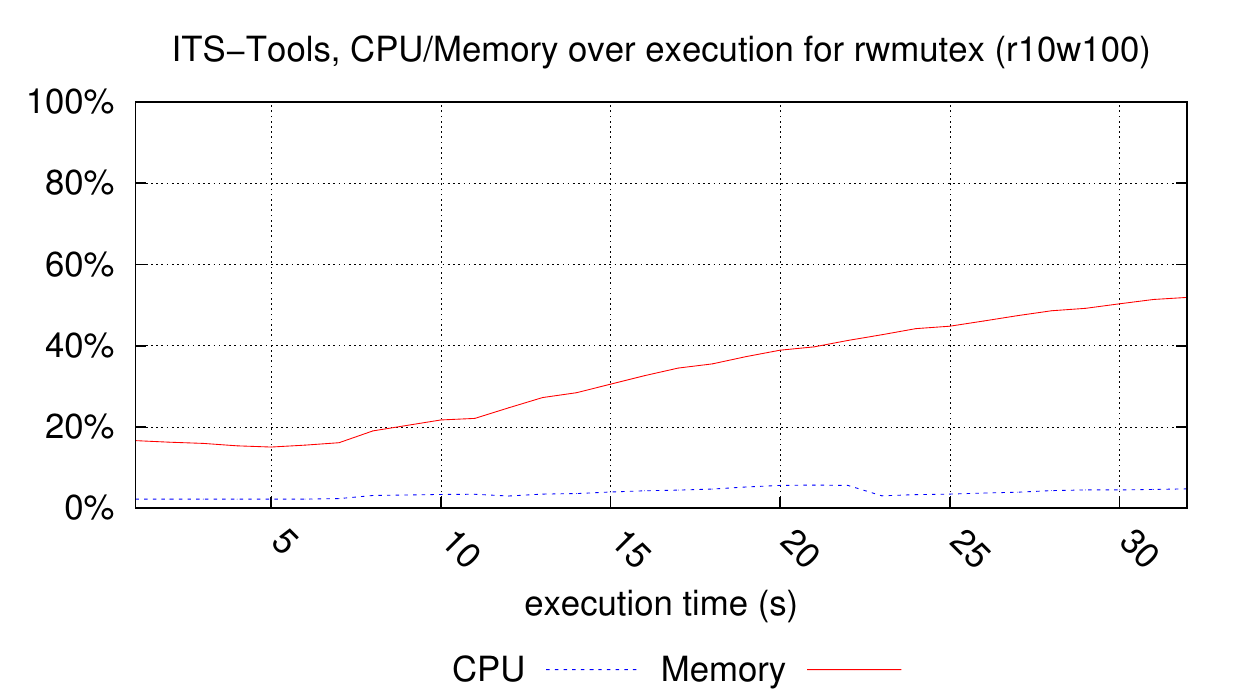}

\noindent\includegraphics[width=.5\textwidth]{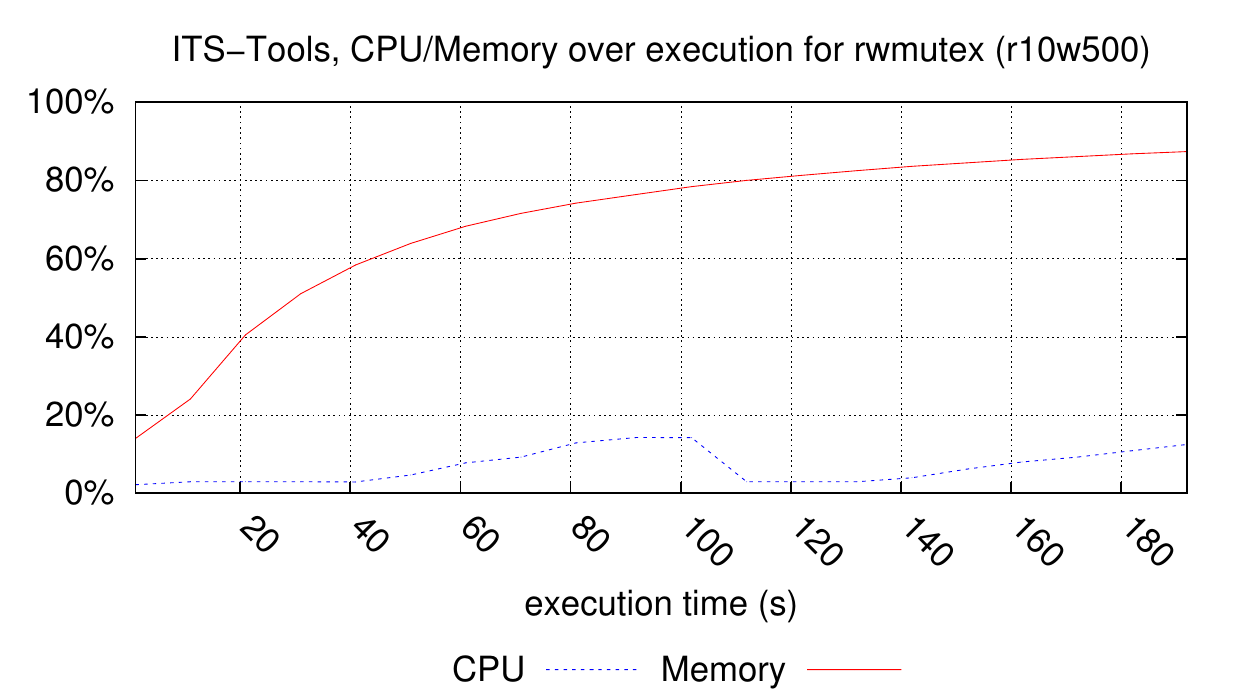}

\vfill\eject
\subsubsection{Executions for SharedMemory}
4 charts have been generated.
\index{Execution (by tool)!ITS-Tools}
\index{Execution (by model)!SharedMemory!ITS-Tools}

\noindent\includegraphics[width=.5\textwidth]{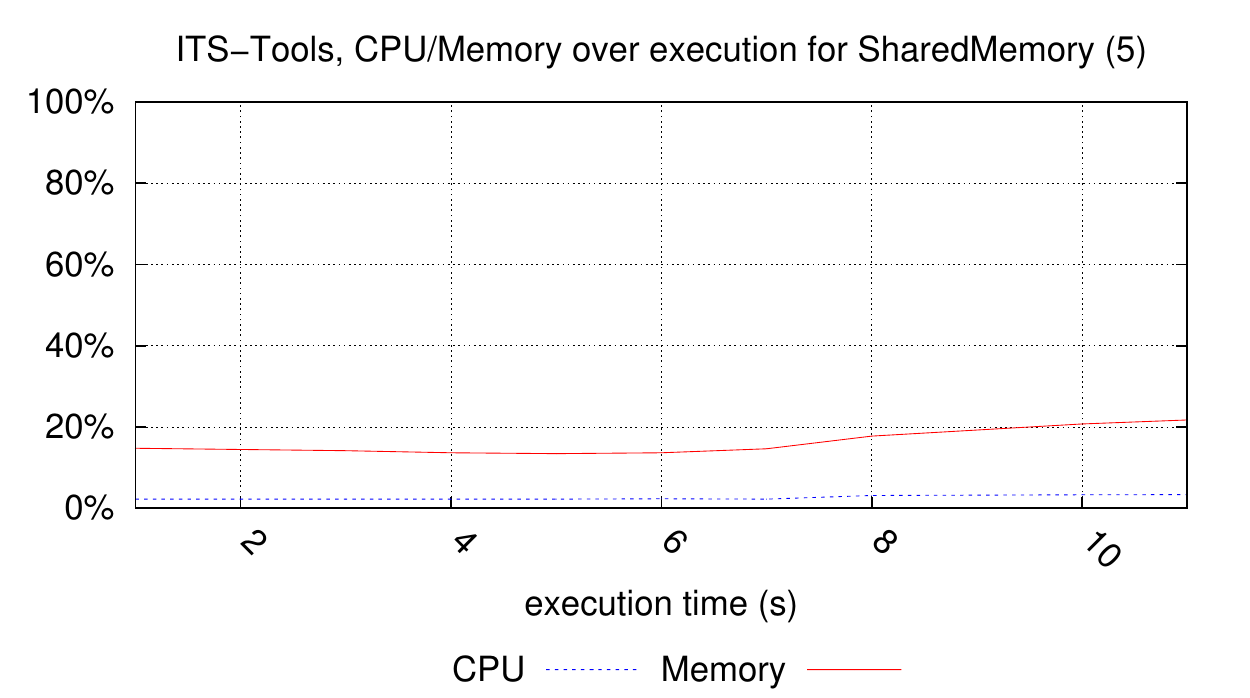}
\includegraphics[width=.5\textwidth]{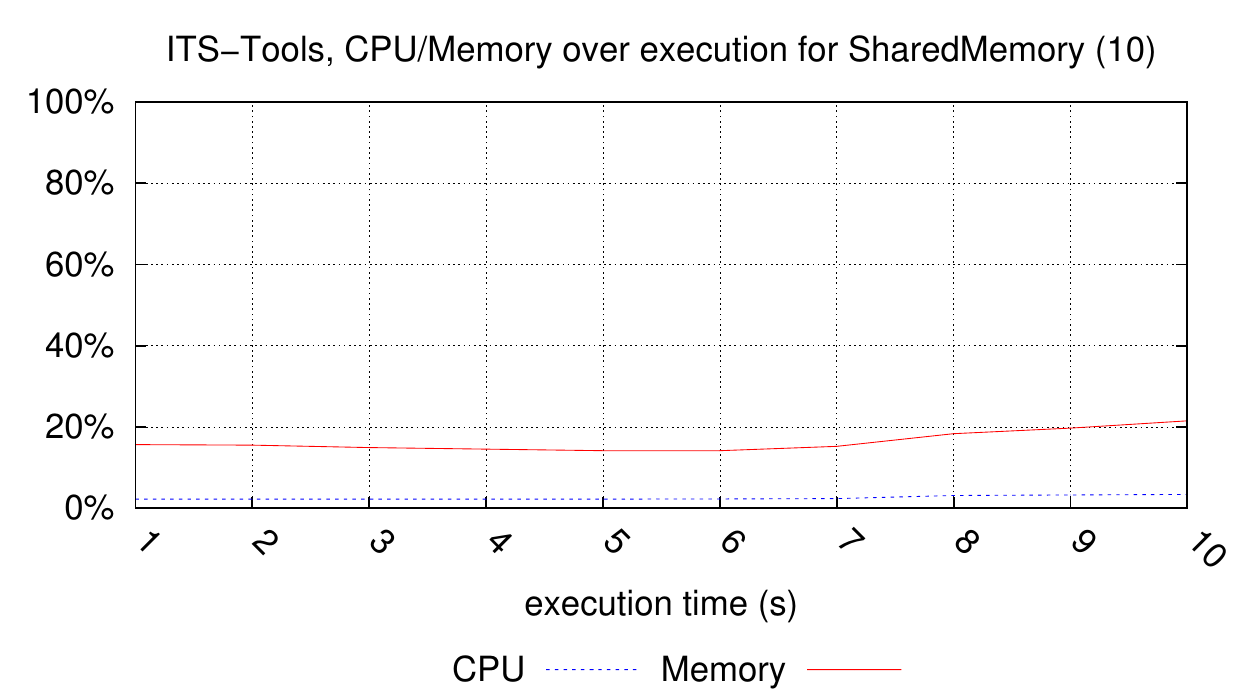}

\noindent\includegraphics[width=.5\textwidth]{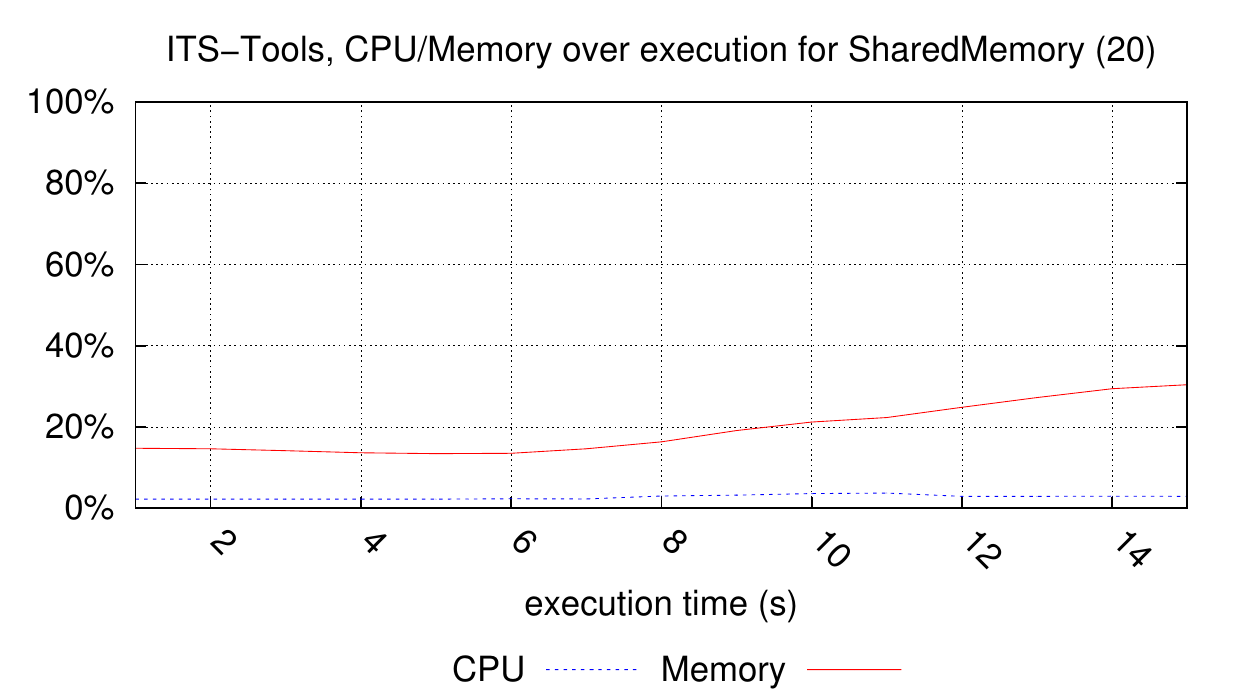}
\includegraphics[width=.5\textwidth]{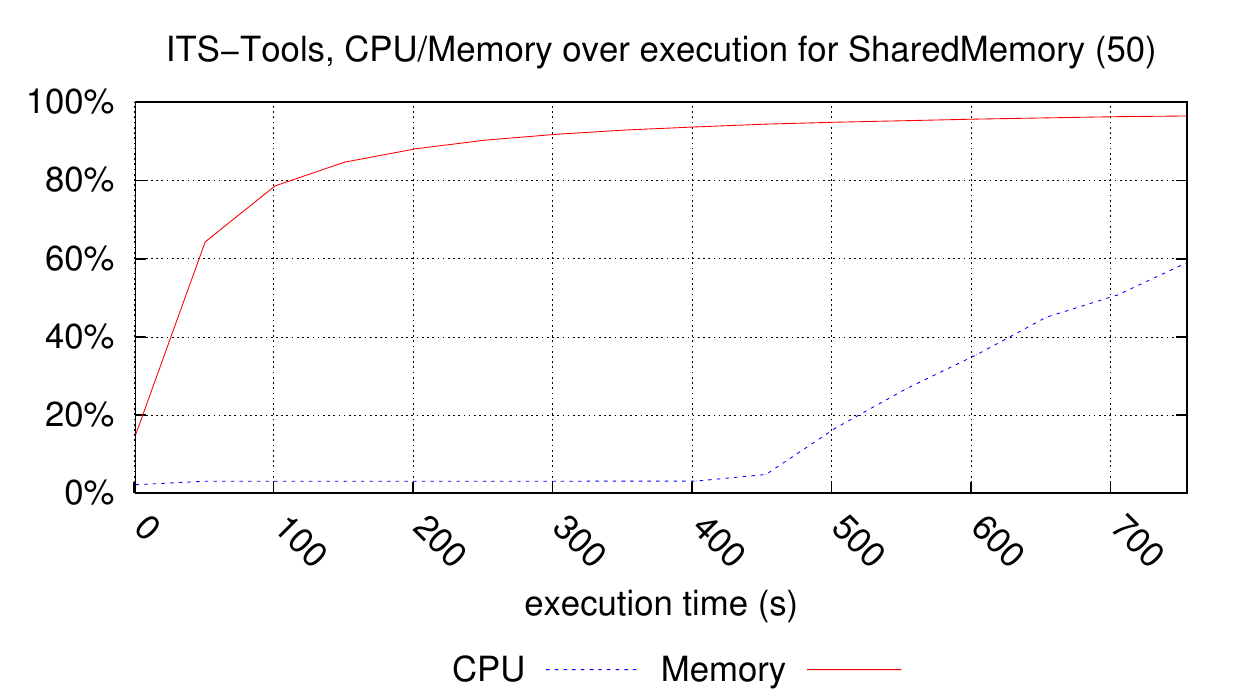}

\subsubsection{Executions for simple\_lbs}
5 charts have been generated.
\index{Execution (by tool)!ITS-Tools}
\index{Execution (by model)!simple\_lbs!ITS-Tools}

\noindent\includegraphics[width=.5\textwidth]{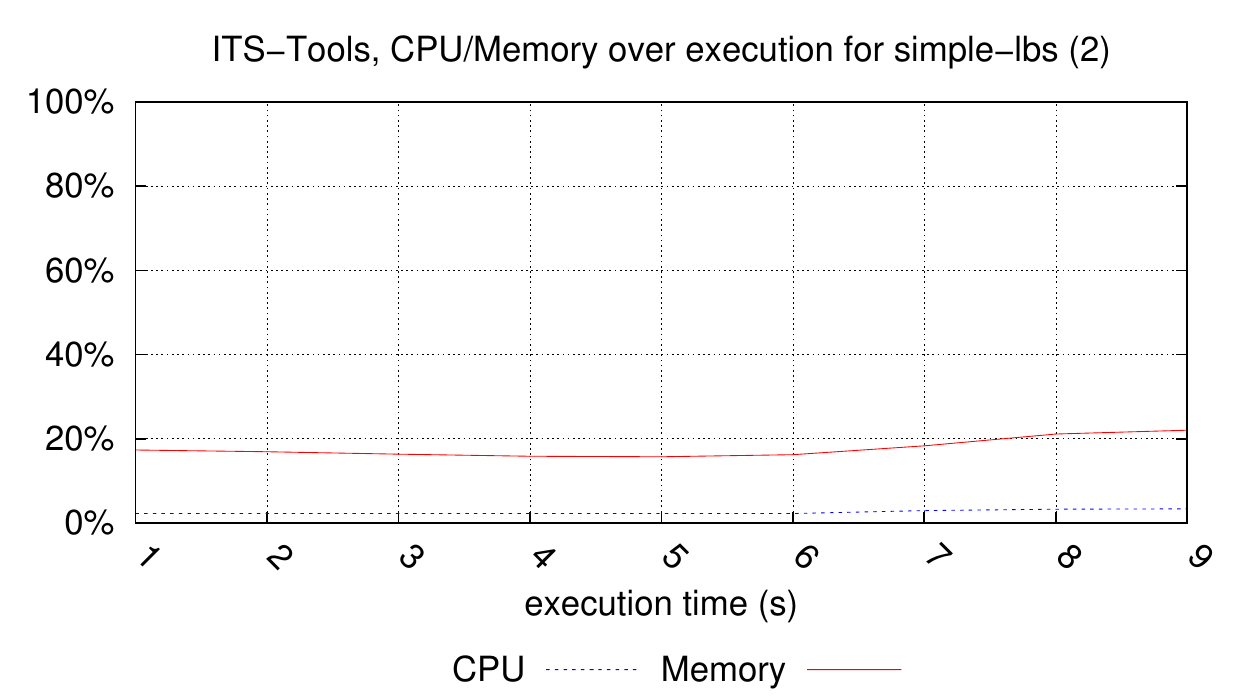}
\includegraphics[width=.5\textwidth]{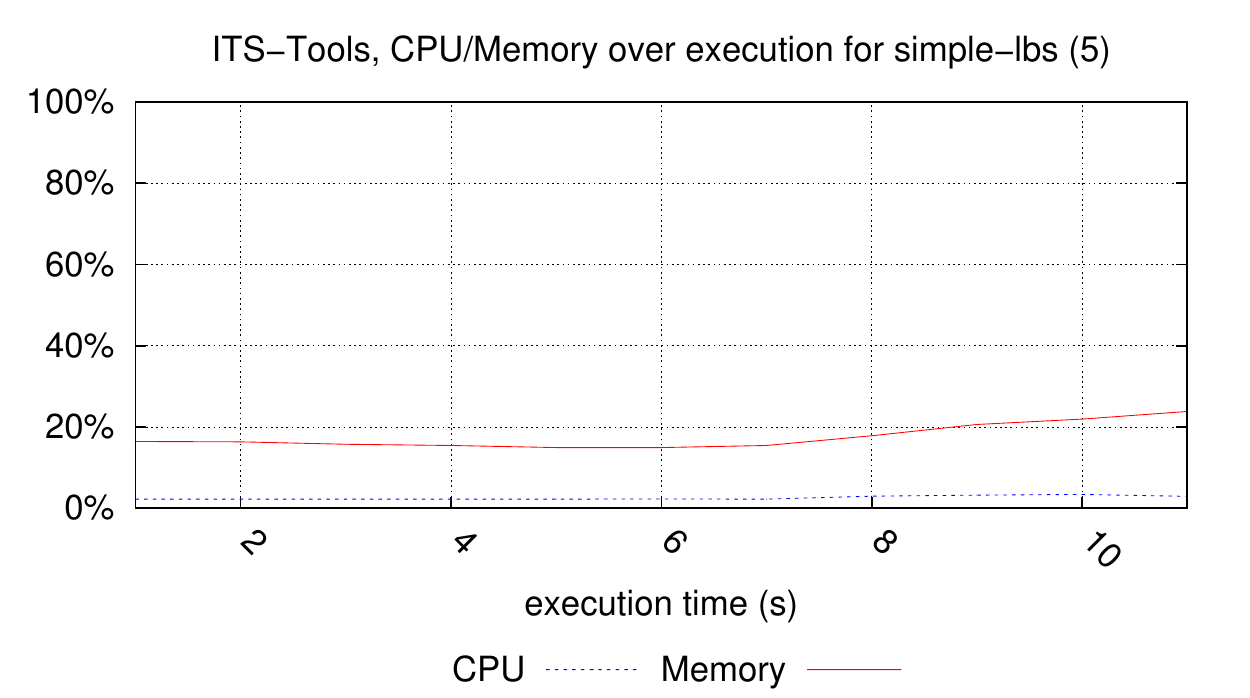}

\noindent\includegraphics[width=.5\textwidth]{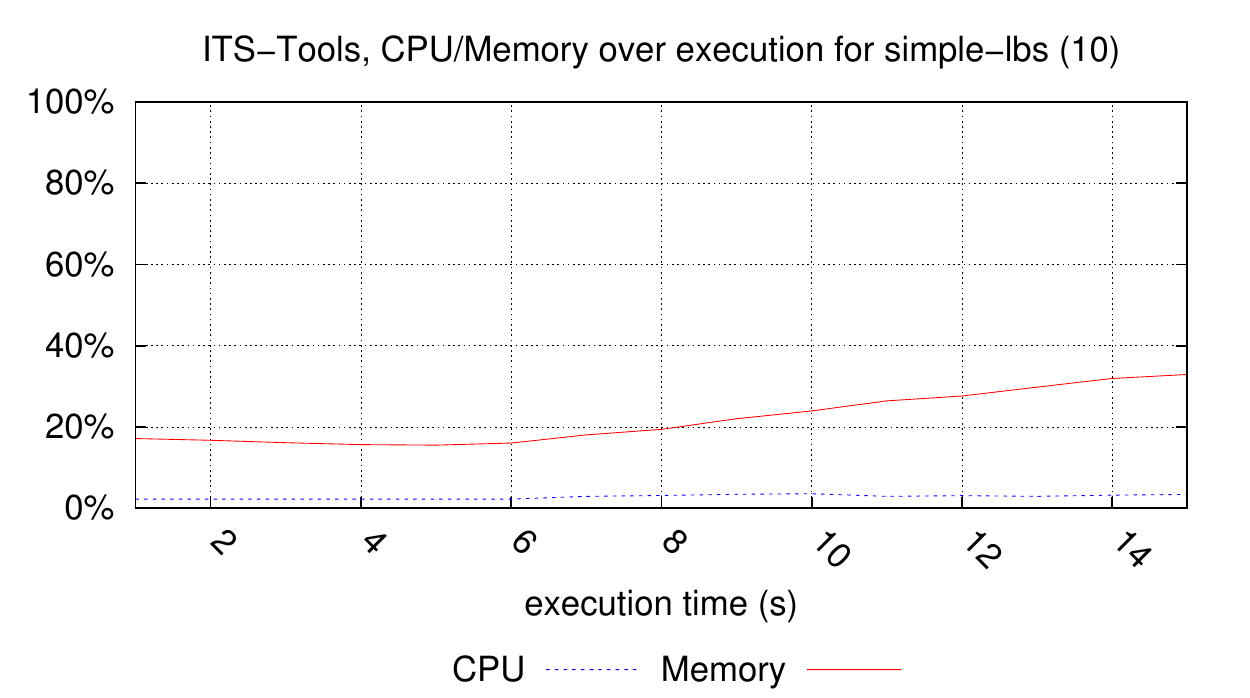}
\includegraphics[width=.5\textwidth]{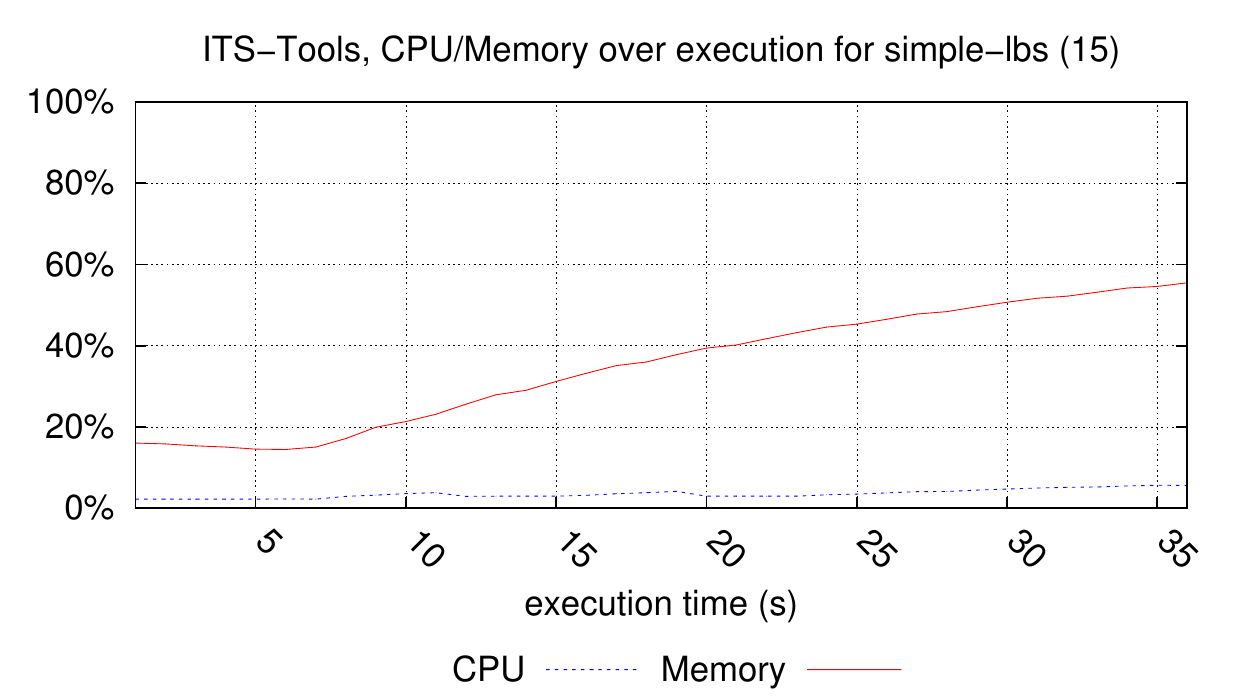}

\noindent\includegraphics[width=.5\textwidth]{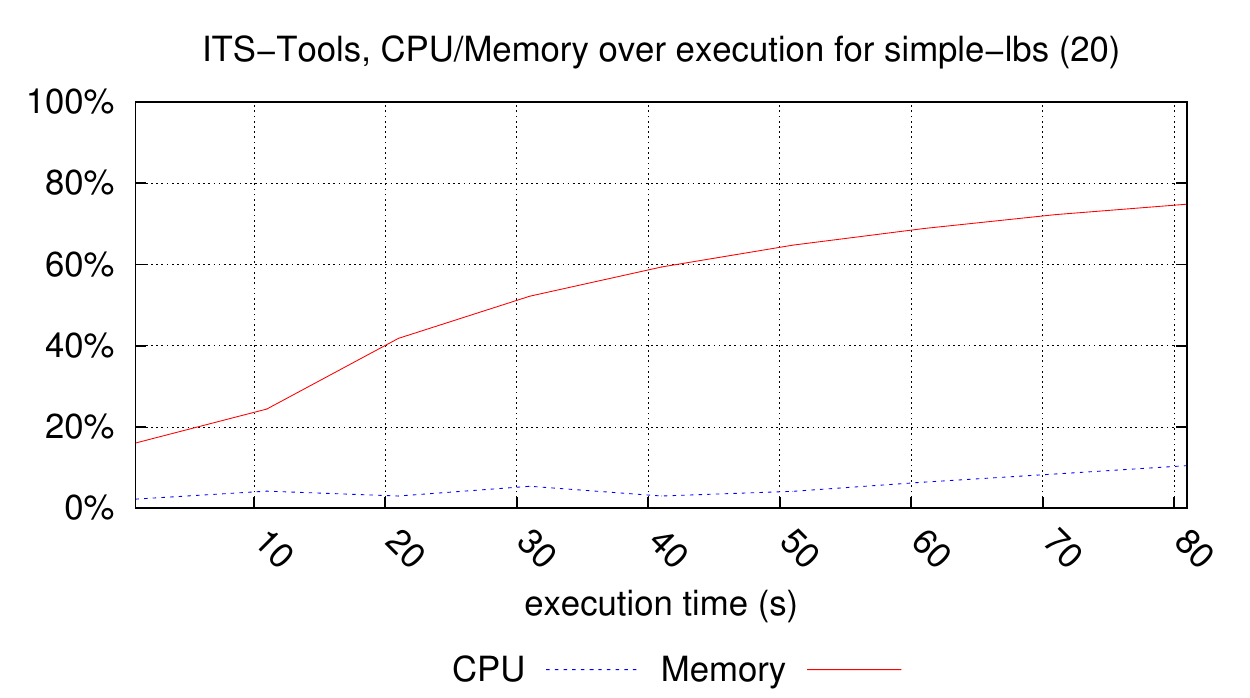}

\subsubsection{Executions for TokenRing}
4 charts have been generated.
\index{Execution (by tool)!ITS-Tools}
\index{Execution (by model)!TokenRing!ITS-Tools}

\noindent\includegraphics[width=.5\textwidth]{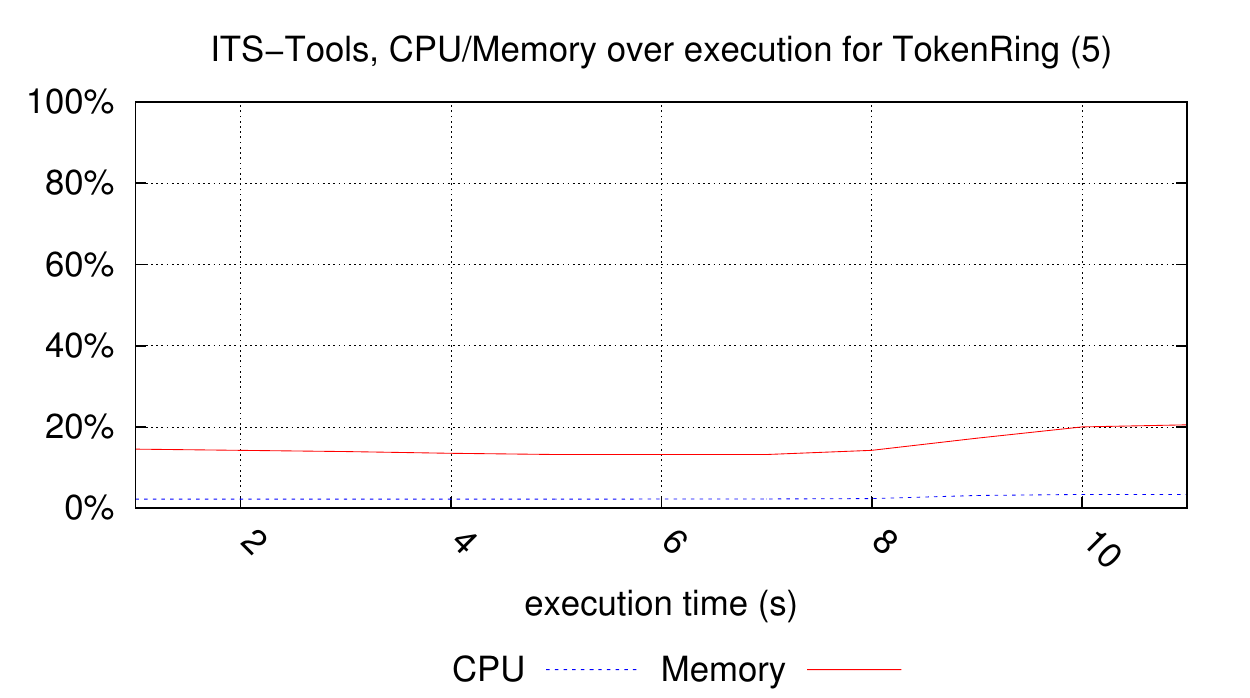}
\includegraphics[width=.5\textwidth]{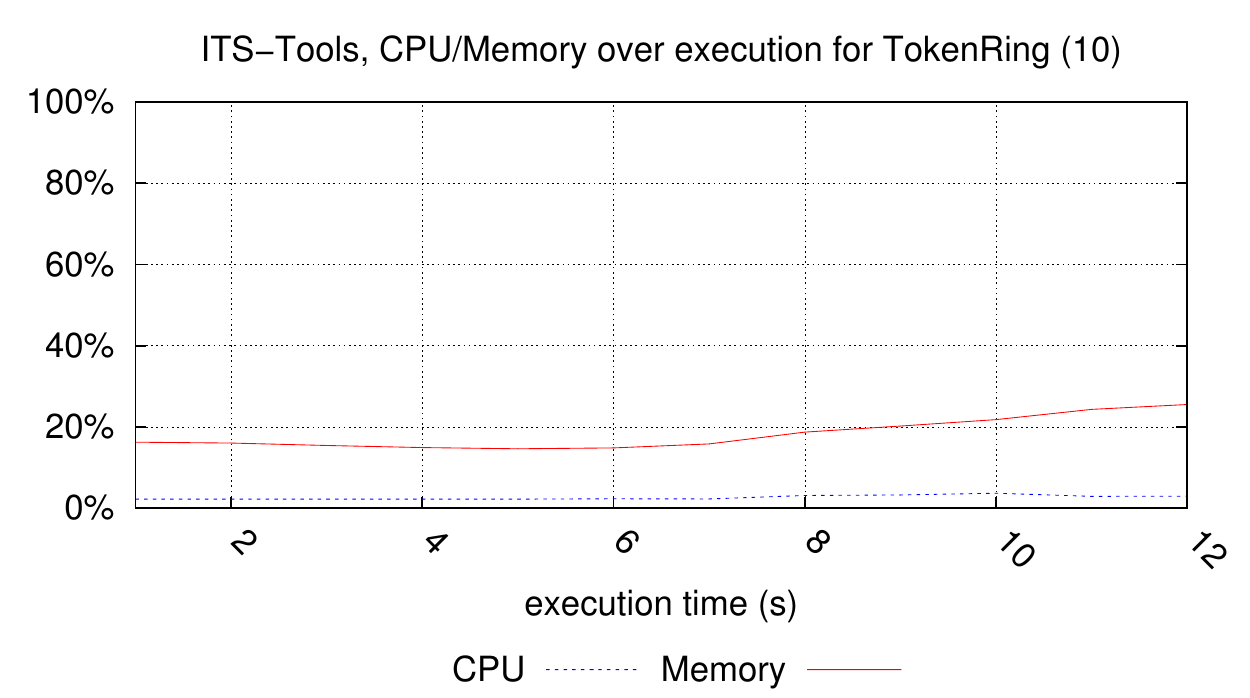}

\noindent\includegraphics[width=.5\textwidth]{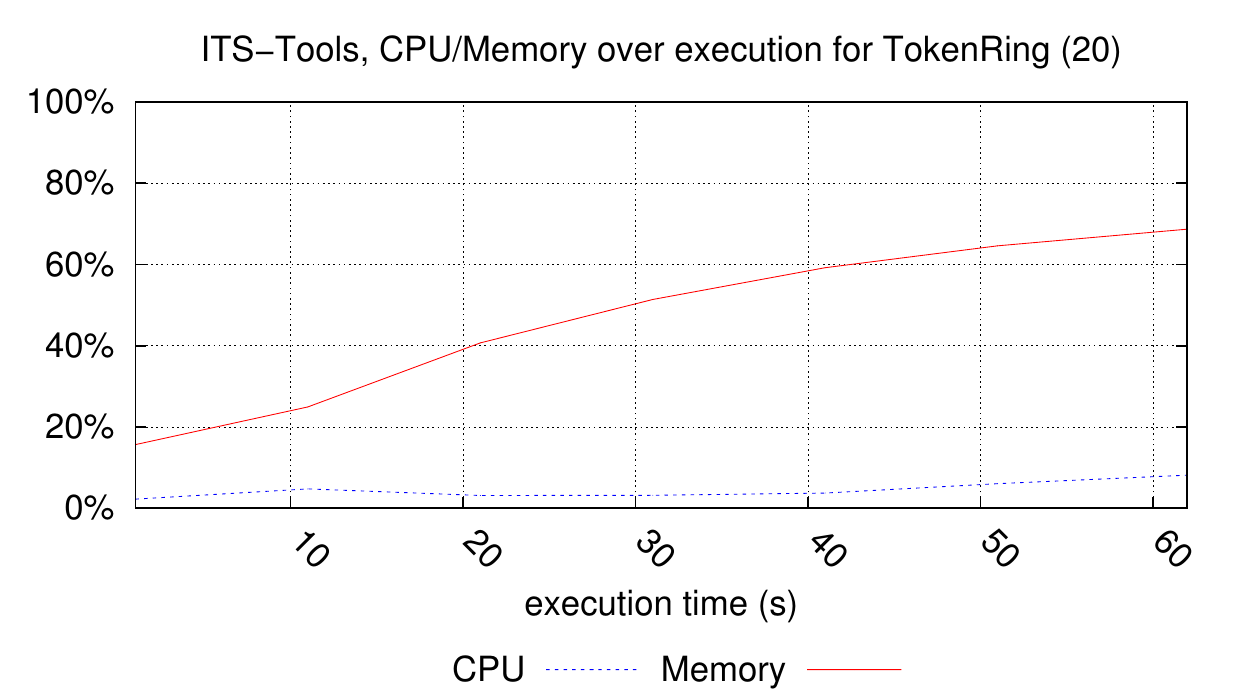}
\includegraphics[width=.5\textwidth]{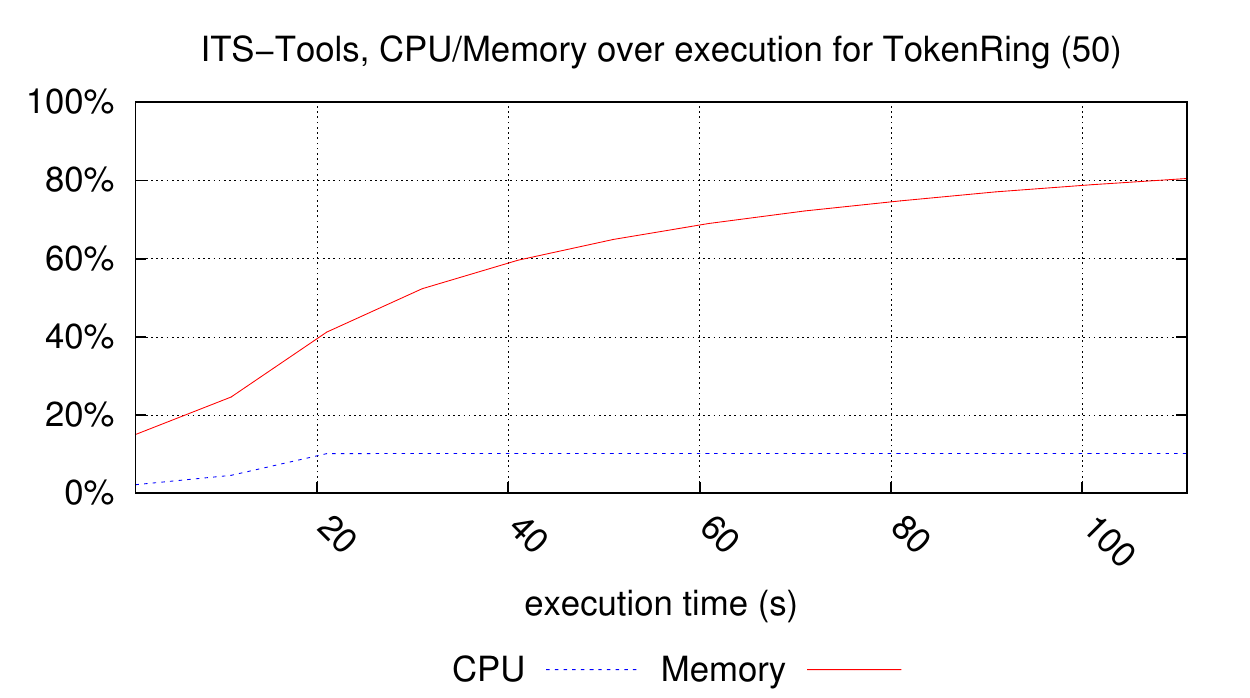}

\subsection{Execution Charts for Marcie}

We provide here the execution charts observed for Marcie over
the models it could compete with.

\subsubsection{Executions for cs\_repetitions}
1 chart has been generated.
\index{Execution (by tool)!Marcie}
\index{Execution (by model)!cs\_repetitions!Marcie}

\noindent\includegraphics[width=.5\textwidth]{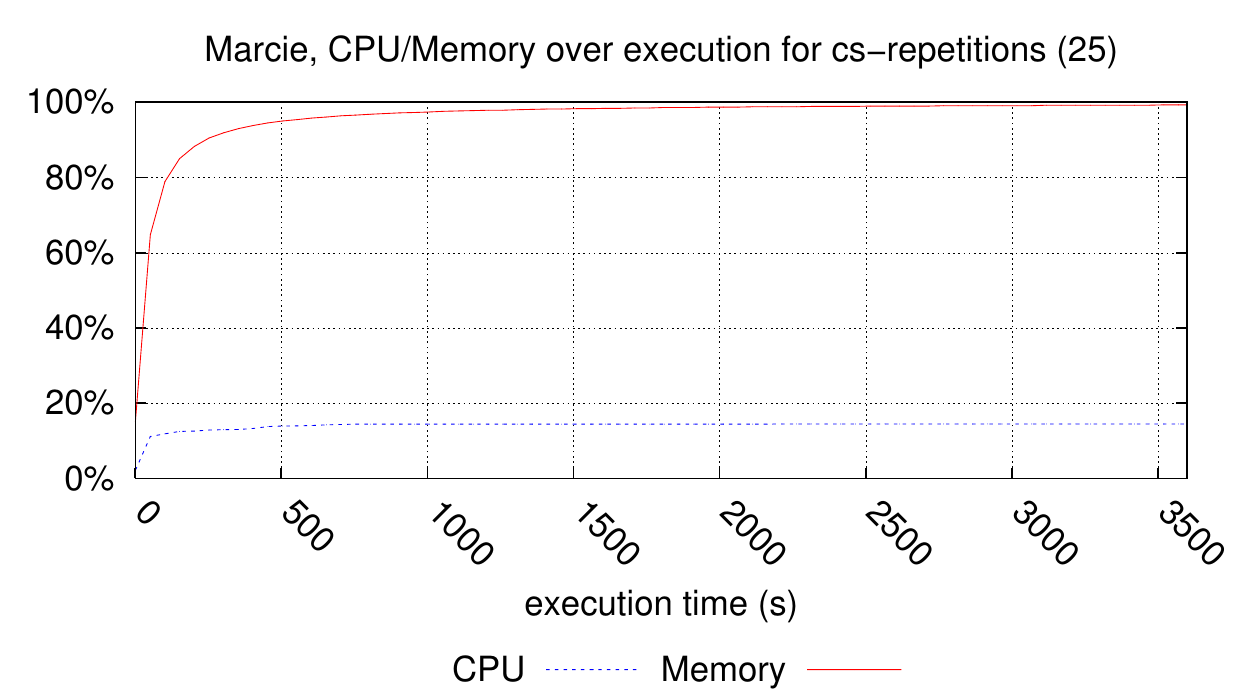}

\vfill\eject
\subsubsection{Executions for echo}
2 charts have been generated.
\index{Execution (by tool)!Marcie}
\index{Execution (by model)!echo!Marcie}

\noindent\includegraphics[width=.5\textwidth]{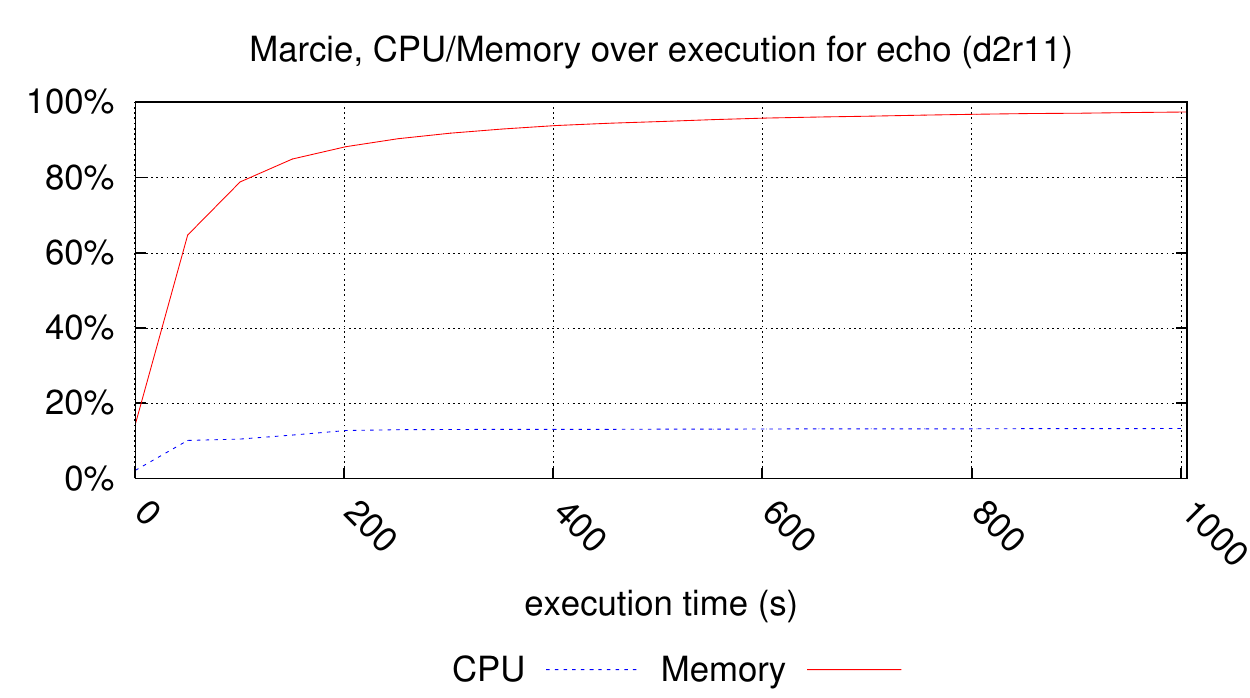}
\includegraphics[width=.5\textwidth]{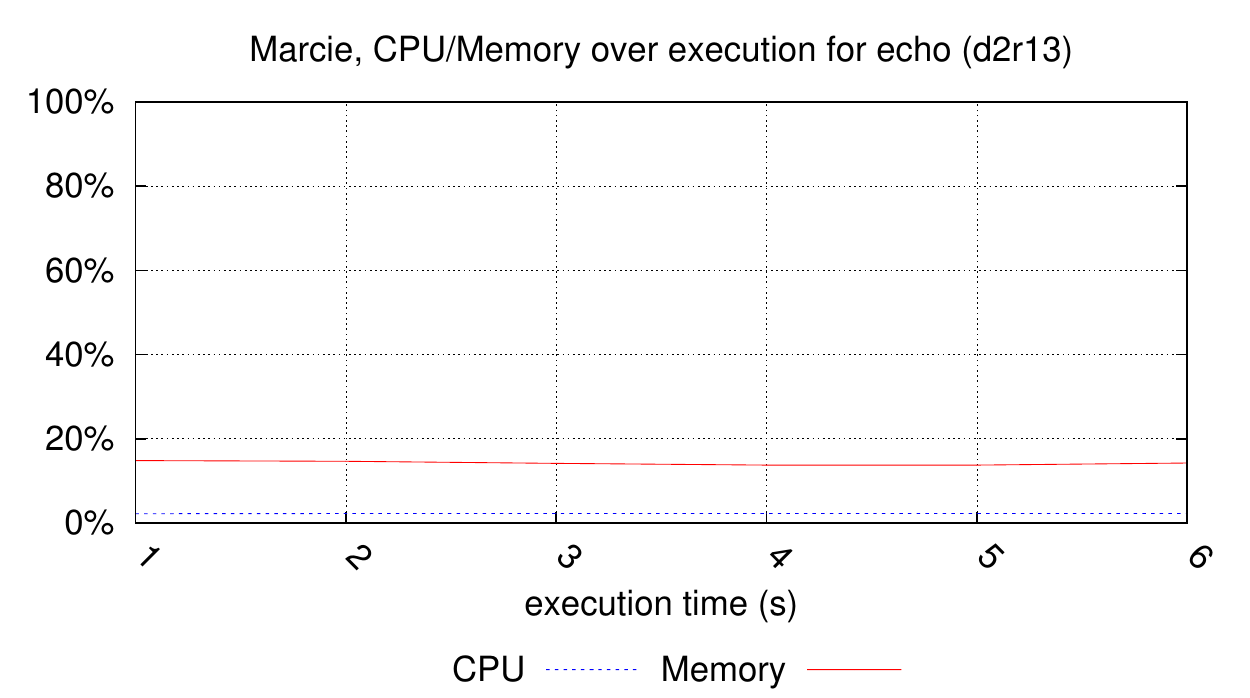}

\subsubsection{Executions for eratosthenes}
7 charts have been generated.
\index{Execution (by tool)!Marcie}
\index{Execution (by model)!eratosthenes!Marcie}

\noindent\includegraphics[width=.5\textwidth]{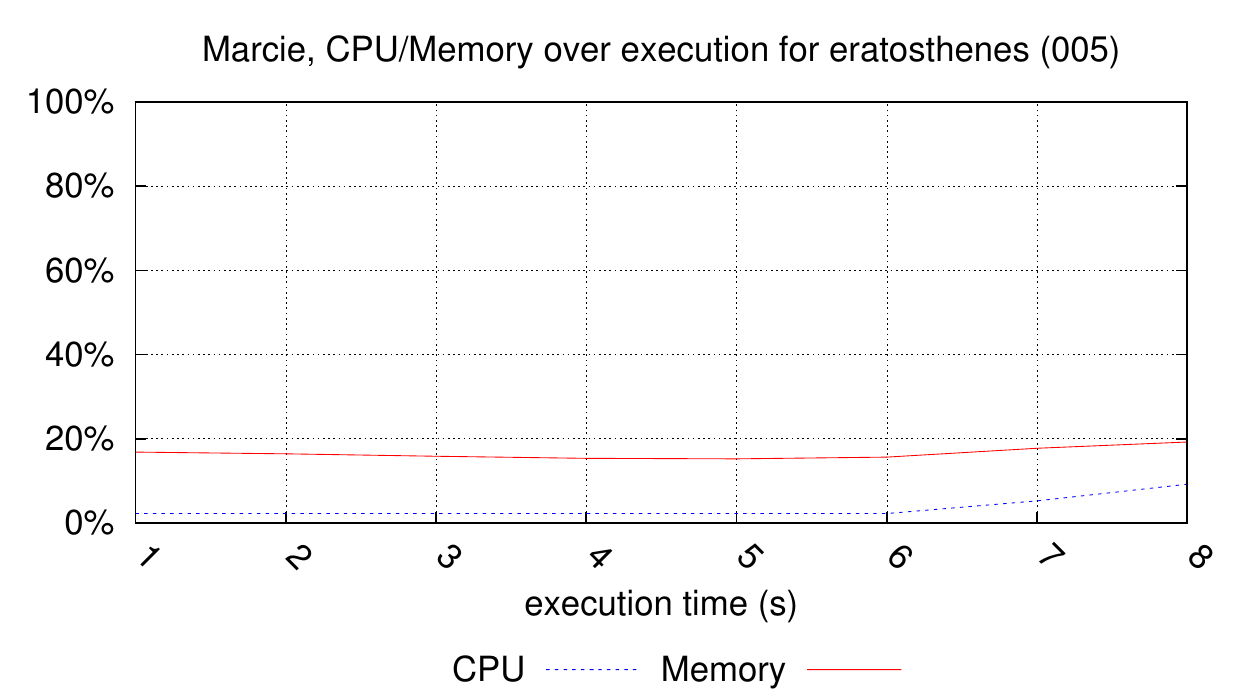}
\includegraphics[width=.5\textwidth]{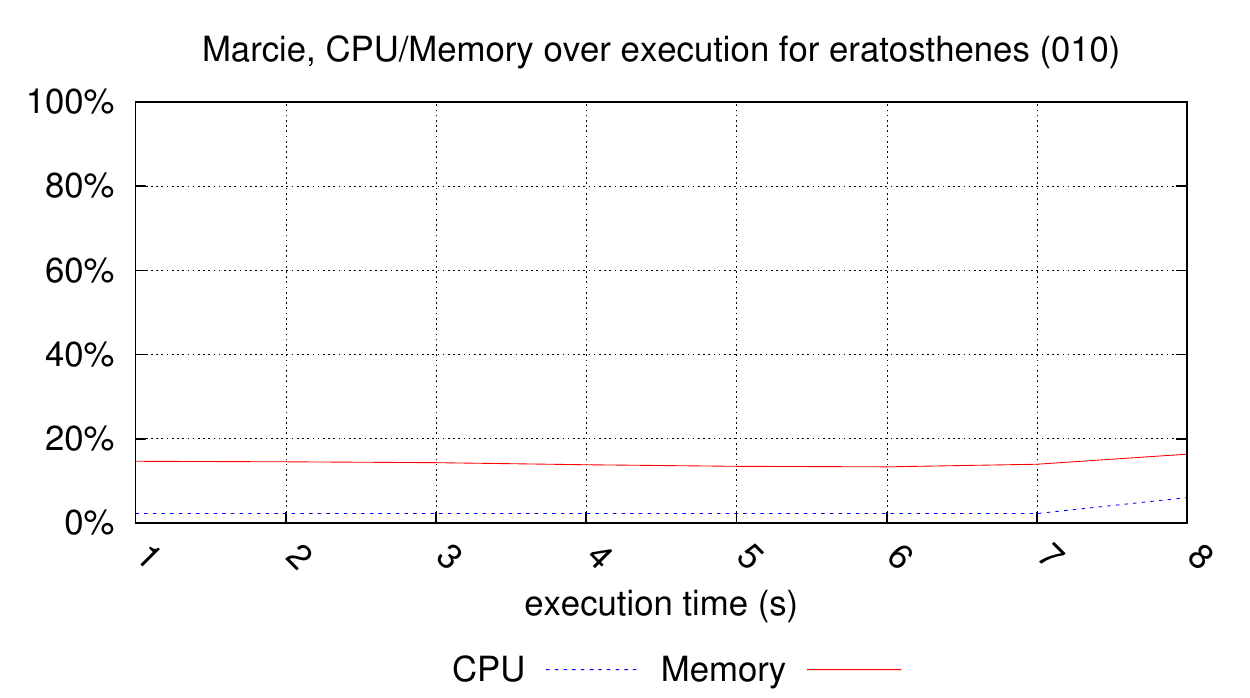}

\noindent\includegraphics[width=.5\textwidth]{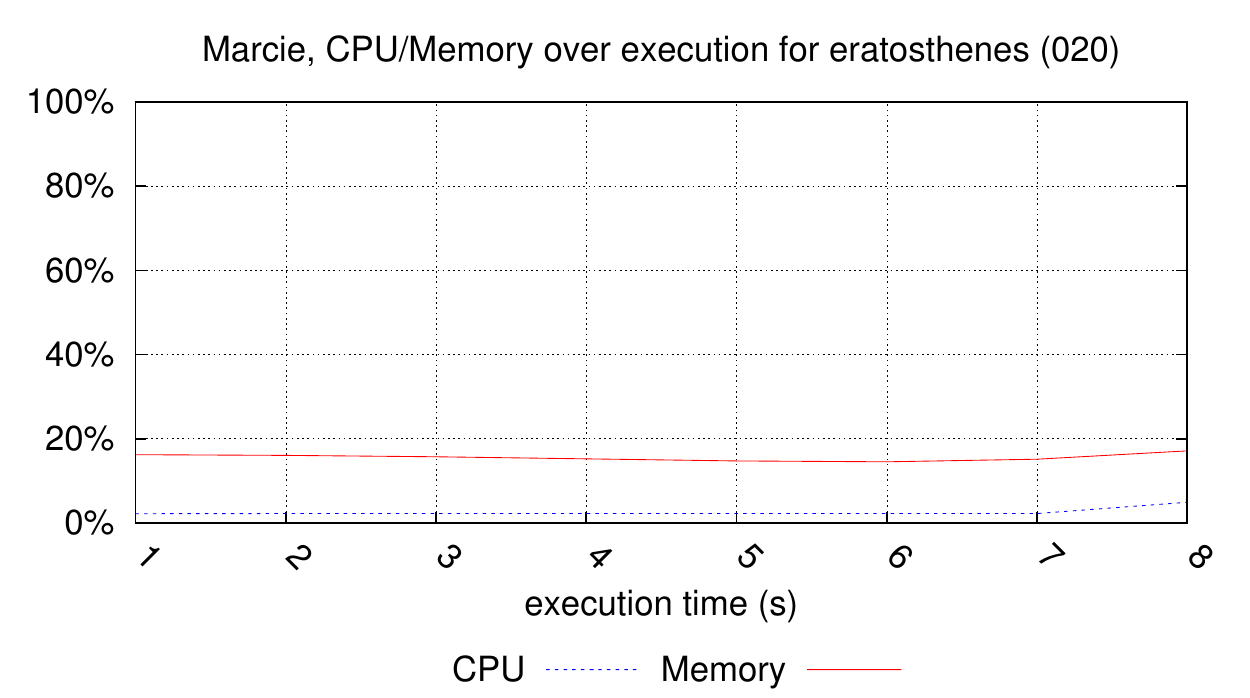}
\includegraphics[width=.5\textwidth]{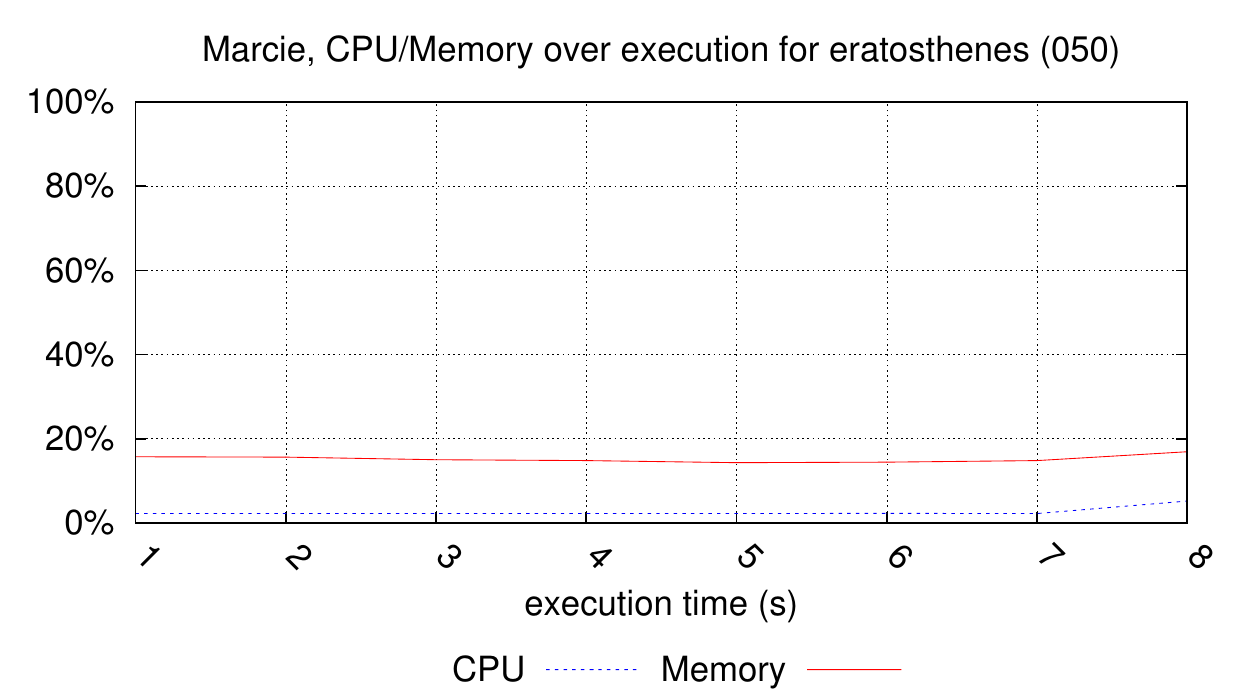}

\noindent\includegraphics[width=.5\textwidth]{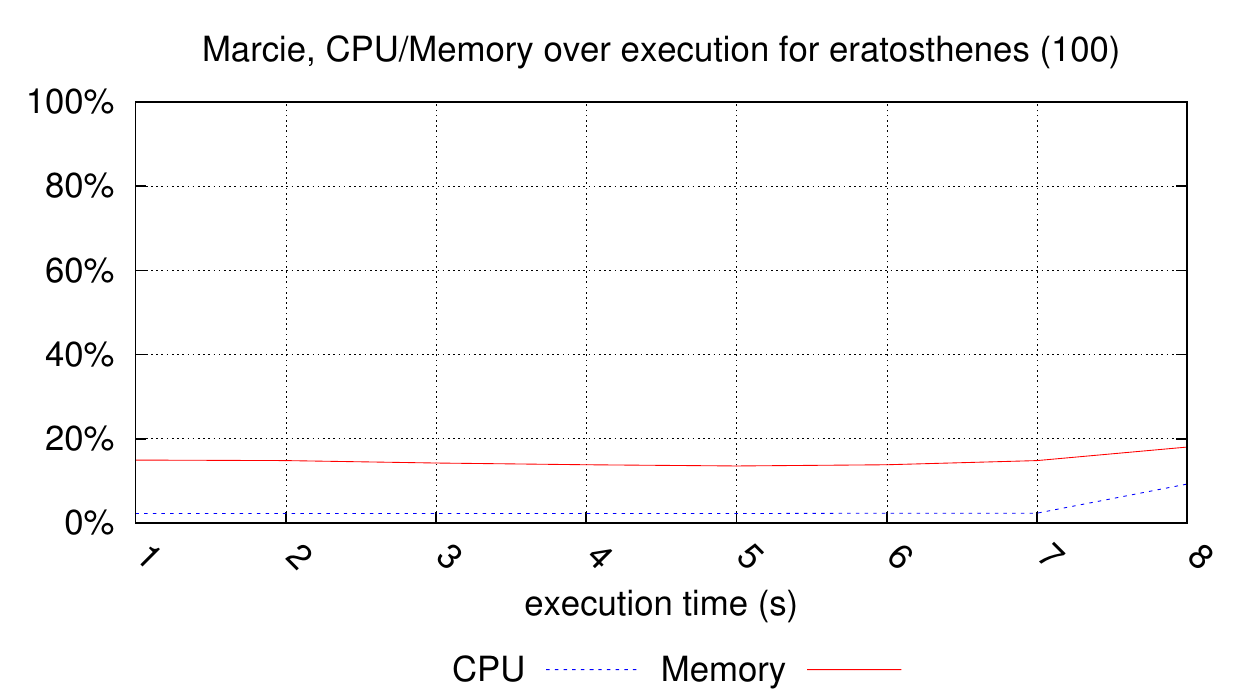}
\includegraphics[width=.5\textwidth]{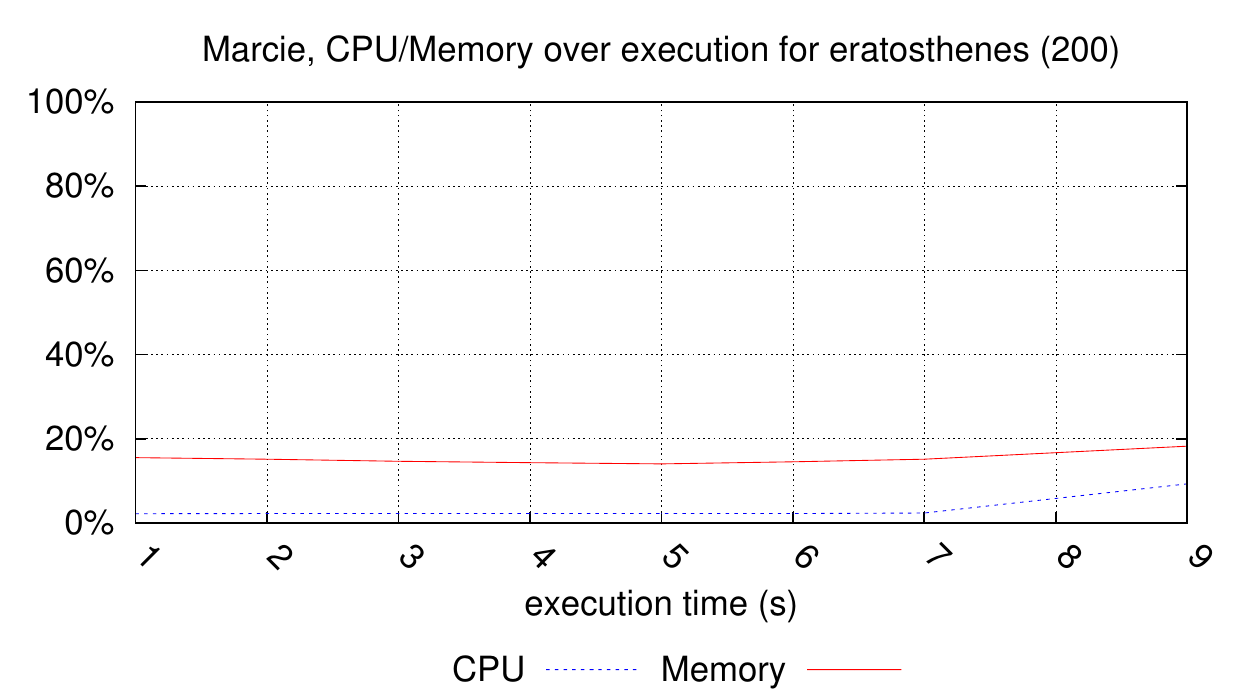}

\noindent\includegraphics[width=.5\textwidth]{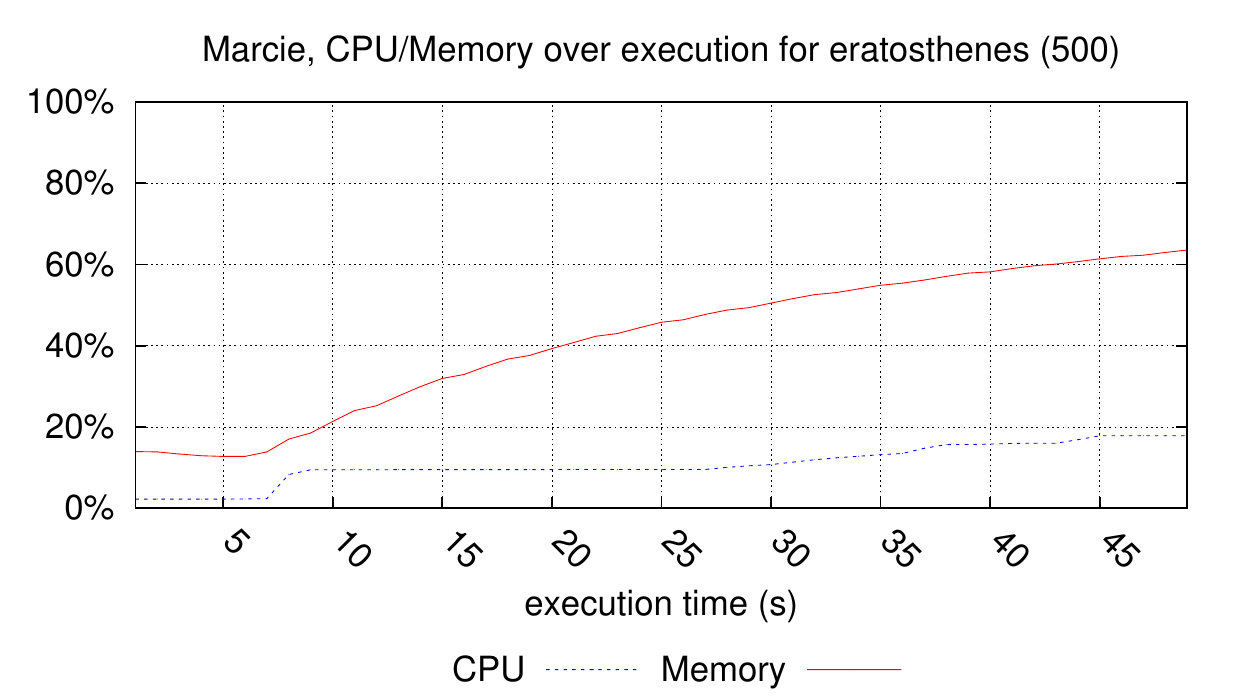}
\subsubsection{Executions for FMS}
7 charts have been generated.
\index{Execution (by tool)!Marcie}
\index{Execution (by model)!FMS!Marcie}

\noindent\includegraphics[width=.5\textwidth]{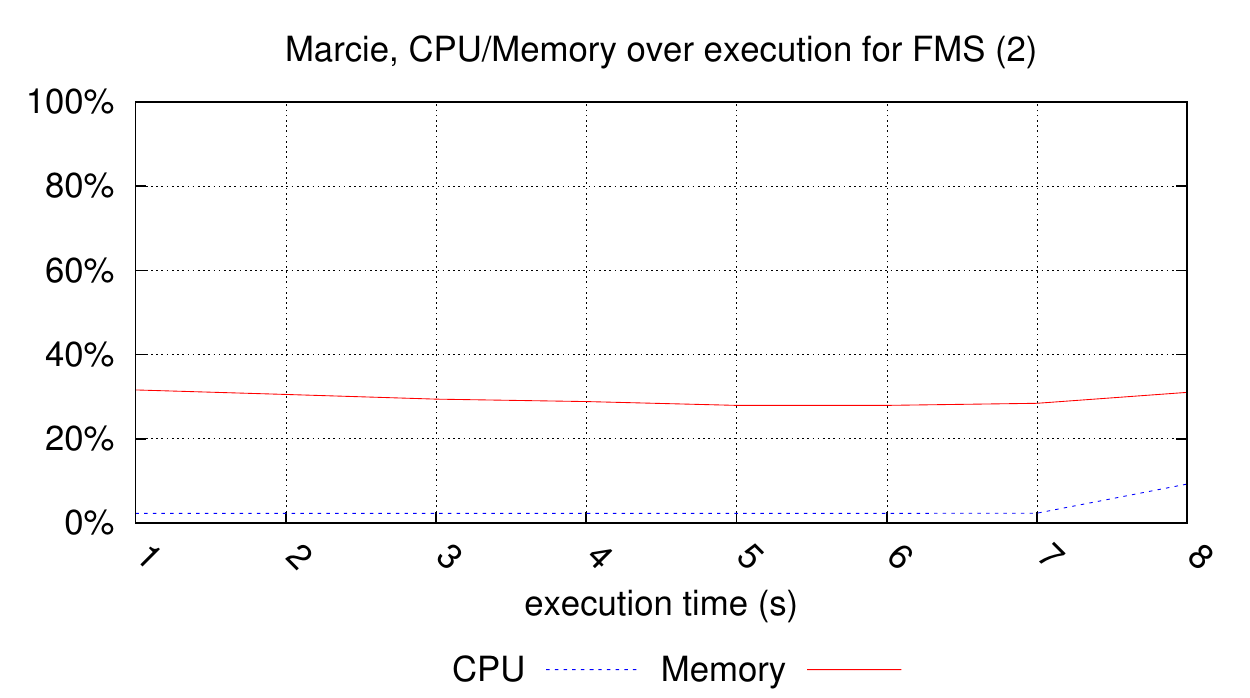}
\includegraphics[width=.5\textwidth]{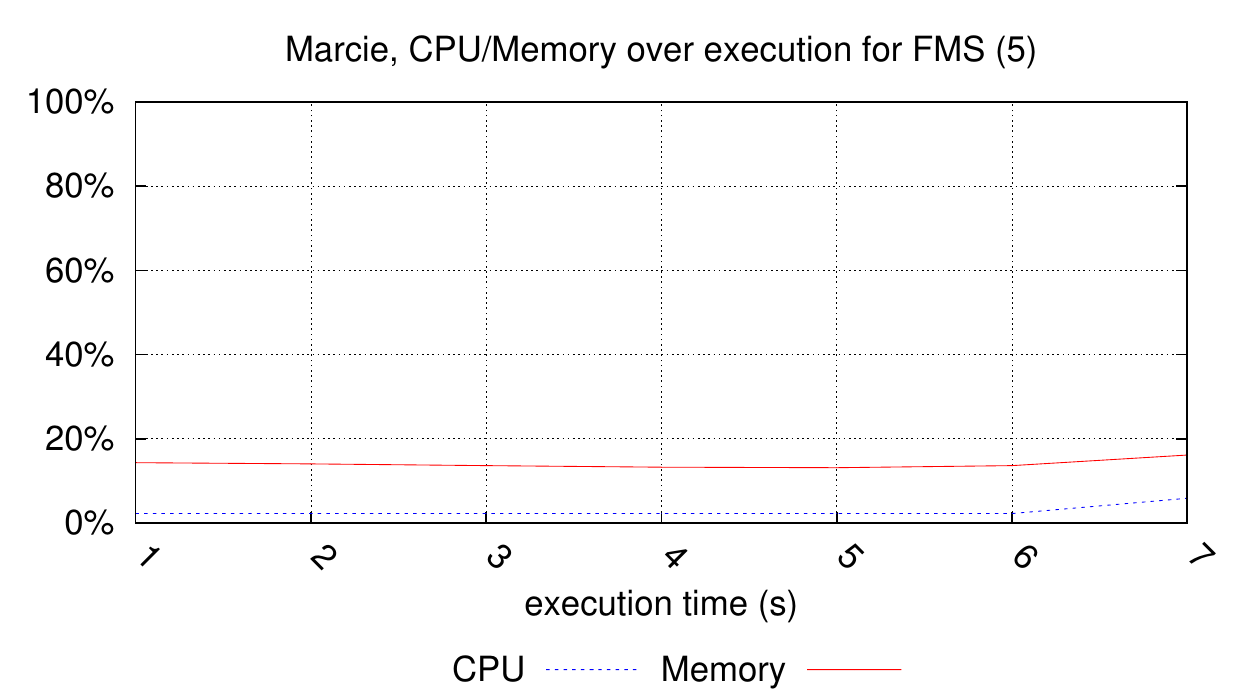}

\noindent\includegraphics[width=.5\textwidth]{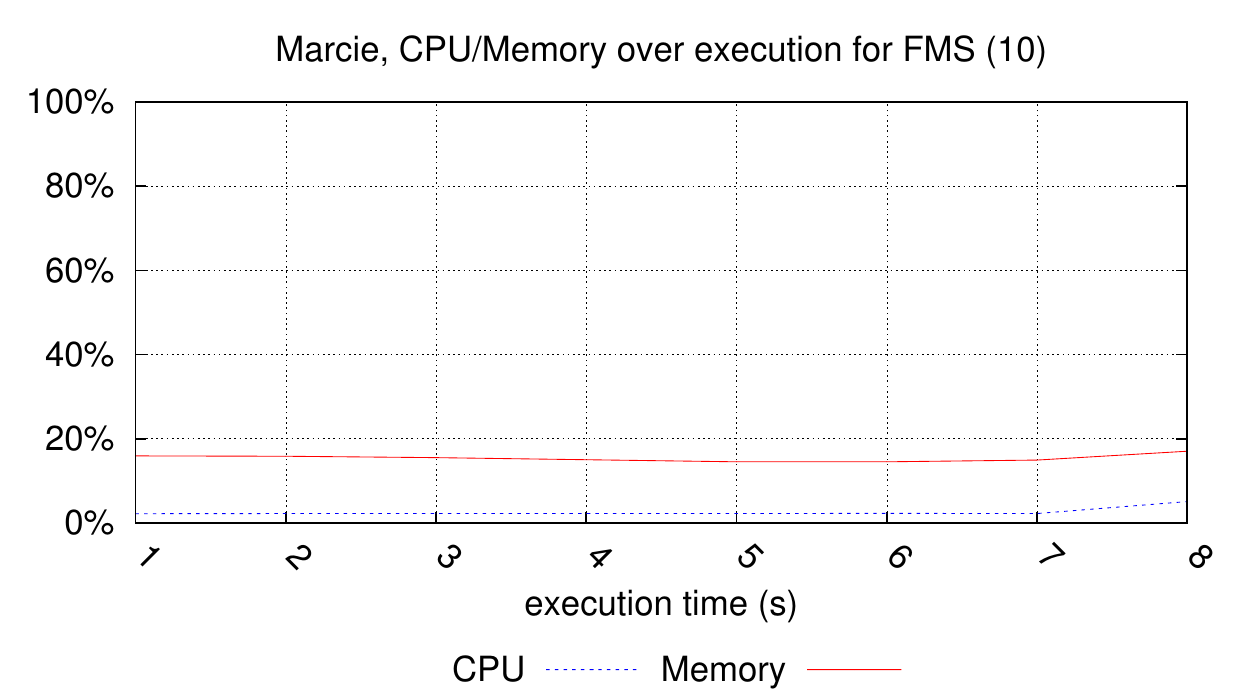}
\includegraphics[width=.5\textwidth]{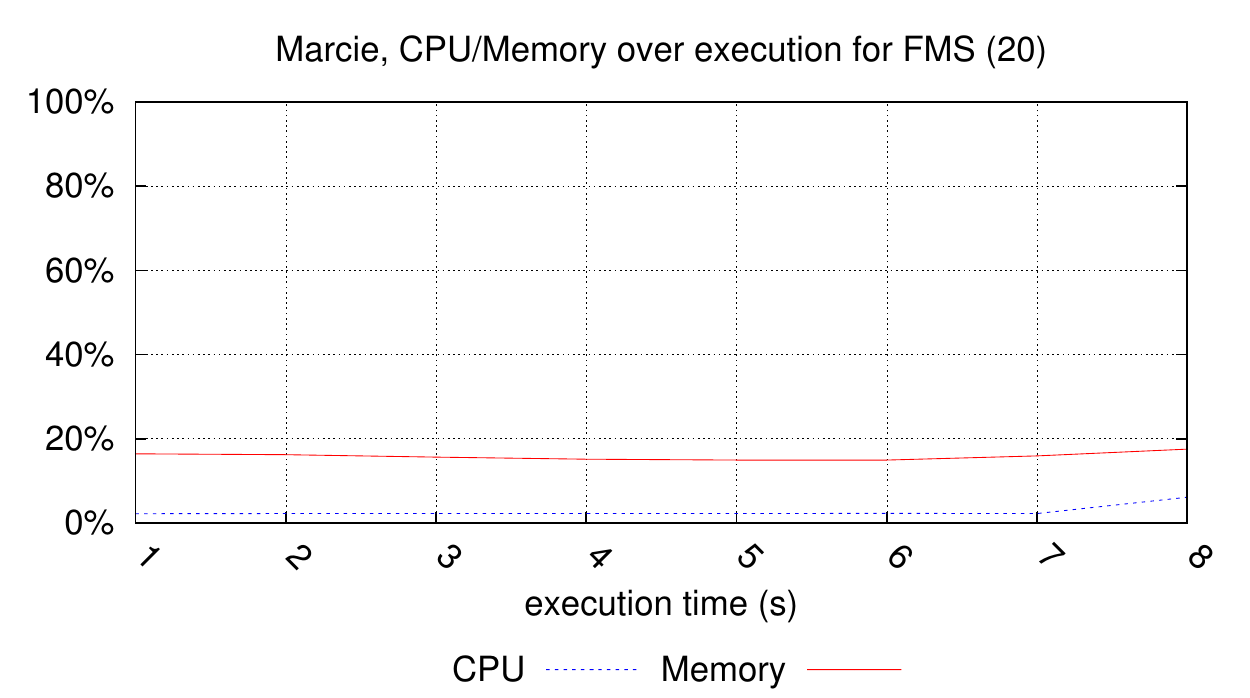}

\noindent\includegraphics[width=.5\textwidth]{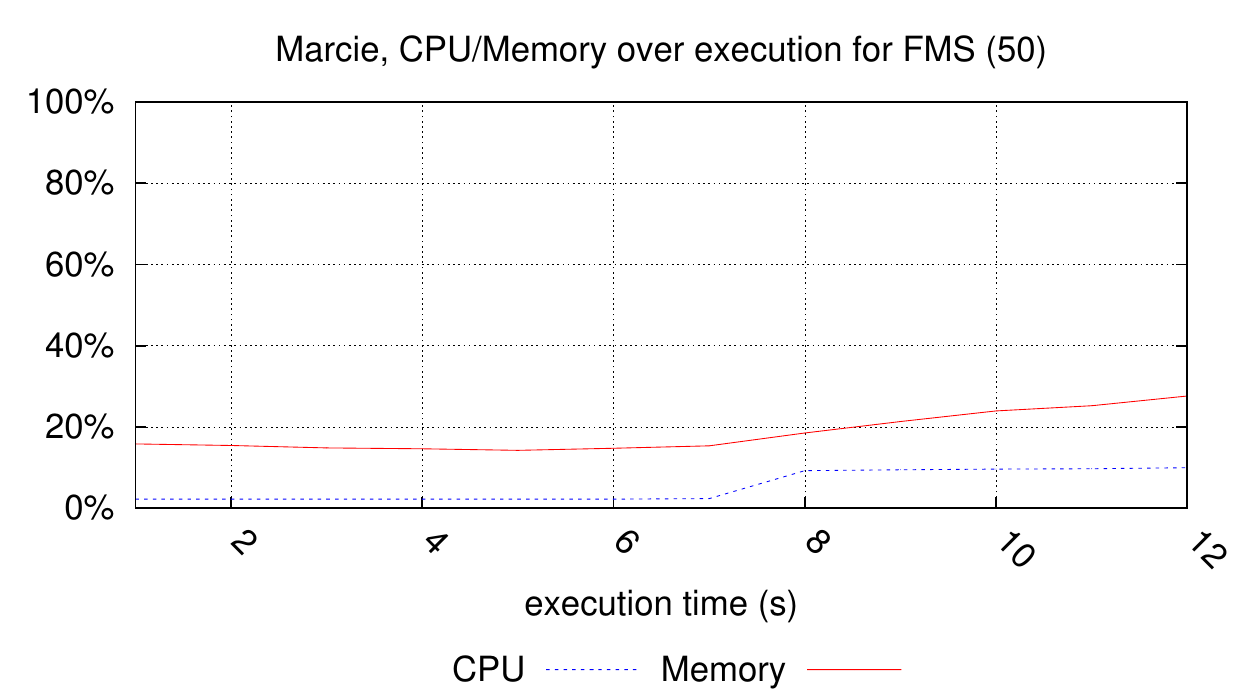}
\includegraphics[width=.5\textwidth]{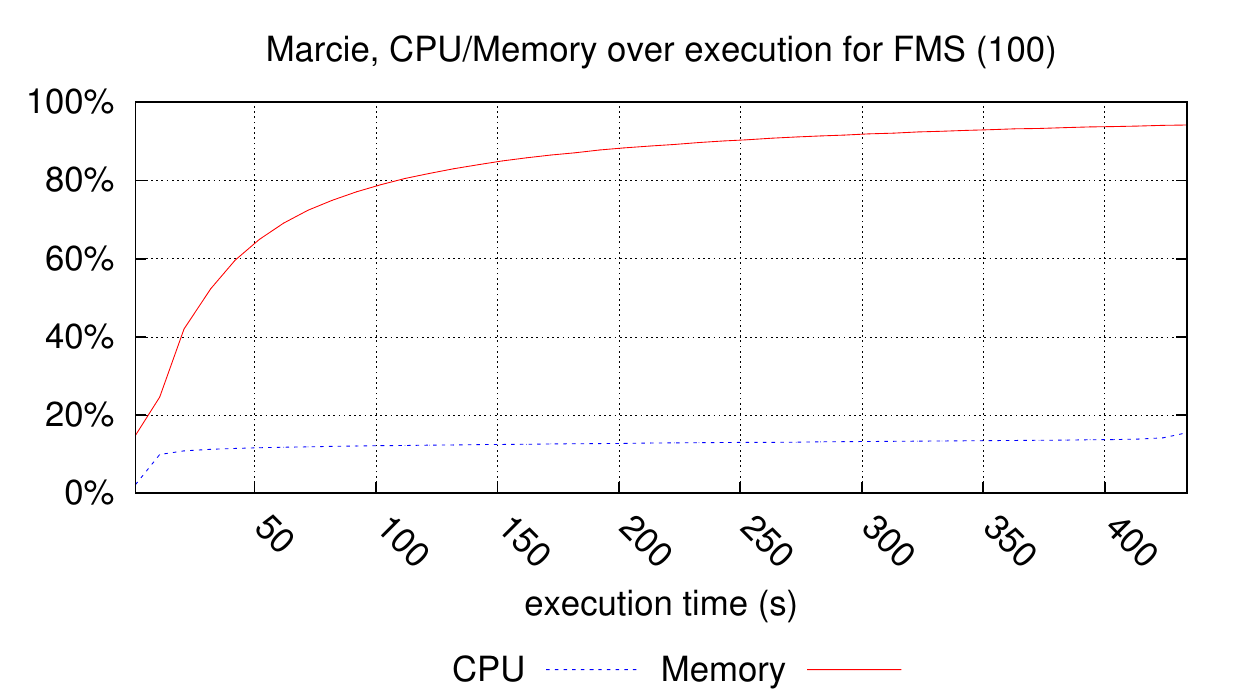}

\noindent\includegraphics[width=.5\textwidth]{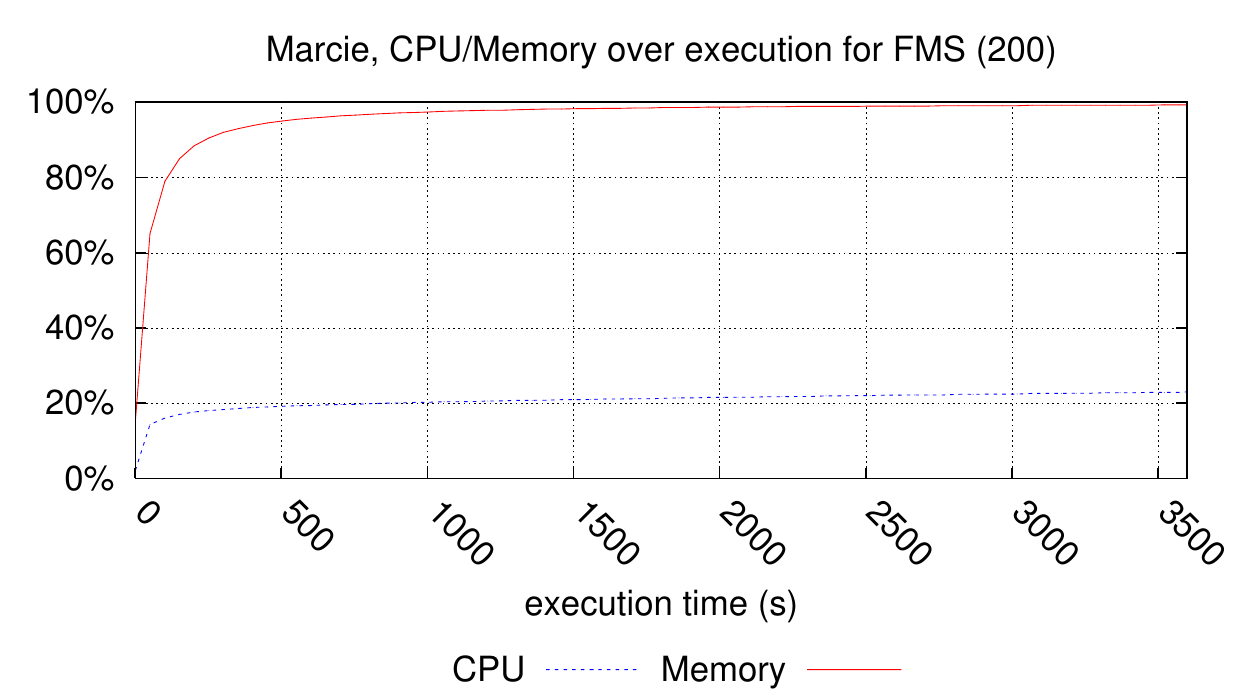}

\subsubsection{Executions for galloc\_res}
2 charts have been generated.
\index{Execution (by tool)!Marcie}
\index{Execution (by model)!galloc\_res!Marcie}

\noindent\includegraphics[width=.5\textwidth]{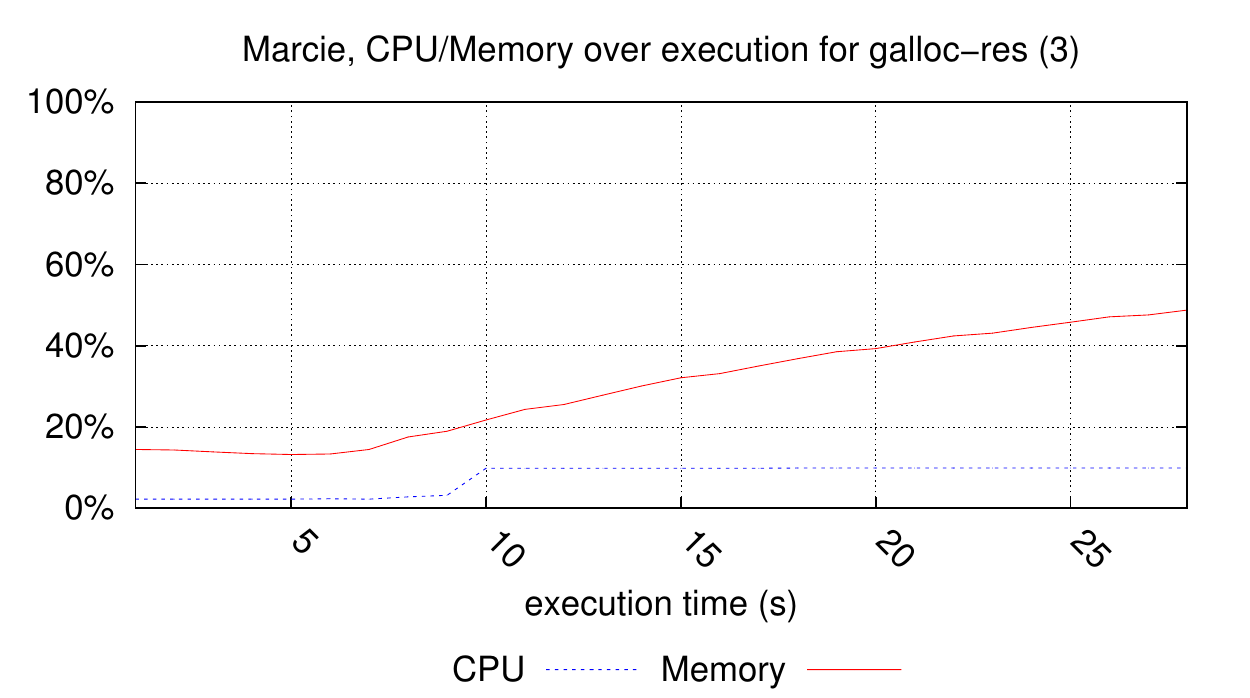}
\includegraphics[width=.5\textwidth]{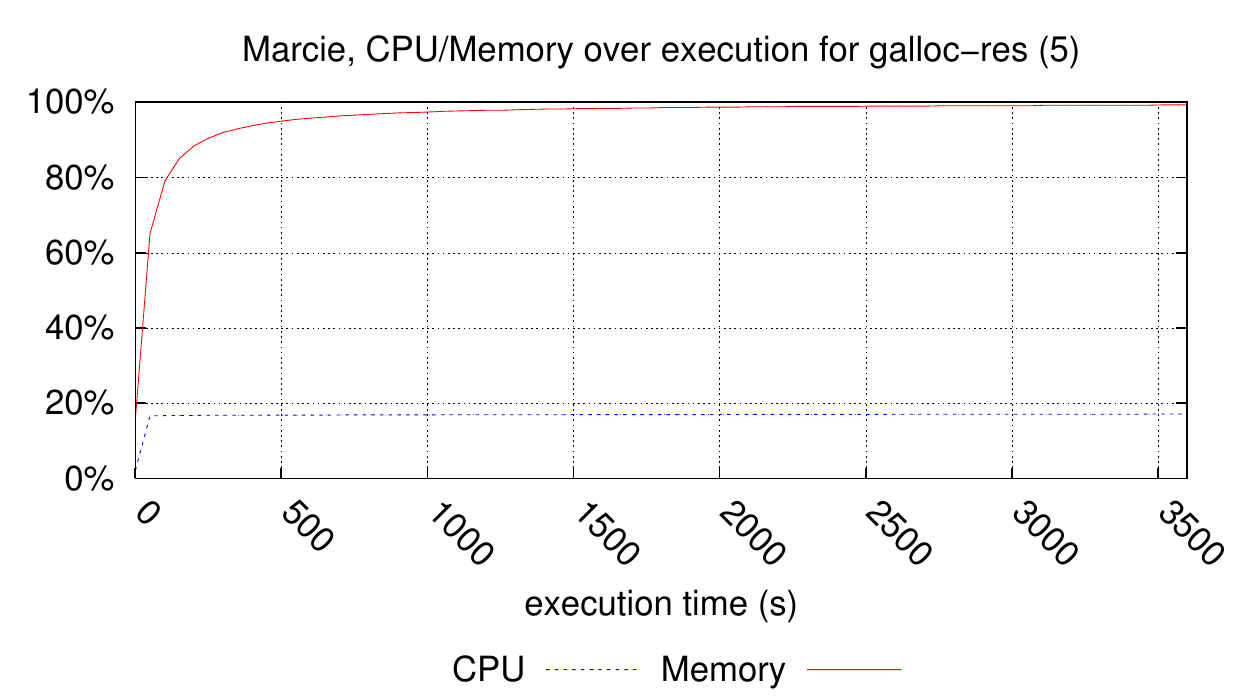}

\subsubsection{Executions for Kanban}
6 charts have been generated.
\index{Execution (by tool)!Marcie}
\index{Execution (by model)!Kanban!Marcie}

\noindent\includegraphics[width=.5\textwidth]{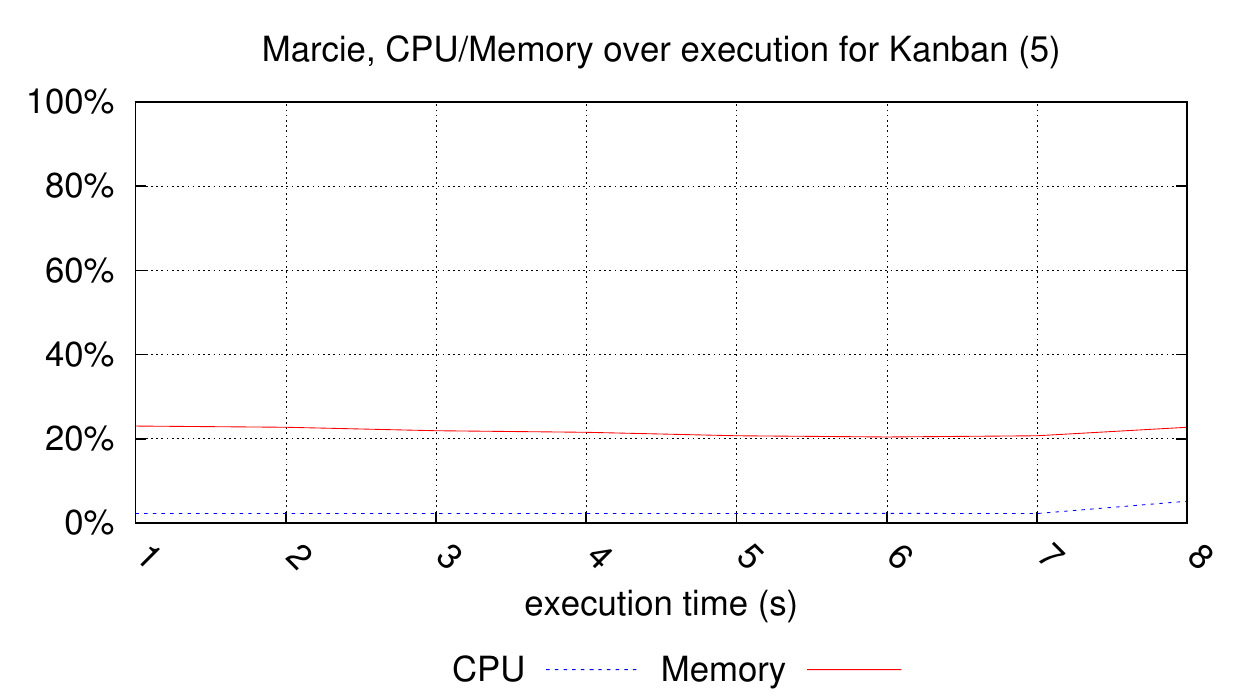}
\includegraphics[width=.5\textwidth]{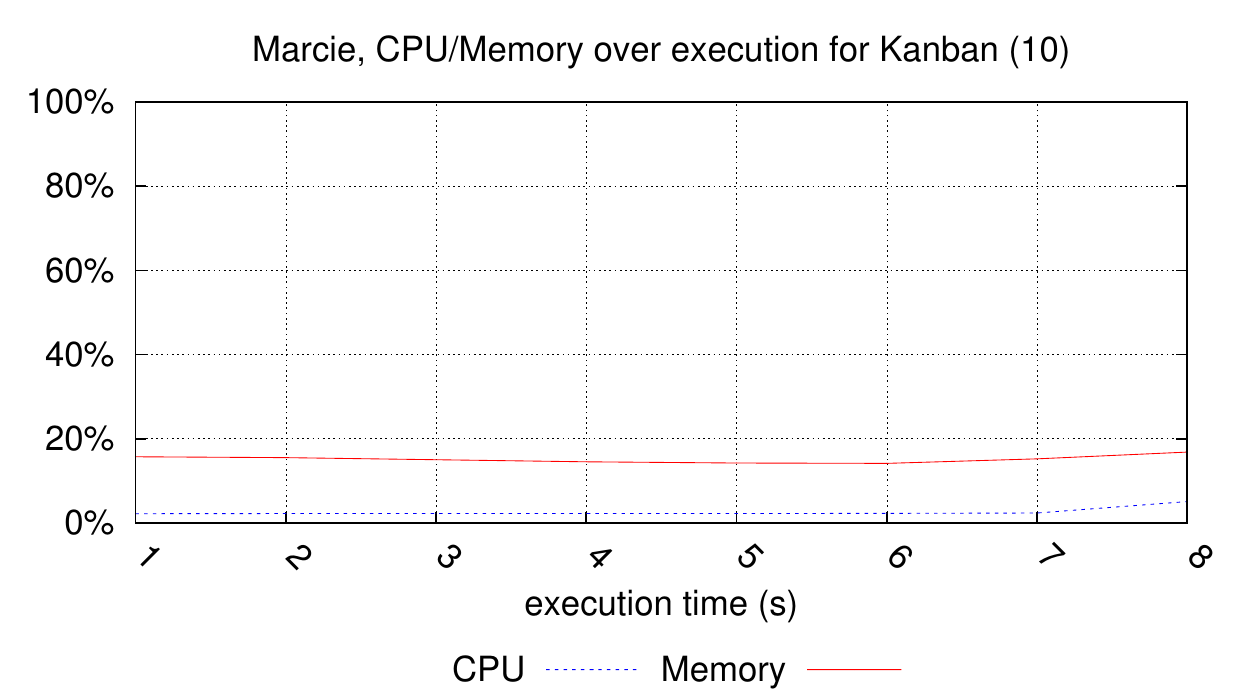}

\noindent\includegraphics[width=.5\textwidth]{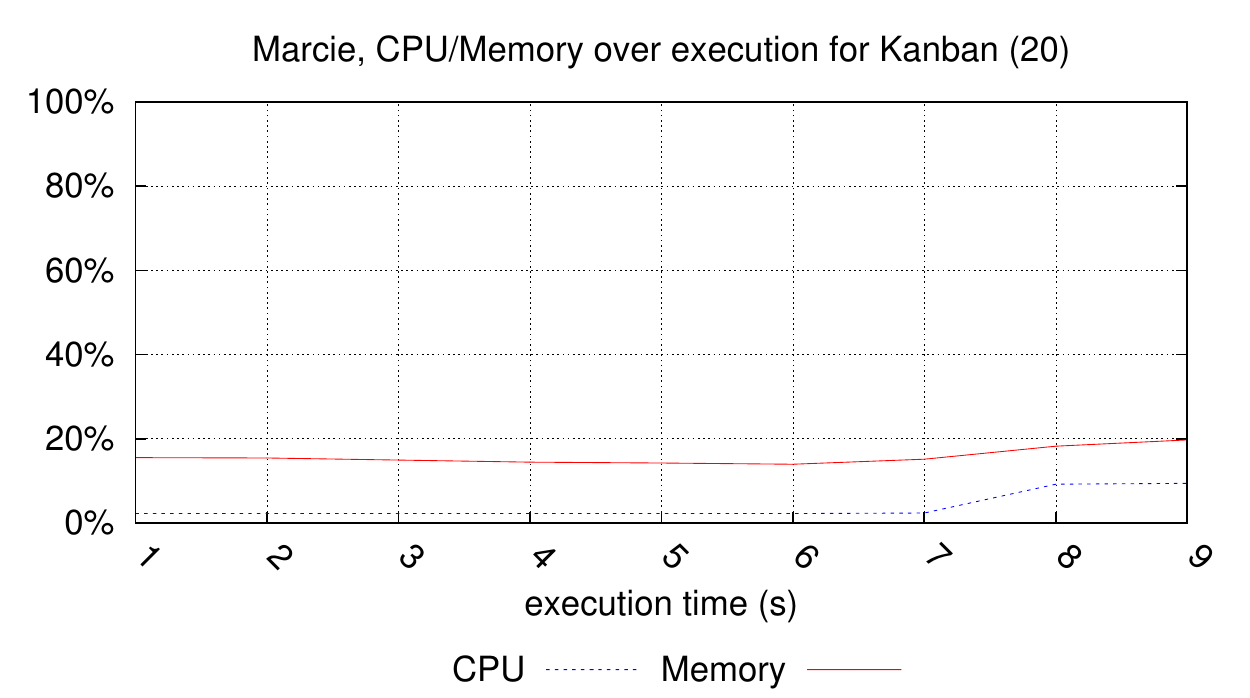}
\includegraphics[width=.5\textwidth]{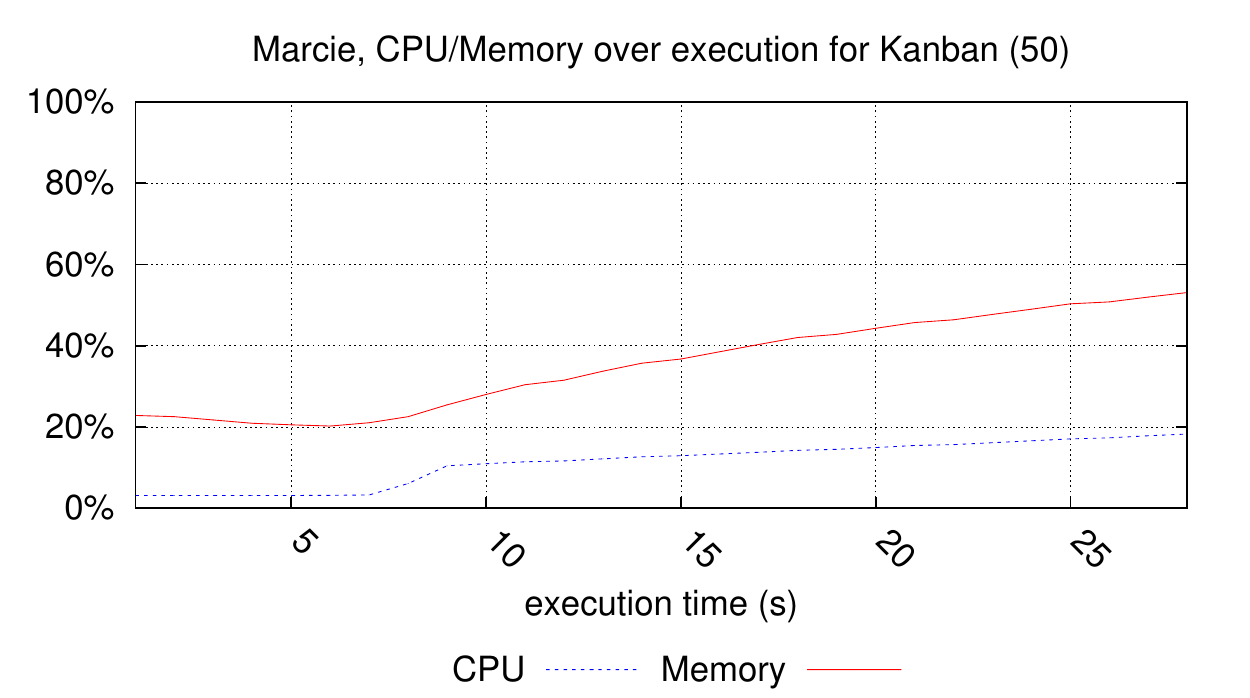}

\noindent\includegraphics[width=.5\textwidth]{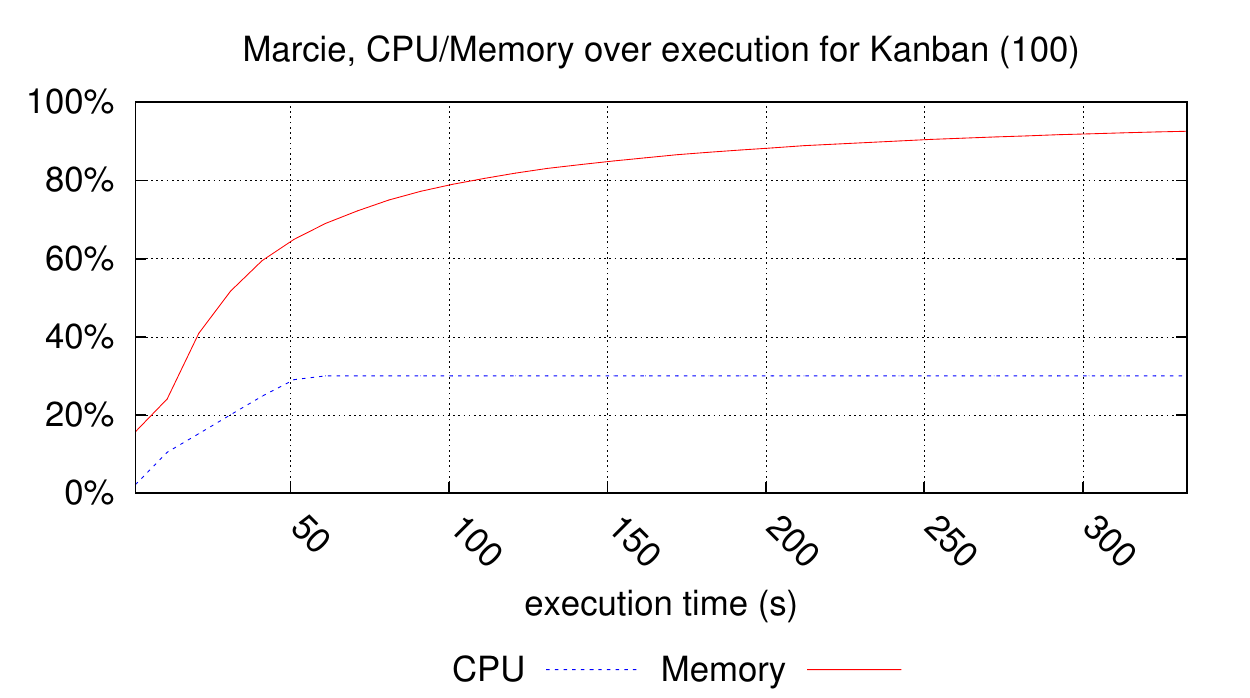}
\includegraphics[width=.5\textwidth]{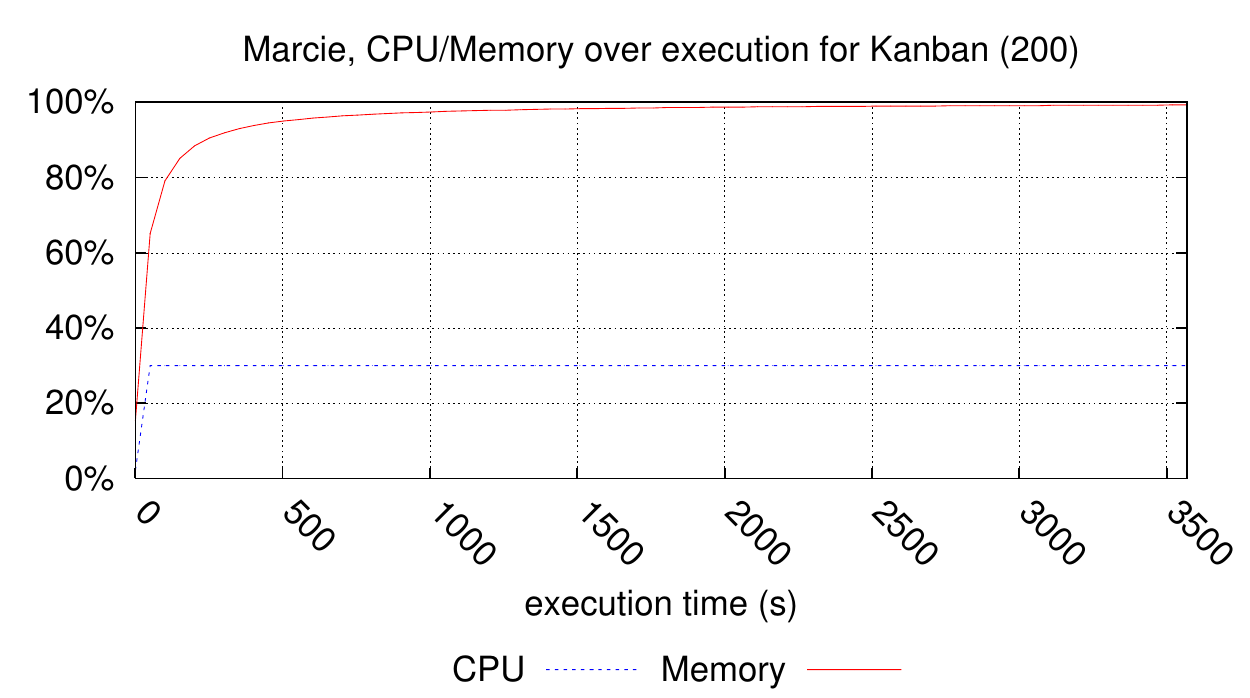}

\vfill\eject
\subsubsection{Executions for lamport\_fmea}
3 charts have been generated.
\index{Execution (by tool)!Marcie}
\index{Execution (by model)!lamport\_fmea!Marcie}

\noindent\includegraphics[width=.5\textwidth]{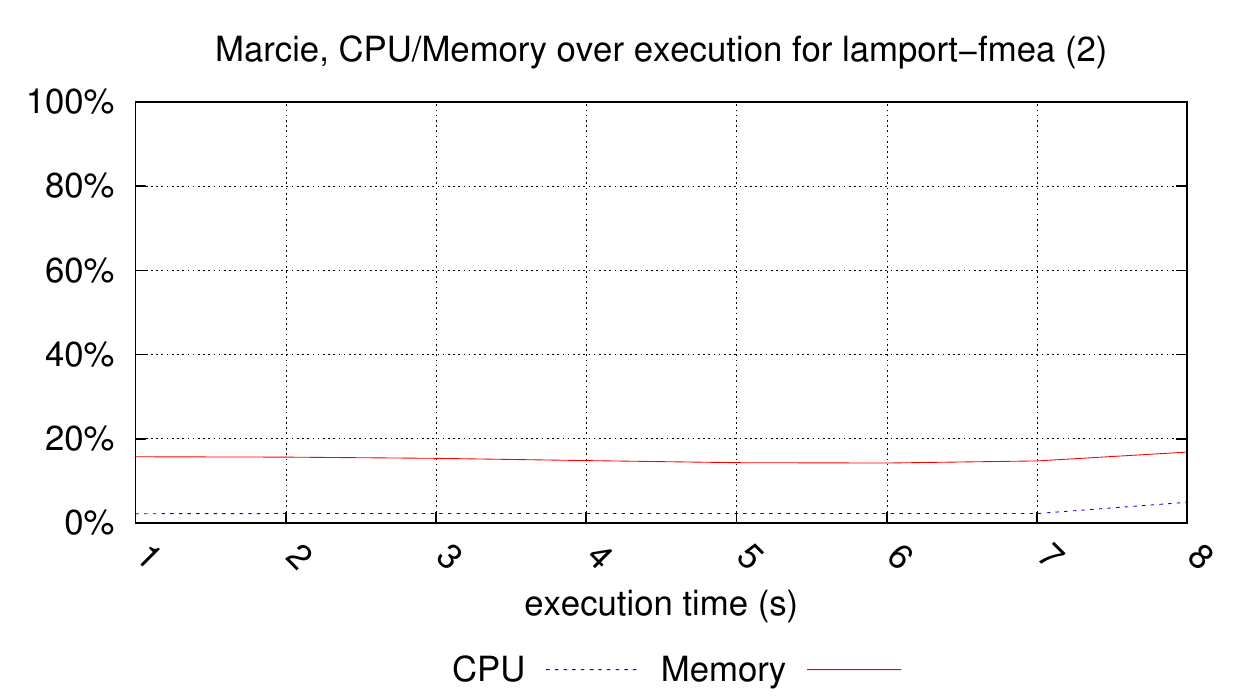}
\includegraphics[width=.5\textwidth]{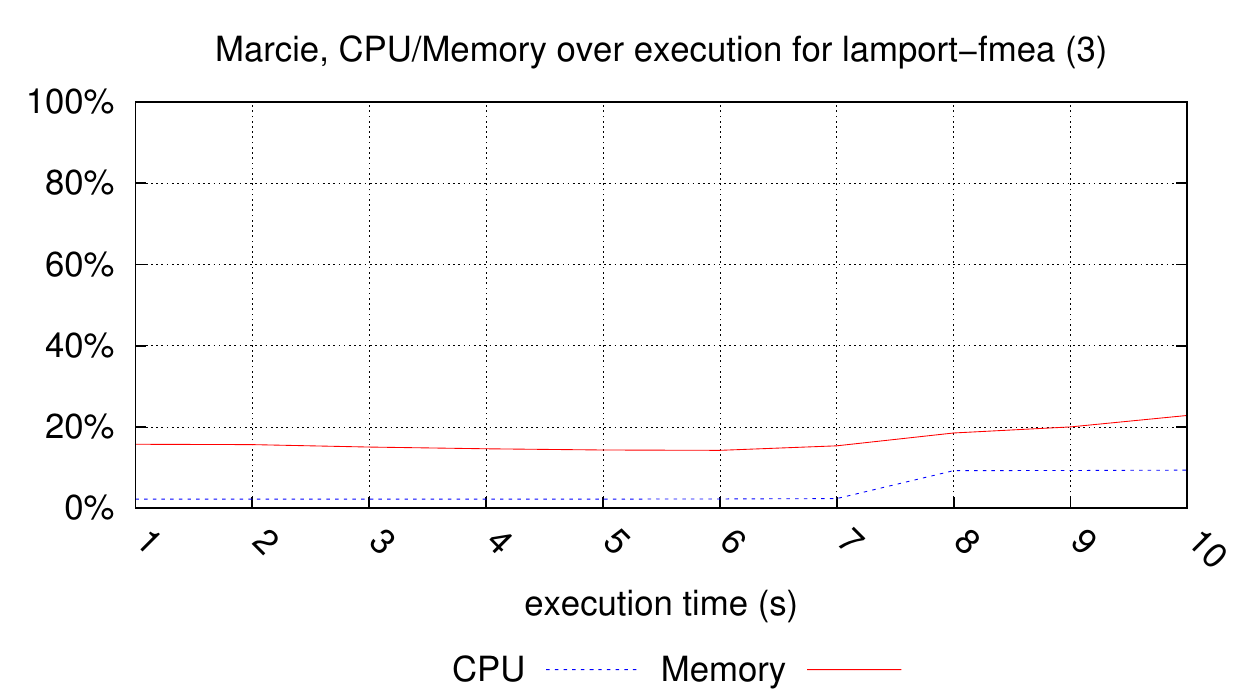}

\noindent\includegraphics[width=.5\textwidth]{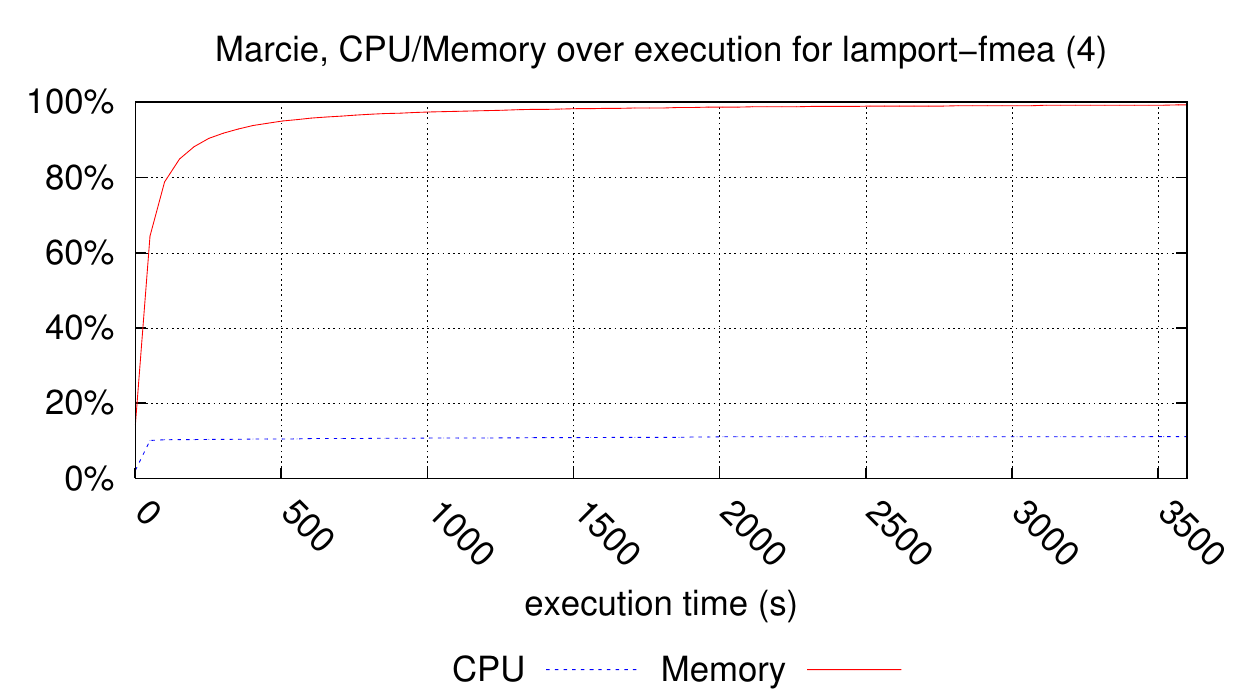}

\subsubsection{Executions for neo-election}
1 chart has been generated.
\index{Execution (by tool)!Marcie}
\index{Execution (by model)!neo-election!Marcie}

\noindent\includegraphics[width=.5\textwidth]{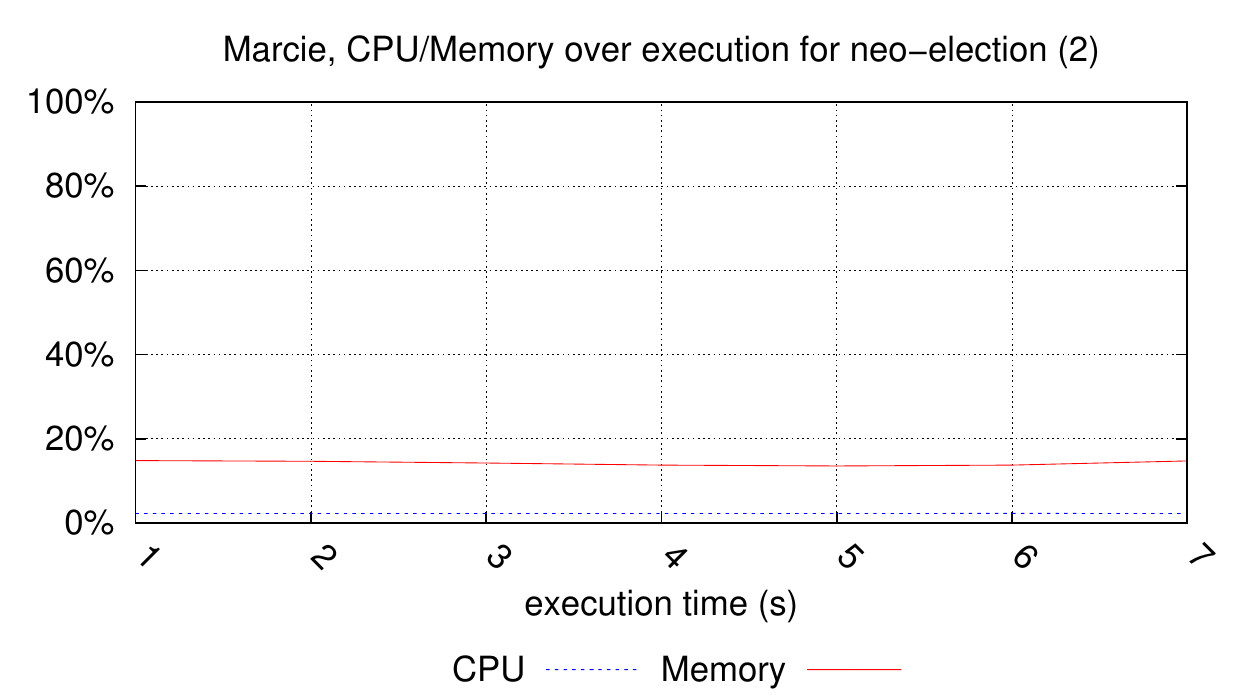}

\subsubsection{Executions for Peterson}
2 charts have been generated.
\index{Execution (by tool)!Marcie}
\index{Execution (by model)!Peterson!Marcie}

\noindent\includegraphics[width=.5\textwidth]{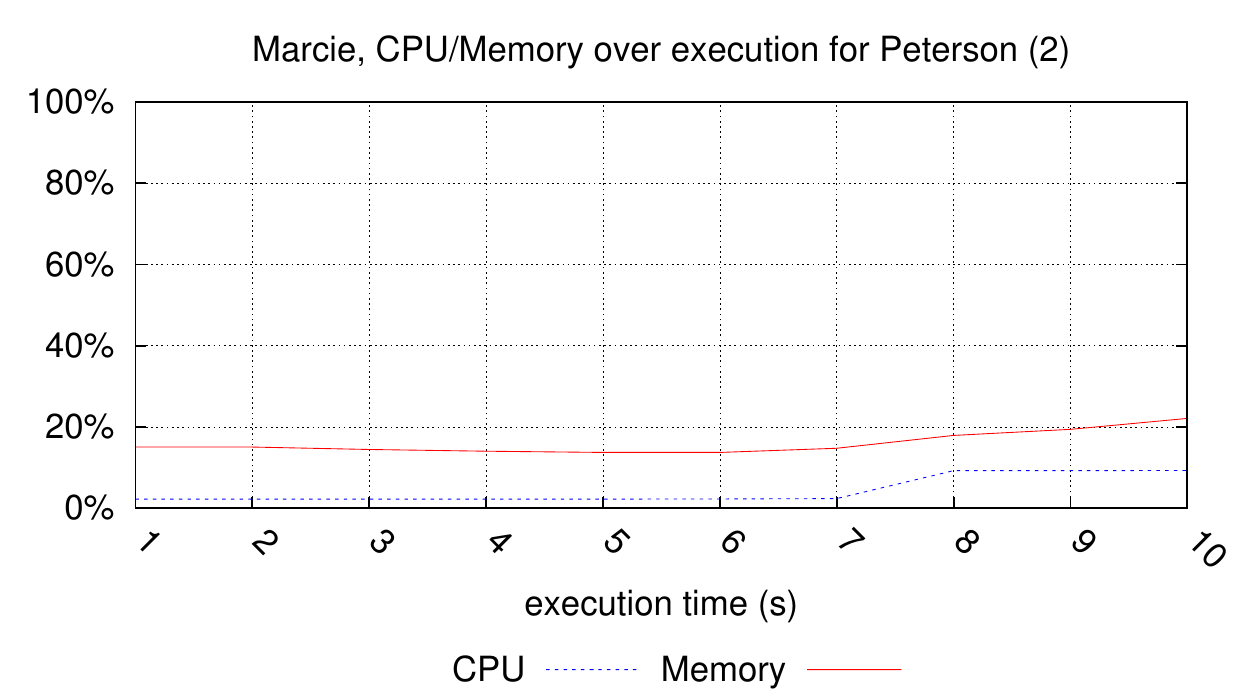}
\includegraphics[width=.5\textwidth]{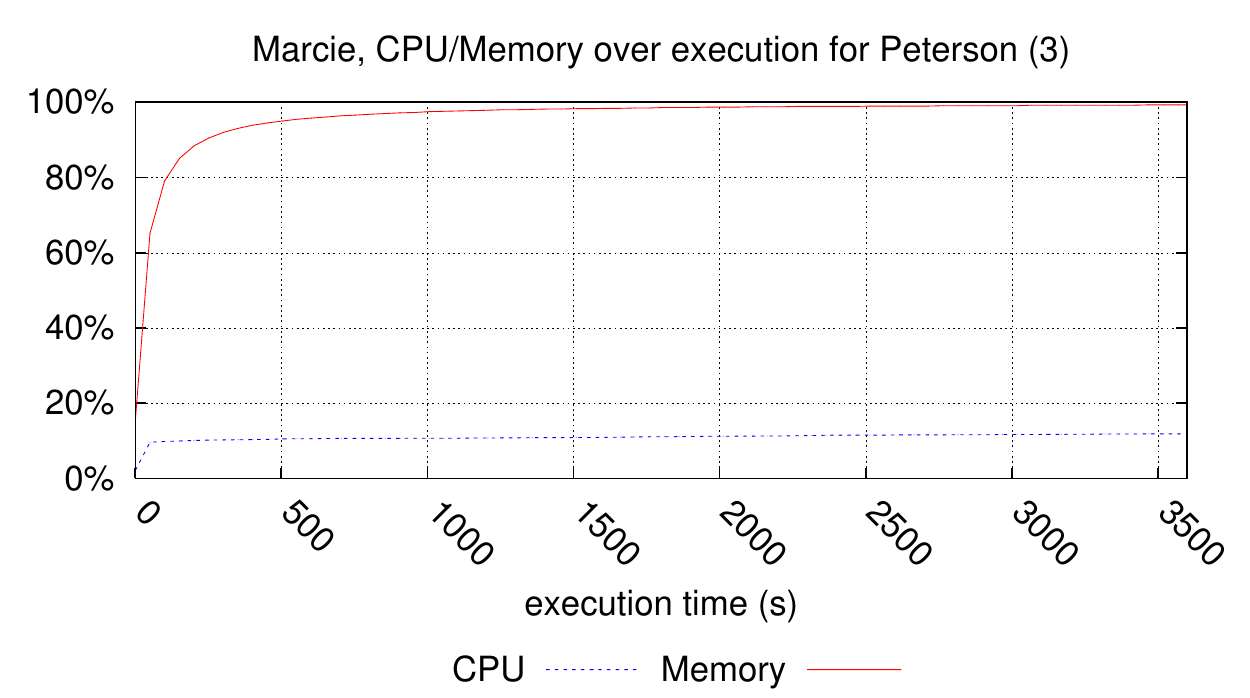}

\vfill\eject
\subsubsection{Executions for philo\_dyn}
3 charts have been generated.
\index{Execution (by tool)!Marcie}
\index{Execution (by model)!philo\_dyn!Marcie}

\noindent\includegraphics[width=.5\textwidth]{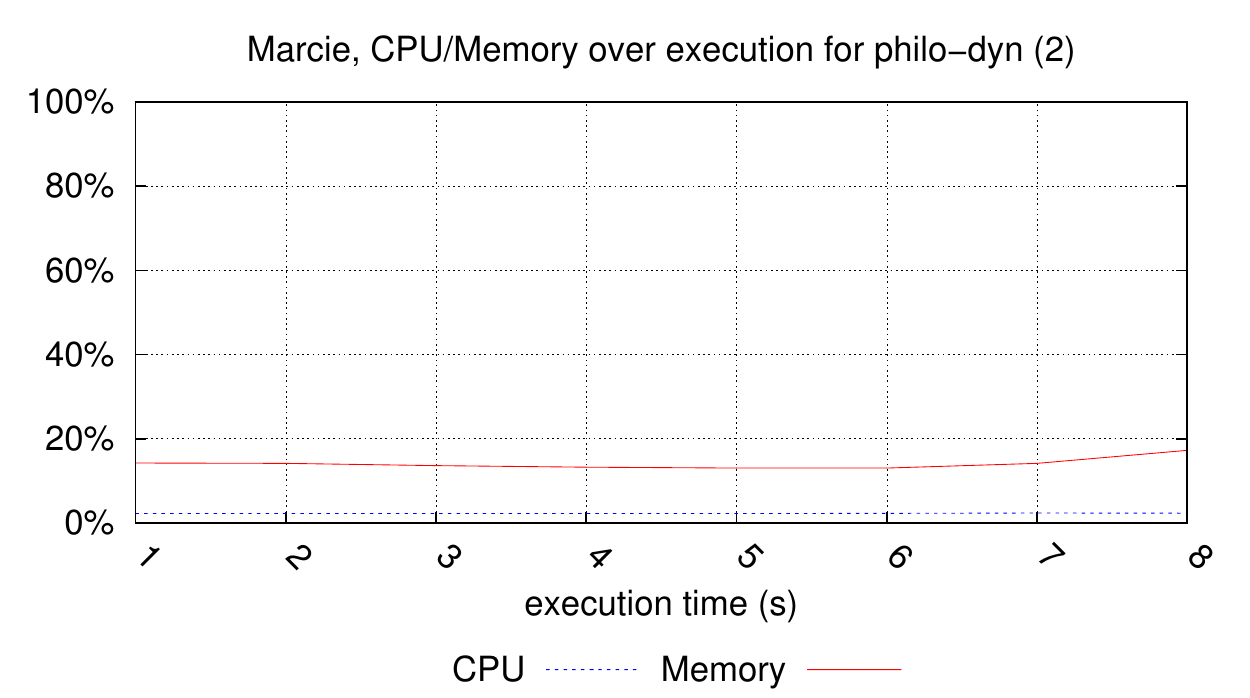}
\includegraphics[width=.5\textwidth]{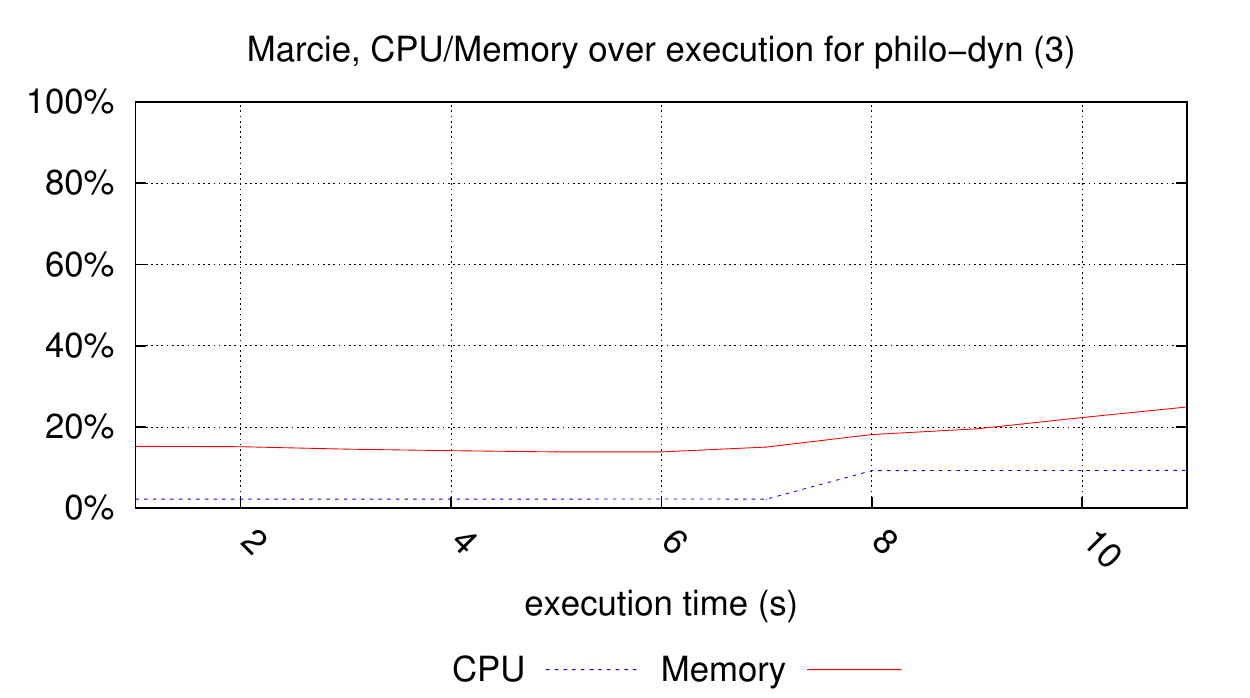}

\noindent\includegraphics[width=.5\textwidth]{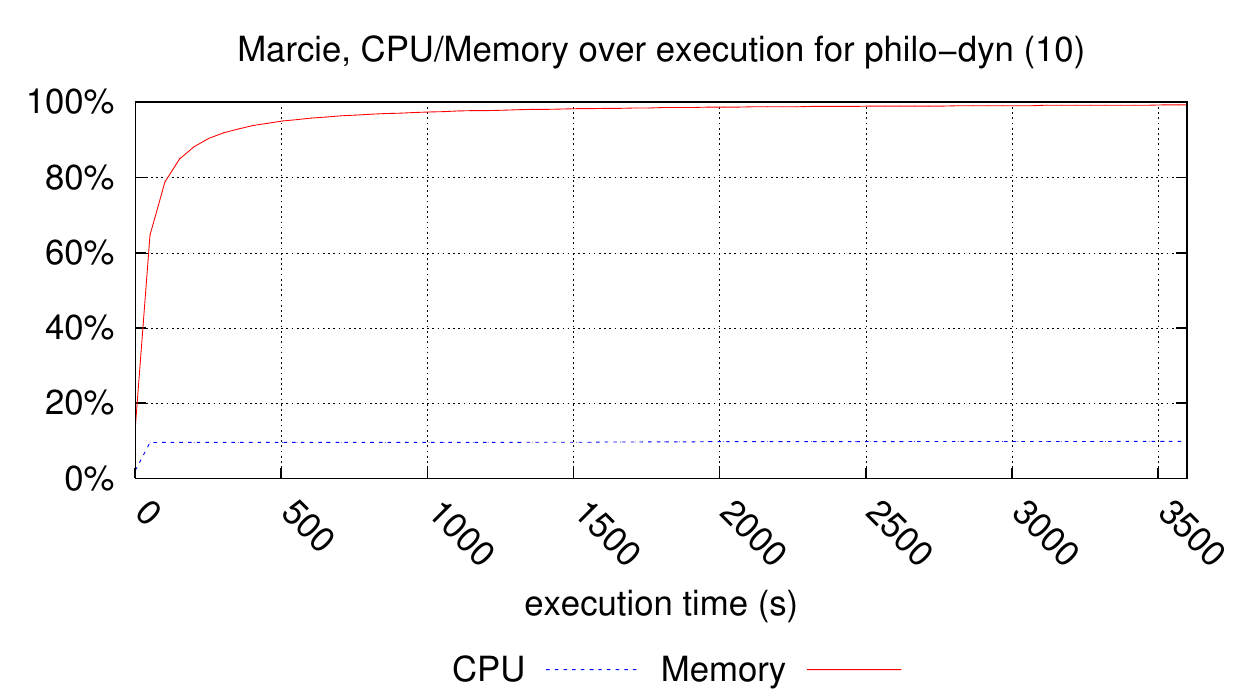}

\subsubsection{Executions for Philosophers}
10 charts have been generated.
\index{Execution (by tool)!Marcie}
\index{Execution (by model)!Philosophers!Marcie}

\noindent\includegraphics[width=.5\textwidth]{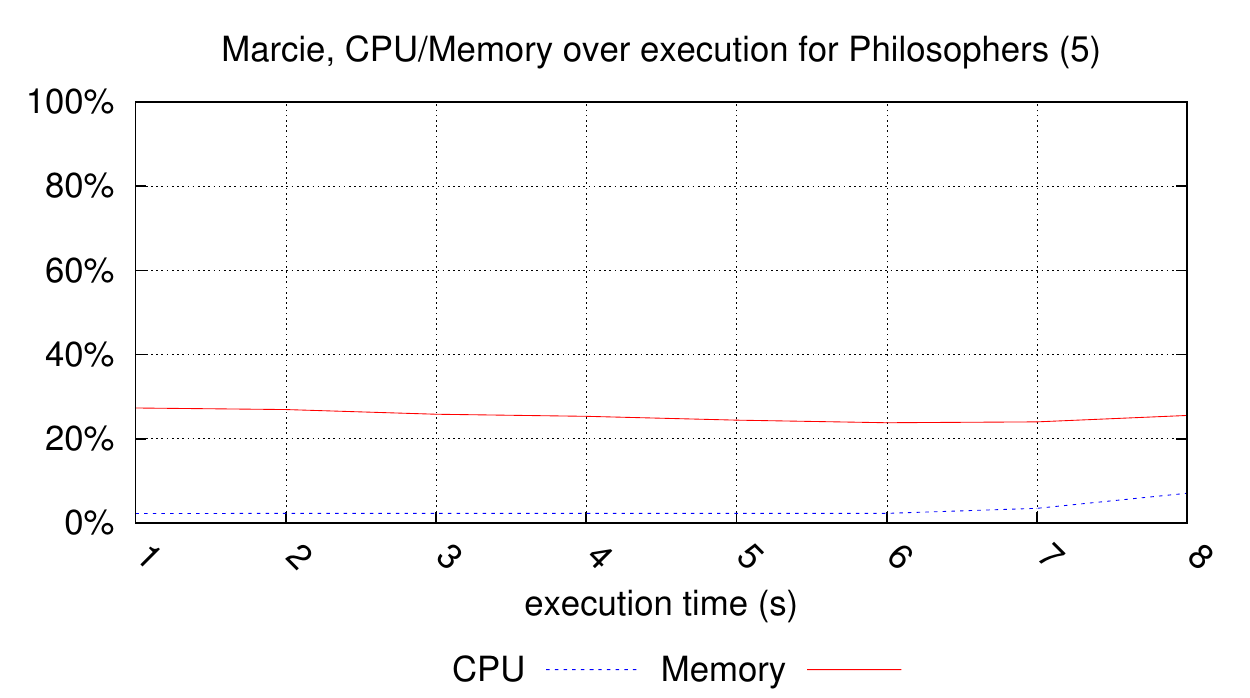}
\includegraphics[width=.5\textwidth]{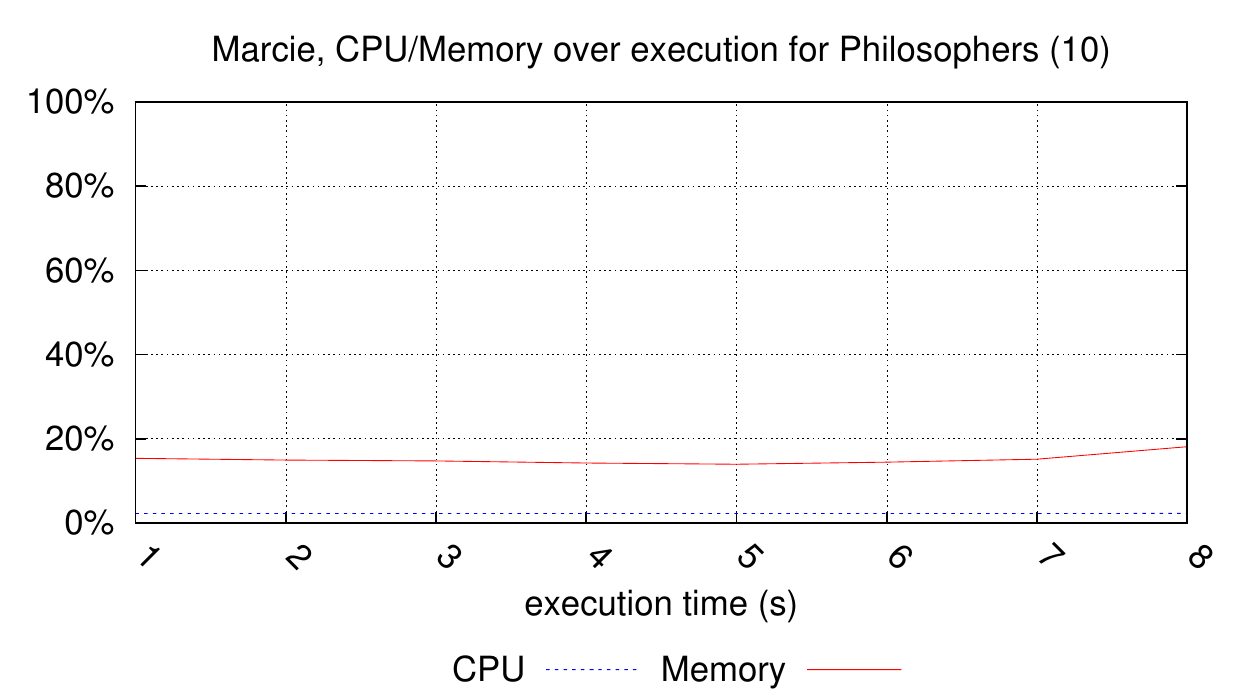}

\noindent\includegraphics[width=.5\textwidth]{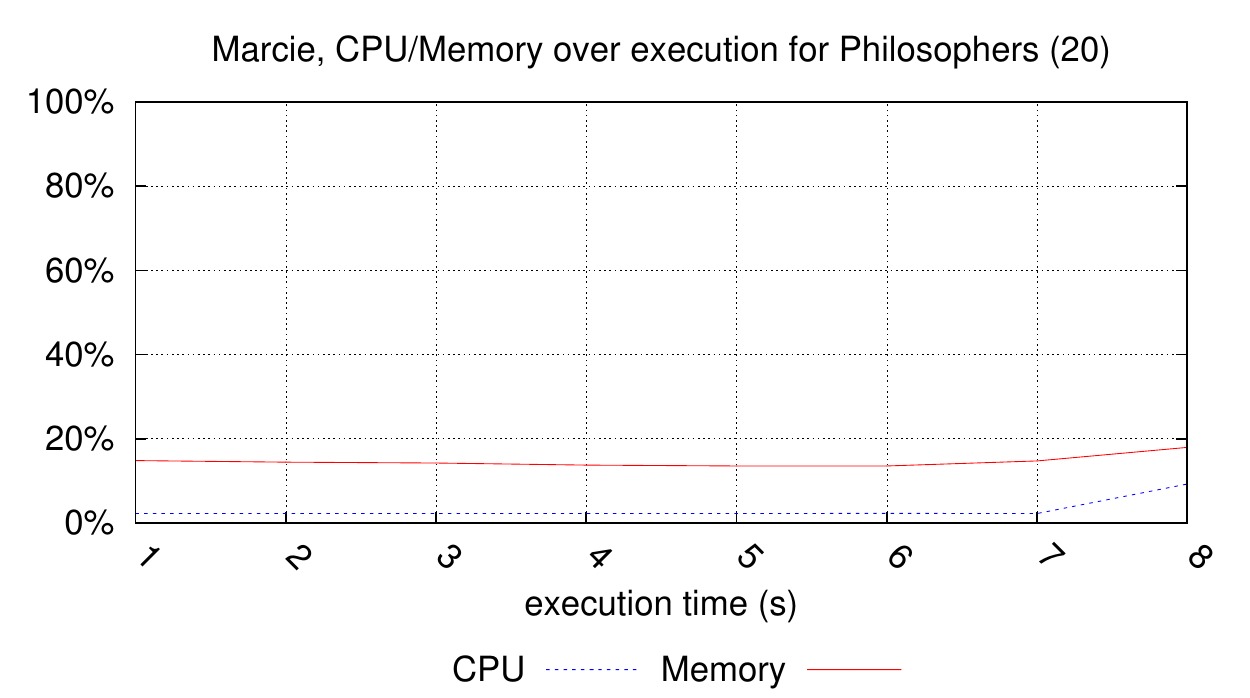}
\includegraphics[width=.5\textwidth]{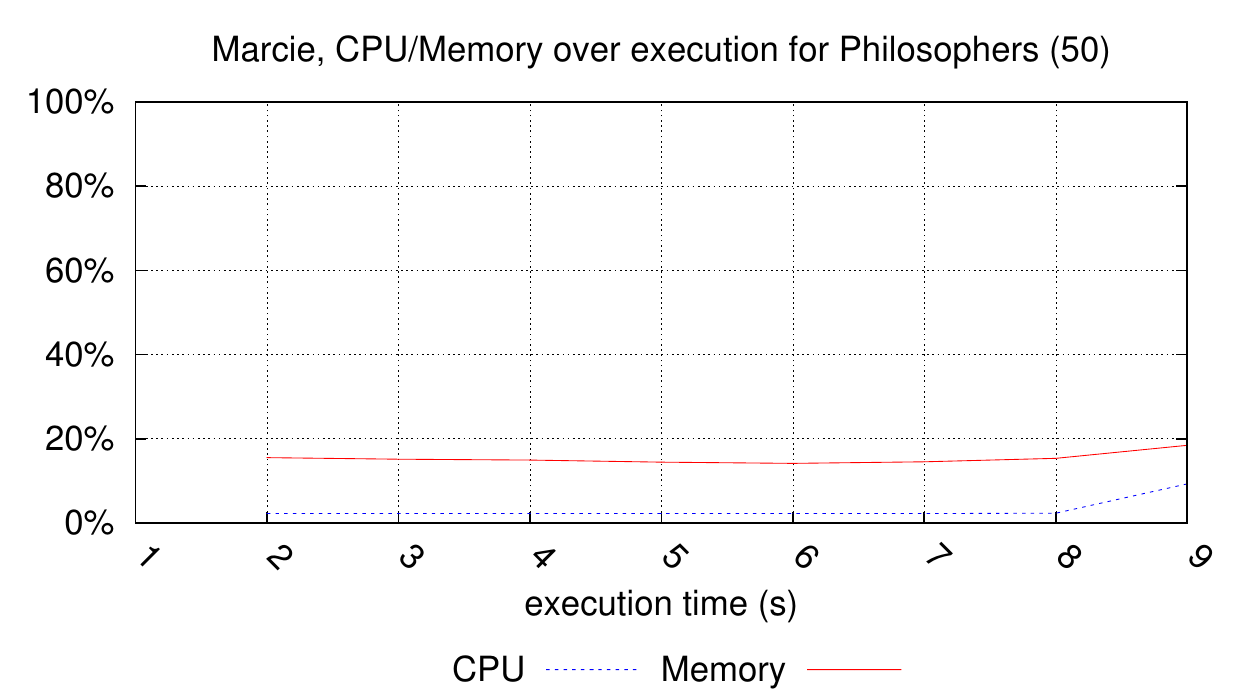}

\noindent\includegraphics[width=.5\textwidth]{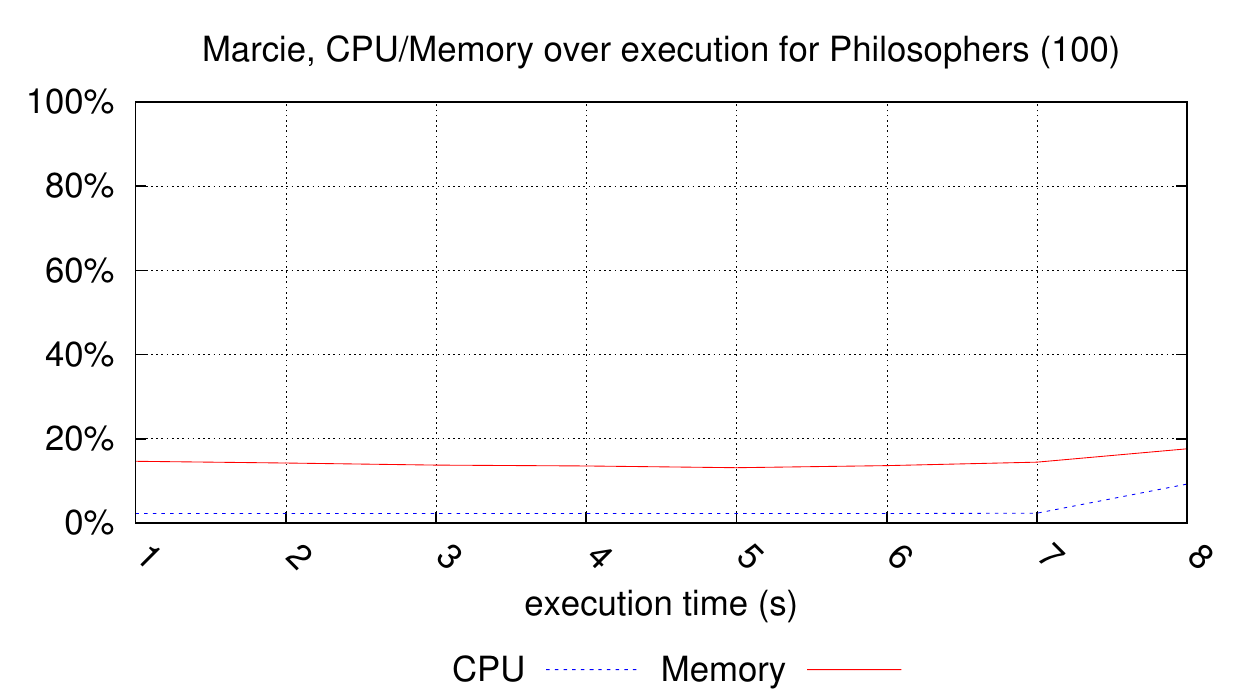}
\includegraphics[width=.5\textwidth]{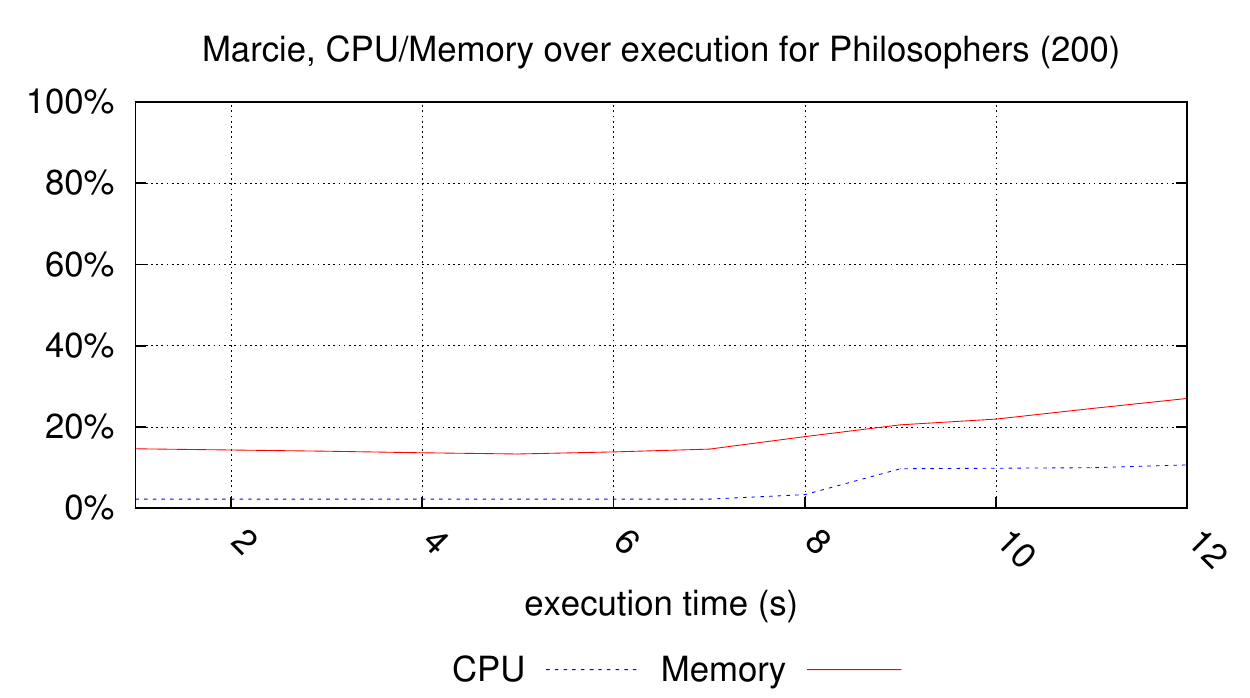}

\noindent\includegraphics[width=.5\textwidth]{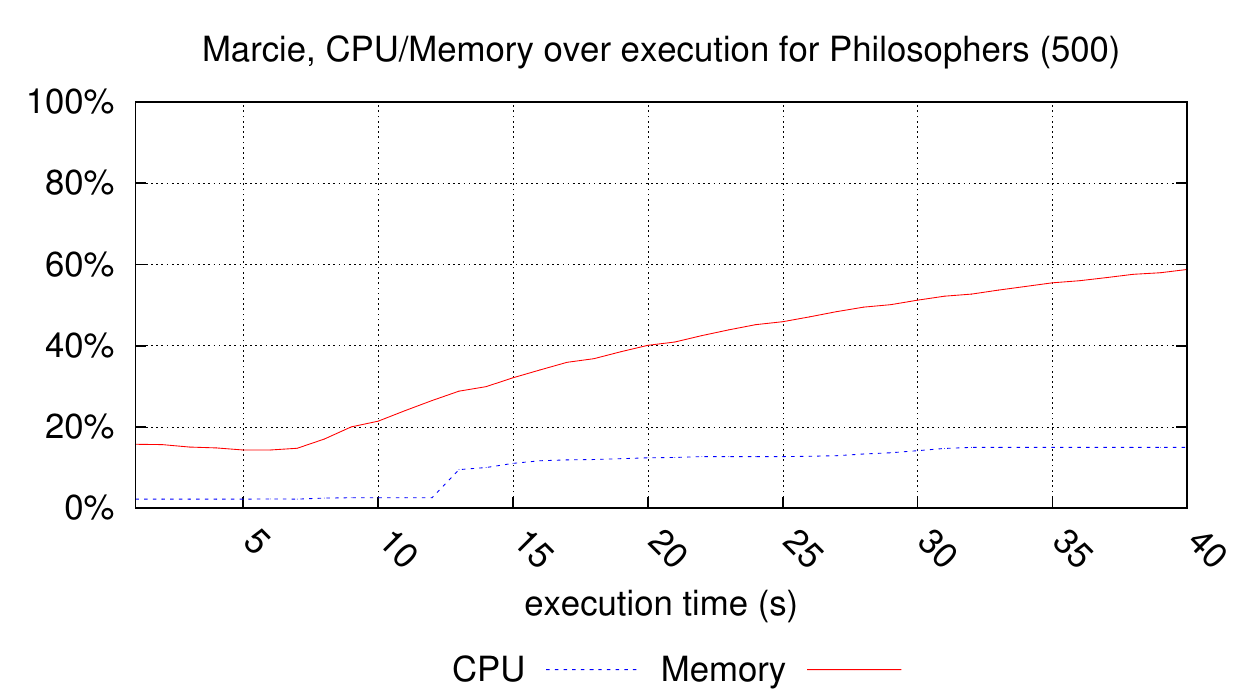}
\includegraphics[width=.5\textwidth]{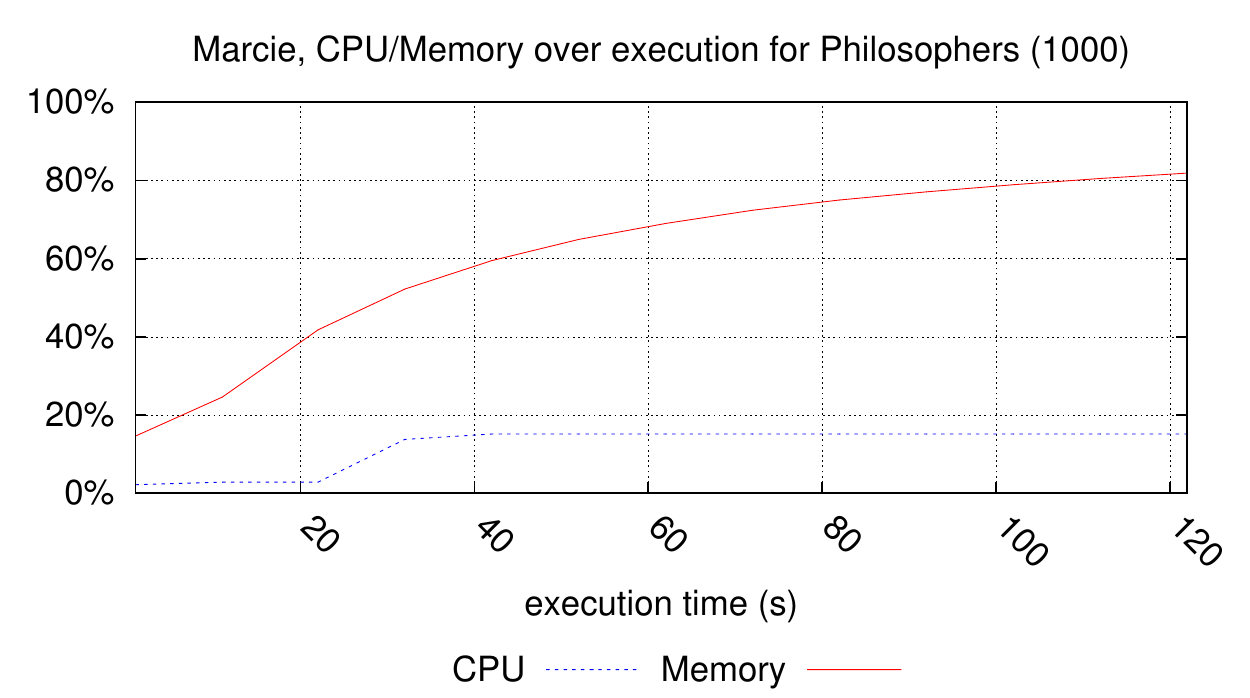}

\noindent\includegraphics[width=.5\textwidth]{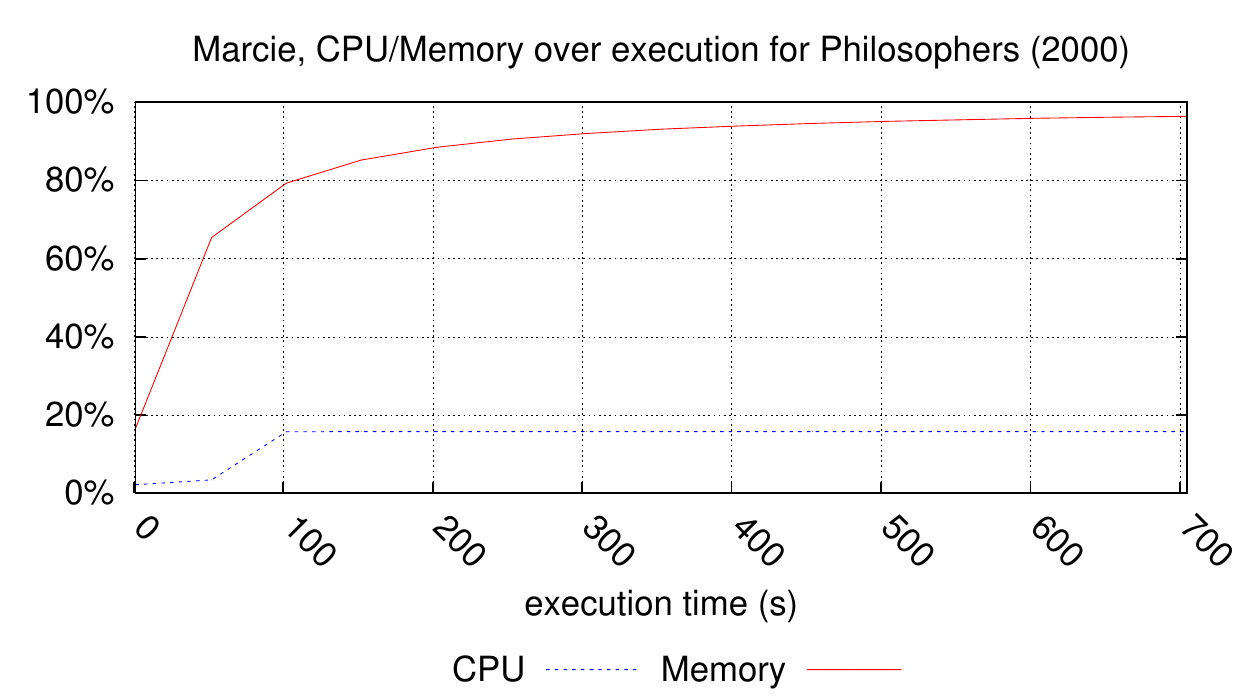}
\includegraphics[width=.5\textwidth]{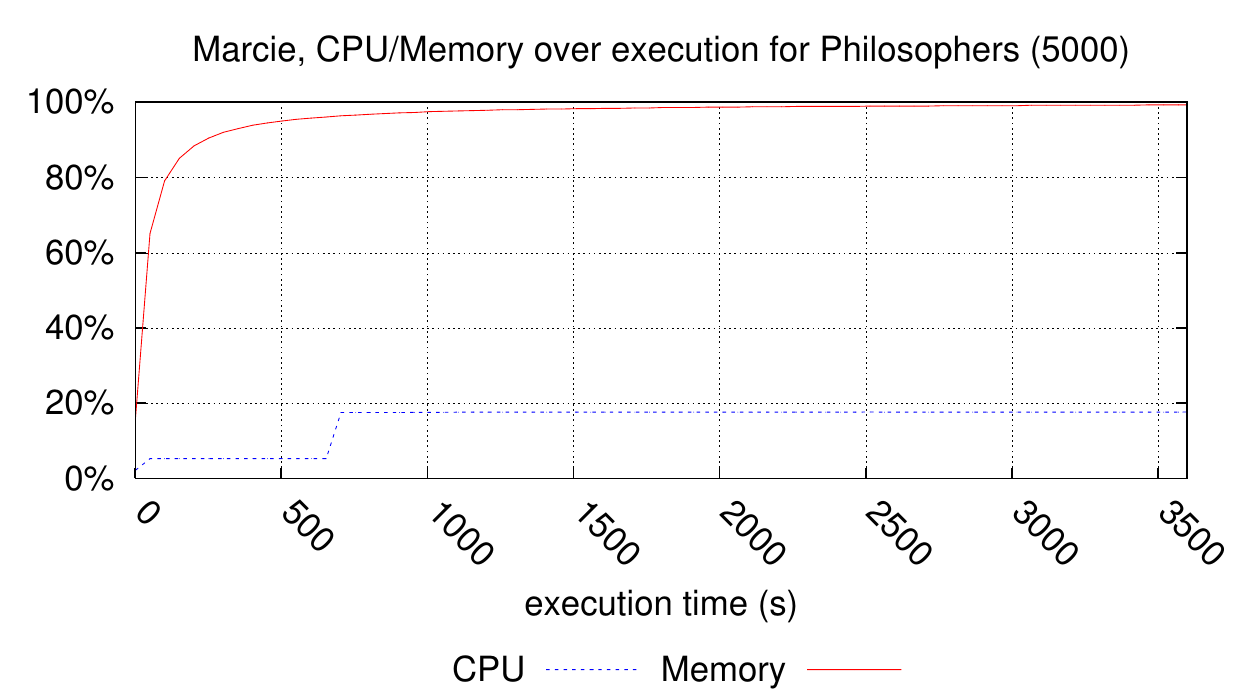}

\subsubsection{Executions for railroad}
3 charts have been generated.
\index{Execution (by tool)!Marcie}
\index{Execution (by model)!railroad!Marcie}

\noindent\includegraphics[width=.5\textwidth]{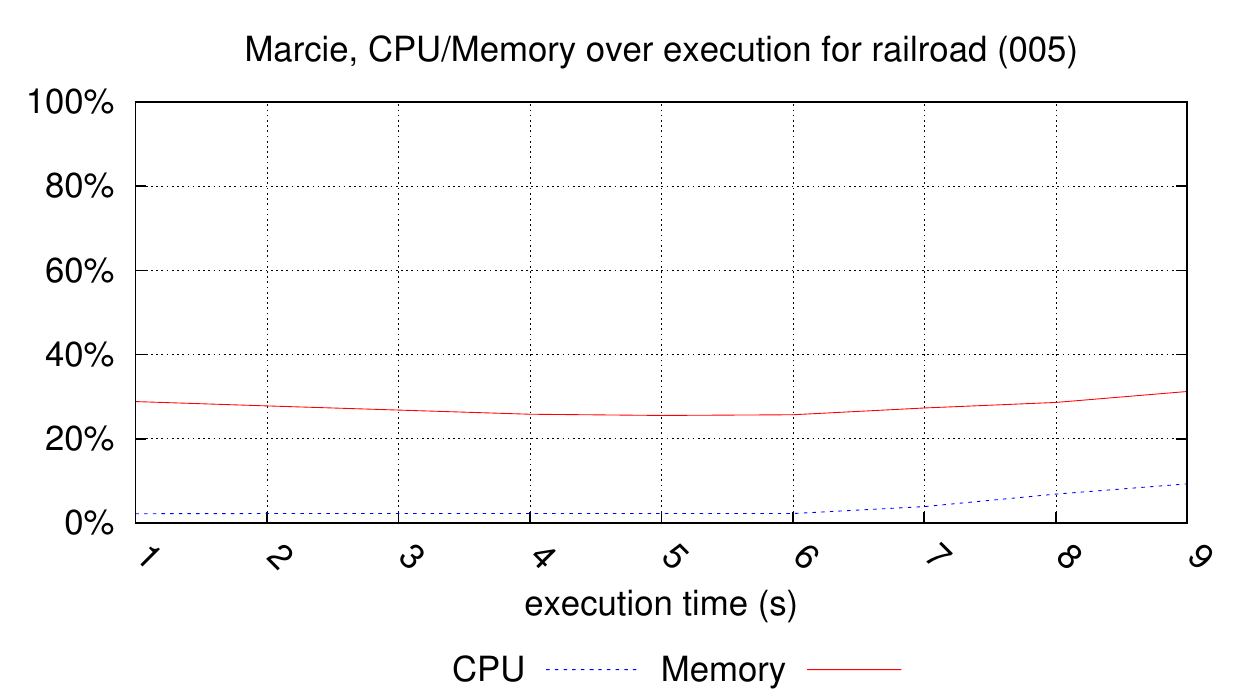}
\includegraphics[width=.5\textwidth]{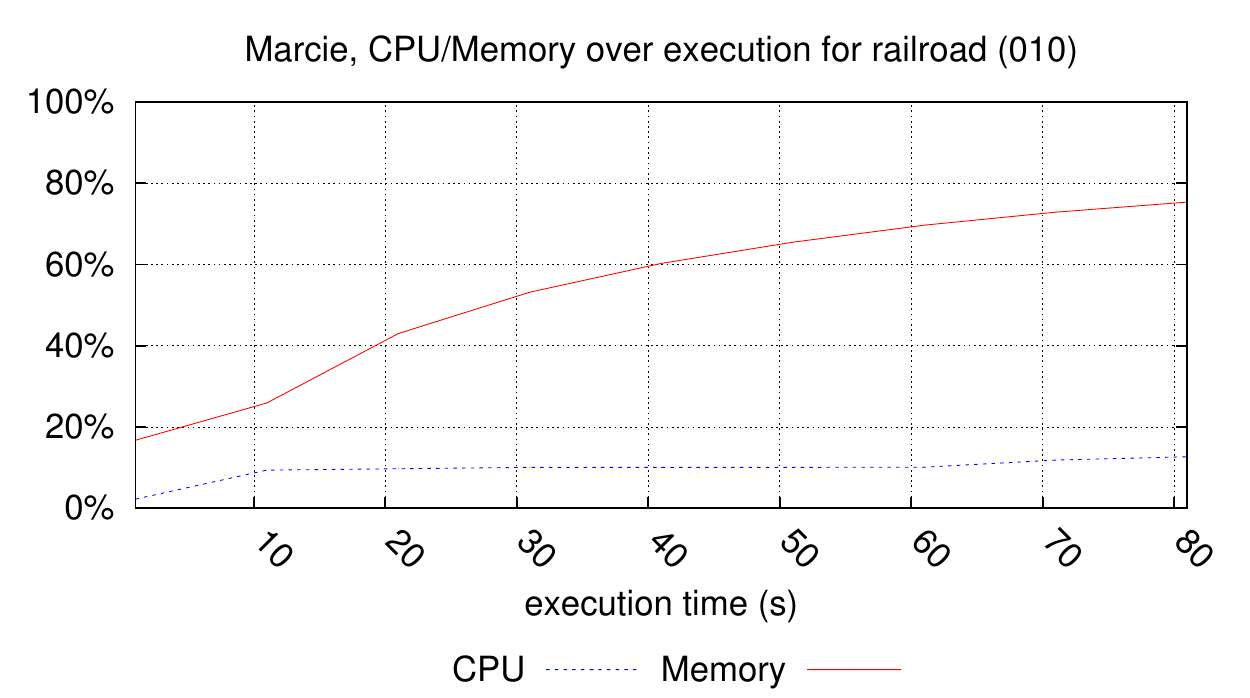}

\noindent\includegraphics[width=.5\textwidth]{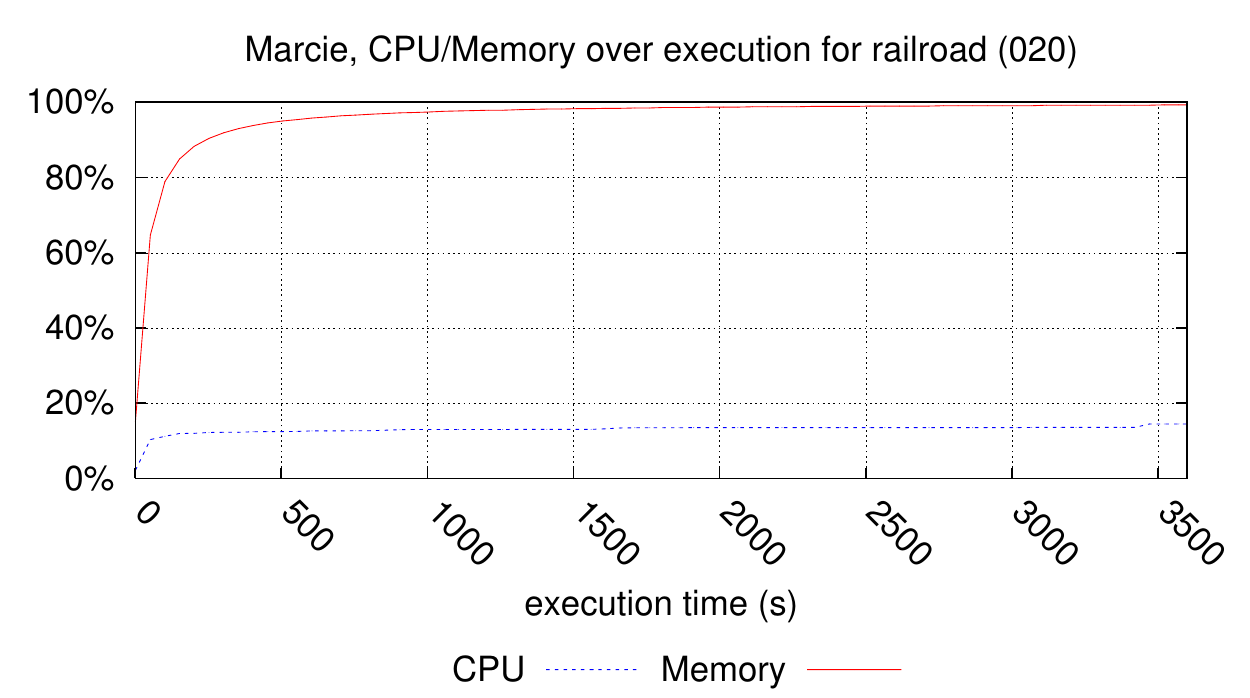}

\subsubsection{Executions for ring}
1 chart has been generated.
\index{Execution (by tool)!Marcie}
\index{Execution (by model)!ring!Marcie}

\noindent\includegraphics[width=.5\textwidth]{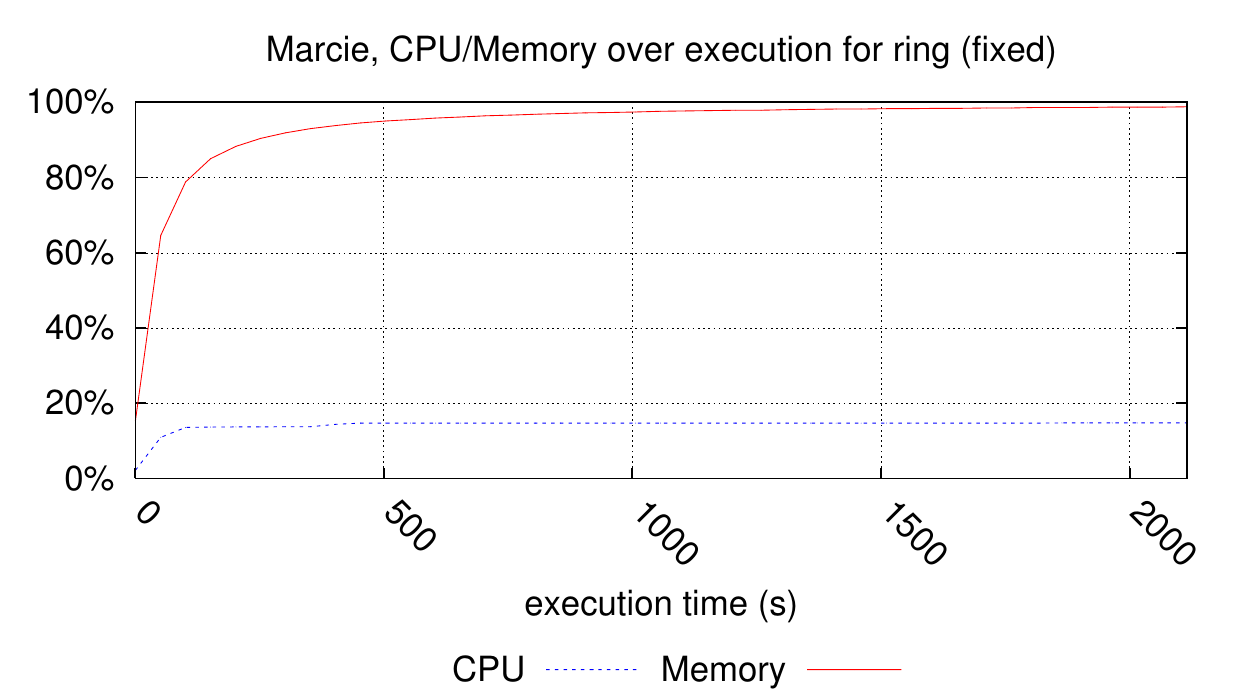}

\subsubsection{Executions for rw\_mutex}
5 charts have been generated.
\index{Execution (by tool)!Marcie}
\index{Execution (by model)!rw\_mutex!Marcie}

\noindent\includegraphics[width=.5\textwidth]{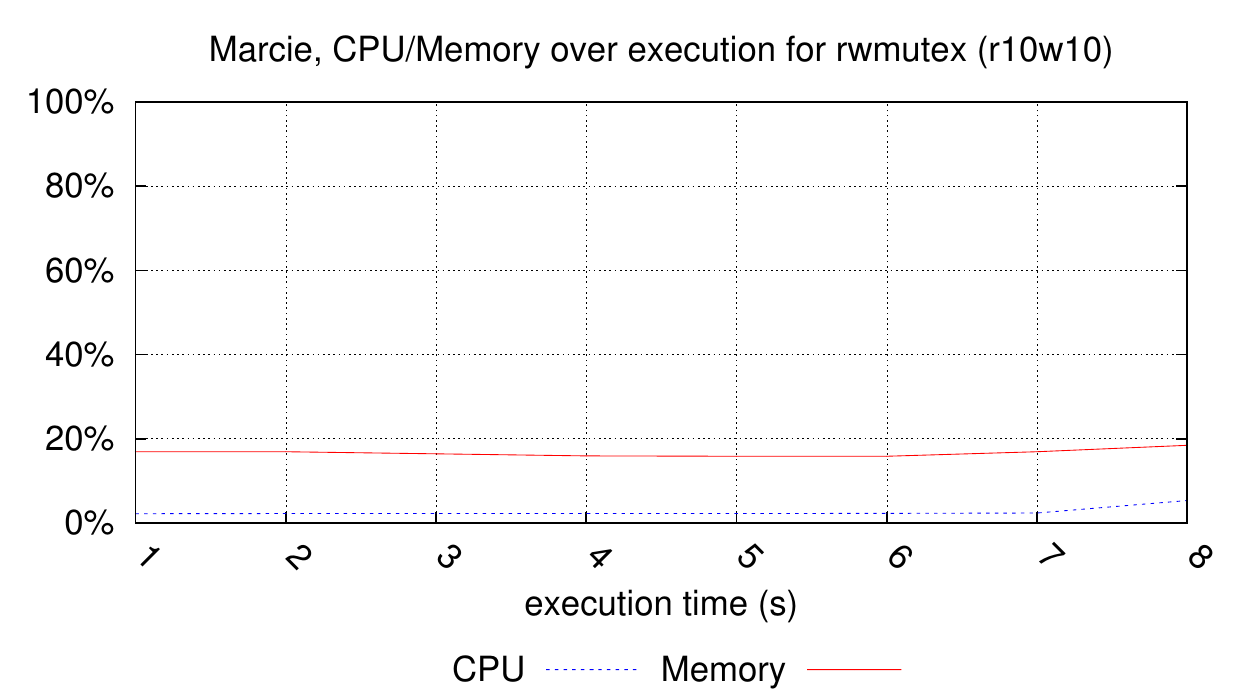}
\includegraphics[width=.5\textwidth]{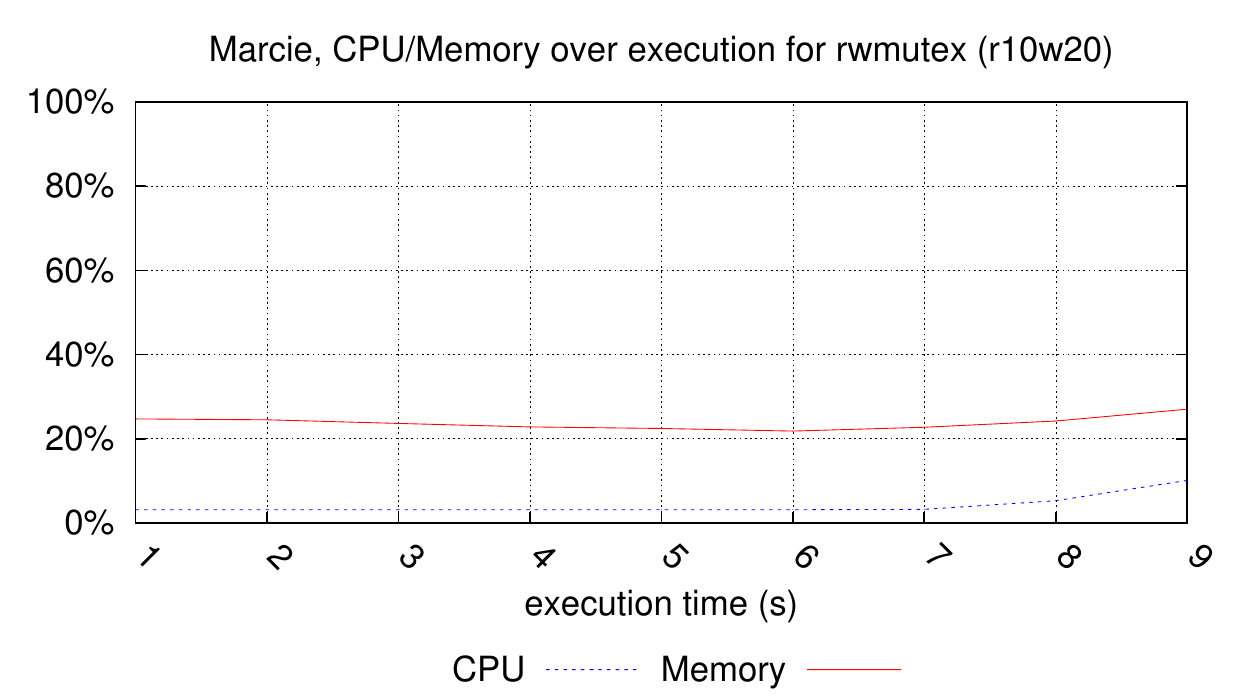}

\noindent\includegraphics[width=.5\textwidth]{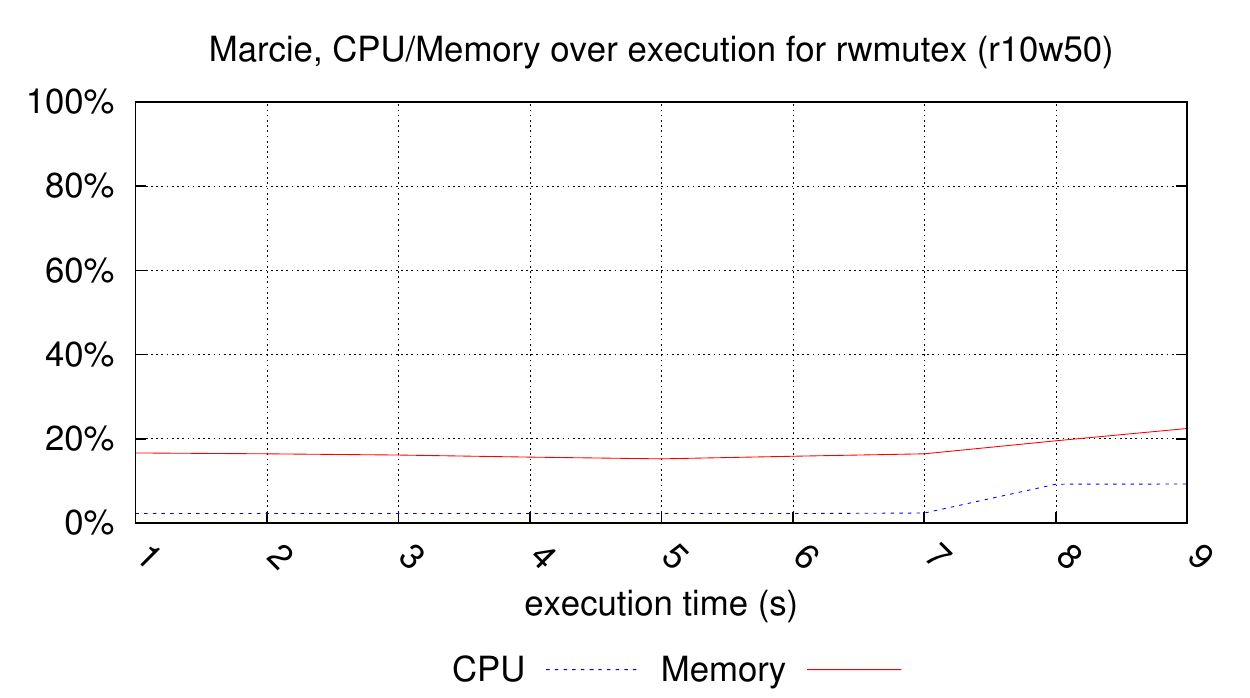}
\includegraphics[width=.5\textwidth]{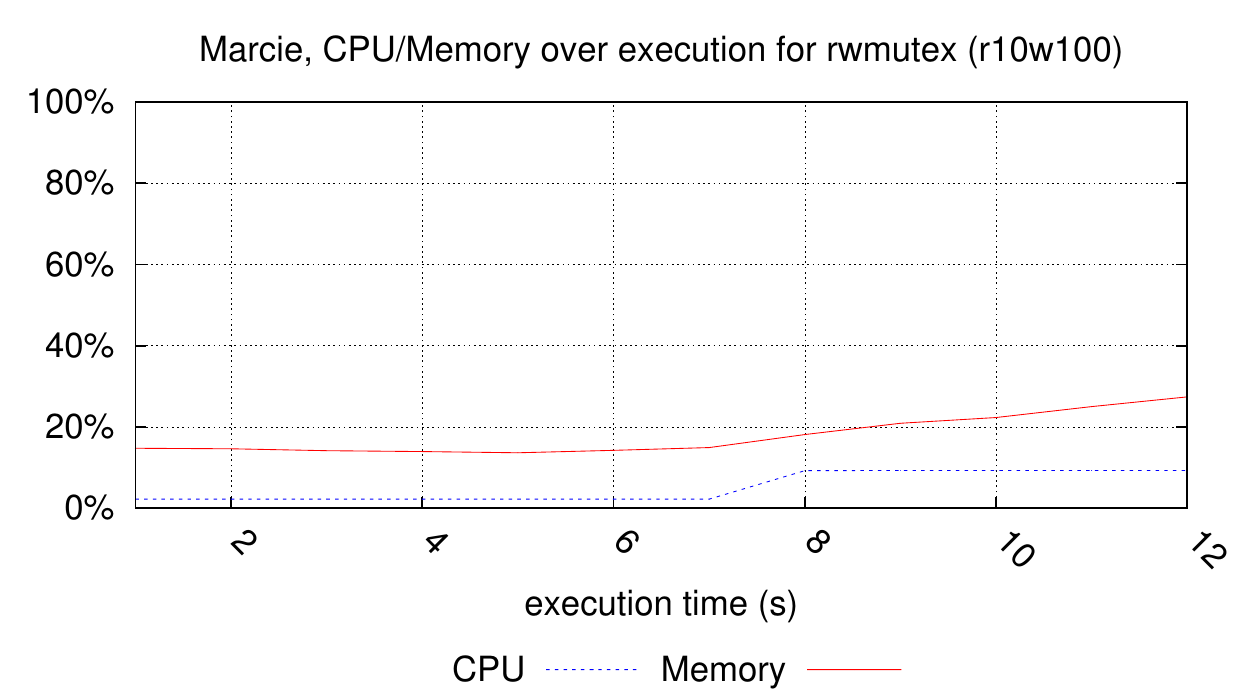}

\noindent\includegraphics[width=.5\textwidth]{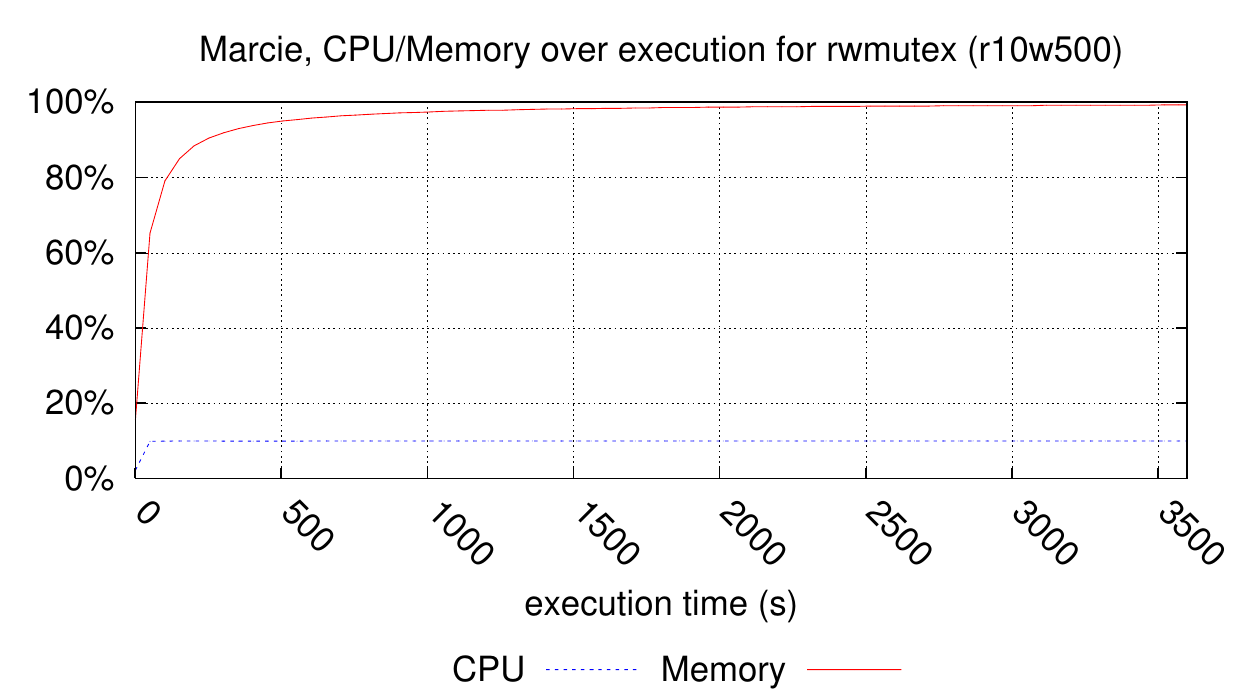}

\vfill\eject
\subsubsection{Executions for SharedMemory}
7 charts have been generated.
\index{Execution (by tool)!Marcie}
\index{Execution (by model)!SharedMemory!Marcie}

\noindent\includegraphics[width=.5\textwidth]{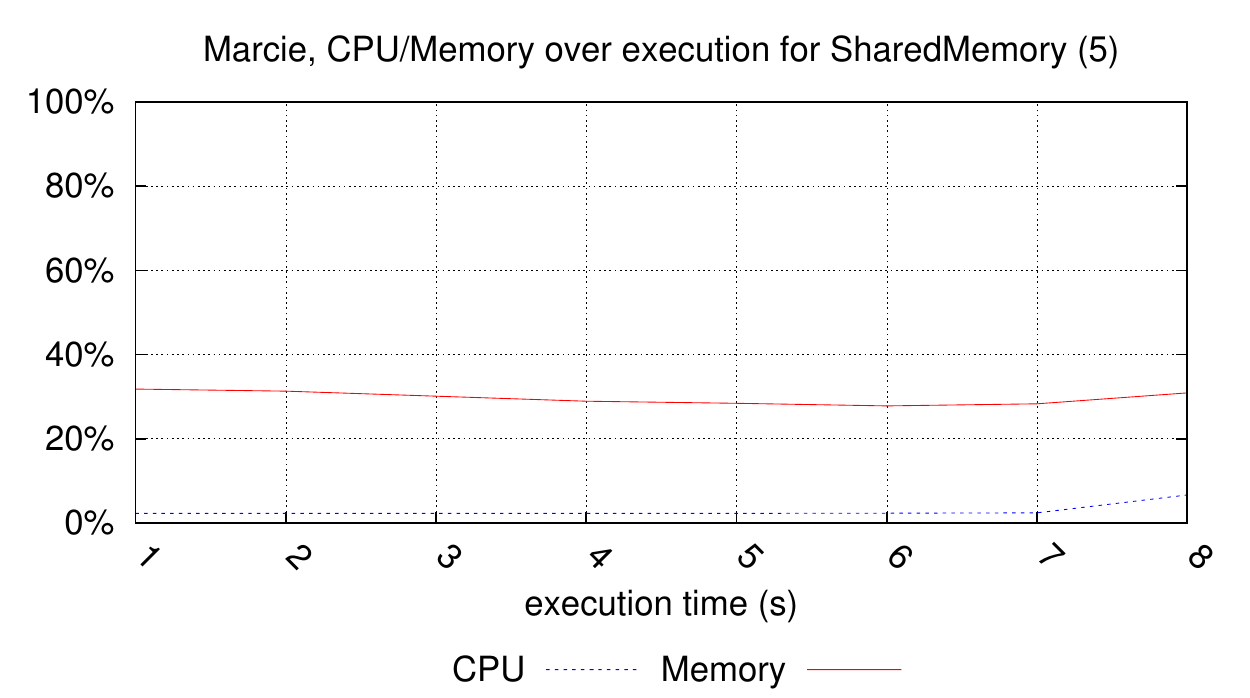}
\includegraphics[width=.5\textwidth]{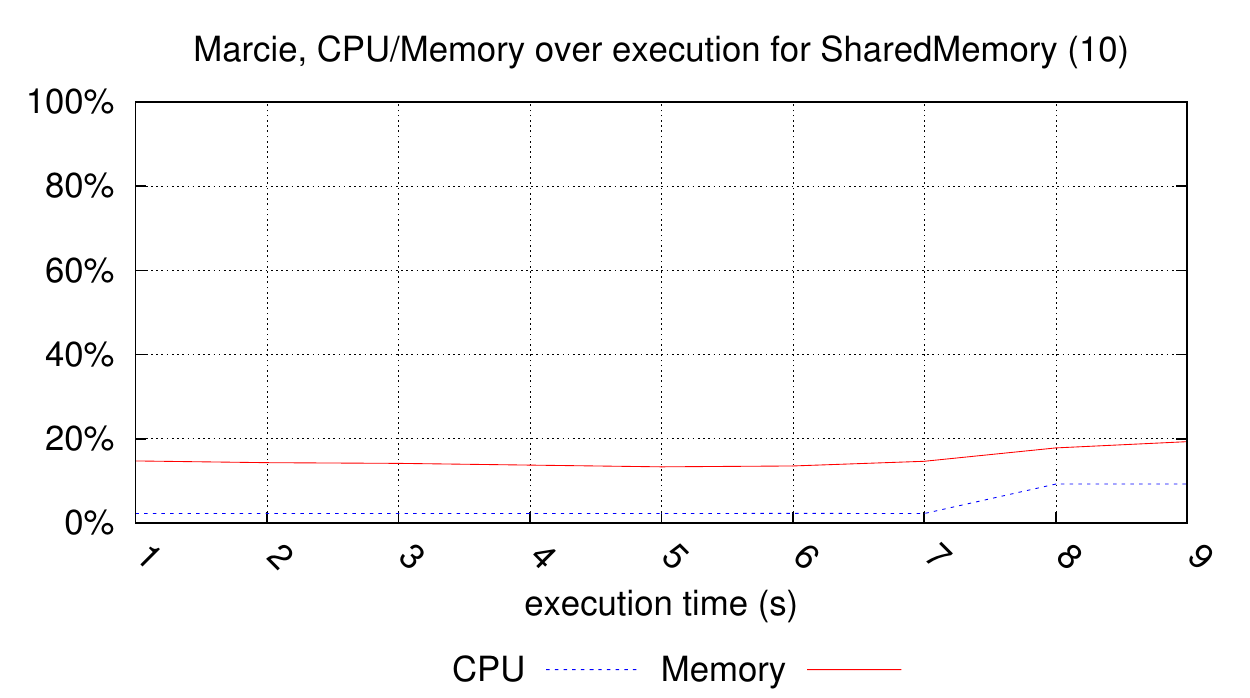}

\noindent\includegraphics[width=.5\textwidth]{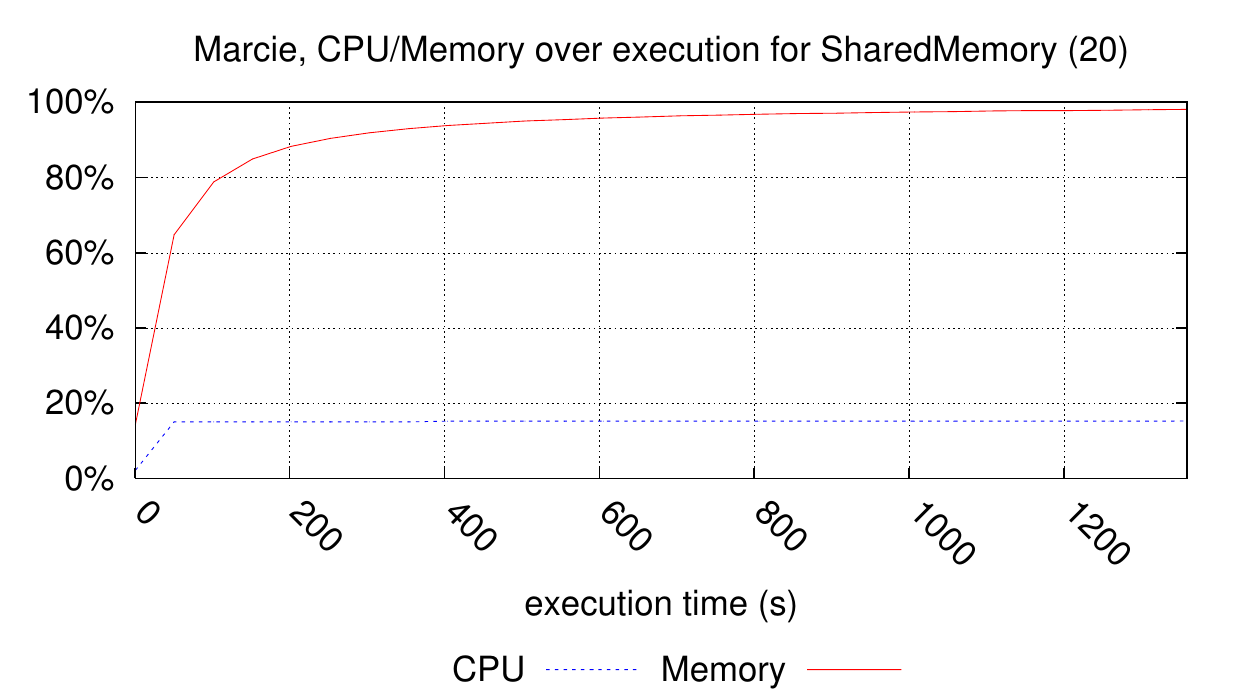}
\includegraphics[width=.5\textwidth]{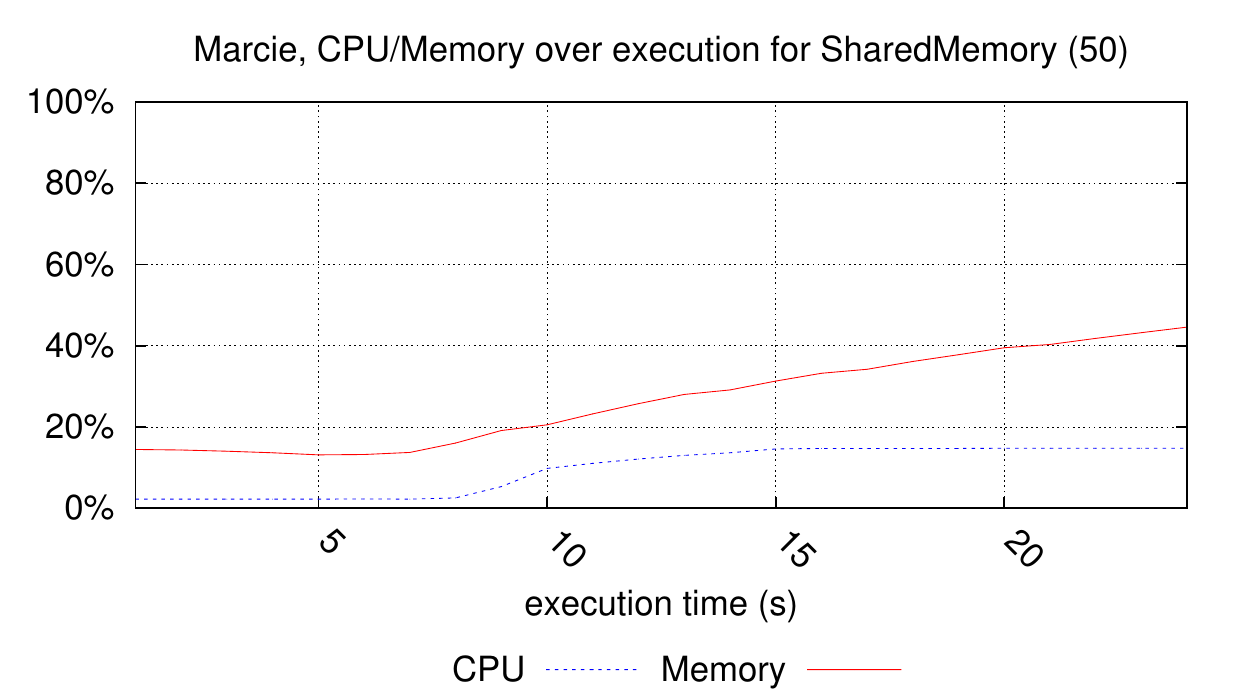}

\noindent\includegraphics[width=.5\textwidth]{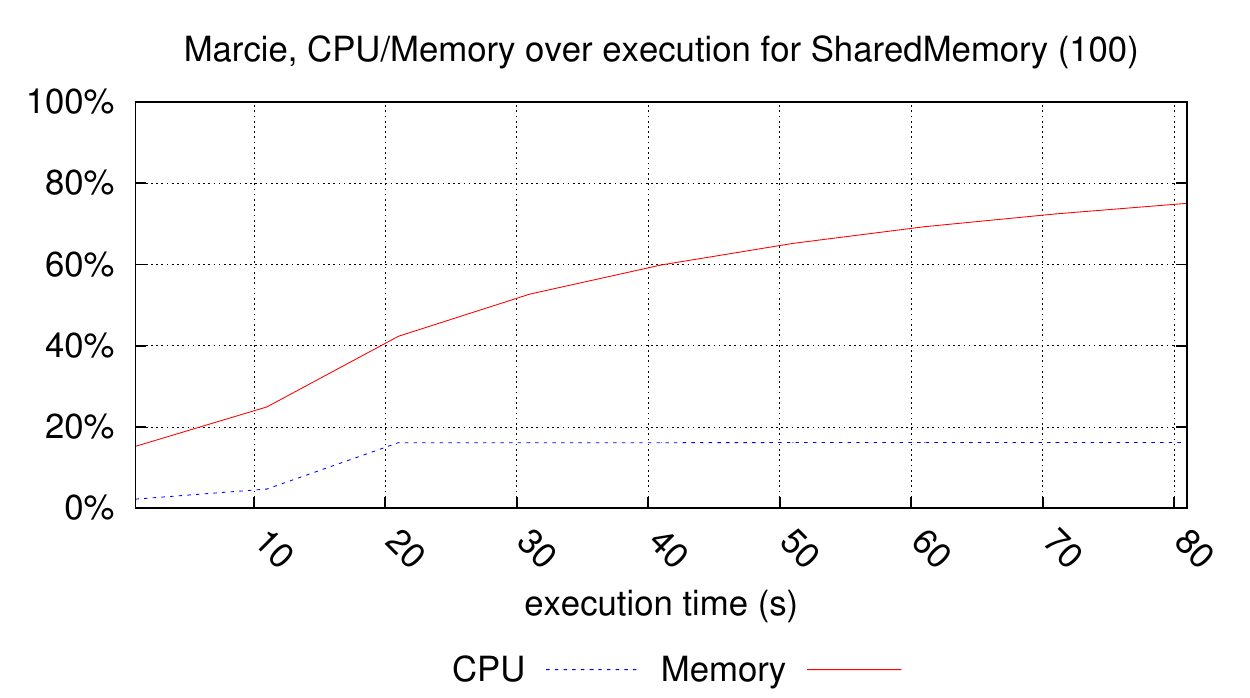}
\includegraphics[width=.5\textwidth]{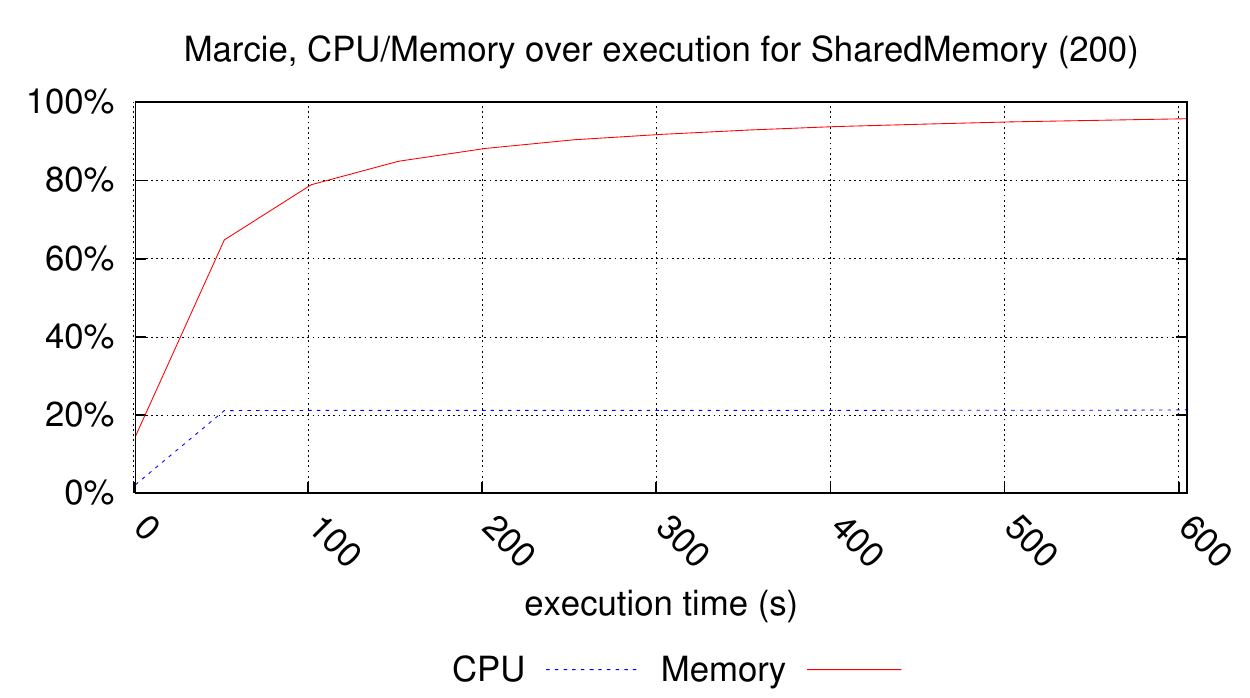}

\noindent\includegraphics[width=.5\textwidth]{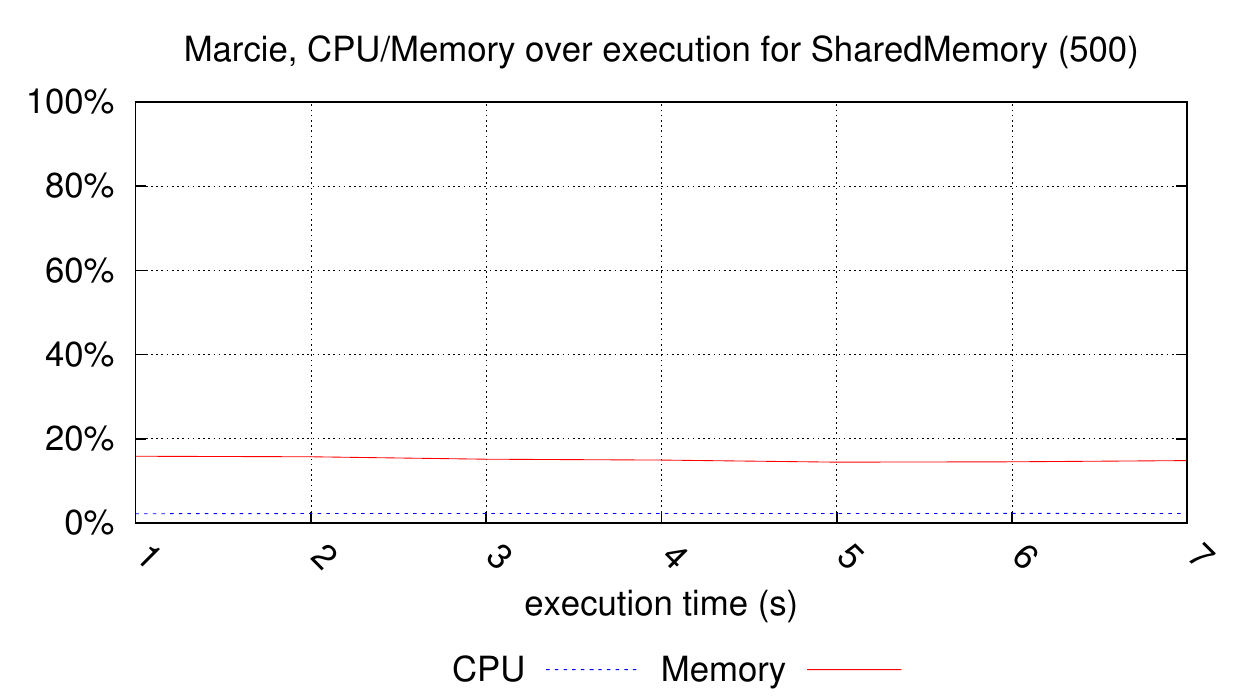}

\vfill\eject
\subsubsection{Executions for simple\_lbs}
3 charts have been generated.
\index{Execution (by tool)!Marcie}
\index{Execution (by model)!simple\_lbs!Marcie}

\noindent\includegraphics[width=.5\textwidth]{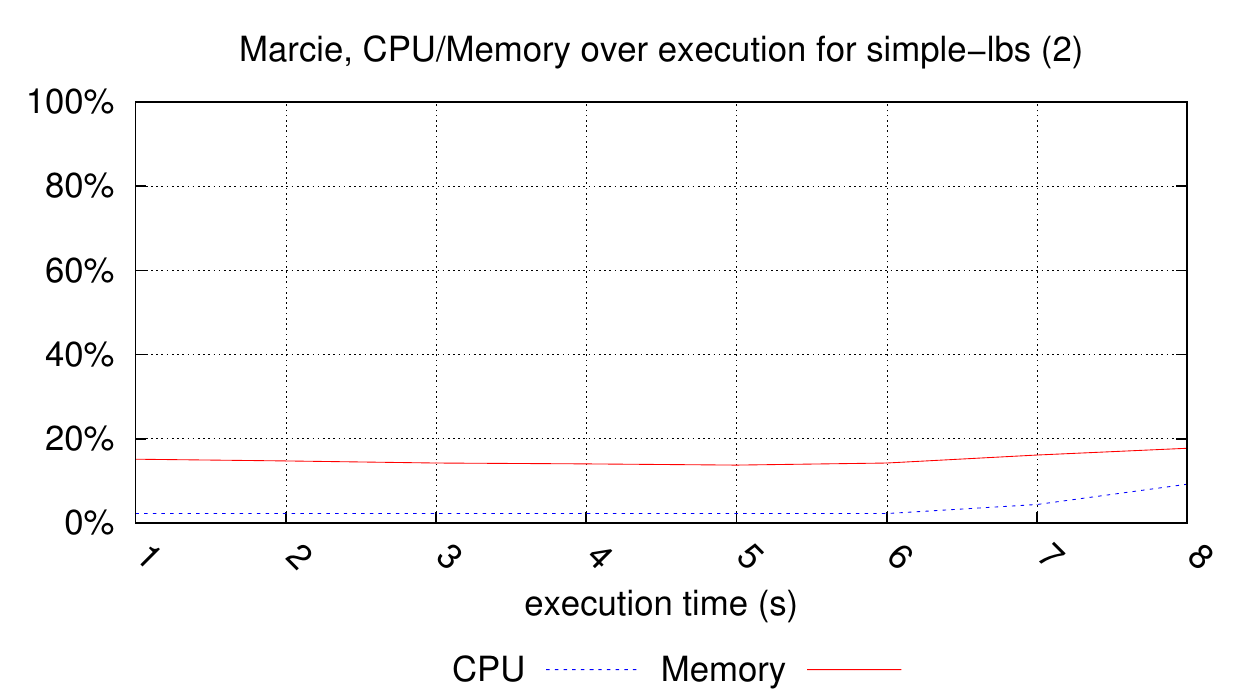}
\includegraphics[width=.5\textwidth]{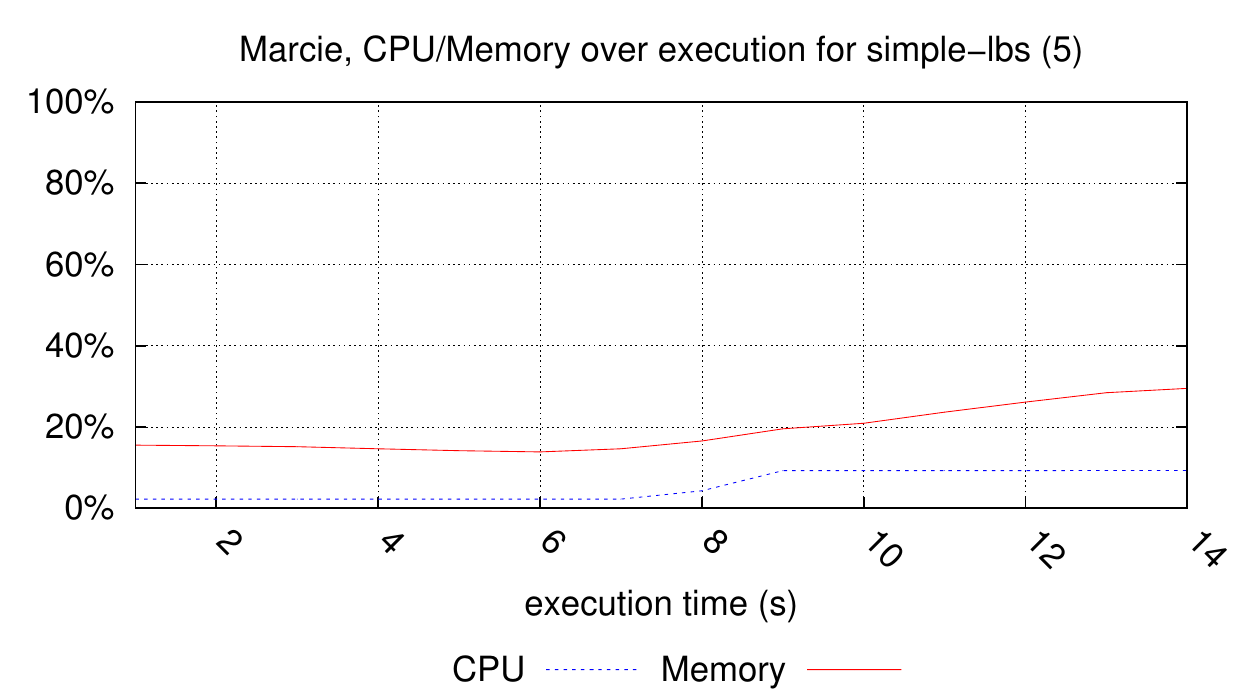}

\noindent\includegraphics[width=.5\textwidth]{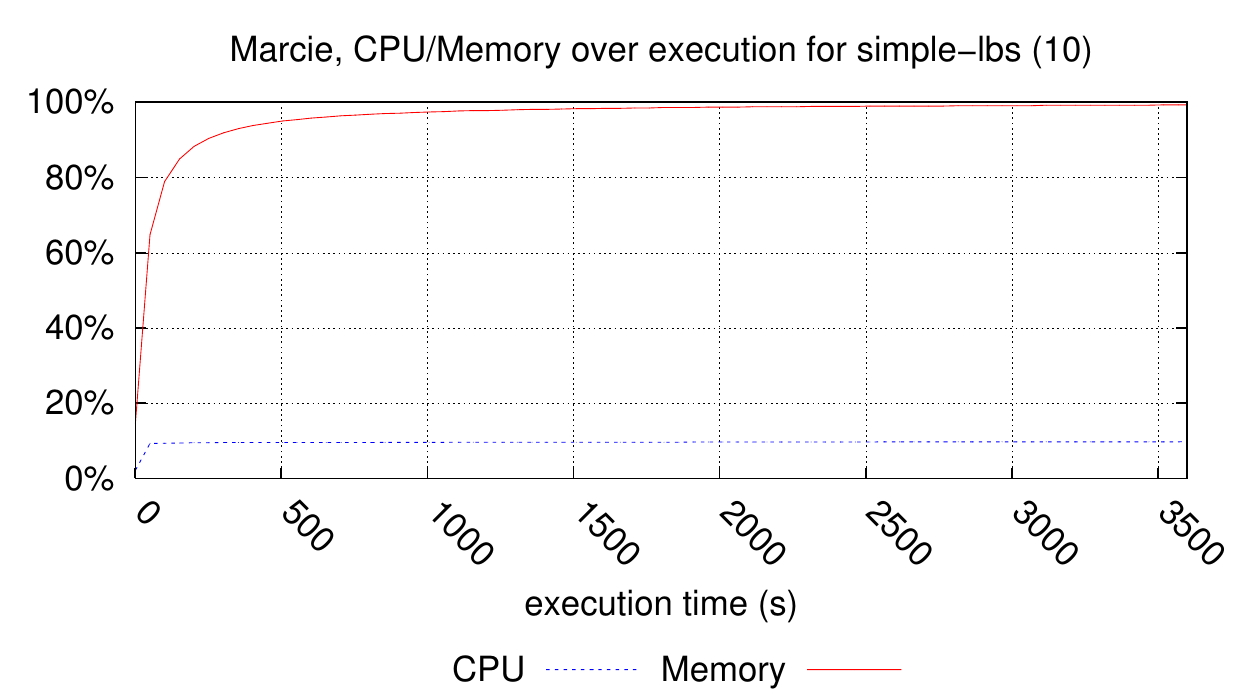}

\subsubsection{Executions for TokenRing}
4 charts have been generated.
\index{Execution (by tool)!Marcie}
\index{Execution (by model)!TokenRing!Marcie}

\noindent\includegraphics[width=.5\textwidth]{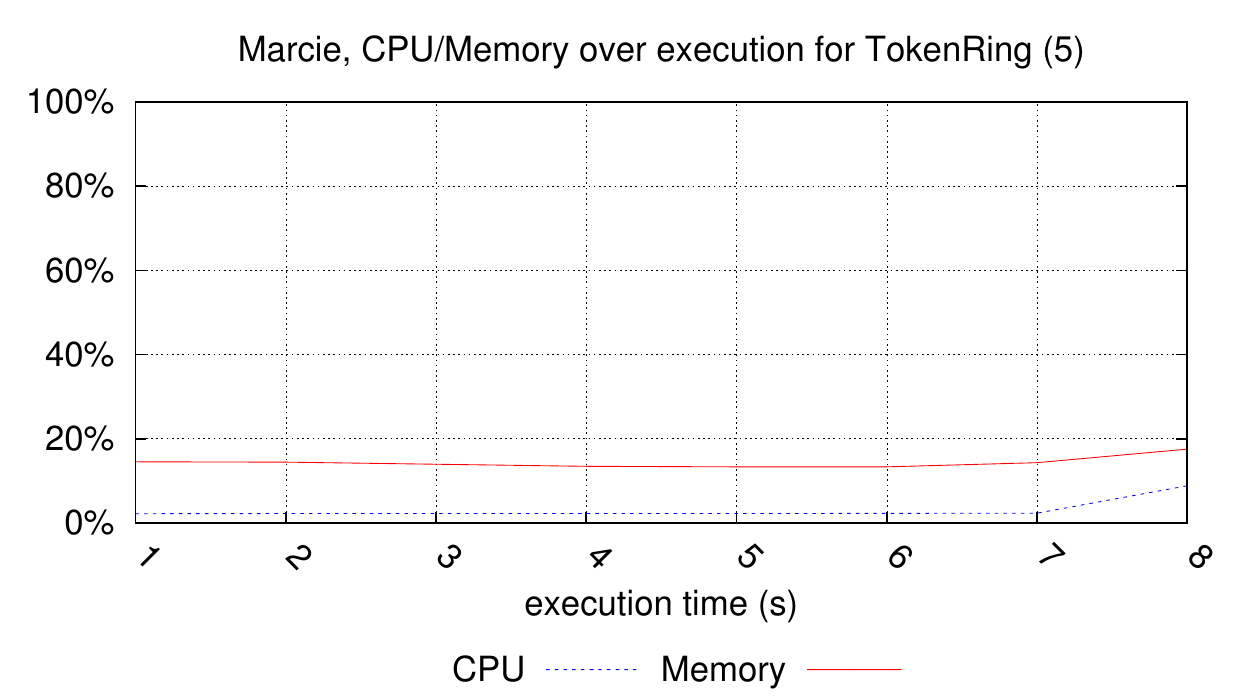}
\includegraphics[width=.5\textwidth]{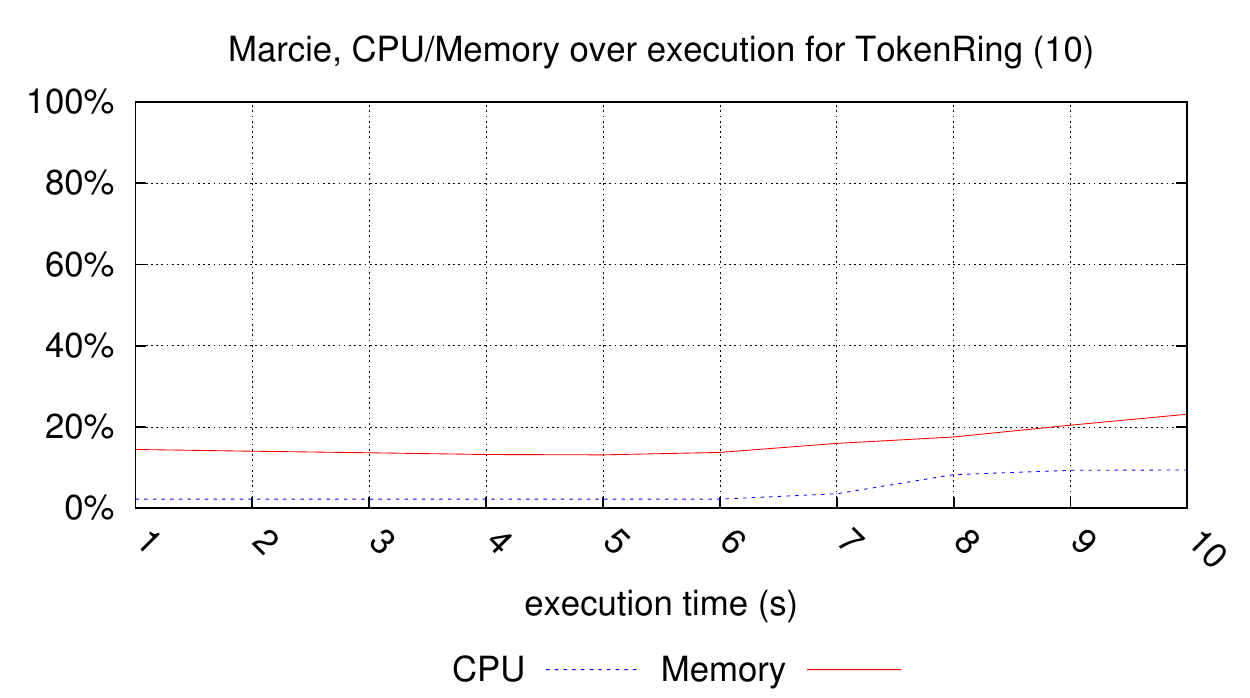}

\noindent\includegraphics[width=.5\textwidth]{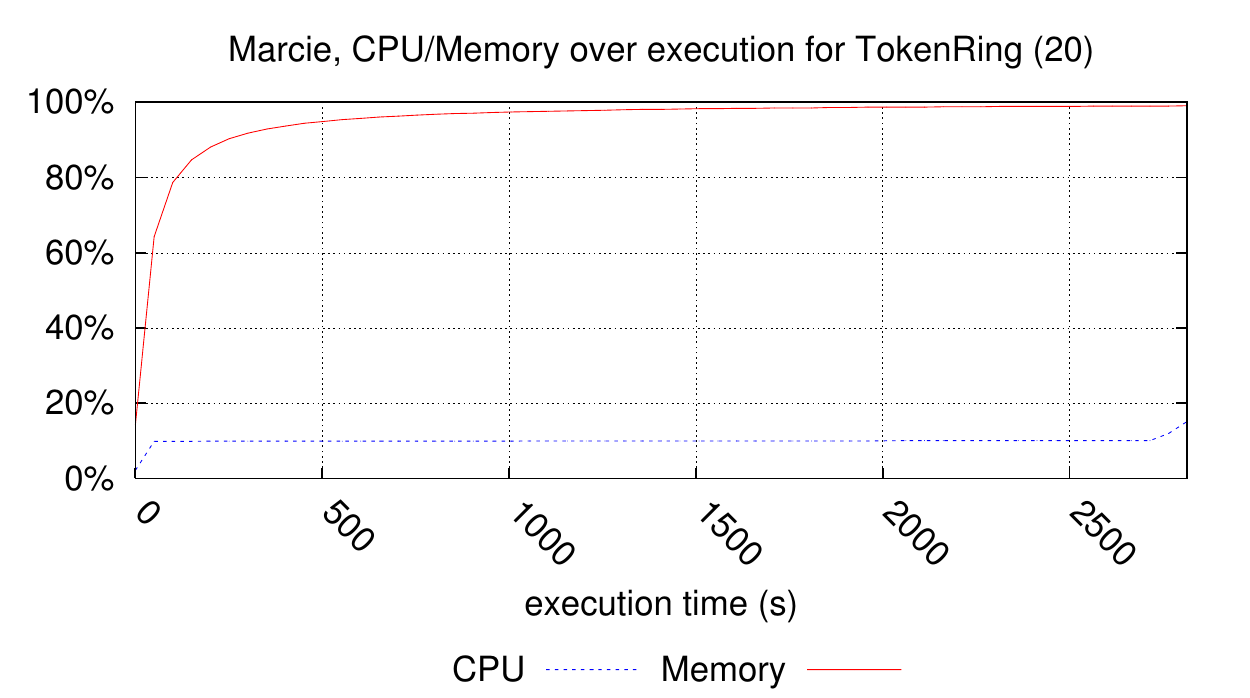}
\includegraphics[width=.5\textwidth]{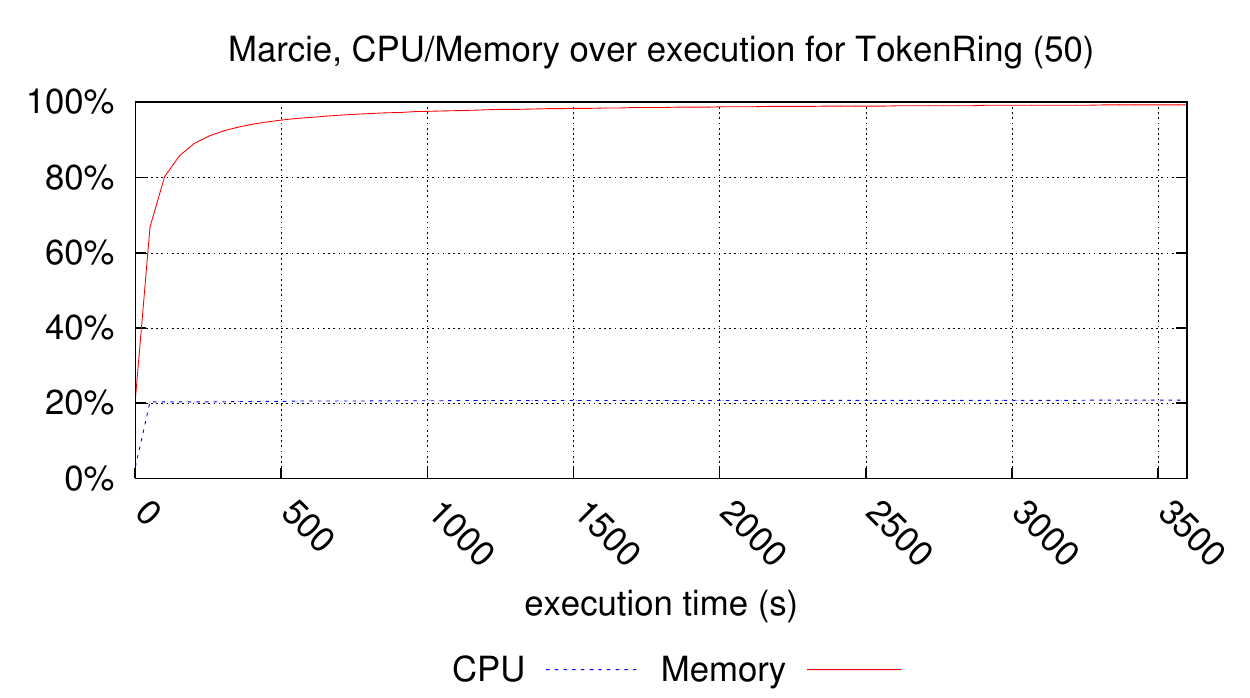}

\subsection{Execution Charts for Neco}

We provide here the execution charts observed for Neco over
the models it could compete with.

\vfill\eject
\subsubsection{Executions for eratosthenes}
4 charts have been generated.
\index{Execution (by tool)!Neco}
\index{Execution (by model)!eratosthenes}

\noindent\includegraphics[width=.5\textwidth]{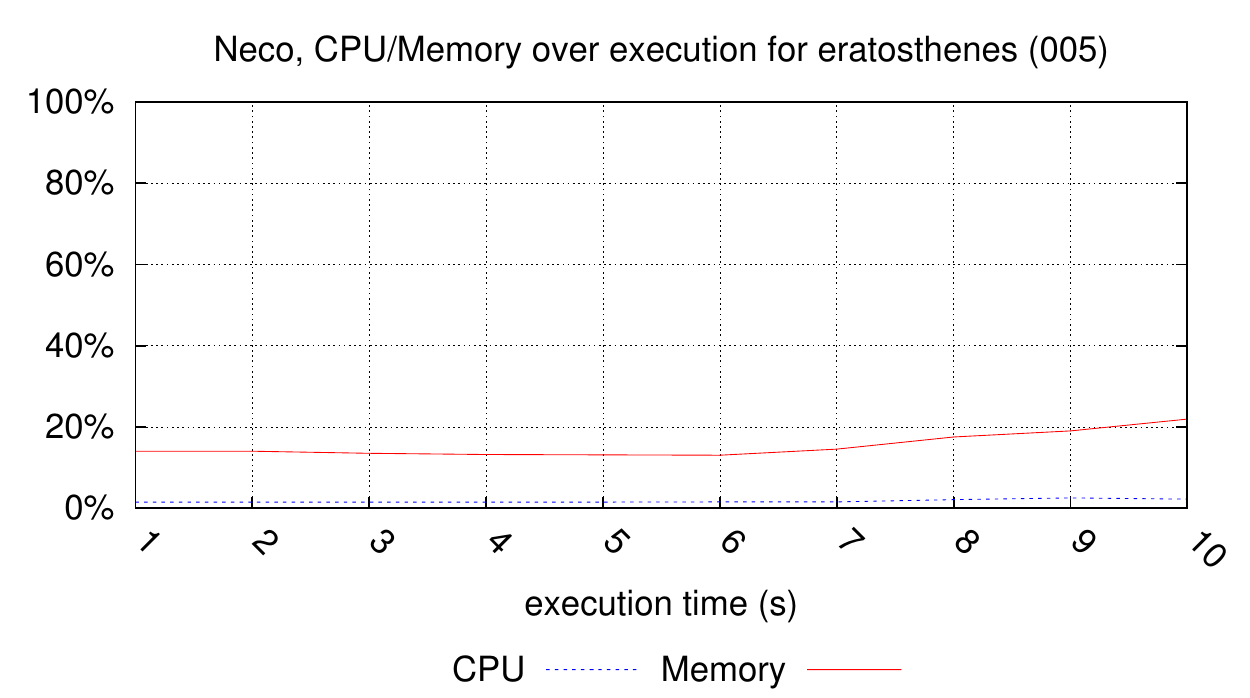}
\includegraphics[width=.5\textwidth]{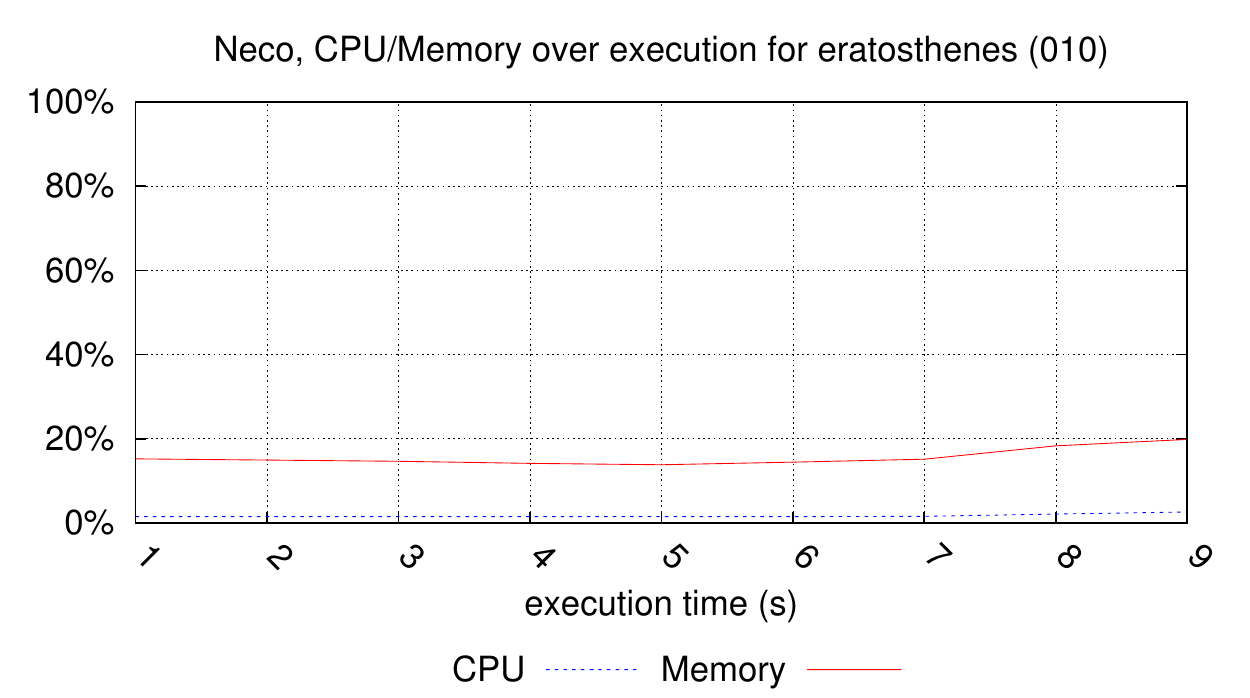}

\noindent\includegraphics[width=.5\textwidth]{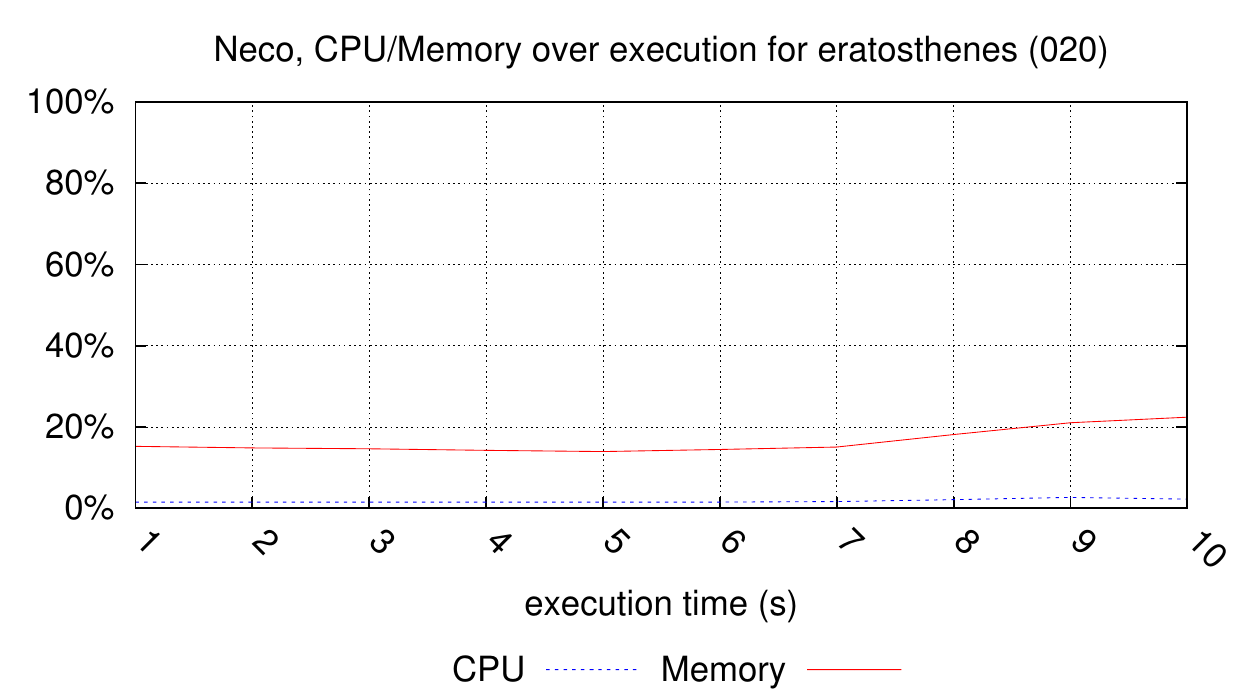}
\includegraphics[width=.5\textwidth]{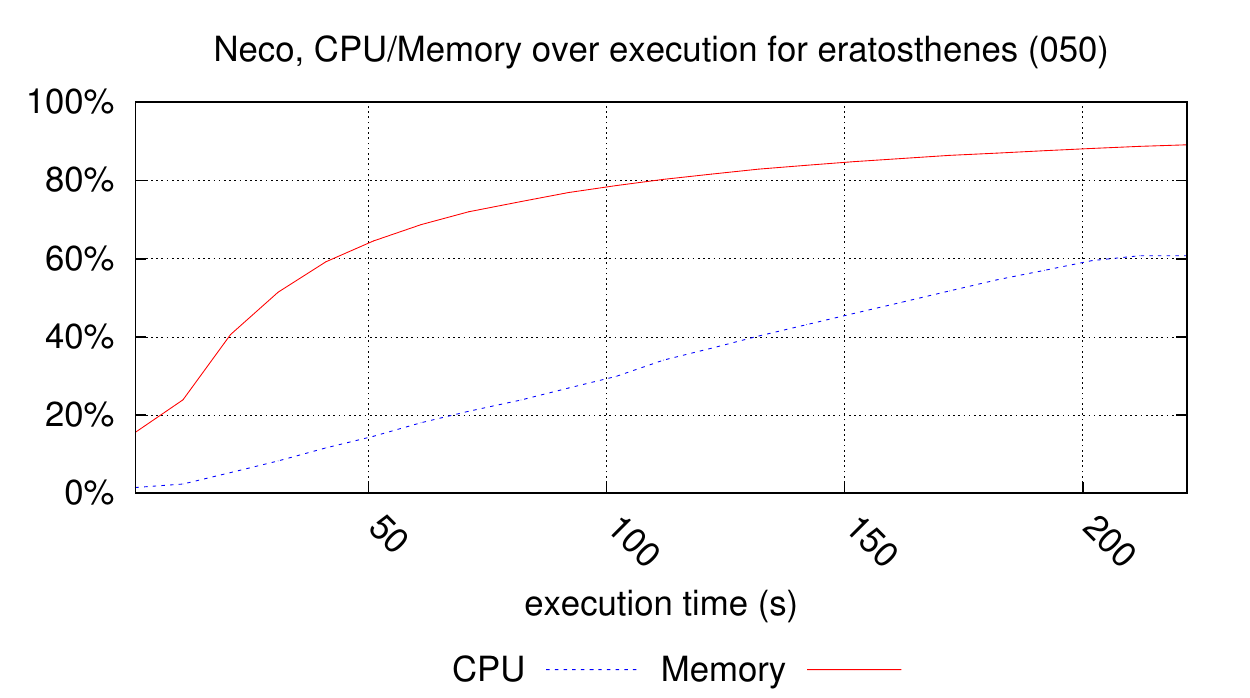}

\subsubsection{Executions for Kanban}
2 charts have been generated.
\index{Execution (by tool)!Neco}
\index{Execution (by model)!Kanban!Neco}

\noindent\includegraphics[width=.5\textwidth]{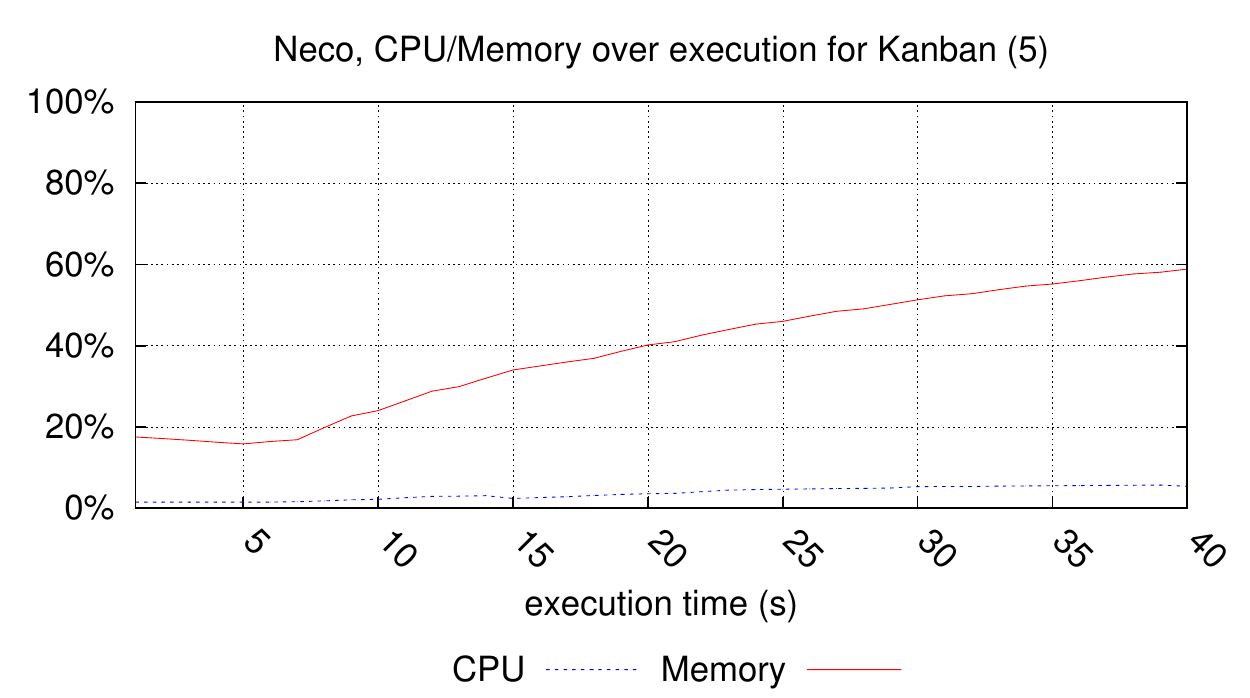}
\includegraphics[width=.5\textwidth]{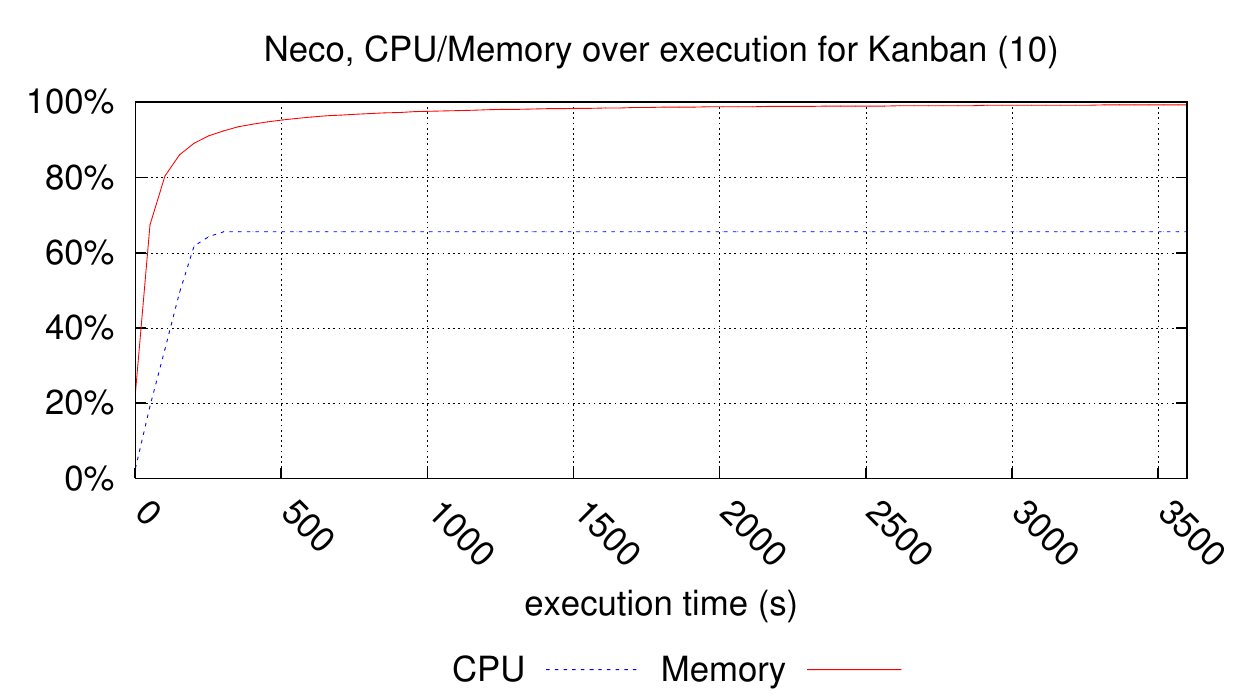}

\subsubsection{Executions for MAPK}
2 charts have been generated.
\index{Execution (by tool)!Neco}
\index{Execution (by model)!MAPK!Neco}

\noindent\includegraphics[width=.5\textwidth]{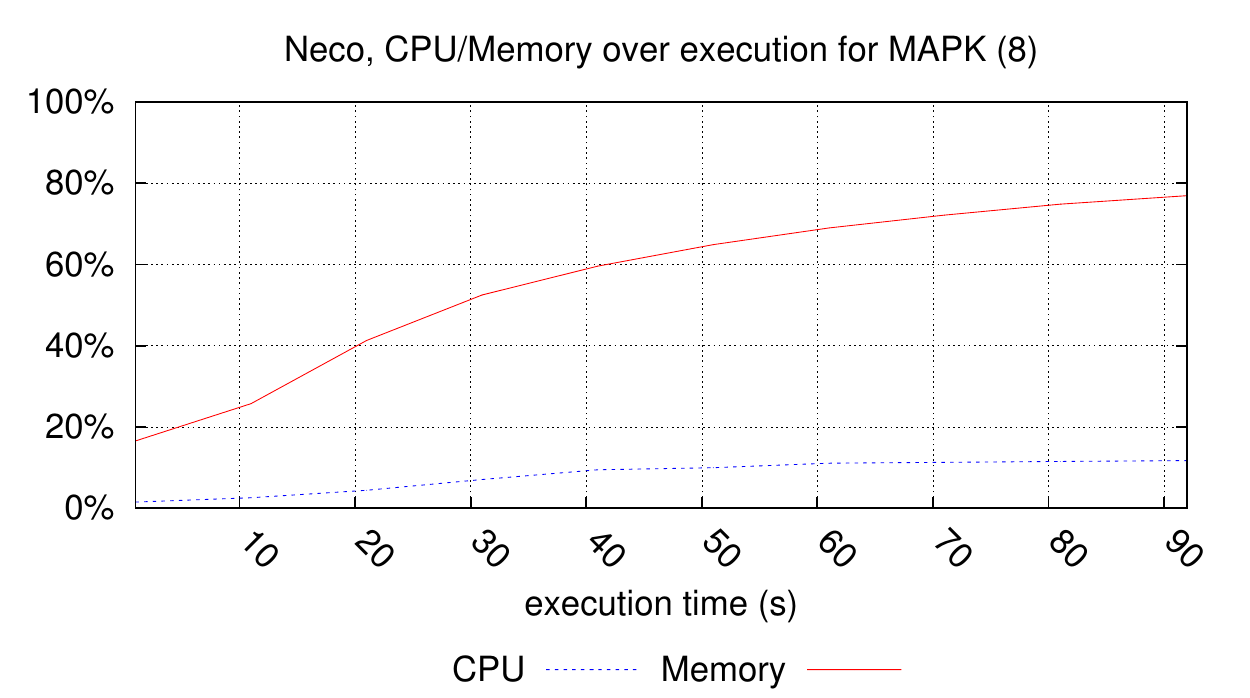}
\includegraphics[width=.5\textwidth]{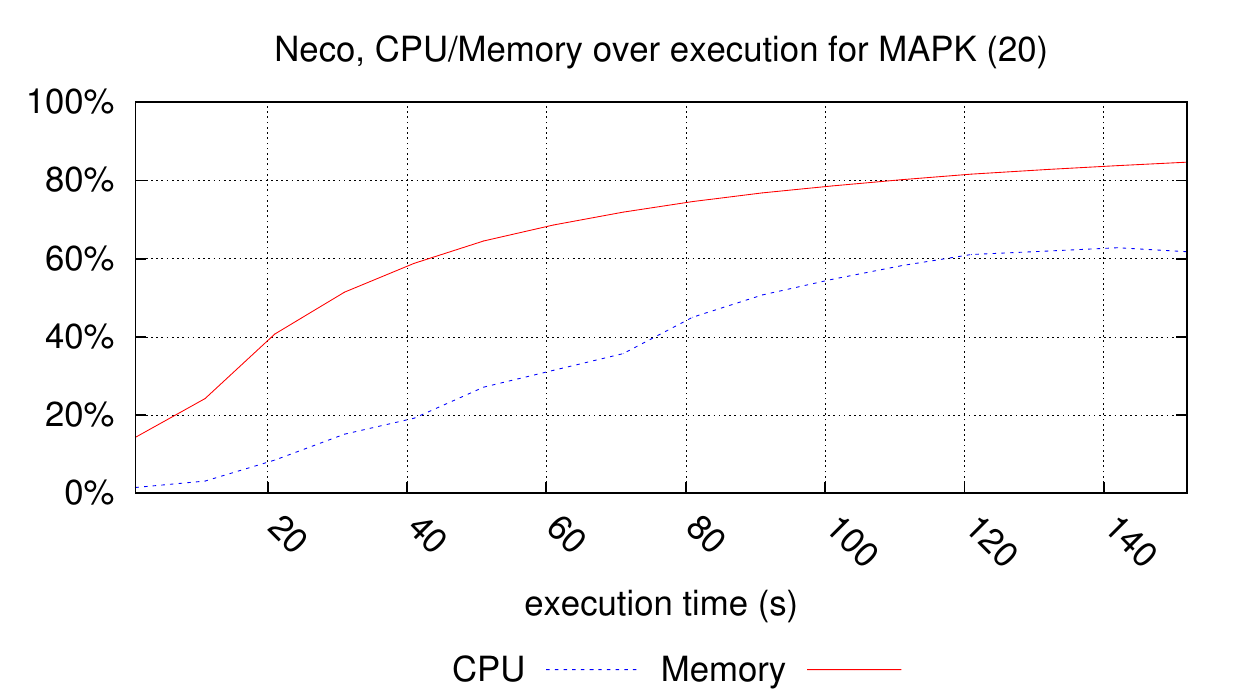}

\vfill\eject
\subsubsection{Executions for Peterson}
2 charts have been generated.
\index{Execution (by tool)!Neco}
\index{Execution (by model)!Peterson!Neco}

\noindent\includegraphics[width=.5\textwidth]{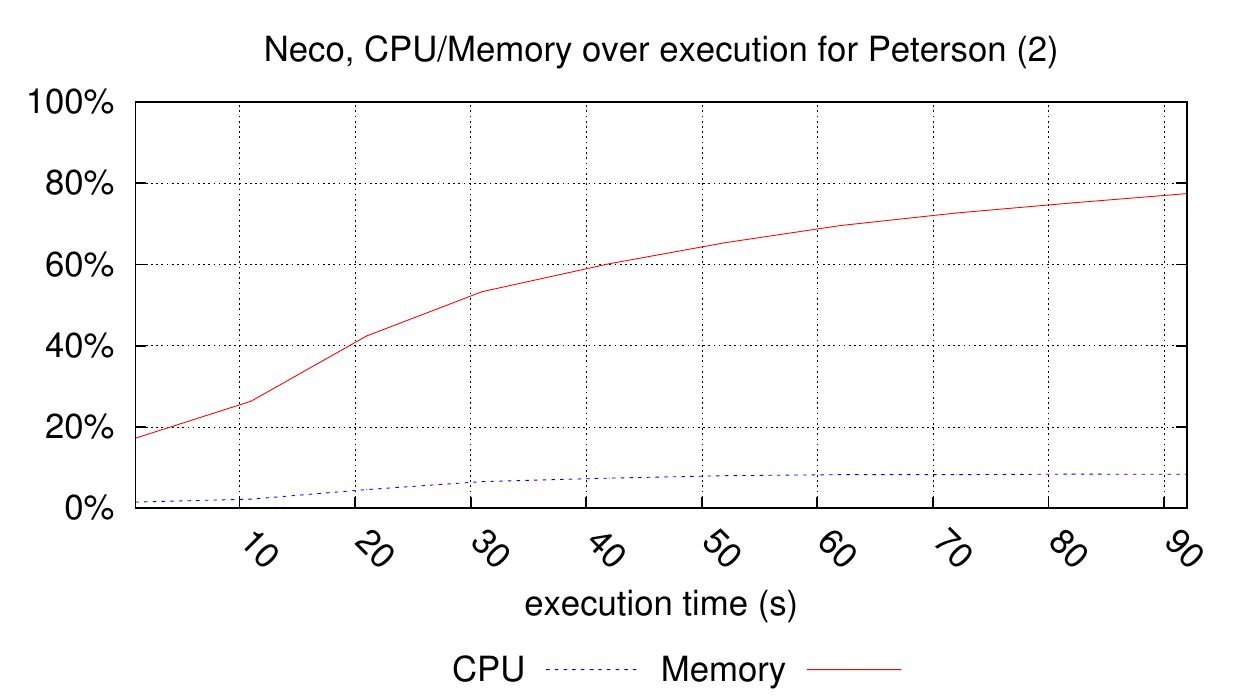}
\includegraphics[width=.5\textwidth]{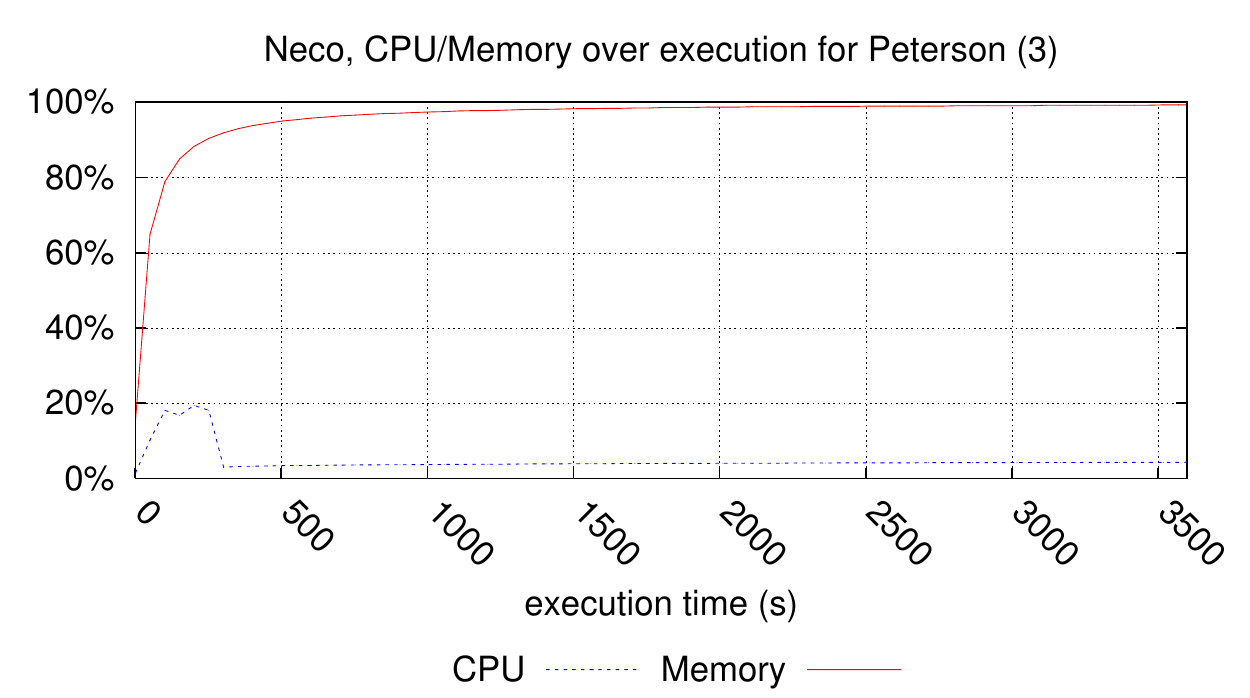}

\subsubsection{Executions for philo\_dyn}
3 charts have been generated.
\index{Execution (by tool)!Neco}
\index{Execution (by model)!philo\_dyn!Neco}

\noindent\includegraphics[width=.5\textwidth]{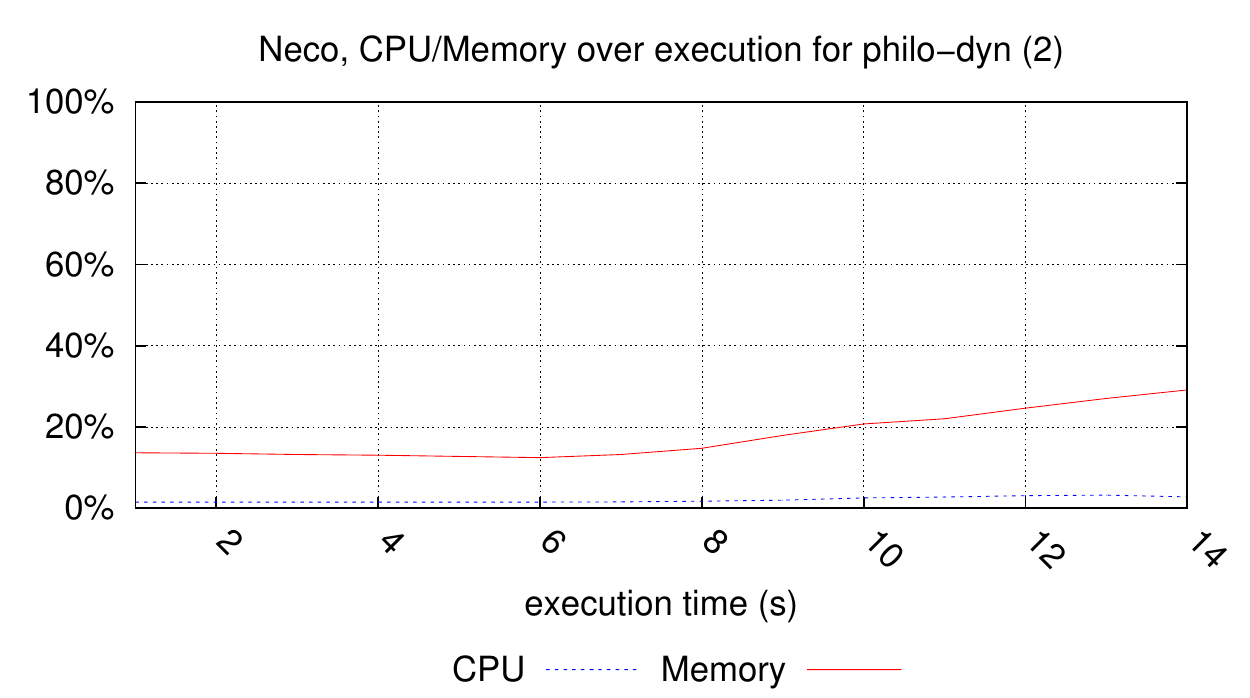}
\includegraphics[width=.5\textwidth]{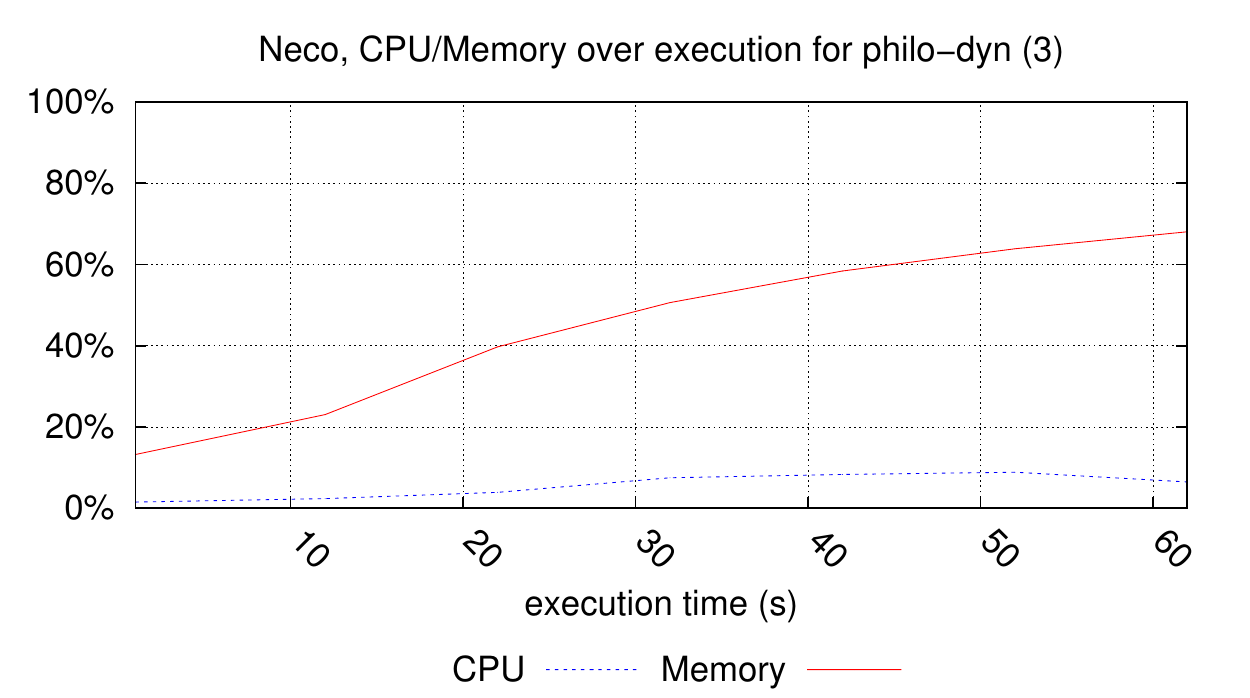}

\noindent\includegraphics[width=.5\textwidth]{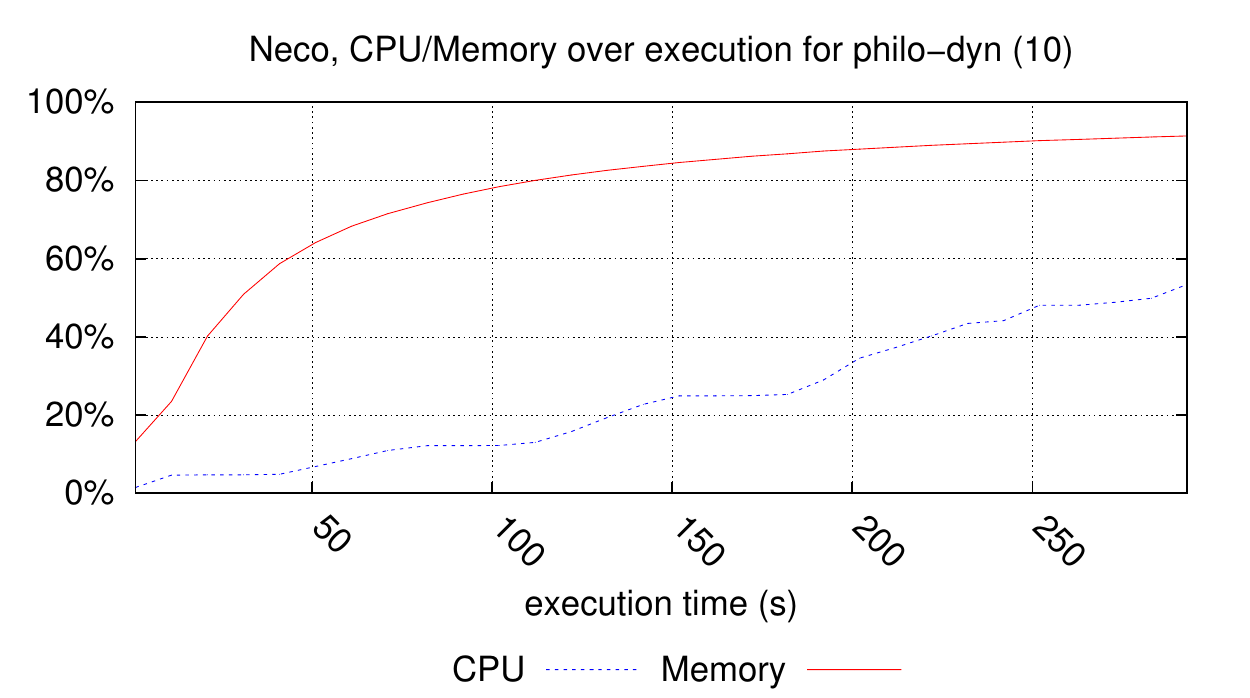}

\subsubsection{Executions for rw\_mutex}
5 charts have been generated.
\index{Execution (by tool)!Neco}
\index{Execution (by model)!rw\_mutex!Neco}

\noindent\includegraphics[width=.5\textwidth]{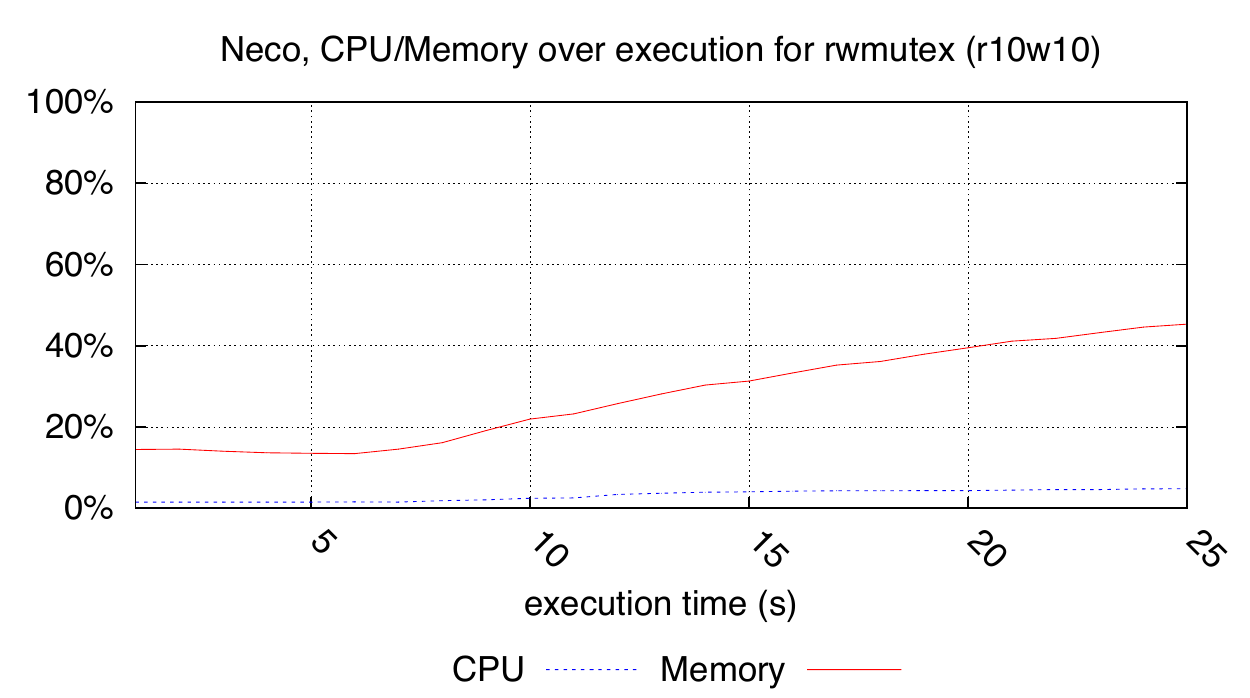}
\includegraphics[width=.5\textwidth]{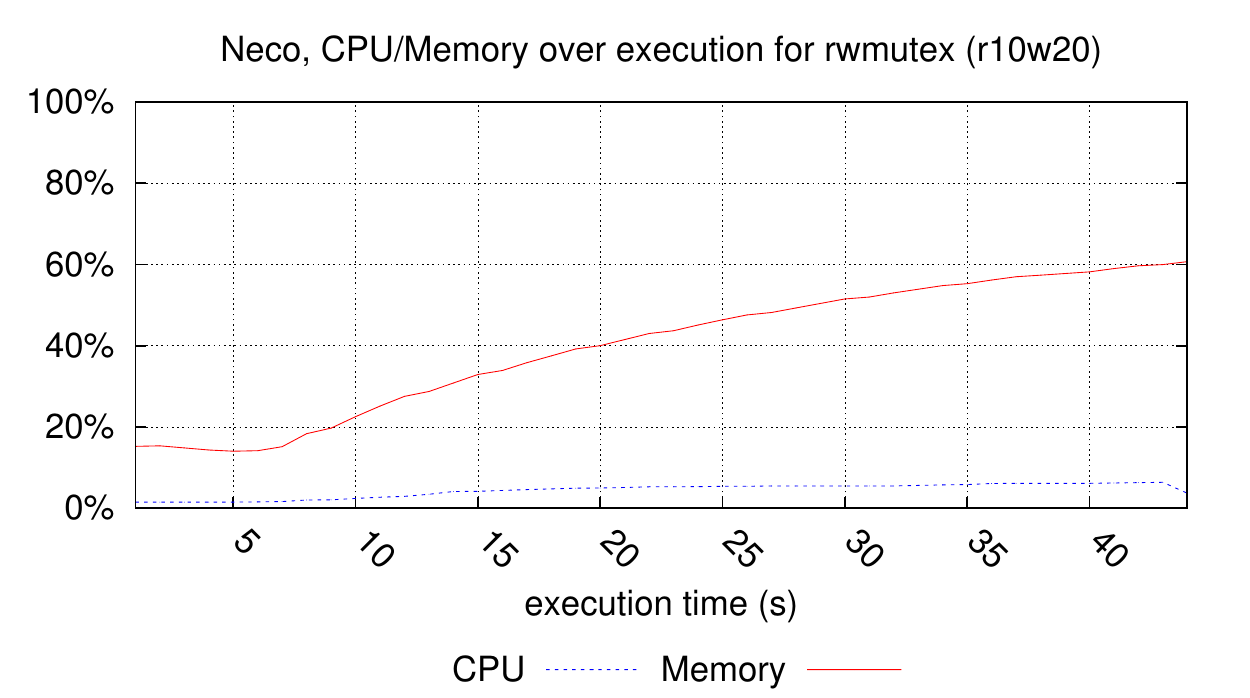}

\noindent\includegraphics[width=.5\textwidth]{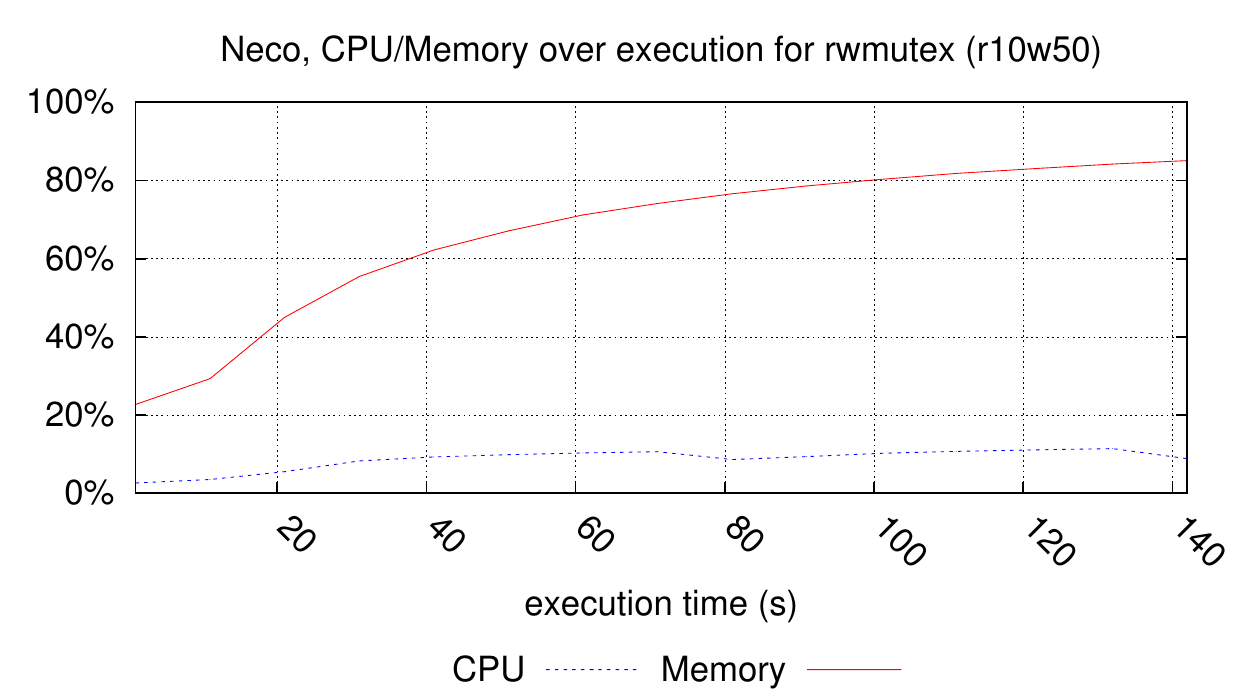}
\includegraphics[width=.5\textwidth]{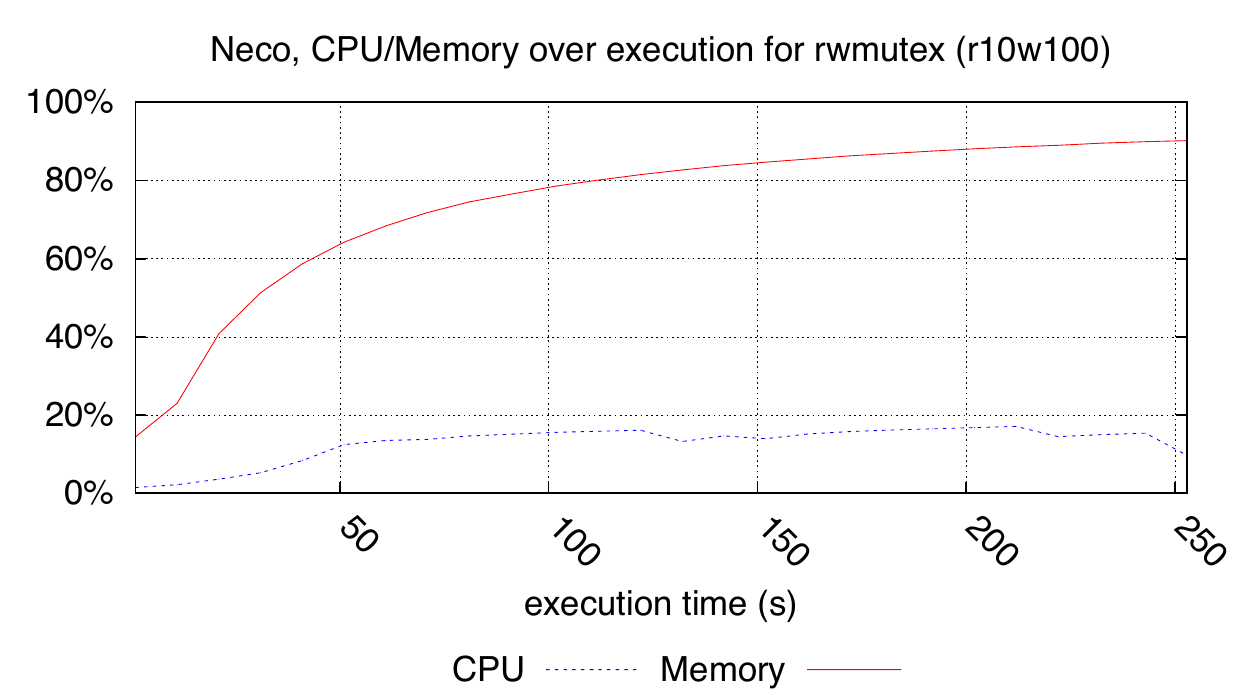}

\noindent\includegraphics[width=.5\textwidth]{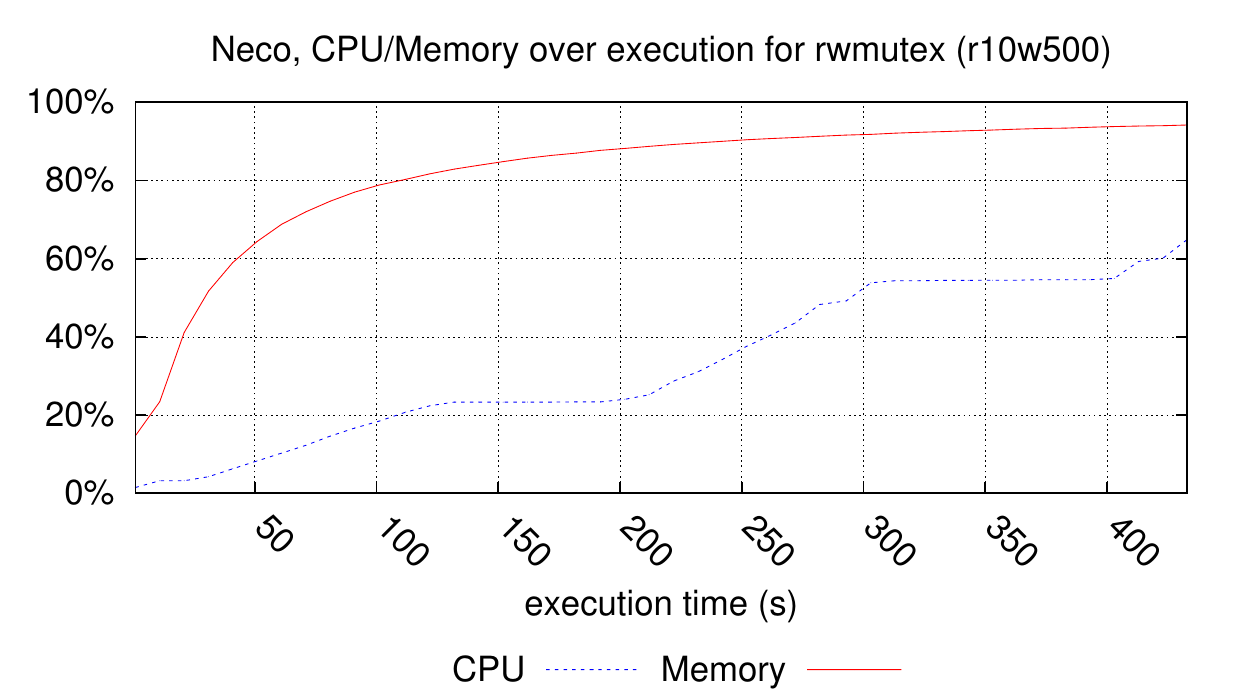}

\subsubsection{Executions for simple\_lbs}
3 charts have been generated.
\index{Execution (by tool)!Neco}
\index{Execution (by model)!simple\_lbs!Neco}

\noindent\includegraphics[width=.5\textwidth]{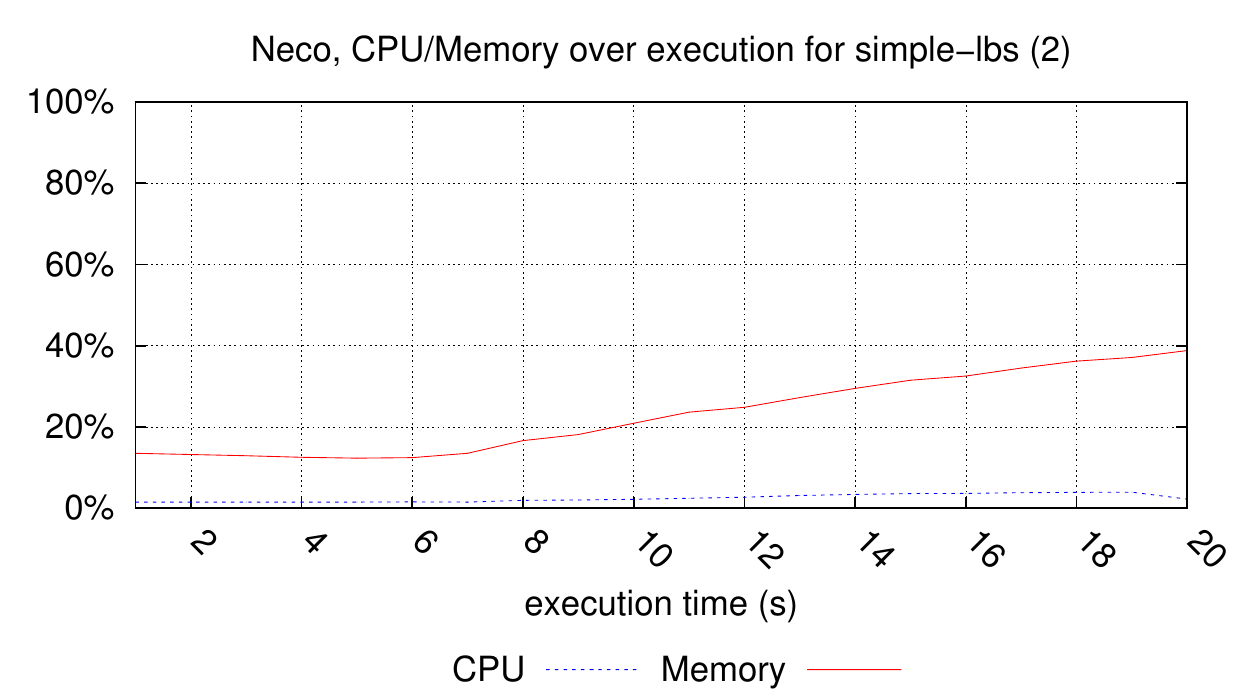}
\includegraphics[width=.5\textwidth]{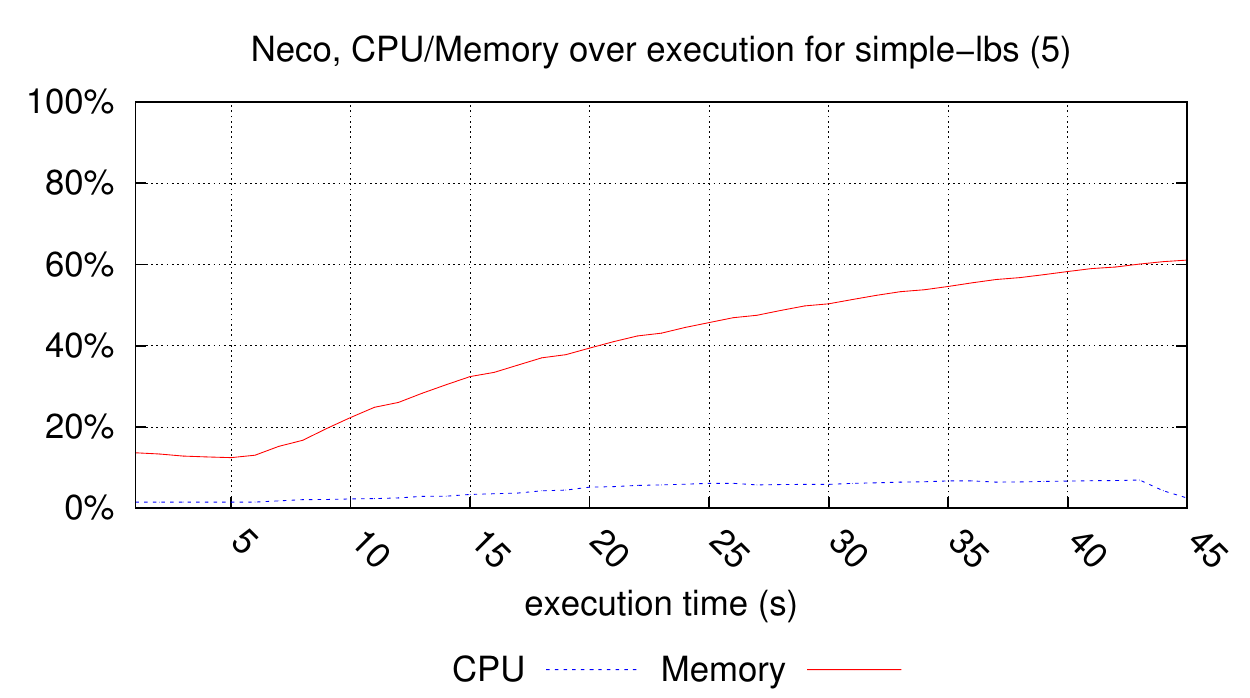}

\noindent\includegraphics[width=.5\textwidth]{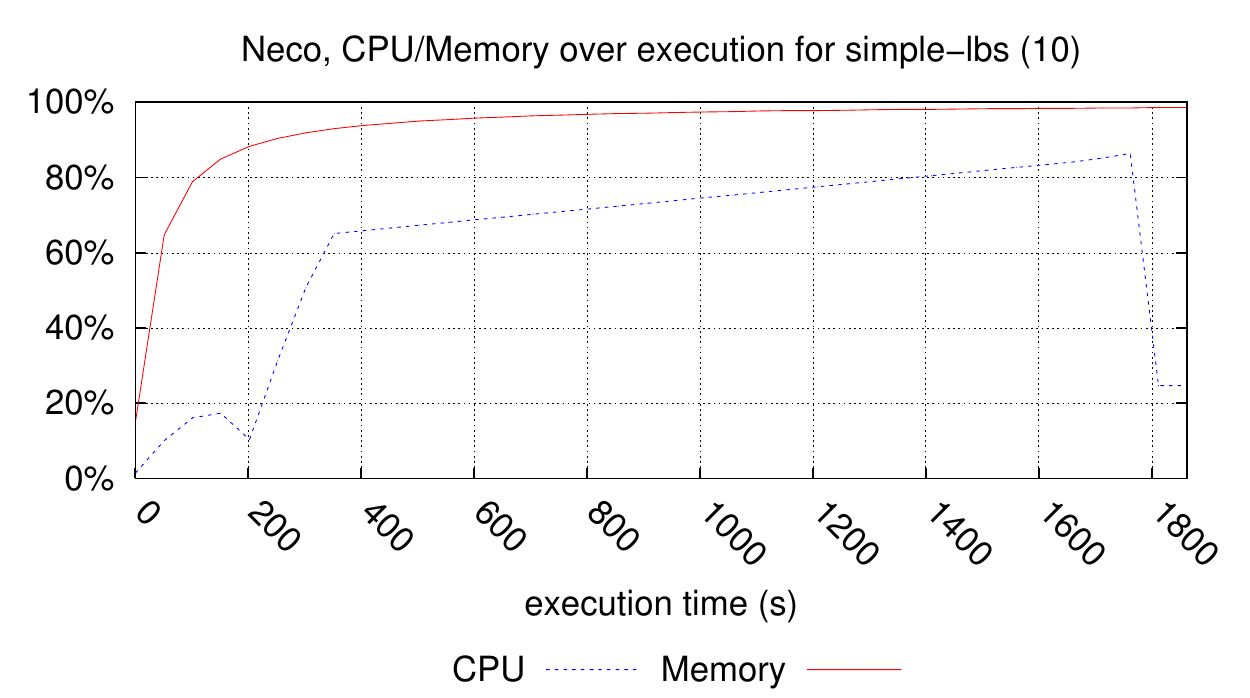}

\subsection{Execution Charts for PNXDD}
\label{sec:sstoollast}

We provide here the execution charts observed for PNXDD over
the models it could compete with.

\vfill\eject
\subsubsection{Executions for cs\_repetitions}
1 chart has been generated.
\index{Execution (by tool)!PNXDD}
\index{Execution (by model)!cs\_repetitions!PNXDD}

\noindent\includegraphics[width=.5\textwidth]{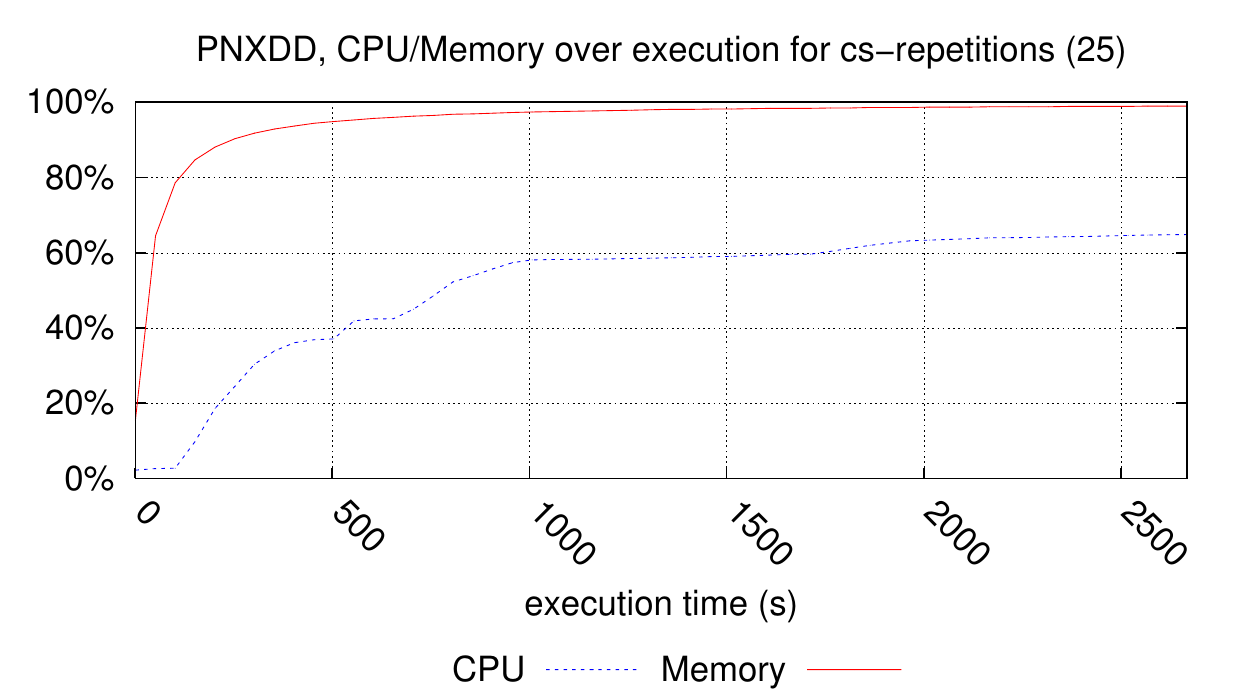}

\subsubsection{Executions for FMS}
6 charts have been generated.
\index{Execution (by tool)!PNXDD}
\index{Execution (by model)!FMS!PNXDD}

\noindent\includegraphics[width=.5\textwidth]{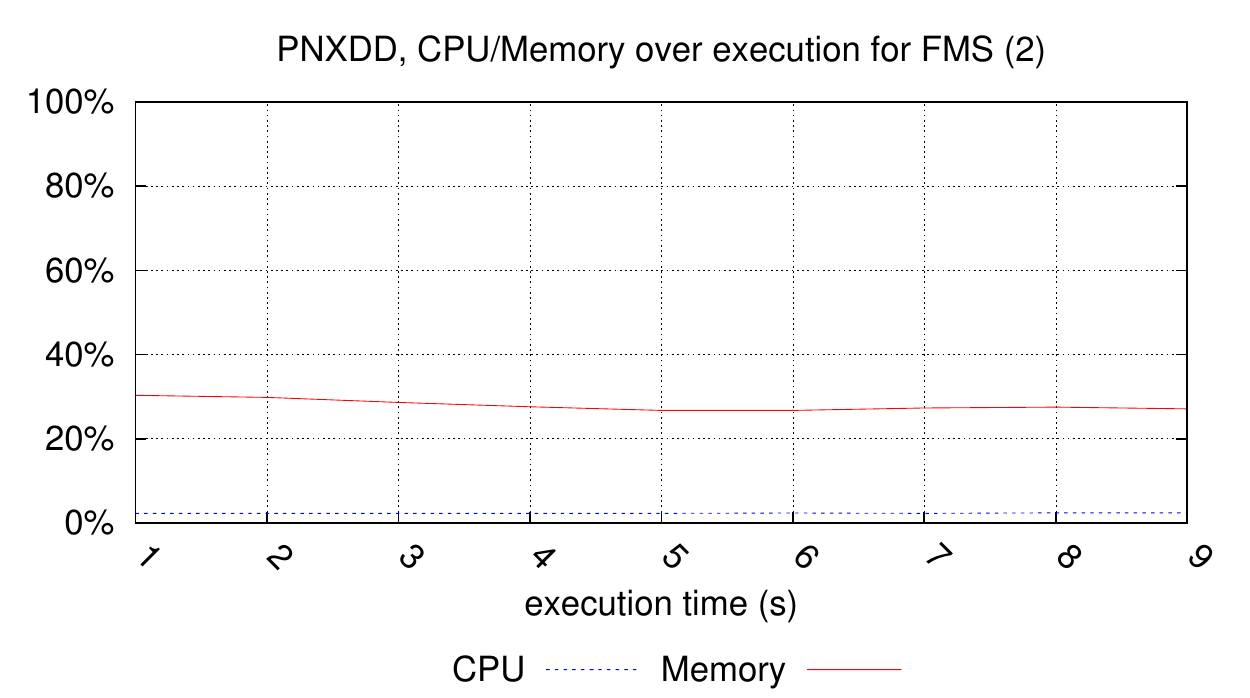}
\includegraphics[width=.5\textwidth]{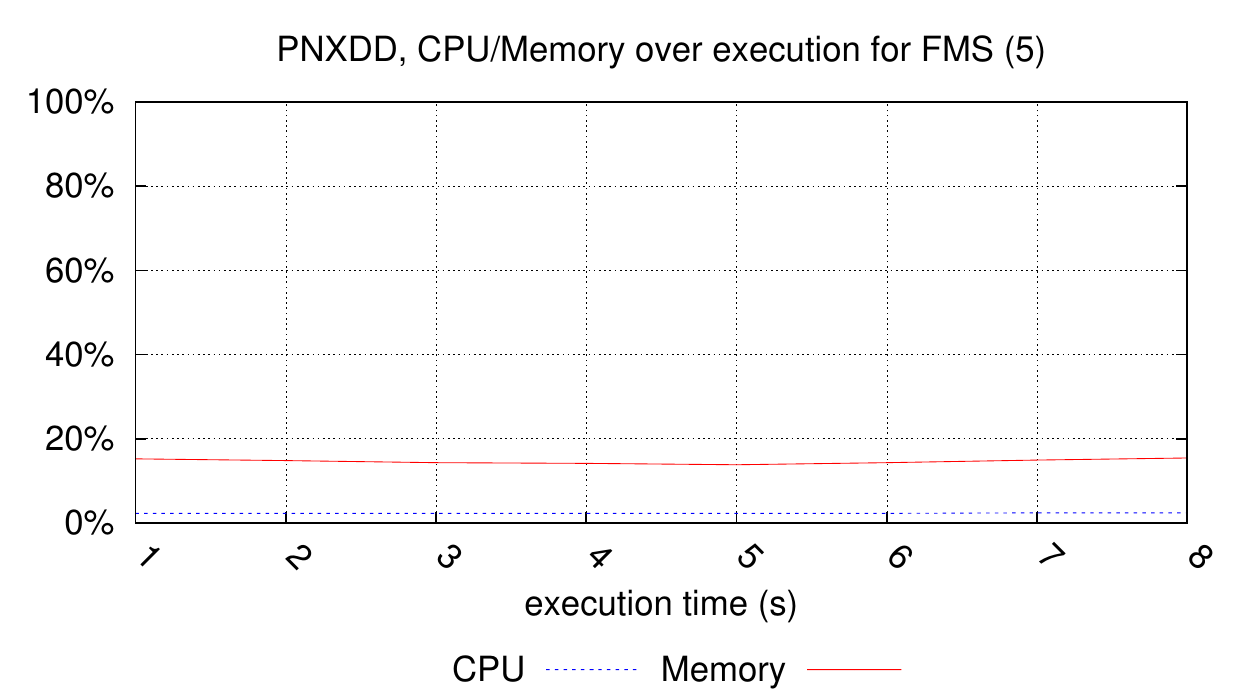}

\noindent\includegraphics[width=.5\textwidth]{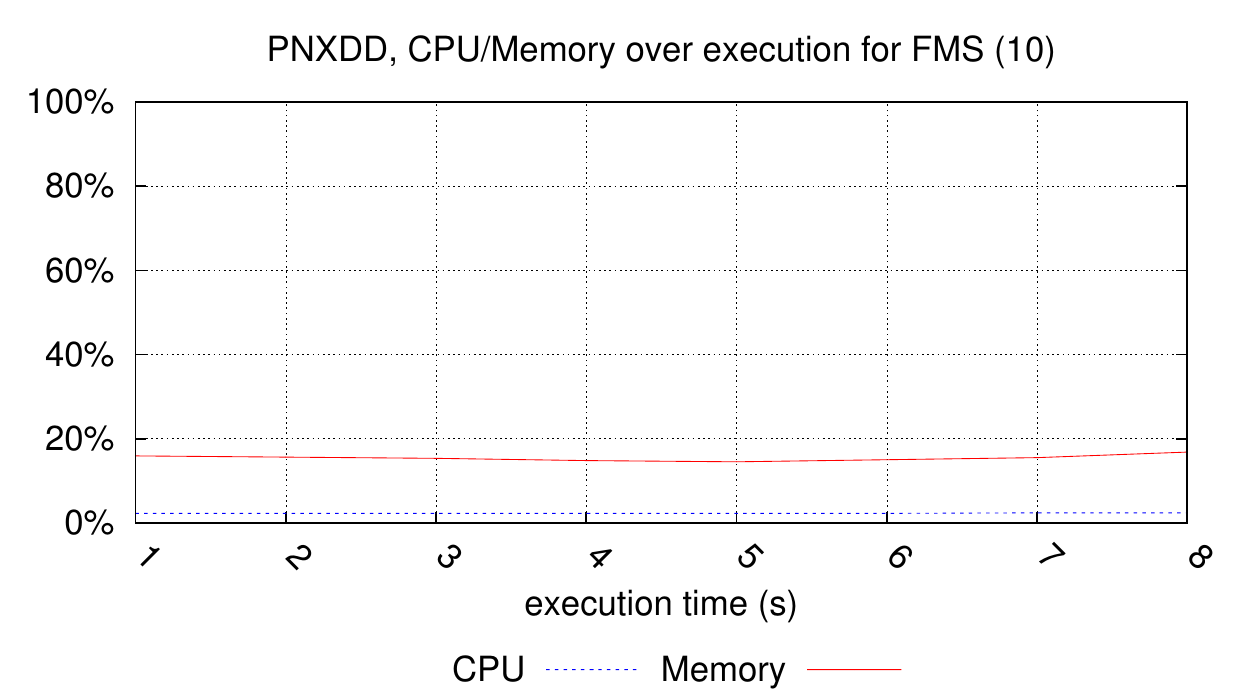}
\includegraphics[width=.5\textwidth]{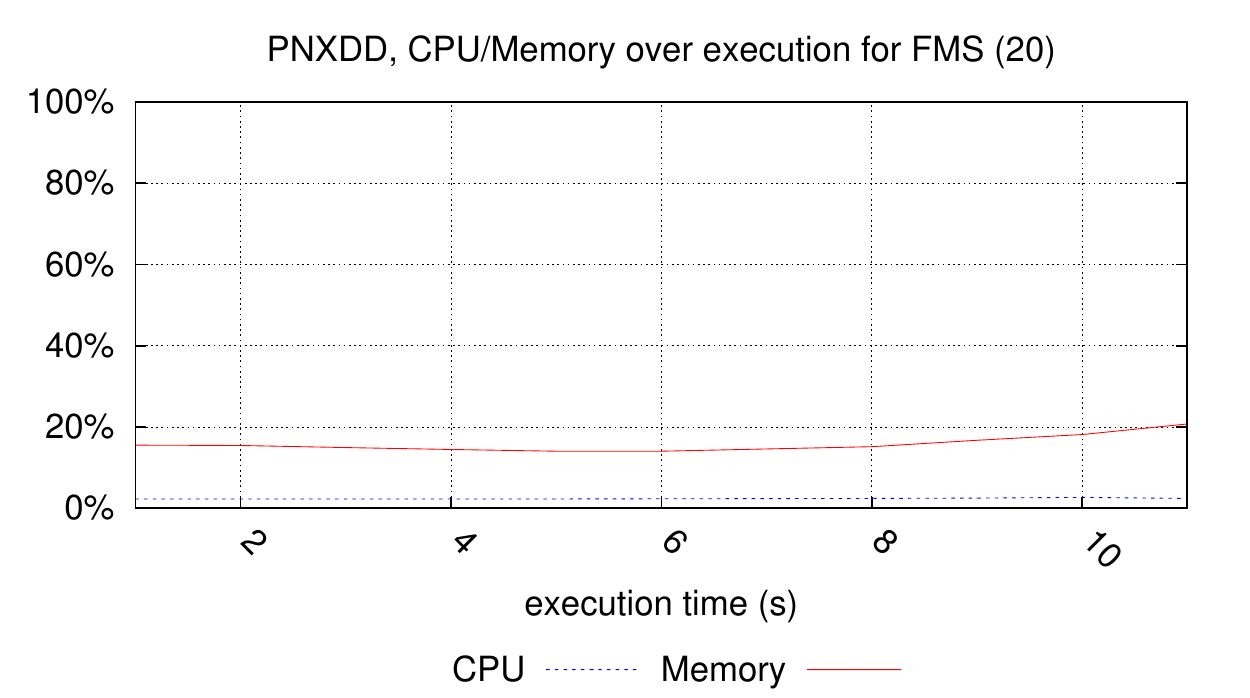}

\noindent\includegraphics[width=.5\textwidth]{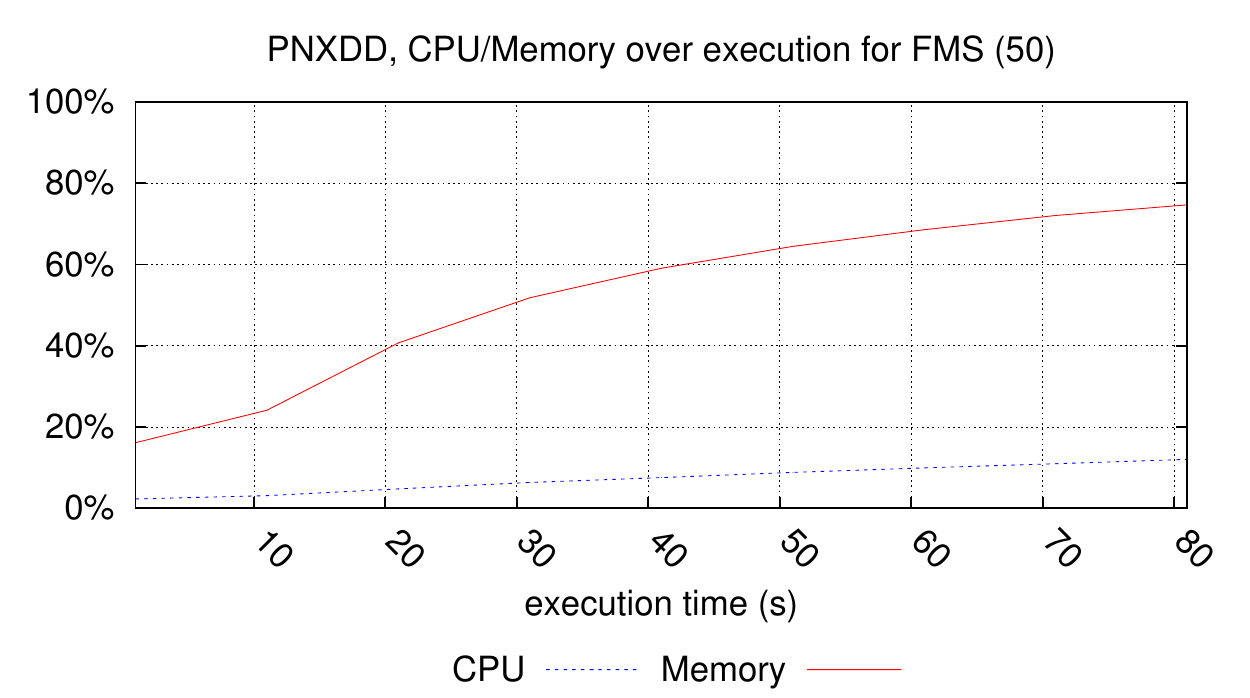}
\includegraphics[width=.5\textwidth]{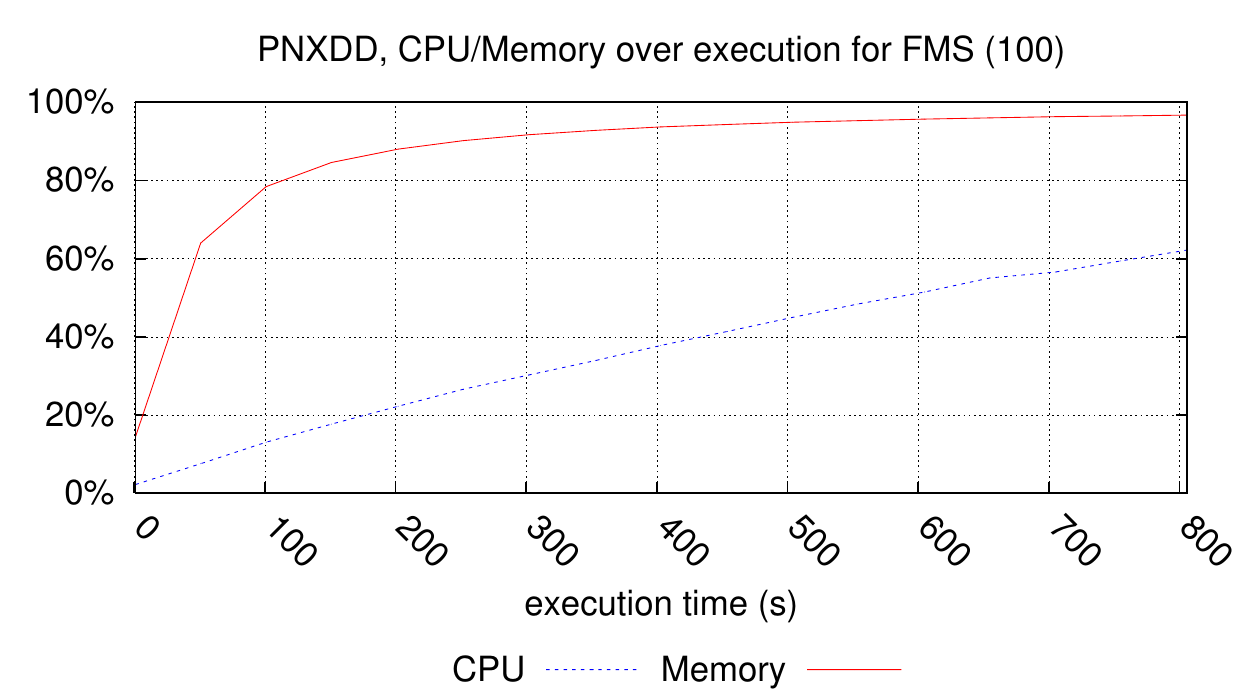}

\vfill\eject
\subsubsection{Executions for galloc\_res}
2 charts have been generated.
\index{Execution (by tool)!PNXDD}
\index{Execution (by model)!galloc\_res!PNXDD}

\noindent\includegraphics[width=.5\textwidth]{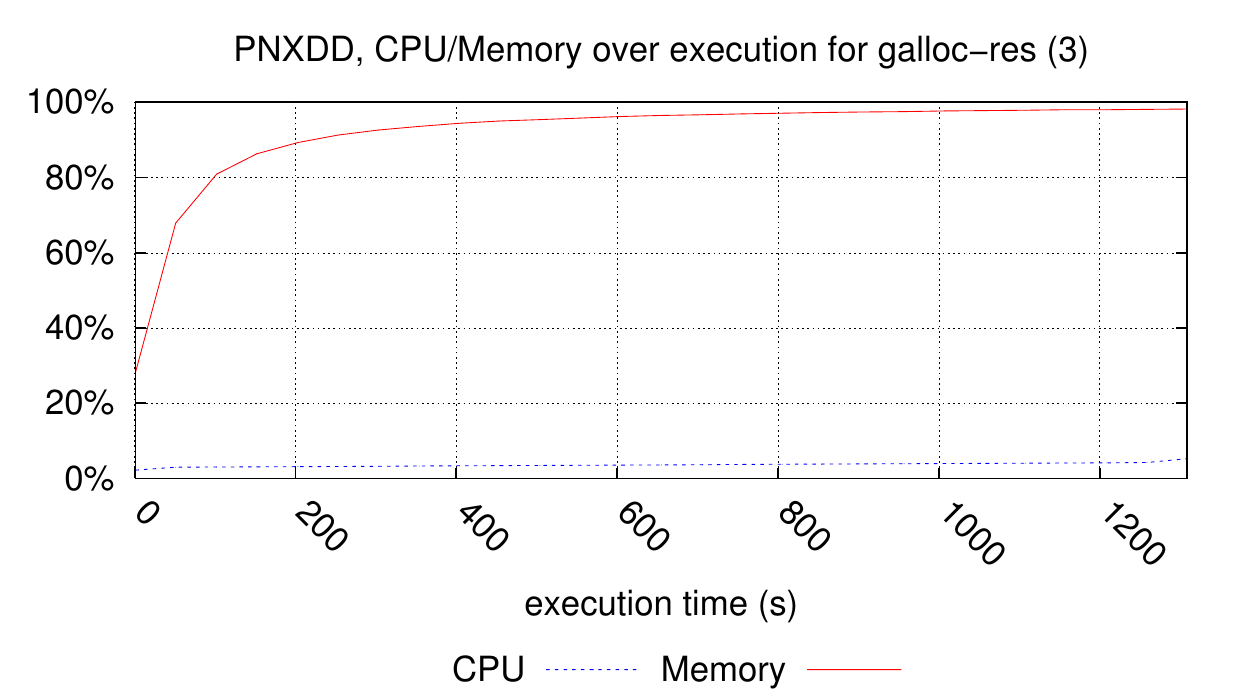}
\includegraphics[width=.5\textwidth]{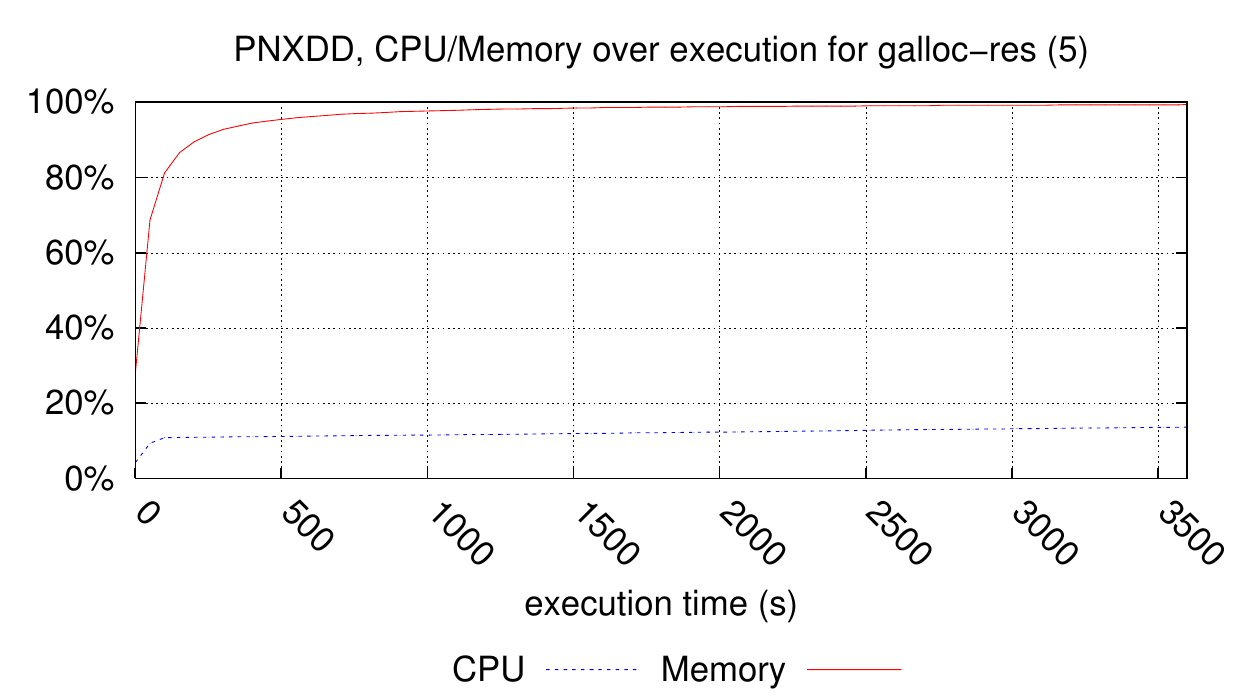}

\subsubsection{Executions for Kanban}
5 charts have been generated.
\index{Execution (by tool)!PNXDD}
\index{Execution (by model)!Kanban!PNXDD}

\noindent\includegraphics[width=.5\textwidth]{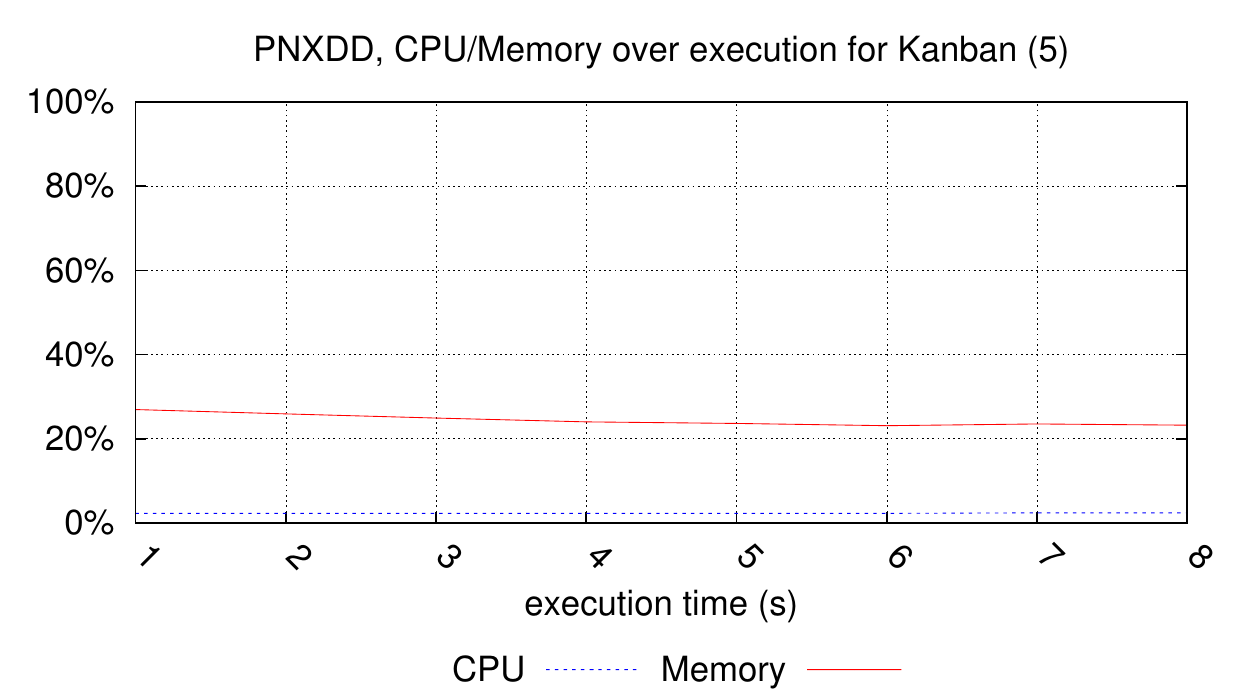}
\includegraphics[width=.5\textwidth]{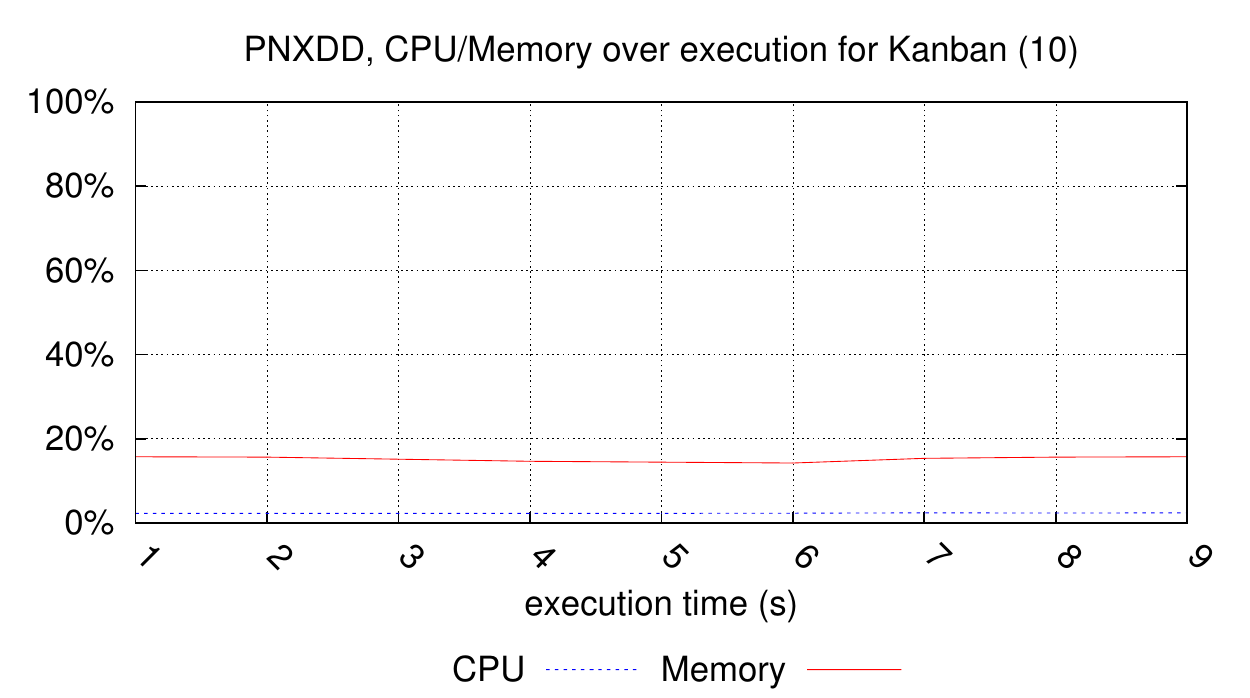}

\noindent\includegraphics[width=.5\textwidth]{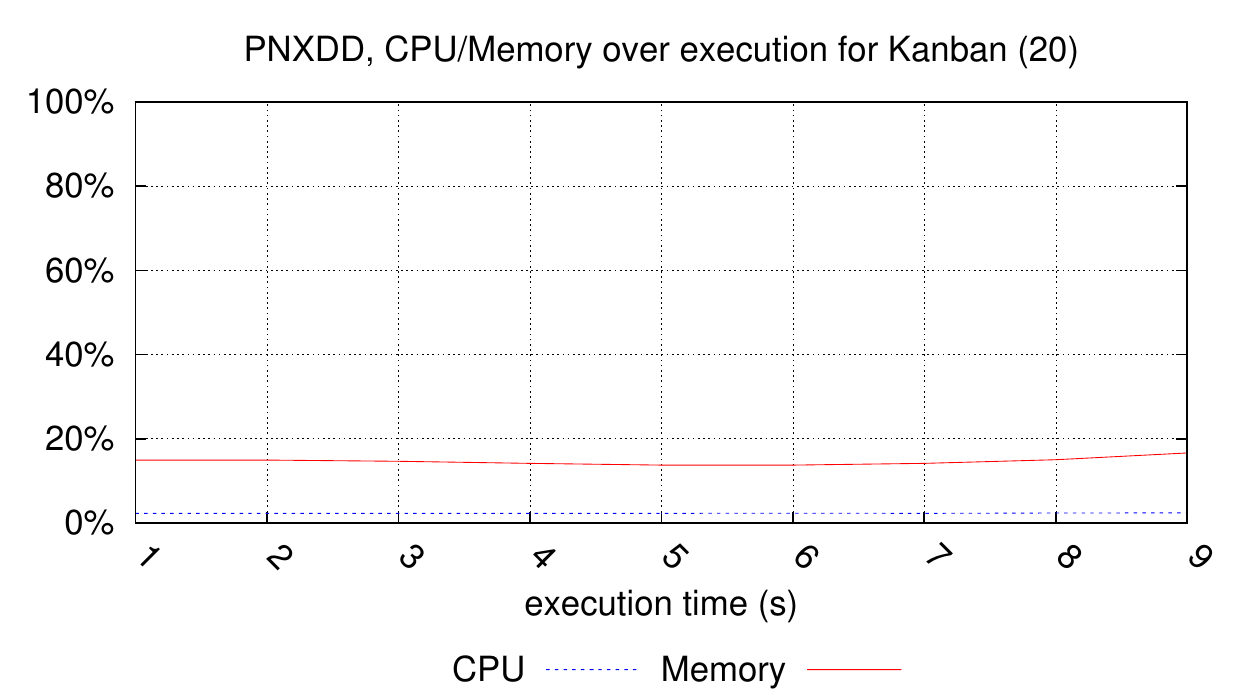}
\includegraphics[width=.5\textwidth]{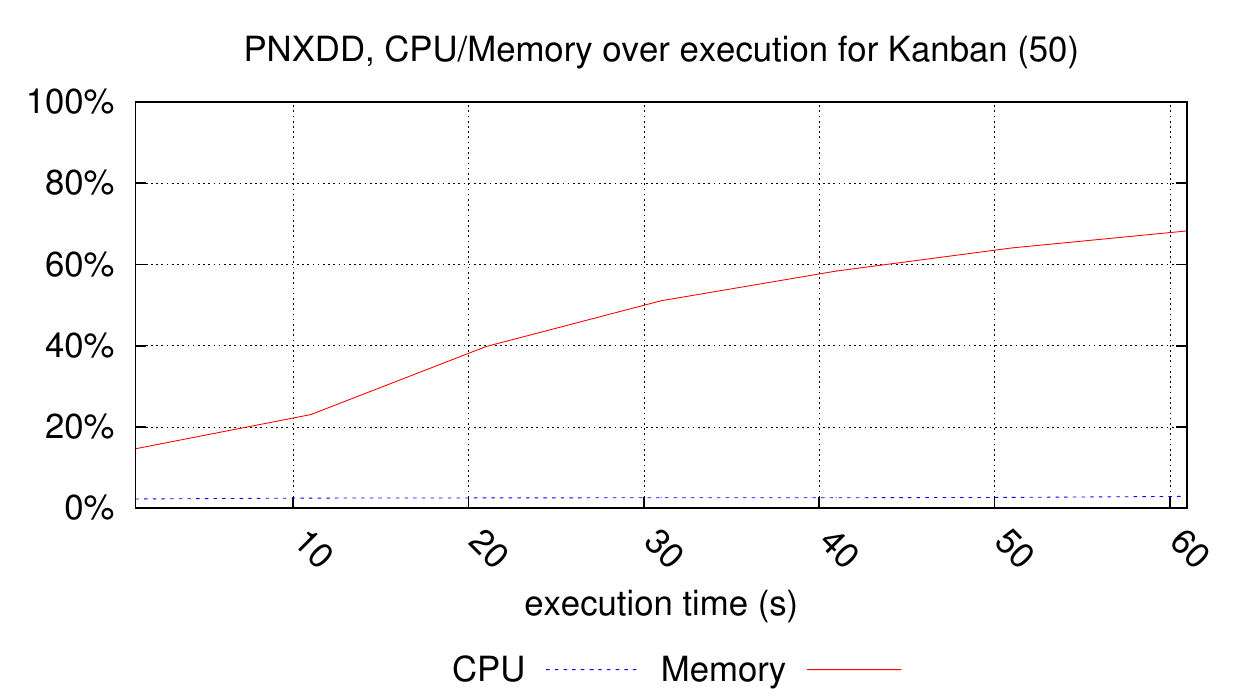}

\noindent\includegraphics[width=.5\textwidth]{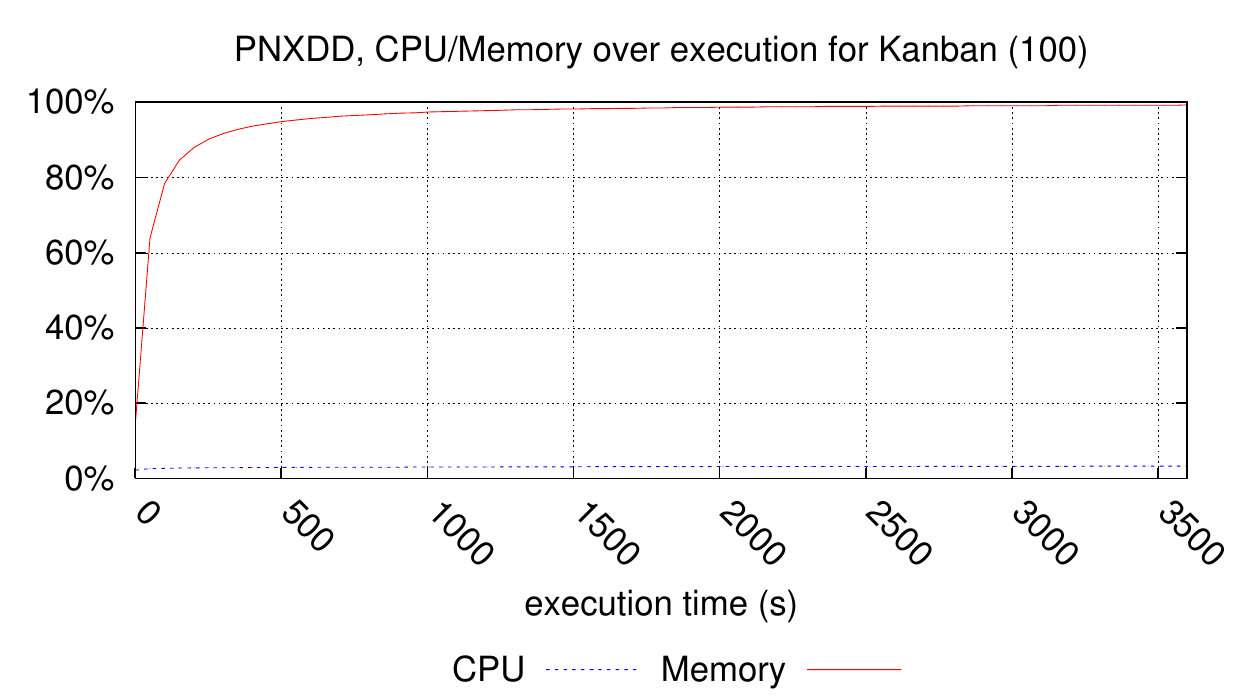}

\vfill\eject
\subsubsection{Executions for MAPK}
5 charts have been generated.
\index{Execution (by tool)!PNXDD}
\index{Execution (by model)!MAPK!PNXDD}

\noindent\includegraphics[width=.5\textwidth]{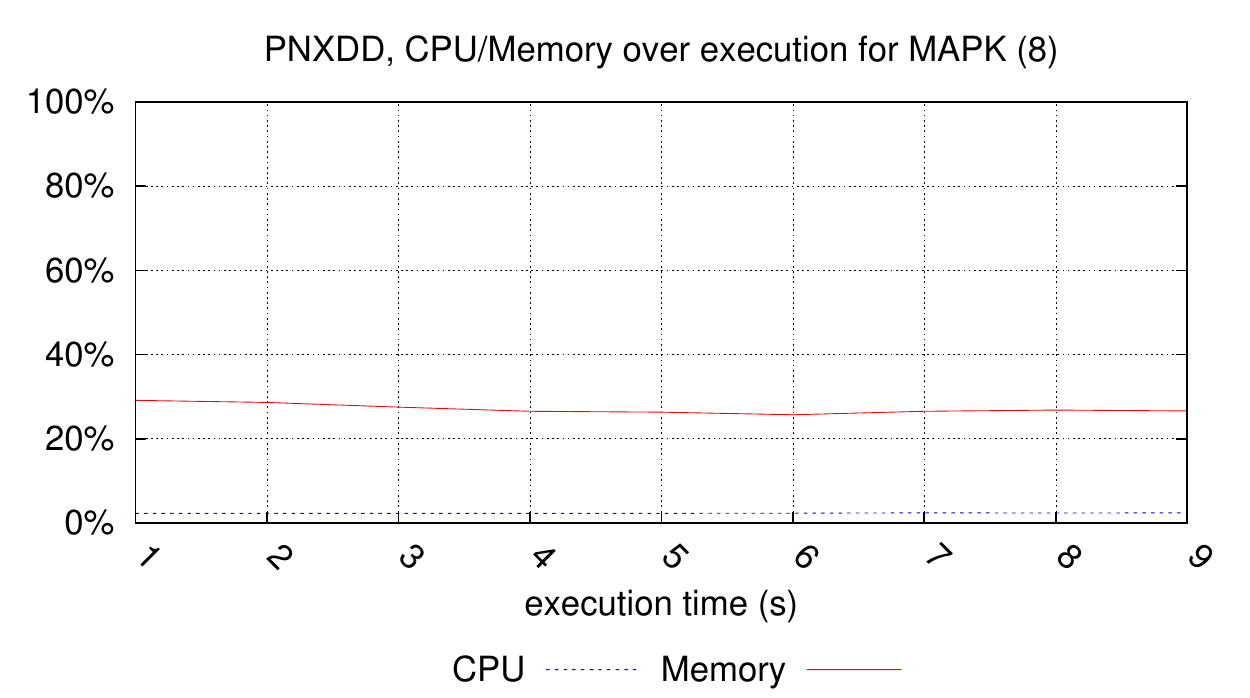}
\includegraphics[width=.5\textwidth]{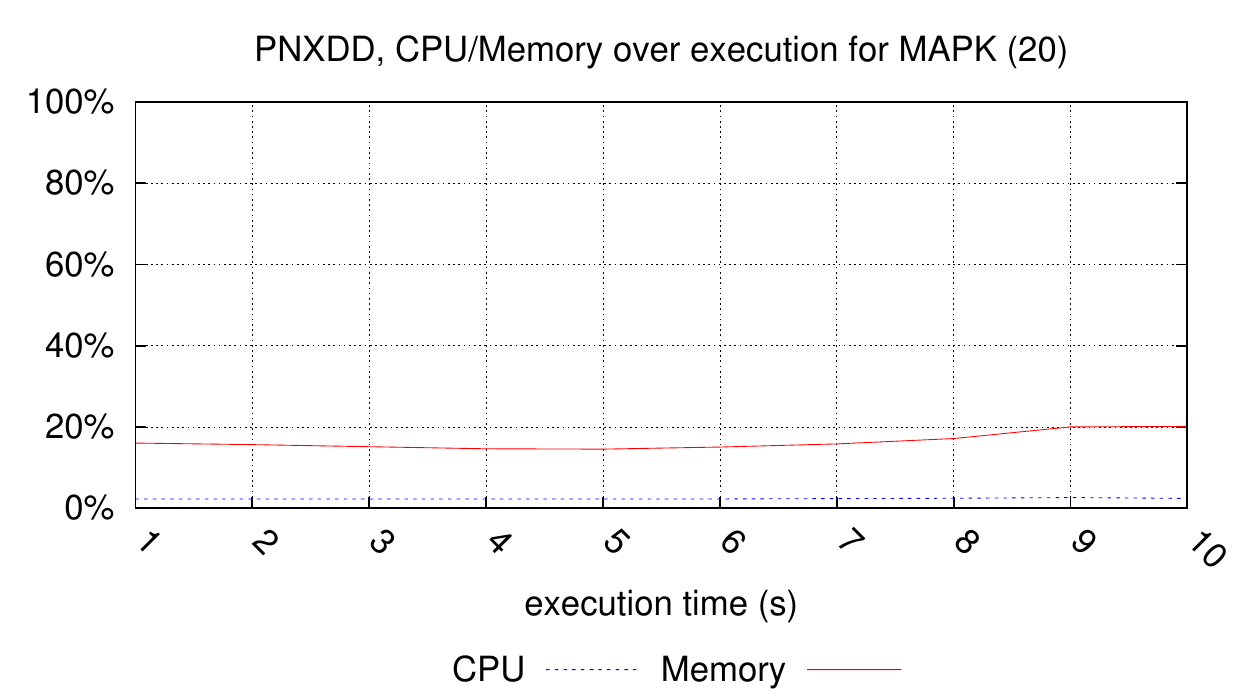}

\noindent\includegraphics[width=.5\textwidth]{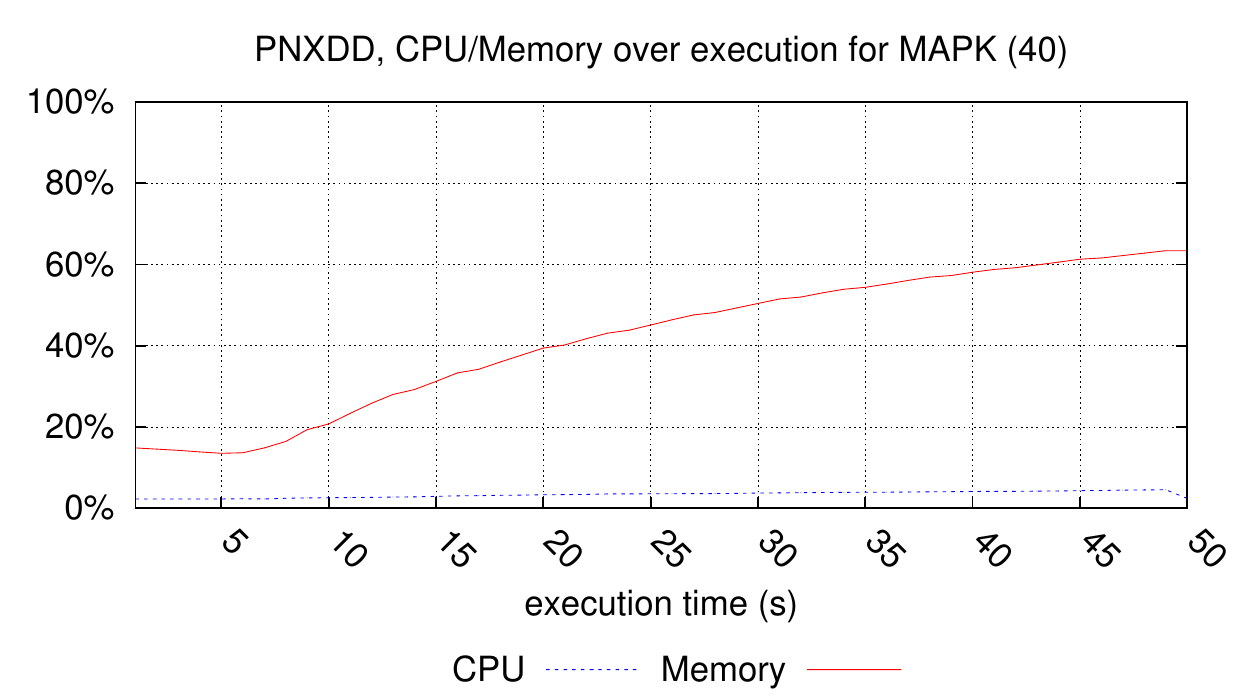}
\includegraphics[width=.5\textwidth]{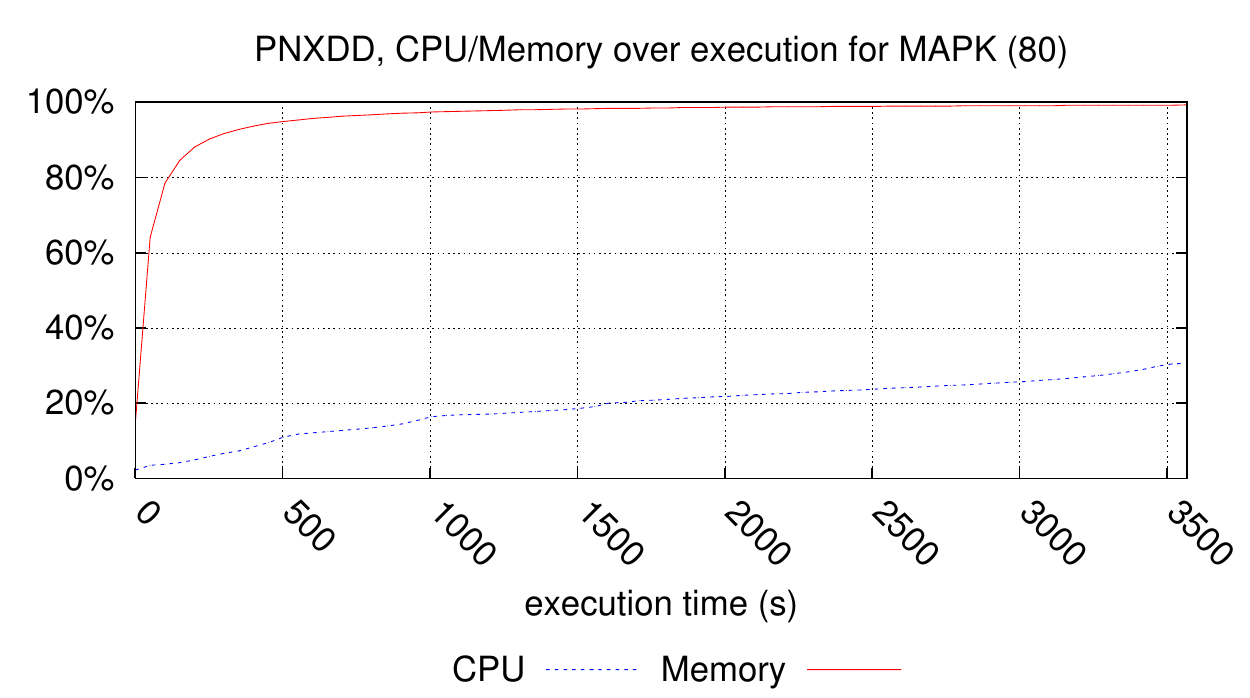}

\noindent\includegraphics[width=.5\textwidth]{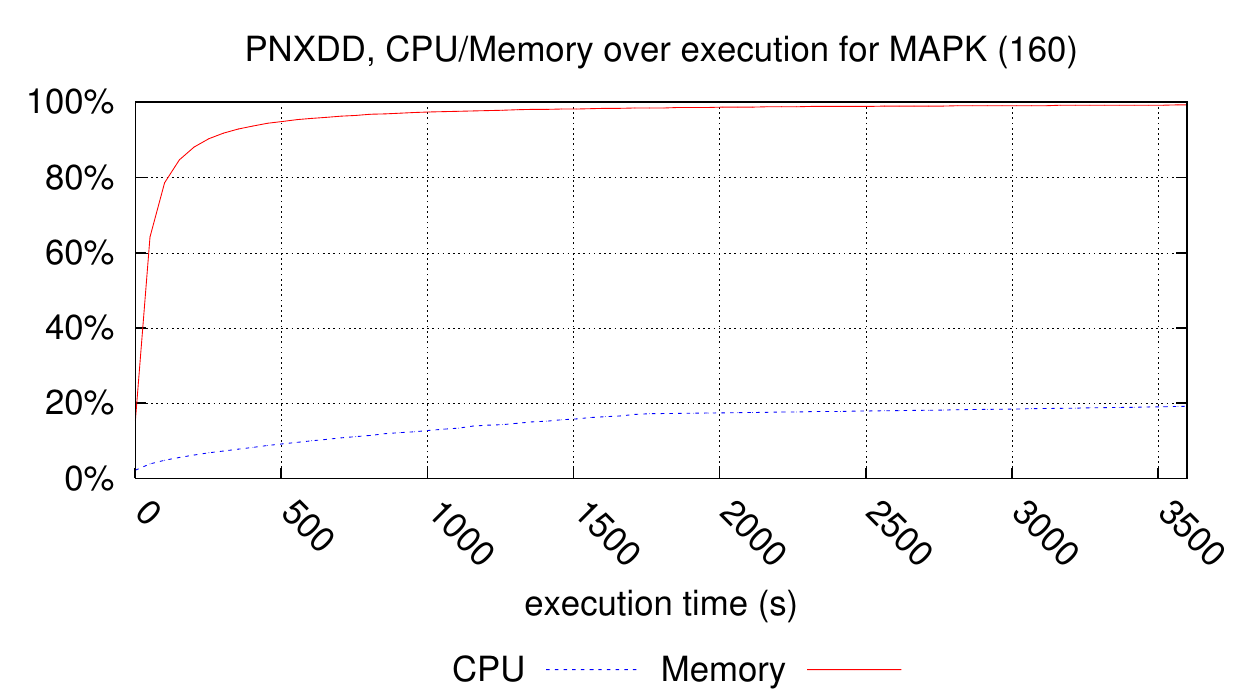}

\subsubsection{Executions for Peterson}
4 charts have been generated.
\index{Execution (by tool)!PNXDD}
\index{Execution (by model)!Peterson!PNXDD}

\noindent\includegraphics[width=.5\textwidth]{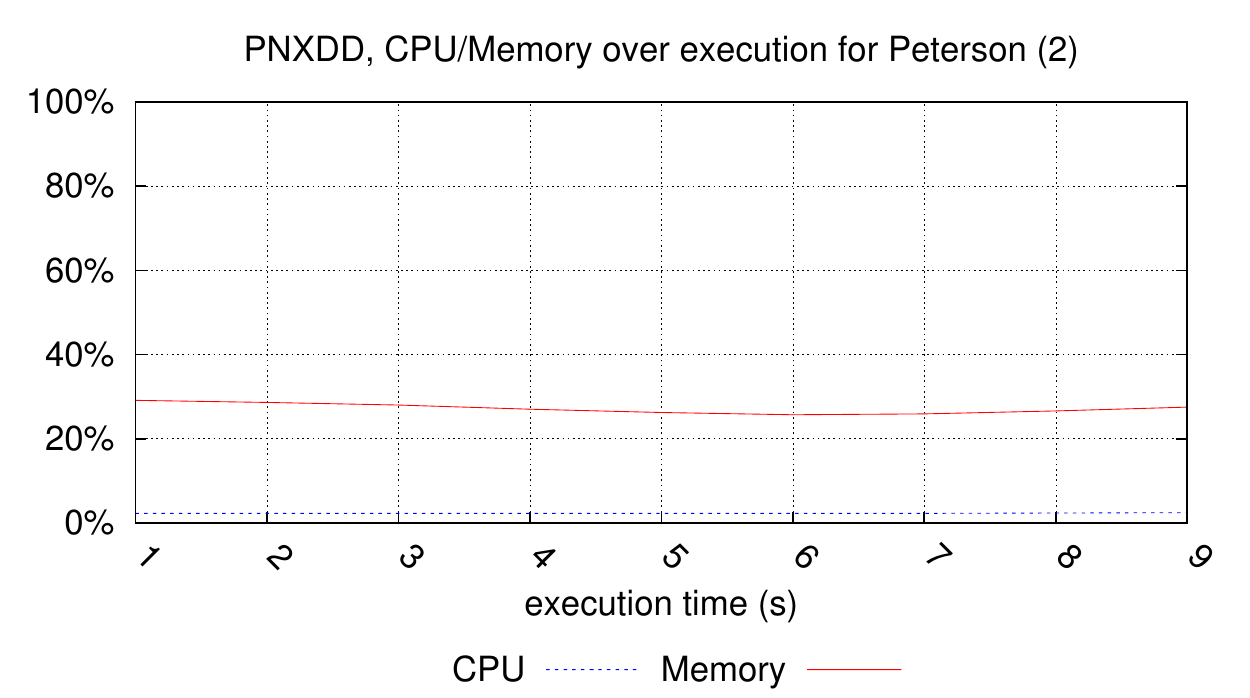}
\includegraphics[width=.5\textwidth]{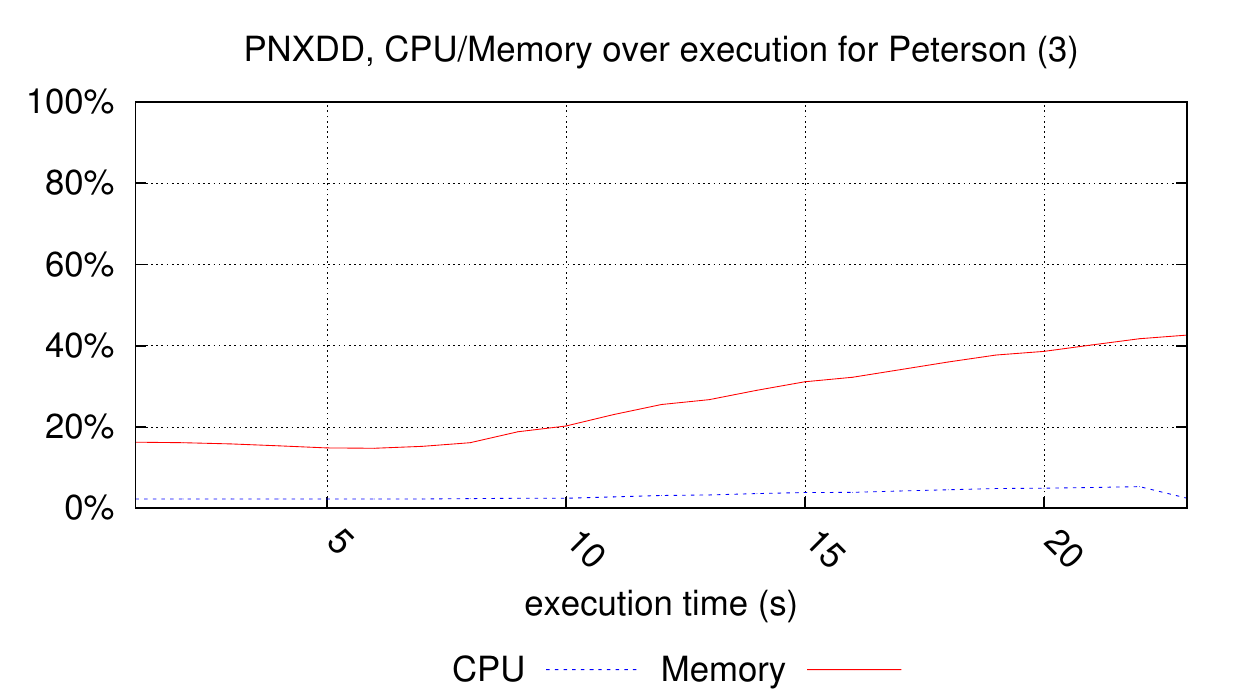}

\noindent\includegraphics[width=.5\textwidth]{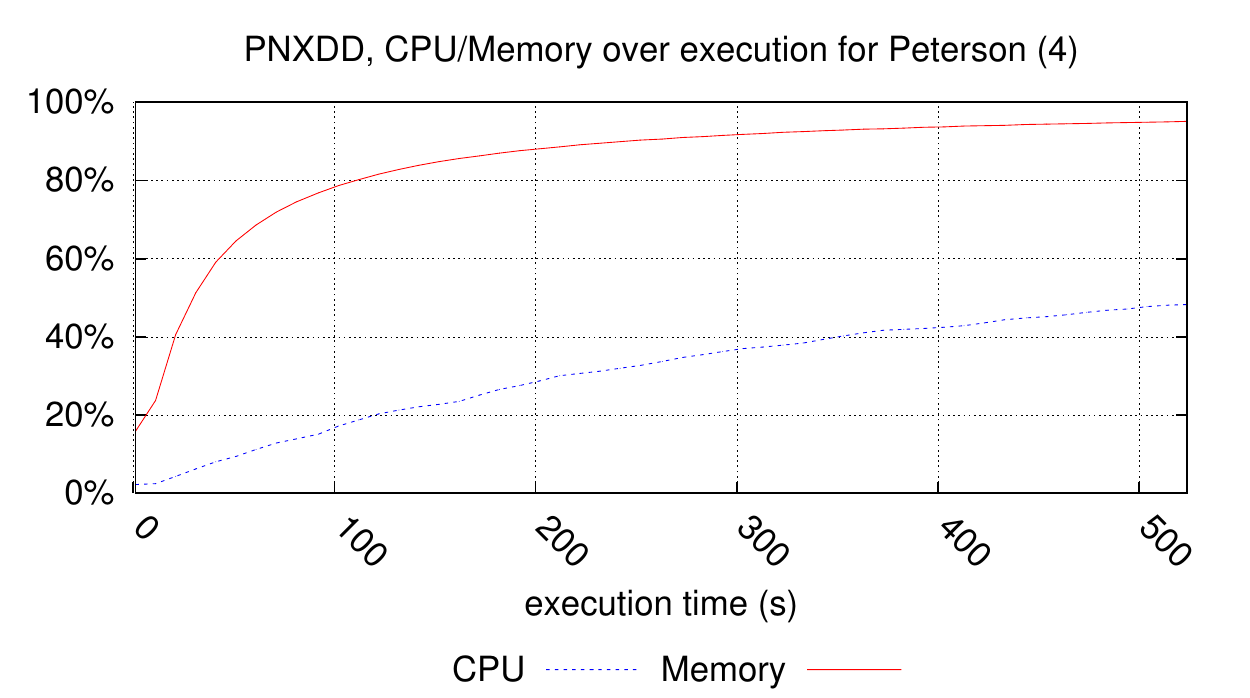}
\includegraphics[width=.5\textwidth]{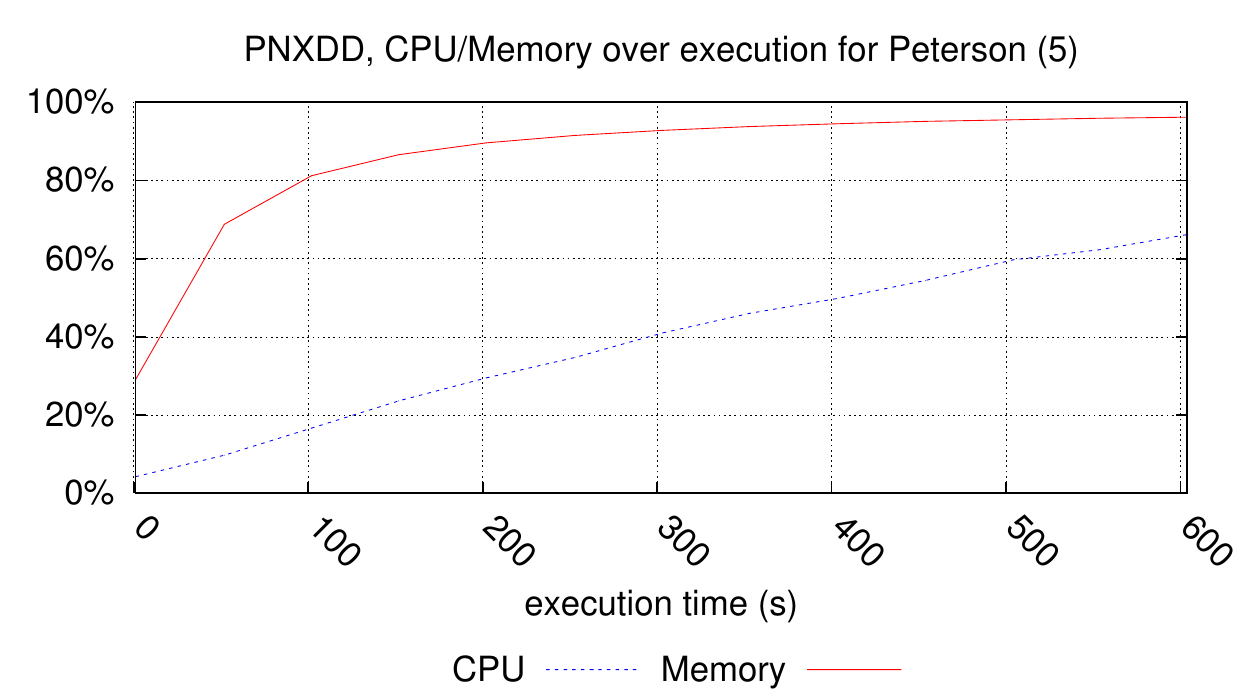}

\subsubsection{Executions for philo\_dyn}
3 charts have been generated.
\index{Execution (by tool)!PNXDD}
\index{Execution (by model)!philo\_dyn!PNXDD}

\noindent\includegraphics[width=.5\textwidth]{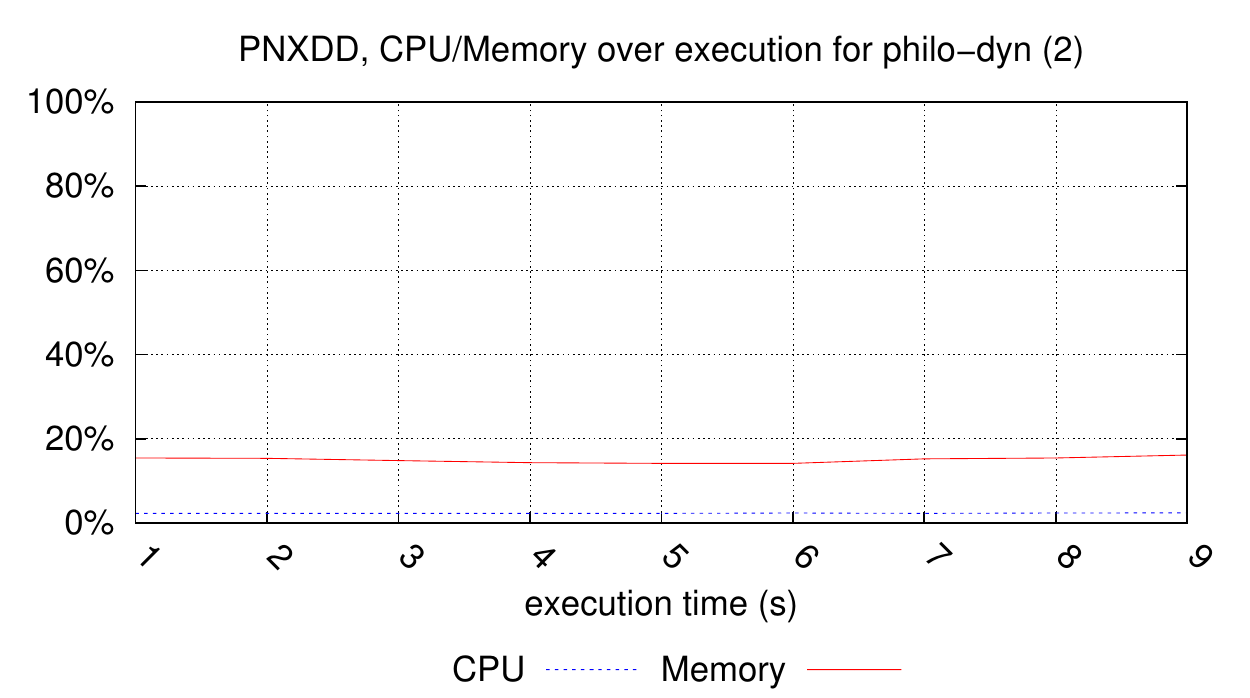}
\includegraphics[width=.5\textwidth]{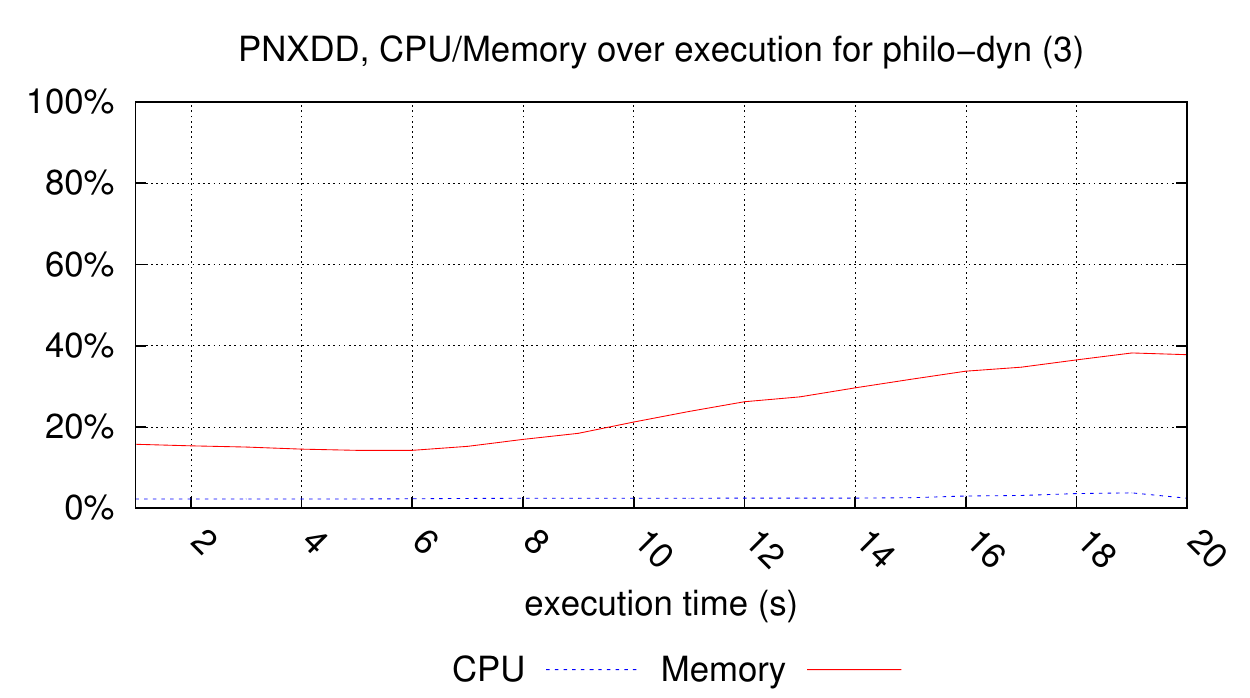}

\noindent\includegraphics[width=.5\textwidth]{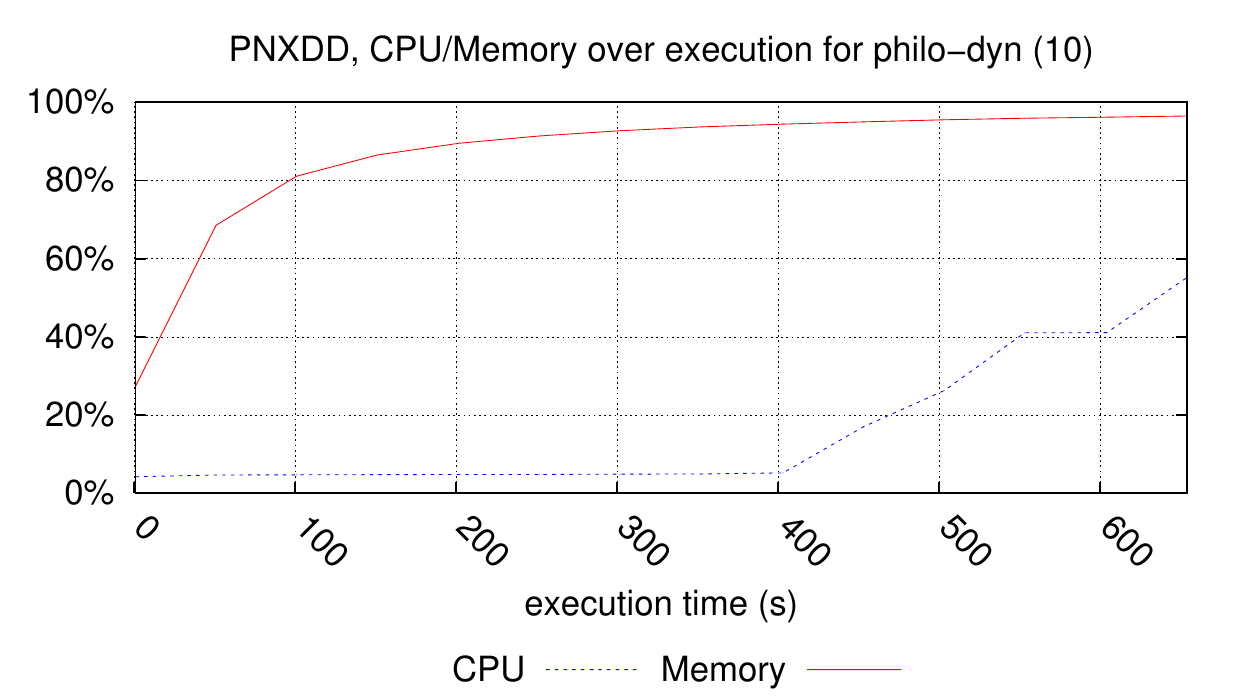}

\subsubsection{Executions for Philosophers}
6 charts have been generated.
\index{Execution (by tool)!PNXDD}
\index{Execution (by model)!Philosophers!PNXDD}

\noindent\includegraphics[width=.5\textwidth]{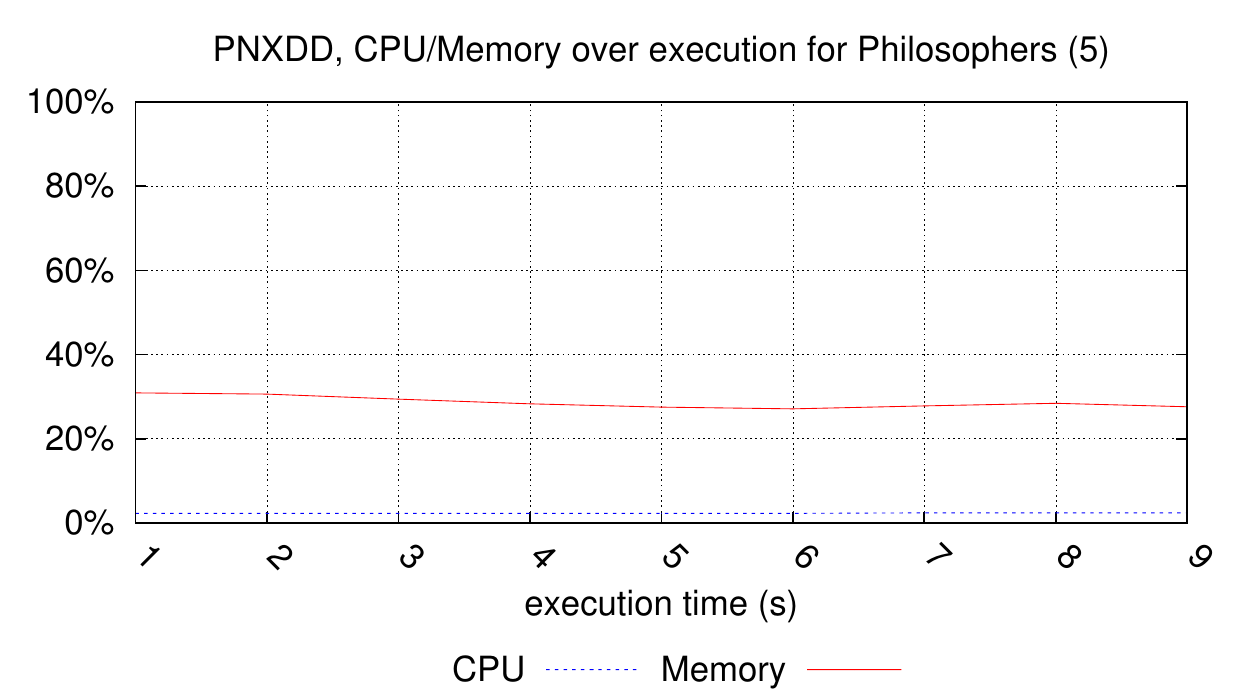}
\includegraphics[width=.5\textwidth]{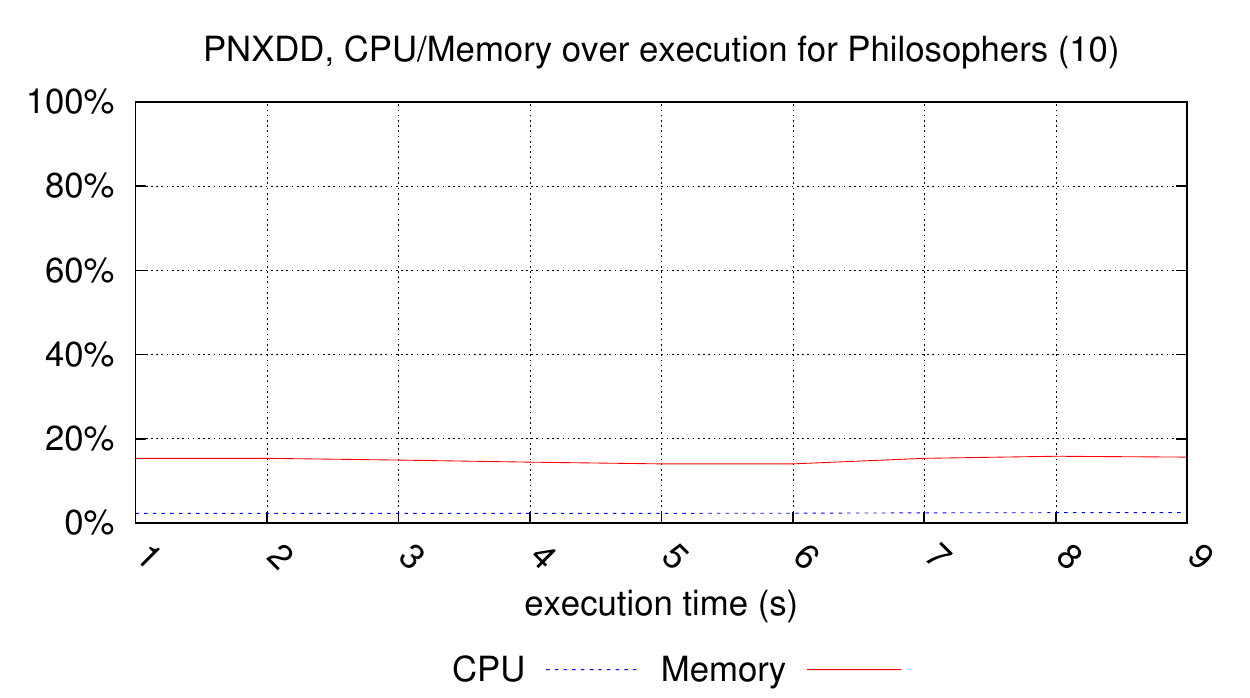}

\noindent\includegraphics[width=.5\textwidth]{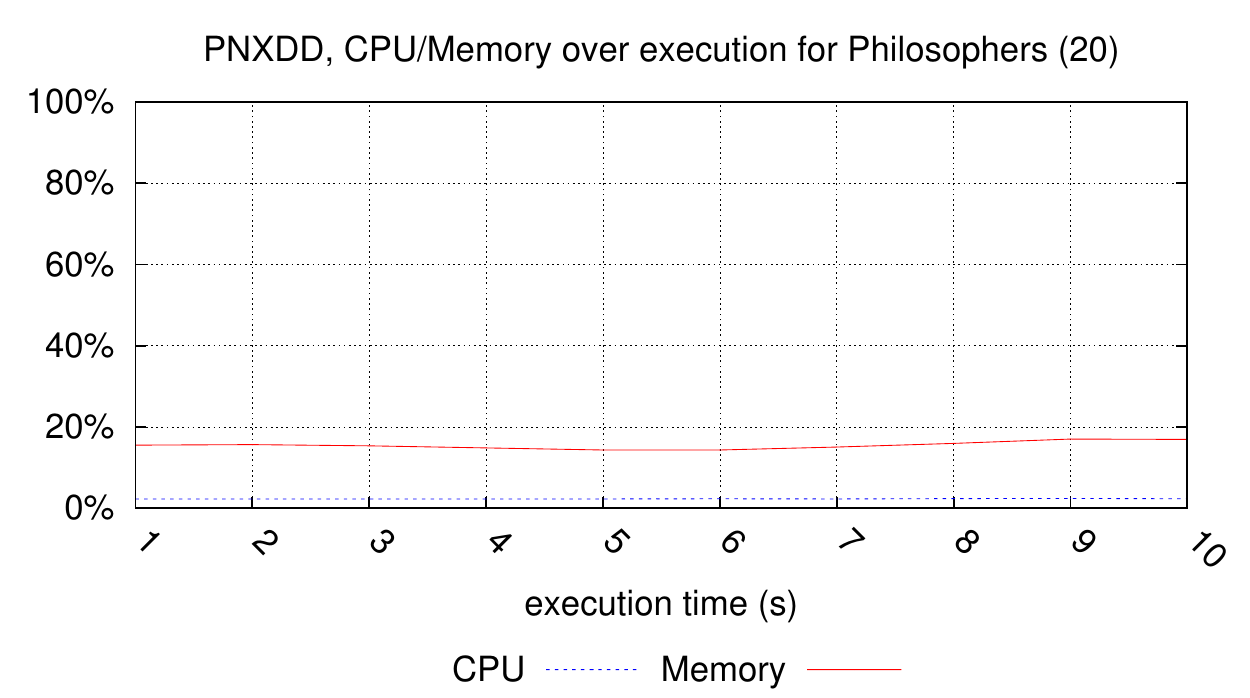}
\includegraphics[width=.5\textwidth]{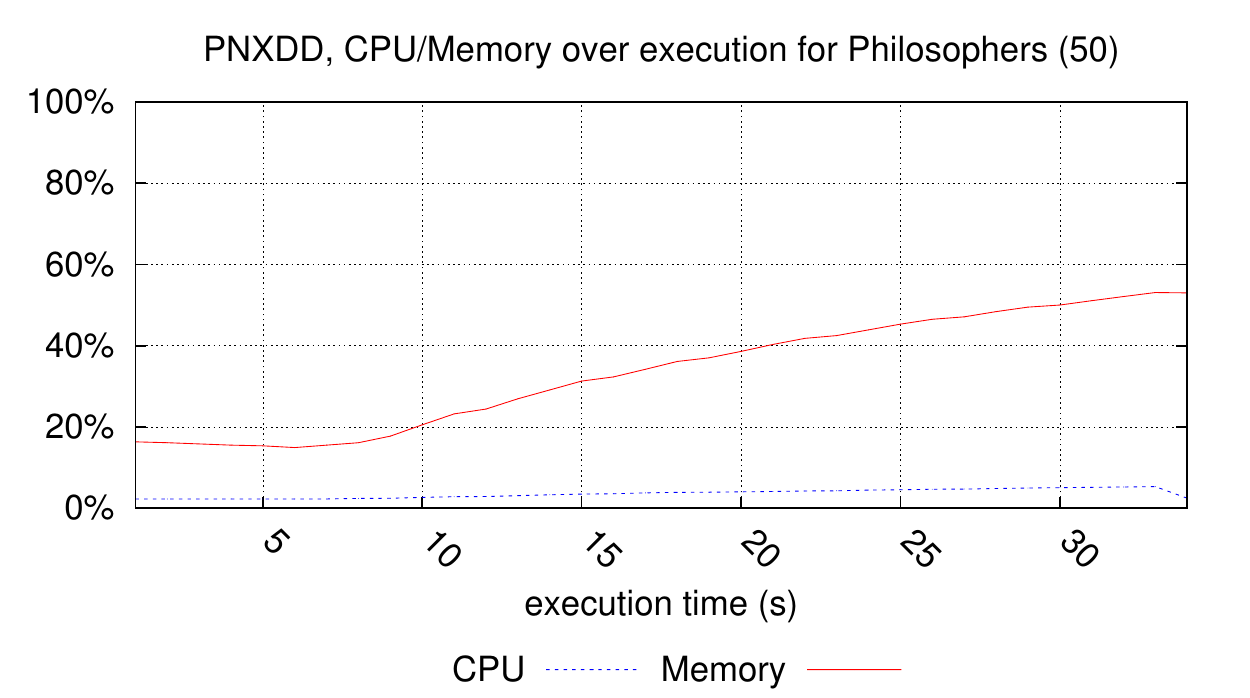}

\noindent\includegraphics[width=.5\textwidth]{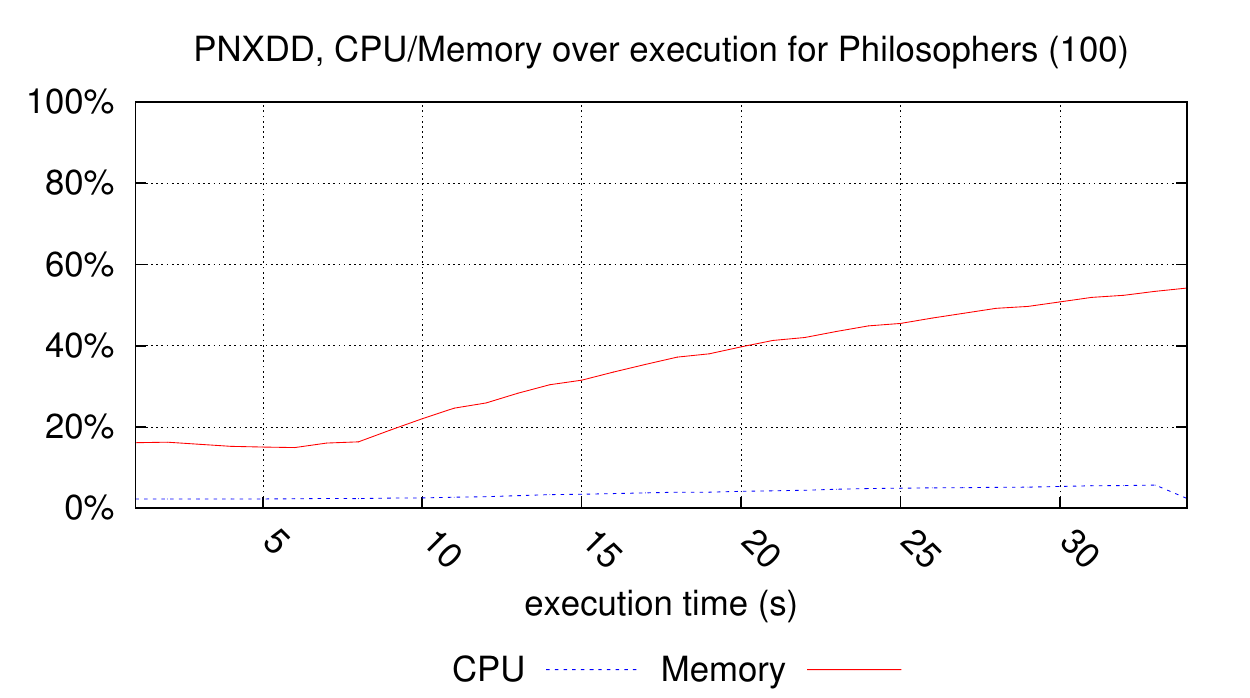}
\includegraphics[width=.5\textwidth]{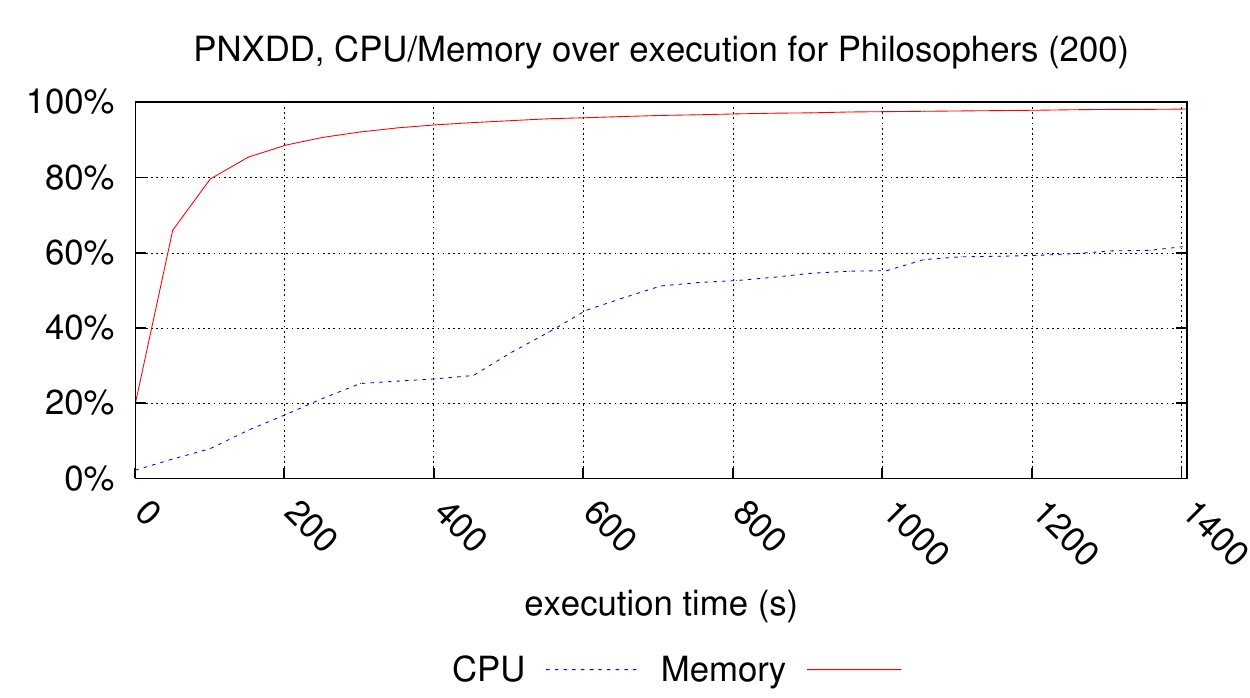}

\subsubsection{Executions for SharedMemory}
4 charts have been generated.
\index{Execution (by tool)!PNXDD}
\index{Execution (by model)!SharedMemory!PNXDD}

\noindent\includegraphics[width=.5\textwidth]{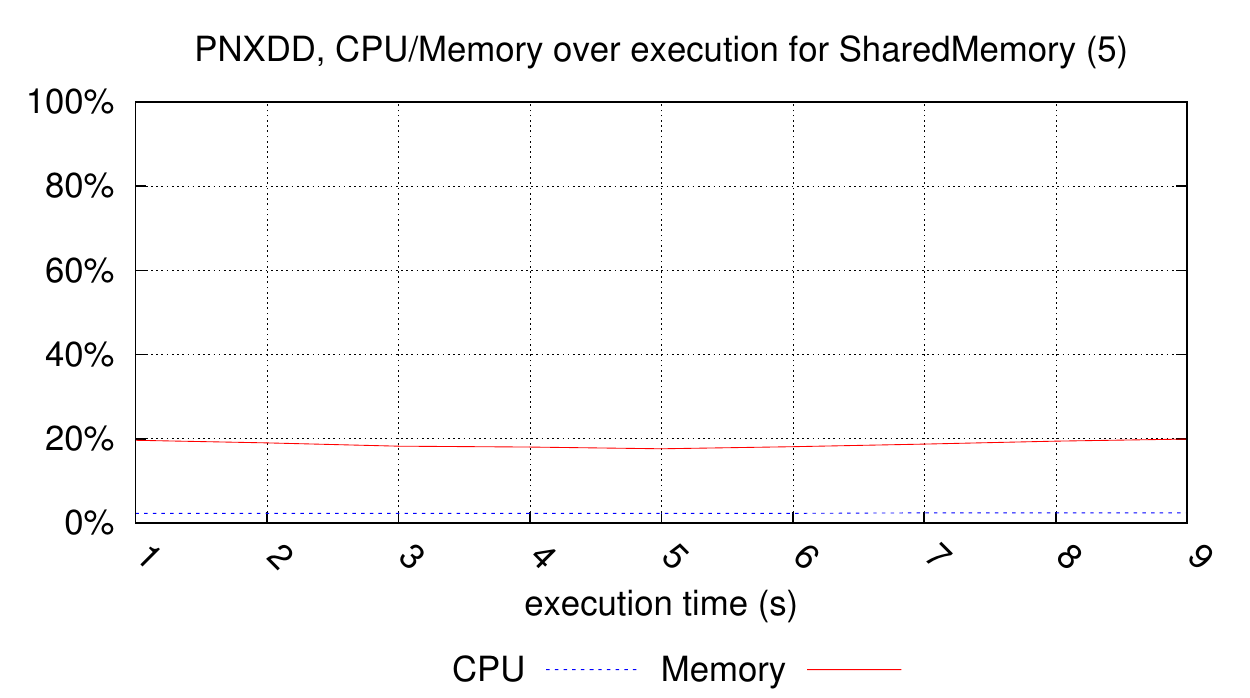}
\includegraphics[width=.5\textwidth]{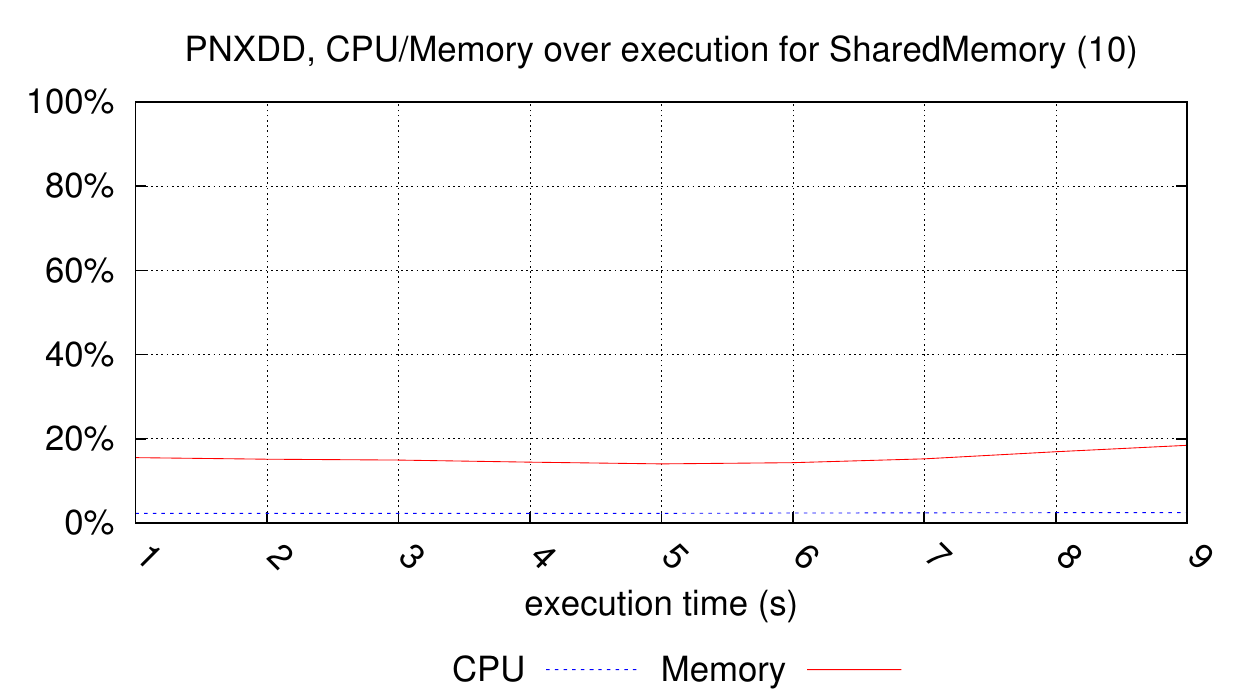}

\noindent\includegraphics[width=.5\textwidth]{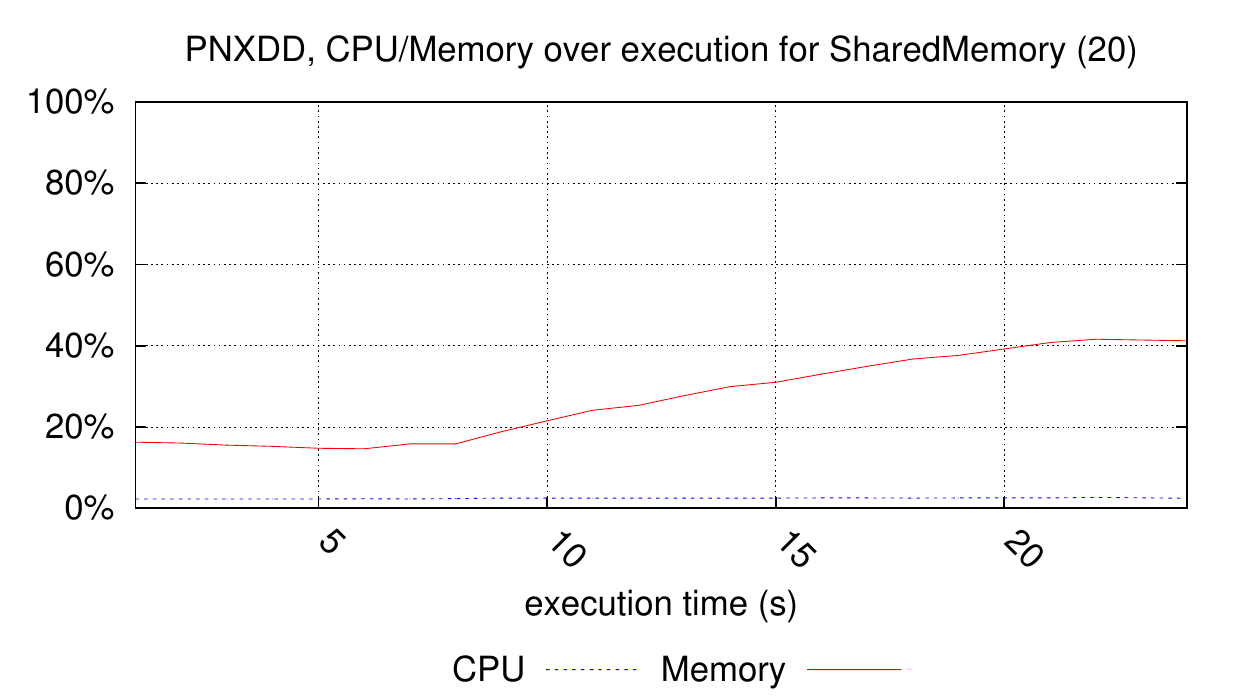}
\includegraphics[width=.5\textwidth]{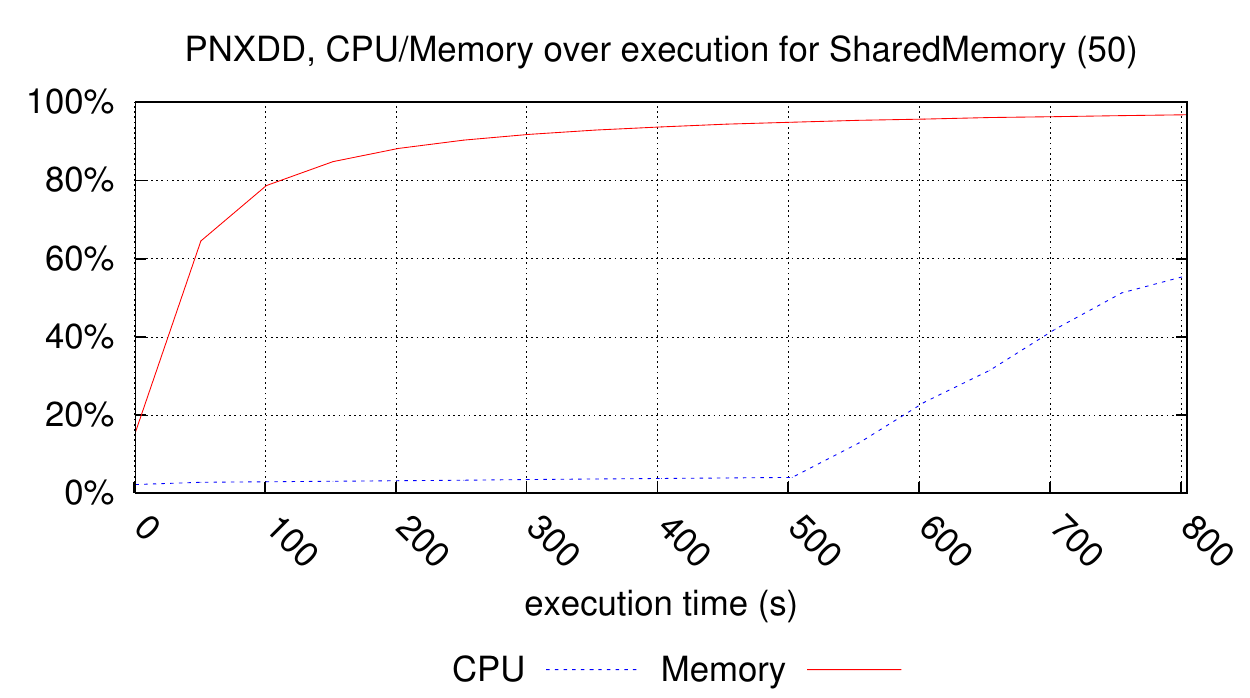}

\vfill\eject
\subsubsection{Executions for simple\_lbs}
5 charts have been generated.
\index{Execution (by tool)!PNXDD}
\index{Execution (by model)!simple\_lbs!PNXDD}

\noindent\includegraphics[width=.5\textwidth]{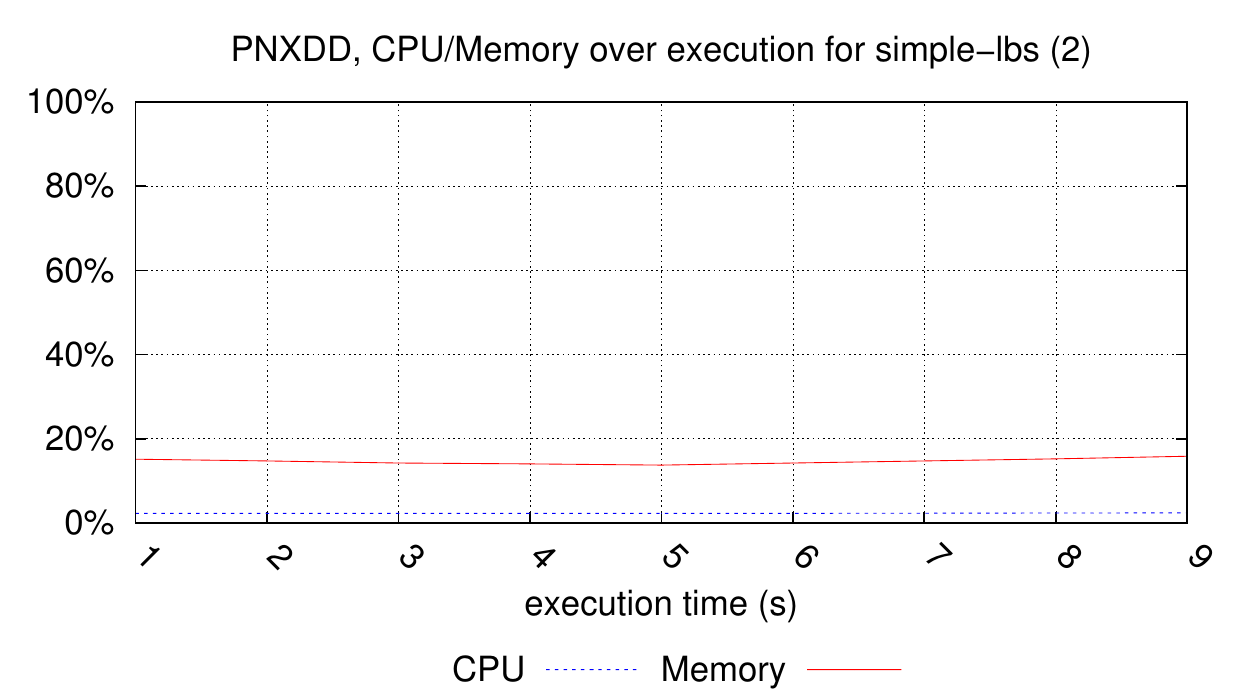}
\includegraphics[width=.5\textwidth]{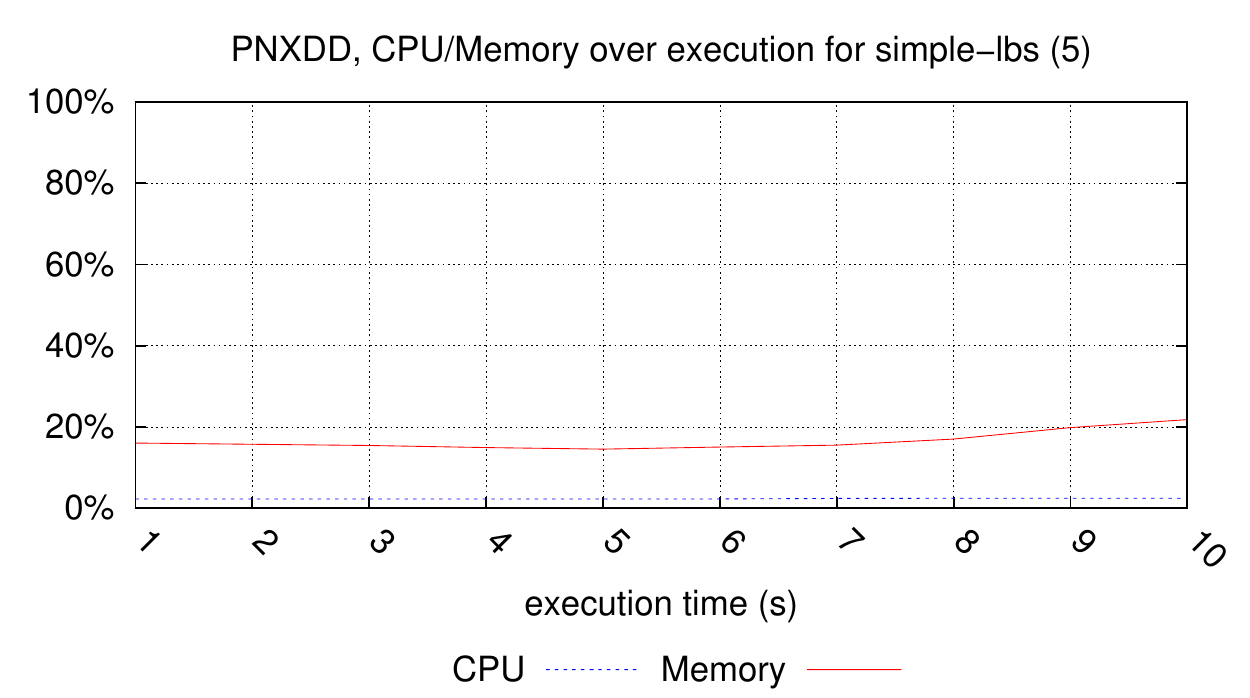}

\noindent\includegraphics[width=.5\textwidth]{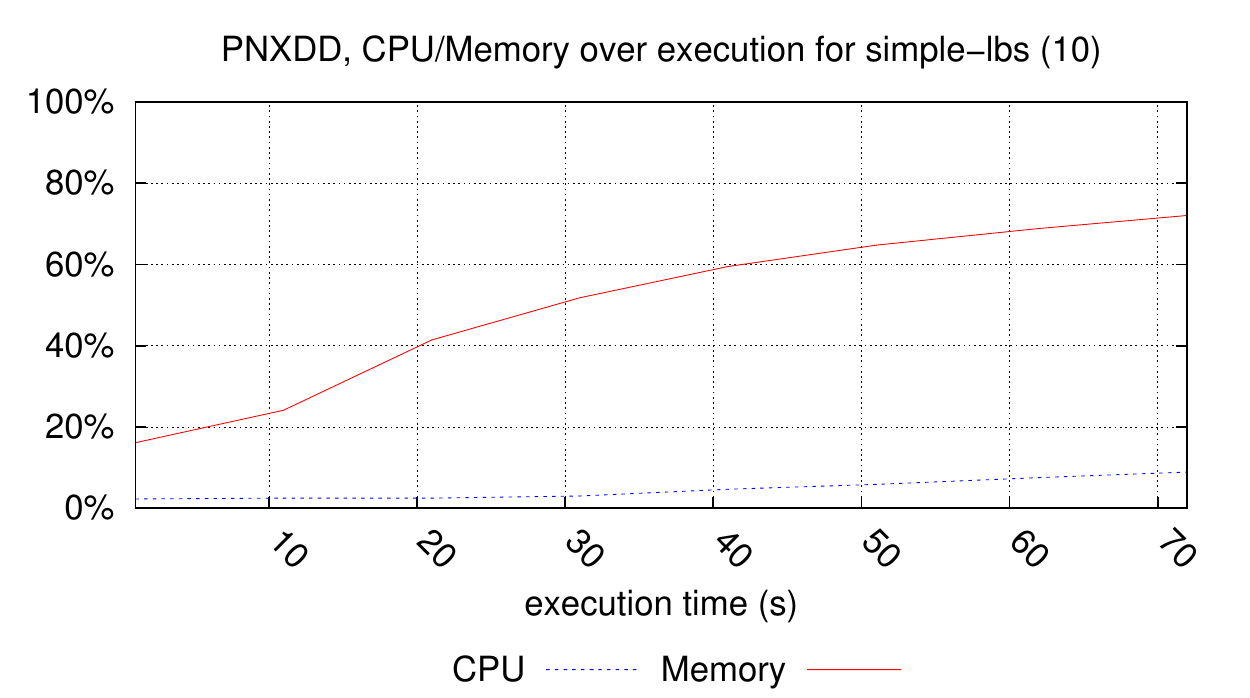}
\includegraphics[width=.5\textwidth]{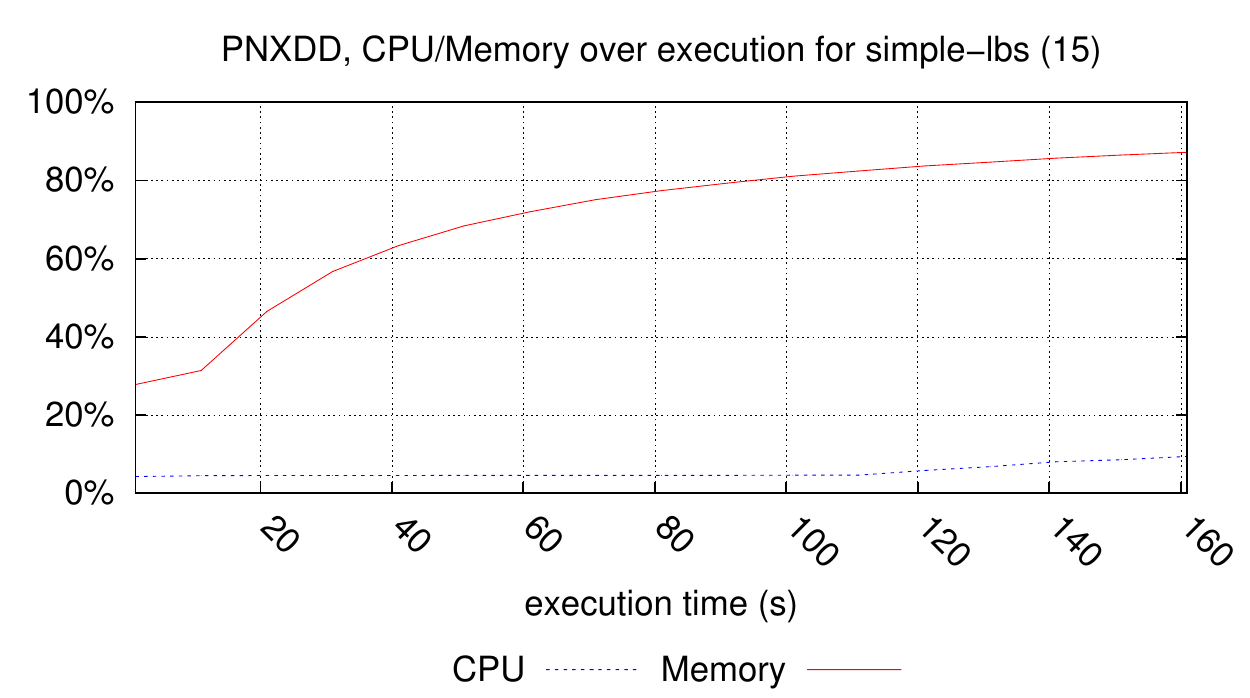}

\noindent\includegraphics[width=.5\textwidth]{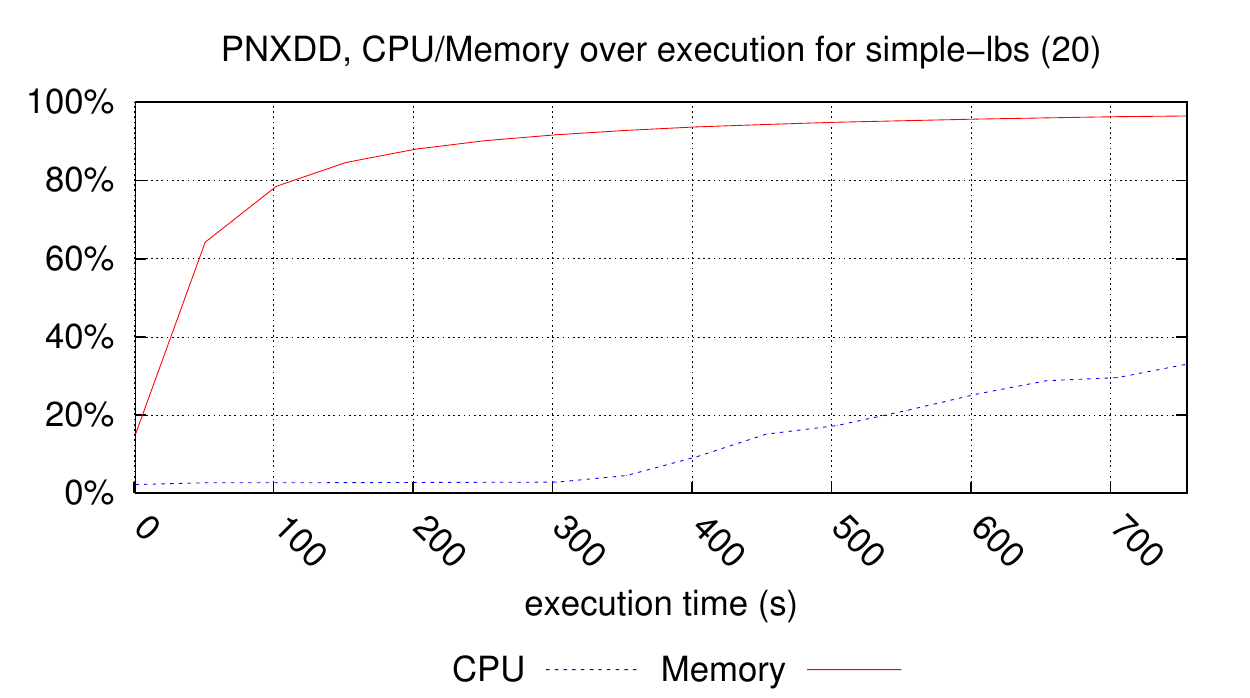}

\subsubsection{Executions for TokenRing}
3 charts have been generated.
\index{Execution (by tool)!PNXDD}
\index{Execution (by model)!TokenRing!PNXDD}

\noindent\includegraphics[width=.5\textwidth]{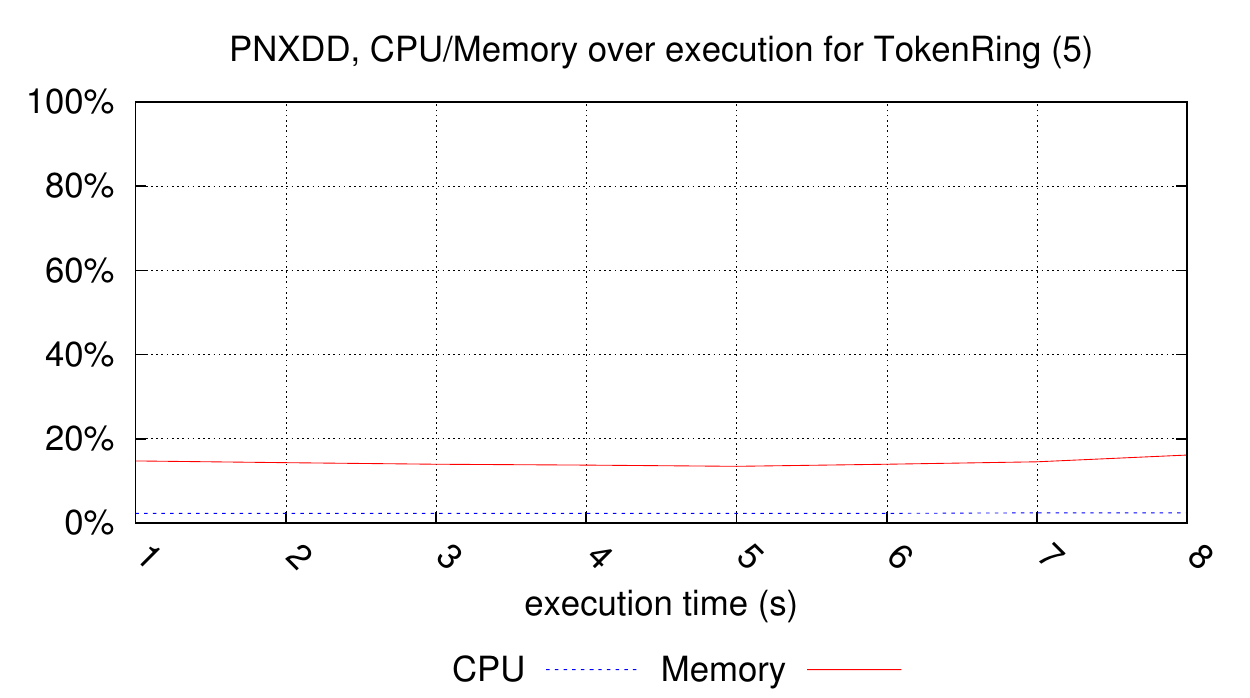}
\includegraphics[width=.5\textwidth]{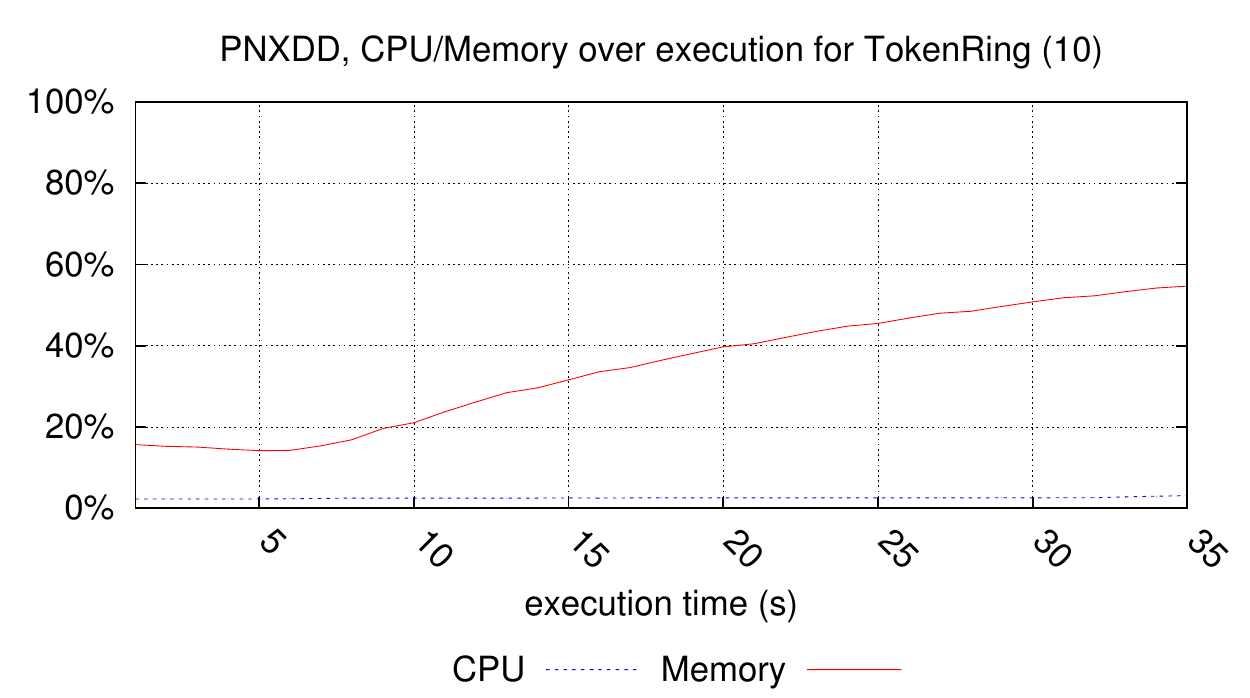}

\noindent\includegraphics[width=.5\textwidth]{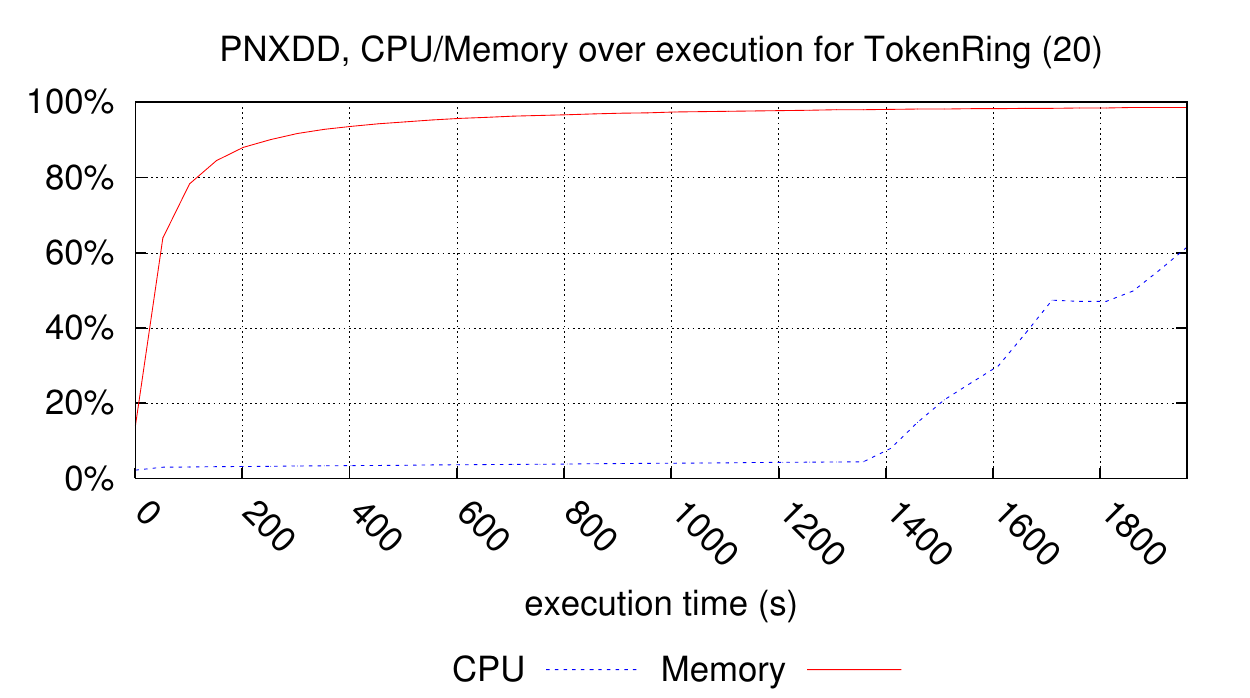}

\clearpage
\section{Raw Data for Structural Formul{\ae} Evaluation}
\label{sec:structformulae}

This section shows the raw results of the structural formul{\ae} examination.
\Cref{tab:fs}~summarizes the highest scaling parameter reached by the tools
for each model. Then Charts generated from the data collected for this
examination are provided. This table should be interpreted using the legend
displayed in~\Cref{position:legend}, page~\pageref{position:legend}.
Let us not that only two tools did participate in the evaluation of structural
formul{\ae}: \acs{AlPiNA} and \acs{Helena}. \acs{LoLA-binstore},
\acs{LoLA-bloom}
and \acs{Sara} could not
compete because of the combination of structural assertions.

\index{Structural Formul{\ae}!Tool results (summary)}
\begin{table}[h]
	\centering
  \lstset{
    basicstyle=\scriptsize\fontfamily{fvm}\selectfont
  }
  \footnotesize
  \renewcommand\cellalign{cc}
  \setlength\rotheadsize{5em}
  \hspace{-1em}
  \begin{tabular}{|c|c||m{5em}|m{1.5em}|m{5em}|m{1.5em}|m{2.5em}|m{2.5em}|m{1.5em}|m{1.5em}|m{1.5em}|m{1.5em}|}
    \cline{3-12}
    \multicolumn{2}{c|}{}
  & \rothead{\textbf{AlPiNA}}
  & \rothead{\textbf{Crocodile}}
  & \rothead{\textbf{Helena}}
  & \rothead{\textbf{ITS-Tools}}
  & \rothead{\textbf{LoLA (binstore)}}
  & \rothead{\textbf{LoLA (bloom)}}
  & \rothead{\textbf{Marcie}}
  & \rothead{\textbf{Neco}}
  & \rothead{\textbf{PNXDD}}
  & \rothead{\textbf{Sara}}
    \lstset{
      basicstyle=\footnotesize\fontfamily{fvm}\selectfont
    }
  \\
  \hline
  \hline
    \multirow{7}{*}{\begin{sideways}\textbf{MCC 2011 models}\end{sideways}}
  & \textbf{FMS}
  & \result[reached=best,typepn=pt]{10} 
  & \result[reached=nc]{} 
  & \result[reached=nc]{} 
  & \result[reached=nc]{} 
  & \result[reached=nc]{} 
  & \result[reached=nc]{} 
  & \result[reached=nc]{} 
  & \result[reached=nc]{} 
  & \result[reached=nc]{} 
  & \result[reached=nc]{} 
  \\
  & \textbf{Kanban}
  & \result[reached=best,typepn=pt]{20} 
  & \result[reached=nc]{} 
  & \result[reached=nc]{} 
  & \result[reached=nc]{} 
  & \result[reached=nc]{} 
  & \result[reached=nc]{} 
  & \result[reached=nc]{} 
  & \result[reached=nc]{} 
  & \result[reached=nc]{} 
  & \result[reached=nc]{} 
  \\
  & \textbf{MAPK}
  & \result[reached=none]{} 
  & \result[reached=nc]{} 
  & \result[reached=nc]{} 
  & \result[reached=nc]{} 
  & \result[reached=nc]{} 
  & \result[reached=nc]{} 
  & \result[reached=nc]{} 
  & \result[reached=nc]{} 
  & \result[reached=nc]{} 
  & \result[reached=nc]{} 
  \\
  & \textbf{Peterson}
  & \result[reached=best,typepn=cn]{2} 
  & \result[reached=nc]{} 
  & \result[reached=none]{} 
  & \result[reached=nc]{} 
  & \result[reached=nc]{} 
  & \result[reached=nc]{} 
  & \result[reached=nc]{} 
  & \result[reached=nc]{} 
  & \result[reached=nc]{} 
  & \result[reached=nc]{} 
  \\
  & \textbf{Philosophers}
  & \result[reached=best,typepn=cn]{10} 
  & \result[reached=nc]{} 
  & \result[reached=nc]{} 
  & \result[reached=nc]{} 
  & \result[reached=nc]{} 
  & \result[reached=nc]{} 
  & \result[reached=nc]{} 
  & \result[reached=nc]{} 
  & \result[reached=nc]{} 
  & \result[reached=nc]{} 
  \\
  & \textbf{Shared-Memory}
  & \result[reached=some,typepn=cn]{10} 
  & \result[reached=nc]{} 
  & \result[reached=best,typepn=cn]{20} 
  & \result[reached=nc]{} 
  & \result[reached=nc]{} 
  & \result[reached=nc]{} 
  & \result[reached=nc]{} 
  & \result[reached=nc]{} 
  & \result[reached=nc]{} 
  & \result[reached=nc]{} 
  \\
  & \textbf{TokenRing}
  & \result[reached=none]{} 
  & \result[reached=nc]{} 
  & \result[reached=best,typepn=cn]{20} 
  & \result[reached=nc]{} 
  & \result[reached=nc]{} 
  & \result[reached=nc]{} 
  & \result[reached=nc]{} 
  & \result[reached=nc]{} 
  & \result[reached=nc]{} 
  & \result[reached=nc]{} 
  \\
  \hline
  \hline
    \multirow{12}{*}{\begin{sideways}\textbf{new models from MCC 2012}\end{sideways}}
  & \textbf{Cs\_repetitions}
  & \result[reached=none]{} 
  & \result[reached=nc]{} 
  & \result[reached=nc]{} 
  & \result[reached=nc]{} 
  & \result[reached=nc]{} 
  & \result[reached=nc]{} 
  & \result[reached=nc]{} 
  & \result[reached=nc]{} 
  & \result[reached=nc]{} 
  & \result[reached=nc]{} 
  \\
  & \textbf{Echo}
  & \result[reached=none]{} 
  & \result[reached=nc]{} 
  & \result[reached=nc]{} 
  & \result[reached=nc]{} 
  & \result[reached=nc]{} 
  & \result[reached=nc]{} 
  & \result[reached=nc]{} 
  & \result[reached=nc]{} 
  & \result[reached=nc]{} 
  & \result[reached=nc]{} 
  \\
  & \textbf{Eratosthenes}
  & \result[reached=none]{} 
  & \result[reached=nc]{} 
  & \result[reached=nc]{} 
  & \result[reached=nc]{} 
  & \result[reached=nc]{} 
  & \result[reached=nc]{} 
  & \result[reached=nc]{} 
  & \result[reached=nc]{} 
  & \result[reached=nc]{} 
  & \result[reached=nc]{} 
  \\
  & \textbf{Galloc\_res}
  & \result[reached=none]{} 
  & \result[reached=nc]{} 
  & \result[reached=nc]{} 
  & \result[reached=nc]{} 
  & \result[reached=nc]{} 
  & \result[reached=nc]{} 
  & \result[reached=nc]{} 
  & \result[reached=nc]{} 
  & \result[reached=nc]{} 
  & \result[reached=nc]{} 
  \\
  & \textbf{Lamport\_fmea}
  & \result[reached=some,typepn=cn]{3} 
  & \result[reached=nc]{} 
  & \result[reached=best,typepn=cn]{5} 
  & \result[reached=nc]{} 
  & \result[reached=nc]{} 
  & \result[reached=nc]{} 
  & \result[reached=nc]{} 
  & \result[reached=nc]{} 
  & \result[reached=nc]{} 
  & \result[reached=nc]{} 
  \\
  & \textbf{NEO-election}
  & \result[reached=nc]{} 
  & \result[reached=nc]{} 
  & \result[reached=nc]{} 
  & \result[reached=nc]{} 
  & \result[reached=nc]{} 
  & \result[reached=nc]{} 
  & \result[reached=nc]{} 
  & \result[reached=nc]{} 
  & \result[reached=nc]{} 
  & \result[reached=nc]{} 
  \\
  & \textbf{Philo\_dyn}
  & \result[reached=some,typepn=cn]{3} 
  & \result[reached=nc]{} 
  & \result[reached=best,typepn=cn]{20} 
  & \result[reached=nc]{} 
  & \result[reached=nc]{} 
  & \result[reached=nc]{} 
  & \result[reached=nc]{} 
  & \result[reached=nc]{} 
  & \result[reached=nc]{} 
  & \result[reached=nc]{} 
  \\
  & \textbf{Planning}
  & \result[reached=nc]{} 
  & \result[reached=nc]{} 
  & \result[reached=nc]{} 
  & \result[reached=nc]{} 
  & \result[reached=nc]{} 
  & \result[reached=nc]{} 
  & \result[reached=nc]{} 
  & \result[reached=nc]{} 
  & \result[reached=nc]{} 
  & \result[reached=nc]{} 
  \\
  & \textbf{Railroad}
  & \result[reached=nc]{} 
  & \result[reached=nc]{} 
  & \result[reached=nc]{} 
  & \result[reached=nc]{} 
  & \result[reached=nc]{} 
  & \result[reached=nc]{} 
  & \result[reached=nc]{} 
  & \result[reached=nc]{} 
  & \result[reached=nc]{} 
  & \result[reached=nc]{} 
   \\
  & \textbf{Ring}
  & \result[reached=none]{} 
  & \result[reached=nc]{} 
  & \result[reached=nc]{} 
  & \result[reached=nc]{} 
  & \result[reached=nc]{} 
  & \result[reached=nc]{} 
  & \result[reached=nc]{} 
  & \result[reached=nc]{} 
  & \result[reached=nc]{} 
  & \result[reached=nc]{} 
  \\
  & \textbf{Rw\_mutex}
  & \result[reached=best,typepn=pt]{r10w100} 
  & \result[reached=nc]{} 
  & \result[reached=nc]{} 
  & \result[reached=nc]{} 
  & \result[reached=nc]{} 
  & \result[reached=nc]{} 
  & \result[reached=nc]{} 
  & \result[reached=nc]{} 
  & \result[reached=nc]{} 
  & \result[reached=nc]{} 
  \\
  & \textbf{Simple\_lbs}
  & \result[reached=some,typepn=pt]{2} 
  & \result[reached=nc]{} 
  & \result[reached=best,typepn=cn]{15} 
  & \result[reached=nc]{} 
  & \result[reached=nc]{} 
  & \result[reached=nc]{} 
  & \result[reached=nc]{} 
  & \result[reached=nc]{} 
  & \result[reached=nc]{} 
  & \result[reached=nc]{} 
  \\
  \hline
  \end{tabular}
  \caption{Results for the structural formul{\ae} examination}
  \label{tab:fs}
\end{table}

\Cref{sec:fsresults}~presents the data computed by tools. Then,
\Cref{sec:fsprocessedmodels}~shows how models have been handled by tools
and \Cref{sec:fs:bytools,sec:fs:bymodels}~summarize how tools did cope with
models.

\Cref{sec:fsresults}~shows that no tool computes the same set of formul{\ae}.
This is why we do not display the charts that have been extracted from the
executions since they have no meaning at all.

\subsection{Computed results for the formul\ae}
\label{sec:fsresults}
\index{Structural Formul{\ae}!Computed data}

Since only two tools did compete and had sometimes different results, we could
not consolidate the values and decided to show the results as they were
produced in~\Cref{tab:result:FS}. For each scaling parameter, we proposed a
set of formul{\ae}. Outputs from the tools were analyzed and formula were
sorted by their identifier in order to build a vector where the $i^{th}$
element corresponds to the $i^{th}$ formula of the set (value \texttt{F} or
\texttt{T}). Sometimes, the tool cannot compute a formula. We then display a
``.'' instead. For this examination, it appears that, the problem is often due
to formula evaluation itself (the grammar was published late and thus). "?"
means that the tool participated but did not compute any
result.\label{explain:fs}

\begin{table}
\centering
\begin{tabular}{|c||c|c|}
	\hline
	\textbf{Scale} & \multirow{2}{*}{\acs{AlPiNA}} & \multirow{2}{*}{\acs{Helena}} \\
	\textbf{value} &  & \\
	\hline
	\hline
	\multicolumn{3}{|c|}{\acs{CS-Repetitions}} \\
	25 & {\scriptsize ?} & \cellcolor{gray!50}\\
	\hline
	\multicolumn{3}{|c|}{\acs{Echo}} \\
	d2r11 & {\scriptsize ?} & \cellcolor{gray!50}\\
	\hline
	\multicolumn{3}{|c|}{\acs{Eratosthenes}} \\
	5 & {\scriptsize ?} & \cellcolor{gray!50}\\
	\hline
	\multicolumn{3}{|c|}{\acs{FMS}} \\
	2 & {\scriptsize .......................FFFFFTTTT.FTTTTTTTTT.TFFFFF.....} & \cellcolor{gray!50}\\
	5 & {\scriptsize .......................FFFFFFFFF.FFFFFFFFFF.FTTTTFFTFT.T....} & \cellcolor{gray!50}\\
	10 & {\scriptsize .......................FFFFFFFFF.FFTFTFFTFT.TFTFFFFTFT.TFTFTFFFFF.F...}& \cellcolor{gray!50}\\
	\hline
	\multicolumn{3}{|c|}{\acs{Galloc}} \\
	3 & {\scriptsize ?} & \cellcolor{gray!50}\\
	\hline
	\multicolumn{3}{|c|}{\acs{Kanban}} \\
	5 & {\scriptsize .......................FFFFFFFFF.FFFFFFFFFF.FTTTTFFFFF.F....} & \cellcolor{gray!50}\\
	10 & {\scriptsize .......................FFFFFFFFF.FFFFFFFFFF.FFFFFFFFFF.FFFFFFFFFF.F...} & \cellcolor{gray!50}\\
	20 & {\scriptsize .......................FFFFFFFFF.FFFFFFFFFF.FFFFFFFFFF.FFFFFFFFFF.FFFFFFFFFF.F..}& \cellcolor{gray!50}\\
	\hline
	\multicolumn{3}{|c|}{\acs{Lamport}} \\
	2 & {\scriptsize TTTTTTTTFTFTTTFFTTTFFTTTTTTTTTFTFTTTTTTT} & {\scriptsize ..........T.............................} \\
	3 & {\scriptsize FTTFFFTTFTFFTTTTFTTFTTTTTTTTTTFTFTTTTTTTTFTTTFFTFT} & {\scriptsize ..................................................} \\
	4 & {\scriptsize ?} & {\scriptsize ..................................................} \\
	5 & {\scriptsize ?} & {\scriptsize ..................................................} \\
	\hline
	\multicolumn{3}{|c|}{\acs{MAPK}} \\
	8 & {\scriptsize ?} & \cellcolor{gray!50}\\
	\hline
	\multicolumn{3}{|c|}{\acs{Peterson}} \\
	2 & {\scriptsize TFTFFFTTTTFTTTFTTTTFFFFTTTTFTTTTFTTTFTFFTFTFFFFTFF} & \cellcolor{gray!50}\\
	\hline
	\multicolumn{3}{|c|}{\acs{Dynamic-Philosophers}} \\
	2 & {\scriptsize FTTTTTTTFFFFTTTTTFTFTFFFFFFFFF} & {\scriptsize ..........T...................}\\
	3 & {\scriptsize FTTTTTTTFFFFTTTTTFTFTFFFFFFFFF} & {\scriptsize ........................................}\\
	10 & {\scriptsize ?} & {\scriptsize .......................................................}\\
	20 & {\scriptsize ?} & {\scriptsize .......................................................}\\
	\hline
	\multicolumn{3}{|c|}{\acs{Philosophers}} \\
	5 & {\scriptsize FFTFFFFTFFFFFTFFFTTFTFTFFFFFFF} & \cellcolor{gray!50}\\
	10 & {\scriptsize FFFFFFFFFFFFFFFFFFTFTFFFTFTFFTFTFFFFFFFF} & \cellcolor{gray!50}\\
	\hline
	\multicolumn{3}{|c|}{\acs{Planning}} \\
	--- & {\scriptsize ?} & \cellcolor{gray!50}\\
	\hline
	\multicolumn{3}{|c|}{\acs{Ring}} \\
	--- & {\scriptsize ?} & \cellcolor{gray!50}\\
	\hline
	\multicolumn{3}{|c|}{\acs{RW-Mutex}} \\
	r10w10 & {\scriptsize .......................FFFFFTTTT.FTTFTFFTFT.F.....} & \cellcolor{gray!50}\\
	r10w20 & {\scriptsize .......................FFFFTTTFT.TTTTTTTTTT.T.....} & \cellcolor{gray!50}\\
	r10w50 & {\scriptsize .......................FFFFFTTTT.FTTFTTTTTT.F.....} & \cellcolor{gray!50}\\
	r10w100 & {\scriptsize .......................FFFFFTTTT.FTTTTTFFFF.F.....} & \cellcolor{gray!50}\\
	\hline
	\multicolumn{3}{|c|}{\acs{Shared-Memory}} \\
	5 & {\scriptsize TFTFTFFTFTFTFTFTFTTFTFFTTTTTTTFTFTTTFTFFFFFFFFFTFF} & {\scriptsize ..................................................}\\
	10 & {\scriptsize FFTFTFFTFFFFFTFTFFTFFFTTTTTTTTTTFFTTTTTTTFTFFFTFTTFFTFT} & {\scriptsize .......................................................}\\
	20 & {\scriptsize ?} & {\scriptsize .......................................................}\\
	\hline
	\multicolumn{3}{|c|}{\acs{Simple-LBS}} \\
	2 & {\scriptsize TTTTTTTTFTFFTTFTFTTFFFFFFFFTFF} & {\scriptsize ..........T...................}\\
	5 & {\scriptsize ?} & {\scriptsize ........................................}\\
	10 & {\scriptsize ?} & {\scriptsize ..................................................}\\
	15 & {\scriptsize ?} & {\scriptsize ..................................................}\\
	\hline
	\multicolumn{3}{|c|}{\acs{Token-Ring}} \\
	2 & {\scriptsize ?} & {\scriptsize ..........T.......................................}\\
	10 & {\scriptsize ?} & {\scriptsize .......................................................}\\
	20 & {\scriptsize ?} & {\scriptsize .......................................................}\\
	\hline
\hline
\end{tabular}
\caption{Results of structural formul{\ae} evaluation for the models where at least one tool competed\label{tab:result:FS}}
\end{table}

\subsection{Processed Models}
\label{sec:fsprocessedmodels}
\index{Structural Formul{\ae}!Processed models}

This section summarizes how models were processed by tools. Let us first note
that no tool succeeded in this examination for the following models:

\begin{itemize}
	\item \acs{CS-Repetitions},
	\item \acs{Echo},
	\item \acs{Eratosthenes},
	\item \acs{Galloc},
	\item \acs{MAPK},
	\item \acs{Neo-Election},
	\item \acs{Planning},
	\item \acs{Railroad},
	\item \acs{Ring}.
\end{itemize}

These models constitute challenges for the next edition of the \acs{MCC}.

\subsection{Radars by models}
\label{sec:fs:bymodels}
\begin{figure}[p]
\centering
\begin{adjustwidth}{-2em}{-2em}
\noindent
\includegraphics[scale=.35]{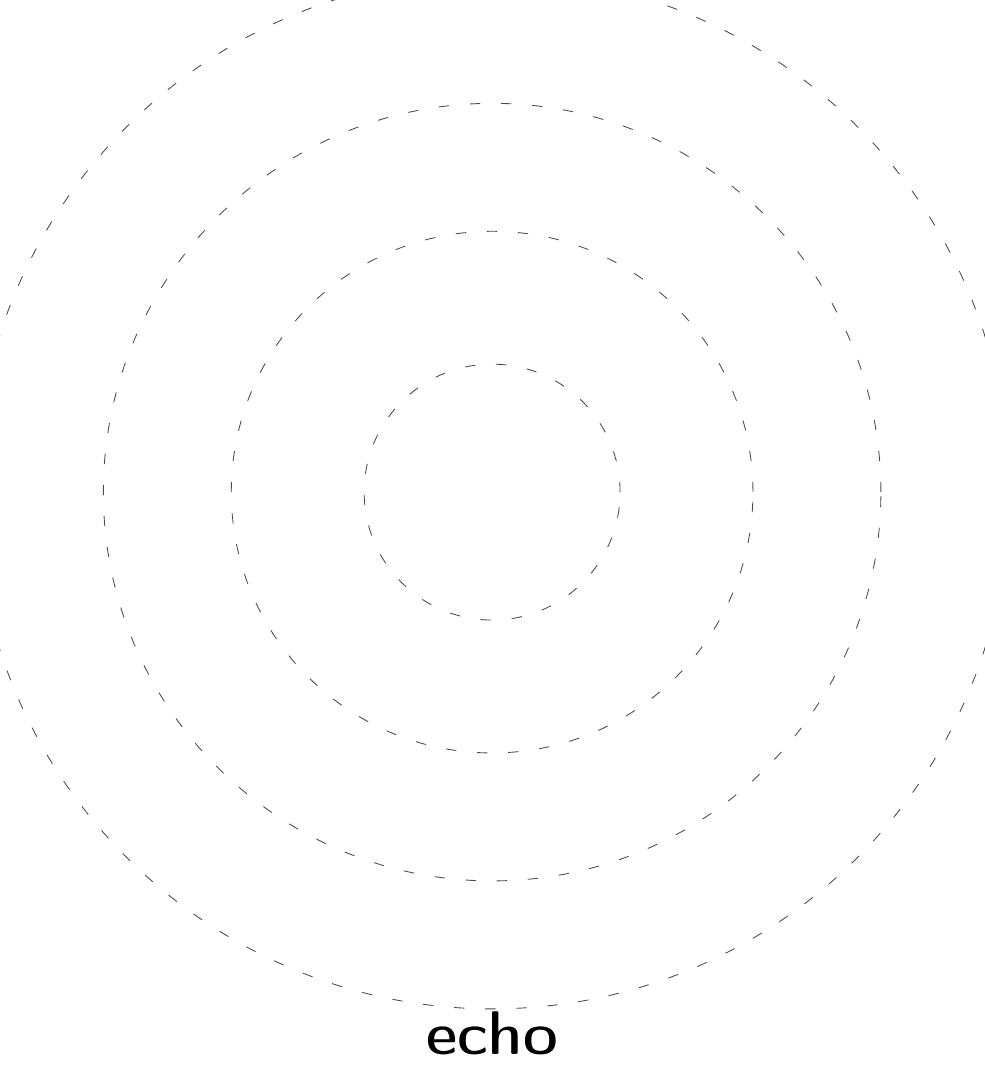}
\hfill
\includegraphics[scale=.35]{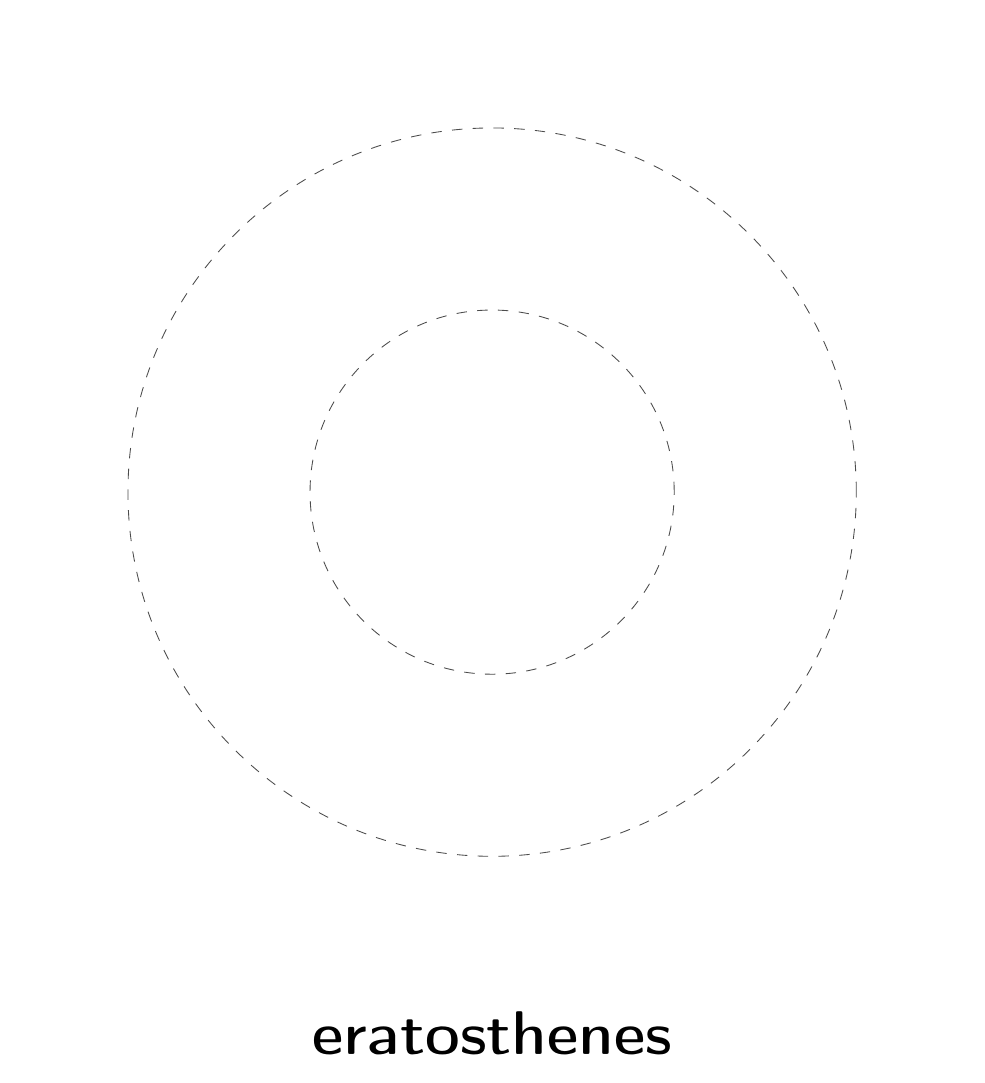}
\hfill
\includegraphics[scale=.35]{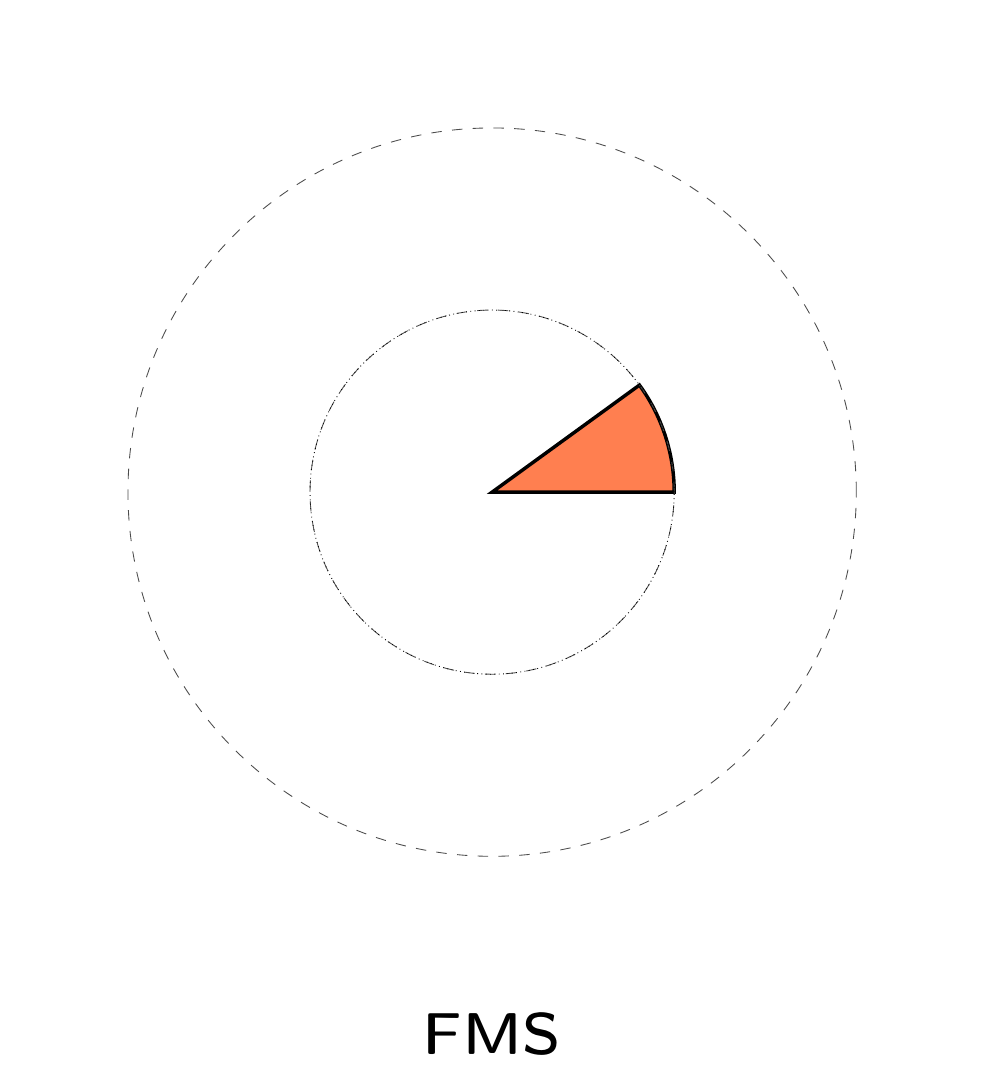}
\hfill
\includegraphics[scale=.35]{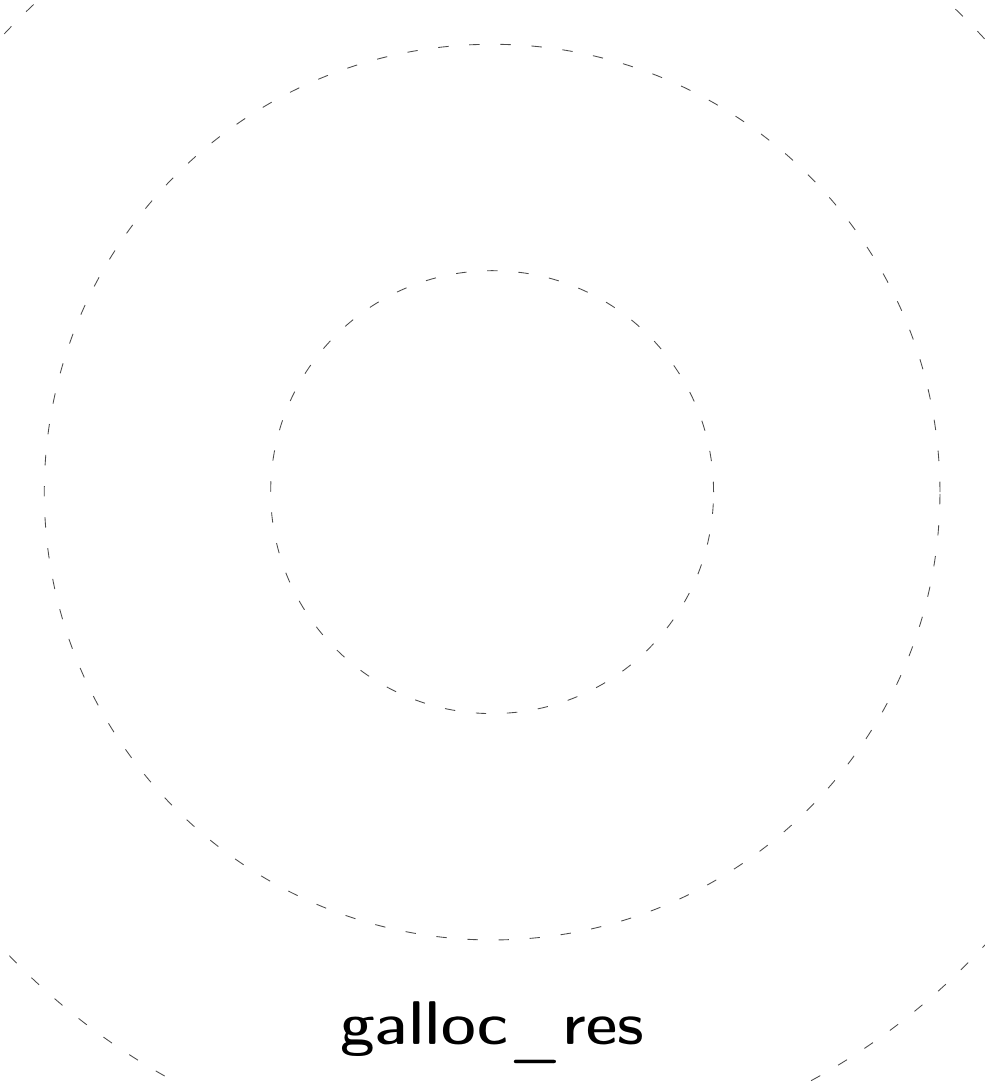}
\\
\medskip
\includegraphics[scale=.35]{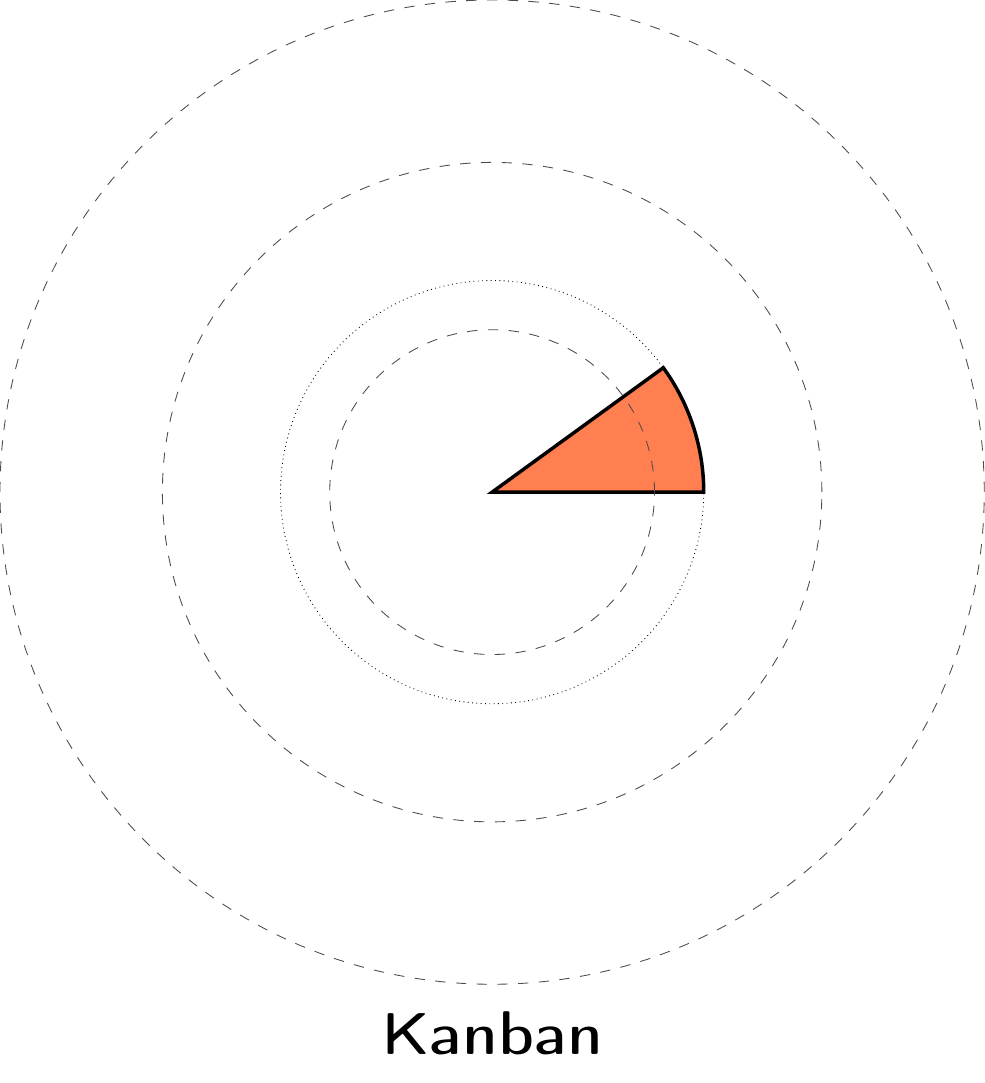}
\hfill
\includegraphics[scale=.35]{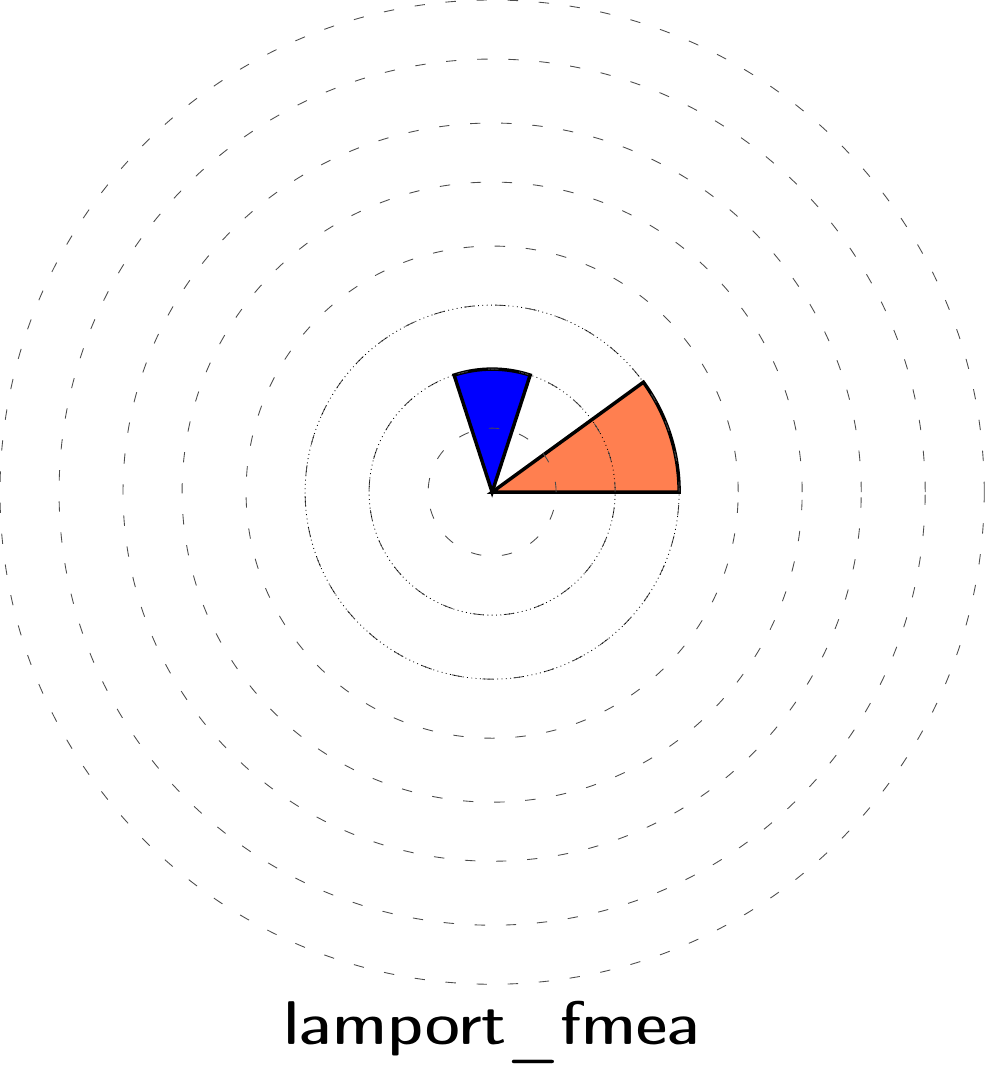}
\hfill
\includegraphics[scale=.35]{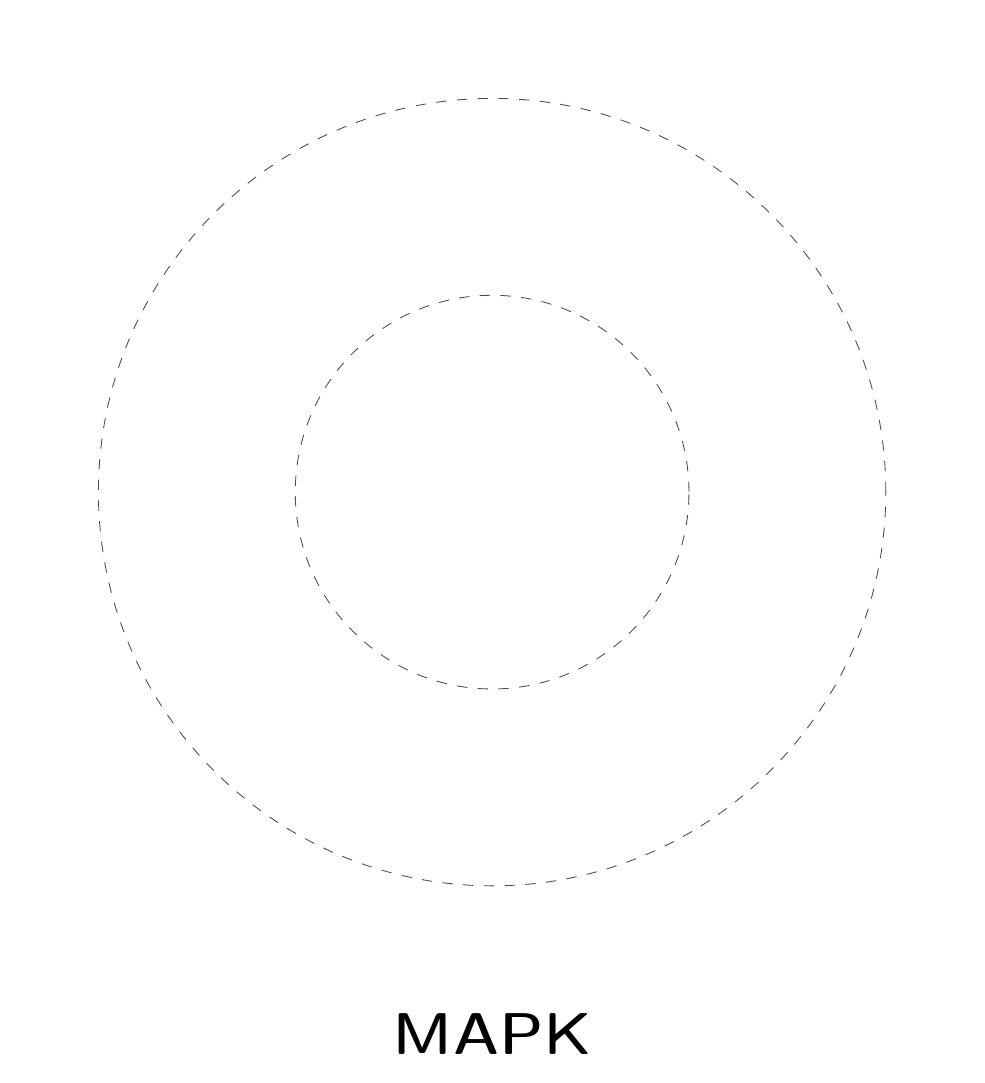}
\hfill
\includegraphics[scale=.35]{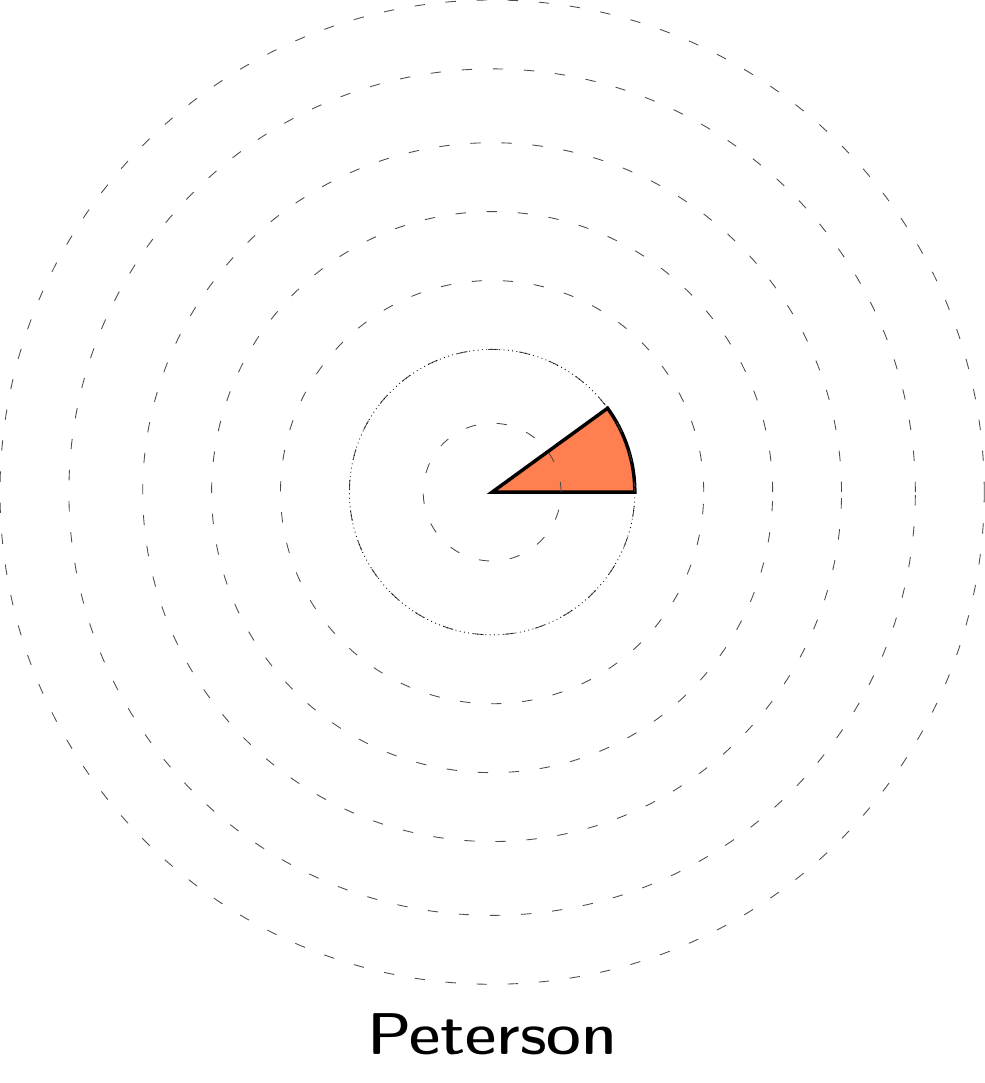}
\\
\medskip
\includegraphics[scale=.35]{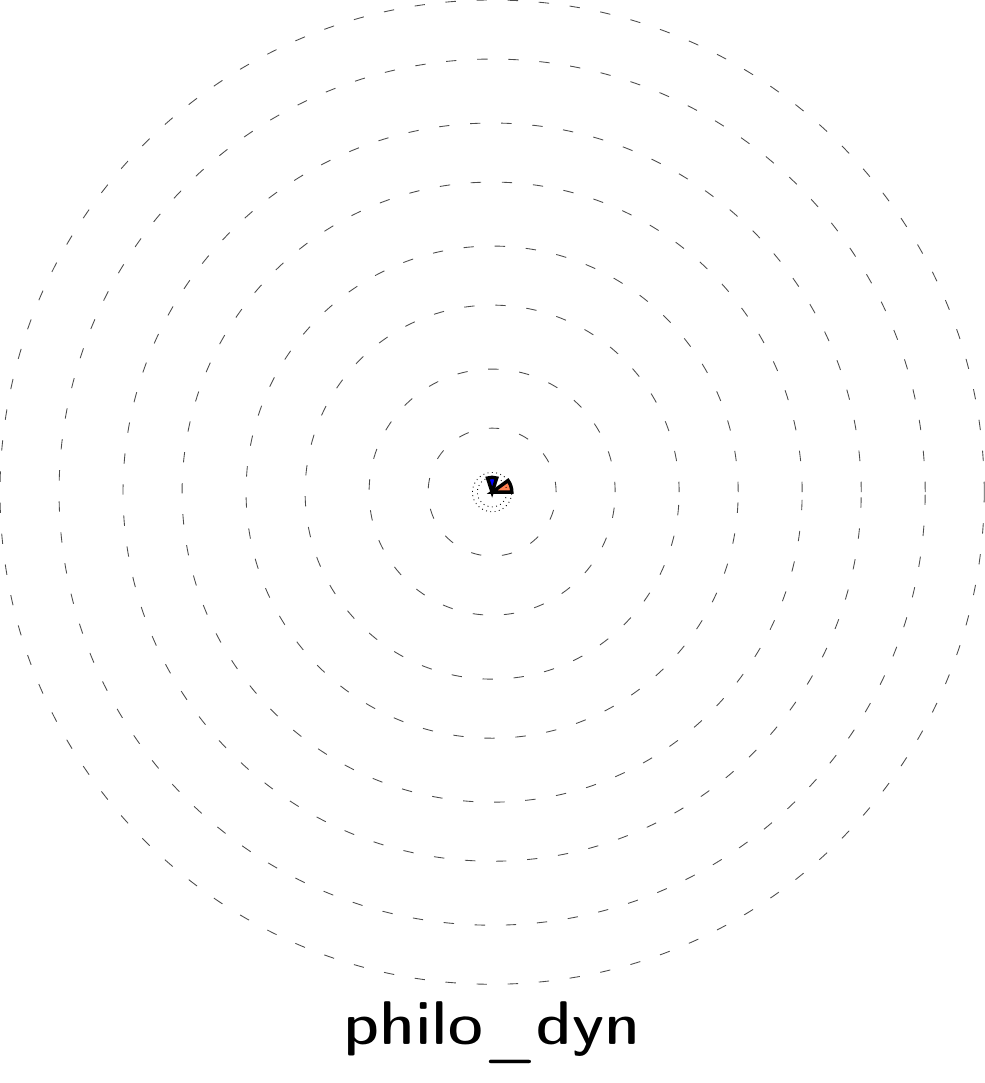}
\hfill
\includegraphics[scale=.35]{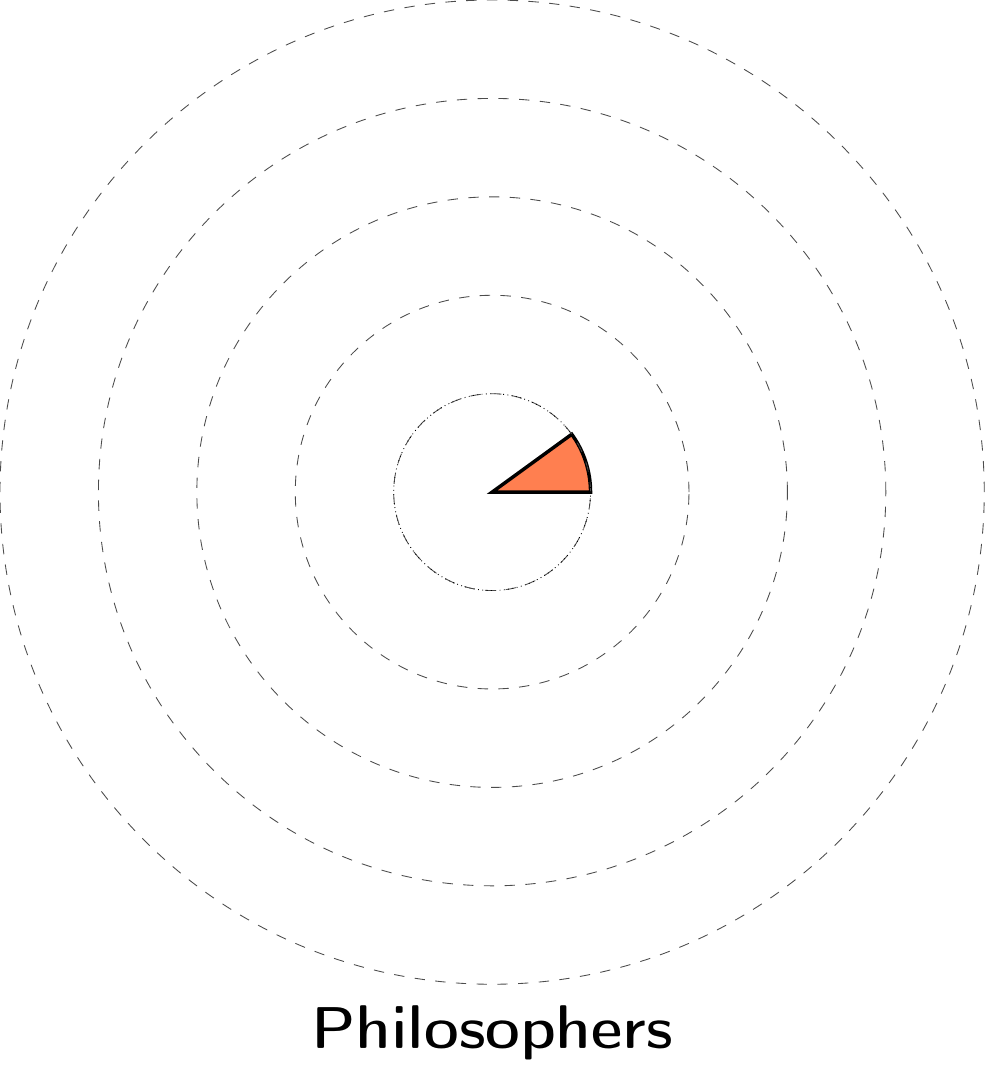}
\hfill
\includegraphics[scale=.35]{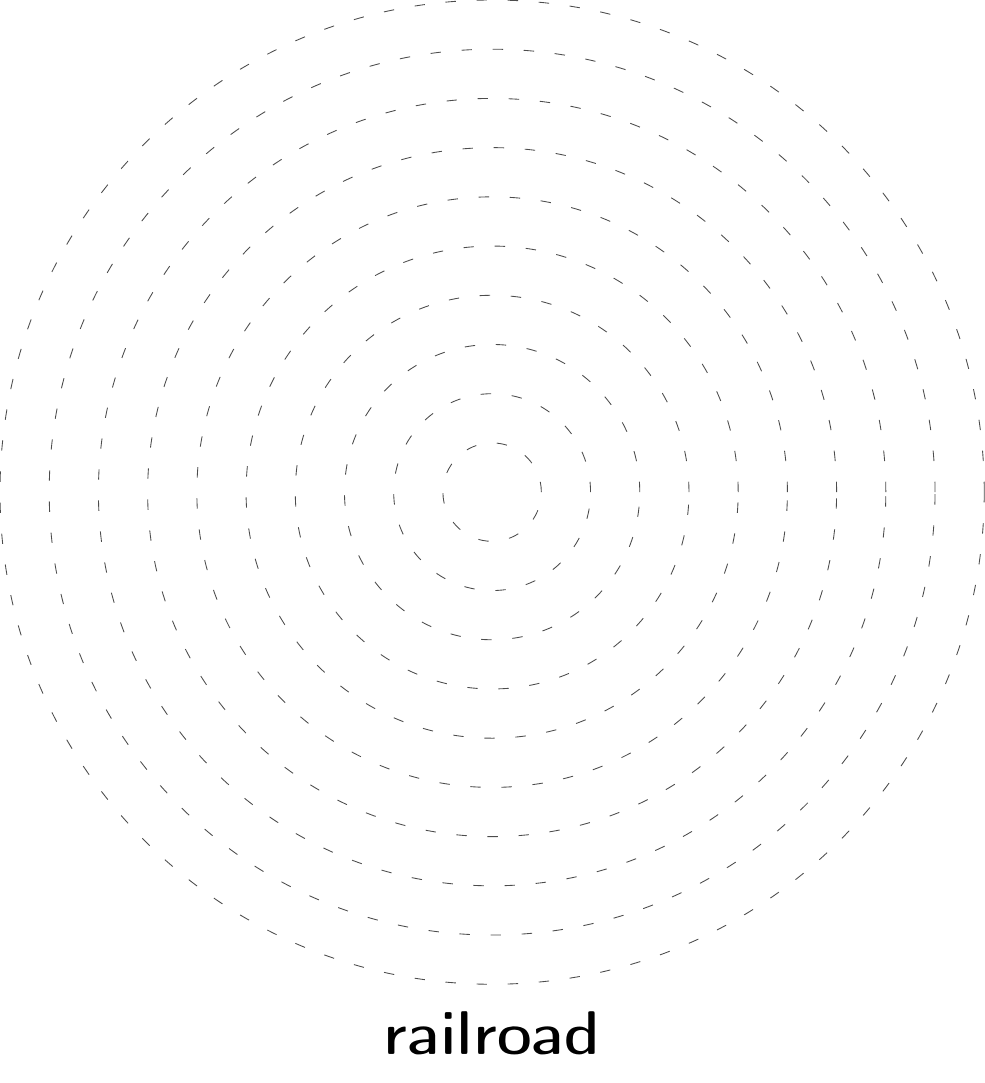}
\hfill
\includegraphics[scale=.35]{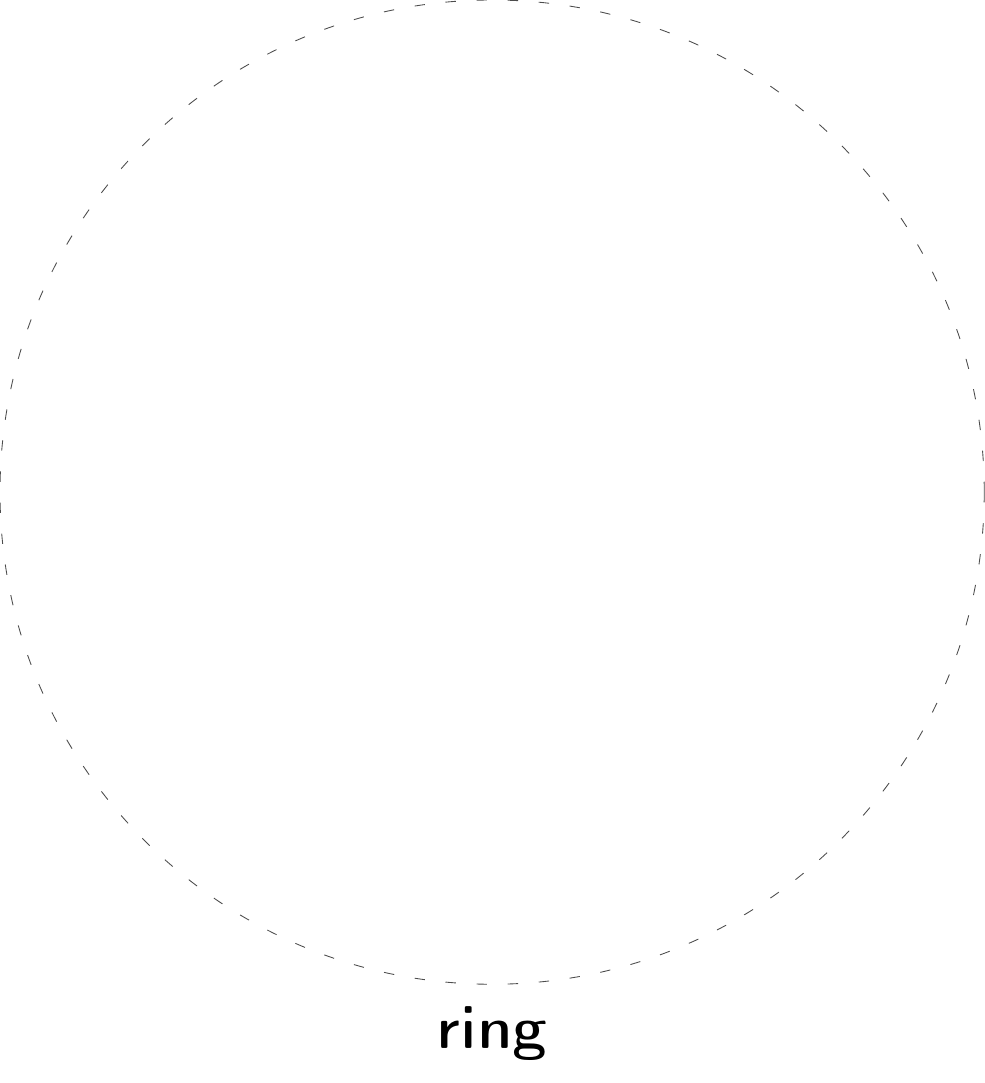}
\\
\medskip
\includegraphics[scale=.35]{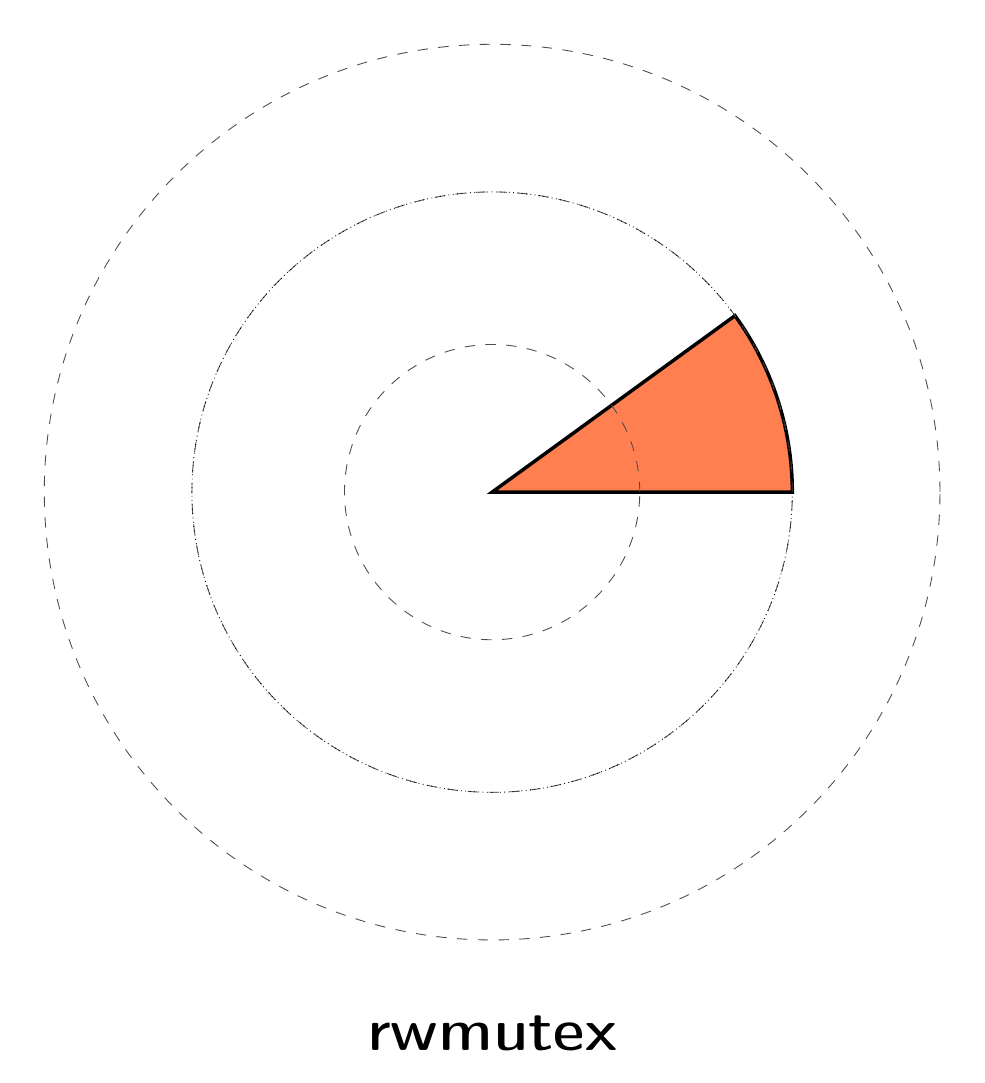}
\hfill
\includegraphics[scale=.35]{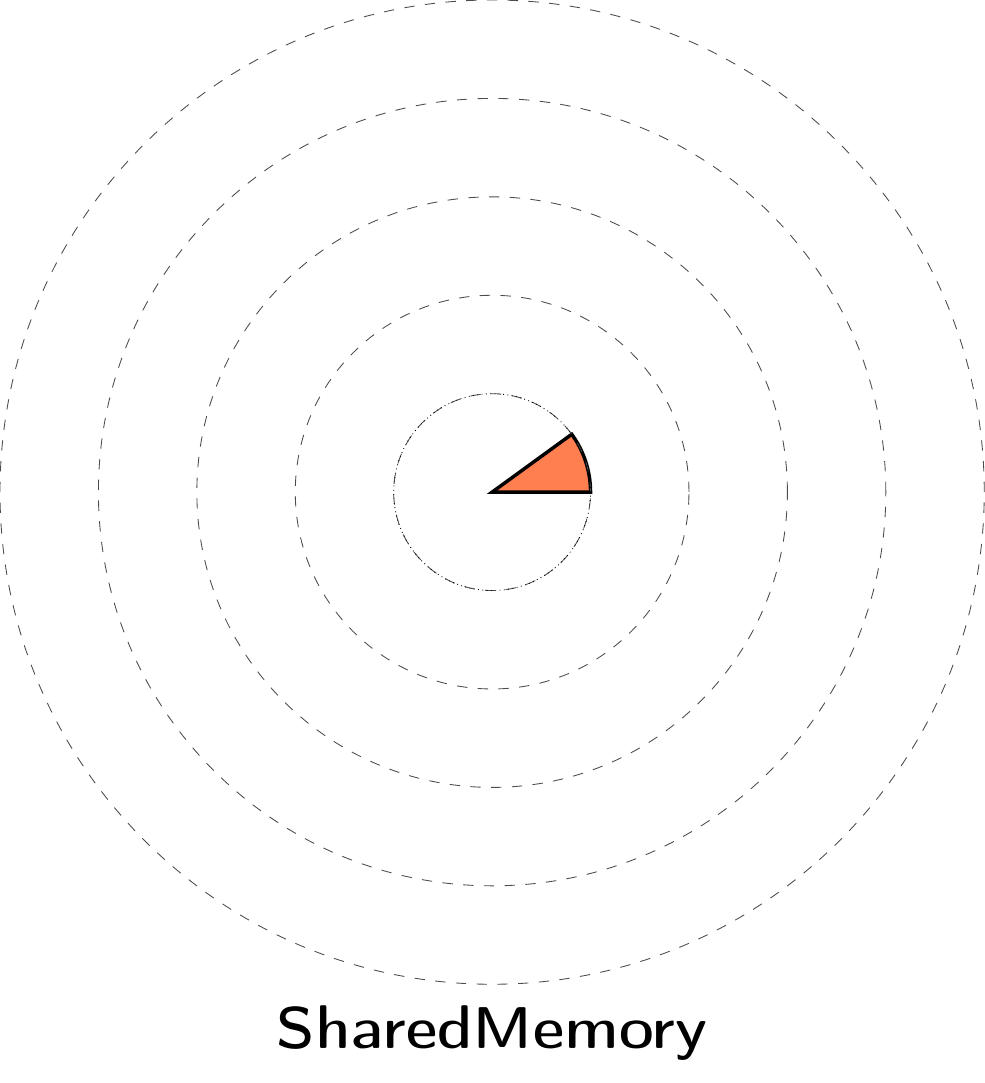}
\hfill
\includegraphics[scale=.35]{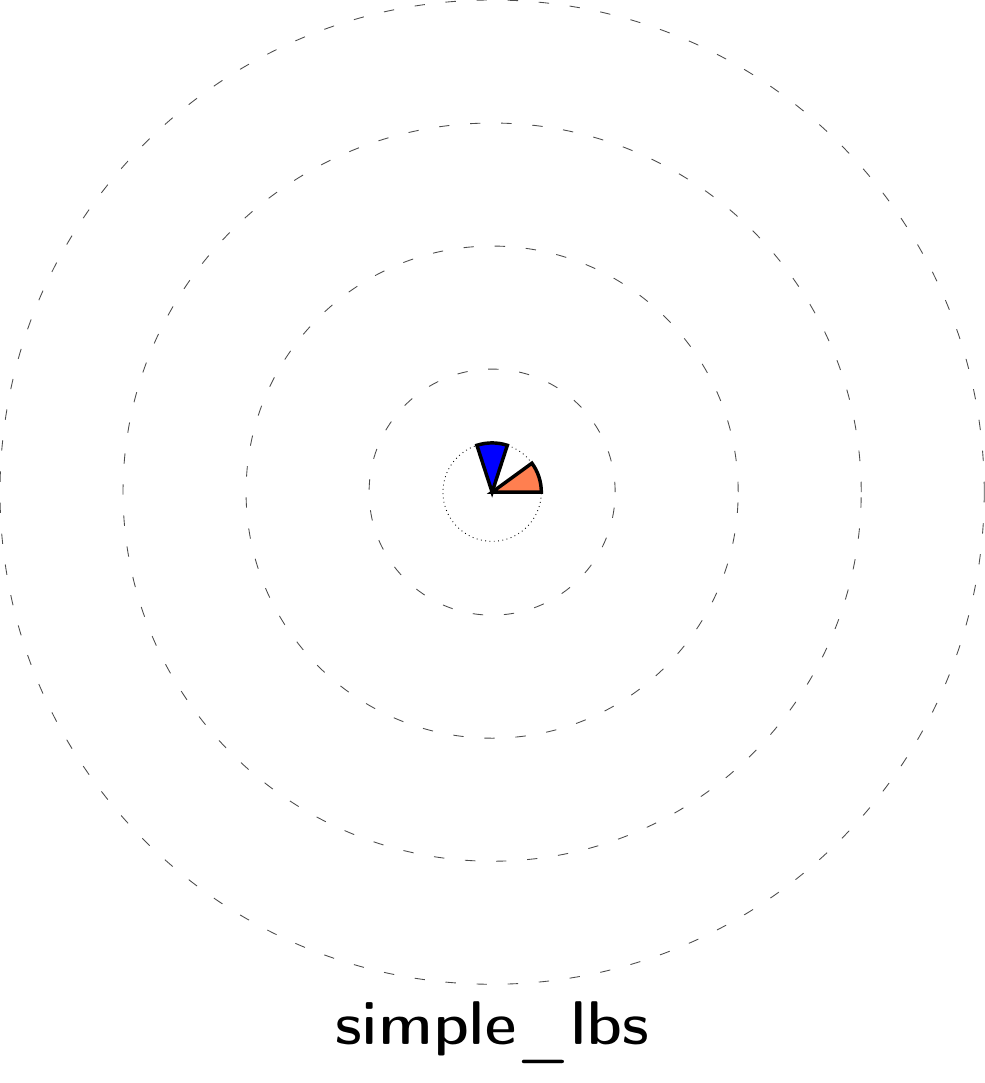}
\hfill
\includegraphics[scale=.35]{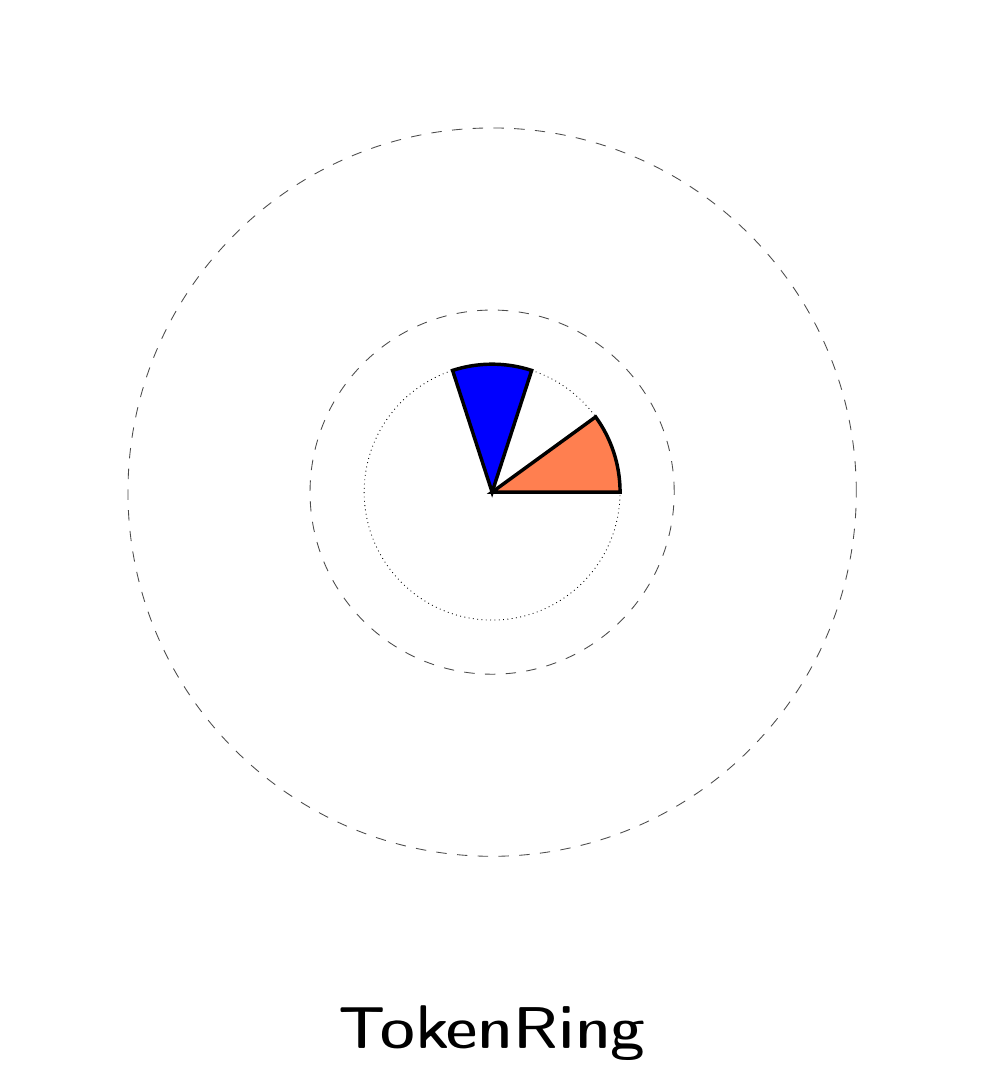}
\\
\bigskip
\mbox{}
\hfill
\includegraphics[scale=.7]{figures/alban-legend-tools.pdf}
\hfill
\mbox{}
\end{adjustwidth}
\caption{Highest parameter reached for each model,
         in structural formul{\ae} evaluation}
\label{fig:fs:radar:models}
\end{figure}

\Cref{fig:fs:radar:models}~represents graphically
through a set of radar diagrams the highest parameter reached by the tools,
for each model.
Each diagram corresponds to one model, \emph{e.g.}, \acs{Echo} or
\acs{Kanban}.
Each diagram is divided in ten slices, one for each competing tool,
always at the same position.

The length of the slice corresponds to the highest parameter reached
by the tool.
When a slice does not appear,
the tool could not process even the smallest parameter.
For instance, \acs{Helena} handles some parameters
for \acs{Lamport} and \acs{Token-Ring},
but is not able to handle the smallest parameter for \acs{Eratosthenes}.
The figure also shows that only two tools (\acs{AlPiNA} and \acs{Helena})
did compete for this examination.
None of them achieves the highest parameter for any model.

Note that the scale depends on the model :
when the parameters of a model vary within a small range (less than~$100$),
a linear scale is used (showed using loosely dashed circles as in the
\acs{Peterson} model),
whereas a logarithmic scale is used for larger parameter values
(showed using densely dashed circles as in the \acs{Philosophers} model).
We also show dotted circles for the results of the tools,
in order to allow easier comparison.

Some models (\acs{RW-Mutex} and \acs{Echo}) have complex parameters,
built from two values.
Tools were only able to handle variation of the second parameter,
so we only represent it in the figure,
in order to show an integer value.

\subsection{Radars by tools}
\label{sec:fs:bytools}
\index{Structural Formul{\ae}!Handling of models}

\begin{figure}[p]
\centering
\begin{adjustwidth}{0em}{0em}
\noindent
\includegraphics[scale=.4]{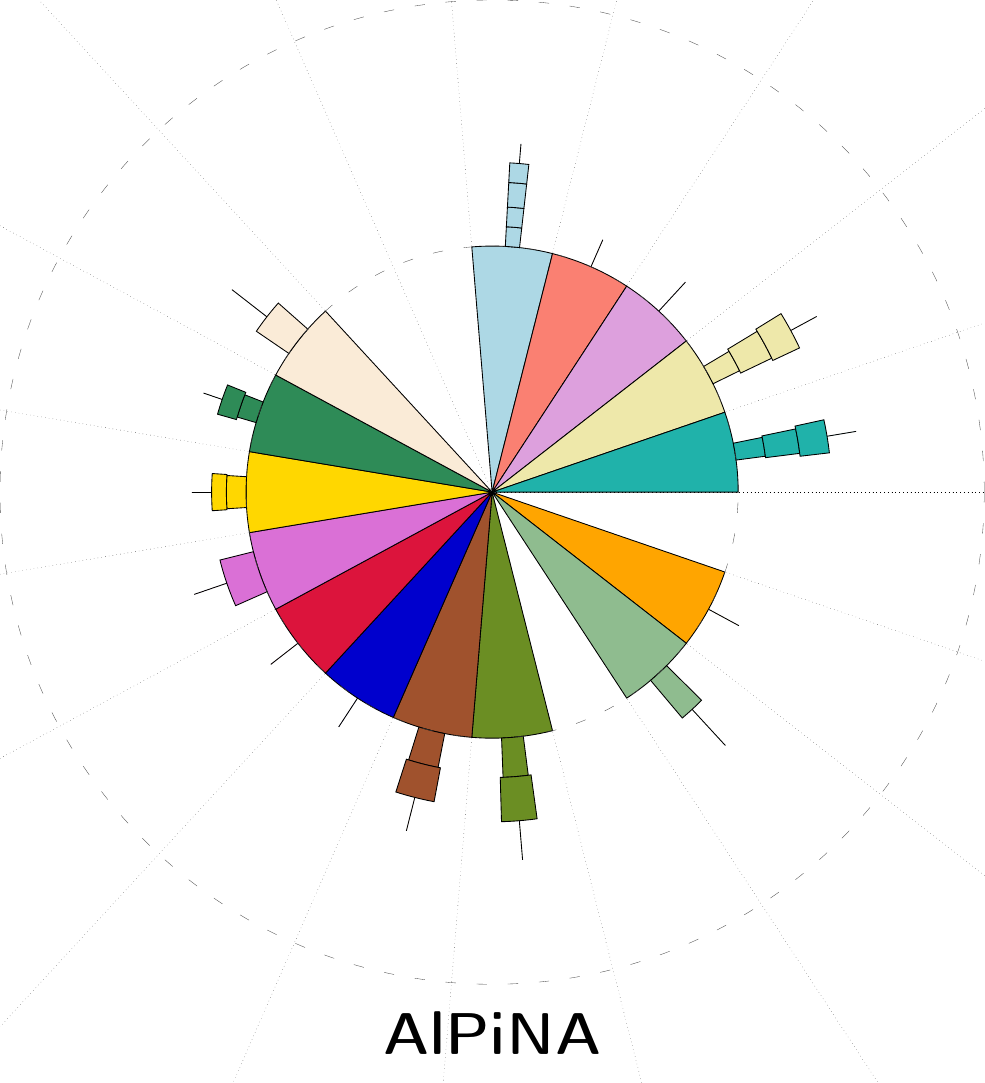}
\hfill
\includegraphics[scale=.4]{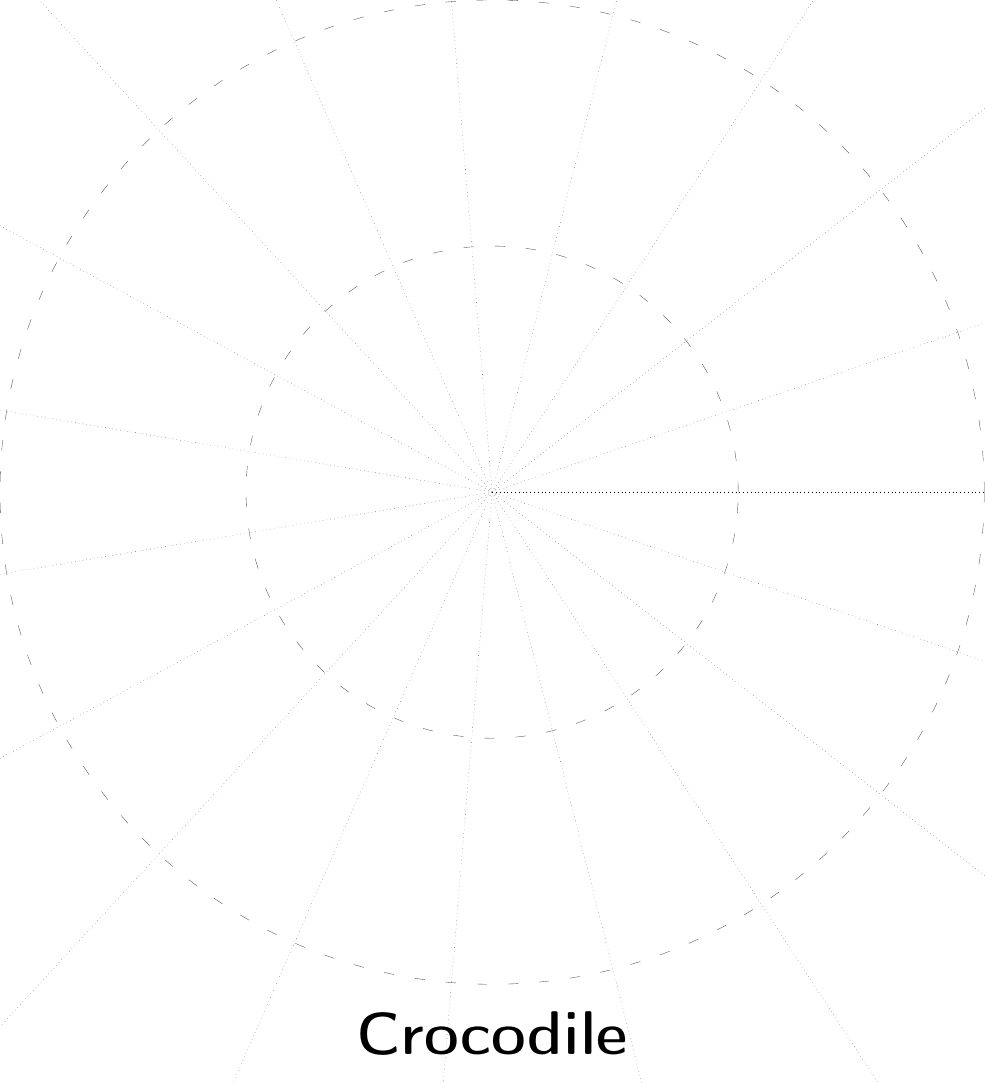}
\hfill
\includegraphics[scale=.4]{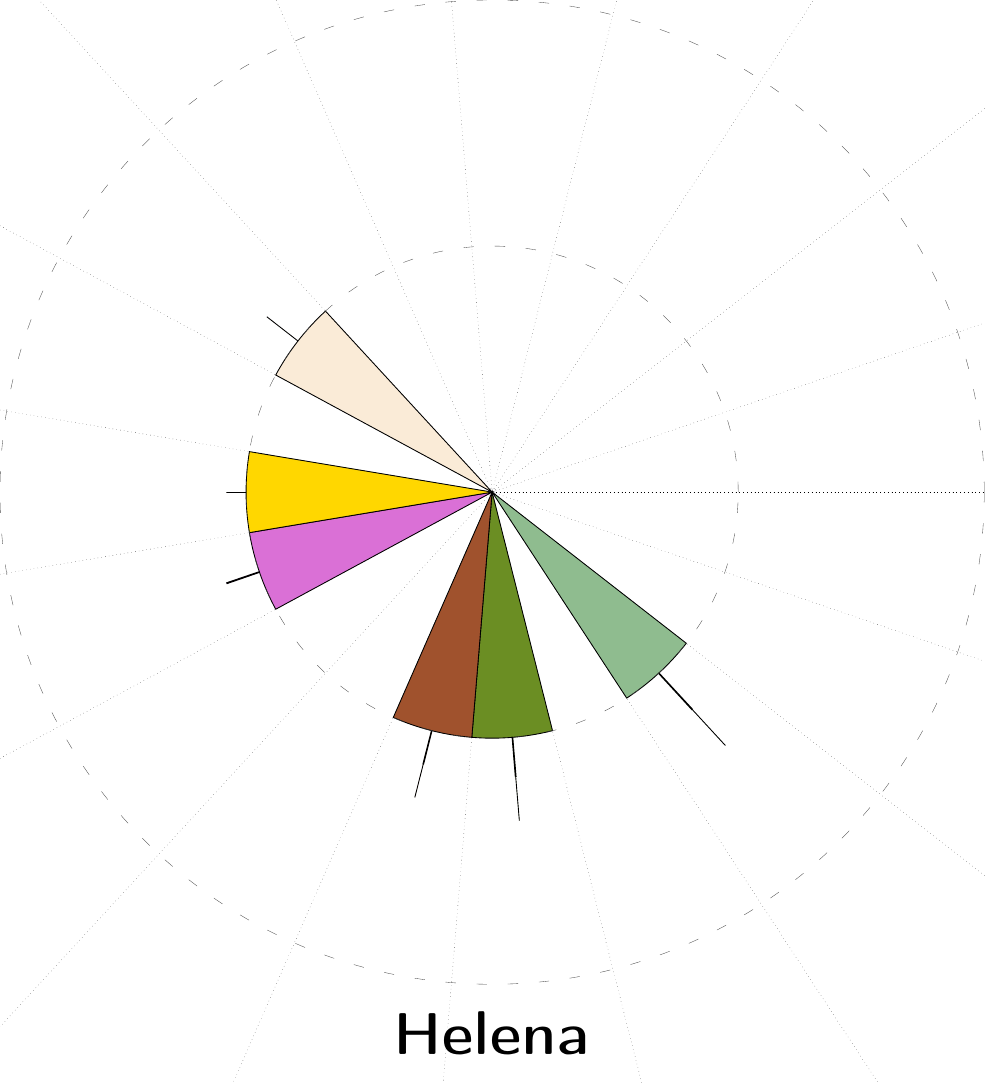}
\\
\medskip
\includegraphics[scale=.4]{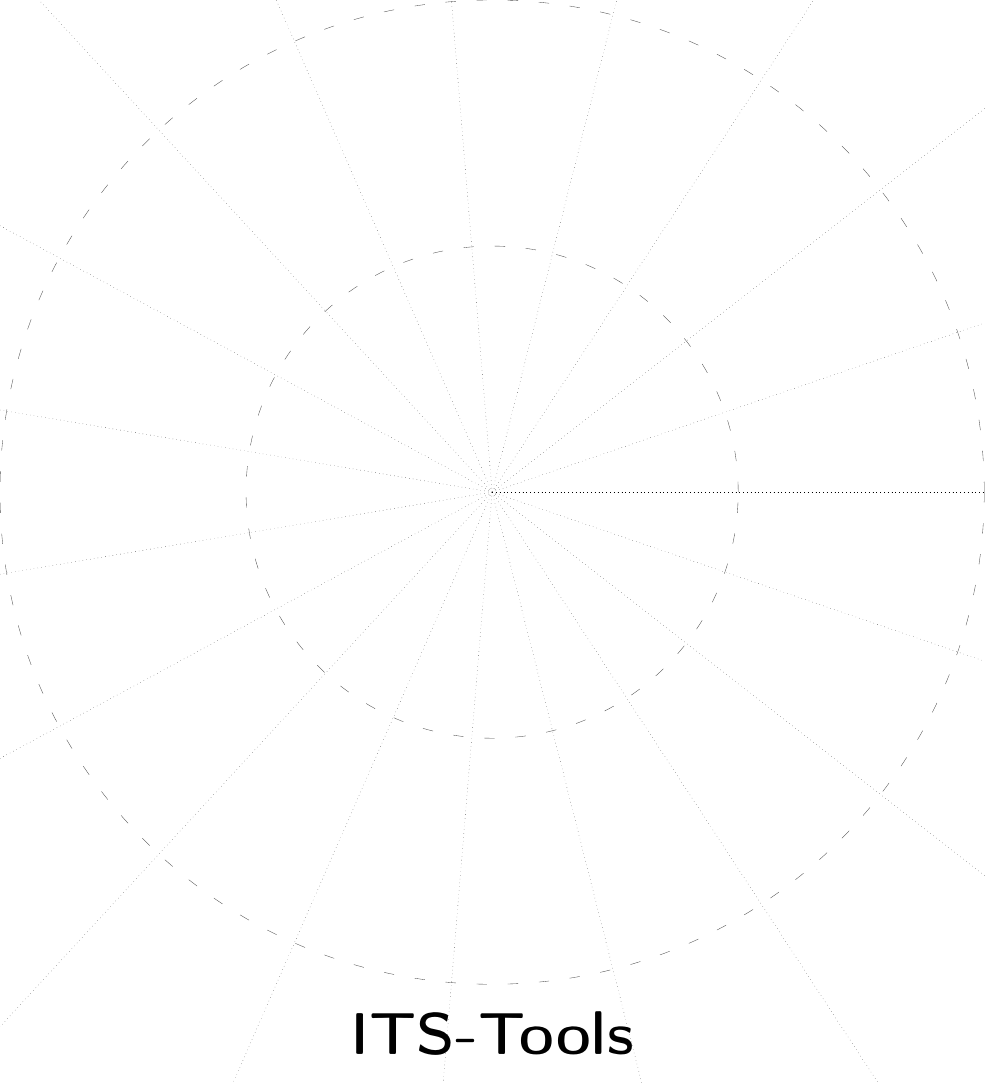}
\hfill
\includegraphics[scale=.4]{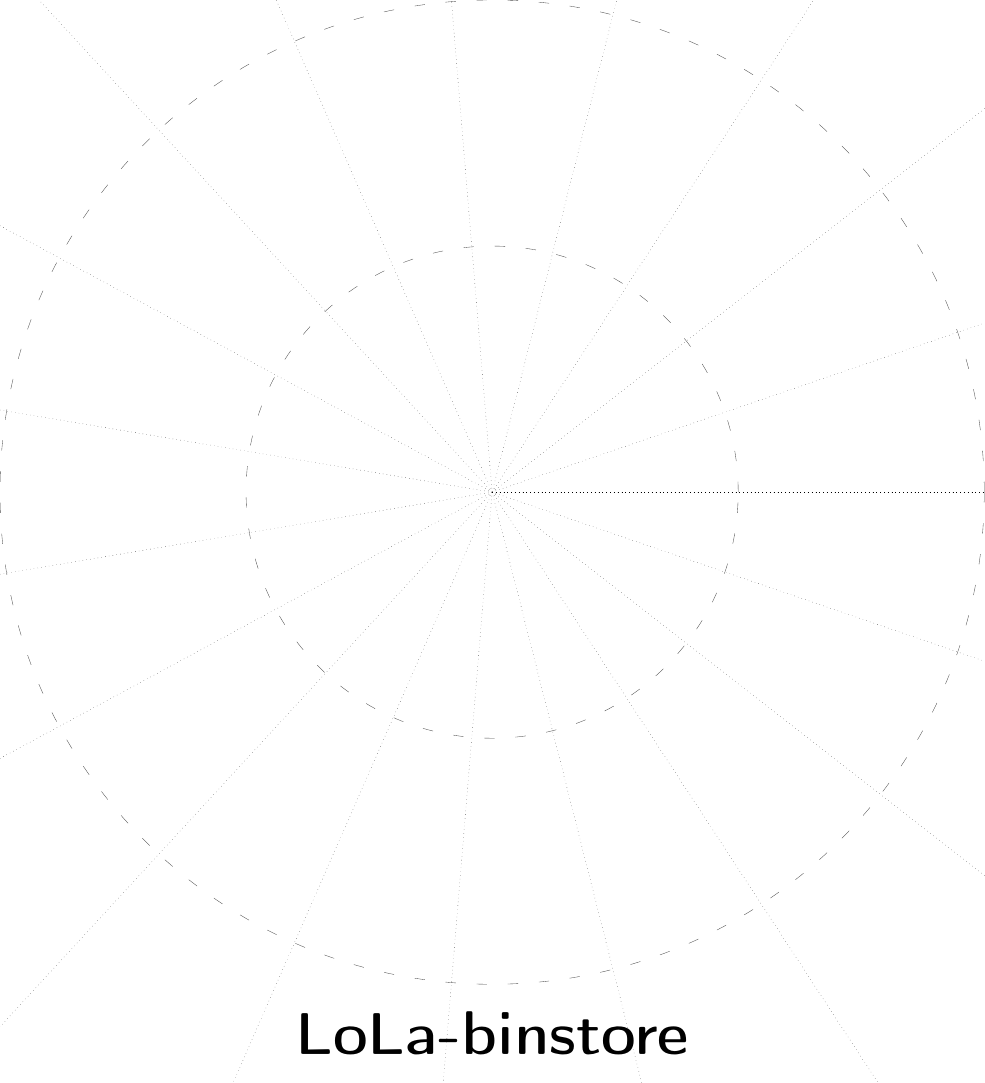}
\hfill
\includegraphics[scale=.4]{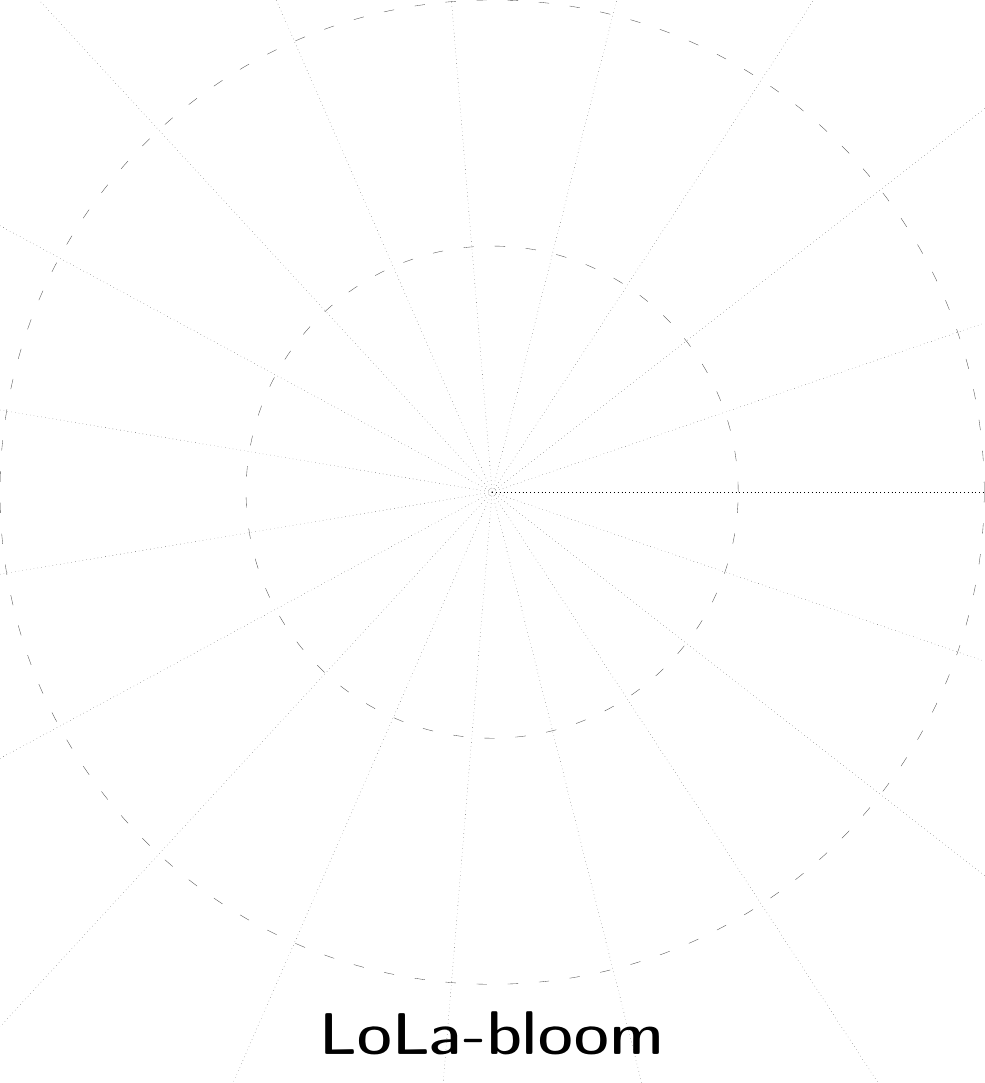}
\\
\medskip
\includegraphics[scale=.4]{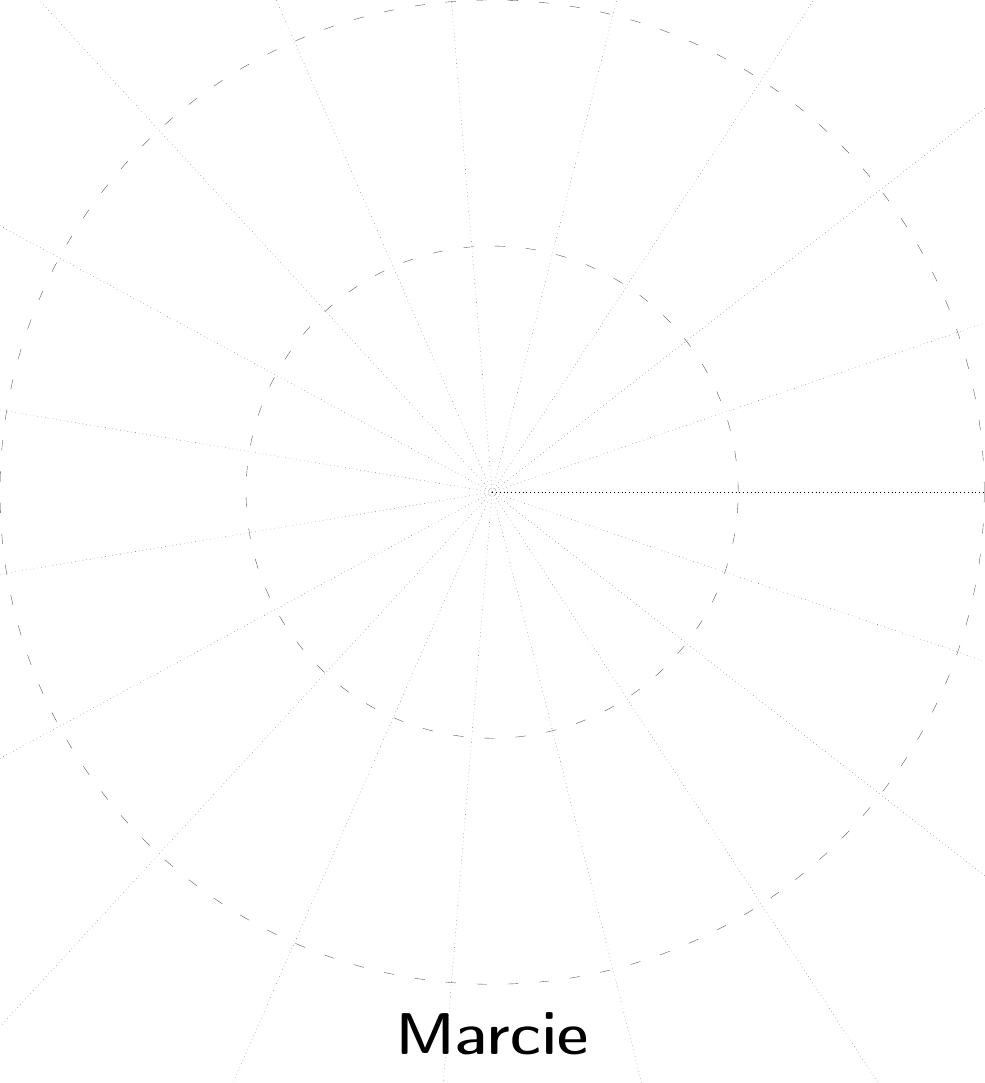}
\hfill
\includegraphics[scale=.4]{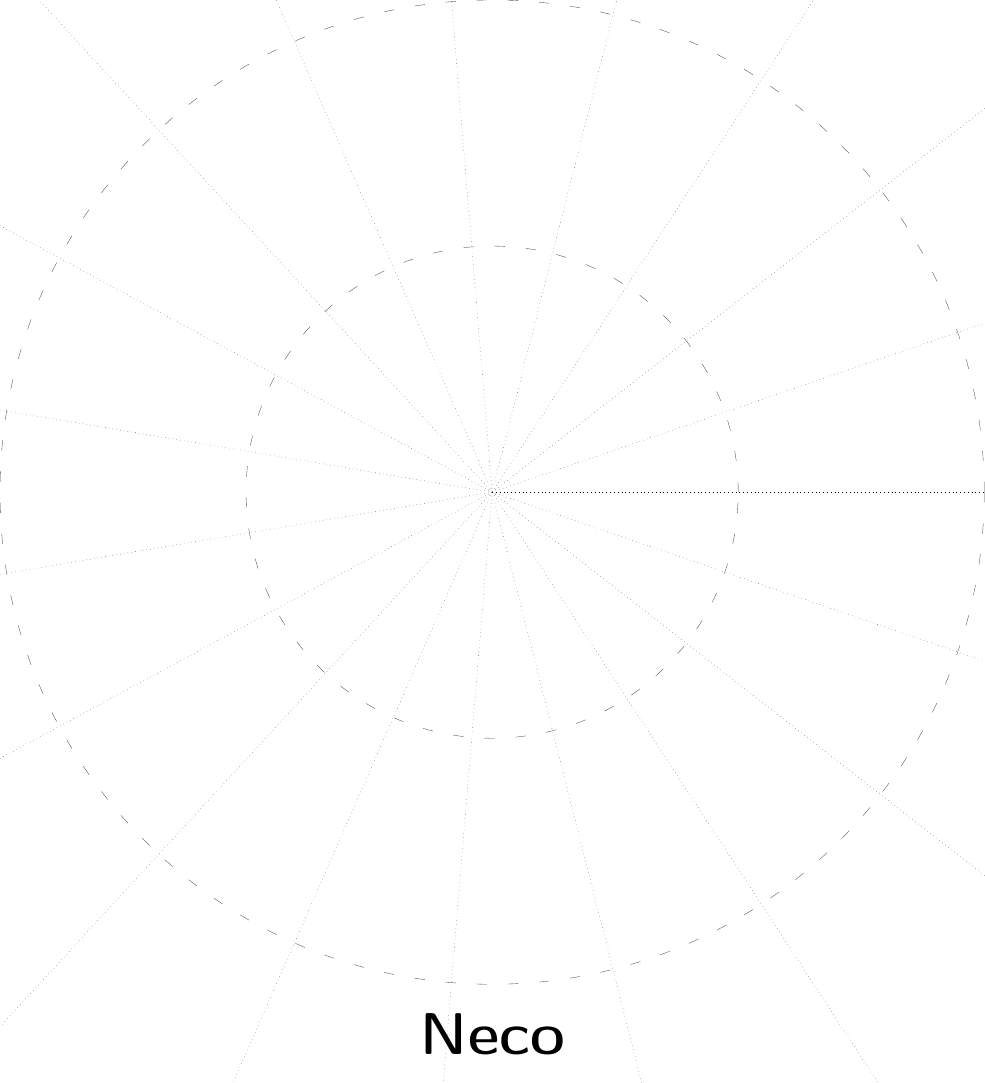}
\hfill
\includegraphics[scale=.4]{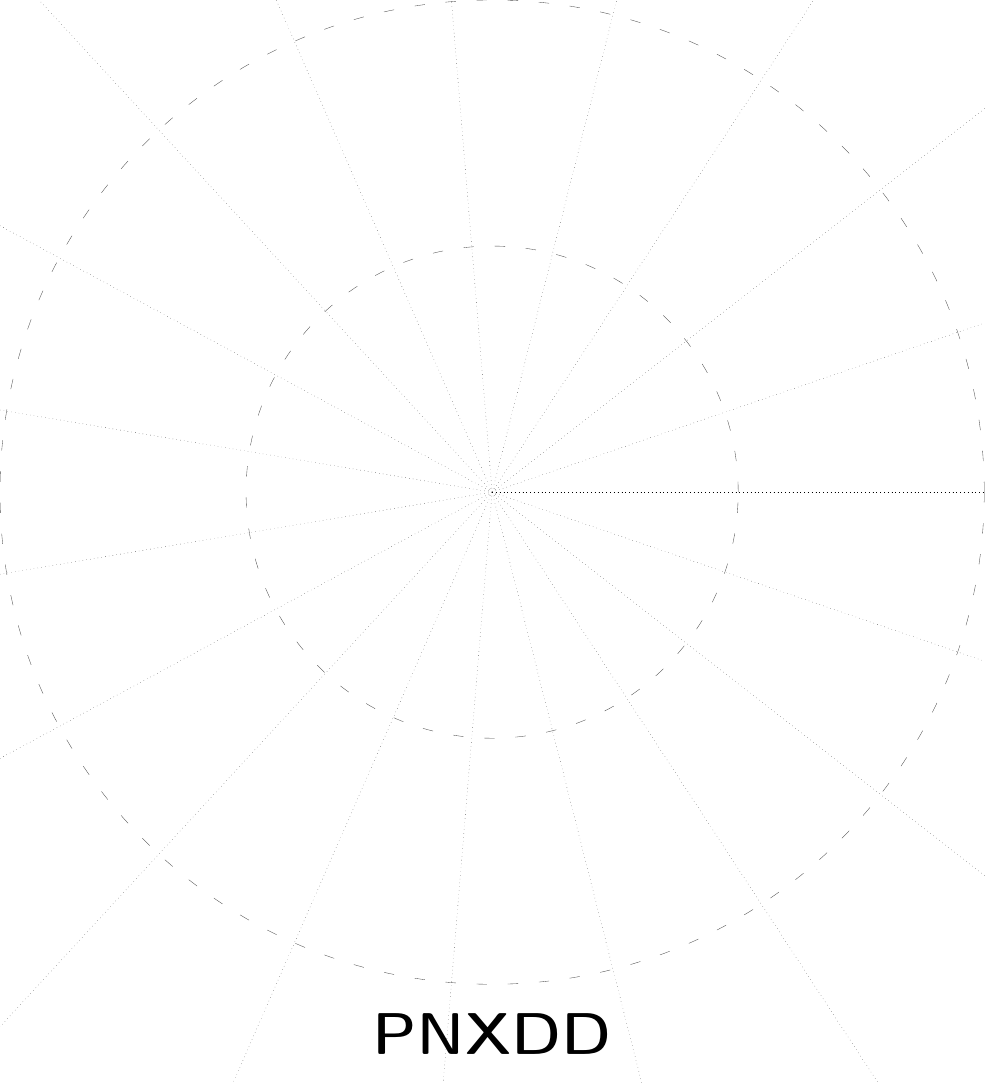}
\\
\medskip
\mbox{}
\hfill
\includegraphics[scale=.4]{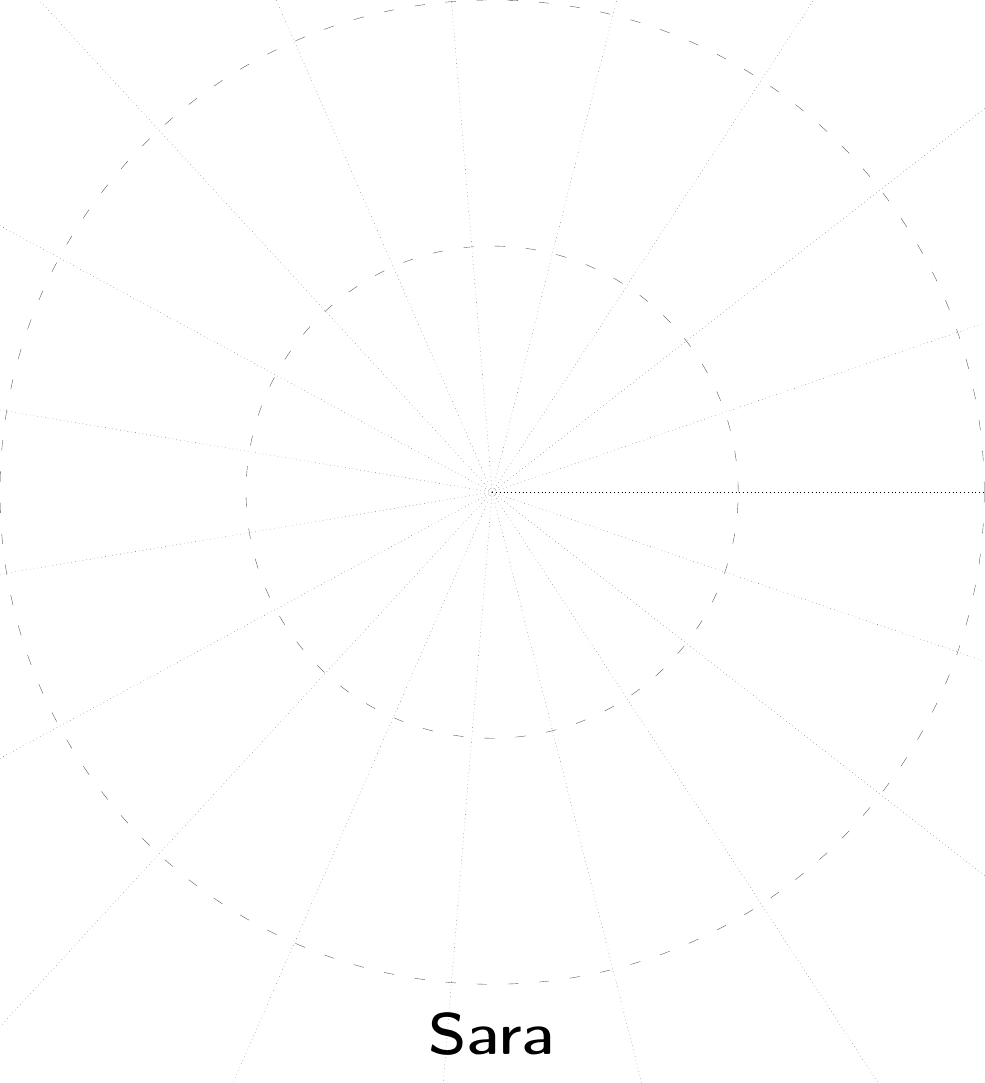}
\hfill
\mbox{}
\\
\bigskip
\mbox{}
\hfill
\includegraphics[scale=.7]{figures/alban-legend-models.pdf}
\hfill
\mbox{}
\end{adjustwidth}
\caption{Handled parameters for each tool,
         in structural formul{\ae} evaluation}
\label{fig:fs:radar:tools}
\end{figure}

\Cref{fig:fs:radar:tools}~presents also radar diagrams showing
graphically the participation and results by tool.
Each slice in the radars represents a model, always at the same position.

These diagrams differ from both those in~\Cref{fig:fs:radar:models}
and those in~\Cref{fig:ss:radar:models}.
Each diagram contains two circles, as in~\Cref{fig:ss:radar:models}:
a slice reaches the inner circle if the tool participates
to the state space competition for this model, but fails.
The number of subslices between the inner and the outer circles represents
how many parameters are handled.
The angle covered by each subslice shows the ratio between
the computed formul{\ae} and the total number of formul{\ae}
in the examination.
For instance, \acs{AlPiNA} has wider subslices than \acs{Helena}:
it is able to handle more formul{\ae}.
But its subslices do not cover the whole angle dedicated to each model:
\acs{AlPiNA} could not handle all the formul{\ae} proposed in this
examination.

The colored surface in the inner circle shows
if the tool could at least handle one formula for the model.
For instance, \acs{AlPiNA} and \acs{Helena}
could handle formul{\ae} for \acs{Simple-LBS},
but not for \acs{Railroad}.

This figure is not sufficient as we do not clearly distinguish tools
that try to handle formul{\ae} from tools that do not compete.
For instance, \acs{AlPiNA} tries for all models,
but has four ``blank'' models in the figure.

\clearpage
\section{Raw Data for Reachability Formul{\ae} Evaluation}
\label{sec:reachformulae}

This section shows the raw results of the reachability formul{\ae}
examination.
\Cref{tab:fr}~summarizes the highest scaling parameter reached by
the tools for each model. Then Charts generated from the data collected for
this examination are provided. This table should be interpreted using the
legend displayed in page~\pageref{position:legend}.

\index{Reachability Formul{\ae}!Tool results (summary)}
\begin{table}[h]
	\centering
  \lstset{
    basicstyle=\scriptsize\fontfamily{fvm}\selectfont
  }
  \footnotesize
  \renewcommand\cellalign{cc}
  \setlength\rotheadsize{5em}
  \hspace{-1em}
  \begin{tabular}{|c|c||m{5em}|m{1.5em}|m{5em}|m{1.5em}|m{5.5em}|m{5.5em}|m{1.5em}|m{1.5em}|m{1.5em}|m{5em}|}
    \cline{3-12}
    \multicolumn{2}{c|}{}
  & \rothead{\textbf{AlPiNA}}
  & \rothead{\textbf{Crocodile}}
  & \rothead{\textbf{Helena}}
  & \rothead{\textbf{ITS-Tools}}
  & \rothead{\textbf{LoLA (binstore)}}
  & \rothead{\textbf{LoLA (bloom)}}
  & \rothead{\textbf{Marcie}}
  & \rothead{\textbf{Neco}}
  & \rothead{\textbf{PNXDD}}
  & \rothead{\textbf{Sara}}
    \lstset{
      basicstyle=\footnotesize\fontfamily{fvm}\selectfont
    }
  \\
  \hline
  \hline
    \multirow{7}{*}{\begin{sideways}\textbf{MCC 2011 models}\end{sideways}}
  & \textbf{FMS}
  & \result[reached=some,typepn=pt]{10} 
  & \result[reached=nc]{} 
  & \result[reached=nc]{} 
  & \result[reached=nc]{} 
  & \result[reached=max,typepn=pt]{500} 
  & \result[reached=max,typepn=pt]{500} 
  & \result[reached=nc]{} 
  & \result[reached=nc]{} 
  & \result[reached=nc]{} 
  & \result[reached=nc]{} 
  \\
  & \textbf{Kanban}
  & \result[reached=some,typepn=pt]{20} 
  & \result[reached=nc]{} 
  & \result[reached=nc]{} 
  & \result[reached=nc]{} 
  & \result[reached=max,typepn=pt]{1\,000} 
  & \result[reached=max,typepn=pt]{1\,000} 
  & \result[reached=nc]{} 
  & \result[reached=nc]{} 
  & \result[reached=nc]{} 
  & \result[reached=nc]{} 
  \\
  & \textbf{MAPK}
  & \result[reached=none]{} 
  & \result[reached=nc]{} 
  & \result[reached=nc]{} 
  & \result[reached=nc]{} 
  & \result[reached=max,typepn=pt]{320} 
  & \result[reached=max,typepn=pt]{320} 
  & \result[reached=nc]{} 
  & \result[reached=nc]{} 
  & \result[reached=nc]{} 
  & \result[reached=nc]{} 
  \\
  & \textbf{Peterson}
  & \result[reached=some,typepn=cn]{2} 
  & \result[reached=nc]{} 
  & \result[reached=none]{} 
  & \result[reached=nc]{} 
  & \result[reached=some,typepn=pt]{4} 
  & \result[reached=nc]{} 
  & \result[reached=nc]{} 
  & \result[reached=nc]{} 
  & \result[reached=nc]{} 
  & \result[reached=best,typepn=pt]{5} 
  \\
  & \textbf{Philosophers}
  & \result[reached=some,typepn=cn]{10} 
  & \result[reached=nc]{} 
  & \result[reached=nc]{} 
  & \result[reached=nc]{} 
  & \result[reached=best,typepn=pt]{50\,000} 
  & \result[reached=nc]{} 
  & \result[reached=nc]{} 
  & \result[reached=nc]{} 
  & \result[reached=nc]{} 
  & \result[reached=nc]{} 
  \\
  & \textbf{Shared-Memory}
  & \result[reached=some,typepn=cn]{5} 
  & \result[reached=nc]{} 
  & \result[reached=best,typepn=cn]{20} 
  & \result[reached=nc]{} 
  & \result[reached=nc]{} 
  & \result[reached=nc]{} 
  & \result[reached=nc]{} 
  & \result[reached=nc]{} 
  & \result[reached=nc]{} 
  & \result[reached=nc]{} 
  \\
  & \textbf{TokenRing}
  & \result[reached=nc]{} 
  & \result[reached=nc]{} 
  & \result[reached=best,typepn=cn]{20} 
  & \result[reached=nc]{} 
  & \result[reached=nc]{} 
  & \result[reached=nc]{} 
  & \result[reached=nc]{} 
  & \result[reached=nc]{} 
  & \result[reached=nc]{} 
  & \result[reached=nc]{} 
  \\
  \hline
  \hline
    \multirow{12}{*}{\begin{sideways}\textbf{new models from MCC 2012}\end{sideways}}
  & \textbf{Cs\_repetitions}
  & \result[reached=none]{} 
  & \result[reached=nc]{} 
  & \result[reached=nc]{} 
  & \result[reached=nc]{} 
  & \result[reached=some,typepn=pt]{225} 
  & \result[reached=nc]{} 
  & \result[reached=nc]{} 
  & \result[reached=nc]{} 
  & \result[reached=nc]{} 
  & \result[reached=max,typepn=pt]{900} 
  \\
  & \textbf{Echo}
  & \result[reached=none,typepn=no]{} 
  & \result[reached=nc]{} 
  & \result[reached=nc]{} 
  & \result[reached=nc]{} 
  & \result[reached=best,typepn=pt]{d2r11} 
  & \result[reached=best,typepn=pt]{d2r11} 
  & \result[reached=nc]{} 
  & \result[reached=nc]{} 
  & \result[reached=nc]{} 
  & \result[reached=best,typepn=pt]{d2r11} 
  \\
  & \textbf{Eratosthenes}
  & \result[reached=none,typepn=no]{} 
  & \result[reached=nc]{} 
  & \result[reached=nc]{} 
  & \result[reached=nc]{} 
  & \result[reached=max,typepn=pt]{500} 
  & \result[reached=nc]{} 
  & \result[reached=nc]{} 
  & \result[reached=nc]{} 
  & \result[reached=nc]{} 
  & \result[reached=max,typepn=pt]{500} 
  \\
  & \textbf{Galloc\_res}
  & \result[reached=none,typepn=no]{} 
  & \result[reached=nc]{} 
  & \result[reached=nc]{} 
  & \result[reached=nc]{} 
  & \result[reached=some,typepn=pt]{3} 
  & \result[reached=nc]{} 
  & \result[reached=nc]{} 
  & \result[reached=nc]{} 
  & \result[reached=nc]{} 
  & \result[reached=best,typepn=pt]{5} 
  \\
  & \textbf{Lamport\_fmea}
  & \result[reached=some,typepn=cn]{3} 
  & \result[reached=nc]{} 
  & \result[reached=best,typepn=cn]{5} 
  & \result[reached=nc]{} 
  & \result[reached=some,typepn=pt]{4} 
  & \result[reached=nc]{} 
  & \result[reached=nc]{} 
  & \result[reached=nc]{} 
  & \result[reached=nc]{} 
  & \result[reached=nc]{} 
  \\
  & \textbf{NEO-election}
  & \result[reached=nc]{} 
  & \result[reached=nc]{} 
  & \result[reached=nc]{} 
  & \result[reached=nc]{} 
  & \result[reached=max,typepn=pt]{8} 
  & \result[reached=nc]{} 
  & \result[reached=nc]{} 
  & \result[reached=nc]{} 
  & \result[reached=nc]{} 
  & \result[reached=nc]{} 
  \\
  & \textbf{Philo\_dyn}
  & \result[reached=some,typepn=cn]{3} 
  & \result[reached=nc]{} 
  & \result[reached=best,typepn=cn]{20} 
  & \result[reached=nc]{} 
  & \result[reached=best,typepn=pt]{20} 
  & \result[reached=nc]{} 
  & \result[reached=nc]{} 
  & \result[reached=nc]{} 
  & \result[reached=nc]{} 
  & \result[reached=nc]{} 
  \\
  & \textbf{Planning}
  & \result[reached=none]{} 
  & \result[reached=nc]{} 
  & \result[reached=nc]{} 
  & \result[reached=nc]{} 
  & \result[reached=nc]{} 
  & \result[reached=nc]{} 
  & \result[reached=nc]{} 
  & \result[reached=nc]{} 
  & \result[reached=nc]{} 
  & \result[reached=nc]{} 
  \\
  & \textbf{Railroad}
  & \result[reached=nc]{} 
  & \result[reached=nc]{} 
  & \result[reached=nc]{} 
  & \result[reached=nc]{} 
  & \result[reached=nc]{} 
  & \result[reached=nc]{} 
  & \result[reached=nc]{} 
  & \result[reached=nc]{} 
  & \result[reached=nc]{} 
  & \result[reached=nc]{} 
   \\
  & \textbf{Ring}
  & \result[reached=none]{} 
  & \result[reached=nc]{} 
  & \result[reached=nc]{} 
  & \result[reached=nc]{} 
  & \result[reached=nc]{} 
  & \result[reached=nc]{} 
  & \result[reached=nc]{} 
  & \result[reached=nc]{} 
  & \result[reached=nc]{} 
  & \result[reached=nc]{} 
  \\
  & \textbf{Rw\_mutex}
  & \result[reached=some,typepn=pt]{r10w100} 
  & \result[reached=nc]{} 
  & \result[reached=nc]{} 
  & \result[reached=nc]{} 
  & \result[reached=max,typepn=pt]{r2000w10} 
  & \result[reached=max,typepn=pt]{r2000w10} 
  & \result[reached=nc]{} 
  & \result[reached=nc]{} 
  & \result[reached=nc]{} 
  & \result[reached=some,typepn=pt]{r20w10} 
  \\
  & \textbf{Simple\_lbs}
  & \result[reached=some,typepn=pt]{2} 
  & \result[reached=nc]{} 
  & \result[reached=some,typepn=cn]{15} 
  & \result[reached=nc]{} 
  & \result[reached=some,typepn=pt]{5} 
  & \result[reached=max,typepn=pt]{20} 
  & \result[reached=nc]{} 
  & \result[reached=nc]{} 
  & \result[reached=nc]{} 
  & \result[reached=max,typepn=pt]{20} 
  \\
  \hline
  \end{tabular}
  \caption{Results for the reachability formul{\ae} examination}
  \label{tab:fr}
\end{table}

\Cref{sec:frresults}~presents the data computed by tools.
Then, \Cref{sec:frprocessedmodels}~shows how models have been handled
by tools and \Cref{sec:fr:bymodels}~summarizes how tools did cope
with models.

\Cref{sec:frresults}~shows that no tool compute the same set
of formul{\ae}.
This is why we do not display the charts that have been extracted
from the executions since they have no meaning at all.

\subsection{Computed results for the formul\ae}
\label{sec:frresults}
\index{Reachability Formul{\ae}!Computed data}

Since the participating tools did produce various values, we could not
consolidate the results and decided to show them as they were produced in
\Cref{tab:result:FR1,tab:result:FR2}. We use the notations already presented
for~\Cref{tab:result:FS}, page~\pageref{tab:result:FS} (use of \texttt{T},
\texttt{F} or ``.'' in the issued vector).

\begin{table}
\centering
\begin{tabular}{|c||c|c|c|c|c|}
	\hline
	\textbf{Scale} & \acs{AlPiNA} &
\acs{Helena}  & \acs{LoLA-binstore} & \acs{LoLA-bloom} &
\acs{Sara}\\
	\hline
	\hline
	\multicolumn{6}{|c|}{\acs{CS-Repetitions}} \\
	25 & \cellcolor{gray!50} & \cellcolor{gray!50} & {\scriptsize T...........................} & \cellcolor{gray!50} & {\scriptsize ....FFFTTT..................} \\
	49 & \cellcolor{gray!50} & \cellcolor{gray!50} & {\scriptsize T...........................} & \cellcolor{gray!50} & {\scriptsize ....TTT..................FFT} \\
	100 & \cellcolor{gray!50} & \cellcolor{gray!50} & {\scriptsize T...........................} & \cellcolor{gray!50} & {\scriptsize ....TTT...TTT...............} \\
	225 & \cellcolor{gray!50} & \cellcolor{gray!50} & {\scriptsize T...........................} & \cellcolor{gray!50} & {\scriptsize ....TTT.....................} \\
	400 & \cellcolor{gray!50} & \cellcolor{gray!50} & {\scriptsize ?} & \cellcolor{gray!50} & {\scriptsize ..........FFF...............} \\
	625 & \cellcolor{gray!50} & \cellcolor{gray!50} & {\scriptsize ?} & \cellcolor{gray!50} & {\scriptsize .............TTT.........FFF} \\
	900 & \cellcolor{gray!50} & \cellcolor{gray!50} & {\scriptsize ?} & \cellcolor{gray!50} & {\scriptsize ....TTT...TTT...............} \\
	\hline
	\multicolumn{6}{|c|}{\acs{Echo}} \\
	s2r11 & {\scriptsize ?} & \cellcolor{gray!50} & {\scriptsize T...FFFFFFFFFFFFFFFFFF} & {\scriptsize ....F.FF.FF..FF.F..FFF} & {\scriptsize .............FFF......}\\
	\hline
	\multicolumn{6}{|c|}{\acs{Eratosthenes}} \\
	5 & {\scriptsize ?} & \cellcolor{gray!50} & {\scriptsize ..............} & {\scriptsize ..............} & {\scriptsize .FFFF...FFFF..}\\
	10 & {\scriptsize ?} & \cellcolor{gray!50} & {\scriptsize ..............} & \cellcolor{gray!50} & {\scriptsize .FFFF...FFFF..}\\
	20 & {\scriptsize ?} & \cellcolor{gray!50} & {\scriptsize ..............} & \cellcolor{gray!50} & {\scriptsize .FFFF...FFFF..}\\
	50 & {\scriptsize ?} & \cellcolor{gray!50} & {\scriptsize ..............} & \cellcolor{gray!50} & {\scriptsize .FFFF...FFFF..}\\
	100 & {\scriptsize ?} & \cellcolor{gray!50} & {\scriptsize ..............} & \cellcolor{gray!50} & {\scriptsize .FFFF...FFFF..}\\
	200 & {\scriptsize ?} & \cellcolor{gray!50} & {\scriptsize ..............} & \cellcolor{gray!50} & {\scriptsize .FFFF...FFFF..}\\
	500 & {\scriptsize ?} & \cellcolor{gray!50} & {\scriptsize ..............} & \cellcolor{gray!50} & {\scriptsize .FFFF...FFFF..}\\
	\hline
	\multicolumn{6}{|c|}{\acs{FMS}} \\
	2 & {\scriptsize ....FFFFFFFTFFFFFFFFFF} & \cellcolor{gray!50} & {\scriptsize F...FFFFFFFTFFFFFFFFFF} & {\scriptsize ....FFFFF.F.FFFFF..F.F} & {\scriptsize ................TTT...}\\
	5 & {\scriptsize ....FFFFFFFFFFFFFFFFFF} & \cellcolor{gray!50} & {\scriptsize F...FFFFFFFFFFFFFFFFFF} & {\scriptsize ....F..FFF...........F} & {\scriptsize ................TTT...}\\
	10 & {\scriptsize ....FFFFFFFFFFFFFFFFFF} & \cellcolor{gray!50} & {\scriptsize F...FFFFFFFFFFFFFFFFFF} & {\scriptsize ....F..F..F....FFFF...} & \cellcolor{gray!50}\\
	20 & {\scriptsize ?} & \cellcolor{gray!50} & {\scriptsize FFFF...FFFFFFFFFFFFFFF} & {\scriptsize FFF....FFFFFF...FF...F} & \cellcolor{gray!50}\\
	50 & {\scriptsize ?} & \cellcolor{gray!50} & {\scriptsize FFFF...FFFFFFFFFFFFFFF} & {\scriptsize ..F....FF....FF...FF.F} & \cellcolor{gray!50}\\
	100 & {\scriptsize ?} & \cellcolor{gray!50} & {\scriptsize FFFFFFFFF...FFFFFFFFFF} & {\scriptsize F.FFFF.F....F..F.FF..F} & \cellcolor{gray!50}\\
	200 & {\scriptsize ?} & \cellcolor{gray!50} & {\scriptsize FFFFFFFFF...FFFFFFFFFF} & {\scriptsize .F..FF.........F....FF} & \cellcolor{gray!50}\\
	500 & {\scriptsize ?} & \cellcolor{gray!50} & {\scriptsize FFFFFFFFF...FFFFFFFFFF} & {\scriptsize ..F..FF.....F..FFFF.FF} & \cellcolor{gray!50}\\
	\hline
	\multicolumn{6}{|c|}{\acs{Galloc}} \\
	3 & {\scriptsize ?} & \cellcolor{gray!50} & {\scriptsize F...........................} & \cellcolor{gray!50} & {\scriptsize .......FFF...TTF............}\\
	5 & {\scriptsize ?} & \cellcolor{gray!50} & \cellcolor{gray!50} & \cellcolor{gray!50} & {\scriptsize ..........TTT...............}\\
	\hline
	\multicolumn{6}{|c|}{\acs{Kanban}} \\
	5 & {\scriptsize ....FFFFFFFFFFFFFFFFFF} & \cellcolor{gray!50} & {\scriptsize F...FFFFFFFFFFFFFFFFFF} & {\scriptsize ....F.FFFFF..F..FF.F..} & {\scriptsize ....TTT.........TTF...}\\
	10 & {\scriptsize ....FTFFFFFFFFFFFFFFFF} & \cellcolor{gray!50} & {\scriptsize F...FTFFFFFFFFFFFFFFFF} & {\scriptsize ....F..FFFF..F.FF.FF.F} & \cellcolor{gray!50}\\
	20 & {\scriptsize FFF....FFFFFFFFFFFFFFF} & \cellcolor{gray!50} & {\scriptsize FFFF...FFFFFFFFFFFFFFF} & {\scriptsize FF.....FFFF.FFFFF..FFF} & \cellcolor{gray!50}\\
	50 & {\scriptsize ?} & \cellcolor{gray!50} & {\scriptsize FFFF...FFFFFFFFFFFFFFF} & {\scriptsize F........FF....F..FF.F} & \cellcolor{gray!50}\\
	100 & {\scriptsize ?} & \cellcolor{gray!50} & {\scriptsize FFFFFFFF....FFFFFFFFFF} & {\scriptsize ..FFF..........FFFF...} & \cellcolor{gray!50}\\
	200 & {\scriptsize ?} & \cellcolor{gray!50} & {\scriptsize FFFFFFFF....FFFFF.FFFF} & {\scriptsize ....FFF.....F.FFF.F.FF} & \cellcolor{gray!50}\\
	500 & {\scriptsize ?} & \cellcolor{gray!50} & {\scriptsize FFFFFFFF....FFFFFFFFFF} & {\scriptsize .F..........FFFF..FF.F} & \cellcolor{gray!50}\\
	1\,000 & {\scriptsize ?} & \cellcolor{gray!50} & {\scriptsize FFFFFFFF....FFFFFFFFFF} & {\scriptsize ..FFFFFF......FF..F...} & \cellcolor{gray!50}\\
	\hline
	\multicolumn{6}{|c|}{\acs{Lamport}} \\
	2 & {\scriptsize ......................FFF...} & {\scriptsize T...........................} & {\scriptsize F...........................} & \cellcolor{gray!50} & \cellcolor{gray!50}\\
	3 & {\scriptsize .........................FFF} & {\scriptsize T...........................} & {\scriptsize F...........................} & \cellcolor{gray!50} & \cellcolor{gray!50}\\
	4 & {\scriptsize ?} & {\scriptsize T...........................} & {\scriptsize F...........................} & \cellcolor{gray!50} & \cellcolor{gray!50}\\
	5 & {\scriptsize ?} & {\scriptsize T...........................} & {\scriptsize ?} & \cellcolor{gray!50} & \cellcolor{gray!50}\\
	\hline
	\multicolumn{6}{|c|}{\acs{MAPK}} \\
	8 & {\scriptsize ?} & \cellcolor{gray!50} & {\scriptsize F...FFFFFFFFFFTFFFFFFF} & {\scriptsize ....FF.F.FF..F.FFFFFF.} & {\scriptsize ....TTT.........FTT...} \\
	20 & {\scriptsize ?} & \cellcolor{gray!50} & {\scriptsize FFF....FFFFFFFFFFFFFFF} & {\scriptsize F...........FFFFF..F.F} & \cellcolor{gray!50} \\
	40 & {\scriptsize ?} & \cellcolor{gray!50} & {\scriptsize FFF....FFFFFFFFFFFFFFF} & {\scriptsize F......F.FFFFF........} & \cellcolor{gray!50} \\
	80 & {\scriptsize ?} & \cellcolor{gray!50} & {\scriptsize FFFFFFFF....FFFFFFFFFF} & {\scriptsize F...FF......FF.FF.F..F} & \cellcolor{gray!50} \\
	160 & {\scriptsize ?} & \cellcolor{gray!50} & {\scriptsize FFFFFFFF....FFFFFFFFFF} & {\scriptsize .FFF.FFF.......FF.FFFF} & \cellcolor{gray!50} \\
	320 & {\scriptsize ?} & \cellcolor{gray!50} & {\scriptsize FFFFFFFF....FFFFFFFF.F} & {\scriptsize .FF.F.......F.F..FF..F} & \cellcolor{gray!50} \\
	\hline
	\multicolumn{6}{|c|}{\acs{Neo-Election}} \\
	2 & \cellcolor{gray!50} & \cellcolor{gray!50} & {\scriptsize T...........................} & \cellcolor{gray!50} & \cellcolor{gray!50}\\
	3 & \cellcolor{gray!50} & \cellcolor{gray!50} & {\scriptsize T...........................} & \cellcolor{gray!50} & \cellcolor{gray!50}\\
	4 & \cellcolor{gray!50} & \cellcolor{gray!50} & {\scriptsize T...........................} & \cellcolor{gray!50} & \cellcolor{gray!50}\\
	5 & \cellcolor{gray!50} & \cellcolor{gray!50} & {\scriptsize T...........................} & \cellcolor{gray!50} & \cellcolor{gray!50}\\
	6 & \cellcolor{gray!50} & \cellcolor{gray!50} & {\scriptsize T...........................} & \cellcolor{gray!50} & \cellcolor{gray!50}\\
	7 & \cellcolor{gray!50} & \cellcolor{gray!50} & {\scriptsize T...........................} & \cellcolor{gray!50} & \cellcolor{gray!50}\\
	8 & \cellcolor{gray!50} & \cellcolor{gray!50} & {\scriptsize T...........................} & \cellcolor{gray!50} & \cellcolor{gray!50}\\
	\hline
\end{tabular}
\caption{Results of reachability formul{\ae} evaluation for the models where at least one tool produced a result\label{tab:result:FR1}}
\end{table}

\begin{table}
\centering
\begin{tabular}{|c||c|c|c|c|c|}
	\hline
	\textbf{Scale} & \acs{AlPiNA} &
\acs{Helena}  & \acs{LoLA-binstore} & \acs{LoLA-bloom} &
\acs{Sara}\\
	\hline
	\hline
	\multicolumn{6}{|c|}{\acs{Peterson}} \\
	2 & {\scriptsize ....FFFFFFFFFFFF......FFFFFF} & {\scriptsize ?} & {\scriptsize F...........................} & \cellcolor{gray!50} & {\scriptsize ..........FFT...............}\\
	3 & {\scriptsize ?} & {\scriptsize ?} & {\scriptsize F...........................} & \cellcolor{gray!50} & {\scriptsize ......................TTTFFT}\\
	4 & {\scriptsize ?} & {\scriptsize ?} & {\scriptsize F...........................} & \cellcolor{gray!50} & {\scriptsize ......................TTTTTT}\\
	5 & {\scriptsize ?} & {\scriptsize ?} & {\scriptsize ?} & \cellcolor{gray!50} & {\scriptsize ......................TTT...}\\
	\hline
	\multicolumn{6}{|c|}{\acs{Dynamic-Philosophers}} \\
	2 & {\scriptsize ....FFFFFFFFFFFF......FFFFFF} & {\scriptsize T...........................} & {\scriptsize T...........................} & \cellcolor{gray!50} & \cellcolor{gray!50}\\
	3 & {\scriptsize ....FTTFFFFFFFTF......FFFFFF} & {\scriptsize F...........................} & {\scriptsize T...........................} & \cellcolor{gray!50} & \cellcolor{gray!50}\\
	10 & {\scriptsize ?} & {\scriptsize F...........................} & {\scriptsize T...........................} & \cellcolor{gray!50} & \cellcolor{gray!50}\\
	20 & {\scriptsize ?} & {\scriptsize F...........................} & {\scriptsize T...........................} & \cellcolor{gray!50} & \cellcolor{gray!50}\\
	\hline
	\multicolumn{6}{|c|}{\acs{Philosophers}} \\
	5 & {\scriptsize ....FFFFFFFFFFFF......FFFFFF} & \cellcolor{gray!50} & {\scriptsize T...........................} & \cellcolor{gray!50} & {\scriptsize ..........TTT...............}\\
	10 & {\scriptsize ....FFFFFFFFFFFF......FFFFFF} & \cellcolor{gray!50} & {\scriptsize T...........................} & \cellcolor{gray!50} & \cellcolor{gray!50}\\
	20 & {\scriptsize ?} & \cellcolor{gray!50} & {\scriptsize T...........................} & \cellcolor{gray!50} & \cellcolor{gray!50}\\
	50 & {\scriptsize ?} & \cellcolor{gray!50} & {\scriptsize T...........................} & \cellcolor{gray!50} & \cellcolor{gray!50}\\
	100 & {\scriptsize ?} & \cellcolor{gray!50} & {\scriptsize T...........................} & \cellcolor{gray!50} & \cellcolor{gray!50}\\
	200 & {\scriptsize ?} & \cellcolor{gray!50} & {\scriptsize T...........................} & \cellcolor{gray!50} & \cellcolor{gray!50}\\
	500 & {\scriptsize ?} & \cellcolor{gray!50} & {\scriptsize T...........................} & \cellcolor{gray!50} & \cellcolor{gray!50}\\
	1\,000 & {\scriptsize ?} & \cellcolor{gray!50} & {\scriptsize T...........................} & \cellcolor{gray!50} & \cellcolor{gray!50}\\
	2\,000 & {\scriptsize ?} & \cellcolor{gray!50} & {\scriptsize T...........................} & \cellcolor{gray!50} & \cellcolor{gray!50}\\
	5\,000 & {\scriptsize ?} & \cellcolor{gray!50} & {\scriptsize T...........................} & \cellcolor{gray!50} & \cellcolor{gray!50}\\
	10\,000 & {\scriptsize ?} & \cellcolor{gray!50} & {\scriptsize T...........................} & \cellcolor{gray!50} & \cellcolor{gray!50}\\
	50\,000 & {\scriptsize ?} & \cellcolor{gray!50} & {\scriptsize T...........................} & \cellcolor{gray!50} & \cellcolor{gray!50}\\
	\hline
	\multicolumn{6}{|c|}{\acs{RW-Mutex}} \\
	r10w10 & {\scriptsize ....FFFFTFFFFFFFFFFFFF} & \cellcolor{gray!50} & {\scriptsize F...FFFFFFFFFFFFFFFFFF} & {\scriptsize ....F.FFF.F..FFFFFF...} & {\scriptsize ..........FFT...FTT...}\\
	r10w20 & {\scriptsize ....FFFFFFFFFFTFFFFFFF} & \cellcolor{gray!50} & {\scriptsize F...FFFFFFFFFFTFFTFFFF} & {\scriptsize ....FFF...F..F.FF.F...} & {\scriptsize ..........TTF......TTT}\\
	r10w50 & {\scriptsize ....FFFFTTFFFFFFFFFFTF} & \cellcolor{gray!50} & {\scriptsize F...FFTFFFFTFFFFFTFFTF} & {\scriptsize ....FF.FFF..FF.F...F..} & {\scriptsize ................TTT...}\\
	r10w100 & {\scriptsize ....FTFFFFFFFFFTFFFFFT} & \cellcolor{gray!50} & {\scriptsize F...FFFFFFFFFFTFFFFFTF} & {\scriptsize ....F..FF.F.....FFF...} & {\scriptsize ....TTT...............}\\
	r10w500 & {\scriptsize ?} & \cellcolor{gray!50} & {\scriptsize F...FTFFFFFFFFFFFFFFTF} & {\scriptsize .......FF.FF.FF.F.FF.F} & {\scriptsize ..........TTT.........}\\
	r10w1000 & {\scriptsize ?} & \cellcolor{gray!50} & {\scriptsize F...FTFFTTFTFFFTFFFFFF} & {\scriptsize ......FF....FF..F..F..} & {\scriptsize ....TTT...TTT.........}\\
	r10w2000 & {\scriptsize ?} & \cellcolor{gray!50} & {\scriptsize F...FTFFFFFFFFFFFFTFFF} & {\scriptsize ......FF.FF.FFFFF..FFF} & {\scriptsize ..........TTT.........}\\
	r20w10 & {\scriptsize ?} & \cellcolor{gray!50} & {\scriptsize F...FFFFFFFFFFFFFFFFFF} & {\scriptsize ....F.....FF.FF......F} & {\scriptsize ....FFT...............}\\
	r100w10 & {\scriptsize ?} & \cellcolor{gray!50} & {\scriptsize F...FFFFFFFFFFFFFFFFFF} & {\scriptsize ....FF.F..FFFF.FF.FFF.} & \cellcolor{gray!50}\\
	r500w10 & {\scriptsize ?} & \cellcolor{gray!50} & {\scriptsize F...FFFFFFFFFFFFFFFFFF} & {\scriptsize ......F...F..F.FF..FF.} & \cellcolor{gray!50}\\
	r1000w10 & {\scriptsize ?} & \cellcolor{gray!50} & {\scriptsize F...FFFFFFFFFFFFFFFFFF} & {\scriptsize .......F.FF.FFF.FF...F} & \cellcolor{gray!50}\\
	r2000w10 & {\scriptsize ?} & \cellcolor{gray!50} & {\scriptsize F...FFFFFFFFFFFFFFFFFF} & {\scriptsize ....F.FF.FFFFF.....FF.} & \cellcolor{gray!50}\\
	\hline
	\multicolumn{6}{|c|}{\acs{Shared-Memory}} \\
	5 & {\scriptsize ....FFFFFFFTFFFF......FFFFFF} & {\scriptsize T...........................} & \cellcolor{gray!50} & \cellcolor{gray!50} & \cellcolor{gray!50}\\
	10 & {\scriptsize ?} & {\scriptsize T...........................} & \cellcolor{gray!50} & \cellcolor{gray!50} & \cellcolor{gray!50}\\
	20 & {\scriptsize ?} & {\scriptsize T...........................} & \cellcolor{gray!50} & \cellcolor{gray!50} & \cellcolor{gray!50}\\
	\hline
	\multicolumn{6}{|c|}{\acs{Simple-LBS}} \\
	2 & {\scriptsize ..........FFFFFF............} & {\scriptsize T...........................} & {\scriptsize F...FTFFTFFFFFFF......FFFFFF} & {\scriptsize ....F.F...F.FF.F.........F.F} & {\scriptsize ................FT..........}\\
	5 & {\scriptsize ?} & {\scriptsize T...........................} & {\scriptsize F...FFFFFFFFFFFF......FFFFFF} & {\scriptsize ....F..F..F..F.F......FF.FFF} & {\scriptsize ................TTT.........}\\
	10 & {\scriptsize ?} & {\scriptsize T...........................} & {\scriptsize ?} & {\scriptsize ....F.FF..F..F.F......FFFF..} & {\scriptsize ................F.F.........}\\
	15 & {\scriptsize ?} & {\scriptsize T...........................} & {\scriptsize ?} & {\scriptsize .......F.FF.FFFF......F..FF.} & {\scriptsize ................F.....FFT...}\\
	20 & {\scriptsize ?} & \cellcolor{gray!50} & {\scriptsize ?} & {\scriptsize ....F..F.FF..F.F......F..FFF} & {\scriptsize .......TT.......F...........}\\
	\hline
	\multicolumn{6}{|c|}{\acs{Token-Ring}} \\
	5 \cellcolor{gray!50} & & {\scriptsize T...............} & \cellcolor{gray!50} & \cellcolor{gray!50} & \cellcolor{gray!50} \\
	10 \cellcolor{gray!50} & & {\scriptsize T...............} & \cellcolor{gray!50} & \cellcolor{gray!50} & \cellcolor{gray!50} \\
	20 \cellcolor{gray!50} & & {\scriptsize T...............} & \cellcolor{gray!50} & \cellcolor{gray!50} & \cellcolor{gray!50} \\
	\hline
\end{tabular}
\caption{Results of reachability formul{\ae} evaluation for the models where at least one tool produced a result (continued)\label{tab:result:FR2}}
\end{table}

\subsection{Processed Models}
\label{sec:frprocessedmodels}
\index{Reachability Formul{\ae}!Processed models}

This section summarizes how models were processed by tools. Let us first note
that no tool succeeded in this examination for the following models:

\begin{itemize}
	\item \acs{Planning},
	\item \acs{Railroad},
	\item \acs{Ring}.
\end{itemize}

These models constitute challenges for the next edition of the \acs{MCC}.

\subsection{Radars by models}
\label{sec:fr:bymodels}
\begin{figure}[p]
\centering
\begin{adjustwidth}{-2em}{-2em}
\noindent
\includegraphics[scale=.35]{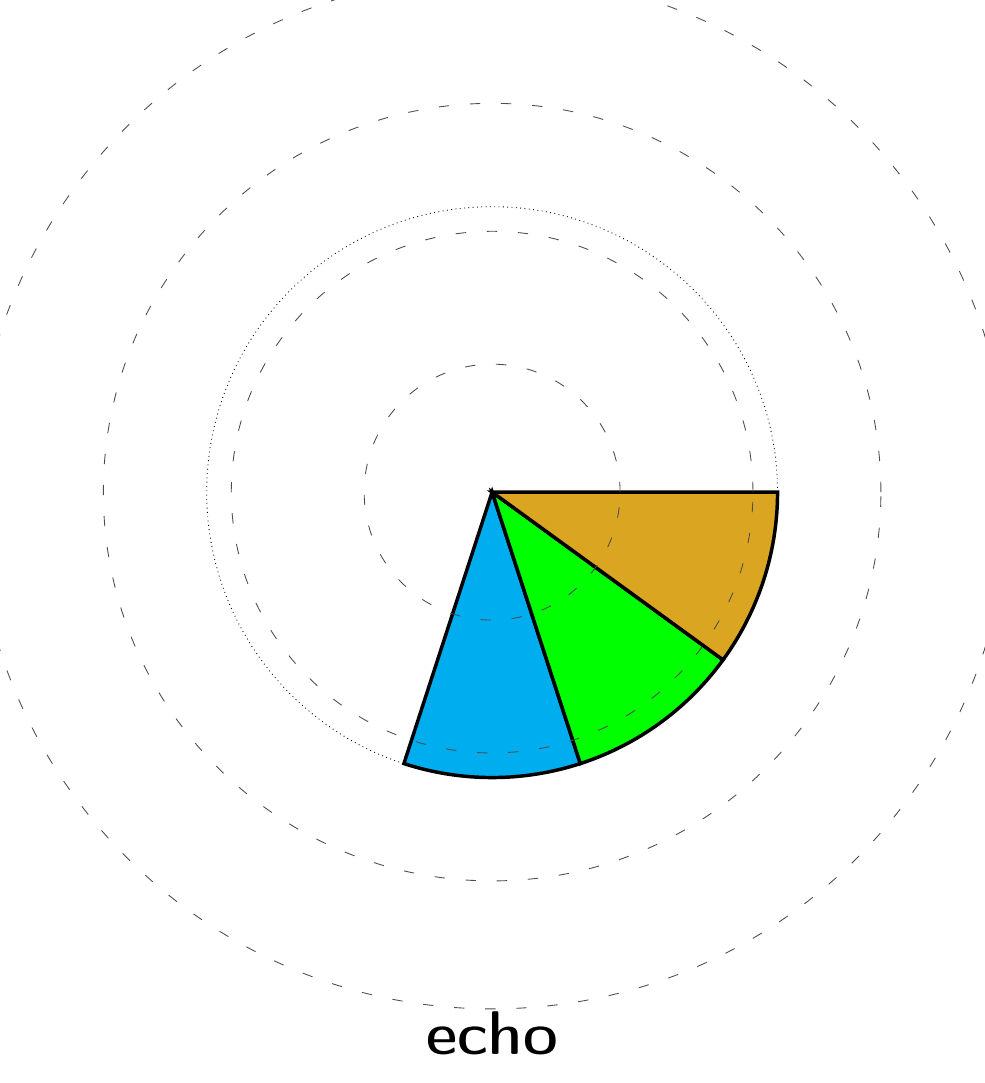}
\hfill
\includegraphics[scale=.35]{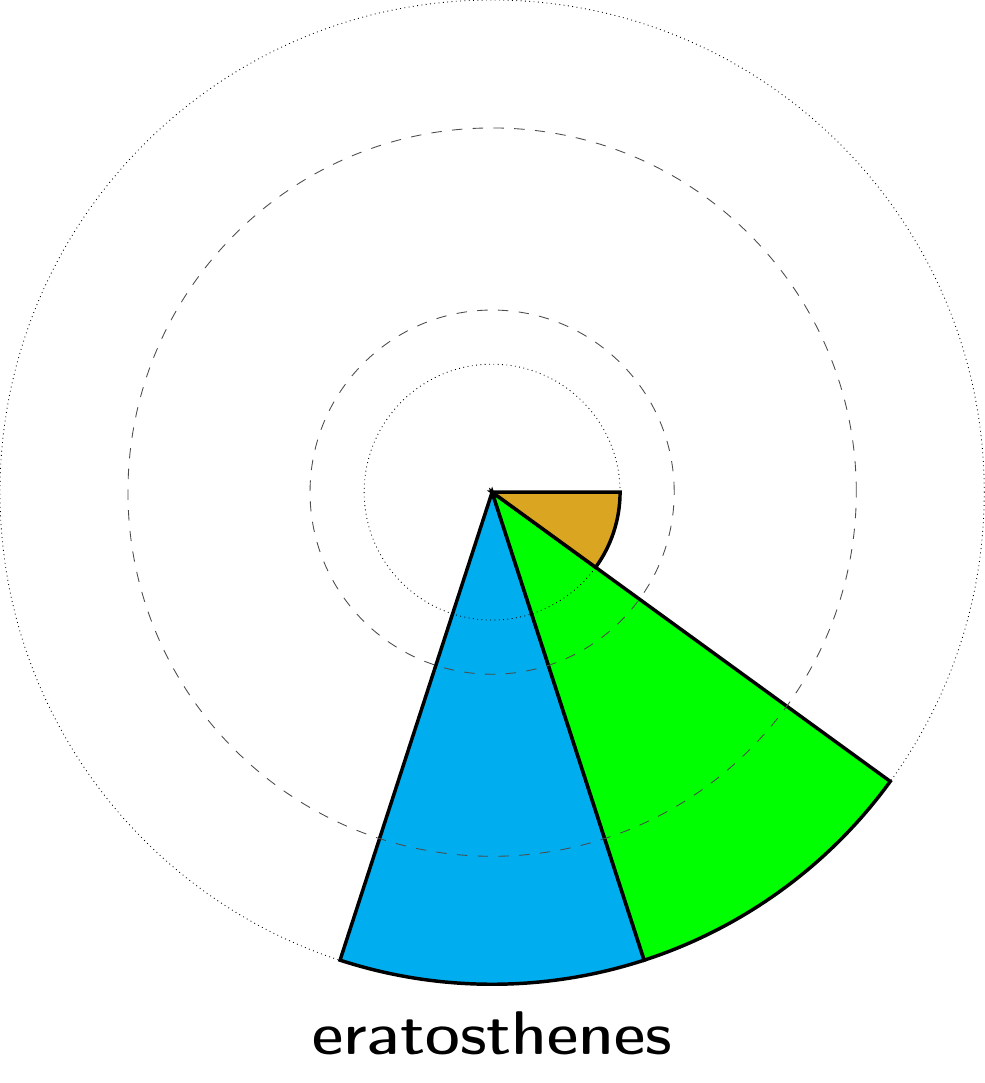}
\hfill
\includegraphics[scale=.35]{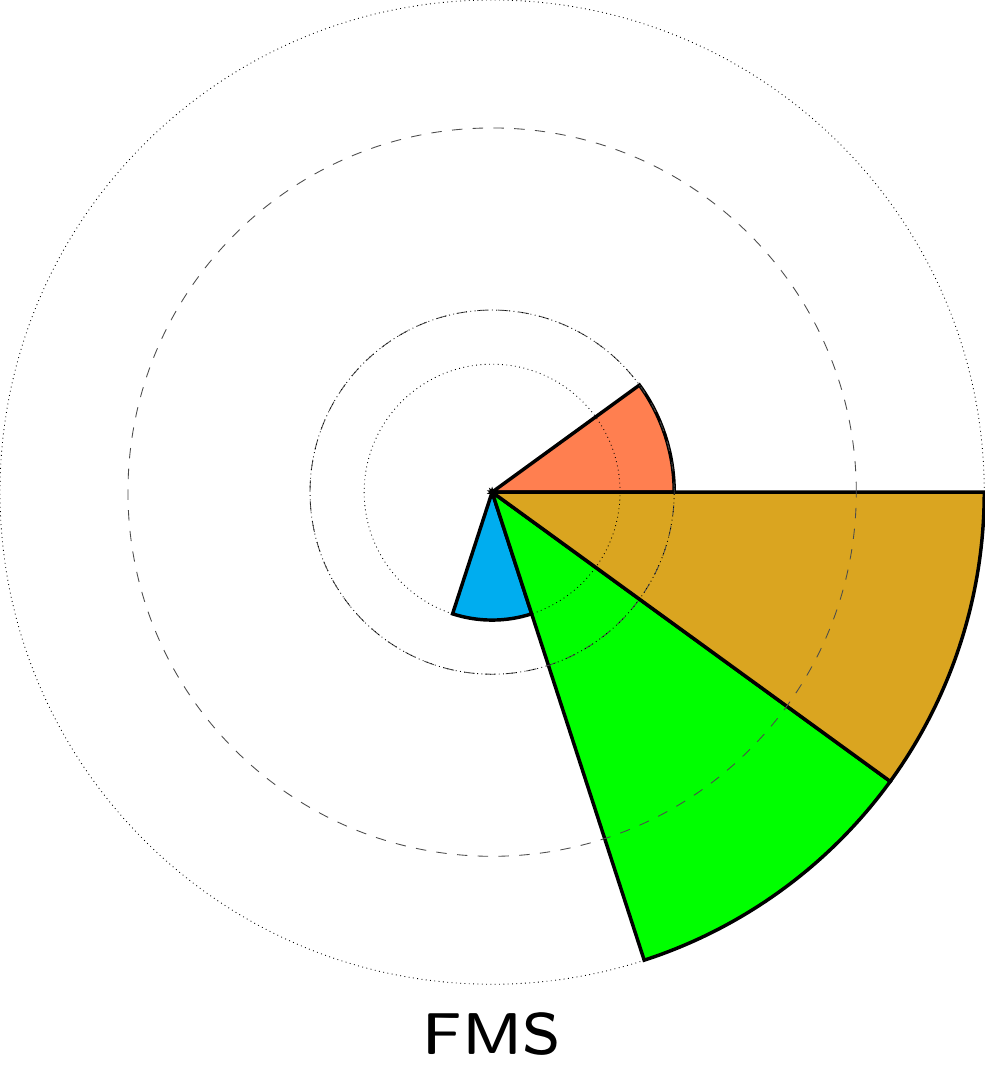}
\hfill
\includegraphics[scale=.35]{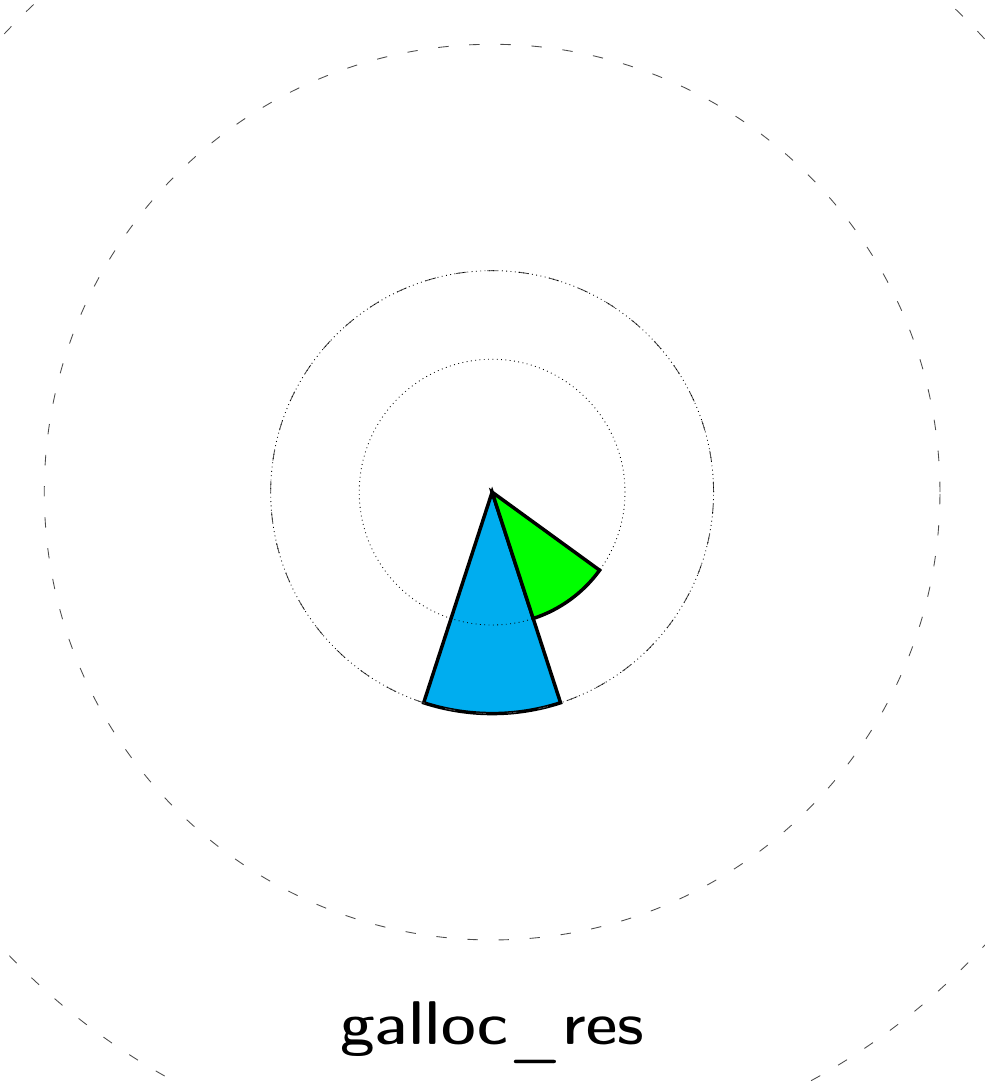}
\\
\medskip
\includegraphics[scale=.35]{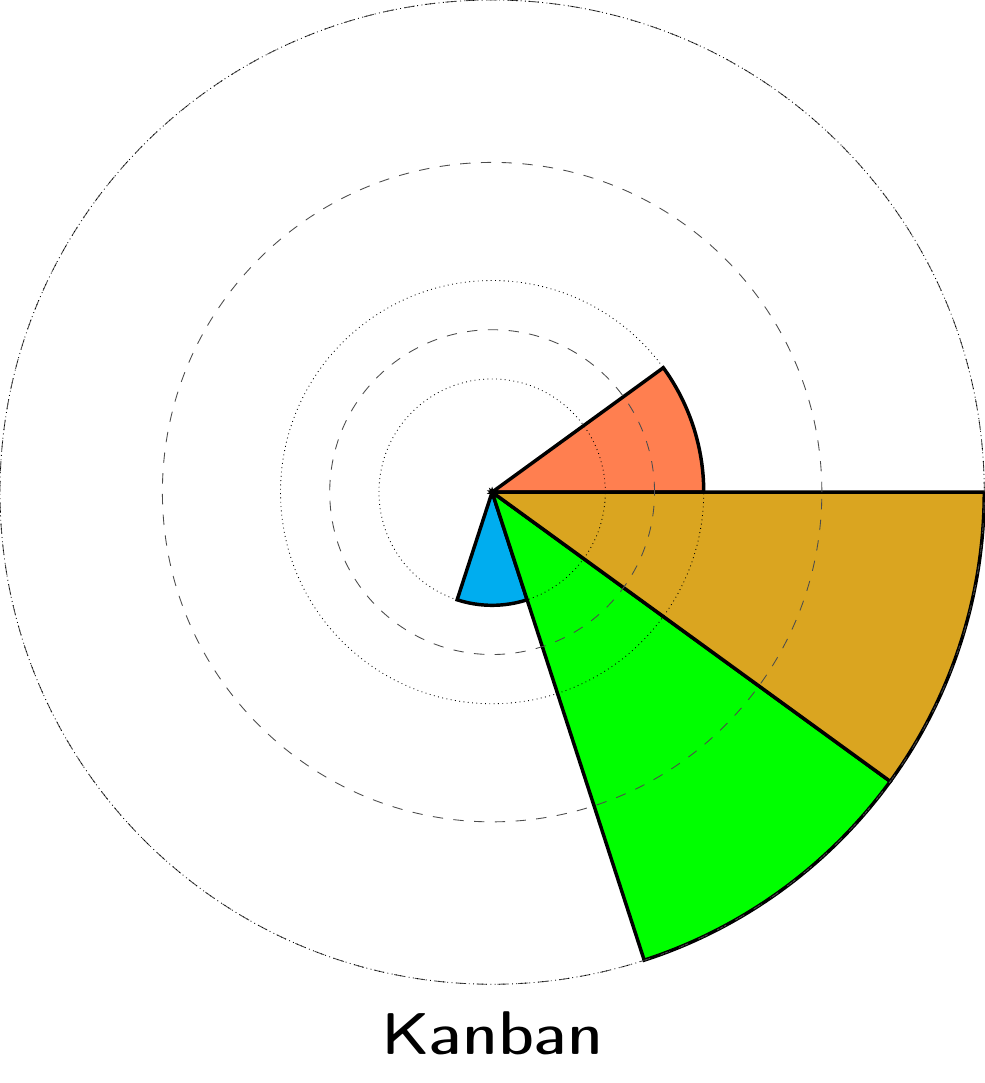}
\hfill
\includegraphics[scale=.35]{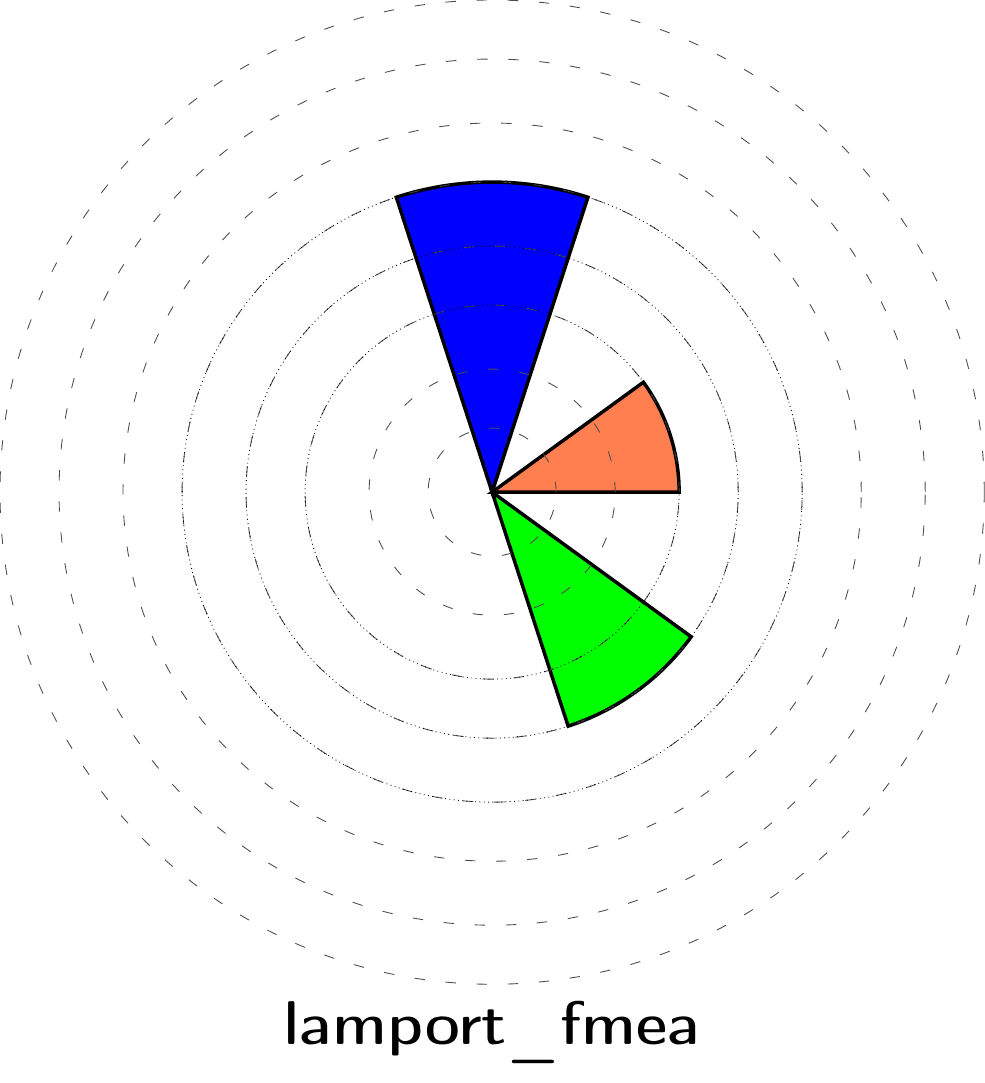}
\hfill
\includegraphics[scale=.35]{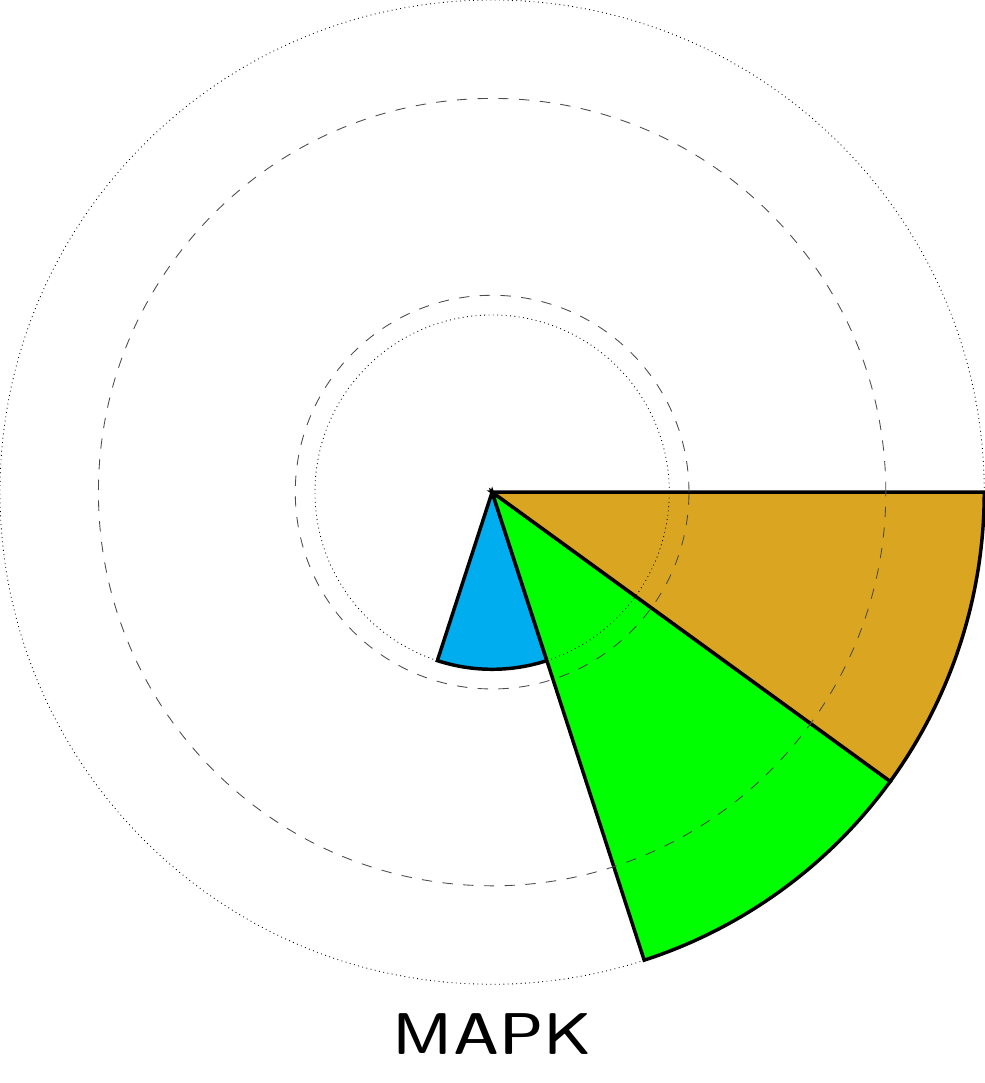}
\hfill
\includegraphics[scale=.35]{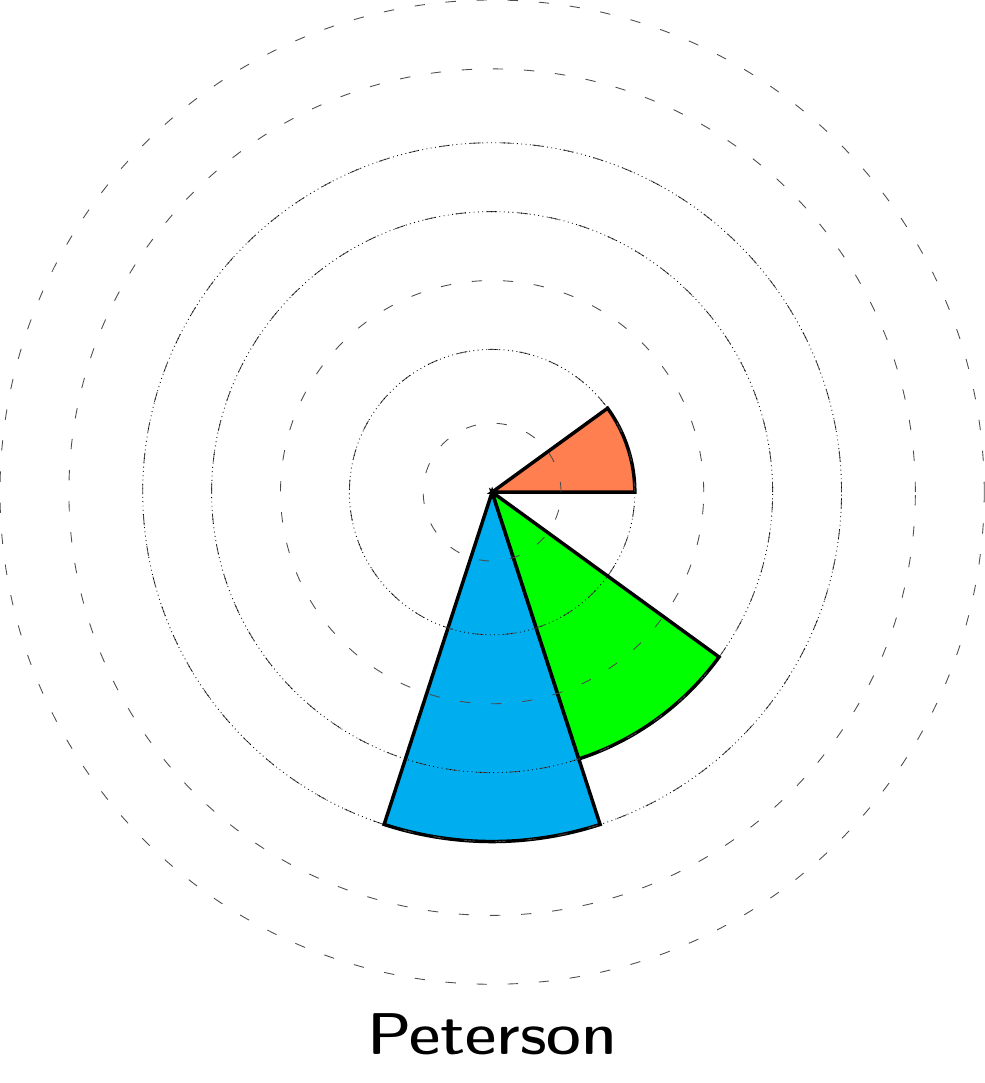}
\\
\medskip
\includegraphics[scale=.35]{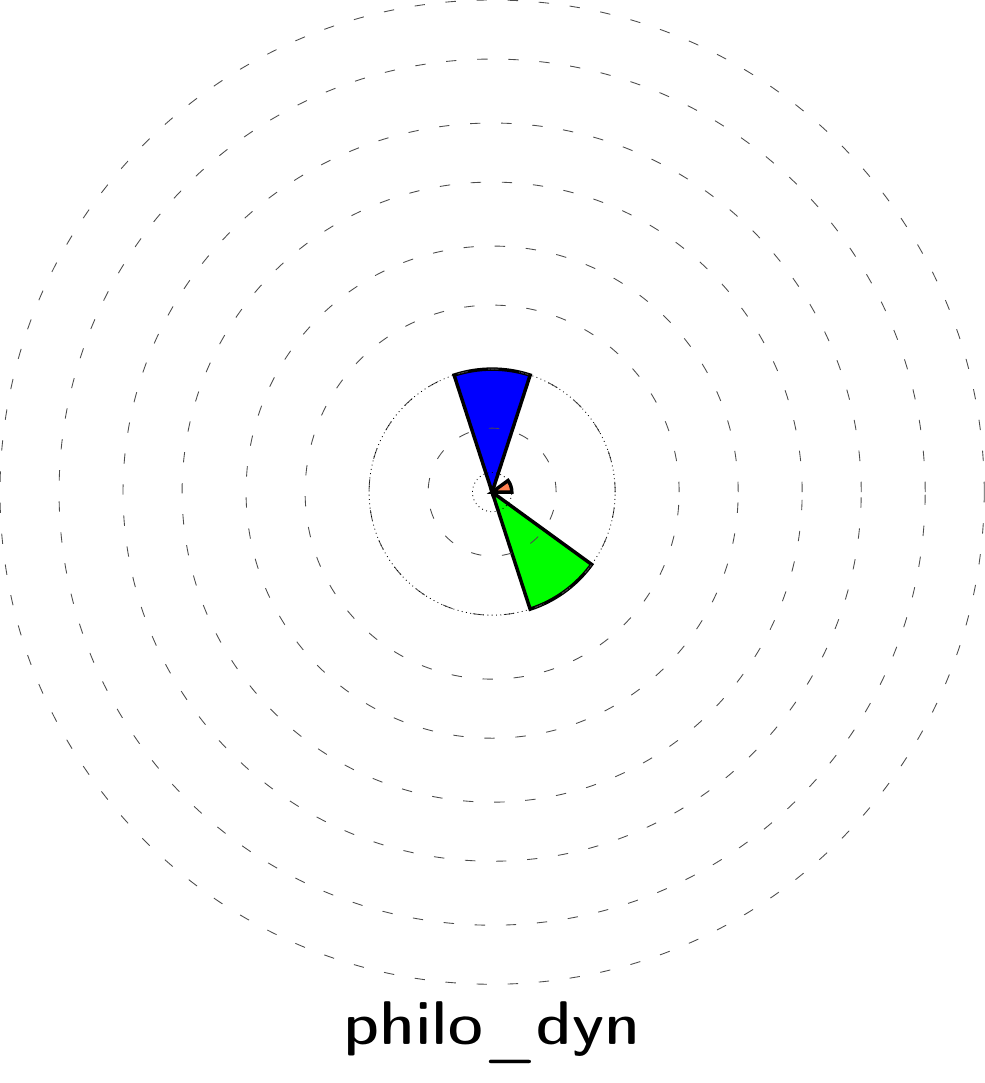}
\hfill
\includegraphics[scale=.35]{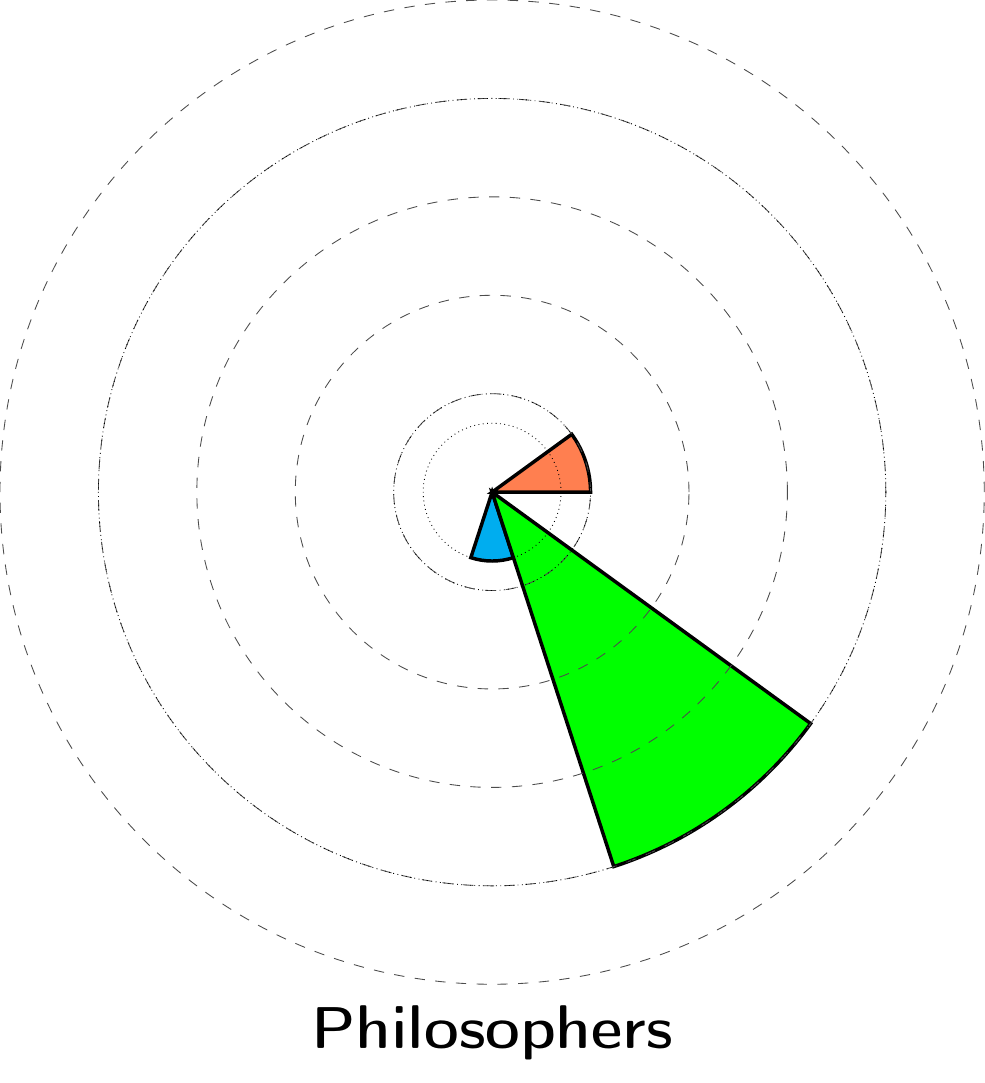}
\hfill
\includegraphics[scale=.35]{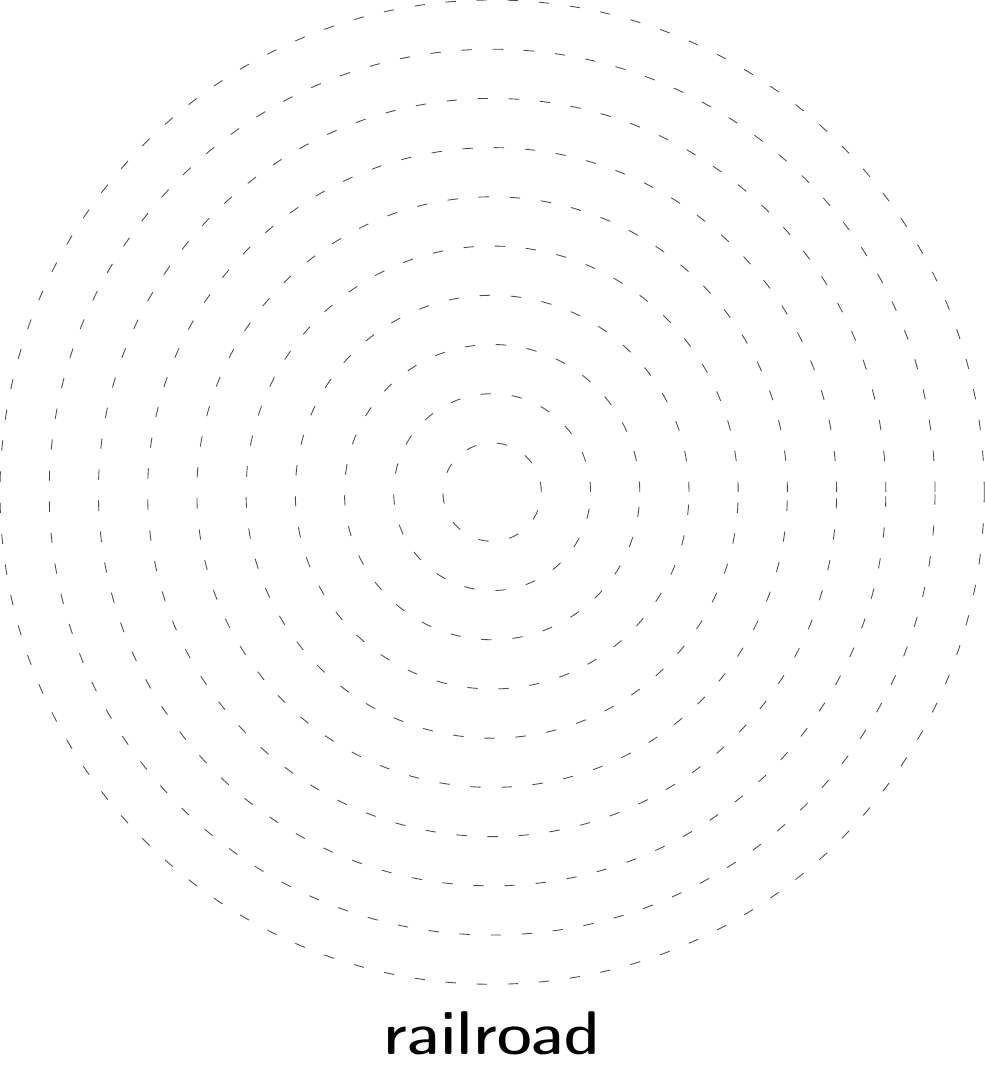}
\hfill
\includegraphics[scale=.35]{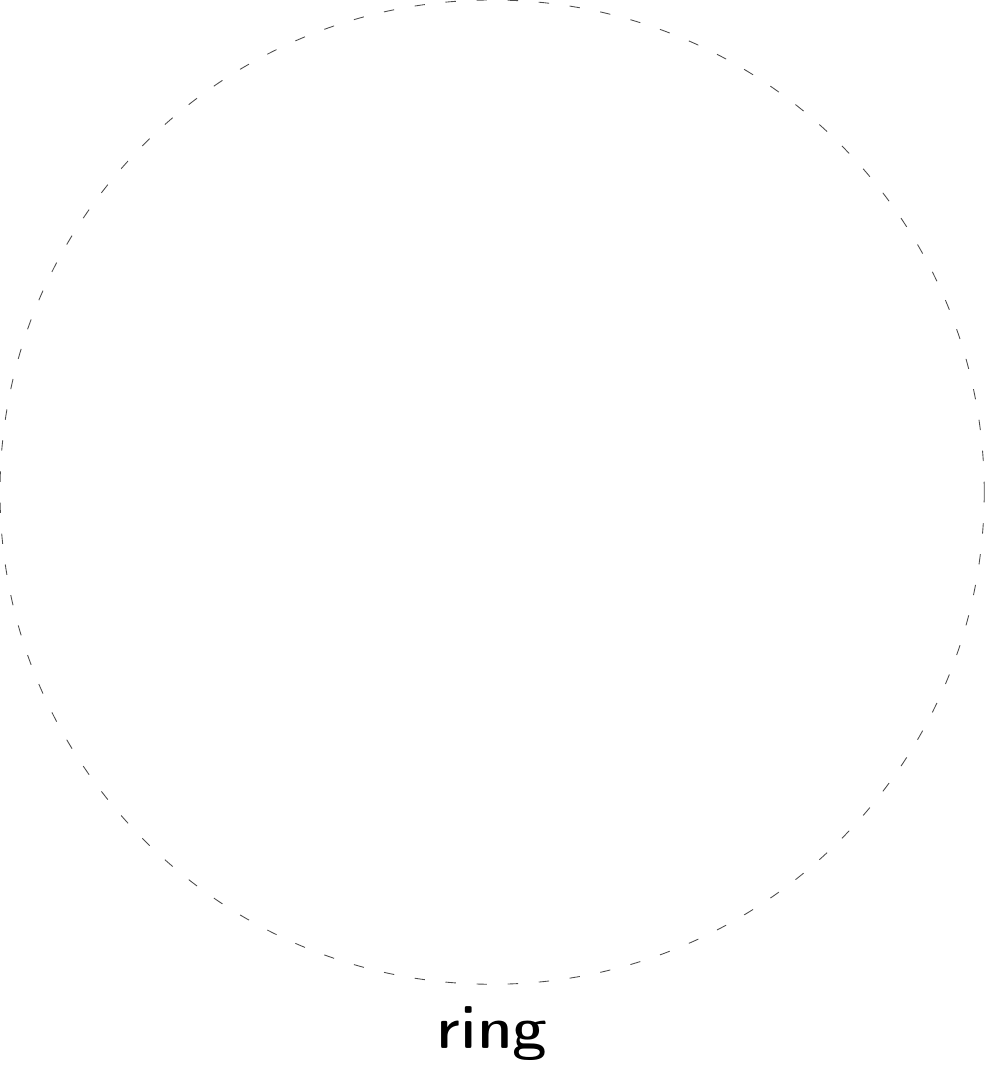}
\\
\medskip
\includegraphics[scale=.35]{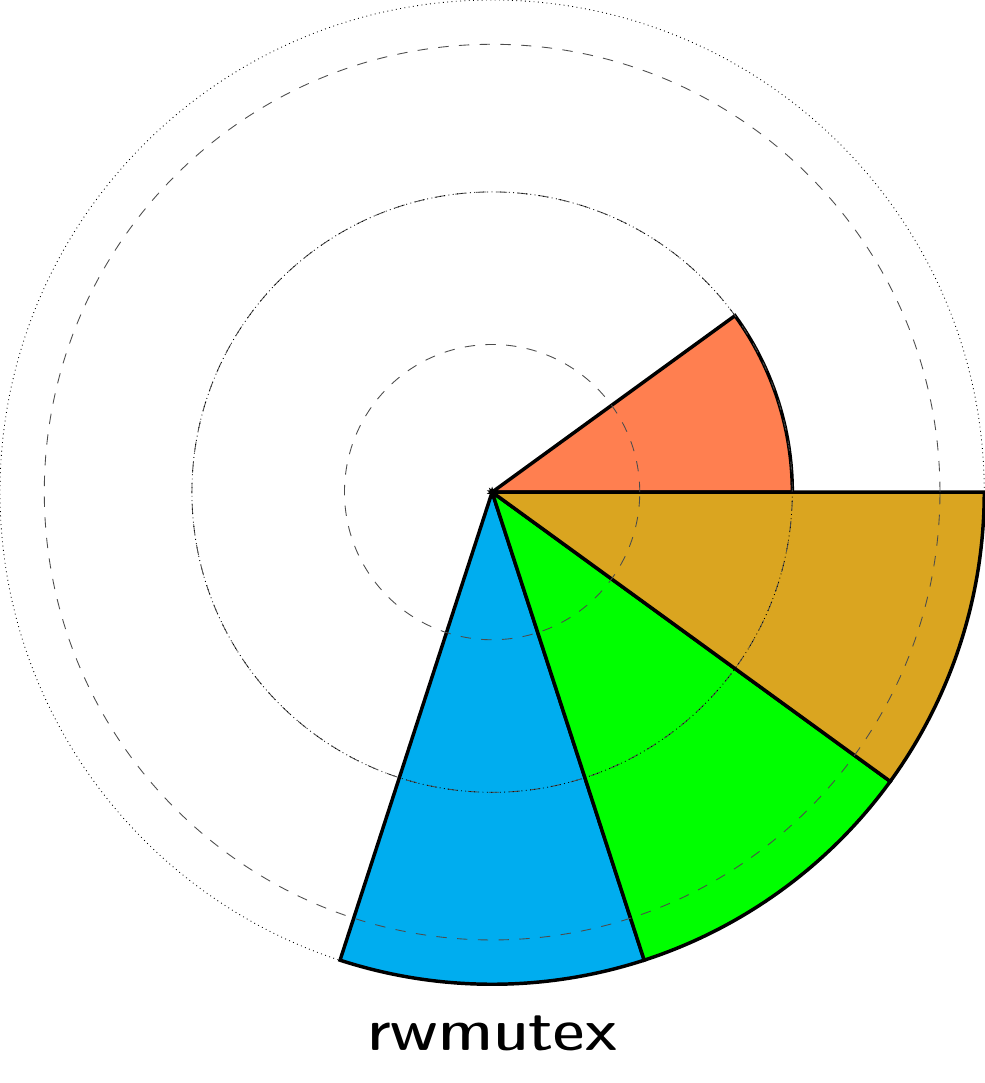}
\hfill
\includegraphics[scale=.35]{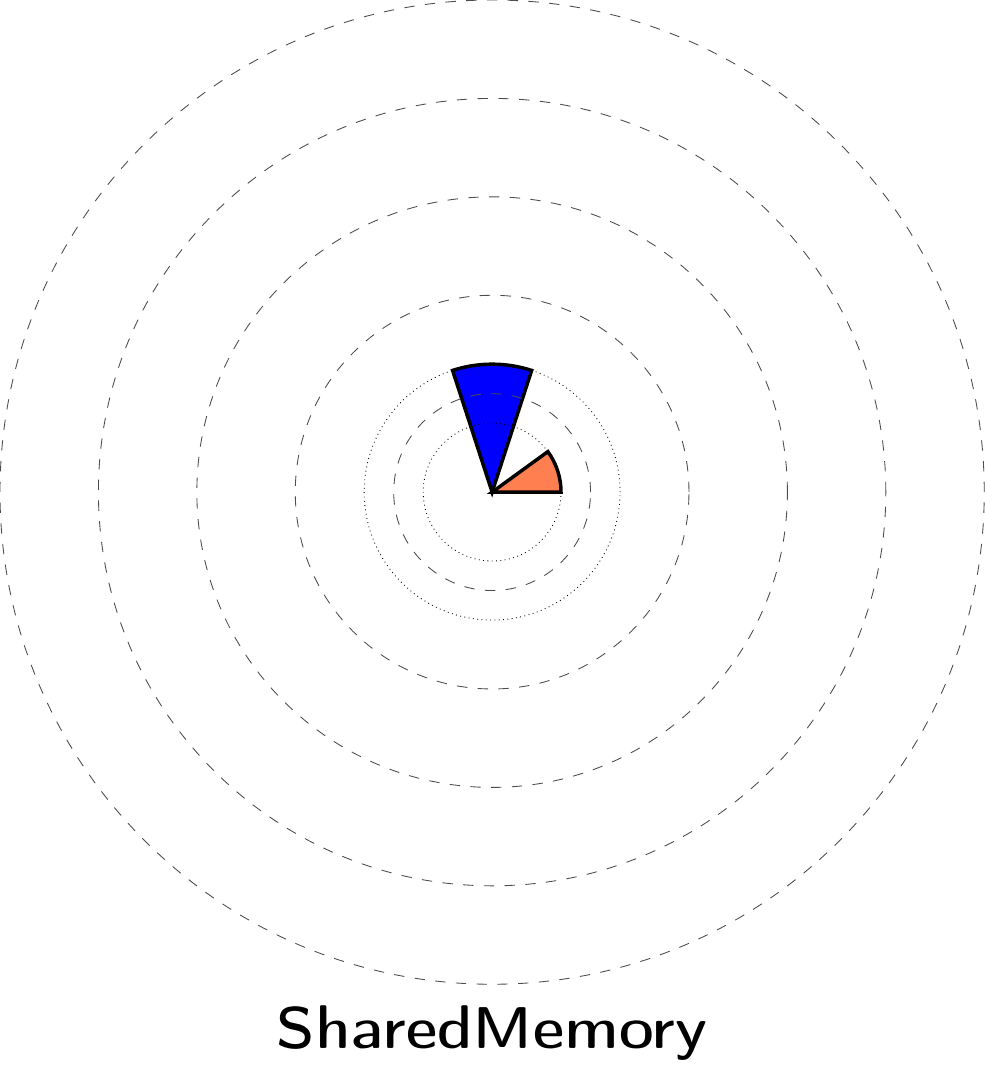}
\hfill
\includegraphics[scale=.35]{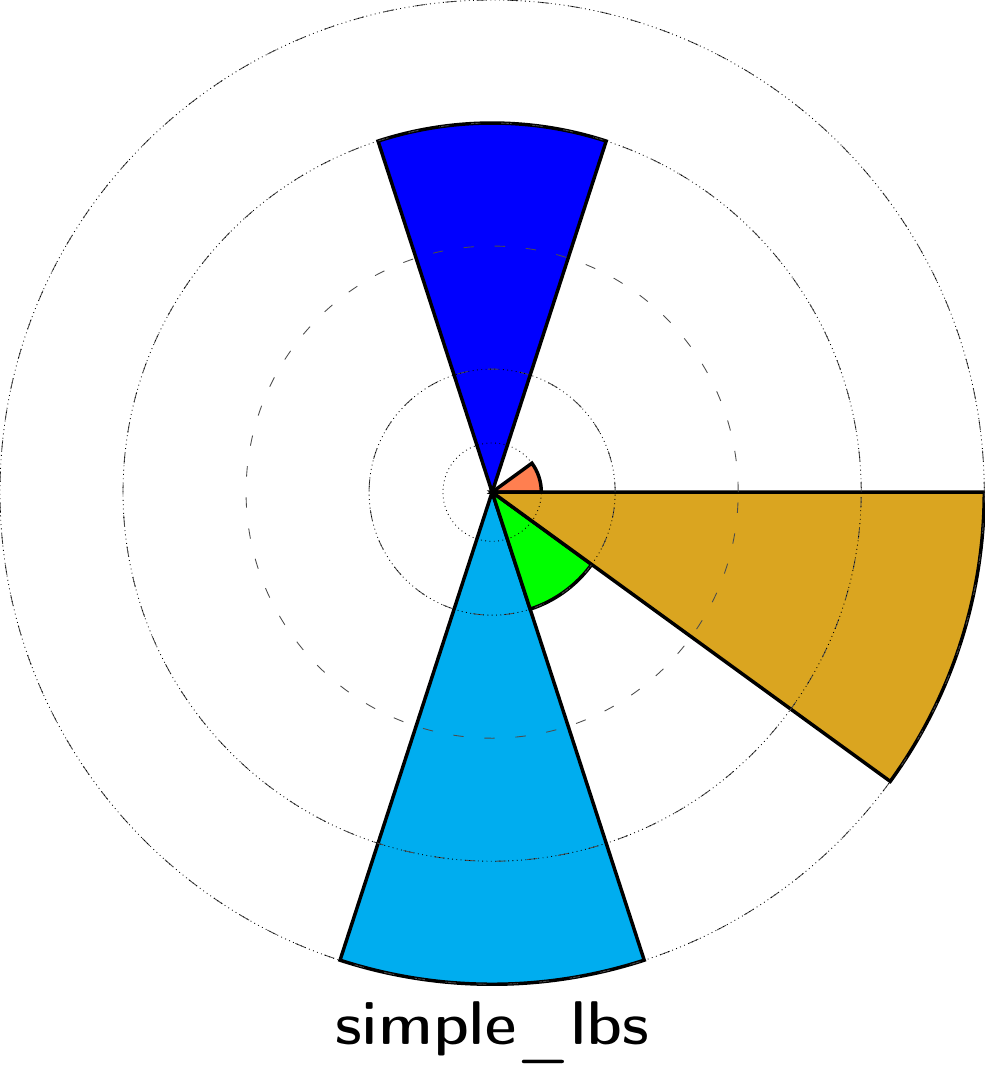}
\hfill
\includegraphics[scale=.35]{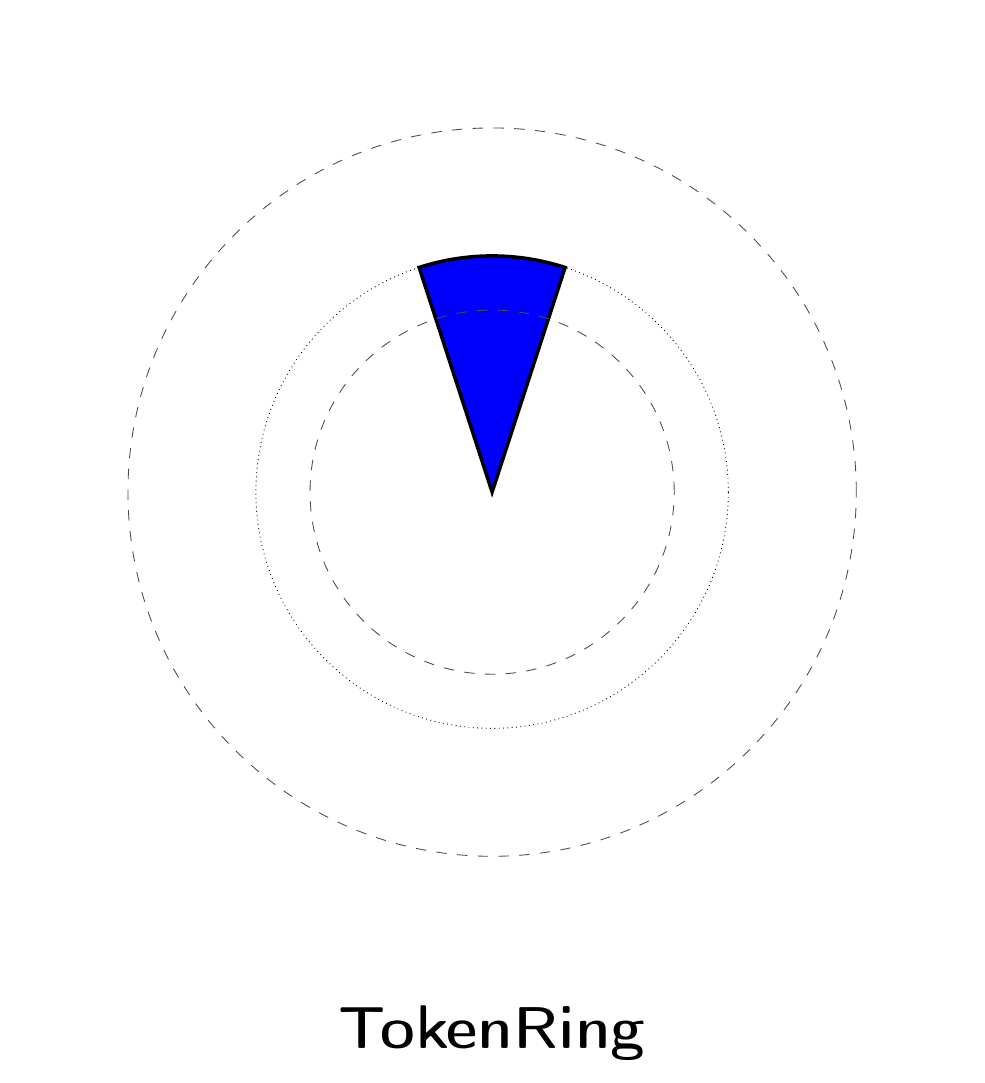}
\\
\bigskip
\mbox{}
\hfill
\includegraphics[scale=.7]{figures/alban-legend-tools.pdf}
\hfill
\mbox{}
\end{adjustwidth}
\caption{Highest parameter reached for each model,
         in reachability formul{\ae} evaluation}
\label{fig:fr:radar:models}
\end{figure}

\Cref{fig:fr:radar:models}~represents graphically
through a set of radar diagrams the highest parameter reached by the tools,
for each model.
Each diagram corresponds to one model, \emph{e.g.}, \acs{Echo} or
\acs{Kanban}.
Each diagram is divided in ten slices, one for each competing tool,
always at the same position.

The length of the slice corresponds to the highest parameter reached
by the tool.
When a slice does not appear,
the tool could not process even the smallest parameter.
For instance, \acs{LoLA-binstore} handles some parameters
for almost all models (and the highest parameter for most of them),
but is not able to handle the smallest parameter for \acs{Shared-Memory}.

Note that the scale depends on the model :
when the parameters of a model vary within a small range (less than~$100$),
a linear scale is used (showed using loosely dashed circles as in the
\acs{Peterson} model),
whereas a logarithmic scale is used for larger parameter values
(showed using densely dashed circles as in the \acs{Philosophers} model).
We also show dotted circles for the results of the tools,
in order to allow easier comparison.

Some models (\acs{RW-Mutex} and \acs{Echo}) have complex parameters,
built from two values.
Tools were only able to handle variation of the second parameter,
so we only represent it in the figure,
in order to show an integer value.

\subsection{Radars by tools}
\label{sec:fr:bytools}
\index{Reachability Formul{\ae}!Handling of models}

\begin{figure}[p]
\centering
\begin{adjustwidth}{0em}{0em}
\noindent
\includegraphics[scale=.4]{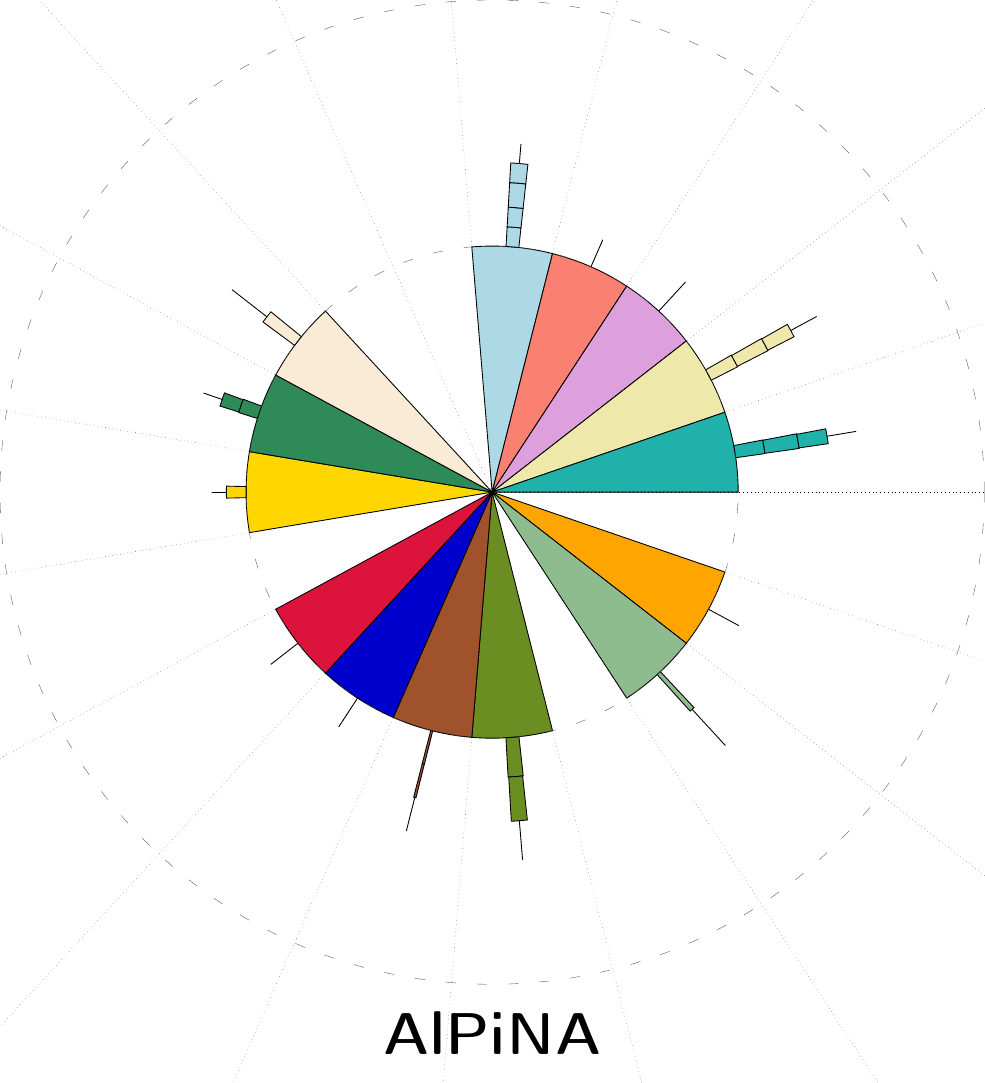}
\hfill
\includegraphics[scale=.4]{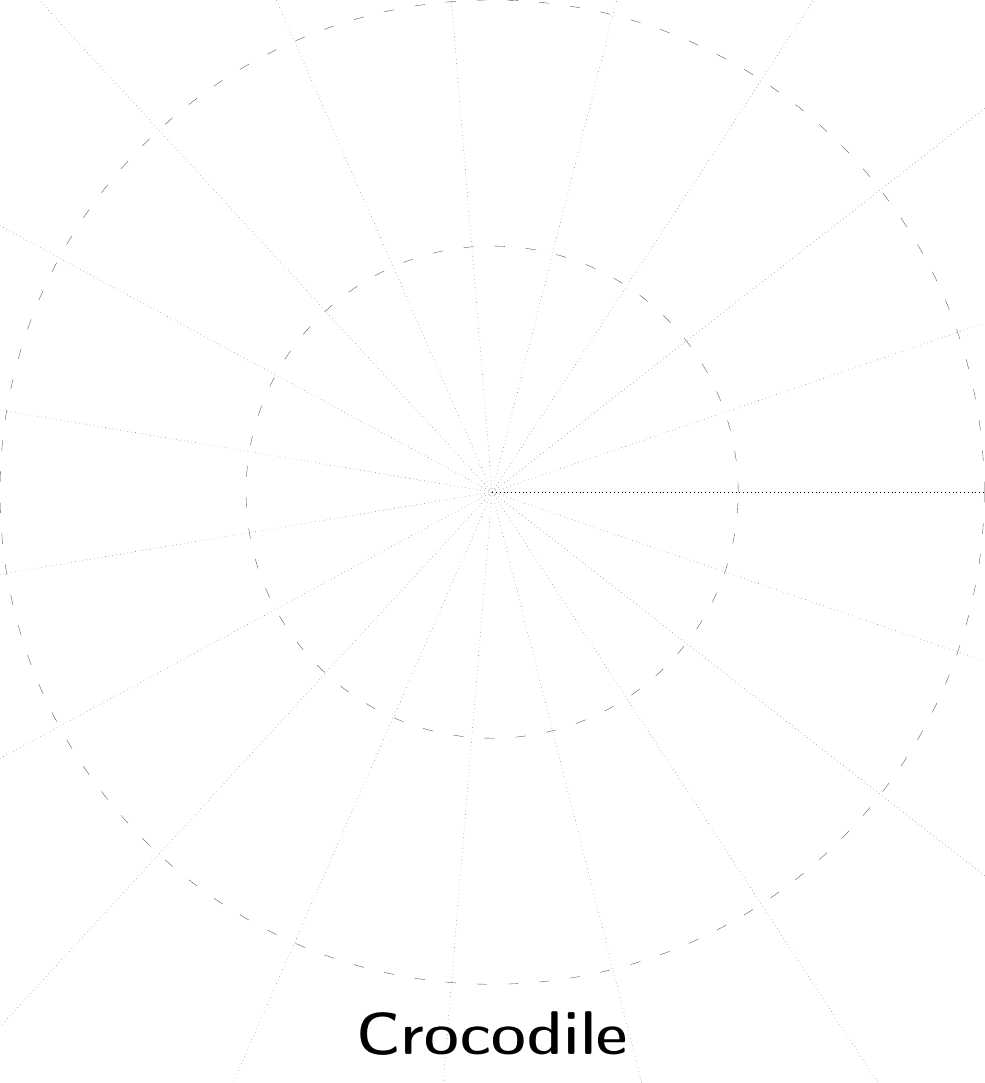}
\hfill
\includegraphics[scale=.4]{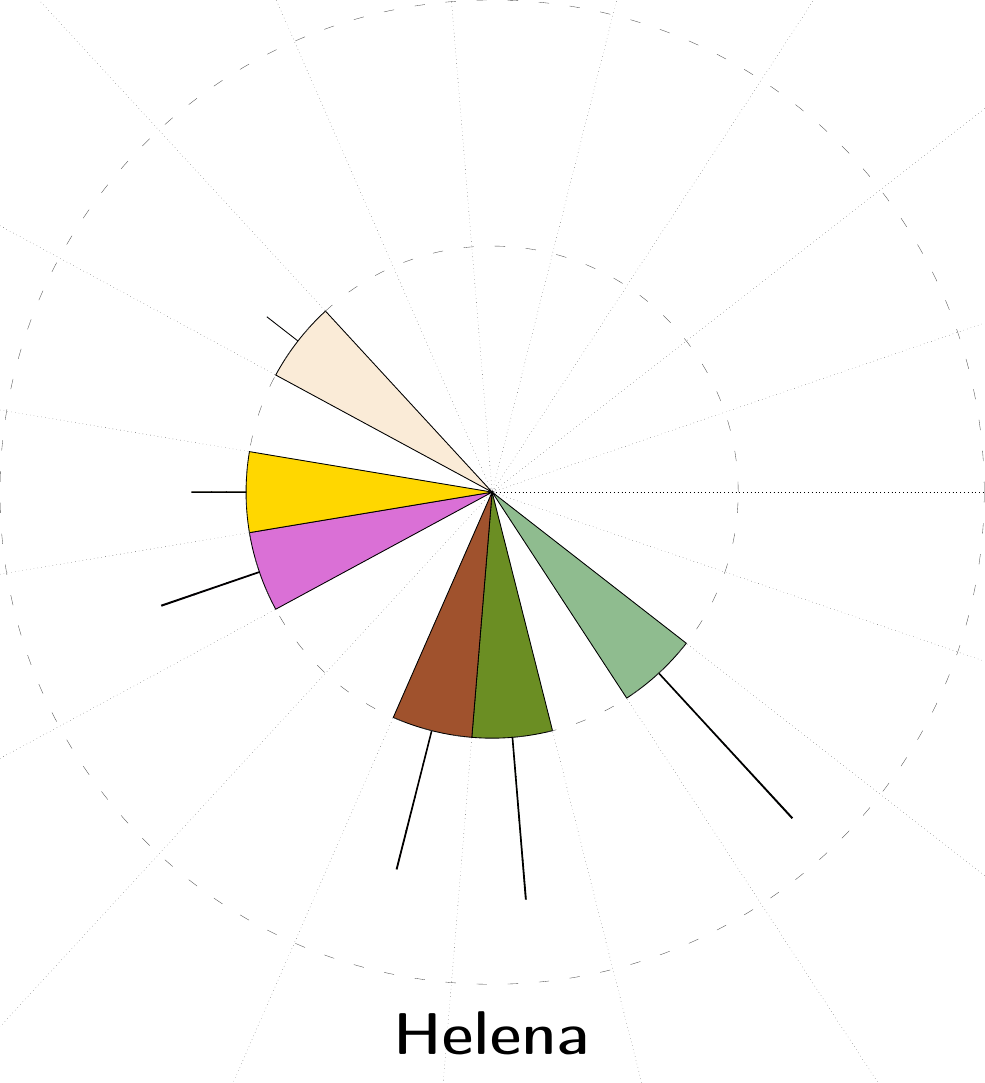}
\\
\medskip
\includegraphics[scale=.4]{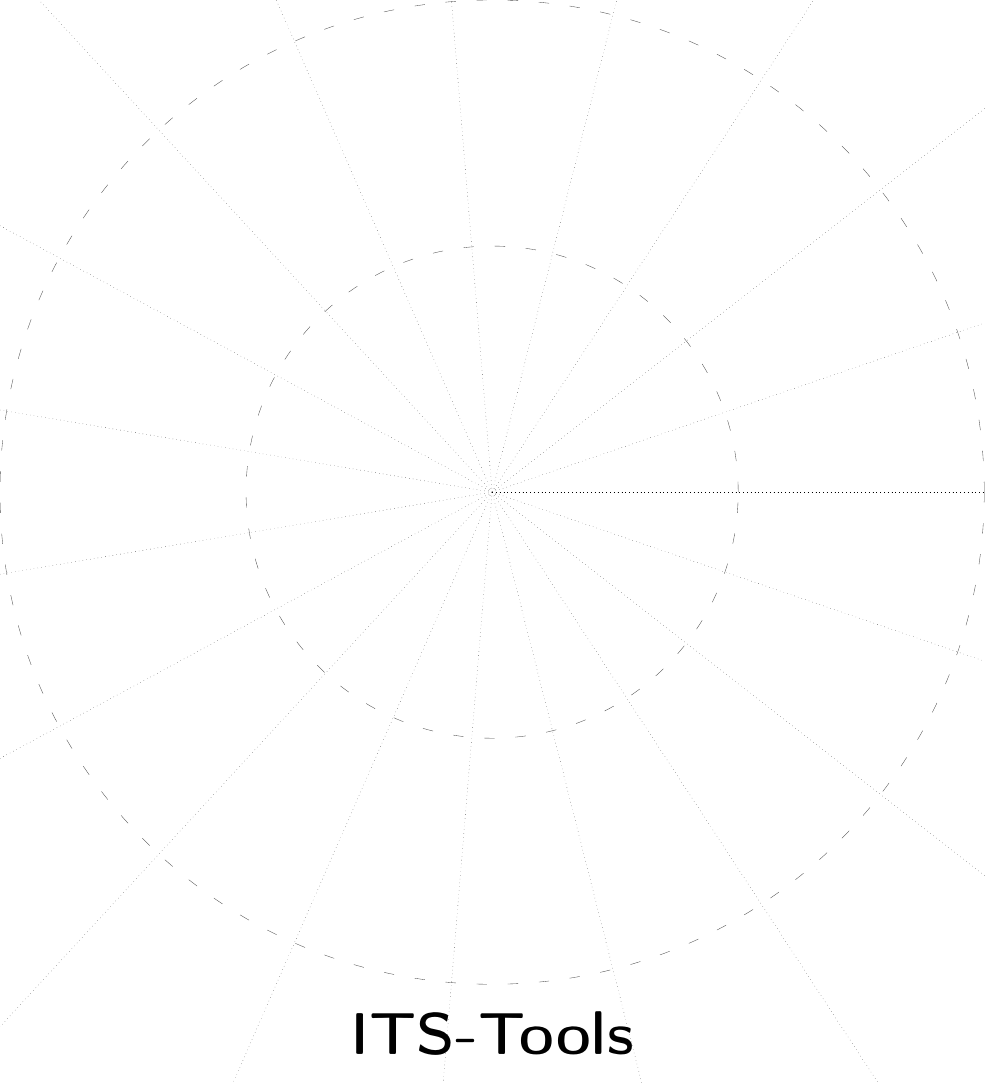}
\hfill
\includegraphics[scale=.4]{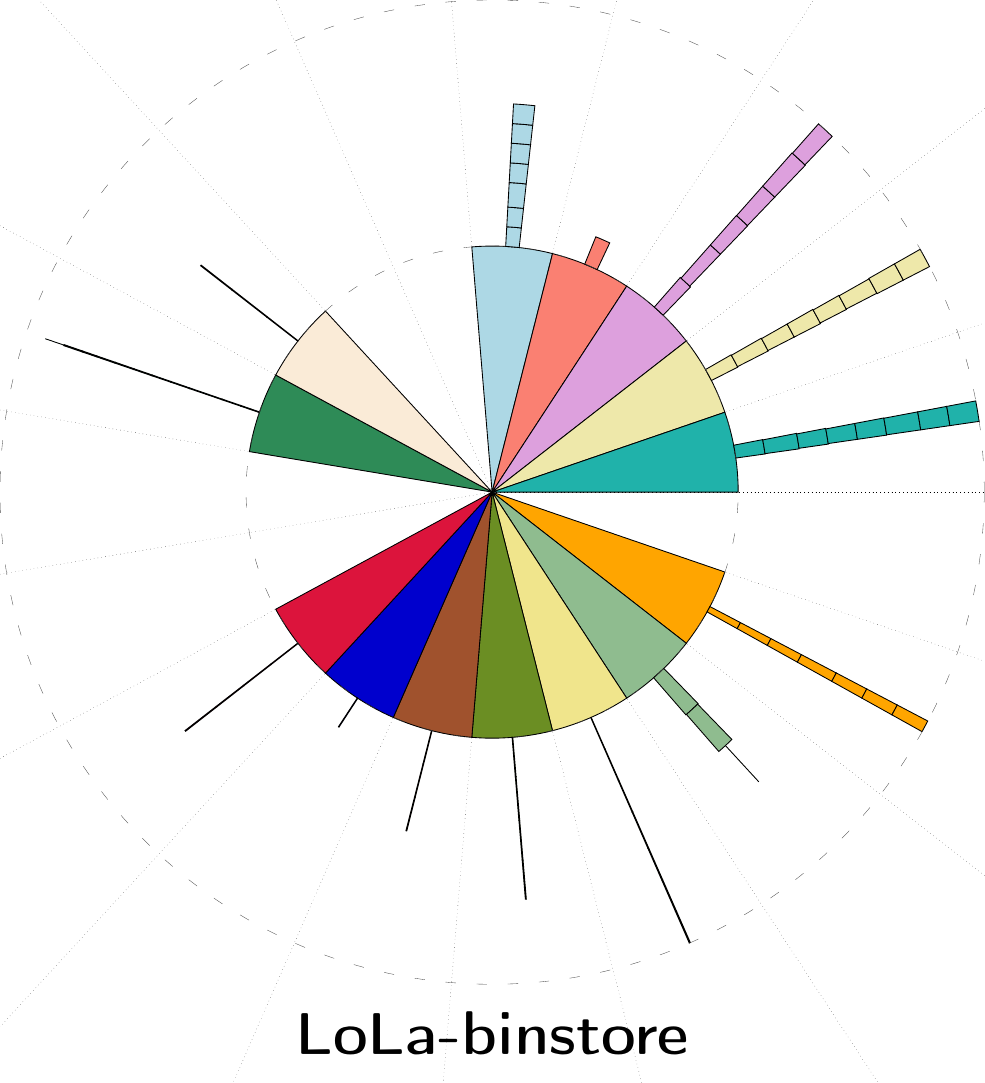}
\hfill
\includegraphics[scale=.4]{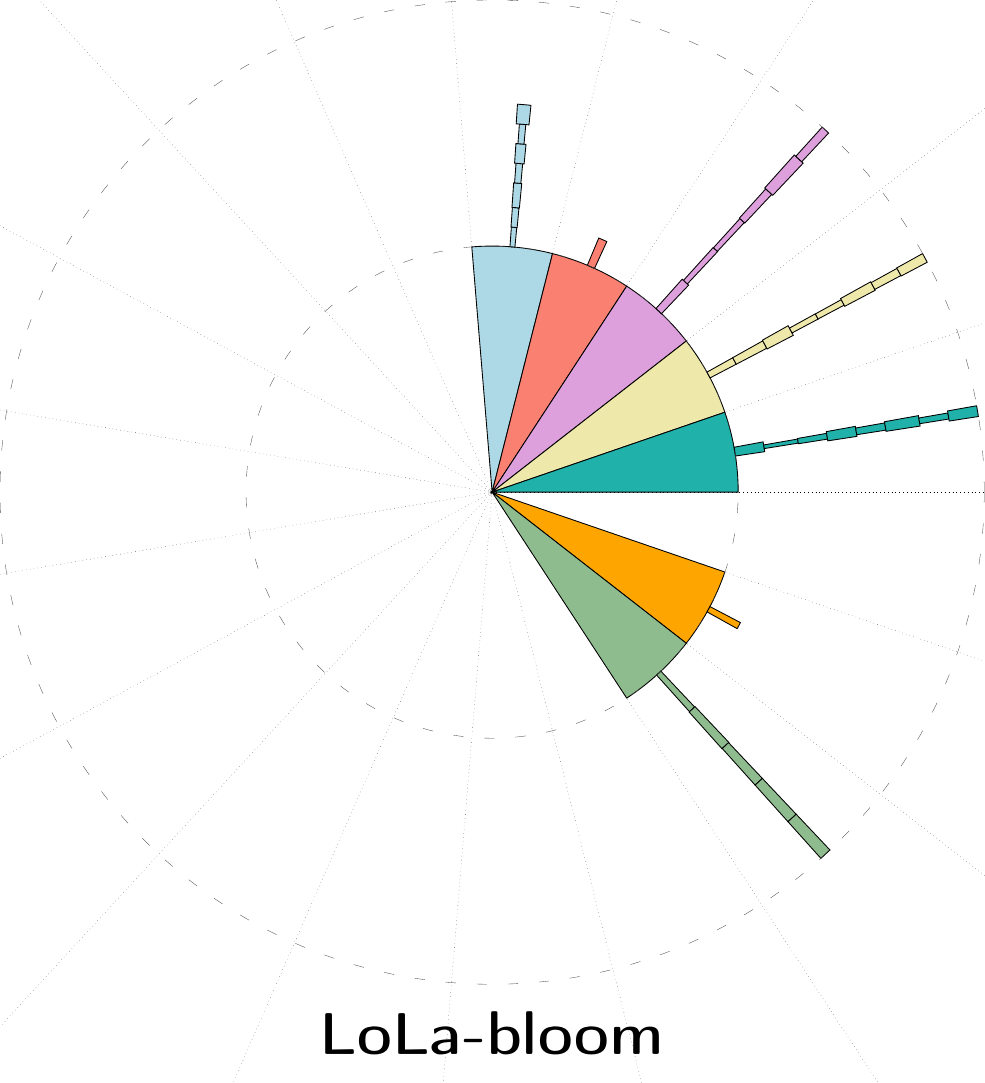}
\\
\medskip
\includegraphics[scale=.4]{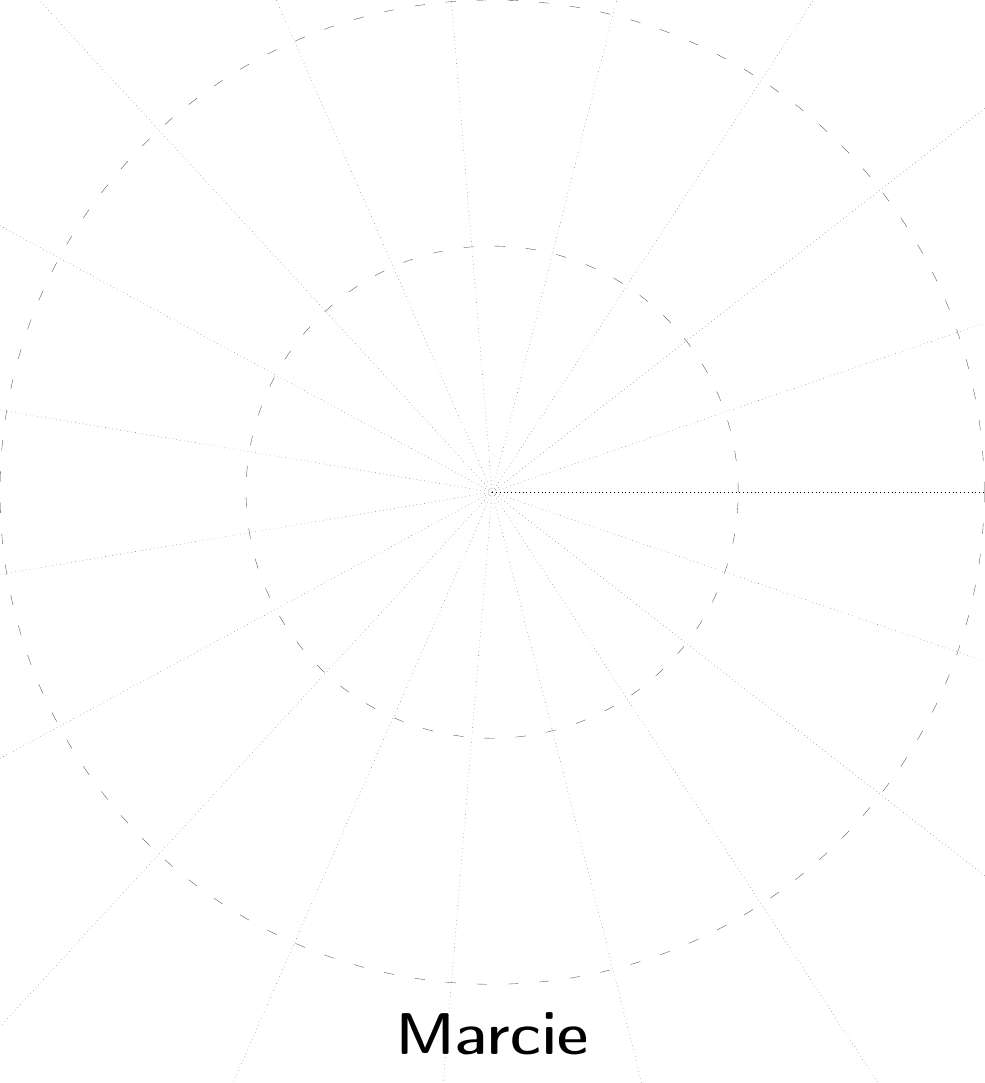}
\hfill
\includegraphics[scale=.4]{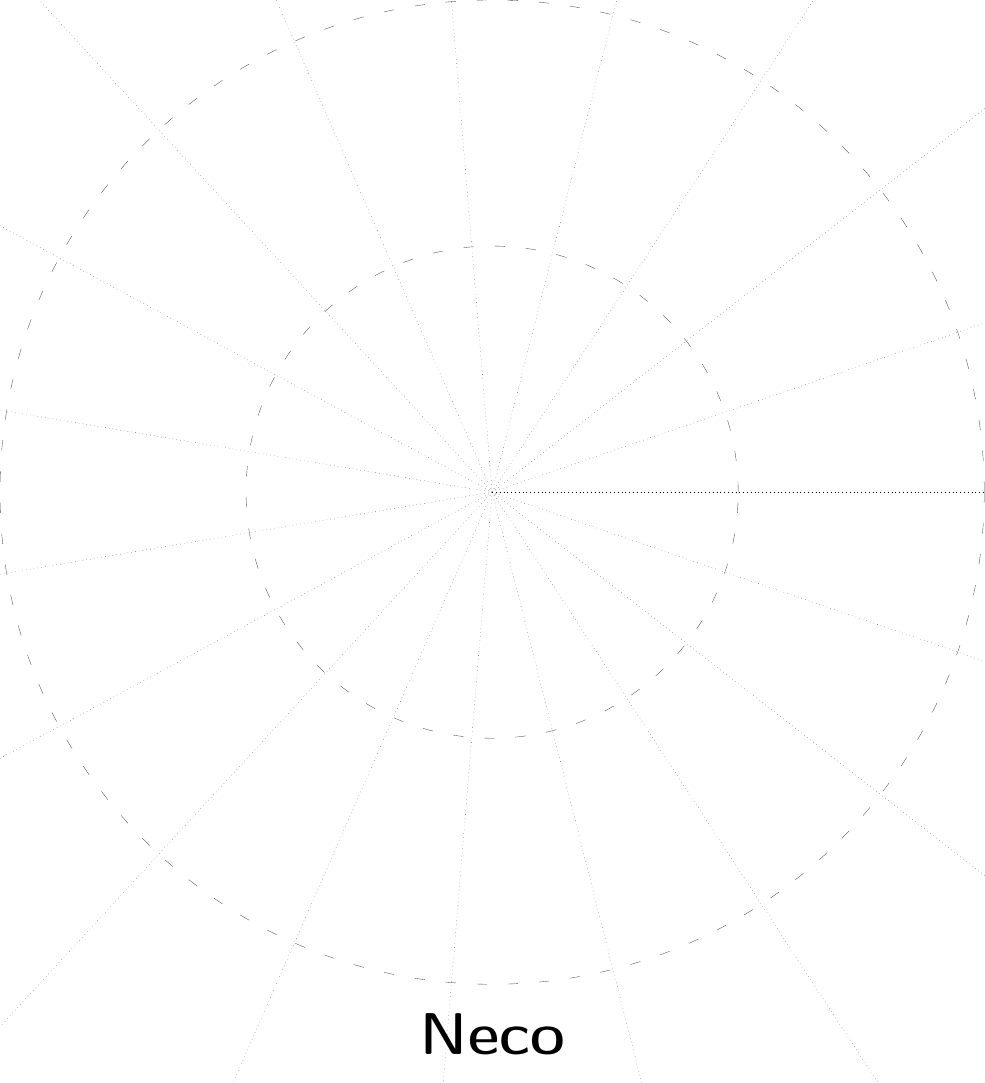}
\hfill
\includegraphics[scale=.4]{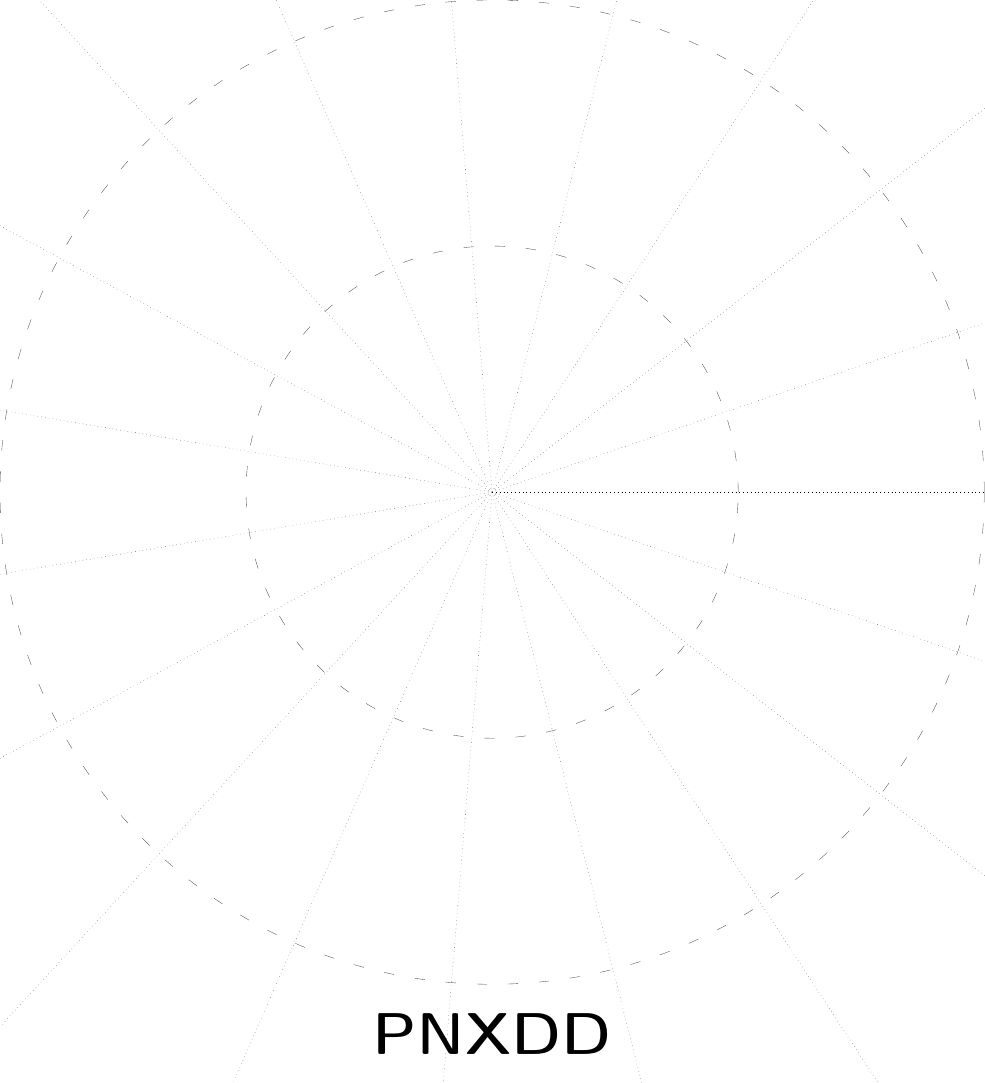}
\\
\medskip
\mbox{}
\hfill
\includegraphics[scale=.4]{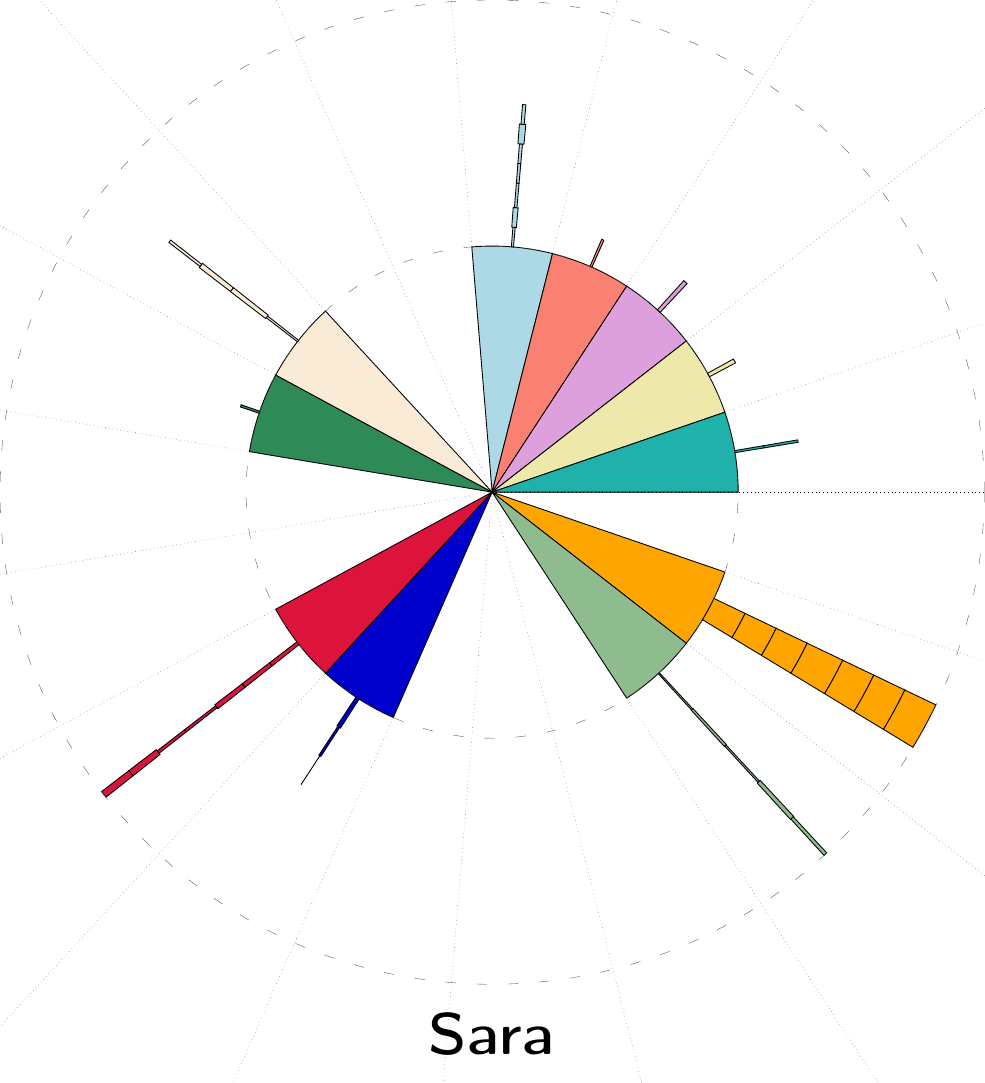}
\hfill
\mbox{}
\\
\bigskip
\mbox{}
\hfill
\includegraphics[scale=.7]{figures/alban-legend-models.pdf}
\hfill
\mbox{}
\end{adjustwidth}
\caption{Handled parameters for each tool,
         in reachability formul{\ae} evaluation}
\label{fig:fr:radar:tools}
\end{figure}

\Cref{fig:fr:radar:tools}~presents also radar diagrams showing
graphically the participation and results by tool.
Each slice in the radars represents a model, always at the same position.

These diagrams differ from both those in~\Cref{fig:fr:radar:models}
and those in~\Cref{fig:ss:radar:models}.
Each diagram contains two circles, as in~\Cref{fig:ss:radar:models}:
a slice reaches the inner circle if the tool participates
to the state space competition for this model, but fails.
The number of subslices between the inner and the outer circles represents
how many parameters are handled.
The angle covered by each subslice shows the ratio between
the computed formul{\ae} and the total number of formul{\ae}
in the examination.
For instance,
\acs{LoLA-binstore} has wider subslices than \acs{LoLA-bloom}:
it is able to handle more formul{\ae}.
But its subslices do not cover the whole angle dedicated to each model:
\acs{LoLA-binstore} could not handle all the formul{\ae} proposed in this
examination.
For \acs{Eratosthenes}, \acs{Sara} has a very wide angle,
so it handles far more formul{\ae} than for the other models.

The colored surface in the inner circle shows
if the tool could at least handle one formula for the model.
For instance, \acs{AlPiNA} and \acs{Helena}
could handle formul{\ae} for \acs{Simple-LBS},
but not for \acs{Railroad}.

This figure is not sufficient as we do not clearly distinguish tools
that try to handle formul{\ae} from tools that do not compete.
For instance, \acs{AlPiNA} tries for all models,
but has four ``blank'' models in the figure.

\clearpage
\section{RowData for CTL and LTL Formulae Evaluation}
\label{sec:ctlltlformulae}
\index{CTL Formul{\ae}}
\index{LTL Formul{\ae}}

No tool participated in these two examinations. 

\clearpage
\section{Conclusion}
\label{sec:conclusion}

This paper reported our experience with the second Model Checking Contest @
Petri nets 2012.

From the tool developers' point of view, such an event allows to compare tools
on a common benchmark that could become a public repository. Also, some
mechanisms established for the contest, such as a language to elaborate the
formula to be verified could become, over the years, a common way to provide
formul{\ae} to the various tools developed by the community.

Some difficulties in the generation of formul{\ae} for the contest led to a
very late publication of the grammar. Moreover, it appeared that some
difficulties were underestimated for tool integration. This explain why, while
the analysis of the state space generation examination is fruitful, this is
not the case for the formul{\ae} evaluation ones.

Thus, we provide in this report raw results as they were computed, and with no
interpretation, in order to emphasize the effort performed by both the
organizers and the tool submitters.

\subsubsection{Lesson Learned from \acs{MCC2012}} This edition was greatly
improved from the lesson learned during the previous edition
(see~\cite{mcc2011}). However, some remarks about the complexity of the
submission procedure led to the following suggestions:

\begin{enumerate}[i]

\item structural formul{\ae} examination will be split in two subcategories: a
single request one (asking for deadlock, bound, etc.) and a combined one where
these single requests can be combined.

\item for formula{\ae} examinations, we would like to distinguish satisfied
properties from unsatisfied ones. This would provide a better feedback
for tool developers (due to the difficulties encountered this year, it was
impossible to complete this for the 2012 edition).

\item the structure of a model$\times$scaling value will be simplified, thus
avoiding the complex naming problems we encountered this year (especially for
fomul{\ae}). This should simplify the integration of model checkers in the
image disk.

\item we will provide, as for last year, a pre-installed disk image with a
dummy model checker, just to let tool developers see how their tool can be
integrated.

\item the ``surprise model'' examination will be processed as well, it is of
interest to evaluate the capability of tools with their default settings. We
had no time to do so this year.

\end{enumerate}

Finally, we would like to set-up an on-line repository that would help tool
developers to submit more models as well as to perform tests on their tools.
However, this task requiring more manpower, we do not know if it will be
operated for \acs{MCC2013}.

\subsubsection{Acknowledgements} The \acl{MCC} organizers would like to thank
the following people for the help they provided in setting up this event:
Nicolas Gibelin (infrastructure and cluster management), Emmanuel Paviot-Adet
and Alexis Marechal (description of selected models), and Steve Hostettler
(definition of properties).

\smallskip\noindent The \acs{MCC} organizers would also like to thank the tool
developers who made possible such a contest. They are:

\begin{itemize}
\item \acs{AlPiNA}: Steve Hostettler, Alexis Marechal, and Edmundo Lopez;
\item \acs{Crocodile}: Maximilien Colange;
\item \acs{Helena}: Sami Evangelista and Jean-Fran\c{c}ois Pradat-Peyre.
\item \acs{ITS-Tools}: Yann Thierry-Mieg;
\item \acs{LoLA-binstore} and \acs{LoLA-bloom}: Karsten Wolf;
\item \acs{Marcie}: Alexey Tovchigrechko, Martin Schwarick, and Christian Rohr;
\item \acs{Neco}: Lukasz Fronc;
\item \acs{PNXDD}: Silien Hong and Emmanuel Paviot-Adet;
\item \acs{Sara}: Harro Wimmel and Karsten Wolf.
\end{itemize} 

\cleardoublepage
\bibliographystyle{splncs03}
\bibliography{MCC-report}

\cleardoublepage
\printindex

\cleardoublepage
\thispagestyle{empty}
\mbox{}
\clearpage
\thispagestyle{empty}
\renewcommand\CoverPicture{
\put(-5,-100){
\parbox[b][\paperheight]{\paperwidth}{%
\vfill
\centering
\includegraphics[width=23.5cm,keepaspectratio]{cover/cover-MCC.jpg}%
\vfill
}}}
\AddToShipoutPicture{\CoverPicture}
\mbox{}
\end{document}